%% file: dissertation.tex
\documentclass[10pt,twoside,openany,a4paper]{report}
\usepackage[utf8]{inputenc}

\usepackage[margin=2.5cm]{geometry}
\usepackage{graphicx, helvet, hyperref, setspace}
\usepackage[english]{babel}
\usepackage[nomain, acronym,nonumberlist]{glossaries}
\usepackage[backend=biber,style=numeric,natbib=true,sortcites=true,sorting=none,maxbibnames=99]{biblatex} 

\usepackage{notoccite}
\usepackage{tabularx}
\usepackage[all]{xy}
\usepackage{graphicx}
\usepackage{slashed}
\usepackage{amsmath,amssymb,bm,mathtools,slashed,tensor,bbm}
\usepackage{epsfig}
\usepackage{epstopdf}
\usepackage{float} 
\usepackage{tensor}
\usepackage{booktabs,tabularx}
\usepackage{multirow}

\input{extra_stuff.tex}

\DeclareSourcemap{
  \maps[datatype=bibtex]{
    \map{
      \step[fieldsource=note, final]
      \step[fieldset=addendum, origfieldval, final]
      \step[fieldset=note, null]
    }
  }
}

\usepackage[autostyle=true]{csquotes} 
\defbibheading{secbib}[References]{%
  \section*{#1}%
  \markboth{#1}{#1}}

\makeglossaries

\newcommand*\NewPage{\newpage\null\thispagestyle{empty}\cleardoublepage}

\def\FontLLb{
  \usefont{T1}{phv}{b}{n}\fontsize{16pt}{16pt}\selectfont}

\def\FontLb{
  \usefont{T1}{phv}{b}{n}\fontsize{14pt}{14pt}\selectfont}

\def\FontMb{
  \usefont{T1}{phv}{b}{n}\fontsize{12pt}{12pt}\selectfont}
\def\FontSn{
  \usefont{T1}{phv}{m}{n}\fontsize{10pt}{10pt}\selectfont}

\input{acronyms/acronyms}

\begin{document}
\pagenumbering{gobble}
\clearpage
\thispagestyle{empty}
\input{variables.tex}

\includegraphics[width=5cm]{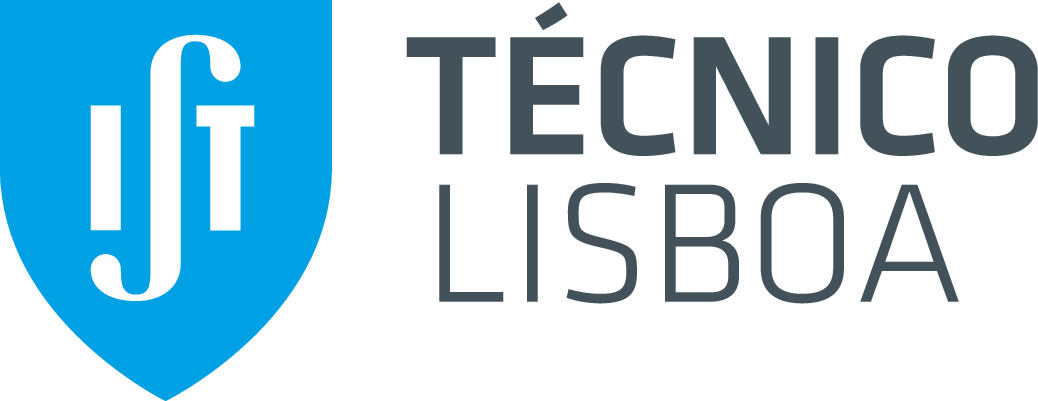}

\begin{center}

\begin{spacing}{1.5}

{\FontLLb UNIVERSIDADE DE LISBOA} \\
{\FontLLb INSTITUTO SUPERIOR TÉCNICO} \\

\vspace{2.5cm}
{\FontLb Constraining Multi-scalars models with colliders and Dark Matter} \\

\vspace{2.5cm}
{\FontLb Rafael Filipe Teixeira Boto} \\

\vspace{2.5cm}
{\FontSn %
\begin{tabular}{ll}
 \textbf{Supervisor}: & Doctor João Paulo Ferreira da Silva\\
 \textbf{Co-Supervisor}:  & Doctor Jorge Manuel Rodrigues Crispim Romão
\end{tabular} } \\

\vspace{3.0cm}
{\FontSn {Thesis approved in public session to obtain the PhD degree in}} \\
\vspace{0.3cm}
{\FontSn \textbf{Physics}} \\
\vspace{1.3cm}
{\FontSn {Jury final classification:  \textbf{Pass with Distinction and Honour}}}

\vspace*{\fill}
{\FontMb 2025}

\end{spacing}
\end{center}

\newpage
\thispagestyle {empty}

\includegraphics[width=5cm]{images/ist_logo}

\begin{center}

\begin{spacing}{1.5}

{\FontLLb UNIVERSIDADE DE LISBOA} \\
{\FontLLb INSTITUTO SUPERIOR TÉCNICO} \\

\vspace{1.0cm}
{\FontLb Constraining Multi-scalars models with colliders and Dark Matter} \\

\vspace{1.0cm}
{\FontMb Rafael Filipe Teixeira Boto} \\

\vspace{1.0cm}
{\FontSn %
\begin{tabular}{ll}
 \textbf{Supervisor}: & Doctor João Paulo Ferreira da Silva\\
 \textbf{Co-Supervisor}:  & Doctor Jorge Manuel Rodrigues Crispim Romão
\end{tabular} } \\
\vspace{0.5cm}
{\FontSn {Thesis approved in public session to obtain the PhD degree in}} \\
\vspace{0.3cm}
{\FontSn \textbf{Physics}} \\
\vspace{0.3cm}
{\FontSn {Jury final classification: \textbf{Pass with Distinction and Honour}}}

\vspace{0.5cm}
{\FontSn {\textbf{Jury}}}\\
{\FontSn %
 \flushleft\textbf{Chairperson}: Doctor Mário João Martins Pimenta, Instituto Superior Técnico,\\ \hspace*{116pt} Universidade de Lisboa \\
 \flushleft\textbf{Members of the Committee:}\\
 \vspace{0.2cm}
 \hspace*{9.3pt}{\setlength{\tabcolsep}{1.75pt}\renewcommand{\arraystretch}{1}\begin{tabular}{lll}  
 & Doctor& Howard Eli Haber, University of California, Santa Cruz, EUA\\
 & Doctor& Francisco Jose Botella Olcina, Institut de Física Corpuscular,\\ 
 & & Universitat de València, Espanha\\
 & Doctor& Mário João Martins Pimenta, Instituto Superior Técnico, Universidade de Lisboa\\
 & Doctor&  João Paulo Ferreira da Silva, Instituto Superior Técnico, Universidade de Lisboa\\
 & Doctor& José Guilherme Teixeira de Almeida Milhano, Instituto Superior Técnico,\\
 & & Universidade de Lisboa\\
 & Doctor & Rui Alberto Serra Ribeiro dos Santos, Instituto Superior de Engenharia de Lisboa\\ 
\end{tabular} }} \\

\vspace{0.5cm}
{\FontSn {\textbf{Funding Institution}}}\\
{\FontSn {FCT: Fundação para a Ciência e a Tecnologia}}

\vspace*{\fill}
{\FontMb 2025}
\end{spacing}
\end{center}

\vspace*{5cm}  

Declaration

\vspace{0.2cm} 

I declare that this document is an original work of my own authorship and that it fulfills all the requirements of the Code of Conduct and Good Practices of the Universidade de Lisboa.
\NewPage

\pagenumbering{roman}

\input{acknowledgments/acknowledgments.tex} 
\NewPage

\input{abstract/abstract-pt}   
\cleardoublepage

\input{abstract/abstract-en} 
\cleardoublepage

%
\tableofcontents
\cleardoublepage 

%
\phantomsection
\addcontentsline{toc}{section}{\listtablename}
\listoftables
\cleardoublepage 

%
\phantomsection
\addcontentsline{toc}{section}{\listfigurename}
\listoffigures
\cleardoublepage

\phantomsection
\addcontentsline{toc}{section}{List of Abreviations}
\printglossaries
\cleardoublepage

\pagenumbering{arabic}

\input{chapters/main.tex}


\printbibliography[heading=bibintoc]

\appendix
\input{appendix/main.tex}

\pagenumbering{gobble}
\NewPage

\end{document}

%% file: extra_stuff.tex
\addto\captionsenglish{}
\addto\captionsenglish{}
\addto\captionsenglish{}

\def\FontSn{
  \usefont{T1}{phv}{m}{n}\fontsize{12pt}{12pt}\selectfont}

\makeatletter
\renewcommand*\env@matrix[1][\arraystretch]{%
  \edef\arraystretch{#1}%
  \hskip -\arraycolsep
  \let\@ifnextchar\new@ifnextchar
  \array{*\c@MaxMatrixCols c}}
\makeatother

\newcolumntype{C}[1]{>{\centering\arraybackslash}p{#1}}

\def\ddel{\!\!\mathrel{\raise1.5ex\hbox{$\leftrightarrow$\kern-.85em
\lower1.7ex\hbox{$\partial$}}}}

\def\ben{\begin{enumerate}}
\def\een{\end{enumerate}}
\def\beq{\begin{equation}}
\def\eeq{\end{equation}}
\def\beqa{\begin{eqnarray}}
\def\eeqa{\end{eqnarray}}

\def\ifmath#1{\relax\ifmmode #1\else $#1$\fi}
\def\lsim{\mathrel{\raise.3ex\hbox{$<$\kern-.75em\lower1ex\hbox{$\sim$}}}}
\def\gsim{\mathrel{\raise.3ex\hbox{$>$\kern-.75em\lower1ex\hbox{$\sim$}}}}

\def\eq#1{Eq.~(\ref{#1})}

\def\eqs#1#2{Eqs.~(\ref{#1}) and (\ref{#2})}
\def\eqsthree#1#2#3{Eqs.~(\ref{#1}), (\ref{#2}) and (\ref{#3})}

\def\Eqs#1#2{Eqs.~(\ref{#1}) and (\ref{#2})}

\def\anti{\overline}

\def\vev#1{\langle #1 \rangle}

\def\beq{\begin{equation}}
\def\eeq{\end{equation}}

\def\Eqs#1#2{Eqs.\ (\ref{#1}) and (\ref{#2})}

\renewcommand{\Re}{{\rm Re}}
\renewcommand{\Im}{{\rm Im}}

\def\lam{\lambda}

\def\half{\ifmath{{\textstyle{\frac{1}{2}}}}}

\usepackage{xparse}

\usepackage{xcolor}

\def\lasup#1{^{\lower 2pt\hbox{$\scriptstyle#1$}}}

\newcommand{\be}{\begin{equation}}
\newcommand{\ee}{\end{equation}}
\newcommand{\ba}{\begin{eqnarray}}
\newcommand{\ea}{\end{eqnarray}}

\newcommand{\Z}[1]{\ensuremath{\mathbbm{Z}_{#1}}} 

\newcommand{\SU}[1]{\ensuremath{\mathrm{SU}(#1)}}

\newcommand{\U}[1]{\ensuremath{\mathrm{U}(#1)}}

\makeatletter
\renewcommand*\env@matrix[1][\arraystretch]{%
  \edef\arraystretch{#1}%
  \hskip -\arraycolsep
  \let\@ifnextchar\new@ifnextchar
  \array{*\c@MaxMatrixCols c}}
\makeatother

\def\sbi  {s_{\beta_1}}
\def\cbi  {c_{\beta_1}}
\def\sbii  {s_{\beta_2}}
\def\cbii  {c_{\beta_2}}
\def\sgi  {s_{\gamma_1}}
\def\cgi  {c_{\gamma_1}}
\def\sgii  {s_{\gamma_2}}
\def\cgii  {c_{\gamma_2}}
\def\sai  {s_{\alpha_1}}
\def\cai  {c_{\alpha_1}}
\def\saii  {s_{\alpha_2}}
\def\caii  {c_{\alpha_2}}
\def\saiii  {s_{\alpha_3}}
\def\caiii  {c_{\alpha_3}}

\def\lsup#1{^{\lower 6pt\hbox{$\scriptstyle#1$}}}
\def\llsup#1{^{\lower 3pt\hbox{$\scriptstyle#1$}}}
\def\lasup#1{^{\lower 2pt\hbox{$\scriptstyle#1$}}}
\def\ls#1{\ifmath{_{\lower1.5pt\hbox{$\scriptstyle #1$}}}}

\usepackage{mathrsfs}
\def\mw{m_W}

\counterwithout{footnote}{chapter}
\newcommand{\one}{\mathbbm{1}}

\usepackage{feynmp-auto}
\usepackage{caption}
\usepackage{subcaption}

\DeclareSourcemap{
  \maps[datatype=bibtex]{
    \map[overwrite=false]{
      \step[fieldsource=note]
      \step[fieldset=addendum, origfieldval, final]
      \step[fieldset=note, null]
    }
  }
}

\renewcommand{\Re}{\mathrm{Re }}
\renewcommand{\Im}{\mathrm{Im }}
\newcommand{\doublet}[2]{ \left( \begin{array}{c}#1 \\ #2 \end{array}\right) }
\newcommand{\triplet}[3]{ \left( \begin{array}{c}#1 \\ #2 \\ #3 \end{array}\right) }

\newcommand{\lr}[1]{ \langle #1 \rangle}
\newcommand{\mmatrix}[4]{ \left(\! \begin{array}{ccc}#1 & #2 \\ #3 & #4 \end{array}\!\right) }

\providecommand{\mtrx}[1]{\begin{pmatrix} #1 \end{pmatrix}}

\newcommand{\id}{\mathbbm{1}}
\newcommand{\lrpartial}{\,\partial^{\hspace{-7pt}\raise3pt\hbox{\small $\leftrightarrow$}}\!}
\newcommand{\scr}[1]{\mbox{\scriptsize #1}}
\newcommand{\VCKM}{V_{\mbox{\scriptsize CKM}}}
\newcommand{\hSM}{h_{\mbox{\scriptsize SM}}}
\newcommand{\mSM}{m_{\mbox{\scriptsize SM}}}

\newcommand{\Rrefst}[2]{Refs.~\cite{#1}--\cite{#2}}

%% file: acronyms/acronyms.tex
\newglossaryentry{2HDM}
{
    name=2HDM,
    description={Two Higgs Doublet Model}
}

\newglossaryentry{3HDM}
{
    name=3HDM,
    description={Three Higgs Doublet Model}
}

\newglossaryentry{NHDM}
{
    name=NHDM,
    description={N Higgs Doublet Model}
}

\newglossaryentry{C2HDM}
{
    name=C2HDM,
    description={Complex 2 Higgs Doublet Model}
}

\newglossaryentry{C3HDM}
{
    name=C3HDM,
    description={Complex 3 Higgs Doublet Model}
}

\newglossaryentry{QCD}
{
    name=QCD,
    description={Quantum Chromodynamics}
}

\newglossaryentry{GeV}
{
    name=GeV,
    description={Giga electron Volt}
}

\newglossaryentry{ATLAS}
{
    name=ATLAS,
    description={A Toroidal LHC ApparatuS}
}

\newglossaryentry{LHC}
{
    name=LHC,
    description={Large Hadron Collider}
}

\newglossaryentry{CMS}
{
    name=CMS,
    description={Compact Muon Solenoid}
}

\newglossaryentry{QFT}
{
    name=QFT,
    description={Quantum Field Theory}
}

\newglossaryentry{BSM}
{
    name=BSM,
    description={Beyond the Standard Model}
}

\newglossaryentry{BAU}
{
    name=BAU,
    description={Baryon Asymmetry in the Universe }
}

\newglossaryentry{LEP}
{
    name=LEP,
    description={Large Electron–Positron Collider}
}

\newglossaryentry{HF}
{
    name=HF,
    description={Higgs Family}
}

\newglossaryentry{BFB}
{
    name=BFB,
    description={bounded from below}
}

\newglossaryentry{FCNCs}
{
    name=FCNCs,
    description={Flavor-changing neutral couplings}
}

\newglossaryentry{SM}
{
    name=SM,
    description={Standard Model}
}

\newglossaryentry{ML}
{
    name=ML,
    description={Machine Learning}
}

\newglossaryentry{CKM}
{
    name=CKM,
    description={Cabibbo-Kobayashi-Maskawa}
}

\newglossaryentry{h.c.}
{
    name=h.c.,
    description={hermitian conjugate}
}

\newglossaryentry{QED}
{
    name=QED,
    description={Quantum Electrodynamics}
}

\newglossaryentry{vev}
{
    name=vev,
    description={vacuum expectation value}
}

\newglossaryentry{CP}
{
    name=CP,
    description={Charge-Parity}
}

\newglossaryentry{Ref.}
{
    name=Ref.,
    description={Reference}
}

\newglossaryentry{IDM}
{
    name=IDM,
    description={Inert Doublet Model}
}

\newglossaryentry{eq.}
{
    name=eq.,
    description={Equation}
}

\newglossaryentry{DM}
{
    name=DM,
    description={Dark matter}
}

\newglossaryentry{SSB}
{
    name=SSB,
    description={Spontaneous Symmetry Breaking}
}

\newglossaryentry{WIMP}
{
    name=WIMP,
    description={Weakly interacting massive particle}
}

\newglossaryentry{CMB}
{
    name=CMB,
    description={Cosmic Microwave Background}
}

\newglossaryentry{SI}
{
    name=SI,
    description={spin-independent}
}

\newglossaryentry{EDM}
{
    name=EDM,
    description={Electric Dipole Moment}
}

\newglossaryentry{LAT}
{
    name=LAT,
    description={Large Area Telescope}
}

\newglossaryentry{AMS}
{
    name=AMS,
    description={Alpha Magnetic Spectrometer}
}

\newglossaryentry{LZ}
{
    name=LZ,
    description={LUX-ZEPLIN}
}

\newglossaryentry{NFW}
{
    name=NFW,
    description={Navarro-Frenk-White}
}

\newglossaryentry{H.E.S.S.}
{
    name=H.E.S.S.,
    description={High Energy Stereoscopic System}
}

%% file: variables.tex

\newcommand {\Title} { Constraining Multi-scalars models with colliders and Dark Matter}
\newcommand {\Subtitle} {My Subtitle}
\newcommand {\StudentName} {Rafael Filipe Teixeira Boto}
\newcommand {\DegreeName} {Physics}
\def\FontSn{
  \usefont{T1}{phv}{m}{n}\fontsize{12pt}{12pt}\selectfont}

\newcommand {\Advisor} {{\large Dr. }}
\newcommand {\CoAdvisor} {{\large Dr. Co Advisor}}

\newcommand {\CommitteeMembers} {
{\large Dr. Doutor Howard Eli Haber, Distinguished Emeritus Professor, University of California, Santa Cruz,
EUA \\ Relator
Doutor Francisco Jose Botella Olcina, Catedrático de Universidad, Institut de Física Corpuscular,
Universitat de València, Espanha; - \\
Doutor Mário João Martins Pimenta, Professor Catedrático do Instituto Superior Técnico da
Universidade de Lisboa;\\
Doutor João Paulo Ferreira da Silva, Professor Catedrático do Instituto Superior Técnico da
Universidade de Lisboa; \\
Doutor José Guilherme Teixeira de Almeida Milhano, Professor Associado do Instituto Superior
Técnico da Universidade de Lisboa;\\
Doutor Rui Alberto Serra Ribeiro dos Santos, Professor Coordenador (com Agregação) do
Instituto Superior de Engenharia de Lisboa. }
}
\newcommand {\Chairperson} {{\large  Doutor Mário João Martins Pimenta, Professor Catedrático do Instituto Superior
Técnico da Universidade de Lisboa. }}

\def \IsFinalVersion{1}

\newcommand {\Month} {June}
\newcommand {\Year} {2025}

\def \acknowledgmentsPage{1}

\def \abstractEnglishPage{3}

\def \abstractPortuguesePage{5}

\def \HasCoAdvisor{0}

\def \finalLogoSpacing{-1.0cm}
\def \draftLogoSpacing{2.0cm}

\def \finalAdvisorsSpacing{1.0cm}
\def \draftAdvisorsSpacing{10.0cm}

\def \dateSpacing{1.5cm}

%% file: acknowledgments/acknowledgments.tex


\section*{Acknowledgments}

\addcontentsline{toc}{section}{Acknowledgments}
This thesis marks the completion of my main goal when I set to study Physics. For most of the journey I always had two official supervisors, João P. Silva and Jorge Romão.  I would like to express my highest respect and eternal gratitude for the invaluable guidance and mentorship throughout the years. Their expertise and values were fundamental to my development, both professionally and personally. With Prof. João P. Silva I want to highlight the commitment, honour and rigour.  Prof. Jorge Romão taught me perseverance, patience and resilience.

I would like to follow with acknowledging José Correia, my Physics highschool professor who played a big part in my decision to follow this career. I am also indebted to Tiago Fernandes who was my first professional collaborator, helping my decision of meeting with my supervisors. I am proudly finishing my PhD in the same year. I want to thank the direct collaborators I have had since starting my career, in chronological order: Howard Haber, Andreas Trautner, Miguel Bento, Dipankar Das, Ipsita Saha, Luis Lourenço, Pedro Figueiredo, Igor Ivanov, Hong Deng, Fernando Souza and Miguel Romão.

I am thankful for the resources I received both from FCT, with the PhD grant PRT/BD/152268/2021, as well as from CFTP and Instituto Superior Técnico. The academic and administrative support were fundamental for the success and the execution of the research plan. I want to highlight the support from Sandra Oliveira. Many thanks to my colleagues here, to Henrique Câmara for the motivation as a classmate from the same year, to the younger who will complete their journeys in the future -
 José Rocha, Aditya Batra, André Santos, João Matos, João Belas, Carolina Lopes, Duarte Correia and Tiago Rebelo - I wish you the very best of luck and success.

 To my friends, who have offered emotional support and helped my growth in countless avenues of life. I especially want to thank Filipe Mendes who, years ago, forced me along a journey that has proven to have changed my perspective on life forever. Along the way, I have been fortunate to meet countless friends from around the world, and I want to thank Pedro Rodrigues, Vitor Velosa and Tiago Vieira, with whom I have had the pleasure of discovering the world together. Finally, I want to show my appreciation and admiration for the role models who I have met, with Jacob, Ethan, Devin, Saurya and many others. I vow to never stop being the kid that wanted to touch the stars. 

Lastly, and most importantly, I want to express my deepest gratitude to my parents and brother for the support and allowing me to follow my dreams. To my father for the understanding and guidance to handle the challenges along the way. To my mother for the encouragement and steadfast belief in my capabilities, ensuring I always knew someone had faith in me.

%% file: abstract/abstract-pt.tex

\begin{otherlanguage}{portuguese}

\section*{Resumo}

\addcontentsline{toc}{section}{Resumo}

Após a observação em 2012 de uma nova partícula escalar muito semelhante ao bosão de Higgs do Modelo Padrão da física de partículas, existe um consenso geral de que deve haver Física Além do Modelo Padrão, com as experiências atuais agora dedicadas à sua descoberta. A extensão do sector escalar é motivada por problemas chave não resolvidos na física de partículas, incluindo a necessidade de novas fontes de violação da simetria Carga Paridade, fornecendo uma explicação para a assimetria bariónica no universo, ou para explicar a Matéria Escura, que constitui cerca de $85\%$ da matéria do Universo. 

Nesta tese, focamo-nos em modelos de três dubletos de Higgs e nas restrições que precisam de ser impostas. Adicionamos contribuições teóricas para a consistência do potencial escalar, com a condição de limite inferior e confirmação do  mínimo global. Consideramos todas as restrições para estudos fenomenológicos completos em modelos com diferentes simetrias, estudando o seu impacto individual e tentando distinguir os modelos com base em dados.

Propomos um modelo com coeficientes de violação de Carga Paridade, levando a acoplamentos de Higgs que se desviam significativamente dos valores do Modelo Padrão e permanecem permitidos. Para explorar o espaço de parâmetros do modelo, utilizamos um algoritmo eficiente de Aprendizagem Automática que encontra novas regiões do espaço de parâmetros e consequências observáveis, não encontradas com técnicas anteriores. As novas técnicas são aplicáveis a qualquer cenário de Física Além do Modelo Padrão.

Conectamos as extensões escalares com soluções experimentalmente viáveis para o problema da Matéria Escura. Ao construir modelos, frequentemente impõe-se simetrias discretas conservadas para estabilizar os candidatos a Matéria Escura. Consideramos esta possibilidade com dois candidatos e procuramos uma possibilidade alternativa de um grupo não Abeliano conservado que leva a uma solução viável, numa tentativa de traçar os limites do que os modelos Multi-Higgs podem acomodar.

\vfill

\textbf{\Large Palavras-chave:} bosão de Higgs, modelos Multi-Higgs, violação de CP, Matéria Escura, Aprendizagem Automática

\end{otherlanguage}

%% file: abstract/abstract-en.tex

\begin{otherlanguage}{english}

\section*{Abstract}

\addcontentsline{toc}{section}{Abstract}
After the observation in 2012 of a new scalar particle closely resembling the Higgs boson of the Standard Model of particle physics, there is a general consensus that there must be Physics Beyond the Standard Model, with present experiments now dedicated to its discovery. Extending the scalar sector is motivated by key unresolved issues in particle physics including the need of new sources of Charge Parity violation, providing an explanation of the baryon asymmetry in the universe, or to explain Dark Matter, which comprises of order $85\%$ of the matter content of the Universe.

In this thesis, we focus on three Higgs doublets models (3HDM) and the constraints that need to be imposed. We add theoretical contributions for the consistency of the scalar potential, with boundedness from below and the global minimum. We consider all constraints for full phenomenological studies in models with different symmetries, studying their individual impact and attempting to distinguish the models based on data. 

We propose a model with Charge Parity violating coefficients, leading to Higgs couplings that significantly deviate from Standard Model values and remain allowed. To explore the parameter space of the model, we employ an efficient Machine Learning algorithm that finds new regions of parameter space and observable consequences, not found with previous techniques we developed and applied. The new techniques are applicable to any Physics Beyond the
Standard Model scenario.

We connect the scalar extensions with experimentally viable solutions to the Dark Matter problem. When building Dark Matter models, one often imposes conserved discrete symmetries to stabilize DM candidates. We consider a possibility with two candidates, and an alternative possibility of a conserved non-Abelian group leading to a viable DM, in an attempt to chart the limits of what Multi-Higgs Models can accommodate.


\vfill

\textbf{\Large Keywords:} Higgs boson, Multi-Higgs Models, CP violation, Dark Matter, Machine Learning

\end{otherlanguage}

%% file: chapters/main.tex


\include{chapters/preface}
\cleardoublepage 
\include{chapters/introduction}
\cleardoublepage 
\include{chapters/3hdm}
\cleardoublepage 
\include{chapters/Constraints}
\cleardoublepage 
\include{chapters/Alignment}

\cleardoublepage 
\include{chapters/c3hdm}
\cleardoublepage 
\include{chapters/ai}

\cleardoublepage 
\include{chapters/DarkMatter_problem}

\cleardoublepage

\include{chapters/DarkMatter_model}

\cleardoublepage 
\include{chapters/DarkMatter_multi}

\cleardoublepage 
\include{chapters/conclusion}

%% file: chapters/preface.tex
\chapter*{Preface}
\addcontentsline{toc}{chapter}{Preface}
\label{chapter:preface}
\hspace*{0.3cm}
The research presented in this thesis has been carried out at Centro de Física Teórica de Partículas (CFTP) of the Physics Department at Instituto Superior Técnico, University of Lisbon, and was supported by Fundação para a Ciência e Tecnologia (FCT) with the PhD grant PRT/BD/152268/2021.

To the present date, I have contributed to the following works:
\begin{itemize}
\item~\cite{Boto:2020wyf} R. Boto, T. V. Fernandes, H. E. Haber, J. C. Romão, and J. P. Silva, 
“Basis-independent treatment of the complex 2HDM”, Phys. Rev. D 101, 055023, March 2020 [arXiv:2001.01430].

\item~\cite{Bento:2020jei} M. P. Bento, R. Boto, J. P. Silva, and A. Trautner, 
“A fully basis invariant Symmetry Map of the 2HDM”, JHEP 21, 229, February 2021 [arXiv:2009.01264].

\item~\cite{Boto:2021qgu} R. Boto, J. C. Romão, and J. P. Silva, 
“Current bounds on the Type-Z Z3 three-Higgs-doublet model”, Phys. Rev. D 104, 095006, November 2021 [arXiv:2106.11977].

\item~\cite{Boto:2022uwv} R. Boto, J. C. Romão, and J. P. Silva, 
“Bounded from below conditions on a class of symmetry constrained 3HDM”, Phys. Rev. D 106, 115010, December 2022 [arXiv:2208.01068].

\item~\cite{Boto:2023nyi} R. Boto, D. Das, L. Lourenco, J. C. Romão, and J. P. Silva, 
“Fingerprinting the Type-Z three-Higgs-doublet models”, Phys. Rev. D 108, 015020 , July 2023, [arXiv:2304.13494]. 

\item~\cite{Boto:2023bpg} R. Boto, D. Das, J. C. Romão, I. Saha, and J. P. Silva, 
“New physics interpretations for nonstandard values of $h\to Z\gamma$”, Phys. Rev. D 109, 095002, May 2024 [arXiv:2312.13050].

\item~\cite{Boto:2024tzp} R. Boto, P. N. Figueiredo, J. C. Romão, and J. P. Silva, 
“Novel two component dark matter features in the $\Z2\times\Z2$ 3HDM”, JHEP 11, 108, November 2024 [arXiv:2407.15933].

\item~\cite{Boto:2024jgj} R. Boto, L. Lourenco, J. C. Romão, and J. P. Silva, 
“Large pseudoscalar Yukawa couplings in the complex 3HDM”, JHEP 11, 106, November 2024 [arXiv:2407.19856].

\item~\cite{Deng:2025dcq} H. Deng, R. Boto, I. P. Ivanov, and J. P. Silva, 
“Dark matter stabilized by a non-abelian group: lessons from the $\Sigma$(36) 3HDM”, Phys.Rev. D 111, 055006, March 2025 [arXiv:2501.05929].

\item~\cite{deSouza:2025bpl} F. Abreu de Souza, R. Boto, M. Crispim Romão, P. N. Figueiredo, J. C. Romão, and J. P. Silva,
“Unearthing large pseudoscalar Yukawa couplings with Machine Learning”, In review, May 2025 [arXiv:2505.10625]
\end{itemize}

Refs.~\cite{Boto:2020wyf,Bento:2020jei} are prior to this thesis and have not been included in it. \Rrefst{Boto:2021qgu}{deSouza:2025bpl} have a significant portion of content included in this thesis.

In this thesis, we develop and study aspects of scalar extensions of the Standard Model (\gls{SM}), with a focus on three Higgs doublet models (\gls{3HDM}). The first Chapter contains basic description of the content of the Standard Model of Particle Physics. The second Chapter, containing part of Refs.~\cite{Boto:2021qgu,Boto:2022uwv}, is dedicated to defining the different scalar extensions studied, with a Section dedicated to the novel contributions done in the topic of ensuring consistency in their formulation. 

In Chapter three we present an extensive description of all the theoretical and experimental constraints to be considered in a full phenomenological study of the models considered. The calculations shown are included in Refs.~\cite{Boto:2021qgu,Boto:2023bpg}. 

In the fourth Chapter we confronts the models with all the bounds using traditional sampling methods. The individual exclusions are studied and we show how the different models can be distinguished. We pioneer the study of the models proposed away from a special region named the alignment limit, to get new predictions for observables. The results shown contain parts of Refs.~\cite{Boto:2021qgu,Boto:2022uwv,Boto:2023nyi}.

The fifth and sixth Chapters are dedicated to the first study of a 3HDM with a $\Z2\times\Z2$ symmetry and explicit charge parity (\gls{CP}) violation, which we name complex 3HDM (\gls{C3HDM}). We start by introducing a parametrization and comparing the results obtained from sampling the parameter space with traditional methods, in Ref.~\cite{Boto:2024jgj}, and new Machine Learning algorithms, in Ref.~\cite{deSouza:2025bpl}.    

The seventh Chapter addresses the Dark Matter (\gls{DM}) problem with experimental evidence and searches. In Chapter~\ref{chapter:Darkmatter_model},  we follow with a construction of 
a $\Sigma(36)$ symmetric model with DM candidates. Chapter~\ref{chapter:Darkmatter_multi} follows with a study of the possibility of having two dark matter particle candidates in a $\Z2\times\Z2$ 3HDM for the complete ${\rm GeV}$ mass range. The main results are on estabilishing the conditions necessary for the models to have viable solutions to the Dark Matter open problem, with published work in Refs.~\cite{Boto:2024tzp,Deng:2025dcq}. 

In the tenth and final Chapter, we present our conclusions.

%% file: chapters/introduction.tex

\chapter{Introduction}
\label{chapter:introduction}
\hspace*{0.3cm}
The Glashow-Salam-Weinberg model~\cite{Weinberg:1967tq,Salam:1968rm,GLASHOW1961579}, known as the Electroweak Standard Model, describes the electromagnetic and the weak interactions through a gauge theory with the symmetry group $\text{\SU2}_\text{L}\times\text{\U1}_Y$ that is spontaneously broken into $\text{\U1}_{\rm EM}$ through the Higgs mechanism~\cite{HIGGS1964132,Englert:1964et,PhysRevLett.13.585}. Combined with an $\text{\SU3}_c$ gauge group responsible for Quantum Chromodynamics (\gls{QCD}), the Standard Model (\gls{SM}) of Particle Physics was constructed and remains as the most precise theory of Nature known. Over the decades, it has been confirmed and culminated with the discovery in July of 2012 of the predicted Higgs boson by the \gls{LHC} collaborations \gls{ATLAS}~\cites{Aad:2012tfa} and \gls{CMS}~\cite{Chatrchyan:2012xdj}. And yet, there is a general consensus that there must be Physics Beyond the Standard Model (\gls{BSM}), with experiments now dedicated to its discovery. 

The main open problems consist of the need for new sources of \gls{CP}-violation, a necessary ingredient for a successful explanation of the baryon asymmetry of the universe (\gls{BAU})~\cite{Sakharov:1967dj}, the existence of Dark Matter (\gls{DM}), which comprises of order 85 \% of the matter content of the Universe~\cite{Planck:2013oqw}, and the origin of the observed tiny neutrino masses, with the SM not including a mechanism for generating massive neutrinos. 
Motivated by the lack of a fundamental reason why the scalar sector should be limited to a single Higgs doublet, multi-Higgs extensions are required for many of the  viable explanations to these problems. Models featuring multiple spin-zero particles stabilized by a symmetry can, by prohibiting tree-level neutrino masses, provide a loop-level mechanism for generating tiny neutrino masses~\cite{Zee:1980ai,Ma:2006km,Kanemura:2011jj}. Dark Matter, consisting of heavy, stable, electrically neutral particles of non-baryonic nature, can be obtained from a scalar sector equipped with a new global symmetry that remains unbroken in the vacuum~\cite{Deshpande:1977rw,LopezHonorez:2006gr,Barbieri:2006dq}. 

In the possible scalar extensions of the SM, the two-Higgs doublet model (\gls{2HDM}) is the simplest one that can provide a new source of CP violation, by imposing a softly-broken $\Z2$ symmetry and allowing the scalar sector to be in general CP-violating. This model, now known by the name complex 2HDM (\gls{C2HDM}), was first introduced in Ref.~\cite{Ginzburg:2002wt}, with only one independent CP-phase. The simplest extensions of the SM, by construction, can only partly provide a solution to the main open questions and are already significantly constrained experimentally. We thus motivate the central topic of this thesis - the study of extended Higgs sectors and their viability.  

The thesis is structured with Chapter~\ref{chapter:3HDM} defining models with multiple Higgs fields in order to study the requirements of consistency, with the development of a method of deriving sufficient conditions for a bounded from below scalar potential for Three Higgs Doublet Models (\gls{3HDM}). Next, in Chapter~\ref{chapter:Constraints} and~\ref{chapter:Alignment}, we perform full phenomenological studies of multiple 3HDM, by computing all relevant observables and comparing them with the most recent experimental data. In doing so, we arrive at how specific enhancements and proposed new decays can also distinguish currently viable models. Additionally, preliminary measurements of a signal may be indicative of an enhancement compared to the SM expectation. We developed a theoretical preparation for such an eventuality in the particular case of the loop-induced decay of the Higgs boson into a photon and Z boson. Previous studies pointed out that sampling the multiple free parameters is easier in simplified scenarios, such as near the alignment limit. We attempt to soften such simplifications to obtain new predictions.  In Chapter~\ref{chapter:C3HDM} and~\ref{chapter:ml}, we define and perform the first studies of an extension of the C2HDM with an additional Higgs doublet and multiple unremovable complex phases, which we name the complex 3HDM (\gls{C3HDM}) with a softly-broken $\Z2\times\Z2$ symmetry. In doing so, we apply novel Machine Learning (\gls{ML}) techniques and compare with traditional sampling near the real and alignment limits we define, highlighting both severe improvements in efficiency and important new physical consequences. 

Starting in Chapter~\ref{chapter:Darkmatter_problem}, we switch focus to the Dark Matter (\gls{DM}) problem for the remainder of the thesis. We review experimental evidence, current exclusion bounds and future sensitivity of experiments. In Chapter~\ref{chapter:Darkmatter_model} we follow with a study on how multi-Higgs doublet models can have viable Dark Matter particle candidates, having the first attempt at stable DM candidates in
a $\Sigma(36)$ symmetric model and discussing the obstacles that appear in model building. Chapter~\ref{chapter:Darkmatter_multi} follows with considering the previous real $\Z2\times\Z2$ 3HDM for two DM particle candidates with different masses including regions already excluded in simpler extensions. 

Each Chapter ends with a short summary and, in Chapter~\ref{chapter:conclusion}, we give an overall Conclusion. Several appendices follow, leading with Appendix~\ref{app:lambdas} giving the full list of the parameterization of the terms in the scalar potential in terms of physical quantities for the symmetries considered. In Appendix~\ref{app:Types}, we complete the definitions shown in Chapter~\ref{chapter:3HDM} for the possible Types of Yukawa couplings in 3HDMs. In Appendix~\ref{a:comparison_bfb}, we compare our method of deriving sufficient conditions for a bounded from below scalar potential with alternative methods. Appendix~\ref{a:appoffdiag} presents a general approach on how to suppress trilinear diagonal
couplings of additional charged scalars with Z-boson when  compared to the corresponding off-diagonal couplings. Appendix~\ref{a:appScattering} follows with a calculation of the scattering amplitudes of a general Lagrangian with such off-diagonal couplings and their high-energy limit. Appendix~\ref{app:parametrization} shows the explicit forms of the matrices defined in Chapter~\ref{chapter:C3HDM} to parameterize the C3HDM. In Appendix~\ref{appendix-group}, we discuss the irreducible representations of the non-abelian group $\Sigma(36\varphi)$, followed by attempting all possibilities at extending to the Yukawa sector in Appendix~\ref{appendix-Yukawa-exact}. In our last Appendix~\ref{app:masses}, we present the mass formulas for minima other than the most general one for the real $\Z2\times\Z2$ symmetric 3HDM.




We will continue this introduction by briefly summarizing the particle content and the Lagrangian for the current formulation of the Standard Model.

\section{Standard Model}
\label{chapter:Standard Model}

The Standard Model is a Quantum Field Theory (\gls{QFT}) based on a Yang-Mills theory~\cite{Darvas:2011zz} defined by the requirement of both Poincaré invariance and invariance under the local gauge symmetry $\text{\SU3}_\text{c}\times\text{\SU2}_\text{L}\times\text{\U1}_Y$, where the subscripts C, L and Y represent color, left-handedness and hypercharge, respectively. The subscript L means that  for a generic fermionic field $\psi$ the left-handed component, $\psi_\text{L}=(1-\gamma_5)\psi/2$, transforms under the fundamental representation (doublet, $\pmb{\underline{2}}$) of $\text{\SU2}$ while the right-handed, $\psi_\text{R}=(1-\gamma_5)\psi/2$, as a singlet, $\pmb{\underline{1}}$. The hypercharge Y is given by 
\beq
Y=Q-T_3\,;
\eeq
where Q is the electric charge and $T_3$ the third component of weak isospin. 

The theory’s particle interactions are encoded in its Lagrangian, $\mathcal{L}_{SM}$, determined by the field content and their transformation properties under the gauge group.  This mathematical framework  describes measurable physical processes, enabling predictions that can be rigorously tested against experimental data to validate the theory. The matter content of the SM is, 
\beqa
&\begin{aligned}&\text{Fermions}\\&(\alpha=1,2,3)\end{aligned}\left\{\begin{array}{l}
\text{\textbf{Quarks}} \left(\qquad\begin{array}{l}q_{\text{L}\alpha}=   \begin{pmatrix}[1.5] u\,_{\text{L}\alpha}   \\ d\,_{\text{L}\alpha} \end{pmatrix} \sim (\,\pmb{\underline{3}}\,,\,\pmb{\underline{2}}\,,\,Y=1/6) \\
u\,_{\text{R}\alpha} \sim (\,\pmb{\underline{3}}\,,\,\pmb{\underline{1}}\,,\,Y=2/3)\quad,\quad d\,_{\text{R}\alpha} \sim (\,\pmb{\underline{3}}\,,\,\pmb{\underline{1}}\,,\,Y=-1/3)

\end{array}\right.\\
\text{\textbf{Leptons}}\left(\qquad\begin{aligned}&l_{\text{L}\alpha}=   \begin{pmatrix}[1.5] \nu_{\text{L}\alpha}   \\ c\,_{\text{L}\alpha} \end{pmatrix} \sim (\,\pmb{\underline{1}}\,,\,\pmb{\underline{2}}\,,\,Y=-1/2) \\
&c\,_{\text{R}\alpha} \sim (\,\pmb{\underline{1}}\,,\,\pmb{\underline{1}}\,,\,Y=-1)\end{aligned}\right.
\end{array}\right.\label{fermionfield} \\
&\text{\textbf{Higgs}}\qquad \qquad\,\phi=\begin{pmatrix}[1.5] \phi^+ \\ \phi^0\end{pmatrix} \sim (\,\pmb{\underline{1}}\,,\,\pmb{\underline{2}}\,,\,Y=1/2)\qquad\qquad    \label{higgsfield}
\eeqa
where the numbers in brackets indicate how the fields transform under the gauge groups $\text{\SU3}_\text{c}$, $\text{\SU2}_\text{L}$ and $\text{\U1}_Y$, respectively. 


 Local gauge invariance is achieved by replacing ordinary derivatives $\partial_\mu$ of the fields \eqref{fermionfield} and \eqref{higgsfield} by
the corresponding covariant derivatives. For a doublet field $\psi_\text{L}$, with hypercharge $Y$, the covariant derivative is given, in the notation of~\cite{Romao:2012pq}, by,

\beq\label{doubletD}
D_\mu \psi_\text{L}=\left(\partial^\mu-i\frac{g}{2}\tau^aW_\mu^a-i g'Y B_\mu\right) \psi_\text{L}\,,
\eeq
where $\tau^a\,(a=1,2,3)$, the Pauli matrices, are the generators of the $\pmb{\underline{2}}$ representation of  $\text{\SU2}_\text{L}$. $W_\mu^a$ and $B_\mu$ are the $\SU2$ and $\U1$ gauge fields. For a quark field q in the $\pmb{\underline{3}}$ representation  of $\text{\SU3}_\text{c}$, with the Gell-Mann matrices $\lambda_a$ as generators, we have the covariant derivative 

\beq
D_\mu q=\left(\partial^\mu-i\frac{g_s}{2}G_\mu^a\lambda^a-i\frac{g}{2}\tau^aW_\mu^a-i g'Y B_\mu\right)q\,,
\eeq
where the 8 gluons $G_\mu^a$ are introduced as Lorentz fields in the adjoint representation.

This work focuses exclusively on the electroweak sector, defined by  $\text{\SU2}_\text{L}\times\text{\U1}_Y$, which governs the interactions of the weak and electromagnetic forces. The full Lagrangian of the theory is:
\beq
\mathcal{L}_{SM}= \mathcal{L}_\text{gauge}+\mathcal{L}_\text{fermions}+\mathcal{L}_\text{Higgs}+\mathcal{L}_\text{Yukawa}+\mathcal{L}_\text{GF}+\mathcal{L}_\text{ghosts}\,.
\eeq

The gauge term is given by:
\beq
\mathcal{L}_\text{gauge}=-\frac{1}{4}\left(\tensor{B}{_\mu_\nu}\tensor{B}{^\mu^\nu}+\frac{1}{4}W^a_{\mu_\nu}\tensor{W}{^a^\mu^\nu}\right)\,,\, \tensor{B}{_\mu_\nu}\equiv\partial_\mu\tensor{B}{_\nu}-\partial_\nu\tensor{B}{_\mu}\,,\,\, W^a_{\mu\nu}\equiv\partial_\mu W_\nu^a-\partial_\nu W_\mu^a+g\tensor{\epsilon}{^a^b^c}W_\mu^b W_\nu^c\label{gaugelag}\,,
\eeq
including the kinetic terms of the gauge fields and the three- and four-gauge boson vertices. The fermion term is expressed as: 
\beq
\mathcal{L}_\text{fermions}=i \,\anti q_L \gamma^\mu D_\mu q_L + i \,\anti u_R \gamma^\mu D_\mu u_R + i \,\anti d_R \gamma^\mu D_\mu d_R + i \,\anti l_L \gamma^\mu D_\mu l_L + i \,\anti c_R \gamma^\mu D_\mu c_R\label{SMgaugeferm}\,,
\eeq
where $\overline{\psi}\equiv\psi^\dagger\gamma_0$ and $D_\mu$ is the covariant derivative, with its form dependant on  the properties of the fields in \eq{fermionfield}.
Yang-Mills theories describe massless gauge bosons, as a mass term $A_\mu A^\mu$ violates local invariance under the gauge group. For the weak force, mediated by massive $W^\pm$ and $Z$ bosons, this issue is addressed through the Higgs mechanism, where a new scalar field enables particles to acquire mass via interactions. Rather than introducing terms that explicitly break gauge symmetry, the theory allows the system itself to break the symmetry spontaneously. In this spontaneous symmetry breaking (\gls{SSB}) mechanism, the Lagrangian remains invariant under the gauge group, but its vacuum solutions exhibit reduced symmetry. The term relative to the Higgs field introduced is 
\beq
\mathcal{L}_\text{Higgs}\mathrel{\mathop:}=D^\mu\phi^\dagger D_\mu \phi -V(\phi)=D^\mu\phi^\dagger D_\mu \phi - \mu^2 \phi^\dagger\phi -\lambda (\phi^\dagger\phi)^2\label{higgslag}\,,
\eeq
where $\mu^2$ and $\lambda$ are real parameters. The scalar potential $V(\phi)$ drives spontaneous symmetry breaking:
\beq
\text{\SU3}_\text{c}\times\text{\SU2}_\text{L}\times\text{\U1}_Y\longrightarrow \text{\SU3}_\text{c}\times\text{\U1}_{EM}\,,
\eeq
since the field possesses a vacuum expectation value (\gls{vev}), configuration of $\phi$ which minimizes $V(\phi)$, for $\mu^2<0$,
\beq
 \phi=\vev{\phi}_0+\rho\mathrel{\mathop:}=\frac{1}{\sqrt{2}}\begin{pmatrix}[1.5] 0\\ v\end{pmatrix}+\begin{pmatrix}[1.5] G^+ \\ (h+iG^0)/\sqrt{2}\end{pmatrix}\,\,,\quad v=\sqrt{\frac{-\mu^2}{\lambda}}=246\,\textrm{\gls{GeV}}\,,
\eeq
where $h$ is the physical Higgs field while $G^+$ and $G^0$ are the massless Nambu-Goldstone bosons. Note that, after SSB, the covariant derivatives in \eqref{higgslag} lead to mass terms for the gauge bosons, proportional to $v^2$, that were missing in \eqref{gaugelag} and interactions with the higgs field and Goldstone bosons.

The masses of the fermions come from the introduction of the term 
\beq
-\mathcal{L}_{\text{Yukawa}}=Y^c \overline{l_{L}} \phi \,c_{R}  + Y^U\overline{q_{L}} \Tilde{\phi} \,u_{R}  + Y^D\overline{q_{L}} \phi \,d_{R} + {\rm h.c.}  \,,
\eeq
where $\tilde{\phi}\equiv i \tau_2\phi^*$ and ${\rm h.c.}$ denotes the Hermitian conjugate. $Y^c$, $Y^U$ and $Y^D$ are $3\times 3$ general complex matrices in flavour space. After symmetry breaking, we can always change to a basis where the fermion fields are mass eigenstates by performing rotations of the type~\cite[pp.595-597]{Schwartz:2013pla}
\begin{equation}
\anti c_L = \anti C_L\left(T^{c}_{L}\right)^\dagger,\quad \anti u_L = \anti U_L\left(T^{u}_{L}\right)^\dagger,\quad \anti d_L = \anti D_L\left(T^{d}_{L}\right)^\dagger,\quad c_R = \anti C_R T^{c}_{R},\quad u_R = \anti U_R T^{u}_{R},\quad d_R = \anti D_R T^{d}_{R} \,,
\end{equation}
with the unitary matrices $T^{c,u,d}_{L,R}$ being defined by the bi-diagonalization of the matrices $Y^{c,u,d}$,
\beqa
\mathbf{M}_C \equiv {\rm diag}(m_e,m_\mu,m_\tau) =\frac{v}{\sqrt{2}}\left(T^{c}_{L}\right)^\dagger Y^c \,T^c_R  \,, \\[8pt]
\mathbf{M}_U \equiv {\rm diag}(m_u,m_c,m_t) =\frac{v}{\sqrt{2}}\left(T^{u}_{L}\right)^\dagger Y^c \,T^u_R  \,,\\[8pt]
\mathbf{M}_D\equiv{\rm diag}(m_d,m_s,m_b)= \frac{v}{\sqrt{2}}\left(T^{d}_{L}\right)^\dagger Y^c \,T^d_R \,.
\eeqa

The quark-gauge interaction terms, given previously in \eq{SMgaugeferm}, will now have terms mixing flavor families when employing this basis change. It can be seen that the mixing effects are given by a single matrix,
\begin{equation}
    V=\left(T^u_L\right)^\dagger\,T^d_L\,,
\end{equation}
known as the Cabibbo-Kobayashi-Maskawa (\gls{CKM}) matrix, containing one physical complex phase.

The fifth term corresponds to the gauge fixing terms, needed to properly define the gauge boson propagators. The last term relates to the Faddeev-Popov ghost fields~\cite{Faddeev:1967fc}  that are introduced into the theory to keep the path integral formulation consistent.

%% file: chapters/3hdm.tex
\chapter{Scalar extensions of the Standard Model}
\label{chapter:3HDM}
\hspace*{0.3cm}
While the properties of the boson observed at the LHC are compatible with those predicted for the Higgs boson of the Standard Model (SM), they are also in agreement with many BSM models. There may be more than one spin zero particle and their number must be determined experimentally. One of the simplest scalar extensions of the SM is the two-Higgs doublet model (2HDM), a minimal extension of the SM obtained through the addition of a new scalar doublet.

Multiple theoretical restrictions must be considered when constructing models with extensions of the scalar sector~\cite{Bento:2017eti}. The boundedness from below of the scalar potential has been solved fully for 2HDMs, but it is known to be a very complex obstacle in the development of more general models~\cite{Faro:2019vcd}. Imposing well-chosen symmetries can significantly simplify the problem, while also removing unwanted effects already ruled out by experiment, such as Higgs-mediated flavour changing neutral couplings (FCNCs). In Ref.~\cite{Ferreira:2010xe}, it was explicitly
demonstrated that tree-level FCNCs are absent if and only if there is a basis for the Higgs doublets in which
all the fermions of a given electric charge couple to only one Higgs doublet.
Such an aspect of the model is quite desirable in
view of the flavour data~\cite{ParticleDataGroup:2022pth}. These types of constructions are usually referred to
as models with natural flavour conservation~(NFC)~\cite{Glashow:1976nt,Paschos:1976ay} in the literature,
of which there are five independent possibilities.

The complete classification of physically distinct symmetry-constrained models has been achieved in the 2HDM~\cite{Ivanov:2006yq,Bento:2020jei} and the three-Higgs doublet model (3HDM)~\cite{Ivanov:2012fp}. The first aim of this Chapter is to review the Lagrangian for the possible 2HDMs and 3HDMs, both in the scalar potential and in the possibilities for scalar-fermion Yukawa couplings with natural flavor conservation.

Secondly, we present a novel method to derive bounded from below (BFB) sufficient conditions along neutral and charged directions,
in cases where necessary \textit{and} sufficient conditions are not
available through other techniques.
We derive sufficient conditions for the 
3HDM with the symmetries $\U1\times \U1$,
$\U1\times \Z2$, $\Z2\times \Z2$ and $\Z3$.
The method can be applied to a generality of other cases;
for example in the case of $A_4$~\cite{Carrolo:2022oyg}. In Appendix~\ref{a:comparison_bfb}, we compare the physical consequences of our derived sufficient conditions with the necessary \textit{and} sufficient conditions, in  cases where the latter are available. 



\section{The 2HDM}

In the Standard Model, the Veltman’s $\rho$ parameter~\cite{Ross:1975fq,Veltman:1977kh} defined by 
\beq    
\rho=\frac{m_W^2}{m_Z^2 \cos^2\theta_W},
\eeq
where $m_W$ and $m_Z$ are the masses of the $W^\pm$ and $Z$ gauge bosons, respectively, and $\theta_W$ is the weak mixing angle, relates the strength of the neutral-current and charged-current interactions in four-fermion processes at zero momentum transfer. In the SM, the $\rho$ parameter is equal to $1$ at tree level, which is protected by the custodial $\SU2$ symmetry in the Higgs sector of the SM. Experimentally, $\rho$ is very close to one~\cite{ParticleDataGroup:2010dbb} and provides a strong constraint on extended electroweak sectors. In a $\SU2\times\U1$ gauge theory with $n$ scalar multiplets $\phi_i$ possessing weak isospin $I_i$, weak hypercharge $Y_i$ and vacuum expectation value (\gls{vev}) $v_i$ in the neutral components, the $\rho$ parameter is given at tree level by~\cite{Langacker:1980js}
\beq 
\rho=\frac{\sum_{i=1}^n\left[I_i(I_i+1)-\tfrac{1}{4}Y_i^2\right]v_i}{\sum_{i=1}^n\tfrac{1}{2}Y_i^2 v_i}.
\eeq
Extensions of the scalar sector can keep $\rho=1$ at tree level in different ways \footnote{There are relevant radiative corrections that need to be calculated and compared with experiment~\cite{Baak:2014ora}. These can be parameterized at one loop level with oblique parameters~\cite{Grimus:2008nb}, with the deviation of $\rho$ from unit given as $\delta \rho = \alpha T$, with $\alpha$ the fine-structure constant.}; the simplest corresponding to the addition of $\SU2$ singlets with $Y=0$ and/or $\SU2$ doublets with $Y=\pm 1$, as both have $I(I+1)=\tfrac{3}{4}Y^2$. In the two-Higgs-doublet model (\gls{2HDM})~\cite{Lee:2hdm,Branco:2011iw}, the fields consist of two
$\SU2_\text{L}$  doublet scalar fields 
$\Phi_a(x)\equiv (\Phi^+_a(x)\,,\,\Phi^0_a(x))$, 
where the ``Higgs flavour'' index $a=1,2$ labels the two Higgs doublet fields. The most general scalar potential obeying the requirements of hermiticity, SU(2)$_L\times \U1_Y$ gauge symmetry and renormalizability \footnote{The action, $S=-\int d^dx\mathcal{L}$, in a theory must be dimensionless, hence the Lagrangian has dimension d. From the kinetic terms we read off the mass dimensions for each field. 

For the SM and it's extensions in four dimensions, the quark fields carry mass dimension $3/2$ and the boson fields dimension $1$. The mass dimension of each coupling constant, $d_g$ is then deduced based off the respective field combination. 


In practical terms, a theory is renormalizable only
if the coupling constants have zero or positive mass dimension.} is 
\beqa  \label{pot:2hdm}
\mathcal{V}&=& m_{11}^2\Phi_1^\dagger\Phi_1+m_{22}^2\Phi_2^\dagger\Phi_2
-[m_{12}^2\Phi_1^\dagger\Phi_2+{\rm \gls{h.c.}}]+\half\lambda_1(\Phi_1^\dagger\Phi_1)^2
+\half\lambda_2(\Phi_2^\dagger\Phi_2)^2
+\lambda_3(\Phi_1^\dagger\Phi_1)(\Phi_2^\dagger\Phi_2)\nonumber\\[8pt]
&&\quad 
+\lambda_4(\Phi_1^\dagger\Phi_2)(\Phi_2^\dagger\Phi_1)
+\left\{\half\lambda_5(\Phi_1^\dagger\Phi_2)^2
+\big[\lambda_6(\Phi_1^\dagger\Phi_1)
+\lambda_7(\Phi_2^\dagger\Phi_2)\big]
\Phi_1^\dagger\Phi_2+{\rm h.c.}\right\}\,,
\eeqa
where $m_{11}^2$, $m_{22}^2$, and $\lam_{1\to 4}$ are real parameters
and $m_{12}^2$, $\lambda_{5\to7}$ are
potentially complex parameters.
Then, assuming the remaining $\U1_{\text{\gls{QED}}}$ symmetry is not broken, the scalar field
vacuum expectations values (vevs) are of the form
\beq \label{potmin}
\langle \Phi_1 \rangle={\frac{1}{\sqrt{2}}} \left(
\begin{array}{c} 0\\ v_1\end{array}\right), \qquad \langle
\Phi_2\rangle=
{\frac{1}{\sqrt{2}}}\left(\begin{array}{c}0\\ v_2\, e^{i\xi}
\end{array}\right)\,,
\eeq
where $v_1$ and $v_2$ are real and non-negative, $0\leq \xi< 2\pi$, and $v$ is determined by the Fermi constant,
\beq \label{v246}
v\equiv (v_1^2+v_2^2)^{1/2}=\frac{2\mw}{g}=(\sqrt{2}G_F)^{-1/2}\sim 246.2~{\rm GeV}\,.
\eeq

\subsection{Symmetries}
The general 2HDMs run into the problem of the existence of tree-level flavour-changing neutral couplings (\gls{FCNCs}). Higgs mediated interactions such as $\bar{d}s\phi$ will lead to $K-\bar{K}$ mixing at tree level and require very small couplings or a very high mass of the exchanged scalar~\cite{McWilliams:1980kj,Shanker:1981mj}. We will always consider models with tree level FCNCs forbidden by imposing an additional discrete or continuous symmetry.

 As summarized in~\cite{Branco:2011iw}, the global symmetries of the 2HDM scalar potential can be classified into:
\begin{itemize}
    \item $\Phi_a$ related to some unitary transformation of $\Phi_b$,
    \beq
    \Phi^a\rightarrow(\Phi^S)^a=\sum_{b=1}^2\tensor{S}{^a_b}\Phi^b\,,
    \label{HFsym}
    \eeq
    where S is a unitary matrix. This type of symmetries is known as Higgs Family (\gls{HF}) symmetries.
    \item $\Phi_a$ related to some unitary transformation of $\Phi_b^*$,
    \beq
    \Phi^a\rightarrow\left(\Phi^{GCP}\right)_a:=
    \Phi_b^*\tensor{\left[X^\mathrm{T}\right]}{^b_a}\;,  
    \label{GCPsym}
    \eeq
    where X is a unitary matrix. These are known as generalized CP (GCP) symmetries.
\end{itemize}
Under a basis transformation $\Phi^a\to \tensor{U}{^{a}_{b}} \Phi^b$, for $ U \in \text{U}(2)$, the specific form of the symmetries gets altered accordingly
\beqa
S'\,=\,U\,S\,U^\dagger,\\
X'\,=\,U\,X\,U^\text{T}.
\eeqa

We assume that the scalar potential in \eq{pot:2hdm} has some explicit internal symmetry. That is, we assume that the coefficients of $V_H$ \textit{stay exactly the same} under a specific transformation. This reduces the number of independent parameters.

Ferreira and Silva~\cite{Ferreira:2008zy} have shown that potentials satisfying symmetries of the same
conjugacy class (i.e. $S'=U\,S\,U^\dagger$ where $U$ is an unitary matrix), are related through a basis change. Consequently, one only has to focus on symmetries of different classes, since these are the ones yielding different physics. In Ref.~\cite{Ivanov:2006yq}, Ivanov proved that there are only six distinct classes of potentials, even when combining symmetries of the two types. These are commonly denoted as $\Z2$, \U1, and \SU2 (HF symmetries) as well as CP1, CP2, CP3 (GCP symmetries), and they are schematically related as~\cite{Ferreira:2009wh,Ferreira:2010yh}
\begin{equation}\label{SchemeSymmetries}
 \mathrm{CP1}~\subset~\Z2~\subset~\left\{\begin{array}{cc} \U1 \\ \mathrm{CP2} \end{array}\right\}~\subset~\mathrm{CP3}~\subset~\SU2\;.
\end{equation}

Having a $\Z2$ symmetry corresponds, in a specific basis, to invariance under the transformation $\phi_1\to\phi_1$ and $\phi_2\to-\phi_2$. The model with a softly-broken $\Z2$ symmetry and unremovable complex phases in the scalar potential is known as the complex 2HDM (\gls{C2HDM})~\cite{Ginzburg:2004vp,Davidson:2005cw,Boto:2020wyf}. Soft breaking corresponds to allowing the presence of a quadratic term that violates the symmetry, allowing $m_{12}^2\neq 0$ while $\lambda_6=\lambda_7=0$ in \eq{pot:2hdm}, without affecting the renormalizability of the theory\footnote{Cf. eg. Ref.~\cite{Cheng:1984vwu}, Chapter 6, on the topic of renormalization of theories with spontaneous symmetry breaking.}. 

If $\lambda_5=0$ in a basis where $m_{12}^2=\lambda_6=\lambda_7=0$, then the Higgs potential respects a continuous Peccei-Quinn global $\U1$ symmetry, generated by $\phi_1\to e^{-i\zeta}\phi_1$ and $\phi_2\to e^{i\zeta}\phi_2$ for real values of $\zeta$.

\section{The real 3HDM scalar potential} \label{sec:3hdm_pot}

\subsection{Realizable symmetries}

We now consider a scalar sector with three Higgs doublet scalar fields and define the potential as,
\begin{equation}\label{3hdmpotential}
    V=V_{2}+V_{4}\, .
\end{equation}
The quadratic part of the potential is
\be
V_2 = m_{11}^2\phi_1^\dagger\phi_1 + m_{22}^2\phi_2^\dagger\phi_2 + m_{33}^2\phi_3^\dagger\phi_3
- \left[m_{12}^2(\phi_1^\dagger\phi_2)
 +m_{13}^2(\phi_1^\dagger\phi_3)
+m_{23}^2(\phi_2^\dagger\phi_3)+{\rm h.c.}\right]
\, ,
\label{3hdmquadratic}
\ee
where we also include terms, $m_{12}^2$, $m_{13}^2$ and $m_{23}^2$,
that break the symmetry softly.
When we impose a symmetry group $G$ on \eq{3hdmpotential},
we restrict its coefficients in specific relations. It might result in a potential that becomes symmetric under a larger symmetry group $\tilde{G}$ containing $G$. To avoid vacua with a symmetry group other than a subgroup of $G$~\cite{deAdelhartToorop:2010jxh}, or cases where $G$ is a finite symmetry while $\tilde{G}$ is continuous~\cite{Ferreira:2008zy}, we follow Ref.~\cite{Ivanov:2011ae,Ivanov:2012fp} in only considering groups $G$ that lead to a potential not symmetric under any larger symmetry group containing $G$. Considering non-trivial Higgs-basis transformations that preserve the kinetic terms, Ref.~\cite{Ferreira:2008zy,Ivanov:2011ae} give a method of identifying and classifying the realizable abelian groups in NHDM. For $N=3$, these groups are:
\beq
\Z2,\quad\Z3,\quad\Z4,\quad\Z2\times\Z2,\quad \U1,\quad\U1\times\Z2,\quad\U1\times\U1\, .
\eeq
For all these symmetries, the scalar potentials
have a piece invariant under rephasings; that is, invariant under the maximal abelian group 
$\U1 \times \U1$:~\footnote{Invariance under hypercharge guarantees that requiring
invariance under rephasings of two scalar fields implies automatically invariance under
rephasing of the third field.}
\begin{align}
V_{4, \textrm{RI}}=&
\lambda_1(\phi_1^\dagger\phi_1)^2
+\lambda_2(\phi_2^\dagger\phi_2)^2
+\lambda_3(\phi_3^\dagger\phi_3)^2+
\lambda_4(\phi_1^\dagger\phi_1)(\phi_2^\dagger\phi_2)
+\lambda_5(\phi_1^\dagger\phi_1)(\phi_3^\dagger\phi_3)\nonumber\\[8pt]
& 
+\lambda_6(\phi_2^\dagger\phi_2)(\phi_3^\dagger\phi_3)
+\lambda_7(\phi_1^\dagger\phi_2)(\phi_2^\dagger\phi_1)
+\lambda_8(\phi_1^\dagger\phi_3)(\phi_3^\dagger\phi_1)
+\lambda_9(\phi_2^\dagger\phi_3)(\phi_3^\dagger\phi_2)\, .
\label{U1U1quartic}
\end{align}
The rephasing invariant quartic couplings can be written alternatively as
\begin{equation}
V_{4, \textrm{RI}}=
V_N + V_{CB}\, ,
\label{U1U1quartic_2}
\end{equation}
where
\begin{align}
  V_N=&\frac{\lambda_{11}}{2}(\phi_1^\dagger\phi_1)^2 +
\frac{\lambda_{22}}{2}(\phi_2^\dagger\phi_2)^2+
\frac{\lambda_{33}}{2}(\phi_3^\dagger\phi_3)^2+
\lambda_{12}(\phi_1^\dagger\phi_1) (\phi_2^\dagger\phi_2)+
\lambda_{13}(\phi_1^\dagger\phi_1)
(\phi_3^\dagger\phi_3)
\nonumber\\[4pt]
&+
\lambda_{23}(\phi_2^\dagger\phi_2) (\phi_3^\dagger\phi_3)\, ,
\label{eq:VN}\\[4pt]
V_{CB}=& \lambda'_{12} z_{12}+\lambda'_{13} z_{13}+
\lambda'_{23} z_{23}\, ,
\label{eq:VCB}
\end{align}
and~\cite{Faro:2019vcd}
\begin{equation}
  \label{eq:25}
  z_{ij}= (\phi_i^\dagger\phi_i) (\phi_j^\dagger\phi_j)
  - (\phi_i^\dagger\phi_j) (\phi_j^\dagger\phi_i) \quad \text{(no sum)}\, .
\end{equation}
Notice that we have always
\begin{equation}
  \label{eq:6}
  0\le z_{ij} \le r_i r_j\, ,
\end{equation}
where $r_k = |\phi_k|^2\ (k=1,2,3)$, which will be key for the boundedness frow below in the following Section. With these conventions the relation between the two notations is
\begin{subequations}
  \label{eq:1}
\begin{eqnarray}
 & \lambda_{11}\to 2 \lambda_1,\
  \lambda_{22}\to 2 \lambda_2,\
  \lambda_{33}\to 2 \lambda_3,\
  \lambda_{12}\to \lambda_4+\lambda_7,\
  \lambda_{13}\to \lambda_5+\lambda_8,\\[+2mm]
 & \lambda_{23}\to \lambda_6+\lambda_9,\ 
  \lambda'_{12}\to -\lambda_7,\
  \lambda'_{13}\to -\lambda_8,\
  \lambda'_{23}\to -\lambda_9 .
\end{eqnarray}
\end{subequations}
Given a potential invariant under a group $G$,
its quartic part may be written as
\begin{equation}
V_4 = V_{4, \textrm{RI}} + V_G\, ,
\end{equation}
where $V_{4, \textrm{RI}}$ is the rephasing invariant piece
of \eqs{U1U1quartic}{U1U1quartic_2}, common to all potentials,
while $V_G$ is the rephasing non-invariant part of the quartic potential
that depends on the
group. The groups $G = \U1\times \U1$ (for which, obviously $V_G=0$),
$\U1\times \Z2$, $\Z2\times \Z2$, and $\Z3$,
are discussed in detail in the corresponding Sections below.

\begin{figure}[htbp!]
	\centering
	\includegraphics[width=60mm]{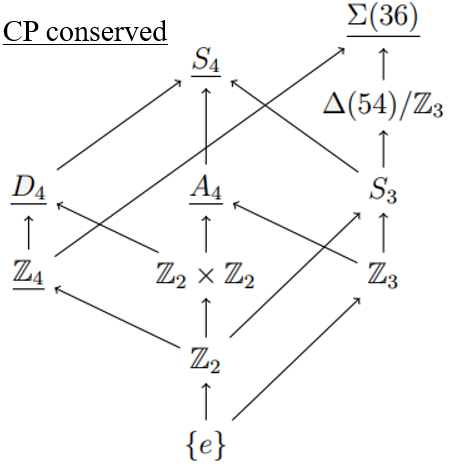}
	\caption{Tree of finite realizable groups of Higgs-family transformations in 3HDM, taken from Ref.~\cite{Ivanov:2014doa}. Groups
leading to automatic explicit CP-conservation are underlined. An arrow from A to B indicates
that $A~\subset~B$.}
	\label{f:symmetrybreaking}
\end{figure}

 The classification of symmetries in the 3HDM also includes non-abelian symmetries. The full list of all discrete non-abelian symmetry groups in the 3HDM was derived in Refs.~\cite{Ivanov:2012ry,Ivanov:2012fp}. Continuous non-abelian groups were later included in Ref.~\cite{deMedeirosVarzielas:2019rrp,Kuncinas:2024zjq} and generalized CP symmetries in Ref.~\cite{Bree:2024edl}. In Figure~\ref{f:symmetrybreaking}, derived in Ref.~\cite{Ivanov:2014doa}, the authors provide a scheme of the finite groups of HF in the 3HDM, similar to \eq{SchemeSymmetries} for the 2HDM, where the groups not underlined may be either $CP$-conserving or $CP$-violating. In this thesis, we will report the first studies of the $\Z2\times\Z2$ scalar potential with complex parameters. We will also consider non-abelian symmetries as the stabilizing symmetry for dark matter models. For Chapters~\ref{chapter:Constraints},~\ref{chapter:Alignment} and the remainder of the present chapter, we focus on abelian symmetries with $CP$ conservation.

\subsection{The physical basis}\label{sec:3hdmphys}

After spontaneous symmetry breaking (SSB), the three doublets can be
parameterized in terms of its component fields as: 
 \begin{equation}
     \phi_i=\begin{pmatrix} w_k^\dagger \\ (v_i+x_i\,+\,i\,z_i)/\sqrt{2}\end{pmatrix} \,\,\qquad (i=1,2,3)\label{fielddefinitions}\, ,
 \end{equation}
where $v_i/\sqrt{2}$ corresponds to the vacuum expectation value
(vev) for the neutral component of $\phi_i$.
It is assumed that the
scalar sector of the
model explicitly and spontaneously conserves
CP.\footnote{Strictly speaking,
it is not advisable to assume a real scalar sector while allowing the
Yukawa couplings to carry the phase necessary for the
CKM matrix.
This is also a problem with the so-called real 2HDM
~\cite{Fontes:2021znm,deLima:2024lfc}.
One can take the view that the complex terms and their counterterms
in the scalar sector exist, with the former set to zero.
}
That is, all the parameters in the scalar potential are real and
the vevs $v_1$, $v_2$, and $v_3$ are also real.
With this assumption, the scalar potential 
contains at most, eighteen parameters. The vevs can be parameterized as follows:
\begin{equation}\label{3hdmvevs}
     v_1=v \cos \beta_1 \cos \beta_2\,,\qquad v_2=v \sin \beta_1 \cos \beta_2\, ,\qquad v_3=v \sin \beta_2,
\end{equation}
with the relation,
\beq \label{v246_c3hdm}
v\equiv (v_1^2+v_2^2+v_3^2)^{1/2}=\frac{2 m_W}{g}=(\sqrt{2}G_F)^{-1/2}\simeq 246.219~{\rm GeV}\,;
\eeq 
and define the Higgs basis ~\cite{Georgi:1978ri,Donoghue:1978cj,Botella:1994cs}
as obtained by the following rotation,
\begin{equation}\label{higgsbasisZ3}
     \begin{pmatrix} H_0 \\ R_1 \\ R_2 \end{pmatrix}
     =
     \mathcal{O}_\beta
     \begin{pmatrix} x_1 \\ x_2 \\ x_3 \end{pmatrix}
     =
     \begin{pmatrix} \cos\beta_2 \cos\beta_1 & \cos\beta_2 \sin\beta_1 & \sin\beta_2 \\ -\sin\beta_1 & \cos\beta_1 & 0 \\ -\cos\beta_1 \sin\beta_2 & -\sin\beta_1 \sin\beta_2 & \cos\beta_2\end{pmatrix}
     \begin{pmatrix} x_1 \\ x_2 \\ x_3 \end{pmatrix} .
\end{equation}
The scalar kinetic Lagrangian is written as
\begin{equation}\label{kinetic3hdm}
    \mathscr{L}_{\text{kin}}=\sum_{k=1}^{n=3}|D_\mu\phi_k|^2 ,
\end{equation}
and contains the terms relevant to the propagators and trilinear couplings of the scalars and gauge bosons.

 We can now define orthogonal matrices which diagonalize the
 squared-mass matrices present in the CP-even scalar, CP-odd scalar
 and charged scalar sectors. These are the transformations that take
 us to the physical basis, with states possessing well-defined
 masses. Following Ref.~\cite{Das:2019yad,Boto:2021qgu,Boto:2022uwv}, the twelve
 quartic couplings for both the $\Z2\times\Z2$ and $\Z3$ can be exchanged for seven
 physical masses (three  
CP-even scalars, two CP-odd scalars and two pairs of charged scalars)
and five mixing angles. For the case of $U(1)\times\Z2$, we have only
10 $\lambda$'s and therefore we can also solve for two of the soft
masses.
Finally, in the case of $\U1\times \U1$ one has only 9
$\lambda$'s, and one can also solve for all the soft masses.
We give all the explicit expressions in Appendix~\ref{app:lambdas}.
The mass terms in the neutral scalar sector
can be extracted through the following rotation,  
\begin{equation}\label{CPevenDiag}
     \begin{pmatrix} h_1 \\ h_2 \\ h_3 \end{pmatrix}
=\mathcal{O}_\alpha \begin{pmatrix} x_1 \\ x_2 \\ x_3 \end{pmatrix},
\end{equation}
where we take $h_1 \equiv h_{125}$ to be the $125~{\rm GeV}$ Higgs particle
found at LHC.
The form chosen for $\mathcal{O}_\alpha$ is
\begin{equation}\label{matrixR}
\textbf{R}\equiv\mathcal{O}_\alpha=\mathcal{R}_3.\mathcal{R}_2.\mathcal{R}_1 ,
\end{equation}
where
\begin{equation}
\mathcal{R}_1=\begin{pmatrix}\cai & \sai & 0\\ -\sai & \cai & 0 \\ 0 & 0 & 1  \end{pmatrix}\,,\quad     \mathcal{R}_2=\begin{pmatrix}\caii & 0 & \saii \\ 0 & 1 & 0 \\ -\saii & 0 & \caii  \end{pmatrix}\,,\quad     \mathcal{R}_3=\begin{pmatrix}1 & 0 & 0\\ 0 & \caiii & \saiii \\ 0 & -\saiii & \caiii  \end{pmatrix}\,.\quad
\end{equation}

For the CP-odd scalar sector, the physical basis is
chosen as $\begin{pmatrix}G^0 & A_1 & A_2\end{pmatrix}^T$
and the transformation to be
\begin{equation}\label{CPoddDiag}
\begin{pmatrix} G^0 \\ A_1 \\ A_2 \end{pmatrix}
=
\mathcal{O}_{\gamma_1} \mathcal{O}_\beta
\begin{pmatrix} z_1 \\ z_2 \\ z_3 \end{pmatrix} ,
\end{equation}
where
\begin{equation}
\mathcal{O}_{\gamma_1}
=
\begin{pmatrix}1  & 0 & 0\\ 0& \cgi & -\sgi \\ 0 & \sgi & \cgi \end{pmatrix}
\label{ogamma1}
\end{equation}
is defined in order to diagonalize the 2x2 submatrix
that remains non-diagonal in the Higgs basis.
For later use, we define the matrix $\textbf{P}$
as the combination
\begin{equation}\label{matrixP}
    \textbf{P}\equiv\mathcal{O}_{\gamma_1} \mathcal{O}_\beta.
\end{equation}
For the charged scalar sector, the physical basis is 
$\begin{pmatrix}G^+ & H_1^+ & H_2^+\end{pmatrix}^T$
and the transformation is
\begin{equation}\label{ChargedDiag}
\begin{pmatrix} G^+ \\ H_1^+ \\ H_2^+ \end{pmatrix}
=\mathcal{O}_{\gamma_2} \mathcal{O}_\beta
\begin{pmatrix} w_1^\dagger \\ w_2^\dagger \\ w_3^\dagger \end{pmatrix},
\end{equation}
where
\begin{equation}
\mathcal{O}_{\gamma_2}= \begin{pmatrix}1  & 0 & 0\\ 0& \cgii & -\sgii \\ 0 & \sgii & \cgii \end{pmatrix} .
\label{ogamma2} 
\end{equation}
We write the masses of $H_1^+$ and $H_2^+$
as $m_{H_1^\pm}$ and $m_{H_2^\pm}$, respectively.
The matrix $\textbf{Q}$ is defined as the combination 
\begin{equation}\label{matrixQ}
\textbf{Q}\equiv\mathcal{O}_{\gamma_2} \mathcal{O}_\beta .
\end{equation}
The matrix $\textbf{Q}$ is relevant for the calculation
of the oblique parameters, shown in Section~\ref{sec:stu},
following the analysis of~\cite{Grimus:2007if}.

Considering that the states in the physical basis have well-defined masses,
we can obtain relations between the set 
\begin{align}
&&\left\{v_1,v_2,v_3,m_{h_1},m_{h_2},m_{h_3},m_{A_1},m_{A_2},
m_{H_1^\pm},m_{H_2^\pm},\alpha_1,\alpha_2,\alpha_3,
\gamma_1,\gamma_2,m^2_{12},m^2_{13},m^2_{23}\right\}
,\label{setphysical}\\[8pt] 
&&\quad 
v_1=v \cos\beta_1 \cos\beta_2\,,\quad\,v_2=v \sin\beta_1 \cos\beta_2\,,\quad\,v_3=v \sin\beta_2 ,
\end{align}
and the parameters of the potential~\footnote{As mentioned above,
for the $U(1)\times U(1)$ and $U(1)\times\Z2$ cases,
since we have less parameters, some or all of
the soft mass squared terms can also be solved for,
as shown explicitly in Appendix~\ref{app:lambdas}.}
as shown in Appendix~\ref{app:lambdas}.

\section{Yukawa Lagrangian and the Type-Z possibility}\label{sec:yukawa}

The most general quark Yukawa Lagrangian of the 3HDM may be written as
\be
\mathcal{L}_\mathrm{Y} =
- \bar q_L \left[
\left( \Gamma_1 \phi_1 + \Gamma_2 \phi_2
+ \Gamma_3 \phi_3\right) n_R
+
\left( \Delta_1 \tilde \phi_1 + \Delta_2 \tilde \phi_2
+ \Delta_3 \tilde \phi_3 \right) p_R
\right]
+ {\rm h.c.},
\label{yuk-q}
\ee
where $\tilde \phi_k \equiv i \tau_2 \phi_k^\ast$,
while $q_L$, $n_R$, and $p_R$
are vectors~\footnote{These vectors are written in a weak basis;
not in the mass basis. For massless neutrinos,
we can take the leptons already in the mass basis.}
in the respective three-dimensional flavour
vector space of left-handed quark doublets,
right-handed down-type quarks,
and right-handed up-type quarks.

Ignoring neutrino masses, the leptonic
Yukawa Lagrangian of the 3HDM may be similarly written as
\be
\mathcal{L}_\mathrm{Y} =
- \bar L_L \left[
\left( \Pi_1 \phi_1 + \Pi_2 \phi_2
+ \Pi_3 \phi_3\right) \ell_R
\right]
+ {\rm h.c.},
\label{yuk-lep}
\ee
where $L_L$ and $\ell_R$
are vectors in the respective three-dimensional flavour
vector space of left-handed leptonic doublets
and right-handed charged leptons.
The $3 \times 3$ matrices
$\Gamma_k$,
$\Delta_k$,
and $\Pi_k$
contain the complex Yukawa couplings
to the right-handed down-type quarks, up-type quarks,
and charged leptons,
respectively.

As is well known,
unless protected by a symmetry,
the Higgs-fermion Yukawa couplings
lead to Higgs-mediated flavor-changing neutral couplings
at a level incompatible with experimental observations.
We choose to remove FCNC by making 
the Yukawa coupling matrices to fermions of a given electric charge
proportional:
\be
\Gamma_1 \propto \Gamma_2 \propto \Gamma_3\, ,
\ \ \ 
\Delta_1 \propto \Delta_2 \propto \Delta_3\, ,
\ \ \ 
\Pi_1 \propto \Pi_2 \propto \Pi_3\, .
\label{propto}
\ee
It has been shown that,
in a general NHDM,
\eq{propto}
remain true (thus removing FCNCs) under the renormalization
group running if and only if there is a basis for the Higgs doublets
in which all the fermions of a given electric charge couple to
only one Higgs doublet~\cite{Ferreira:2010xe}.
This can be imposed in the 2HDM through
a $\Z2$ symmetry~\cite{Glashow:1976nt,Paschos:1976ay},
leading to four Types of
models.
For $N\geq 3$ there are five possible choices~\cite{Ferreira:2010xe},
which~\cite{Yagyu:2016whx} dubbed
Types I, II, X, Y, and Z,
as 
\begin{eqnarray}
\textrm{Type-I:}
&&
\phi_u=\phi_d=\phi_e\, ,
\nonumber\\
\textrm{Type-II:}
&&
\phi_u \neq \phi_d=\phi_e\, ,
\nonumber\\
\textrm{Type-X:}
&&
\phi_u=\phi_d \neq \phi_e\, ,
\nonumber\\
\textrm{Type-Y:}
&&
\phi_u=\phi_e \neq \phi_d\, ,
\nonumber\\
\textrm{Type-Z:}
&&
\phi_u\neq \phi_d;\ \phi_d\neq \phi_e,\ \phi_e\neq \phi_u\, ,
\label{Types}
\end{eqnarray}
with $\phi_{u,d,e}$ being the single scalar fields that couple
exclusively to the up-type quarks, down-type quarks, and charged
leptons, respectively.  

We wish to see how these choices can be implemented, starting with the $U(1)\times U'(1)$ symmetric 3HDM.
Without loss of generality,
we can choose $\phi_u=\phi_3$,
with the scalar fields transforming under
$U(1)$ and $U'(1)$, respectively, as
\begin{align}
U(1) &:   &   \phi_1 &\rightarrow e^{i \theta}\phi_1
&   \phi_2 &\rightarrow \phi_2
&   \phi_3 &\rightarrow \phi_3\, ,
\label{U(1)_scalar-2}
\\
U'(1) &:   &   \phi_1 &\rightarrow \phi_1
&   \phi_2 &\rightarrow e^{i \theta'}\phi_2
&   \phi_3 &\rightarrow \phi_3\, .
\label{U'(1)_scalar-2}
\end{align}
We choose three fields to remain invariant under the two groups:
\be
U(1)\ \textrm{and} \ U'(1):
q_L \rightarrow q_L\, ,
\ \ 
p_R \rightarrow p_R\, ,
\ \
L_L \rightarrow L_L\, ,
\label{U(1)ANDU'(1)}
\ee
under both $U(1)$ and $U'(1)$.
Eqs.~\eqref{U(1)_scalar-2}, \eqref{U'(1)_scalar-2}, and
\eqref{U(1)ANDU'(1)} ensure that $\phi_3=\phi_u$.
The various Types can now be implemented by
choosing the other fields to transform as in
Table~\ref{tab:Types}.
\begin{table}[htb]
   \centering
   \begin{tabular}{|c||c|c|c|c|c|c|}\hline
     &$\phi_1$&$\phi_2$&$\phi_3$&$n_R$&$\ell_R$&$\phi_u$
     $\phi_d$ $\phi_{\ell}$\\\hline\hline
    Type-I &$(e^{\, i \theta},\hspace{7mm})$ &$(\hspace{7mm},e^{\, i \theta'})$ &$(\hspace{7mm},\hspace{7mm})$
    &$(\hspace{7mm},\hspace{7mm})$ &$(\hspace{7mm},\hspace{7mm})$
     &$\phi_3$ $\phi_3$ $\phi_3$\\*[1mm]
     Type-II &$(e^{\, i \theta},\hspace{7mm})$ &$(\hspace{7mm},e^{\, i \theta'})$ &$(\hspace{7mm},\hspace{7mm})$
     &$(\hspace{7mm},e^{-i \theta'})$ & $(\hspace{7mm},e^{-i \theta'})$
     & $\phi_3$ $\phi_2$ $\phi_2$\\*[1mm]
     Type-X &$(e^{\, i \theta},\hspace{7mm})$ &$(\hspace{7mm},e^{\, i \theta'})$ & $(\hspace{7mm},\hspace{7mm})$
     &$(\hspace{7mm},\hspace{7mm})$ & $(\hspace{7mm},e^{-i \theta'})$
     & $\phi_3$ $\phi_3$ $\phi_2$\\*[1mm]
     Type-Y & $(e^{\, i \theta},\hspace{7mm})$&$(\hspace{7mm},e^{\, i \theta'})$ &$(\hspace{7mm},\hspace{7mm})$
     &$(\hspace{7mm},e^{-i \theta'})$ &$(\hspace{7mm},\hspace{7mm})$
     & $\phi_3$ $\phi_2$ $\phi_3$\\*[1mm]
     Type-Z &$(e^{\, i \theta},\hspace{7mm})$ &$(\hspace{7mm},e^{\, i \theta'})$ & $(\hspace{7mm},\hspace{7mm})$
     &$(\hspace{7mm},e^{-i \theta'})$ &$(e^{-i \theta},\hspace{7mm})$
     & $\phi_3$ $\phi_2$ $\phi_1$\\\hline
   \end{tabular}
   \caption{All possible models with natural flavour conservation.
   The transformation
     properties under $U(1)\times U'(1)$ are indicated by
     $(\ ,\ )$. For instance $(\ \ ,e^{i \theta'})$ indicates that the field is
     invariant under the first $U(1)$ but transforms as
     $\psi \rightarrow e^{i \theta'} \psi$ under $U'(1)$.
     For $U(1)\times \mathbbm{Z}'_2$ do
     $e^{\pm i \theta'} \rightarrow -$,
     for $\Z2\times \mathbbm{Z}'_2$ do
     $e^{\pm i \theta}, e^{\pm i \theta'} \rightarrow -$, and for $\Z3$ set
     $\theta=-\theta'=\tfrac{2\pi}{3}$.
     }
     \label{tab:Types}
 \end{table}


The transformations of the fields under
$U(1)\times \mathbbm{Z}'_2$ are obtained from Table~\ref{tab:Types}
by changing $e^{\pm i \theta'} \rightarrow -$.
Similarly,
The transformations of the fields under
$\Z2\times \mathbbm{Z}'_2$ are obtained from Table~\ref{tab:Types}
by changing both $e^{\pm i \theta} \rightarrow -$ and
$e^{\pm i \theta'} \rightarrow -$. $\Z3$ is obtained by
     $\theta=-\theta'=\tfrac{2\pi}{3}$ such that $\phi_1\to e^{2\pi\, i/3}\phi_1$, $\phi_2\to e^{-2\pi\, i/3}\phi_2$ and the corresponding substitutions for the fermionic fields.

We continue this Section with the \textbf{Type-Z} model in detail.
The remaining Types are relegated to Appendix~\ref{app:Types}.
For this case we choose transformation properties in Table~\ref{tab:Types} such that $\phi_1$ only has
interaction terms with the charged leptons, giving them mass; $\phi_3$
and $\phi_2$ are responsible for masses of the up and down type
quarks, respectively.
When taking into account the restrictions imposed by the symmetry, the
Yukawa couplings to fermions can be written in a compact form. For the
couplings of neutral Higgs to fermions, 
\begin{equation}\label{eq:couplingNeutralFerm}
    \mathscr{L}_{\rm Y}\ni -\frac{m_f}{v}\bar{f}(a^f_j+i\, b^f_j\gamma_5)fh_j ,
\end{equation}
where we group the physical Higgs fields in a vector, as
$h_j\equiv(h_1,h_2,h_3,A_1,A_2)_j$. We have
\ba
a_j^f &\to&
\frac{\textbf{R}_{j,1}}{\hat{v_1}},
\qquad\qquad j=1,2,3\qquad \text{for all leptons} ,\nonumber\\[8pt]
b_j^f &\to&
\frac{\textbf{P}_{j-2,1}}{\hat{v_1}},
\qquad\quad j=4,5\quad\qquad \text{for all leptons} ,\nonumber\\[8pt]
a_j^f &\to&
\frac{\textbf{R}_{j,3}}{\hat{v_3}},
\qquad\qquad j=1,2,3\qquad \text{for all up quarks} ,\nonumber\\[8pt]
b_j^f &\to&
-\frac{\textbf{P}_{j-2,3}}{\hat{v_3}},
\quad\quad j=4,5\quad\qquad \text{for all up quarks} ,\nonumber\\[8pt]
a_j^f &\to&
\frac{\textbf{R}_{j,2}}{\hat{v_2}},
\qquad\qquad j=1,2,3\qquad \text{for all down quarks} ,\nonumber\\[8pt]
b_j^f &\to&
\frac{\textbf{P}_{j-2,2}}{\hat{v_2}},
\qquad\quad j=4,5\quad\qquad \text{for all down quarks} ,
\label{coeffNeutralFerm-Type-Z}
\ea
%
where we introduce $\hat{v_i}=v_i/v$, with the vevs in \eq{3hdmvevs}. The couplings of the charged Higgs, $H_1^\pm$ and $H_2^\pm$, to fermions can be expressed as
\begin{eqnarray}
\mathscr{L}_{\rm Y} &\ni& \frac{\sqrt{2}}{v}
\bar{\psi}_{d_i}
\left[m_{\psi_{d_i}} V_{ji}^\ast\, \eta_k^L P_L
+ m_{\psi_{u_j}} V_{ji}^\ast\, \eta_k^R P_R\right] \psi_{u_j} H_k^-
\nonumber\\
&&
+ \frac{\sqrt{2}}{v}\bar{\psi}_{u_i}
\left[m_{\psi_{d_j}} V_{ij}\, \eta_k^L P_R 
+ m_{\psi_{u_i}} V_{ij}\, \eta_k^R P_L \right] \psi_{d_j} H_k^+ ,
\label{eq:couplingChargedFerm}
\end{eqnarray}
where $(\psi_{u_i},\psi_{d_i})$ is $(u_i,d_i)$ for
quarks\footnote{Here, the up-type quarks $u$ and
down-type quarks $d$ are already written in the mass basis.}
or $(\nu_i,\ell_i)$ for leptons, and $P_{R,L}\equiv (1\pm \gamma_5)/2$ are the usual chiral projection operators.
For quarks, $V$ is the CKM matrix, while for leptons,
$V_{ij}=\delta_{ij}$ since we are considering massless neutrinos. The
couplings are,
\begin{equation}
\label{eq:coeffChargedFerm-Type-Z}  
    \eta_k^{l\,L}=-\frac{\textbf{Q}_{k+1,1}}{\hat{v_1}}\,,\quad\eta_k^{l\,R}=0\,,\quad\eta_k^{q\,L}=-\frac{\textbf{Q}_{k+1,2}}{\hat{v_2}}\,,\quad\eta_k^{q\,R}=\frac{\textbf{Q}_{k+1,3}}{\hat{v_3}}\,,\quad \text{k=1,2} \, .
\end{equation}

\section{Bounded from below conditions}\label{sec:bfb}

The issue of having
a potential bounded from below (BFB)  is often ignored. Occasionally, some BFB conditions are included
without stressing whether such conditions are necessary, sufficient or both. And, most
articles addressing this problem concentrate on BFB conditions analyzing only vacua along
the neutral directions; that is, vacua which do not break electric charge.

However,
Faro and Ivanov ~\cite{Faro:2019vcd} showed,
using the specific case of a $U(1) \times U(1)$ symmetric
three Higgs doublet model (3HDM),
that one can have a minimum of the potential which satisfies the condition for
bounded for below along charge preserving directions,
but is still unbounded from below along the charge breaking (CB) directions.
They then proceeded to establish necessary and sufficient conditions for
BFB along both neutral (BFB-n) and charge breaking (BFB-c) directions,
for the specific case of the $U(1) \times U(1)$ 3HDM.
Faro~\cite{Faro:2019} extended this analysis to the $U(1) \times \Z2$ symmetric 3HDM.
Surprisingly,
there are no known necessary and sufficient conditions for BFB
for such a simple and classical model as the $\Z2 \times \Z2$ 3HDM.
This model was first proposed by Weinberg in~\cite{Weinberg:1976hu},
in order to have CP violation in the scalar sector,
without exhibiting flavour changing neutral scalar couplings.
The best result has been derived in~\cite{Grzadkowski:2009bt},
which has the necessary and sufficient conditions for BFB-n and
sufficient conditions for BFB-c.

We show a method of deriving sufficient  BFB-n and BFB-c sufficient conditions. The method hinges on finding a potential which lies lower than the potential
desired, and for which one can apply the copositivity conditions
of Klimenko~\cite{Klimenko:1984qx} and Kannike~\cite{Kannike:2012pe}
in order to find BFB conditions for that new potential.

If one uses necessary (but not sufficient) conditions for BFB, one is basing
the analysis on some potentials which are unphysical. Conversely, if one
uses sufficient (but not necessary) conditions for BFB, one is excluding
perfectly good potentials, running the risk that these have some special
features, potentially ignoring interesting new physics signals or, as an extreme, exclude a perfectly viable model. In Appendix~\ref{a:comparison_bfb}, we address this concern by comparing our method with the complete necessary and sufficient BFB conditions for the $U(1)\times U(1)$ and $U(1)\times \Z2$ 3HDM, where those are avaliable.

\subsection{\texorpdfstring{BFB conditions in the $\U1\times \U1$ case}{BFB conditions in the U1xU1 case}}
\label{sec:BFB-U1xU1}

Let us consider the $\U1\times U'(1)$ transformation\footnote{When convenient to distinguish
the two $\U1$'s, we will denote the second one by a prime.}
\begin{align}
\U1 &:   &   \phi_1 &\rightarrow e^{i \theta}\phi_1
&   \phi_2 &\rightarrow \phi_2
&   \phi_3 &\rightarrow \phi_3\, ,
\label{U(1)_scalar}
\\
U'(1) &:   &   \phi_1 &\rightarrow \phi_1
&   \phi_2 &\rightarrow e^{i \theta'}\phi_2
&   \phi_3 &\rightarrow \phi_3\, .
\label{U'(1)_scalar}
\end{align}
where the transformations are to be implemented for all $\theta$ and $\theta'$.
This is the simplest case because the symmetry forces $V_G=0$, so
\begin{equation}
  \label{eq:15a}
  V_4=V_N + V_{CB}\, ,
\end{equation}
given in \eqs{eq:VN}{eq:VCB}.

\subsubsection{Necessary and sufficient conditions for BFB}
\label{subsec:U1U1_BFB_NS}

The necessary and
sufficient conditions for BFB of the potential
for this case were found by Faro and
Ivanov~\cite{Faro:2019vcd}. They can be enunciated in three steps.
For these,
we use gauge invariance to parameterize the (vevs of the) doublets
as~\cite{Faro:2019vcd},
\begin{equation}
  \label{eq:22}
  \phi_1=\sqrt{r_1}
  \begin{pmatrix}
    0\\
    1
  \end{pmatrix},\quad
  \phi_2=\sqrt{r_2}
  \begin{pmatrix}
    \sin(\alpha_2)\\
    \cos(\alpha_2) e^{i \beta_2}
  \end{pmatrix},\quad
  \phi_3=\sqrt{r_3} e^{i \gamma}
  \begin{pmatrix}
    \sin(\alpha_3)\\
    \cos(\alpha_3) e^{i \beta_3}
  \end{pmatrix}  .
\end{equation}
%


\textbf{\textit{Step 1}}---
The potential along the neutral directions, $V_N$, can be written as
\begin{equation}
  \label{eq:2}
  V_N= \frac{1}{2} \sum_{ij} r_i A_{ij} r_j,\quad\text{with}\quad
  A=
    \begin{pmatrix}
    \lambda_{11}&\lambda_{12}&\lambda_{13}\\
    \lambda_{12}&\lambda_{22}&\lambda_{23}\\
    \lambda_{13}&\lambda_{23}&\lambda_{33}
  \end{pmatrix}  .
\end{equation}
For the potential to be BFB, this quadratic form has to be positive
definite for $r_i\ge 0$. Then we should have the following relations
known as copositivity conditions~\cite{Klimenko:1984qx,Kannike:2012pe},
\begin{align}
  \label{eq:5}
  & A_{11} \ge 0, A_{22} \ge 0, A_{33} \ge 0\, ,\nonumber\\[+2mm]
  &\overline{A}_{12}=\sqrt{A_{11}A_{22}} + A_{12} \ge 0,\quad
  \overline{A}_{13}=\sqrt{A_{11}A_{33}} + A_{13} \ge 0,\quad
  \overline{A}_{23}=\sqrt{A_{22}A_{33}} + A_{23} \ge
  0\, ,\nonumber\\[+2mm]
  &\sqrt{A_{11}A_{22}A_{33}} + A_{12}\sqrt{A_{33}}
  + A_{13}\sqrt{A_{22}}+ A_{23}\sqrt{A_{11}}
  +\sqrt{2 \overline{A}_{12}\overline{A}_{13}\overline{A}_{23}} \ge 0 .
\end{align}
This ensures that $V_N$ is BFB. For $V_{CB}$ we need two extra steps.

\textbf{\textit{Step 2}}---
This step is only necessary if at least one of the $\lambda'_{ij}$ in
\eq{eq:VCB} is negative, otherwise because of \eq{eq:6},
the potential along the charge breaking directions,
$V_{CB}$,  is positive definite. If at least one of the
$\lambda'_{ij}$ is negative we construct the matrices
\begin{equation}
  \label{eq:7}
  \Delta_1=
  \begin{pmatrix}
    0&\lambda'_{12}&0\\
    \lambda'_{12}&0&\lambda'_{23}\\
    0&\lambda'_{23}&0
  \end{pmatrix},\quad
  \Delta_2=
  \begin{pmatrix}
    0&0&\lambda'_{13}\\
    0&0&\lambda'_{23}\\
    \lambda'_{13}&\lambda'_{23}&0
  \end{pmatrix},\quad
  \Delta_3=
  \begin{pmatrix}
    0&\lambda'_{12}&\lambda'_{13}\\
    \lambda'_{12}&0&0\\
    \lambda'_{13}&0&0
  \end{pmatrix}.
\end{equation}
Then form the matrices
\begin{equation}
  \label{eq:8a}
  A_i=A_N + \Delta_i
\end{equation}
where $A_N$ is obtained from $V_N$. Then check the copositivity of all
$A_i$. 

\textbf{\textit{Step 3}}---
If $\lambda'_{12} \lambda'_{13}\lambda'_{23} <0$, a final step is
needed. We form the matrix~\cite{Faro:2019vcd},
\begin{equation}
  \label{eq:9a}
  \Delta_4 =  \frac{1}{2}\, 
  \begin{pmatrix}\displaystyle
   \frac{\lambda'_{12}\lambda'_{13}}{\lambda'_{23}}&
    \displaystyle
    \lambda'_{12} &\displaystyle
    \lambda'_{13} \\
\displaystyle    \lambda'_{12}
& \displaystyle
 \frac{\lambda'_{12}\lambda'_{23}}{\lambda'_{13}}
&\displaystyle
\lambda'_{23}\\
\displaystyle   \lambda'_{13} & \displaystyle
\lambda'_{23}
& \displaystyle
\frac{\lambda'_{13}\lambda'_{23}}{\lambda'_{12}}
  \end{pmatrix}\, ,
\end{equation}
and construct the matrix
\begin{equation}
  \label{eq:10}
  A_4 = A_N + \Delta_4 .
\end{equation}
Now,
this matrix has to be copositive inside a tetrahedron in
the first octant and with one of
the vertices at the origin. To handle this, in Ref.~\cite{Faro:2019vcd}
the authors show that this is equivalent to finding the copositivity of
the matrix
\begin{equation}
  \label{eq:37}
  B=R^T A_4 R
\end{equation}
in the first octant, where
\begin{equation}
  \label{eq:38}
  R =
  \begin{pmatrix}
    |\lambda'_{23}| &0 &0\\
    0&|\lambda'_{13}|&0\\
    0&0&\lambda'_{12}|
  \end{pmatrix}
  \begin{pmatrix}
    0&1&1\\
    1&0&1\\
    1&1&0
  \end{pmatrix}\, .
\end{equation}

In summary, the copositivity of the matrices $A_N,A_1,A_2,A_3,B$ are
the necessary and sufficient conditions for the $\U1 \times \U1$
potential to be BFB.

\subsubsection{\texorpdfstring{The sufficient conditions of Ref.~\cite{Grzadkowski:2009bt}}{The sufficient conditions of Ref.~}}
\label{subsec:U1U1_BFB_Grz}

We now consider the conditions from Ref.~\cite{Grzadkowski:2009bt} that are known
to be sufficient but not necessary~\cite{Faro:2019vcd}. These were
derived for the case of $\Z2\times\Z2$ but our potential for $\U1\times \U1$ in
\eq{U1U1quartic} is a particular case with, in our notation
(see \eq{Z2Z2quartic} below),
\begin{equation}
  \label{eq:39}
  \lambda''_{10}=\lambda''_{11}=\lambda''_{12}=0\, .
\end{equation}
The conditions then read~\cite{Grzadkowski:2009bt},
\begin{subequations}
 \label{eq:u1u1_bfb}
	\begin{eqnarray}
  \bullet &\hskip 2mm
  \lambda_1 >0,\ \lambda_2 > 0,\ \lambda_3 >0,\\[+2mm]
  \bullet &\hskip 2mm
  \lambda_x > -2\sqrt{\lambda_1\lambda_2},
  \ \lambda_y > -2 \sqrt{\lambda_1\lambda_3},\
  \lambda_z >-2 \sqrt{\lambda_2\lambda_3},\\[+2mm]
  \bullet &\hskip 1mm
  \left\{
    \lambda_x \sqrt{\lambda_3} +\lambda_y \sqrt{\lambda_2}
    +\lambda_z \sqrt{\lambda_1} \ge 0
  \right\}
  \cup
  \left\{
  \lambda_1
  \lambda_z^2+\lambda_2\lambda_y^2+\lambda_3\lambda_x^2
  -4\lambda_1\lambda_2\lambda_3  
  - \lambda_x\lambda_y\lambda_z < 0  
  \right\}\, ,
	\end{eqnarray}
\end{subequations}
where
\begin{equation}
  \label{eq:13}
  \lambda_x=\lambda_4+\text{min}(0,\lambda_7),\ \ 
  \lambda_y=\lambda_5+\text{min}(0,\lambda_8),\ \ 
  \lambda_z=\lambda_6+\text{min}(0,\lambda_9)\, .
\end{equation}

\subsubsection{Sufficient conditions for a lower bound}
\label{subsec:U1U1_BFB_us}

In this case we know the necessary and sufficient conditions but in many
other symmetry constrained models we do not. So we can think of a
potential that it is always lower than $V_4$ and for which the
copositivity conditions can be easily applied. This is
will be important in the following. Because of \eq{eq:6},
we should have,
\begin{equation}
  \label{eq:11}
  V_{CB} \ge V_{CB}^{\rm lower}=
  r_1 r_2\ \text{min}(0,\lambda'_{12})
+ r_1 r_3\ \text{min}(0,\lambda'_{13})
+ r_2 r_3\ \text{min}(0,\lambda'_{23})
\end{equation}
and therefore
\begin{equation}
  \label{eq:4}
  V_4 \ge V_4^{\rm lower}= V_N +  V_{CB}^{\rm lower}\, .
\end{equation}
So we just have to check the copositivity of the matrix
\begin{equation}
  \label{eq:12}
  \begin{pmatrix}
    \lambda_{11}&\hat\lambda_{12}&
    \hat\lambda_{13}\\
    \hat\lambda_{12}&\lambda_{22}
    &\hat\lambda_{23}\\
    \hat\lambda_{13}
    &\hat\lambda_{23}
    &\lambda_{33}
  \end{pmatrix}\, ,
\end{equation}
where we have defined
\begin{equation}
  \label{eq:13a}
  \hat\lambda_{12}\equiv\lambda_{12}+\text{min}(0,\lambda'_{12}),\
  \hat\lambda_{13}\equiv\lambda_{13}+\text{min}(0,\lambda'_{13}),\
  \hat\lambda_{23}\equiv\lambda_{23}+\text{min}(0,\lambda'_{23}) .  
\end{equation}
These will ensure sufficient conditions for the potential to be BFB,
but they are not necessary. There will be good points in parameter
space that are discarded by this procedure. We will address this issue when we compare the respective sets of points in Appendix~\ref{a:comparison_bfb}.

\subsection{\texorpdfstring{BFB conditions in the $\U1\times\Z2$ case}{BFB conditions in the U1xZ2 case}}
\label{sec:BFB-U1xZ2}

The quartic part of our $\U1\times\Z2$ invariant potential reads,
\begin{align}
V_{\text{quartic}}=&
\lambda_1(\phi_1^\dagger\phi_1)^2
+\lambda_2(\phi_2^\dagger\phi_2)^2
+\lambda_3(\phi_3^\dagger\phi_3)^2+
\lambda_4(\phi_1^\dagger\phi_1)(\phi_2^\dagger\phi_2)
+\lambda_5(\phi_1^\dagger\phi_1)(\phi_3^\dagger\phi_3)\nonumber\\[8pt]
& 
+\lambda_6(\phi_2^\dagger\phi_2)(\phi_3^\dagger\phi_3)
+\lambda_7(\phi_1^\dagger\phi_2)(\phi_2^\dagger\phi_1)
+\lambda_8(\phi_1^\dagger\phi_3)(\phi_3^\dagger\phi_1)
+\lambda_9(\phi_2^\dagger\phi_3)(\phi_3^\dagger\phi_2)\nonumber\\[8pt]
&
+\left[\lambda''_{12}(\phi_2^\dagger\phi_3)^2
+
{\rm h.c.}\right]\, ,
\label{U1Z2quartic-our}
\end{align}
satisfying
\begin{subequations}
 \label{eq:u1z2_def}
	\begin{eqnarray}
  \U1 :&\ \phi_1 \to e^{i \theta} \phi_1,\quad \phi_2\to\phi_2,\quad
  \phi_3\to   \phi_3\, ,  \\
  \Z2 :&\ \phi_1 \to \phi_1,\quad\phi_2 \to - \phi_2,\quad
  \phi_3 \to \phi_3\, ,
	\end{eqnarray}
\end{subequations}
obtained from \eqref{U'(1)_scalar} by setting $\theta'=\pi$.
In \eqref{U1Z2quartic-our}, 
``h.c.'' stands for Hermitian conjugate.
Also, we use double primes, $\lambda''_{12}$, to distinguish from the definitions in
\eq{eq:1}. 

\subsubsection{The necessary and sufficient conditions for BFB}
\label{subsec:U1Z2_BFB_NS}

The conditions for this potential to be BFB were developed by Faro and
can be found in his Master thesis~\cite{Faro:2019}. In 
an adaptation of his notation\footnote{In Faro's implementation of
$\U1\times \Z2$, $\phi_3$ is the field getting a phase.
In our notation,
this role is played by $\phi_1$.
We get from his to ours with $1 \leftrightarrow 3$.},
the non-rephasing invariant part of the potential reads
\begin{equation}
  \label{eq:14a}
  V_{\U1\times\Z2}= \frac{1}{2}\left[\bar{\lambda}_{23}(\phi_2^\dagger\phi_3)^2
+ {\rm h.c.}\right]\, .
\end{equation}
Therefore, comparing with \eq{U1Z2quartic-our},
we get the relation
\begin{equation}
  \label{eq:18}
  \bar{\lambda}_{23} = 2 \lambda''_{12} .
\end{equation}
Now the BFB conditions are as in the $\U1\times \U1$  case doing the 3
steps mentioned there, with the substitutions
\begin{equation}
  \label{eq:21}
  \lambda_{23} \to \lambda_{23}- |\bar{\lambda}_{23}|,\quad
  \lambda'_{23} \to \lambda'_{23} + |\bar{\lambda}_{23}| .
\end{equation}

\subsubsection{\texorpdfstring{The sufficient conditions of Ref.~\cite{Grzadkowski:2009bt}}{The sufficient conditions of Ref.~}}

We now consider the sufficient conditions from Ref.~\cite{Grzadkowski:2009bt}.
They were
derived for the $\Z2\times\Z2$ case. Comparing our $\U1\times\Z2$
potential in \eq{U1Z2quartic-our} with the $\Z2\times\Z2$ case
in \eq{Z2Z2quartic} we require
\begin{equation}
  \label{eq:40}
  \lambda''_{10}=\lambda''_{11}=0
\end{equation}
The conditions from Ref.~\cite{Grzadkowski:2009bt} then read,
\begin{subequations}
 \label{eq:u1z2_bfb}
	\begin{eqnarray}
  \bullet &\hskip 2mm
  \lambda_1 >0,\ \lambda_2 > 0,\ \lambda_3 >0, \\[+2mm]
  \bullet &\hskip 2mm
  \lambda_x > -2\sqrt{\lambda_1\lambda_2},
  \ \lambda_y > -2 \sqrt{\lambda_1\lambda_3},\
  \lambda_z >-2 \sqrt{\lambda_2\lambda_3}, \\[+2mm]
  \bullet &\hskip 1mm
  \left\{
    \lambda_x \sqrt{\lambda_3} +\lambda_y \sqrt{\lambda_2}
    +\lambda_z \sqrt{\lambda_1} \ge 0
  \right\}
  \cup
  \left\{
  \lambda_1
  \lambda_z^2+\lambda_2\lambda_y^2+\lambda_3\lambda_x^2
  -4\lambda_1\lambda_2\lambda_3  
  - \lambda_x\lambda_y\lambda_z < 0  
  \right\}\, ,
	\end{eqnarray}
\end{subequations}
where
\begin{equation}
  \label{eq:13b}
  \lambda_x=\lambda_4+\text{min}(0,\lambda_7),\ \ 
  \lambda_y=\lambda_5+\text{min}(0,\lambda_8),\ \ 
  \lambda_z=\lambda_6+\text{min}(0,\lambda_9-2|\lambda''_{12}|)\, ,
\end{equation}
or, with the equivalence of \eq{eq:18},
\begin{equation}
  \label{eq:30}
   \lambda_x=\lambda_4+\text{min}(0,\lambda_7),\ \ 
  \lambda_y=\lambda_5+\text{min}(0,\lambda_8),\ \ 
  \lambda_z=\lambda_6+\text{min}(0,\lambda_9-|\bar{\lambda}_{23}|) \, .
\end{equation}

\subsubsection{Sufficient conditions for a lower bound}

Although in this case there are necessary and sufficient BFB conditions,
it is instructive to
find a lower potential like in the previous case. This will serve 
to compare the set of points regarding physical observables. For the
$V_{CB}$ part, the reasoning is the same as in \eq{eq:11}.

Now,
for the $V_{\U1\times \Z2}$ part, 
we note that 
\begin{equation}
(\phi_2^\dagger \phi_3)^2 + {\rm h.c.}
=
2\, \textrm{Re}\left\{ (\phi_2^\dagger \phi_3)^2 \right\}
\geq
- 2\, \left| (\phi_2^\dagger \phi_3)^2 \right|
\geq
- 2\, |\phi_2|^2 |\phi_3|^2 = - 2 r_2 r_3\, ,
\label{simpler}
\end{equation}
where we have used the parameterization \eqref{eq:22} on the last step.
A more complicated route would be to use
\eqref{eq:22} from the start, finding
\begin{equation}
  \label{eq:23}
  V_{\U1\times \Z2} = \bar{\lambda}_{23}\ r_2 r_3\,
f(\alpha_2,\alpha_3,\beta_2,\beta_3,\gamma)\, , 
\end{equation}
where we take  $\bar{\lambda}_{23}$ to be real but not necessarily
positive, and
\begin{align}
  \label{eq:35}
  f(\alpha_2,\alpha_3,\beta_2,\beta_3,\gamma)=&\cos ^2(\alpha_2) \cos
  ^2(\alpha_3) \cos\left[2 (\beta_2-\beta_3-\gamma)\right] +\sin ^2(\alpha_2) \sin ^2(\alpha_3) \cos (2
   \gamma)\nonumber\\[+2mm]
    &+\sin (\alpha_2) \cos
   (\alpha_2) \sin (2 \alpha_3) \cos (\beta_2-\beta_3-2 \gamma)\, .
\end{align}
Now, we can verify that we always have
\begin{equation}
  \label{eq:36}
  -1 \le f(\alpha_2,\alpha_3,\beta_2,\beta_3,\gamma) \le 1\, .
\end{equation}
Thus, using either route,
we have always
\begin{equation}
  \label{eq:24}
   V_{\U1\times \Z2}\ge  V_{\U1\times \Z2}^{\rm lower} = -
   |\bar{\lambda}_{23}| r_2 r_3 .
\end{equation}
Combining with \eq{eq:4} we get
\begin{equation}
  \label{eq:26}
  V_4 \ge V_N + V_{CB}^{\rm lower} + V_{\U1\times \Z2}^{\rm lower} .
\end{equation}
So we have just to look at the copositivity of the matrix
\begin{equation}
  \label{eq:12a}
  \begin{pmatrix}
    \lambda_{11}&\hat\lambda_{12}&
    \hat\lambda_{13}\\
    \hat\lambda_{12}&\lambda_{22}
    &\hat\lambda_{23}\\
    \hat\lambda_{13}
    &\hat\lambda_{23}
    &\lambda_{33}
  \end{pmatrix}\, ,
\end{equation}
where we have defined
\begin{equation}
  \label{eq:13c}
  \hat\lambda_{12}\equiv\lambda_{12}+\text{min}(0,\lambda'_{12})\, ,\
  \hat\lambda_{13}\equiv\lambda_{13}+\text{min}(0,\lambda'_{13})\, ,\
  \hat\lambda_{23}\equiv\lambda_{23}+\text{min}(0,\lambda'_{23})
  -|\bar{\lambda}_{23}|\, .  
\end{equation}
These will ensure sufficient conditions for the potential to be BFB,
but they are not necessary.

\subsection{\texorpdfstring{BFB conditions in the $\Z2\times\Z2$ case}{BFB conditions in the Z2xZ2 case}}
\label{sec:BFB-Z2xZ2}

The $\Z2\times \Z2$ symmetry is given by the transformation:
\begin{subequations}
 \label{eq:z2z2_def}
	\begin{eqnarray}
 \Z2 :&\ \phi_1 \to -\phi_1,\quad \phi_2\to\phi_2,\quad
  \phi_3\to  \phi_3\, ,  \\
  \Z2' :&\ \phi_1 \to \phi_1,\quad\phi_2 \to -\phi_2,\quad
  \phi_3 \to \phi_3\, ,
	\end{eqnarray}
\end{subequations}
which can be obtained
from \eqs{U(1)_scalar}{U'(1)_scalar}
by setting $\theta=\theta'=\pi$. The quartic part of the $\Z2\times\Z2$ invariant potential reads,
\begin{align}
V_{4}=&
\lambda_1(\phi_1^\dagger\phi_1)^2
+\lambda_2(\phi_2^\dagger\phi_2)^2
+\lambda_3(\phi_3^\dagger\phi_3)^2+
\lambda_4(\phi_1^\dagger\phi_1)(\phi_2^\dagger\phi_2)
+\lambda_5(\phi_1^\dagger\phi_1)(\phi_3^\dagger\phi_3)\nonumber\\[8pt]
& 
+\lambda_6(\phi_2^\dagger\phi_2)(\phi_3^\dagger\phi_3)
+\lambda_7(\phi_1^\dagger\phi_2)(\phi_2^\dagger\phi_1)
+\lambda_8(\phi_1^\dagger\phi_3)(\phi_3^\dagger\phi_1)
+\lambda_9(\phi_2^\dagger\phi_3)(\phi_3^\dagger\phi_2)\nonumber\\[8pt]
&
+\left[\lambda''_{10}(\phi_1^\dagger\phi_2)^2 +
  \lambda''_{11}(\phi_1^\dagger\phi_3)^2 +
  \lambda''_{12}(\phi_2^\dagger\phi_3)^2
  +
{\rm h.c.}\right].
\label{Z2Z2quartic}
\end{align}
The potential can be decomposed as
\begin{equation}
  \label{eq:27}
  V_4 = V_N + V_{CB} + V_{\Z2\times\Z2} ,
\end{equation}
where $V_N$ and $V_{CB}$ are given in \eq{eq:VN} and
\eq{eq:VCB}, respectively, and
\begin{align}
  \label{eq:z2z2part}
  V_{\Z2\times\Z2}=&\left[\lambda''_{10}(\phi_1^\dagger\phi_2)^2 +
  \lambda''_{11}(\phi_1^\dagger\phi_3)^2 +
  \lambda''_{12}(\phi_2^\dagger\phi_3)^2
  +
  {\rm h.c.}\right]\nonumber\\[+2mm]
=&\ \frac{1}{2}\left[\bar{\lambda}_{12}(\phi_1^\dagger\phi_2)^2 +
  \bar{\lambda}_{13}(\phi_1^\dagger\phi_3)^2 +
  \bar{\lambda}_{23}(\phi_2^\dagger\phi_3)^2
  +
  {\rm h.c.}\right]\, ,
\end{align}
where
\begin{equation}
  \label{eq:29}
  \bar{\lambda}_{12} =2 \lambda''_{10},\quad
  \bar{\lambda}_{13} =2 \lambda''_{11},\quad
  \bar{\lambda}_{23} =2 \lambda''_{12} .
\end{equation}

\subsubsection{\texorpdfstring{The sufficient conditions of Ref.~\cite{Grzadkowski:2009bt}}{The sufficient conditions of Ref.~}}
\label{sec:SuffCondsOgreid}

We now consider the sufficient conditions from Ref.~\cite{Grzadkowski:2009bt},
as implemented in Ref.~\cite{Hernandez-Sanchez:2020aop}. 
We have verified that there is a misprint in
Ref.~\cite{Hernandez-Sanchez:2020aop} when quoting
\eq{eq:19c3} below, taken here from
Ref.~\cite{Grzadkowski:2009bt} (where it is correct).
We find,
 \begin{subequations}
\label{eq:19c3}
	\begin{eqnarray}
  \bullet &\hskip 2mm
  \lambda_1 >0,\ \lambda_2 > 0,\ \lambda_3 >0, \\[+2mm]
  \bullet &\hskip 2mm
  \lambda_x > -2\sqrt{\lambda_1\lambda_2},
  \ \lambda_y > -2 \sqrt{\lambda_1\lambda_3},\
  \lambda_z >-2 \sqrt{\lambda_2\lambda_3}, \\[+2mm]
  \bullet &\hskip 1mm
  \left\{
    \lambda_x \sqrt{\lambda_3} +\lambda_y \sqrt{\lambda_2}
    +\lambda_z \sqrt{\lambda_1} \ge 0
  \right\}
  \cup
  \left\{
  \lambda_1
  \lambda_z^2+\lambda_2\lambda_y^2+\lambda_3\lambda_x^2
  -4\lambda_1\lambda_2\lambda_3  
  - \lambda_x\lambda_y\lambda_z < 0  
  \right\}\, ,
	\end{eqnarray}
\end{subequations}
where
 \begin{subequations}
\label{eq:13d}
	\begin{eqnarray}
  \lambda_x = \lambda_4+\text{min}(0,\lambda_7-2|\lambda''_{10}|)\, ,\lambda_y = \lambda_5+\text{min}(0,\lambda_8-2|\lambda''_{11}|)\, ,\lambda_z = \lambda_6+\text{min}(0,\lambda_9-2|\lambda''_{12}|)\, ,
	\end{eqnarray}
\end{subequations}
or, with the equivalence of \eq{eq:29},
 \begin{subequations}
  \label{eq:30a}
	\begin{eqnarray}
   \lambda_x = \lambda_4+\text{min}(0,\lambda_7-|\bar{\lambda}_{12}|)\, ,\lambda_y = \lambda_5+\text{min}(0,\lambda_8-|\bar{\lambda}_{13}|)\, ,\lambda_z = \lambda_6+\text{min}(0,\lambda_9-|\bar{\lambda}_{23}|)\, .
	\end{eqnarray}
\end{subequations}

\subsubsection{Sufficient conditions for a lower bound}
\label{sec:SufCondtsLowestBound}

In the $\Z2\times\Z2$, case there are no known necessary and sufficient BFB conditions.
One only has the sufficient conditions of Ref.~\cite{Grzadkowski:2009bt}
described in the previous Section.
Thus,
it is interesting to
find sufficient conditions from a lower potential like in the previous cases. For the
$V_{CB}$ part the reasoning is the same as in \eq{eq:11}. Now
for the $V_{\Z2\times \Z2}$ part,
we can either follow the steps in \eqref{simpler},
or use the parameterization
of \eq{eq:22} to get
\begin{align}
\label{eq:31}
  V_{\Z2\times \Z2} =& \bar{\lambda}_{12}\ r_1 r_2 \cos^2(\alpha_2)
  \cos(2 \beta_2) + \bar{\lambda}_{13}\ r_1 r_3 \cos^2(\alpha_3)
  \cos\left[2 (\beta_3+\gamma)\right] \nonumber\\[+2mm]
  &+
  \bar{\lambda}_{23}\ r_2 r_3\, f(\alpha_2,\alpha_3,\beta_2,\beta_3,\gamma)\, ,
\end{align}
where we take  $\bar{\lambda}_{ij}$ to be real but not necessarily
positive.
In either case,
we have always
\begin{equation}
\label{eq:32}
V_{\Z2\times \Z2}\ge  V_{\Z2\times \Z2}^{\rm lower} =
- |\bar{\lambda}_{12}| r_1 r_2
- |\bar{\lambda}_{13}| r_1 r_3
- |\bar{\lambda}_{23}| r_2 r_3\, .
\end{equation}
Combining with \eq{eq:4} we get
\begin{equation}
  \label{eq:26a}
  V_4 \ge V_N + V_{CB}^{\rm lower} + V_{\Z2\times \Z2}^{\rm lower} .
\end{equation}
So we have just to look at the copositivity of the matrix
\begin{equation}
\label{eq:33}
  \begin{pmatrix}
    \lambda_{11}&\hat\lambda_{12}&
    \hat\lambda_{13}\\
    \hat\lambda_{12}&\lambda_{22}
    &\hat\lambda_{23}\\
    \hat\lambda_{13}
    &\hat\lambda_{23}
    &\lambda_{33}
  \end{pmatrix}\, ,
\end{equation}
where we have now defined
 \begin{subequations}
\label{eq:34}
	\begin{eqnarray}
 & \hat\lambda_{12}\equiv\lambda_{12}+\text{min}(0,\lambda'_{12})
  -|\bar{\lambda}_{12}|,\\[+2mm] 
 & \hat\lambda_{13}\equiv\lambda_{13}+\text{min}(0,\lambda'_{13})
  -|\bar{\lambda}_{13}|, \\[+2mm] 
 & \hat\lambda_{23}\equiv\lambda_{23}+\text{min}(0,\lambda'_{23})
 -|\bar{\lambda}_{23}| .   
	\end{eqnarray}
\end{subequations}
These will ensure sufficient conditions for the potential to be BFB,
but they are not necessary. There will be good points in parameter
space that are discarded by this procedure.

\subsection{\texorpdfstring{Sufficient BFB conditions in the $\Z3$ case}{Sufficient BFB conditions in the Z3 case}}
 The $\Z3$ symmetry is realizable through the
following representation,
\begin{equation}
    S_{\Z3}=\text{diag}(1,e^{i\frac{2\pi}{3}}e^{-i\frac{2\pi}{3}}) ,
\end{equation}
such that 
\begin{equation}
\label{e:Z3sym}
\phi_1 \to e^{2\pi i/3} \, \phi_1 \,, \qquad  \phi_2 \to e^{-2\pi i/3} \phi_2 \,,\qquad \phi_3 \to \phi_3\, ;
\end{equation}
The terms quartic 
invariant under this transformation are given by 
\ba
V_{4}&=&
\lambda_1(\phi_1^\dagger\phi_1)^2
+\lambda_2(\phi_2^\dagger\phi_2)^2
+\lambda_3(\phi_3^\dagger\phi_3)^2+\lambda_4(\phi_1^\dagger\phi_1)(\phi_2^\dagger\phi_2)+\lambda_5(\phi_1^\dagger\phi_1)(\phi_3^\dagger\phi_3)\nonumber\\[8pt]
&&\quad 
+\lambda_6(\phi_2^\dagger\phi_2)(\phi_3^\dagger\phi_3)+\lambda_7(\phi_1^\dagger\phi_2)(\phi_2^\dagger\phi_1)+\lambda_8(\phi_1^\dagger\phi_3)(\phi_3^\dagger\phi_1)
+\lambda_9(\phi_2^\dagger\phi_3)(\phi_3^\dagger\phi_2)\nonumber\\[8pt]
&&\quad 
+\left[\lambda_{10}(\phi_1^\dagger\phi_2)(\phi_1^\dagger\phi_3)
+\lambda_{11}(\phi_1^\dagger\phi_2)(\phi_3^\dagger\phi_2)
+\lambda_{12}(\phi_1^\dagger\phi_3)(\phi_2^\dagger\phi_3)+
{\rm h.c.}\right],
\label{Z3quartic}
\ea
which can be written as
\begin{equation}\label{VquarticSep}
    V_{4}= V_N + V_{CB} + V_{\Z3} ,
\end{equation}
where $V_N$ and $V_{CB}$ are given in \eq{eq:VN} and
\eq{eq:VCB}, respectively, with terms in $\lambda_{1\to 9}$ and $V_{\Z3}$ corresponds to the terms $\lambda_{10\to12}$. The differences between the symmetries in \eqs{eq:z2z2_def}{e:Z3sym} are captured by the following
quartic terms in the scalar potential:
\begin{subequations}
\label{e:VZ2Z3}
\begin{eqnarray}
	V_{\Z3} &=& V_C + \left[\lambda_{10}(\phi_1^\dagger\phi_2)(\phi_1^\dagger\phi_3)
	+\lambda_{11}(\phi_1^\dagger\phi_2)(\phi_3^\dagger\phi_2) +\lambda_{12}(\phi_1^\dagger\phi_3)(\phi_2^\dagger\phi_3)+
	{\rm h.c.}\right] \,,
	\label{e:VZ3} \\
	V_{\Z2\times\Z2} &=& V_C  +\left[\lambda'_{10}(\phi_1^\dagger\phi_2)^2 +
	\lambda'_{11}(\phi_1^\dagger\phi_3)^2 +	\lambda'_{12}(\phi_2^\dagger\phi_3)^2 +	{\rm h.c.}\right]\,.
	\label{e:VZ2}
\end{eqnarray}
\end{subequations}
The $V_{CB}$ part is considered similarly to \eq{eq:11}. The $V_{\Z3}$ follows the previous procedure that led to the matrix in \eq{eq:33}. We consider the copositivity of the matrix with the following change,
 \begin{subequations}
\label{eq:Z3bfbsuff}
	\begin{eqnarray}
 & \hat\lambda_{12}\equiv\lambda_{12}+\text{min}(0,\lambda'_{12})
  -2|\lambda_{10}|-2|\lambda_{11}|,\\[+2mm]
 & \hat\lambda_{13}\equiv\lambda_{13}+\text{min}(0,\lambda'_{13})
  -2|\lambda_{10}|-2|\lambda_{12}|, \\[+2mm] 
 & \hat\lambda_{23}\equiv\lambda_{23}+\text{min}(0,\lambda'_{23})
 -2|\lambda_{10}|-2|\lambda_{12}| ,
	\end{eqnarray}
\end{subequations}
with $\lambda_{1-9}$ replaced by \eq{eq:1}. These will ensure sufficient conditions along both neutral and charged breaking directions. In Ref.~\cite{Boto:2021qgu} we derived only along neutral directions. Using the method shown here and published in Ref.~\cite{Boto:2022uwv}, charge breaking directions are included.

\section{Summary}

We dedicated this Chapter to multi Higgs-doublet models with a review on the physically distinct symmetries of the 2HDM and 3HDM scalar potential and the corresponding possibilities of natural flavour conservation in the Yukawa couplings. We followed with a study focused on bounded from below conditions for the 3HDM scalar potential. 
We developed a strategy to find sufficient BFB conditions.
It hinges on finding a potential which lies below the original potential
and to which the positivity conditions
\cite{Klimenko:1984qx,Kannike:2012pe} can be applied. This method is applicable to cases without necessary and sufficient conditions available, as shown. 
We studied in detail the $U(1) \times U(1)$, $U(1) \times \Z2$, $\Z2\times\Z2$ and $\Z3$ 3HDMs, for both BFB-n and BFB-c.
Then, we adapted the sufficient BFB $\Z2\times\Z2$ results of
Ref.~\cite{Grzadkowski:2009bt} to the $U(1)\times U(1)$
and $U(1)\times \Z2$ 3HDM, comparing our bounds in Appendix~\ref{a:comparison_bfb}. 


After analyzing hundreds of correlations in two-dimensional planes
of experimental observables, we find no evidence that points
allowed by the complete necessary and sufficient BFB conditions but
excluded by our sufficient BFB bounds would yield any new phenomenological
features.
We did this for both $U(1) \times U(1)$ and $U(1) \times \Z2$ with Type-I couplings in Appendix~\ref{a:comparison_bfb}.
Although not an airtight proof,
as is the case in any numerical simulation,
our results provide some reassurance that the sufficient BFB conditions
developed here do not significantly skew the phenomenology in cases
where no complete necessary and sufficient conditions are known,
such as the $\Z2 \times \Z2$ and $\Z3$ 3HMDs.

%% file: chapters/Constraints.tex

\chapter{Current constraints on the real 3HDM}
\label{chapter:Constraints}
\hspace*{0.3cm}
After deciding on a symmetry and establishing the physical parameters, the physical observables can be extracted and compared with experimental data. This Chapter is dedicated to the constraints that must be applied in order to perform a phenomenological study of a multi-Higgs doublet model. We start by completing the theoretical restrictions that still have to be considered after ensuring a bounded from below potential. We follow with all the relevant experimental results and highlight some of the details for the most important calculations, providing references to original works. In addition, preliminary measurements of a signal may be indicative of an enhancement compared to the SM expectation. In this Chapter, we develop a theoretical preparation for such an eventuality, with the particular case of loop-induced decay of the Higgs boson into a photon and Z boson, motivated by its first measurements.
 
\section{Theoretical Constraints}\label{ch:constraints_real}

\subsection{Perturbative Unitarity}
The theoretical bounds for all symmetry-constrained 3HDM arising from the perturbative unitarity of two-to-two scattering amplitudes are computed explicitly in Ref.~\cite{Bento_2022}.  For a given model, we must guarantee that the absolute value of the eigenvalues of the scattering matrices is smaller than $8\pi$.

\subsection{Perturbative Yukawa couplings}

As we want to explore the range of low $\tan\beta_1$ and $\tan\beta_2$
we should prevent the Yukawa couplings becoming non-perturbative. For
the Type-Z Yukawa structure, the top, bottom, and tau Yukawa couplings are given by
\begin{eqnarray}
  \label{eq:perturbativeyuk_req}
y_t = \frac{\sqrt{2}\, m_t }{v \sin\beta_2}\;,
\quad y_b = \frac{\sqrt{2}\, m_b }{v \sin\beta_1 \cos\beta_2} \;,
\quad y_\tau = \frac{\sqrt{2}\, m_\tau }{v \cos\beta_1 \cos\beta_2} \;,
\end{eqnarray}
which follow from our convention that $\phi_3$, $\phi_2$, and $\phi_1$ couple to up-type quarks, down-type quarks, and charged leptons, respectively. To maintain the perturbativity of Yukawa couplings, we impose $\lvert y_t \rvert,\lvert y_b\rvert,\lvert y_\tau\rvert < \sqrt{4\pi}$. 
Throughout our analysis, we used values of $\tan\beta_{1,2}$ which are consistent with this perturbative region. For other Types of Yukawa couplings, \eq{eq:perturbativeyuk_req} has to be adapted accordingly.

\section{Electroweak precision data}\label{sec:stu}

In order to discuss the effect of the $S,T,U$ parameters,
we use the results in~\cite{Grimus:2007if}.
To apply the relevant expressions,
 the matrices $U$ and $V$ used in~\cite{Grimus:2007if} need to be written for the notation choices that we have made when obtaining the mass
eigenstates in Section~\ref{sec:3hdmphys}.
The $3\times 6$ matrix $V$ is defined as
  \begin{equation}
      \begin{pmatrix}x_1+\,i\,z_1 \\ x_2+\,i\,z_2 \\ x_3+\,i\,z_3  \end{pmatrix} = V \begin{pmatrix}G^0 \\ h_1 \\ h_2 \\ h_3 \\ A_1 \\ A_2 \end{pmatrix} \,.
  \end{equation}
We find, by comparison with \eqs{CPevenDiag}{CPoddDiag}, that 
 \begin{equation}
     V= \begin{pmatrix}[1.5] i \textbf{P}^T_{11} & \textbf{R}^T_{11}& \textbf{R}^T_{12}&\textbf{R}^T_{13} &  i \textbf{P}^T_{12} &  i \textbf{P}^T_{13} \\ i \textbf{P}^T_{21} & \textbf{R}^T_{21}& \textbf{R}^T_{22}&\textbf{R}^T_{23} &  i \textbf{P}^T_{22} &  i \textbf{P}^T_{23}\\i \textbf{P}^T_{31} & \textbf{R}^T_{31}& \textbf{R}^T_{32}&\textbf{R}^T_{33} &  i \textbf{P}^T_{32} &  i \textbf{P}^T_{33}  \end{pmatrix} .
 \end{equation}
The $3\times 3$ matrix U for the charged scalar diagonalization is defined as
\begin{equation}
\begin{pmatrix} w_1^\dagger \\ w_2^\dagger \\ w_3^\dagger \end{pmatrix}
=U \begin{pmatrix} G^\dagger \\ H_1^+ \\ H_2^+  \end{pmatrix},
\label{chargedtransfU}
\end{equation}
giving us the correspondence $U=\textbf{Q}^T $ from \eq{ChargedDiag}.
  
Having applied the expressions for $S, T, U $,
the constraints implemented on $S$ and $T$ follow
Fig.~4 of Ref.~\cite{Baak:2014ora}, at $95\%$ confidence level.
This corresponds to only considering points that satisfy the condition
\begin{equation}
      a_1 S^2 + a_2 S T + a_3 T^2 + a_4 S + a_5 T + a_6 > 0 ,
\end{equation}
with $a_1=-0.3422, a_2=0.7760, a_3= -0.5262, a_4=
-0.0320, a_5=0.0528, a_6=0.0014$.
For $U$, the allowed interval is fixed to be
\begin{equation}
    U=0.03\pm0.10 .
\end{equation}

\section{Collider Constraints}

\subsection{\texorpdfstring{Couplings modifiers $\kappa$'s}{Couplings modifiers Ks}}
\label{sec:kappas}

Coupling-strength modifiers $\kappa$'s are introduced to test modifications of the Higgs boson couplings derived from new Physics, within a $\kappa$-framework~\cite{LHCHiggsCrossSectionWorkingGroup:2013rie} based on the leading-order contributions to each production and decay channel. Computing all deviations from the SM assuming a single state at $125~{\rm GeV}$ and neglecting the width, the coupling-strength modifiers $\kappa_j$ are defined for a given cross section $\sigma_{jj}$ or partial decay width $\Gamma_{jj}$ as,
\be
\kappa_j^2=\frac{\sigma_{jj}}{\sigma_{jj}^{\text{SM}}}\quad \text{or}\quad \kappa_j^2=\frac{\Gamma_{jj}}{\Gamma_{jj}^{\text{SM}}};
\ee
such that, for example, $\kappa_W^2$ gives the scaling relative to the SM prediction for the associated production with $W$, $\sigma_{WH}/\sigma_{WH}^{\text{SM}}$, and the decay mode into final state $WW$, $\Gamma_{WW^{(*)}}/\Gamma_{WW^{(*)}}^{\text{SM}}$.
We require all coupling modifiers
to be within 3$\sigma$ of the LHC data~\cite{ATLAS:2019nkf}.
We
list below the expressions for the kappas for the various Types. For
all Types of real 3HDMs considered we have
\begin{equation}
  \label{eq:1a}
  \kappa_V=\cos(\alpha_2) \cos(\alpha_1-\beta_1) \cos(\beta_2) +
  \sin(\alpha_2) \sin(\beta_2)\, ,
\end{equation}
which gives $\kappa_V=1$ when $\alpha_1=\beta_1$ and
$\alpha_2=\beta_2$. For the couplings with fermions,

\textbf{\textit{Type-I}}--- 
\begin{equation}
  \kappa_U= \frac{\sin(\alpha_2)}{\sin(\beta_2)},
\quad
  \kappa_D= \frac{\sin(\alpha_2)}{\sin(\beta_2)},
  \quad
  \kappa_L=  \frac{\sin(\alpha_2)}{\sin(\beta_2)}.
\end{equation}

\textbf{\textit{Type-II}}--- 
\begin{equation}
  \kappa_U= \frac{\sin(\alpha_2)}{\sin(\beta_2)},
\quad
  \kappa_D=\frac{\sin(\alpha_1) \cos(\alpha_2)}{\sin(\beta_1) \cos(\beta_2)},
  \quad
  \kappa_L=  \frac{\sin(\alpha_1) \cos(\alpha_2)}{\sin(\beta_1) \cos(\beta_2)}.
\end{equation}

\textbf{\textit{Type-X}}--- 
\begin{equation}
  \kappa_U= \frac{\sin(\alpha_2)}{\sin(\beta_2)},
\quad
  \kappa_D=\frac{\sin(\alpha_2)}{\sin(\beta_2)},
  \quad
  \kappa_L= \frac{\sin(\alpha_1) \cos(\alpha_2)}{\sin(\beta_1) \cos(\beta_2)}. 
\end{equation}

\textbf{\textit{Type-Y}}--- 
\begin{equation}
  \kappa_U= \frac{\sin(\alpha_2)}{\sin(\beta_2)},
\quad
  \kappa_D=\frac{\sin(\alpha_1) \cos(\alpha_2)}{\sin(\beta_1) \cos(\beta_2)},
  \quad
  \kappa_L=  \frac{\sin(\alpha_2)}{\sin(\beta_2)}.
\end{equation}

\textbf{\textit{Type-Z}}--- 
\begin{equation}
  \kappa_U= \frac{\sin(\alpha_2)}{\sin(\beta_2)},
\quad
  \kappa_D=\frac{\sin(\alpha_1) \cos(\alpha_2)}{\sin(\beta_1) \cos(\beta_2)},
  \quad
  \kappa_L= \frac{\cos(\alpha_1) \cos(\alpha_2)}{\cos(\beta_1) \cos(\beta_2)}.
\end{equation}

\subsection{\texorpdfstring{Signal strengths $\mu$'s}{Signal strengths mus}}

We comply with the most recent collider measurements on the $125~{\rm GeV}$ Higgs, both for production cross sections and branching ratios. For comparison with experiment,  we consider only the contributions of the
lowest non-vanishing order in perturbation theory. 
The decays that require one-loop calculations are those of neutral scalars
into two photons ($h_j\to\gamma\gamma$),
one Z and one photon ($h_j\to Z\gamma$),
and two gluons ($h_j\to gg$).
The final formulas for the first two widths are given in
Ref.~\cite{Fontes:2014xva},
only having to adapt the particles and their couplings to our case.
The formula for the width $h_j\to\gamma\gamma$ reads,
\begin{equation}
\Gamma(h_j\to \gamma\gamma)=\frac{G_F\alpha^2m_h^3}{128\sqrt{2}\pi^3}
(|X_F^{\gamma\gamma}+X_W^{\gamma\gamma}+X_H^{\gamma\gamma}|^2),
\label{eq:gaga}
\end{equation}
where, noticing that for scalars the $Y$ terms in~\cite{Fontes:2014xva} vanish,
 \begin{subequations}
\label{XHformula}
	\begin{eqnarray}
X_F^{\gamma\gamma}
&=&
-\sum_f N_c^f2a_j^fQ_f^2\tau_f[1+(1-\tau_f)f(\tau_f)],
\\[8pt]
X_{W}^{\gamma \gamma}
&=&
C_{j}\left[2+3 \tau_{W}+3 \tau_{W}\left(2-\tau_{W}\right)
f\left(\tau_{W}\right)\right],
\label{need_Cj}
\\[8pt]
X_{H}^{\gamma \gamma}
&=&
-\sum_{k=1}^{2} \frac{\lambda_{h_j H_k^+ H_k^-} v^{2}}{2 m_{H_{k}^{\pm}}^{2}}
\tau_{j k}^{\pm}
\left[1-\tau_{j k}^{\pm} f\left(\tau_{j k}^{\pm}\right)
\right].
	\end{eqnarray}
\end{subequations}
We used
\begin{equation}
\tau_f = 4m_f^2/m_{h_j}^2\, ,
\ \ \ 
\tau_{j k}^{\pm} = 4m_{H_k^\pm}^2/m_{h_j}^2\, ,
\end{equation}
where $m_f$ ($m_{H_k^\pm}$) is the mass of the relevant particle
in the loop, while $m_{h_j}$ is the mass of the decaying
Higgs boson. In the Higgs decays, the masses are calculated at the energy scale of $m_h$, following Ref.~\cite{Djouadi:2005gj}.  The function $f(\tau)$ is defined in
the Higgs Hunter's Guide~\cite{Gunion:1989we},
\begin{equation}
f(\tau)=\left\{\begin{array}{ll}
{\left[\sin ^{-1}(\sqrt{1 / \tau})\right]^{2},} & \text { if } \tau \geq 1 \\
-\frac{1}{4}\left[\ln \left(\frac{1+\sqrt{1-\tau}}{1-\sqrt{1-\tau}}\right)-i \pi\right]^{2}, & \text { if } \tau<1
\end{array}\right. ,
\end{equation}
and the couplings $C_{j}$ and $\lambda_{h_j H_k^+ H_k^-}$
for a given model are always derived with the
help of the 
software \texttt{FeynMaster}~\cite{Fontes:2019wqh,Fontes:2021iue}, that
uses \texttt{QGRAF}~\cite{Nogueira:1991ex},
\texttt{FeynRules}~\cite{Christensen:2008py,Alloul:2013bka} and
\texttt{FeynCalc}~\cite{Mertig:1990an,Shtabovenko:2016sxi}  in an
integrated way.

The decay into gluons can be obtained from the expression for the $\gamma\gamma$ decay,
\begin{equation}
    \Gamma(h_j\to gg)=\frac{G_F\alpha_S^2m_h^3}{64\sqrt{2}\pi^3}(|X_F^{gg}|^2) ,
\end{equation}
where
\begin{equation}
    X_F^{gg}=-\sum_q2a_j^q\tau_q[1+(1-\tau_q)f(\tau_q)] ,
\end{equation}
and the sum runs only over quarks q.

\begin{figure}[htbp!]
	\centering
	\includegraphics[width=150mm]{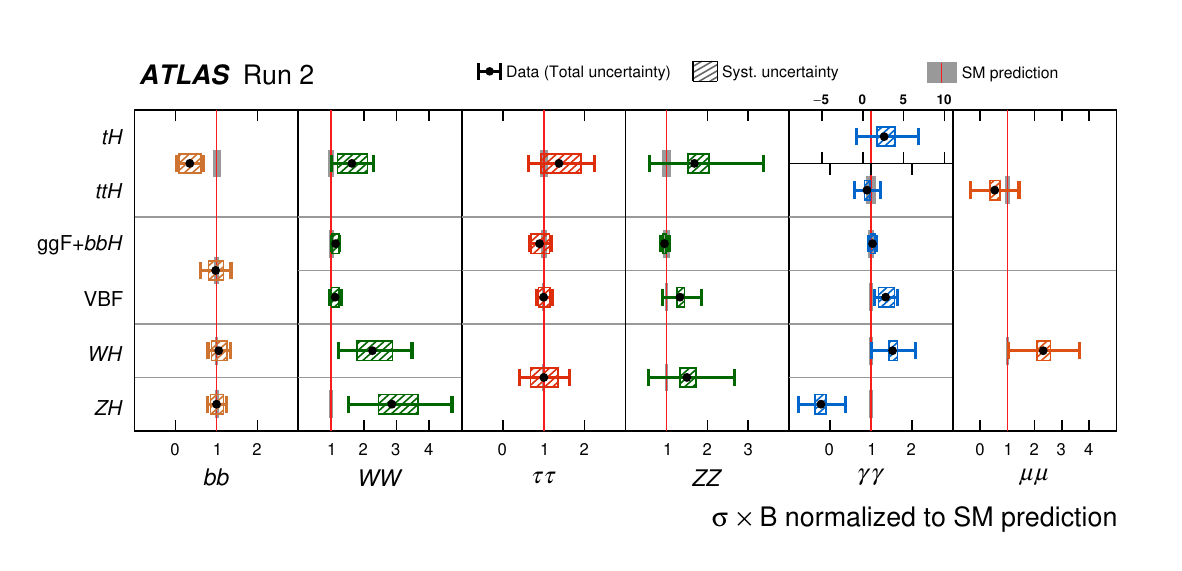}
	\caption{The most recent ATLAS results taken from~\cite{ATLAS:2022vkf} for the ratio of observed rate to predicted SM event rate for different combinations of Higgs boson production
and decay processes. The horizontal bar on each point denotes the 68\% confidence interval.}
	\label{f:mus}
\end{figure}

Having chosen a specific production and decay channel,
the collider event rates can be conveniently  described by the ratios $\mu_{if}^h$,
\begin{equation}
	\mu_{if}^h = \frac{\sigma^{\text{3HDM}}_{i}(pp \rightarrow h_{125})}{\sigma^{\text{SM}}_{i}(pp \rightarrow h_{125})} \times\frac{\text{BR}^{\text{3HDM}}(h_{125} \rightarrow f)}{\text{BR}^{\text{SM}}(h_{125} \rightarrow f)}\,, \label{eq:mus}
\end{equation}
with the subscript `$i$' indicating the production mode,
and `$f$' for the decay channel of the $125~{\rm GeV}$ Higgs. Starting from the collision of two protons, the relevant production
mechanisms include: gluon fusion (ggH), vector boson fusion (VBF),
associated production with a vector boson (VH, V = W or Z), and
associated production with a pair of top quarks (ttH). The SM cross
section for the gluon fusion process is calculated using HIGLU
~\cite{Spira:1995mt}, and for the other production mechanisms we use
the results of Ref.~\cite{deFlorian:2016spz}.  Each of the 3HDM
processes is obtained by rescaling the SM cross sections by the
relevant relative couplings. As for the decay channels, we calculated
the branching rations for final states $f=\,W\,W,\, Z\,Z,\,
b\,\overline{b}, \gamma\,\gamma$ and $\tau^+\tau^-$. Finally, we
require that the $\mu_{if}^h$ for each individual initial state
$\times$ final state combination is consistent, within twice the total
uncertainty, with the best-fit results presented in the most recent
study of data collected at $\sqrt{s}=13\,\text{TeV}$ with ATLAS ~\cite{ATLAS:2022vkf}, shown in Fig.~\ref{f:mus} for completion,
which are also consistent with CMS~\cite{CMS:2022dwd}. The \texttt{HiggsSignals} module
of \texttt{HiggsTools-1.1.3} can also be used to perform a $\Delta \chi^2$ test. With a value corresponding to $2\sigma$ or $3\sigma$, we only identify a quantitative but not a qualitative difference between
the two different methods of including the signal strengths.

\subsection{\texorpdfstring{Direct searches for new particles}{Direct searches for new particles}}

 To comply with the LHC results on the
direct searches for other scalars, either neutral or charged.
we use the software package \texttt{HiggsBounds-5.9.1} in Ref.~\cite{Bechtle:2020pkv} and more recently \texttt{HiggsTools-1.1.3}
~\cite{Bahl:2022igd}, which includes the latest data from the ATLAS and
CMS experiments at CERN.
For the decays allowing for off-shell bosons,
we use the method explained in~\cite{Romao:1998sr}. The production of charged scalars
in association with $t$ quark, $pp\to H^\pm t\,(b)$ , is calculated
with the routine in \texttt{HiggsTools-1.1.3}, for the $13\,\textrm{TeV}$
LHC and mass range $145~{\rm GeV}-2~{\rm TeV}$. For values below $145~{\rm GeV}$, \texttt{HiggsTools-1.1.3} estimates the cross section from the $BR(t \to H^+ b)$ that we give to the program.

\section{Flavour bounds}\label{sec:bsgamma}
The next step is to take into consideration the bounds coming from
flavour data. In the symmetric 3HDMs to be considered there are no FCNCs at the tree-level. Therefore the
NP contribution at 
one-loop order to observables such as $b\to s\gamma$ and the neutral meson mass
differences will come from the charged scalar Yukawa couplings. In
Ref.~\cite{Chakraborti:2021bpy} it was shown that the constraints coming from the meson mass
differences tend to exclude very low values of $\tan\beta_{1,2}$. Therefore, we
only consider
\begin{eqnarray}
	\tan\beta_{1,2} > 0.3 \,.
\end{eqnarray}
It is well known that the experimental bounds on $B\to X_s \gamma$
place stringent restrictions on the parameter space of models with charged
scalars~\cite{Borzumati:1998tg,Borzumati:1998nx,
Misiak:2017bgg,Misiak:2018cec,Akeroyd:2020nfj}.
Most notably,
there is a bound on the mass of the only charged Higgs boson
present in the Type-II 2HDM which, at 95\% CL
(2$\sigma$), is according to~\cite{Misiak:2017bgg} 
\begin{equation}
  \label{eq:16}
  m_{H^+}> 580~{\rm GeV}\, .
\end{equation}
The exact value for this bound depends on both the theoretical 
approximations~\cite{Bernlochner:2020jlt} and the experimental errors.
The experimental average gives~\cite{Amhis:2019ckw}
\begin{equation}
  \label{eq:bsgammaexp}
  \text{BR}^{\rm exp}(B\to X_s \gamma) = (3.32 \pm 0.15) \times 10^{-4}\, ,
\end{equation}
while the NNLO calculation within the SM yields
~\cite{Misiak:2020vlo,Akeroyd:2020nfj}
\begin{equation}
  \label{eq:bsgammasm}
  \text{BR}^{\rm SM}(B\to X_s \gamma) = (3.40 \pm 0.17) \times 10^{-4}\, ,
\end{equation}
with an error of about 5\%. We will take an error of 2.5\% around the central value of the
calculation shown below and, following~\cite{Akeroyd:2020nfj}, consider 99\% CL (3$\sigma$) for the experimental error:
\begin{equation}
  \label{eq:17}
  2.87 \times 10^{-4} < \text{BR}(B\to X_s \gamma) < 3.77 \times 10^{-4}\, .
\end{equation}

\subsection{\texorpdfstring{The $b \to s \gamma$ decay calculation}{The b to s gamma decay calculation}}

We follow closely the calculation by Borzumati and Greub in
Ref.~\cite{Borzumati:1998tg}.
There, the new contributions from the charged Higgs bosons are encoded
in the Wilson coefficients,
\begin{subequations}   
  \label{eq:formulawilson}
\begin{align}
  C^{0,{\rm eff}}_7(\mu_W) =&
  C^{0,{\rm eff}}_{7,\rm SM}(\mu_W) +|Y|^2 C^{0,{\rm eff}}_{7,\rm YY}(\mu_W)
  +(X Y^*)C^{0,{\rm eff}}_{7,\rm XY}(\mu_W)\, , \\[+2mm]
  C^{0,{\rm eff}}_8(\mu_W) =&
  C^{0,{\rm eff}}_{8,\rm SM}(\mu_W) +|Y|^2 C^{0,{\rm eff}}_{8,\rm YY}(\mu_W)
  +(X Y^*)C^{0,{\rm eff}}_{8,\rm XY}(\mu_W)\, , \\[+2mm]
  C^{1,{\rm eff}}_4(\mu_W) =&
  E_0(x) + \frac{2}{3}
  \log\left(\frac{\mu_W^2}{M_W^2}\right) +|Y|^2 E_H(y)\, ,\\[+2mm]  
  C^{1,{\rm eff}}_7(\mu_W) =&
  C^{1,{\rm eff}}_{7,\rm SM}(\mu_W) +|Y|^2 C^{1,{\rm eff}}_{7,\rm YY}(\mu_W)
  +(X Y^*)C^{1,{\rm eff}}_{7,\rm XY}(\mu_W)\, , \\[+2mm]
  C^{1,{\rm eff}}_8(\mu_W) =&
  C^{1,{\rm eff}}_{8,\rm SM}(\mu_W) +|Y|^2 C^{1,{\rm eff}}_{8,\rm YY}(\mu_W)
  +(X Y^*)C^{1,{\rm eff}}_{8,\rm XY}(\mu_W) \, ,
\end{align}
\end{subequations}   
where we are using the notation in Ref.~\cite{Borzumati:1998tg}
which should be consulted for the definitions and also for the procedure
used in evolving the coefficients to the scale $\mu_b=m_b$.
The dependence on the charged Higgs mass
appears because the functions $C^{0,{\rm eff}}_{i,\rm YY},
C^{0,{\rm eff}}_{i,\rm XY}, C^{1,{\rm eff}}_{i,\rm YY}$, and
$C^{1,{\rm eff}}_{i,\rm XY}$ depend on $y=m_t^2/m_{H^+}^2$,
while the SM coefficients depend on $x= m_t^2/M_W^2$.

For models with multiple charged Higgs there is one contribution
(and one parameter $y_k$) for each particle.
A model with two charged Higgs is discussed in~\cite{Akeroyd:2020nfj,Logan:2020mdz},
with interesting earlier work highlighting the possible cancellation
between the two charged Higgs contributions appearing in
Refs.~\cite{Hewett:1994bd,Akeroyd:2016ssd}. In Section~\ref{Z3bsgamma} we show a realization of this possibility, in a specific model with 3 Higgs Doublets and find viable points in parameter space with masses below $400~{\rm GeV}$.
We obtain, for example,
\begin{align}
  \label{eq:wilsoncalc}
  C^{1,{\rm eff}}_7(\mu_W) =&
  C^{1,{\rm eff}}_{7,\rm SM}(\mu_W) +|Y_1|^2 C^{1,{\rm eff}}_{7,\rm  YY}(\mu_W,y_1)
  +|Y_2|^2 C^{1,{\rm eff}}_{7,\rm YY}(\mu_W,y_2)\nonumber\\
& +(X_1 Y_1^*)C^{1,{\rm eff}}_{7,\rm XY}(\mu_W,y_1) 
  +(X_2 Y_2^*)C^{1,{\rm eff}}_{7,\rm XY}(\mu_W,y_2) \, ,
\end{align}
where we wrote explicitly the dependence on the
charged Higgs masses,
\begin{equation}
  \label{eq:wilsoncalc_y}
  y_1= \frac{m_t^2}{m_{H_1^+}^2}, \quad
    y_2= \frac{m_t^2}{m_{H_2^+}^2}\, ,
\end{equation}
and used
\begin{equation}
X_1 = -\frac{\textbf{Q}_{22}}{\cos\beta_2 \sin\beta_1}, \quad
      Y_1 =  \frac{\textbf{Q}_{23}}{\sin\beta_2}, \quad X_2 = -\frac{\textbf{Q}_{32}}{\cos\beta_2\,\sin\beta_1}, \quad Y_2 =  \frac{\textbf{Q}_{33}}{\sin\beta_2}.
\end{equation}
We take the input parameters from Ref.~\cite{Borzumati:1998tg} except for
$\alpha_s(M_Z),m_t,M_Z,M_W$, that were updated to the most recent
values of the Particle Data Group
~\cite{Zyla:2020zbs}:\footnote{If we use exclusively the input values of
Ref.~\cite{Borzumati:1998tg},
we reproduce their SM results.}
\begin{subequations}   
\begin{align}
  \label{eq:alphamz}
 & \alpha_s(M_Z)= 0.1179\pm 0.0010, && m_t=172.76 \pm 0.3 ~{\rm GeV},\\
  &m_c/m_b= 0.29 \pm 0.02,  && m_b-m_c = 3.39 \pm 0.04~{\rm GeV},\\
  &\alpha_{em}^{-1}=137.036, && |V_{ts}^*V_{tb}/V_{cb}|^2=0.95\pm
  0.03,\\
  &\text{BR}_{SL}= 0.1049\pm 0.0046\, .&& &&
\end{align}
\end{subequations}   

\section{\texorpdfstring{Interpretations for nonstandard values of $h\to Z\gamma$}{Interpretations for nonstandard values of h to Z gamma}}\label{studyhzgamma}

The loop-induced decay modes of the Higgs boson~($h$) have been impactful in many
different aspects of Higgs physics. The decay $h\to \gamma\gamma$, calculated using the formula in \eq{eq:gaga},
played a pivotal role in the discovery of the Higgs boson~\cite{ATLAS:2012yve,CMS:2012qbp}.
Such loop-induced Higgs couplings have also been proved useful in sensing the presence of
new physics beyond the Standard Model~(BSM) through new loop
contributions arising from additional nonstandard particles~\cite{Bhattacharyya:2014oka}.
This is essentially how the sequential fermionic fourth generation
models fell out of favor~\cite{Kribs:2007nz,Eberhardt:2012gv,Djouadi:2012ae}.
These loop-induced Higgs couplings can also provide important insights into
the constructional aspects of the scalar extensions of the SM. Measurements of these
couplings can
severely restrict the fraction of nonstandard masses that stems from the electroweak vev,
~\cite{Bhattacharyya:2014oka, Bandyopadhyay:2019jzq}
 thereby providing nontrivial information
about the mechanism of electroweak symmetry breaking.

Now that a preliminary measurement of $h\to Z\gamma$ signal strength has become
available, it opens up new avenues to investigate the nature of new physics that may lie
beyond the SM. The currently measured value stands at~\cite{ATLAS:2020qcv,CMS:2023mku,CMS:2022ahq}
\begin{eqnarray}
	\mu_{Z\gamma} = 2.2 \pm 0.7 \,,
	\label{e:Zgval}
\end{eqnarray}
which, although not statistically significant yet, may be indicative of an enhancement
compared to the corresponding SM expectation~\cite{Buccioni:2023qnt}, $\mu_{Z\gamma} = 1$. This poses a
rather curious question:  if the measurement of $\mu_{Z\gamma}$ settles to a
nonstandard value  while $\mu_{\gamma\gamma}$ is consistent with the SM expectation,
then what kind of new physics would be required to reconcile such observation?

\begin{figure}[htbp!]
	\centering
	\includegraphics[width=60mm]{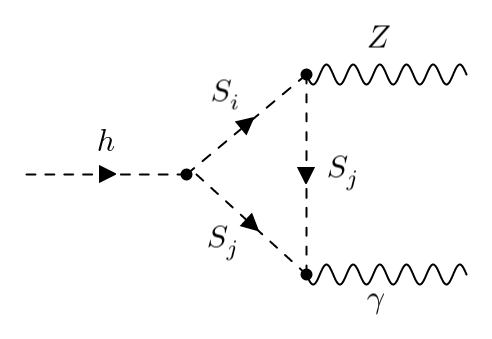}
	\includegraphics[width=60mm]{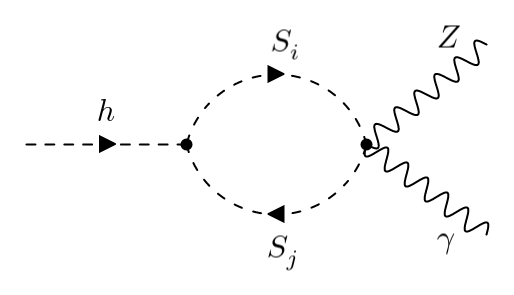}
	\caption{Representative Feynman diagrams that give additional contributions to $\mu_{Z\gamma}$
		exclusively.}
	\label{f:FD}
\end{figure}

\begin{figure}[htbp!]
	\centering
		\includegraphics[width=80mm]{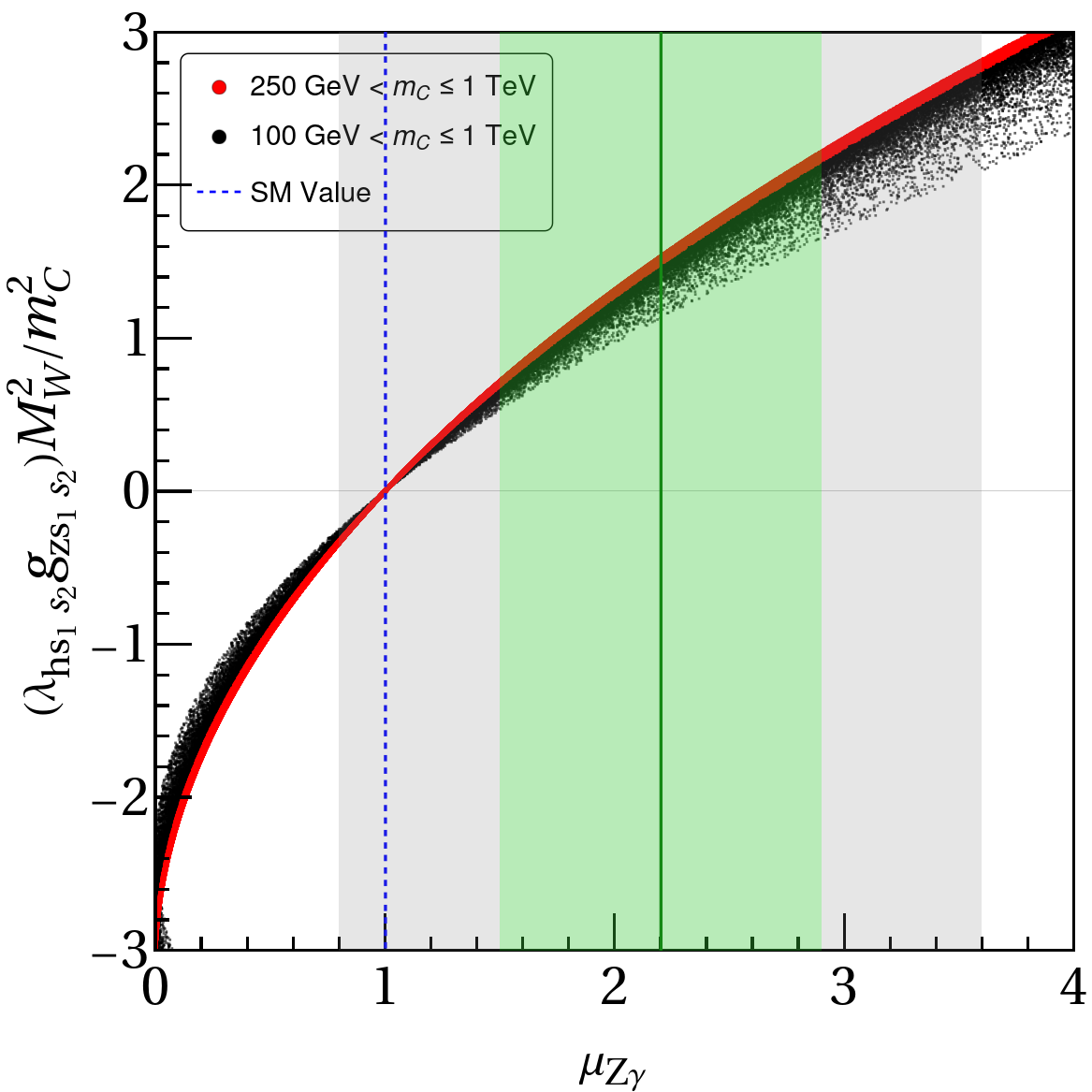}
		\caption{Required values of $f(\mu_{Z\gamma})$ (defined in \eq{e:fmu}) as a function of $\mu_{Z\gamma}$.
			The common charged scalar mass ($m_{C_1}=m_{C_2}=m_C$) has been scanned within the range
			$[100~{\rm GeV}, 1~{\rm TeV}]$ for the black region and within $[250~{\rm GeV}, 1~{\rm TeV}]$ for
			the thin red region.
			The dark-green solid vertical line marks the currently measured central value of $\mu_{Z\gamma}$ and the light-green
			and gray vertical bands around it correspond to the $1\sigma$ and $2\sigma$ ranges respectively. The dashed blue
			vertical line denotes the SM value of $\mu_{Z\gamma}$.
			In making this plot the unitarity conditions of \Eqs{ineq:h}{ineq:dirprod} have been satisfied. 
	}
		\label{f:scalar1}
\end{figure}

%
It is important to realize that, in the usual BSM scenarios, the new physics contributions
affect $\mu_{\gamma\gamma}$ and $\mu_{Z\gamma}$ in a correlated manner~\cite{Gunion:1989we,Djouadi:2005gi}. However, if we
are to keep $\mu_{\gamma\gamma}$ intact at its SM value, we must seek new interactions
that exclusively contribute to $\mu_{Z\gamma}$ without altering $\mu_{\gamma\gamma}$.
A little contemplation reveals that `off-diagonal' couplings of the Higgs and the
$Z$-boson would achieve this goal without much hardship. To illustrate this prescription,
let us assume that there exist new charged scalars with couplings parametrized in
the following manner:\footnote{
A similar exercise can also be done assuming the presence of extra charged fermions or vector bosons
possessing analogous off-diagonal couplings.
}
\begin{eqnarray}
\label{e:scalar}
	{\mathscr L}_S^{\rm int} &=& \lambda_{hs_is_j} M_W \, h\, S_i^{+Q}S_j^{-Q} + i\, g_{zs_is_j}
	Z^\mu \left\{\left(\partial_\mu S_i^{+Q}\right)S_j^{-Q} -\left(\partial_\mu S_j^{-Q}\right)S_i^{+Q} \right\} 
	\nonumber \\
	&& + 	{e Q g_{zs_is_j} A^\mu Z_\mu S_i^{+Q}S_j^{-Q}} + g_{zzs_is_j} Z^\mu Z_\mu S_i^{+Q}S_j^{-Q}
	+ {\rm h.c.} \,, 
\end{eqnarray}
where $M_W$ is the $W$-boson mass, $e$ is the electromagnetic coupling constant
 and $S_i^{+Q}$ denotes the $i$-th charged scalar
with electric charge $+Q$. Note that the correlation between the trilinear and
quartic couplings should follow from the underlying gauge theory.\footnote{
A connection between $g_{zs_is_j}$ and $g_{zzs_is_j}$ has been established
in Appendix~\ref{a:appoffdiag}.} In \eq{e:scalar}
we also assume that the off-diagonal couplings corresponding to $i\ne j$ are
overwhelmingly dominant over the diagonal couplings corresponding to $i=j$,
except for the quartic couplings of the form $ZZSS$. Under
these assumptions only $h\to Z\gamma$ will pick up additional contributions
through the Feynman diagrams shown in Fig.~\ref{f:FD}. These diagrams can not contribute to $h\to \gamma\gamma$ as the photon, in its tree-level
couplings, does not change particle species.
The uncommon interactions of \eq{e:scalar} 
	will become quintessential if $\mu_{Z\gamma}$ settles to a nonstandard value while
the other signal strengths are compatible with the SM.
%
%
The strengths of the couplings required for accommodating nonstandard values
$\mu_{Z\gamma}$ have been presented in Fig.~\ref{f:scalar1}.\footnote{
The general expression for the $h\to Z\gamma$ amplitude may be found in
Ref.~\cite{Hue:2017cph}. Although for simplicity we chose $Q=1$, for
	the general case the quantity on the vertical axis of Fig.~\ref{f:scalar1}
	will be scaled by a factor of $Q$.
	Additionally, we choose to focus on the
		scenario with only two species of charged scalars.
}
As can be observed from the figure, the quantity $g_{z s_1 s_2} \lambda_{hs_1s_2}/m_C^2$ can be almost pinned down
uniquely as a function of $\mu_{Z\gamma}$, $\frac{f(\mu_{Z\gamma})}{M_W^2}$, in the limit $m_{C1}=m_{C2}=m_C$ where 
$m_{Ci}$ denotes the mass of the
$i$-th charged scalar. With this spirit we may approximately write
\begin{eqnarray}
\label{e:fmu}
\frac{\lambda_{hs_1s_2}g_{z s_1 s_2}}{m_C^2} \approx \frac{f(\mu_{Z\gamma})}{M_W^2} \,,
\end{eqnarray}
with the understanding that $f(\mu_{Z\gamma})=0$ for $\mu_{Z\gamma}=1$, as can be confirmed using Fig.~\ref{f:scalar1}.
For a particular value of $\mu_{Z\gamma}$, the thickness of the black plot arises because $m_C$ is scanned from
relatively low values, 
within the range $100~{\rm GeV}$ $< m_C <$ 1~TeV. The thickness of the plot, for practical purposes,
	becomes negligible once we go beyond $m_C \gtrsim 250~{\rm GeV}$, as can be seen from the thin red overlaid region, and
	in this case the equality in \eq{e:fmu} becomes more robust.
%

Now that the essential strategy to accommodate a nonstandard $\mu_{Z\gamma}$ has been
laid out, it might be reasonable to ask whether the couplings of \eq{e:scalar} 
have any additional observable consequences which can potentially falsify such a scenario.
A related study will be to investigate whether the couplings of
\eq{e:scalar} can arise from a more complete gauge theoretical framework.
It is well-known that the off-diagonal charged scalar couplings can emerge
whenever the physical charged scalars are derived from an admixture of two different
$SU(2)_L\times U(1)_Y$ multiplets. The Zee-type~\cite{Zee:1980ai} scalar potential
constitutes a good example
of such a scenario. For the Zee-type set-up, dominant off-diagonal
couplings overpowering the diagonal couplings 
can be achieved when the two charged scalars mix maximally~\footnote{
Even in the presence of diagonal couplings, one may try to keep $\mu_{\gamma\gamma}$ in the
neighborhood of unity by adjusting ${\cal A}^{\rm NP}_{\gamma\gamma} = -2{\cal A}^{\rm SM}_{\gamma\gamma}$
in the $h\to \gamma\gamma$ amplitude. An example of this with fermionic couplings can be
found in a recent work~\cite{Barducci:2023zml}.
A similar effort within the context of a left-right symmetry\cite{Hong:2023mwr} leads to modifications of
	$\mu_{Z\gamma}$ and $\mu_{\gamma\gamma}$ in a correlated manner resulting in a limitation
	to the possible enhancement in $\mu_{Z\gamma}$.
}~\cite{Florentino:2021ybj}.
However, instead of channeling our efforts to construct a specific model, we can follow
a bottom-up method by exploring the high-energy unitarity behaviors of the tree-level
scattering amplitudes~\cite{Gunion:1990kf} involving the couplings of \eq{e:scalar}\footnote{
	Our approach in this regard is different from previous studies. For
	example, the unitarity bounds considered in Ref.~\cite{Abu-Ajamieh:2021egq} arise mostly from
	modifications in the tree-level couplings of the Higgs boson with the SM
	particles.
Of course it is well known that the unitarity of the theory will be compromised if such
couplings in the SM are tinkered with~\cite{Bhattacharyya:2012tj}. We, on the other hand, do
not touch any of the tree-level SM couplings and the new physics interactions we introduce do not
even affect $\mu_{\gamma\gamma}$.
}. Such an analysis is
known to reveal the compatibility of the set of couplings in \eq{e:scalar} with a UV-complete gauge
theory~\cite{LlewellynSmith:1973yud, Cornwall:1974km}. If the interactions in \eq{e:scalar} necessitate additional dynamics accompanying
them, the scattering amplitudes are expected to possess undesirable energy growths which will
lead to violation of tree-unitarity~\cite{Lee:1977eg} at high energies. The energy scale at which
unitarity is violated, can be interpreted as the maximum energy scale before which the effects
of new physics must set in to restore unitarity. Such an exercise provides an alternative
strategy to discover the need for additional effects that should be accompanied by
an enhanced $\mu_{Z\gamma}$. As shown in Appendix~\ref{a:appScattering}, inclusion of proper quartic interaction
of the form $ZZSS$ will neutralize the bad high-energy behaviors.
To demonstrate this explicitly,
we now concentrate on the impact of the $ZZSS$ quartic couplings in \eq{e:scalar}, namely,
\begin{eqnarray}
\label{e:zzss}
{\mathscr L}_{ZZSS}^{\rm int} = g_{zzs_is_j} Z^\mu Z_\mu S_i^{+Q}S_j^{-Q}
+ {\rm h.c.} \,. 
\end{eqnarray}
As mentioned earlier, these couplings will be required
to complement the underlying gauge structure. As we show in Appendix~\ref{a:appoffdiag},
even in the limit when the rest of the couplings of \eq{e:scalar} are purely off-diagonal,
the quartic interactions of \eq{e:zzss} should be diagonal with a specific
relation between $g_{zzs_is_j}$ and $g_{zs_is_j}$\footnote{For off-diagonal
couplings in \eq{e:scalar} and diagonal couplings in \eq{e:zzss} there will
be no additional loop-induced effects for the $hZZ$ vertex as well~\cite{Hernandez-Juarez:2023dor} as long as we work with only
	two flavors of charged scalars.}.
With this information, we can now proceed to calculate the amplitude for the process
$Z_L Z_L \to S_1^+ S_1^-$ where the subscript `$L$' represents longitudinal polarization.
In the high-energy limit, $E_{CM}\gg M$, meaning the CM energy is much larger than all
the masses in our current theory, we obtain
\begin{eqnarray}
{\cal M}_{Z_L Z_L \to S_1^+ S_1^-} \approx \frac{2 g_{z s_1 s_2}^2}{M_Z^2}
\left(m_{C_1}^2 -m_{C_2}^2 \right)
 +{\cal O}\left(\frac{M^2}{E_{CM}^2}\right) \,.
 \label{eq:matproc1}
\end{eqnarray}
Thus it is clear that the splitting between the two charged scalar masses is
constrained from unitarity as 
\begin{eqnarray}
\left|\frac{2 g_{z s_1 s_2}^2}{M_Z^2}
\left(m_{C_1}^2 -m_{C_2}^2\right)\right| < 16 \pi\,.
\label{e:split}
\end{eqnarray}
Therefore our simplified assumption of $m_{C_1}=m_{C_2}=m_C$ is manifestly
consistent with the unitarity requirements irrespective of the magnitude of
$g_{z s_1 s_2}$.
%
Next we consider the process $Z_L Z_L \to S_1^+ S_2^-$.
In the high energy limit, the amplitude is found to be
\begin{eqnarray}
	{\cal M}_{Z_L Z_L \to S_1^+ S_2^-} \approx -\frac{g}{2} \lambda_{hs_1s_2} + {\cal O}\left(\frac{M^2}{E_{\rm CM}^2}\right) \,,
	\label{eq:mth}
\end{eqnarray}
where $g$ is the $SU(2)_L$ gauge coupling.
This puts an upper limit on $\lambda_{hs_1s_2}$ as follows
\begin{eqnarray}
	\left|\frac{g}{2} \lambda_{hs_1s_2}\right| < 16 \pi\,.
	\label{ineq:h}
\end{eqnarray}
Finally we note that a direct upper bound on the charged scalar masses can be
placed by considering the scattering process $Z_L S^+_1 \to h S_1^+$. In the high-energy limit the tree-level amplitude can be written as
\begin{eqnarray}
{\cal M}_{Z_L S^+_1 \to h S_1^+} \approx - g_{zs_1s_2} \lambda_{hs_1s_2} \frac{M_W}{M_Z} + {\cal O}\left(\frac{M^2}{E_{\rm CM}^2}\right)\,. 
\label{eq:matproc3}
\end{eqnarray}
Therefore the unitarity constraint should imply
\begin{eqnarray}
\left|g_{z s_1s_2} \lambda_{hs_1s_2}\right| \frac{M_W}{M_Z} < 16\pi \,.
\label{ineq:dirprod}
\end{eqnarray}
This is where the experimental determination of
$\mu_{Z\gamma}$ becomes relevant. A nonstandard value of $\mu_{Z\gamma}$
exclusively, will necessitate such couplings whose strength can be estimated
using \eq{e:fmu} as follows
\begin{eqnarray}
	\lambda_{hs_1s_2}g_{z s_1 s_2} = \frac{f(\mu_{Z\gamma})\, m_C^2}{M_W^2} \,.
\end{eqnarray}
Plugging this into \eq{ineq:dirprod} we may infer
\begin{eqnarray}
m_C < \sqrt{16 \pi \frac{ M_Z M_W}{\left|f(\mu_{Z\gamma})\right|}} \,.
\label{e:mclim}
\end{eqnarray}
\begin{figure}[htbp!]
	\centering
	\includegraphics[width=0.5\textwidth]{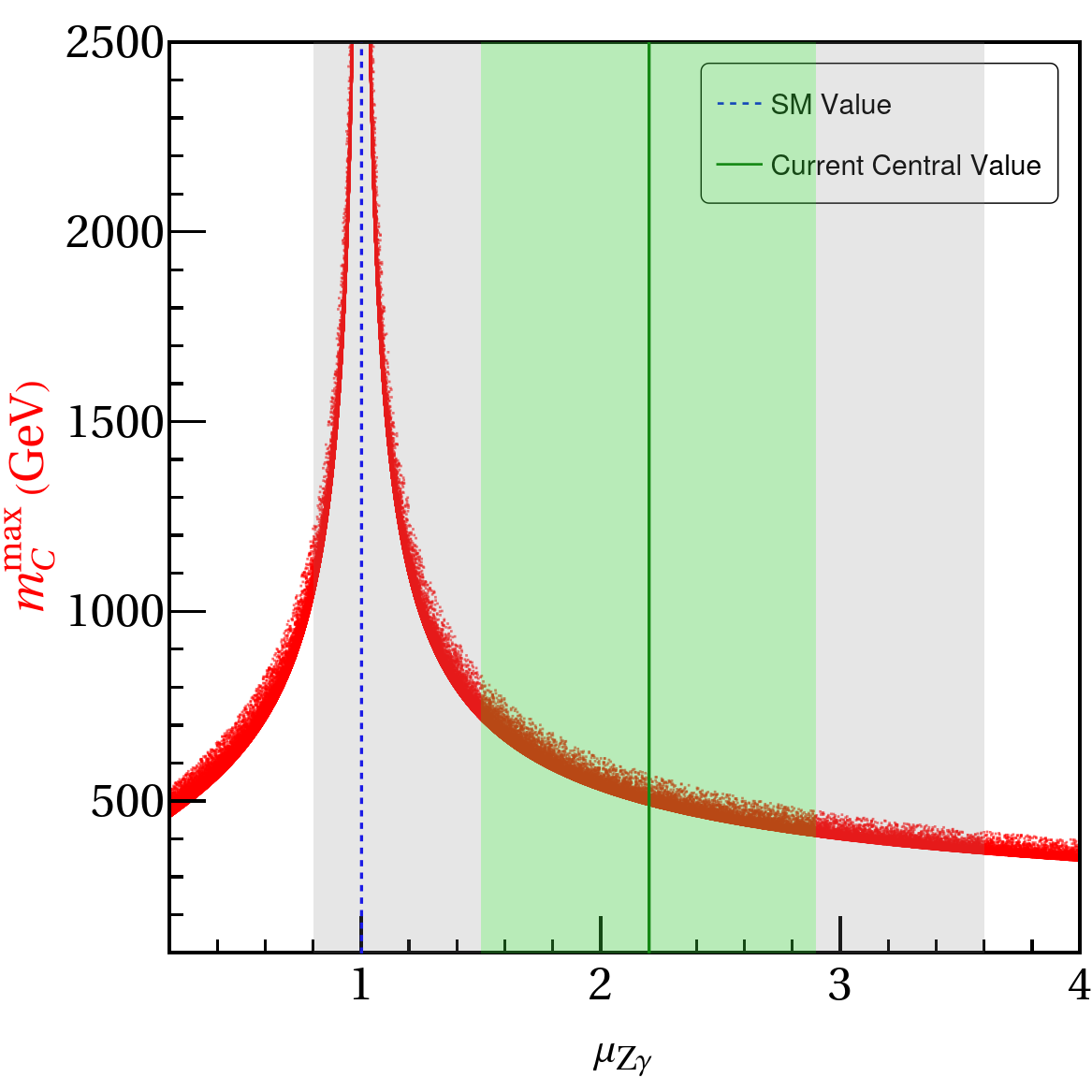}
	\caption{\it The upper limits in \eq{e:mclim} plotted against $\mu_{Z\gamma}$
		as the red region respectively.
		The dark-green solid vertical line marks the currently measured central value of $\mu_{Z\gamma}$ and the light-green
		and gray vertical bands around it correspond to the $1\sigma$ and $2\sigma$ ranges respectively. The dashed blue
		vertical line denotes the SM value of $\mu_{Z\gamma}$.
	}
	\label{f:uniNP}
\end{figure}
It should be noted that, as $f(\mu_{Z\gamma}=1)=0$, the upper limit on $m_C$ can be
infinitely large for $\mu_{Z\gamma}=1$, implying that the new physics effects can be safely decoupled in
the SM-limit as expected. However a more intriguing thing to note will be the fact that any
deviation of $\mu_{Z\gamma}$ from the SM value will mandate the intervention of new physics. To quantify
the required proximity of the new physics scale, we plot the right hand side of \eq{e:mclim} as the
red region in Fig.~\ref{f:uniNP} where
the value of the function $f(\mu_{Z\gamma})$ is mapped from Fig.~\ref{f:scalar1}.
From Fig.~\ref{f:uniNP} we can see that new physics effects in the sub-TeV regime will be expected
even when $\mu_{Z\gamma}$ deviates from unity by only $20\%$. In fact, the current central
value of $\mu_{Z\gamma}=2.2$ (marked by the dark-green vertical solid line)
decrees the common charged scalar mass to be below $500~{\rm GeV}$ which should be well within
the reach of the LHC.


\section{Summary}

In this Chapter we reviewed the current constraints present in real multi-Higgs doublet models. We highlighted the calculation of loop decays measured at colliders and the contributions of the additional charged scalars on the $b \to s \gamma$ decay. We continued with a study motivated by recent experimental data on the $\mu_{Z\gamma}$ signal strength. We consider the possibility that $\mu_{Z\gamma}$ deviates substantially from its SM value while all other Higgs signal strengths remain in excellent agreement with the corresponding SM expectations. Our study is considered a theoretical preparation for such an eventuality in a bottom-up manner. 

We 
provided a general template for the new physics interactions which will exclusively affect $h \to Z \gamma$.
We particularized our strategy with new charged scalars endowed with dominant off-diagonal couplings
with the Higgs and the $Z$ bosons, as exemplified through \eq{e:scalar}. 
However, as we have explicitly shown,
such interactions do not compromise unitarity at high-energies
indicating compatibility with spontaneously broken gauge theories. 
The unitarity constraints have been used to place upper bounds on
the magnitudes of the new couplings. We have then translated them
into an upper
bound on the common charged scalar mass, which can be as low as $500~{\rm GeV}$ for the current central value of
$\mu_{Z\gamma}$. 
This means that, if the current non-standard central value of $\mu_{Z\gamma}$ becomes statistically
significant as more data accumulates, discovery of new physics effects at the LHC should be just around the
corner. LHC can act as a win-win experiment for BSM searches if $\mu_{Z\gamma}$ eventually settles towards a nonstandard value. That would definitely be an exciting future to look forward to.

%% file: chapters/alignment.tex
\chapter{Traditional sampling techniques and the alignment limit}
\label{chapter:Alignment}

\hspace*{0.3cm}
This Chapter is dedicated to phenomenological studies of real 3HDMs with multiple symmetries. We set complete attention on the
Type-Z Yukawa couplings that can only appear for NHDM with $N\geq 3$. The interest is in 
finding the differences that exist in this new Type of model, since it
decouples completely the up quark, down quark and charged lepton
sectors from one-another. For each of the symmetry constrained 3HDM, we built a dedicated
code, which is an extension of our previous
codes~\cite{Fontes:2014xva,Florentino:2021ybj}, and combines the respective BFB conditions in Section~\ref{sec:bfb} with all the constraints described in Chapter~\ref{chapter:Constraints}.
We performed an extensive scan of the parameter space.
Our fixed inputs are $v = 246~{\rm GeV}$ and $m_{h1} = 125~{\rm GeV}$.
We then took random values in the ranges:
 \begin{subequations}
\label{eq:scanparameters}
	\begin{eqnarray}
&\alpha_1,\, \alpha_2,\, \alpha_3,\, \gamma_1,\, \gamma_2\, \in \left[-\frac{\pi}{2},\frac{\pi}{2}\right];\qquad \tan{\beta_1},\,\tan{\beta_2}\,\in \left[0,10\right];\\[8pt]
&m_{H_1}\equiv m_{h_2},\, m_{H_2}\equiv m_{h_3}\,
\in \left[125,1000\right]\,{\rm GeV};\\[8pt]
&
m_{A_1},\,m_{A_2}\,m_{H_1^\pm},\,m_{H_2^\pm}\,
\in \left[100,1000\right]~{\rm GeV};\\[8pt]
&
m^2_{12},m^2_{13},m^2_{23} \in  \left[\pm 10^{-1},\pm 10^{7}\right]\,
~{\rm GeV}^2\, ,
	\end{eqnarray}
\end{subequations}
where the last expression applies only to the soft
masses that are not obtained as derived quantities.
These parameter ranges are used in all scans and figures
presented below.
The lower limits chosen for the masses satisfy the
constraints listed in
Ref.~\cite{Aranda:2019vda}.

There have been implementations of Type-Z in three-Higgs-doublet
models (3HDM) using a $\Z2\times\Z2$ symmetry
~\cite{Akeroyd:2016ssd,Akeroyd:2020nfj,Alves:2020brq,Logan:2020mdz} or $\Z3$
~\cite{Das:2019yad}.  There has also been an analysis of $\Z3$ 3HDM which takes the
exact alignment limit and looks at specific values of the physical
parameters~\cite{Chakraborti:2021bpy}.  It does not seem to consider
the theoretical constraints coming from perturbative unitarity and BFB conditions. We will directly compare our results in light of recent
LHC measurements, for their specific parameter choices.  We
then show that by scanning for a larger range of parameters (away from
exact alignment, but still consistent with all experimental data) we
can obtain viable points corresponding to smaller masses for the
additional particles. We will finish our analysis with a search for observable features which can distinguish between the $\Z2\times\Z2$ and $\Z3$ Type-Z 3HDMs.

\section{The alignment limit for the real 3HDM}
When studying 3HDM, it was
noted~\cite{Chakraborti:2021bpy,Das:2019yad} that in
order to be able to generate good points in an easy way one should not be far away
from alignment, defined as the situation where the lightest Higgs
scalar has the SM couplings. It was shown in Ref.~\cite{Das:2019yad}
that this corresponds to the case when
\begin{equation}
  \label{eq:alignment}
  \alpha_1=\beta_1,\quad \alpha_2=\beta_2, 
\end{equation}
with the remaining parameters allowed to be free, although subject to
the constraints below. It turns out that for $\Z3$ 3HDM~\cite{Boto:2021qgu}, this
constraint alone is not enough to generate a reasonably large set of
good points starting from a completely unconstrained scan as in
\eq{eq:scanparameters}. Ref.~\cite{Chakraborti:2021bpy}
noticed a quite remarkable situation. If, besides the alignment of
\eq{eq:alignment}, one also requires
\begin{equation}
  \label{eq:alignment2}
  \gamma_1=\gamma_2=-\alpha_3,\quad m_{H_1}=m_{A_1}=m_{H_1^\pm},\quad
  m_{H_2}=m_{A_2}=m_{H_2^\pm} ,
\end{equation}
then the potentials of \eq{U1U1quartic},
(\ref{U1Z2quartic-our}) and (\ref{Z2Z2quartic}) all collapse into a
very symmetric form,
\begin{equation}
  \label{eq:scanbsgamma_mass5}
  V_{\rm Sym\, Lim} = \lambda_{\rm SM} \left[ (\phi_1^\dagger \phi_1) +
    (\phi_2^\dagger \phi_2) +(\phi_3^\dagger \phi_3) \right]^2\, ,
\end{equation}
with
\begin{equation}
  \label{eq:scanbsgamma_mass6}
  \lambda_{\rm SM}= \frac{m_h^2}{2 v^2} ,
\end{equation}
being the SM quartic Higgs coupling. This requires that, for the
conditions in \eq{eq:alignment} and \eq{eq:alignment2}, we have
\begin{equation}
  \label{eq:scanbsgamma_mass7}
  \lambda_1=\lambda_2=\lambda_3=\lambda_{\rm SM},\ \ \ 
  \lambda_4=\lambda_5=\lambda_6=2\lambda_{\rm SM},
\end{equation}
with all other $\lambda'$s vanishing. Imposing the validity of \eq{eq:alignment}
and \eq{eq:alignment2} also implies that the soft masses can be explicitly
solved as,
 \begin{subequations}
  \label{eq:softs-sym}
	\begin{eqnarray}
  m^2_{12}&=&c_{\beta_{1}}^2 c_{\gamma_{2}} s_{\beta_{2}} s_{\gamma_{2}}
   \left(m_{H^{+}_{1}}^2-m_{H^{+}_{2}}^2\right)+c_{\beta_{1}} s_{\beta_{1}}
   \left[s_{\beta_{2}}^2 \left(c_{\gamma_{2}}^2 m_{H^{+}_{2}}^2+m_{H^{+}_{1}}^2
   s_{\gamma_{2}}^2\right)-c_{\gamma_{2}}^2 m_{H^{+}_{1}}^2-m_{H^{+}_{2}}^2
 s_{\gamma_{2}}^2\right]\nonumber\\
&&+c_{\gamma_{2}} s_{\beta_{1}}^2 s_{\beta_{2}} s_{\gamma_{2}}
 \left(m_{H^{+}_{2}}^2-m_{H^{+}_{1}}^2\right)\\[+3mm]
 m^2_{13}&=&-c_{\beta_{2}} \left[c_{\beta_{1}} s_{\beta_{2}} \left(c_{\gamma_{2}}^2
   m_{H^{+}_{2}}^2+m_{H^{+}_{1}}^2 s_{\gamma_{2}}^2\right)-c_{\gamma_{2}} s_{\beta_{1}}
   s_{\gamma_{2}} \left(m_{H^{+}_{1}}^2-m_{H^{+}_{2}}^2\right)\right]\\[+3mm]
 m^2_{23}&=& -c_{\beta_{2}} \left[c_{\beta_{1}} c_{\gamma_{2}} s_{\gamma_{2}}
   \left(m_{H^{+}_{1}}^2-m_{H^{+}_{2}}^2\right)+s_{\beta_{1}} s_{\beta_{2}}
   \left(c_{\gamma_{2}}^2 m_{H^{+}_{2}}^2+m_{H^{+}_{1}}^2 s_{\gamma_{2}}^2\right)\right]
	\end{eqnarray}
\end{subequations}
We have verified that this simplification works for the symmetry constrained $U(1)\times U(1)$, $U(1)\times\Z2$, $\Z2\times\Z2$ and $\Z3$. Now it
is easy to understand that all such points are good points. Due to
alignment, the LHC results on the $h_{125}$ are easily obeyed, while
the perturbativity unitarity, STU and the other constraints are
automatically obeyed. In fact we are quite close to the SM. 

\begin{figure}[htb]
\centering
\includegraphics[width = 0.5\textwidth]{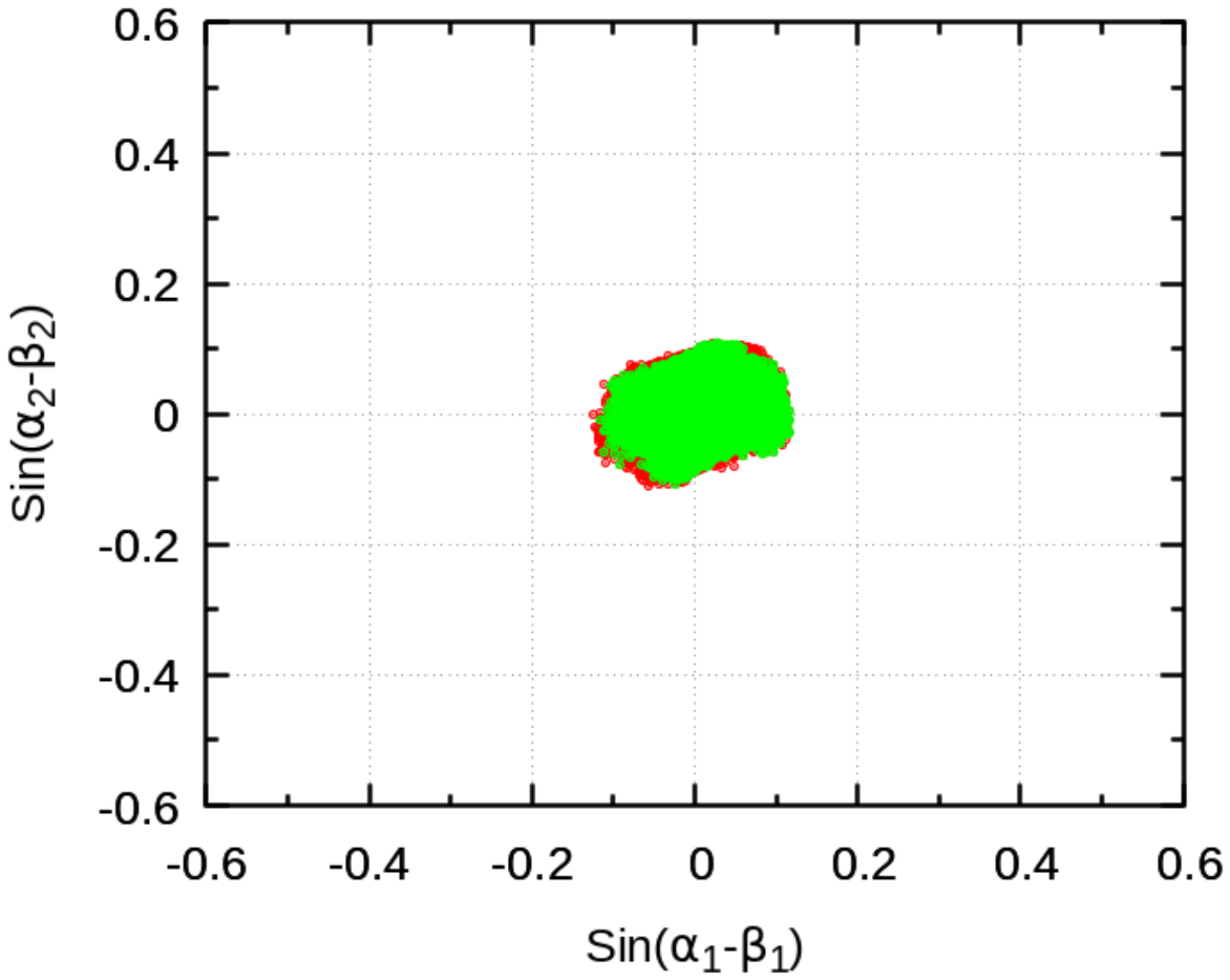}
\caption{Results of the simulation of the $\Z3$ real 3HDM in the $\sin{(\alpha_2-\beta_2)}-\sin{(\alpha_1-\beta_1)}$ plane. The green points passed all constraints including \texttt{HiggsBounds-5.9.1}, while the red points did not pass \texttt{HiggsBounds-5.9.1}. \label{plota1a2}}
\end{figure}

In studying the 3HDM with $\Z3$ we found that we
could go away from the conditions of \eq{eq:alignment}
and \eq{eq:alignment2}
by a given percentage (10\%, 20\%, 50\%) and enhance the possibility
of some signals, while at the same time being able to generate enough
points. The set of points that are consistent with all the bounds is
plotted in the $\sin{(\alpha_2-\beta_2)}-\sin{(\alpha_1-\beta_1)}$
plane as shown in Fig.~\ref{plota1a2}. Comparing with the plot in the
same plane shown in~\cite[Fig.1]{Das:2019yad}, it can be seen that the
use of more recent experimental data for the simulated results leads
to us being closer to the alignment limit, defined by
$\alpha_1=\beta_1$ and $\alpha_2=\beta_2$. 

For the case of $U(1)\times\Z2$,  and
specially for $U(1)\times U(1)$, we could generate a large set of
points just implementing a percentage of 50\% around
\eq{eq:alignment}. That is, we scanned within $x\%=50\%$ of the alignment
	condition of \eq{eq:alignment} by choosing to scan within the following
	range:
	\begin{equation}
	\label{Al-1}
	\frac{\alpha_1}{\beta_1} \,,\ 
	\frac{\alpha_2}{\beta_2}\, \in\, [1-x\%\,,1+x\%]\, ,
	\ \ \ \textbf{(\textrm{Al-1}-x\%)}
	\end{equation}
and setting $x\%=50\%$ without imposing the conditions in \eq{eq:alignment2}.
The second, more stringent alignment condition,
combines $\textrm{Al-1}$ with the six conditions of \eq{eq:alignment2},
\begin{equation}
  \label{Al-2}
  \frac{\alpha_1}{\beta_1},\ 
  \frac{\alpha_2}{\beta_2},\ 
  \frac{\gamma_2}{\gamma_1},\ 
  \frac{-\alpha_3}{\gamma_1},\ 
  \frac{m_{A_1}}{m_{H_1}},\ 
  \frac{m_{H_1^\pm}}{m_{H_1}},\ 
  \frac{m_{A_2}}{m_{H_2}},\ 
    \frac{m_{H_2^\pm}}{m_{H_2}}\, \in\, [1-x\%\,,1+x\%]\, .
    \ \ \ \textbf{(\textrm{Al-2}-x\%)}
\end{equation}
Lastly, we focused on the $\Z2 \times \Z2$ model.
We generated four not overlapping
sets based on \eq{Al-2},
\begin{itemize}
\item $\text{Set A}: [0\%,1\%]\ \text{red}$
\item $\text{Set B}: [1\%,10\%]\ \text{green}$
\item $ \text{Set C}: [10\%,20\%]\ \text{orange}$
\item $\text{Set D}: [20\%,50\%]\  \text{blue}$
\end{itemize}

\begin{figure}[htb]  
  \centering
  \begin{tabular}{cc}
\includegraphics[width=0.47\textwidth]{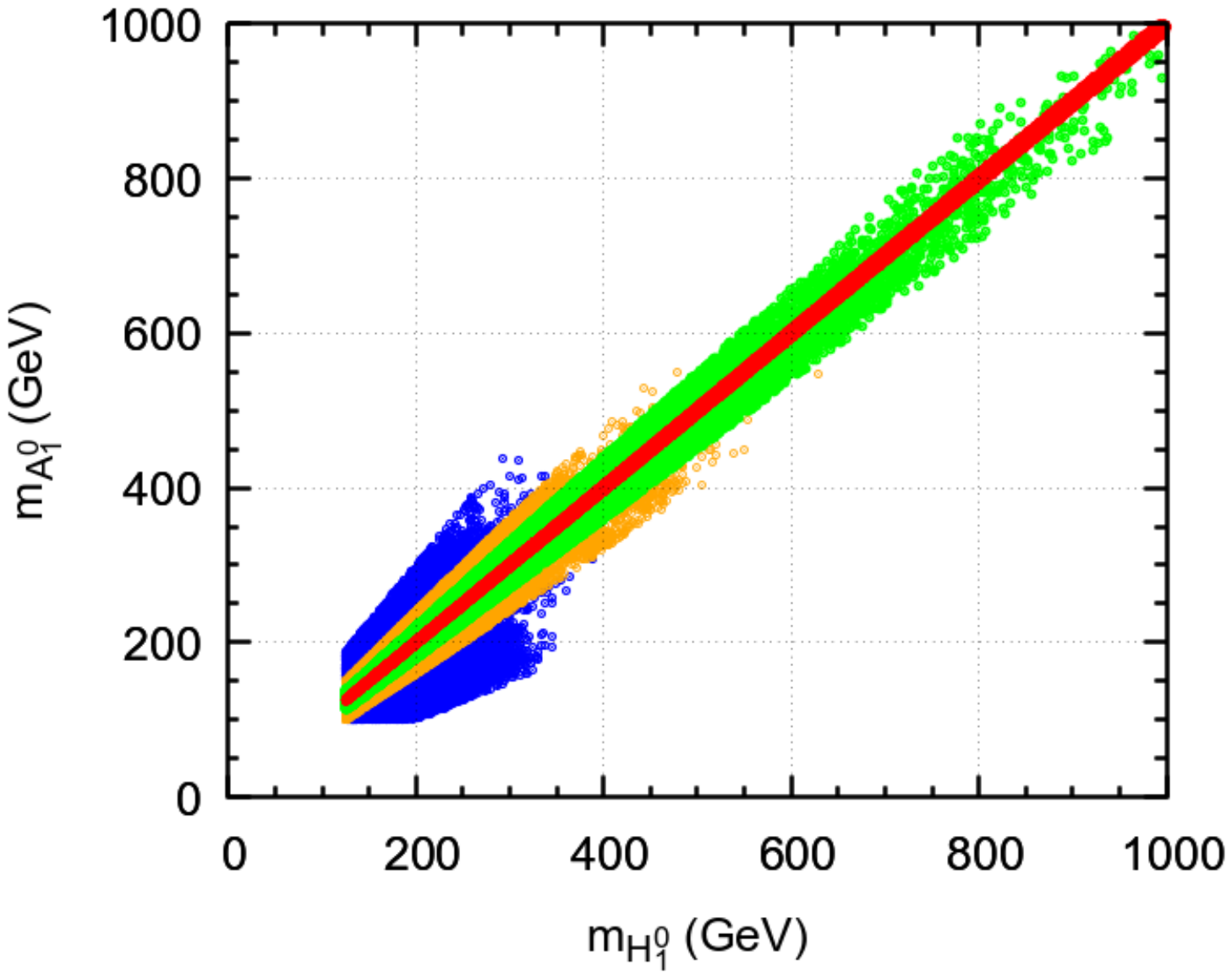}
    &
\includegraphics[width=0.47\textwidth]{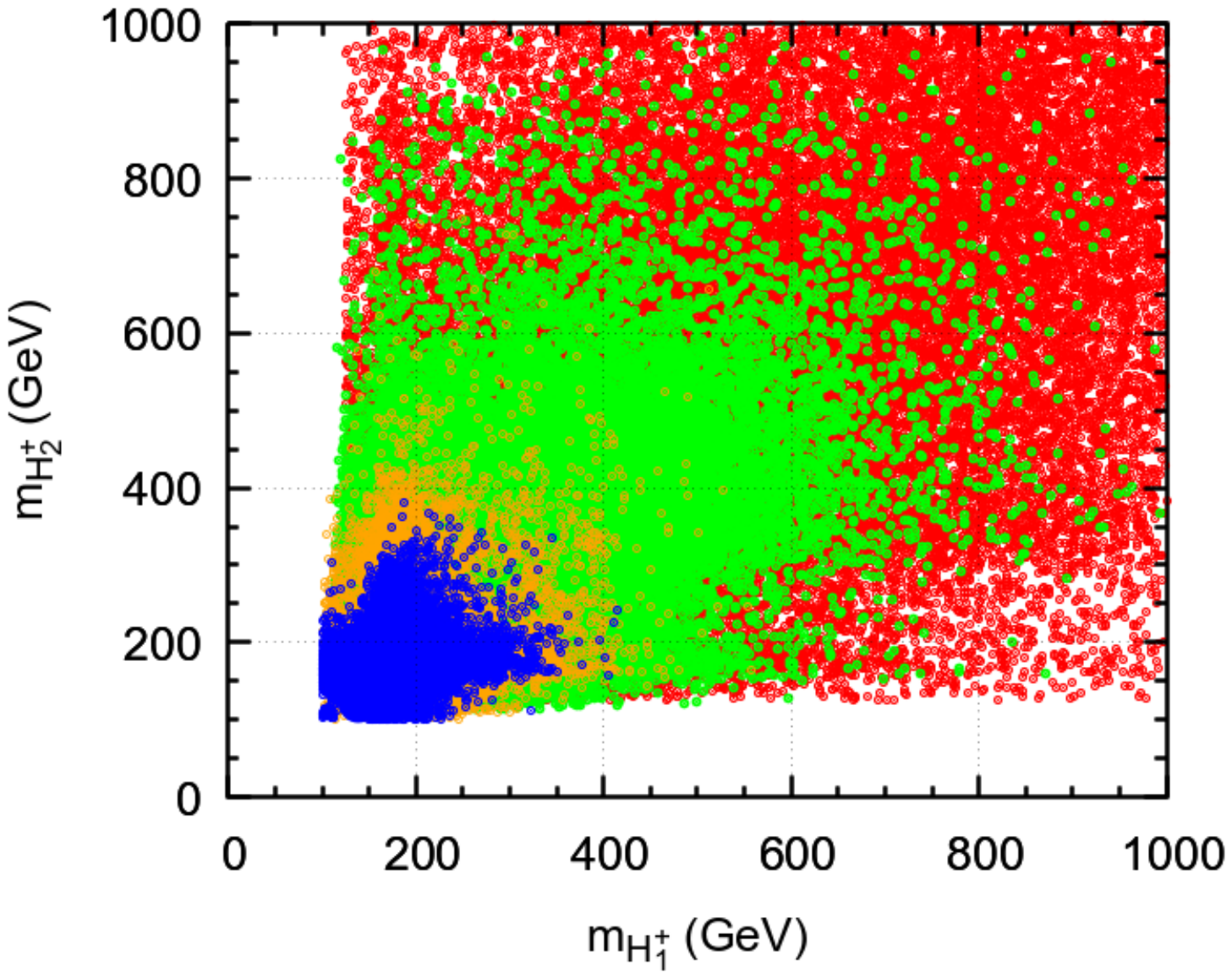}      
  \end{tabular}
  \caption{Left panel: $m_{H_{1}}$ versus $m_{A_{1}}$ for the four
  sets indicated in the text for the  $\Z2 \times \Z2$ real 3HDM.\\ Right panel: Same for $m_{H^+_{1}}$
  versus $m_{H^+_{1}}$.}
\label{fig:masses}
\end{figure}

In the left panel of Fig.~\ref{fig:masses} we plot the mass of $A_1$
versus the mass of $H_1$. In the conditions of \eq{eq:alignment2}, this
would just be a straight line. Here, for Set A, we are very close to
those conditions. As we move away from  \eq{eq:alignment2}, we notice
two things. First, the line moves into a broader band. Second, larger
masses are being cut. This is especially true for Set~D, where we are
more than 20\% away from \eq{eq:alignment2}. This is due to the fact
that we are now far away from the quasi-SM situation.
Then, the
combination of constraints, including the LHC results, makes it
increasingly difficult to generate good points with large masses. We
show this in a different way in the right panel of
Fig.~\ref{fig:masses}. We see that high masses are easy to be
generated for smaller deviations from the symmetric situation of
\eq{eq:alignment2}.

\begin{figure}[htb]
  \centering
  \begin{tabular}{cc}
\includegraphics[width=0.47\textwidth]{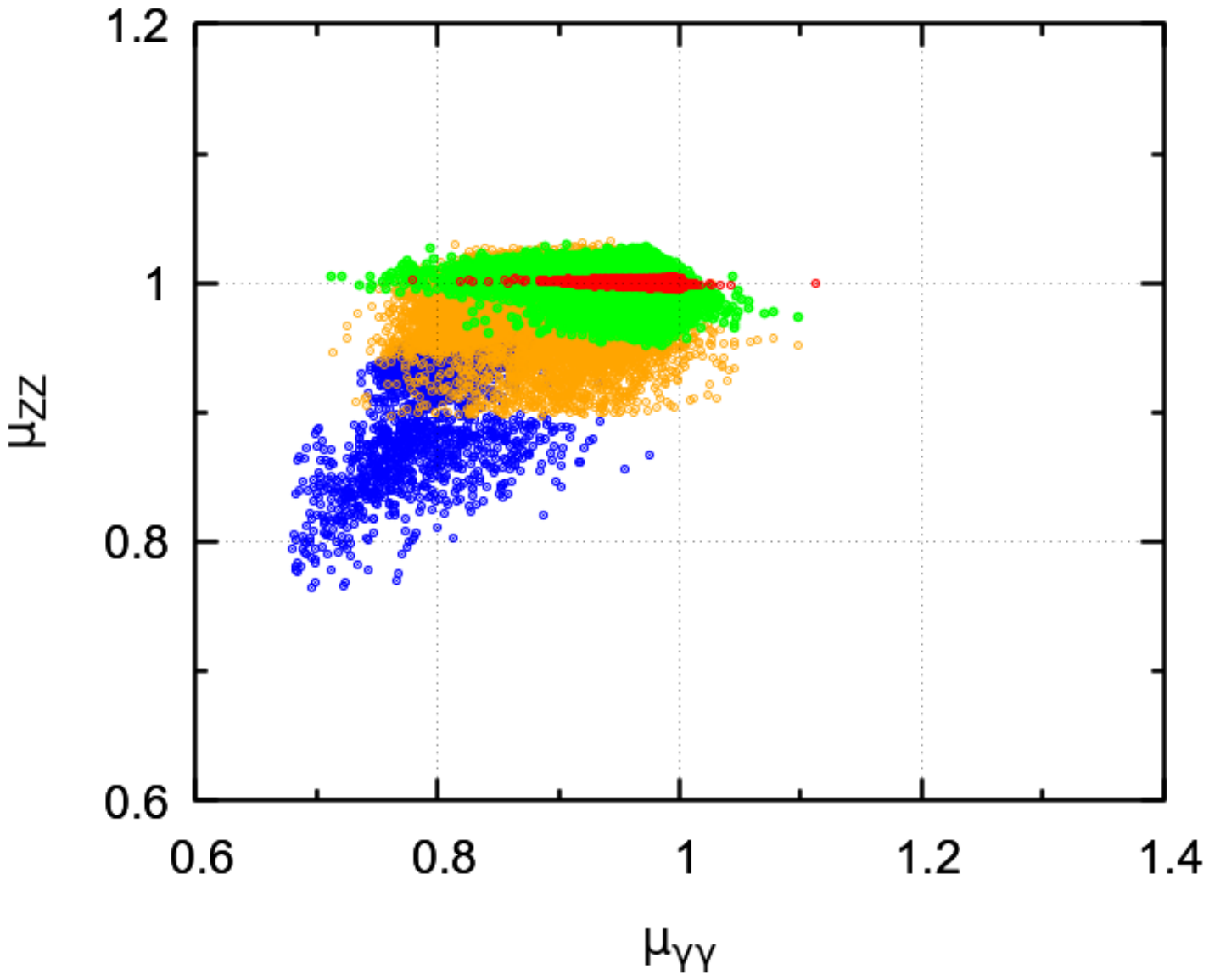}
    &
\includegraphics[width=0.47\textwidth]{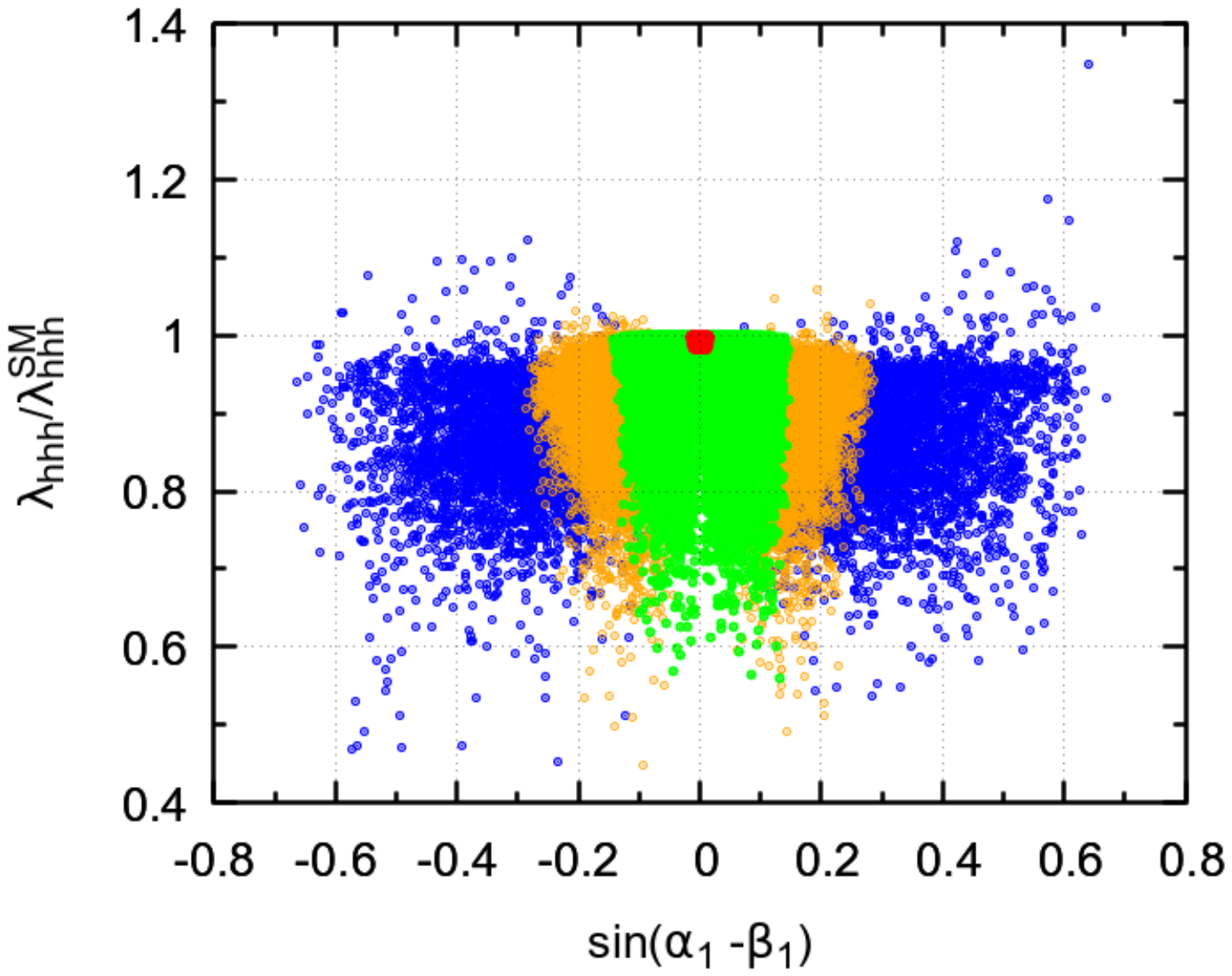}      
  \end{tabular}
  \caption{Left panel: $\mu_{\gamma\gamma}$ versus $\mu_{ZZ}$ for the four
  sets indicated in the text for the  $\Z2 \times \Z2$ real 3HDM.\\ Right panel: Same for
  $\lambda_{hhh}/\lambda_{hhh}^{\rm SM}$
  versus $\sin(\alpha_1-\beta_1)$.}
  \label{fig:observables}
\end{figure}

The reason for including points away from the symmetric limit is that
otherwise we get results that are very close to the SM, with very
little room for new phenomenology.  This is illustrated in Fig.~\ref{fig:observables} where in the left
panel we plot the signal strengths $\mu_{ZZ}$ versus
$\mu_{\gamma\gamma}$ for the same four sets. In the right panel we
plot the ratio of the $\lambda_{hhh}$ coupling to the SM. In both
cases we see that, if we remain too close to the symmetric limit of
\eq{eq:alignment2}, the results are very close to the SM, especially for
the Higgs boson triple coupling. We should try to be as
far away as possible from the symmetric limit (but still compatible
with all constraints) in order to explore a richer beyond the SM phenomenology.
The figures also show that such large deviations are still compatible with all
present theoretical and experimental bounds.

\section{\label{sec:Z3_constraints}\texorpdfstring{Impact of constraints on the $\Z3$ 3HDM parameter space}{Impact of constraints on the Z3 3HDM parameter space}}

We continue with our results for the $\Z3$ 3HDM with a focus on canceling contributions to $h \rightarrow \gamma \gamma$ and  $B\to X_s \gamma$, following the formulas in \eqs{eq:gaga}{eq:formulawilson}, respectively.
For our numerical
analysis, we traded the 18 parameters of the scalar potential in favor of the equivalent set in the physical basis, as defined in Section~\ref{sec:3hdmphys}. The 12 quartic parameters are purposefully
interchanged with the 7 physical masses (two charged scalar masses labeled 
as $m_{C1}$ and $m_{C2}$, two pseudoscalar masses labeled as
$m_{A1}$ and $m_{A2}$, and three CP-even scalar masses labeled as
$m_h$, $m_{H1}$ and $m_{H2}$) and 5 mixing angles, with the ranges in
\eq{eq:scanparameters}.

\subsection{\label{sec:hgaga}\texorpdfstring{Decays of $h_{125}$ in the $\Z3$ 3HDM}{Decays of h125 in the Z3 3HDM}}

\begin{figure}[htb]
\centering
\includegraphics[width = 0.5\textwidth]{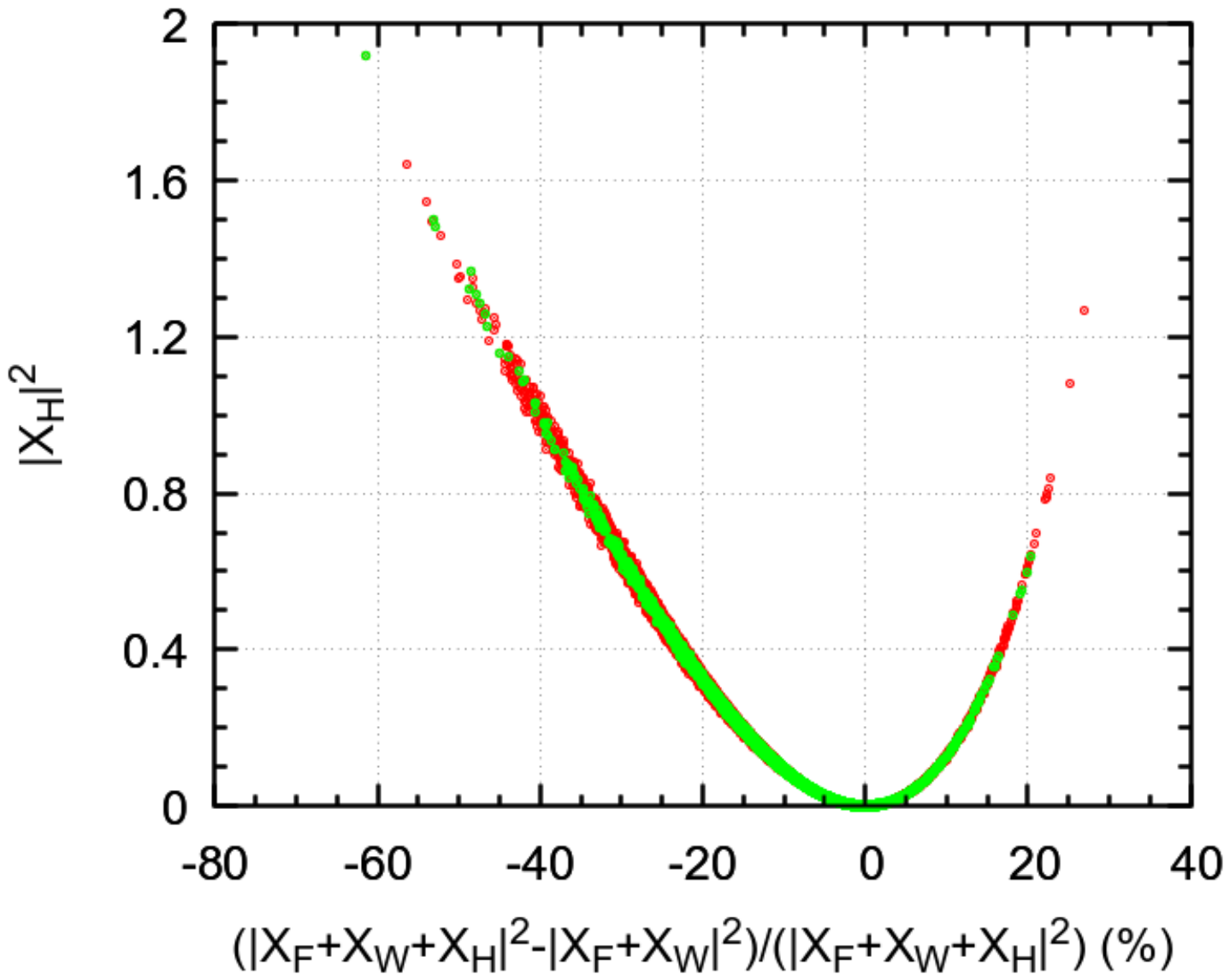}
\caption{\label{HtoGaGa}Effect of the charged Higgs on the
$h_{125}\to\gamma\gamma$ decay, with the definitions of \eq{eq:gaga}.
The green points passed all constraints including \texttt{HiggsBounds-5.9.1},
while the red points did not pass \texttt{HiggsBounds-5.9.1}.}
\end{figure}

The contribution from the two charged scalars to the $h_{125}\to\gamma\gamma$ decay process is shown in Fig.~\ref{HtoGaGa}. There are two interesting regimes.
To the left (right) of the vertical line at coordinate zero,
the two charged Higgs conspire to decrease (increase) the branching
ratio into $\gamma\gamma$.
Most of the points are on the left and correspond
to a significant reduction of the decay width.
However, there are indeed points on the right,
which allow for an increase which could be up by 20\%. 
We have also confirmed the existence of allowed results where
the destructive interference between the two charged Higgs
leads to a null $X_H$, occurring when the signs of the
couplings $\lambda_{h_jH_1^+H_1^-}$ and $\lambda_{h_jH_2^+H_2^-}$
are opposite in \eq{XHformula}.
This means that, barring other constraints,
the charged Higgs masses could be relatively light without
contradicting the observed $h_{125}\to\gamma\gamma$,
as long as their contributions to this decay canceled,
as they may.
\noindent
The points of Fig.~\ref{HtoGaGa} where $|X_H|^2$ is large,
for which the charged Higgs provide a considerable contribution
to the overall $h_{125}\to\gamma\gamma$ decay rate (the latter, still
within current bounds) is only obtained for very fine tuned
points in parameter space with some charged Higgs mass
below $200~{\rm GeV}$.
As we will see in Figs.~\ref{fig:6}-\ref{fig:7} below,
this is a very constrained (fine tuned) region.

For the allowed regions for the other signal strengths $\mu_{if}^h$, our set of points is  shown in Figs.~\ref{fig:gaga-zz} -~\ref{fig:gaga-Zga}. Similar to the complex 2HDM analyzed by Fontes, Rom\~{a}o and Silva
in~\cite{Fontes:2014xva}, there is a strong correlation between $\mu_{Z\gamma}$ and $\mu_{\gamma\gamma}$ in our Type-Z model, as shown in Fig.~\ref{fig:gaga-Zga}. 
\begin{table}[htb]
	\begin{minipage}{0.45\linewidth}
     \centering
    \includegraphics[width = 0.95\textwidth]{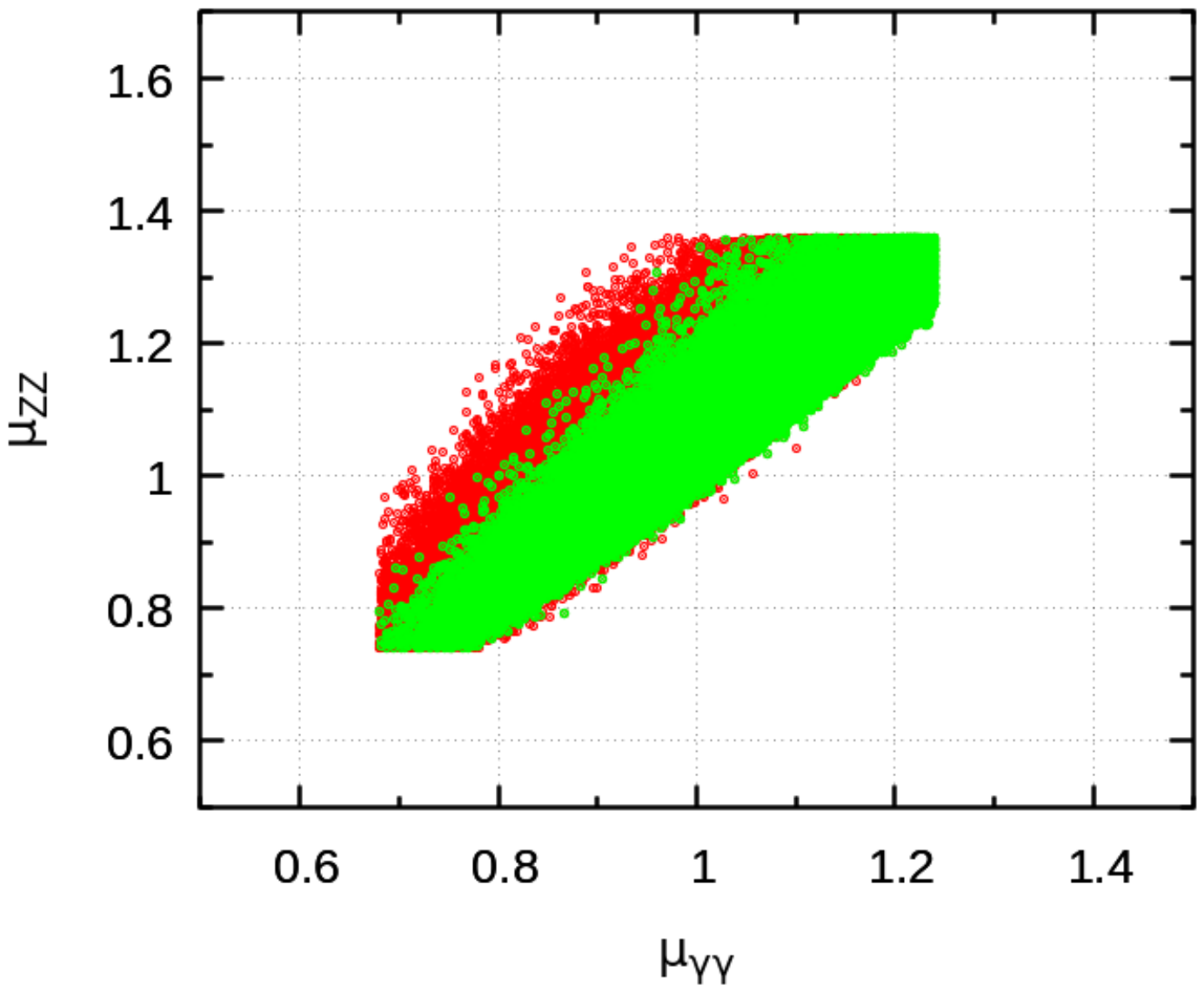}
    \captionof{figure}{Results in the $\mu_{ZZ}-\mu_{\gamma\gamma}$ plane for the gluon fusion production channel. The green points passed all constraints
    including \texttt{HiggsBounds-5.9.1}, while the red points do not.}
    \label{fig:gaga-zz}
	\end{minipage}\hfill
    \begin{minipage}{0.45\linewidth}
     \centering
    \includegraphics[width = 0.95\textwidth]{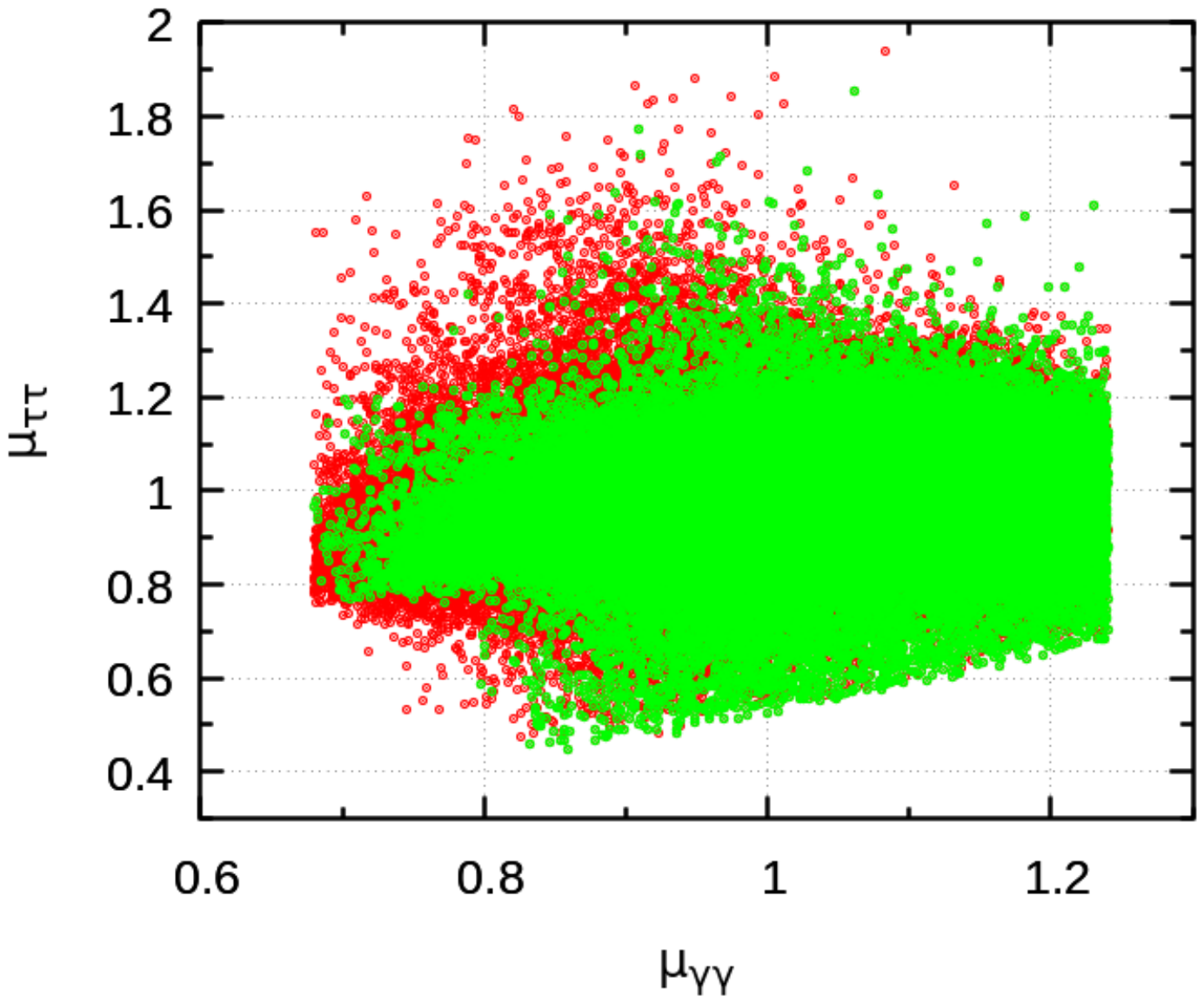}
    \captionof{figure}{Results in the $\mu_{\tau\tau}-\mu_{\gamma\gamma}$ plane for all production channels. The green points passed all constraints
        including \texttt{HiggsBounds-5.9.1}, while the red points do not.}
    \label{fig:gaga-tautau}
	\end{minipage}
\end{table}
\begin{table}[htb]
	\begin{minipage}{0.45\linewidth}
     \centering
    \includegraphics[width = 0.95\textwidth]{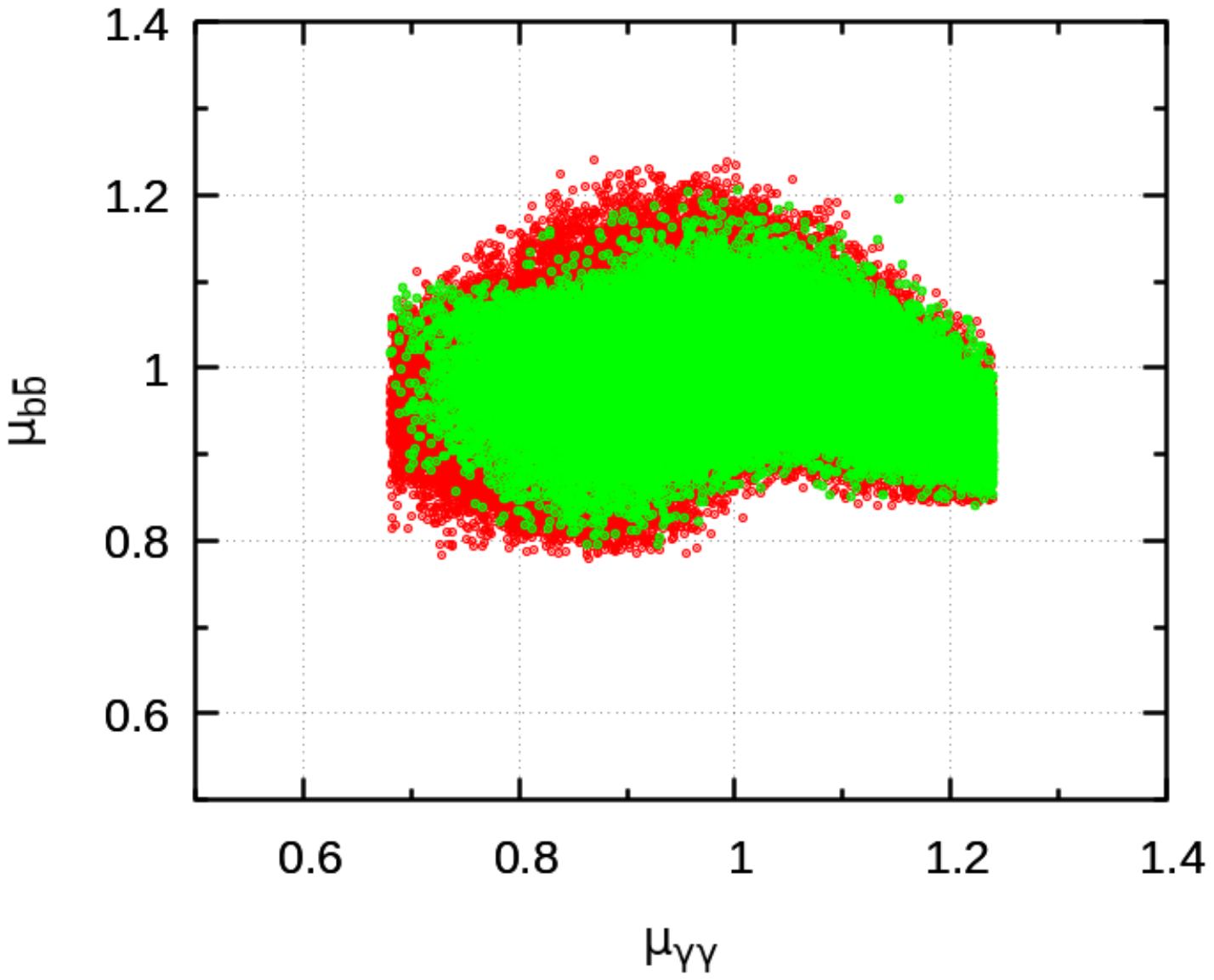}
    \captionof{figure}{Results in the $\mu_{b\overline{b}}-\mu_{\gamma\gamma}$ plane for the gluon fusion production channel. The green points passed all constraints
        including \texttt{HiggsBounds-5.9.1}, while the red points do not.}
    \label{fig:gaga-bb}
	\end{minipage}\hfill
    \begin{minipage}{0.45\linewidth}
     \centering
    \includegraphics[width = 0.95\textwidth]{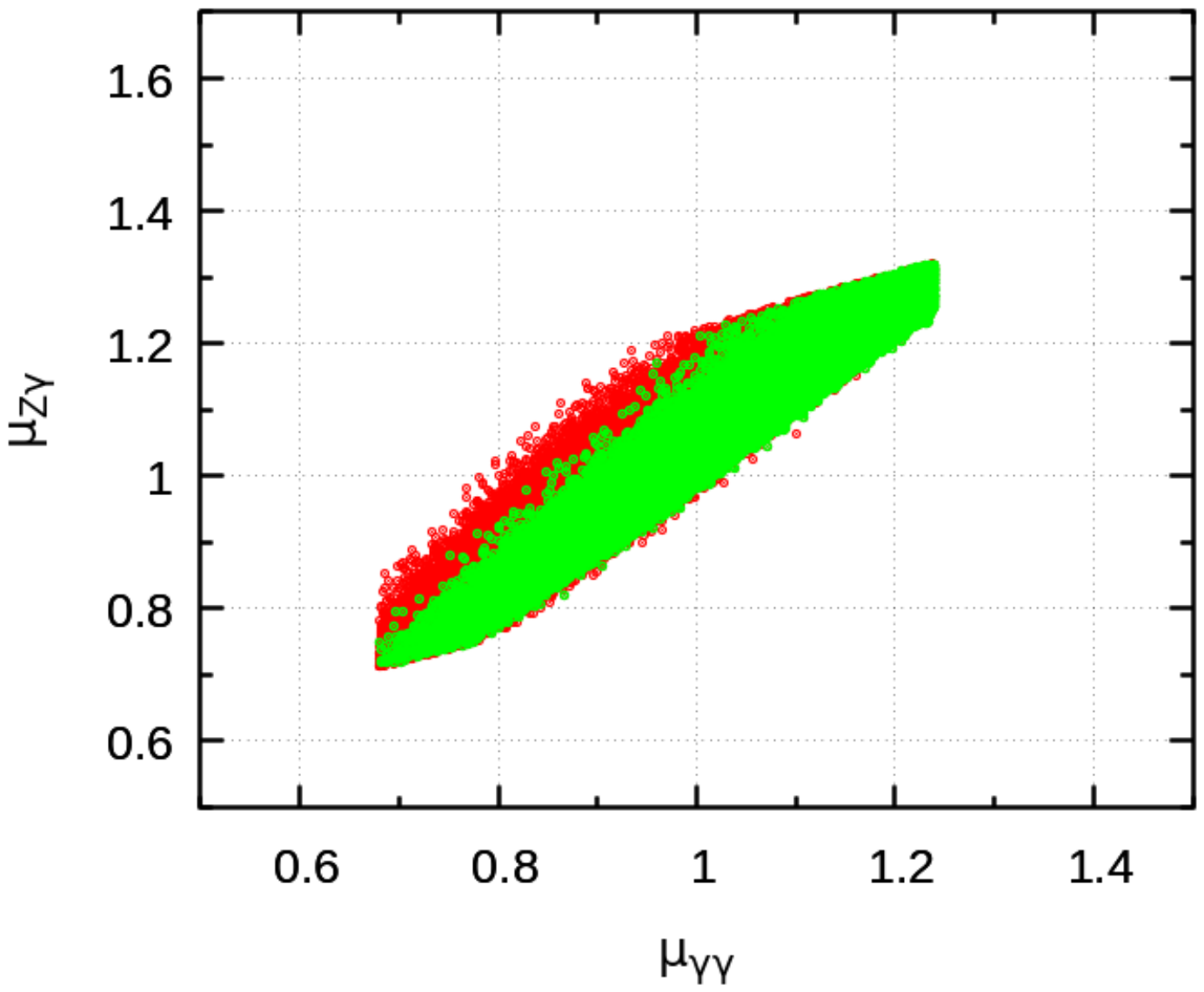}
    \captionof{figure}{Results in the $\mu_{Z\gamma}-\mu_{\gamma\gamma}$ plane for the gluon fusion production channel. The green points passed all constraints
        including \texttt{HiggsBounds-5.9.1}, while the red points do not.}
    \label{fig:gaga-Zga}
	\end{minipage}
\end{table}
Such a correlation is also visible between $\mu_{ZZ}$ and $\mu_{\gamma\gamma}$
in Fig.~\ref{fig:gaga-zz}.
It is less apparent in correlations with $\tau^+ \tau^-$
and $b \bar{b}$,
as shown in Figs.~\ref{fig:gaga-tautau} and~\ref{fig:gaga-bb}.

\subsection{\texorpdfstring{Impact of $b \rightarrow s \gamma$}{Impact of b to s gamma}\label{Z3bsgamma}}

We find that much of the parameter space considered in
Ref.~\cite{Chakraborti:2021bpy} is forbidden. This
is most apparent by considering their Fig.~2,
which we turn to next.

\subsubsection{\texorpdfstring{Only $b\to s \gamma$}{Only b to s gamma}}

\begin{figure}[htb]
    \centering
    \begin{tabular}{cc}
      \includegraphics[width=0.48\textwidth]{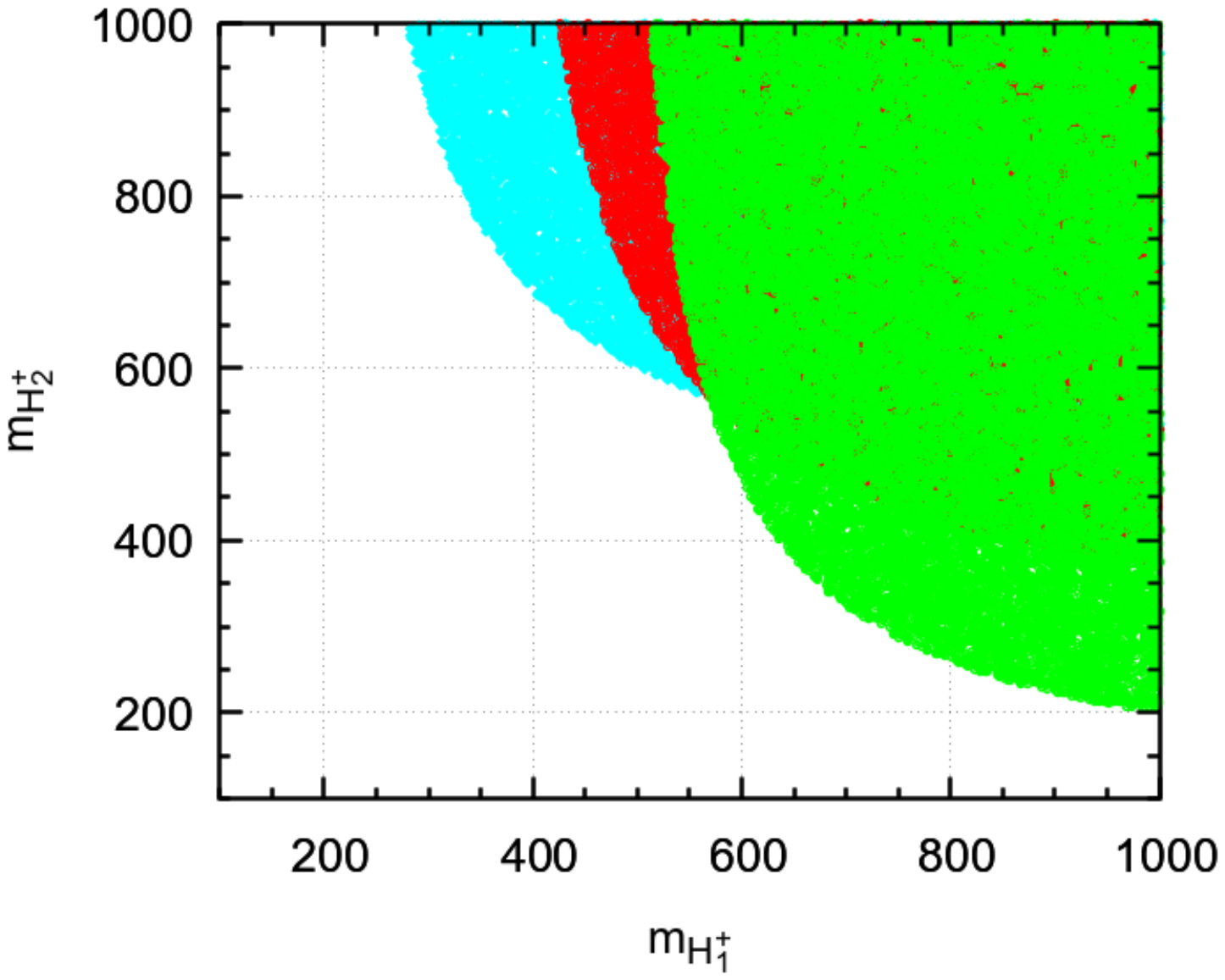}
      &
        \includegraphics[width=0.48\textwidth]{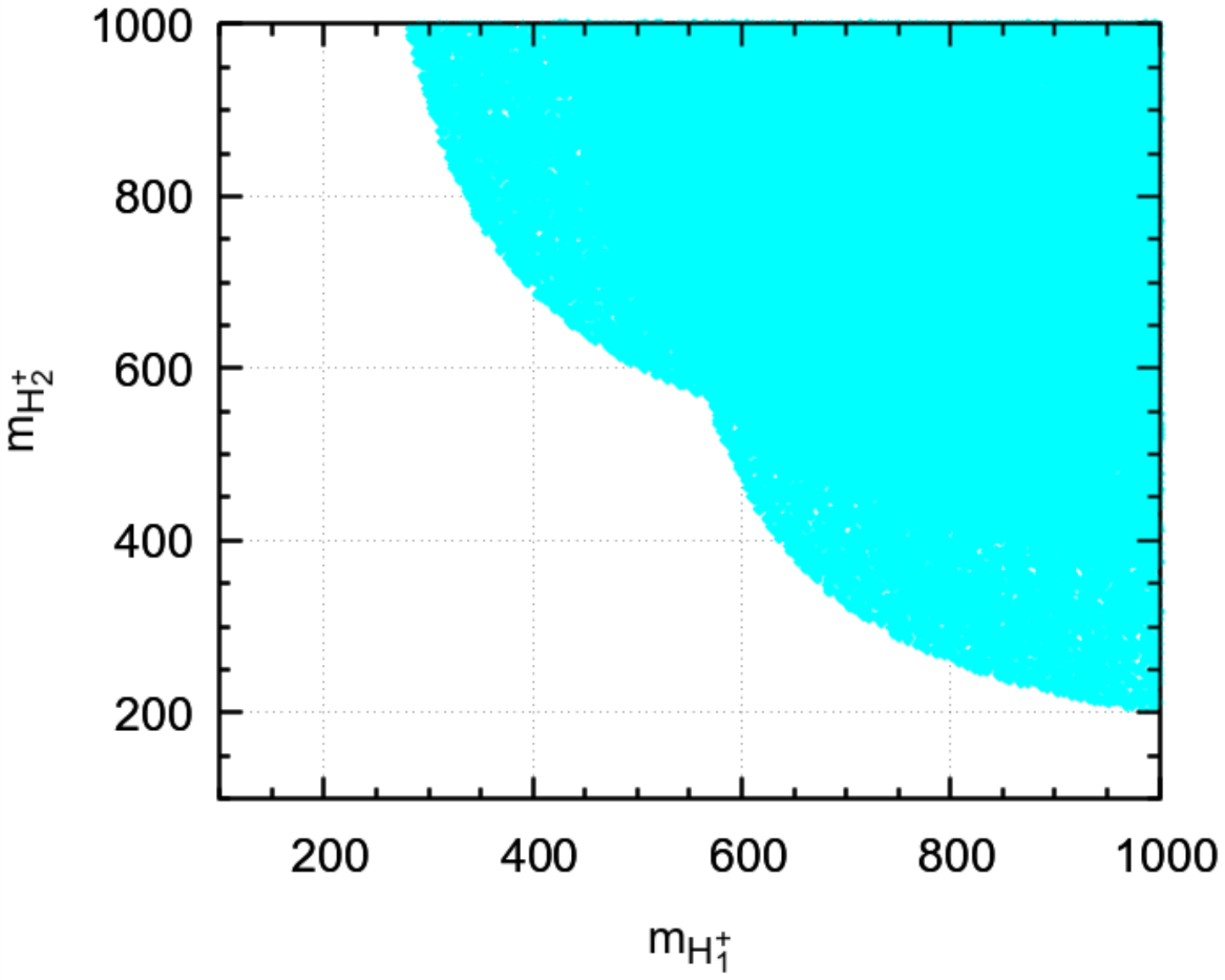}
    \end{tabular}
    
    \caption{Comparison with Fig.~2 of
      Ref.~\cite{Chakraborti:2021bpy}. On the left panel we have
      $\gamma_2= \frac{\pi}{6}$ (cyan),  $\gamma_2= \frac{\pi}{4}$
      (red), $\gamma_2= \frac{\pi}{3}$ (green). On the right panel the 3 regions
    are superimposed.}
    \label{fig:1}
  \end{figure}

On Fig.~2 of Ref.~\cite{Chakraborti:2021bpy} the parameters are fixed as
\begin{equation}
  \label{eq:scanbsgamma}
  \tan\beta_1=10,\quad\tan\beta_2=2,\quad
  \gamma_2=\frac{\pi}{6},\frac{\pi}{4},\frac{\pi}{3} \, ,
\end{equation}
while imposing the exact conditions of \eq{eq:alignment}
and \eq{eq:alignment2}.
Applying only the $b\to s \gamma$ cut we reproduce their Fig.~2
in our Fig.~\ref{fig:1}.
\noindent
Fig.~\ref{fig:1} passes all theoretical constraints,
even including unitarity and BFB.

From the previous plots, the conclusion that we can have
one of the charged Higgs relatively light if the other is sufficiently
heavy seems correct. However we now show that for this choice of
parameters this is not the case. With the choice of
\eqsthree{eq:alignment}{eq:alignment2}{eq:scanbsgamma},
the bounds from the decays of the $125~{\rm GeV}$ Higgs are simply satisfied.
However the same is not true for current bounds on heavier scalars.
Indeed, every single point in Fig.~\ref{fig:1} is excluded by \texttt{HiggsBounds-5.9.1}; not
a single point remains.
This will be explained in detail in the following Section.

\subsubsection{Good points after LHC bounds}

\begin{figure}[htb]
  \centering
  \begin{tabular}{cc}
      \includegraphics[width=0.48\textwidth]{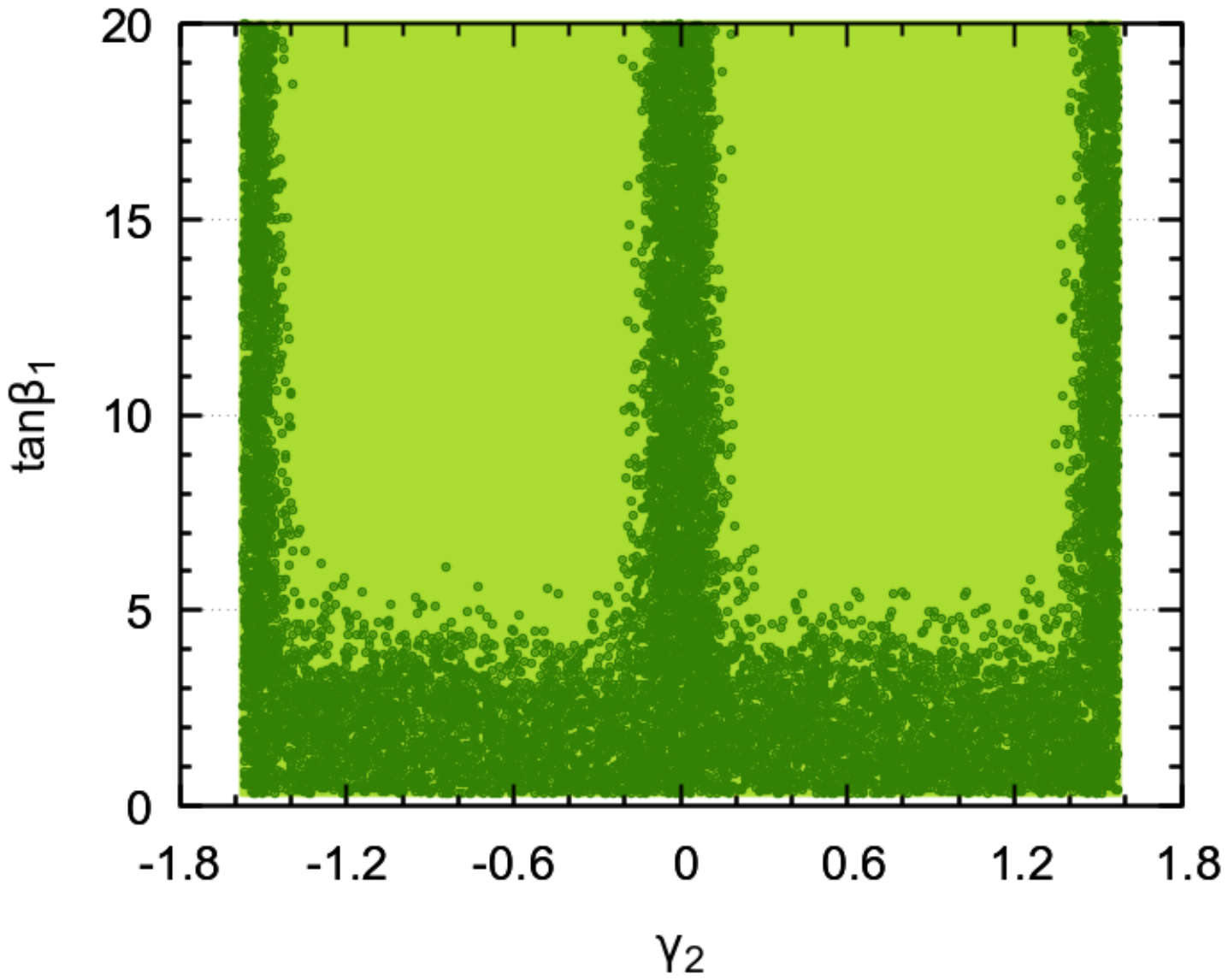}
      &
      \includegraphics[width=0.48\textwidth]{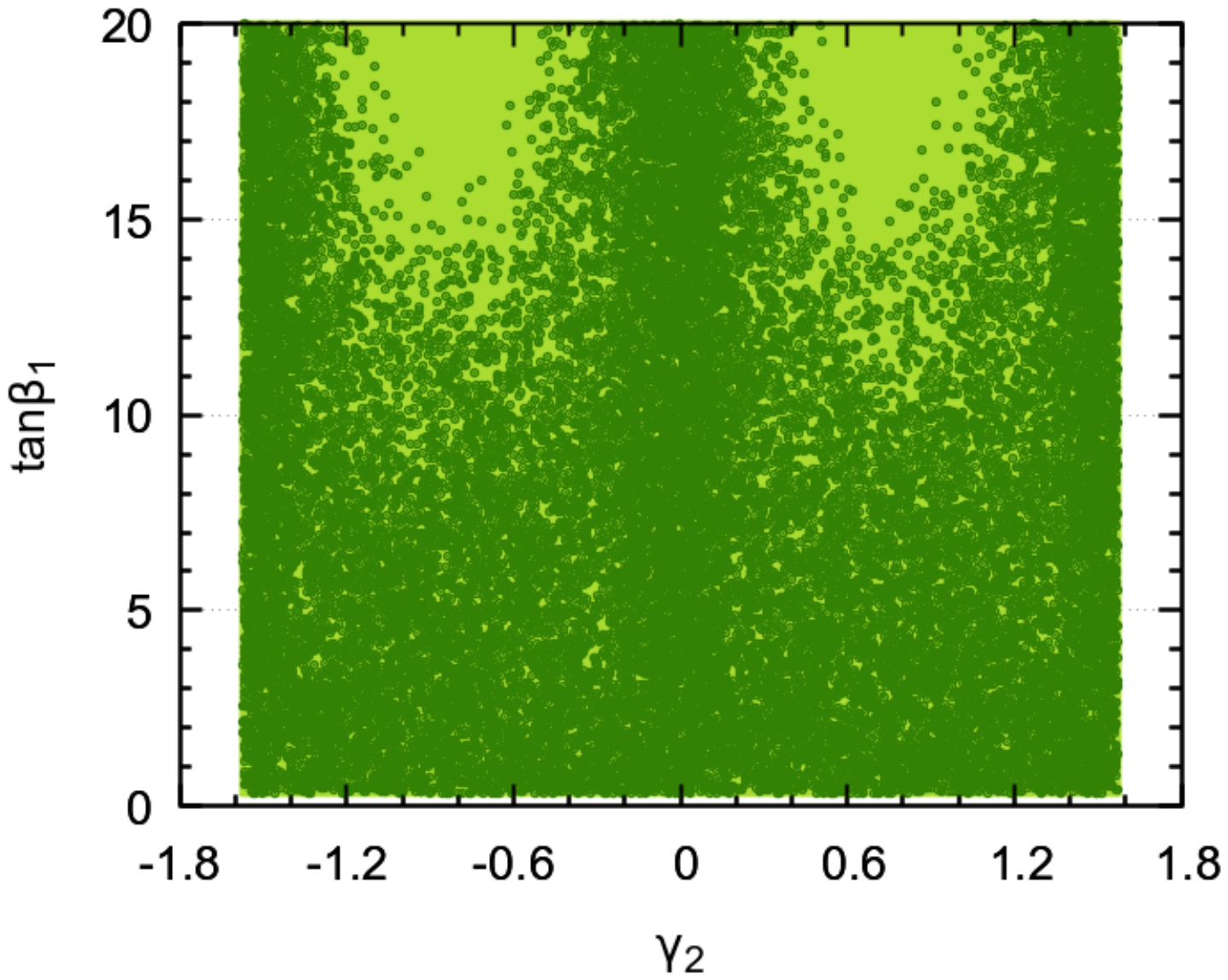}
   \end{tabular}
  \caption{\label{fig:new}Enlarging good points, taking $\gamma_2 \in
      [-\pi/2,\pi/2]$ and varying $\beta_1$, all other conditions in
      \eqsthree{eq:alignment}{eq:alignment2}{eq:scanbsgamma} were
      kept. The dark green points passed all constraints including \texttt{HiggsBounds-5.9.1},
      while the light green points did not. Left panel: All points
      passing \texttt{HiggsBounds-5.9.1}. Right panel: All points
      passing \texttt{HiggsBounds-5.7.1}.}
\end{figure}

We discovered that the situation described is
a consequence of the small range chosen for $\gamma_2$. To illustrate this, we
kept the other conditions in \eqsthree{eq:alignment}{eq:alignment2}{eq:scanbsgamma},
but allowed for
\begin{equation}
  \label{eq:gamma2}
  \gamma_2 \in [-\pi/2,\pi/2] \, ,
\end{equation}
and (for Fig.~\ref{fig:new}) also varied $\tan{\beta_1}$.
The points which survive \texttt{HiggsBounds-5.9.1} are shown in dark green on
the left panel of Fig.~\ref{fig:new}.
\noindent
The allowed points for $\tan{\beta_1}=10$ are concentrated
around $\gamma_2 = 0, \pm\pi/2$,
excluding $\gamma_2 = \pi/6,\pi/4,\pi/3$.
Taking the interval in \eq{eq:gamma2} one can indeed find
regions of good points.\footnote{This it true regardless of whether or not
we vary $\beta_1$, as long as we enlarge the region of $\gamma_2$.}

\begin{figure}[htb]
  \centering
  \begin{tabular}{cc}
    \includegraphics[width=0.48\textwidth]{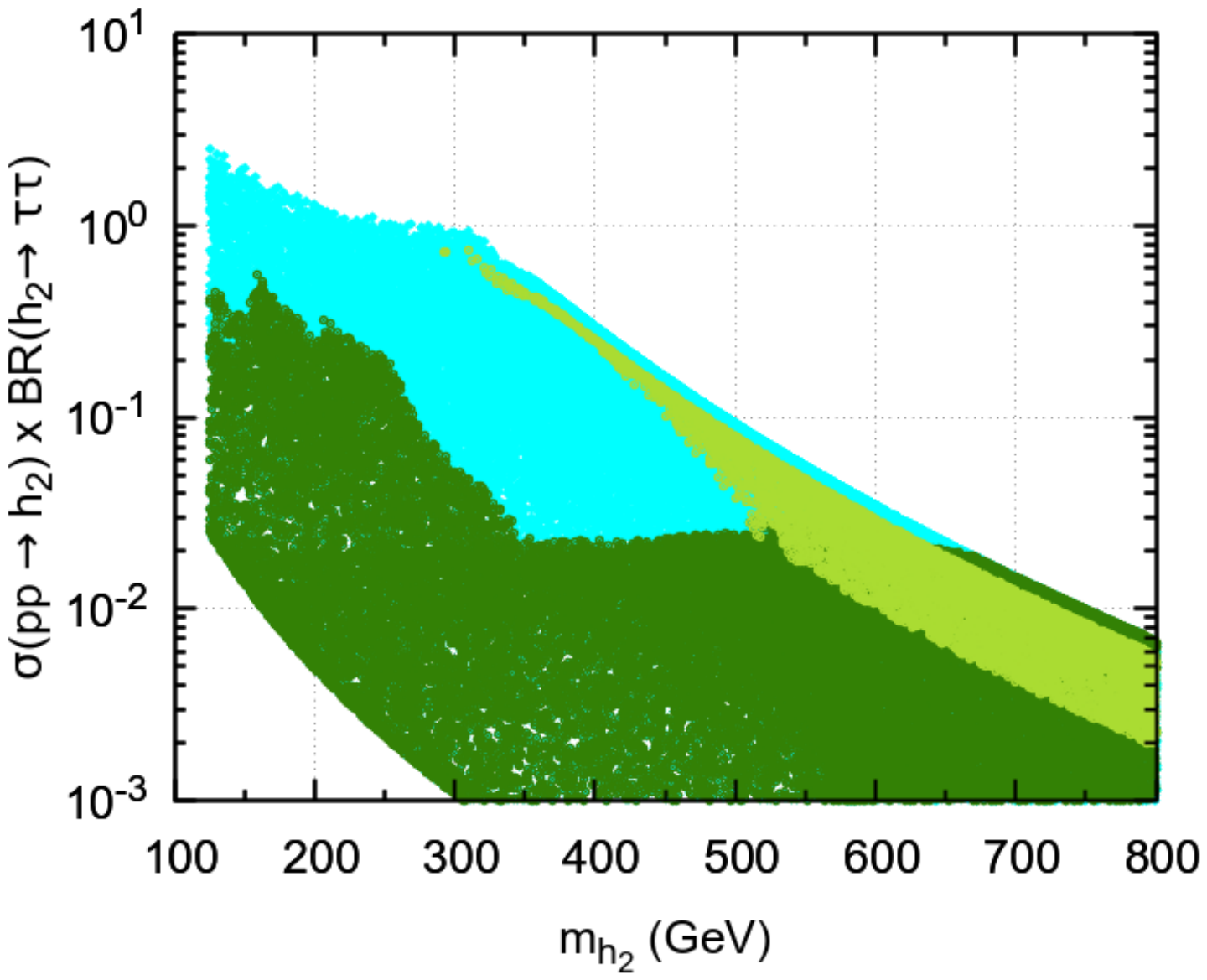}
    &
    \includegraphics[width=0.48\textwidth]{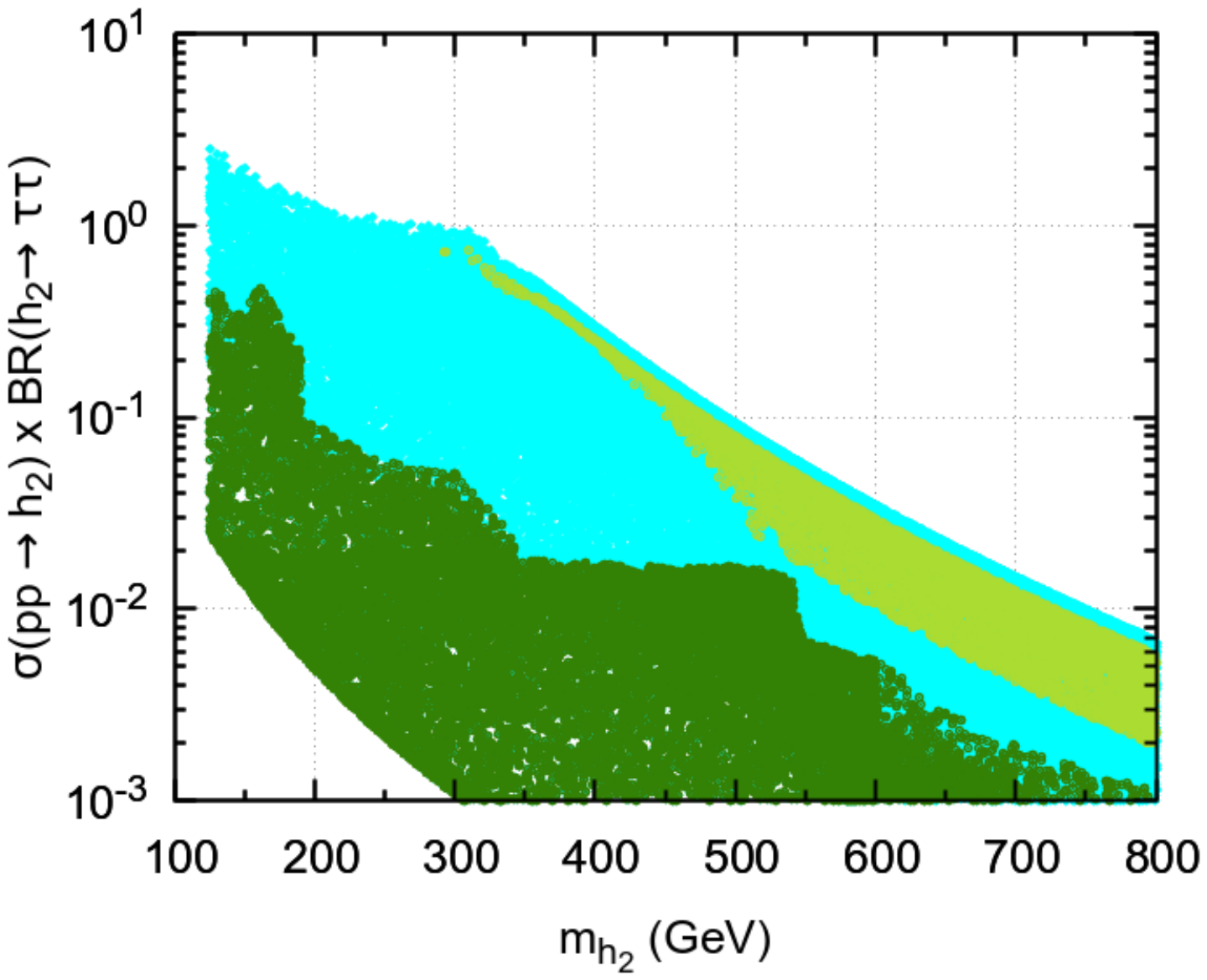}
  \end{tabular}
  \caption{\label{fig:sigBR}
    Left Panel: $\sigma(pp\to h_2) \times \text{BR}(h_2 \to \tau\tau)$
    as function of the $m_{h_2}$. Parameters are as in \eqsthree{eq:alignment}{eq:alignment2}{eq:scanbsgamma},
    except that $\gamma_2 \in [-\pi/2,\pi/2]$. Points in
    cyan are points that pass all constraints before
    \texttt{HiggsBounds}
    and in dark
    green after \texttt{HiggsBounds-5.7.1}. In light green are the points in the
    interval $\gamma_2 
    \in [\pi/6,\pi/3]$. Right Panel: the same but for
    \texttt{HiggsBounds-5.9.1}}  
\end{figure}

It is interesting to compare with what happens with the previous
version of \texttt{HiggsBounds-5.7.1}, shown on
the right panel of Fig.~\ref{fig:new}.
For that case there are many points allowed for all values of $\gamma_2$,
even for $\tan{\beta_1}=10$.
We have found that this is due to the recent bounds on
$h_{2,3} \rightarrow \tau^+ \tau^-$ decay in
Ref.~\cite{ATLAS:2020zms},
included in \texttt{HiggsBounds-5.9.1} but not in
\texttt{HiggsBounds-5.7.1},
which used the previous
bounds~\cite{CMS:2015mca,CMS:2017epy}.\footnote{In
Ref.~\cite{Chakraborti:2021bpy} the strong constraints from neutral
scalar decays 
into $\tau\tau$ still seemed to allow points with the choices in
\eqsthree{eq:alignment}{eq:alignment2}{eq:scanbsgamma}.
}
To better illuminate this point,
we show  $\sigma(pp\to h_2) \times \text{BR}(h_2 \to \tau\tau)$
versus $m_{h_2}$ in Fig.~\ref{fig:sigBR}.
In this figure, the parameters are as in \eqsthree{eq:alignment}{eq:alignment2}{eq:scanbsgamma}, 
except that $\gamma_2 \in [-\pi/2,\pi/2]$.
Points in cyan are points that pass all constraints before
\texttt{HiggsBounds}.
In light green are the points in the restricted
interval $\gamma_2  \in [\pi/6,\pi/3]$.
In the left panel points in  
dark green  are those which survided after
\texttt{HiggsBounds-5.7.1}.
In the right panel we have
the same situation but now we used \texttt{HiggsBounds-5.9.1}. We see that
there were good points in the restricted interval $\gamma_2  \in
[\pi/6,\pi/3]$ in the left panel, but they disappeared with the
newer version \texttt{HiggsBounds-5.9.1}. We have confirmed that
similar plots can be obtained for $h_3$.
This is a good point to stress again the role that the LHC is having in
constraining models with new scalar physics.
One sees the strong impact that the updated LHC results have
in constraining the $\Z3$ 3HDM.
This highlights the importance that the new LHC run will have
in constraining the parameter space of extended scalar sectors.

\begin{figure}[htb]
\centering
\includegraphics[width = 0.5\textwidth]{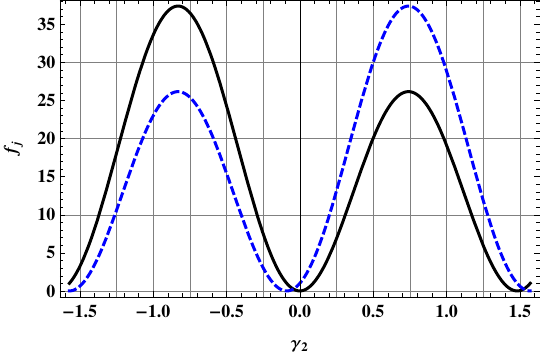}
\caption{\label{fig:fj}Graphic of the functions $f_j$ defined
in \eq{eq:cross_fi} for varying $\gamma_2=-\alpha_3$, with
$\tan{\beta_1} = 10$ and 
$\tan{\beta_2} = 2$.
Function $f_2$ ($f_3$) in black/solid (blue/dashed) line.}
\end{figure}

To better understand the behaviour of 
$\sigma(pp\to h_i) \times \text{BR}(h_i \to \tau\tau)$ ($i=2,3$) we can make
the simplified assumption\footnote{We are neglecting the dependence of
the cross section on the mass.} that this product is proportional to
\begin{equation}
  \label{eq:cross_fi}
  \sigma(pp\to h_i) \times \text{BR}(h_i \to \tau\tau)
  \propto
  g_{h_i \tau \tau}^2\, g_{h_i t t}^2 \equiv f_i\ ,
\end{equation}
where we are assuming that the production occurs mainly
via gluon fusion with the top quark in the loop.
Now, using the assumptions of \eqs{eq:alignment}{eq:alignment2}
in \eq{coeffNeutralFerm-Type-Z}, we have
\begin{align}
  \label{eq:8}
  &g_{h_2 \tau\tau} = - \frac{c_{\alpha_3} t_{\beta_1}}{c_{\alpha_2}}
  - s_{\alpha_3} t_{\beta_2}
= -\frac{t_{\beta_1}}{c_{\beta_2}}\, c_{\gamma_2} +t_{\beta_2}
s_{\gamma_2}\, ,
  && g_{h_2 tt} = \frac{c_{\alpha_2} s_{\alpha_3}}{s_{\beta_2}} = -
  \frac{1}{t_{\beta_2}}\, s_{\gamma_2}
  \ ,\nonumber\\[+2mm]
  &g_{h_3 \tau\tau} = 
-c_{\alpha_3} t_{\beta_2}+
\frac{s_{\alpha_3}t_{\beta_1}}{c_{\alpha_2}}
=- t_{\beta_2} c_{\gamma_2} - \frac{t_{\beta_1}}{c_{\beta_2}}
s_{\gamma_2} \, ,
&& g_{h_3 tt} = \frac{c_{\alpha_2} c_{\alpha_3}}{s_{\beta_2}}
= \frac{1}{t_{\beta_2}}\, c_{\gamma_2}
  \ ,
\end{align}
where, for Fig.~\ref{fig:sigBR}, $\beta_1,\beta_2$ are fixed and
$\gamma_2 \in [-\pi/2,\pi/2]$.
Fig.~\ref{fig:fj} shows the functions in \eq{eq:cross_fi} --
$f_2$ for $h_2$ and $f_3$ for $h_3$ -- for $\tan{\beta_1} = 10$ and
$\tan{\beta_2} = 2$ as in \eq{eq:scanbsgamma},
but keeping $\gamma_2$ free.
\noindent
We see that these functions are largest precisely in the approximate interval
$\pm \gamma_2 \in [\pi/6, \pi/3]$. This explains why these points are the first
to be excluded by the bounds on
$\sigma(pp\to h_2) \times \text{BR}(h_2 \to \tau\tau)$,
and why, going outside such bounds, some points can be
preserved.\footnote{Of course, we have ignored in this simple reasoning the
dependence on $m_{h_i}$, which has been taken into account appropriately
in our scans and \texttt{HiggsBounds-5.9.1} limits.}

\subsection{\texorpdfstring{The effect of $\tan\beta$'s}{The effect of tan beta's}}

\begin{figure}[htb]
  \centering
  \begin{tabular}{cc}
      \includegraphics[width=0.48\textwidth]{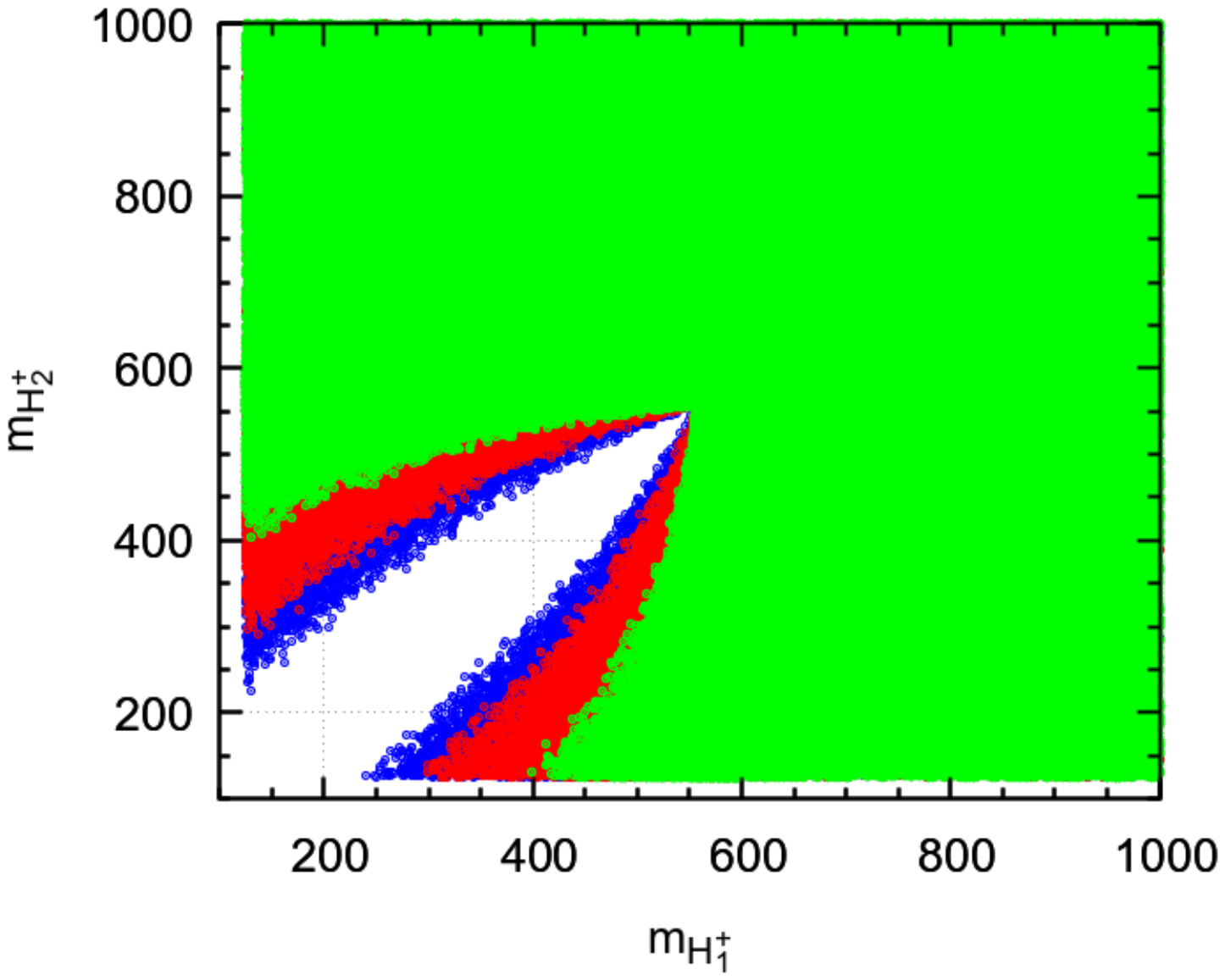}
      &
      \includegraphics[width=0.48\textwidth]{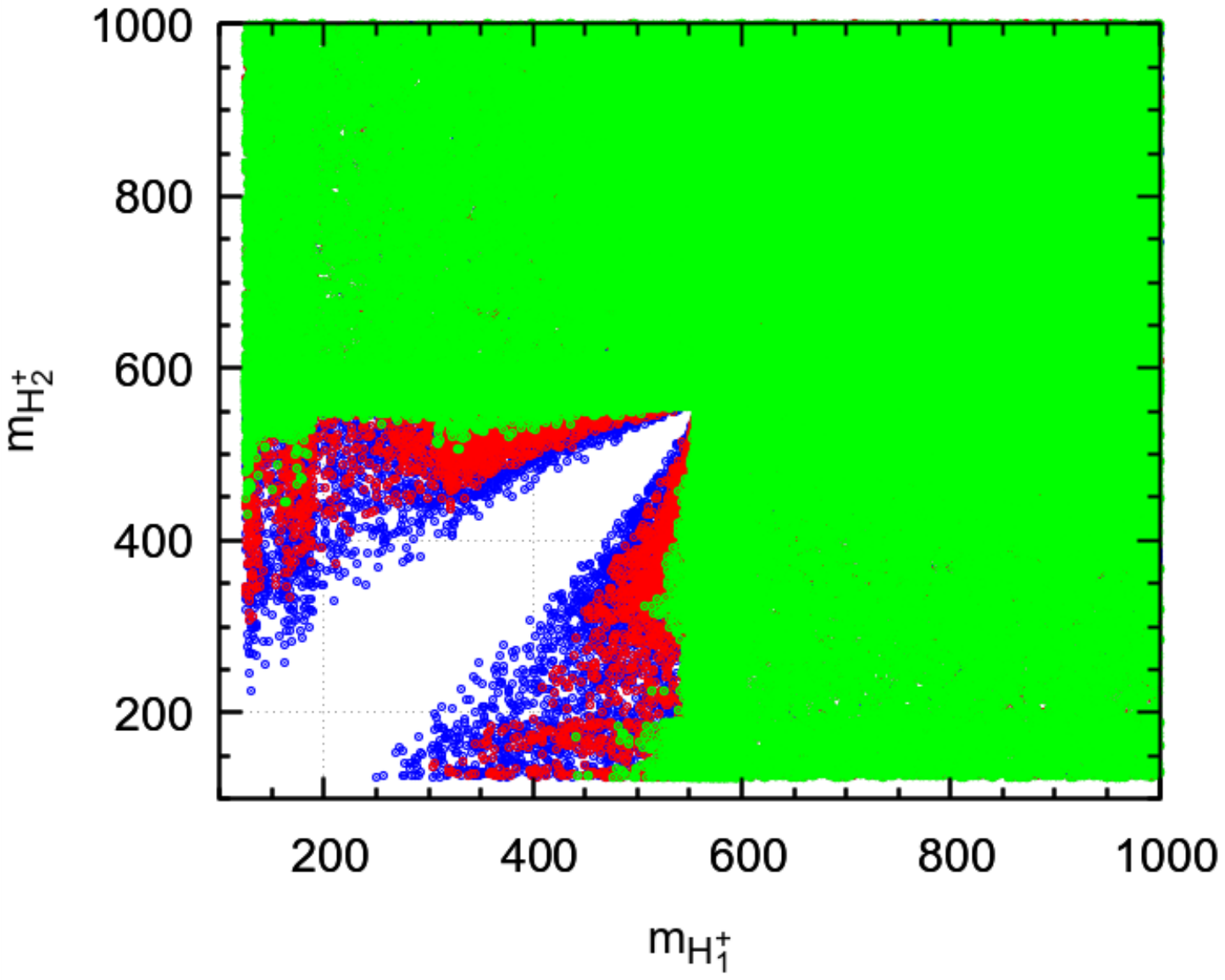}
   \end{tabular}
  \caption{All points satisfy exact alignment by \eqs{eq:alignment}{eq:alignment2}. Left panel: All points
    passed all constraints except for 
    \texttt{HiggsBounds-5.9.1}.
    The blue
    points satisfy \eq{eq:tanbetaeffect}. The red points are for
    $\tan\beta_{1,2}>0.5$ and the green points
    are for $\tan\beta_{1,2}>1$. Right panel: same color code as in
    the left panel but only showing points surviving after requiring
    \texttt{HiggsBounds-5.9.1}.}
  \label{fig:5}
\end{figure}

In the last Section we saw that while maintaining the main features of
\eqsthree{eq:alignment}{eq:alignment2}{eq:scanbsgamma}, but enlarging the range of variation of
$\gamma_2$, 
we could find points allowed by all current experimental constraints.
Here we exploit the variation of both
$\tan\beta$'s in the range
\begin{equation}
  \label{eq:tanbetaeffect}
  \tan\beta_{1,2} \in [10^{-0.5},10] \, ,
\end{equation}
subject to the condition of perturbativity of the Yukawa couplings in
\eq{eq:perturbativeyuk_req}. The result is shown in Fig.~\ref{fig:5}. 
\noindent
We see that by varying the range of $\tan\beta$'s we can have smaller
masses for the charged Higgs bosons. For $\tan\beta< 1$ it is even possible
to have both charged Higgs with masses below $400~{\rm GeV}$.

\subsection{\label{sec:beyondalign}Going beyond exact alignment}

We have performed a completely uniform scan and found out that very
few points survived and those were not too far away from the alignment
condition of \eqs{eq:alignment}{eq:alignment2}. So another strategy can be to scan
points that differ from the perfect alignment by
1\% or 10\%.

\begin{figure}[htb]
  \centering
  \begin{tabular}{cc}
      \includegraphics[width=0.48\textwidth]{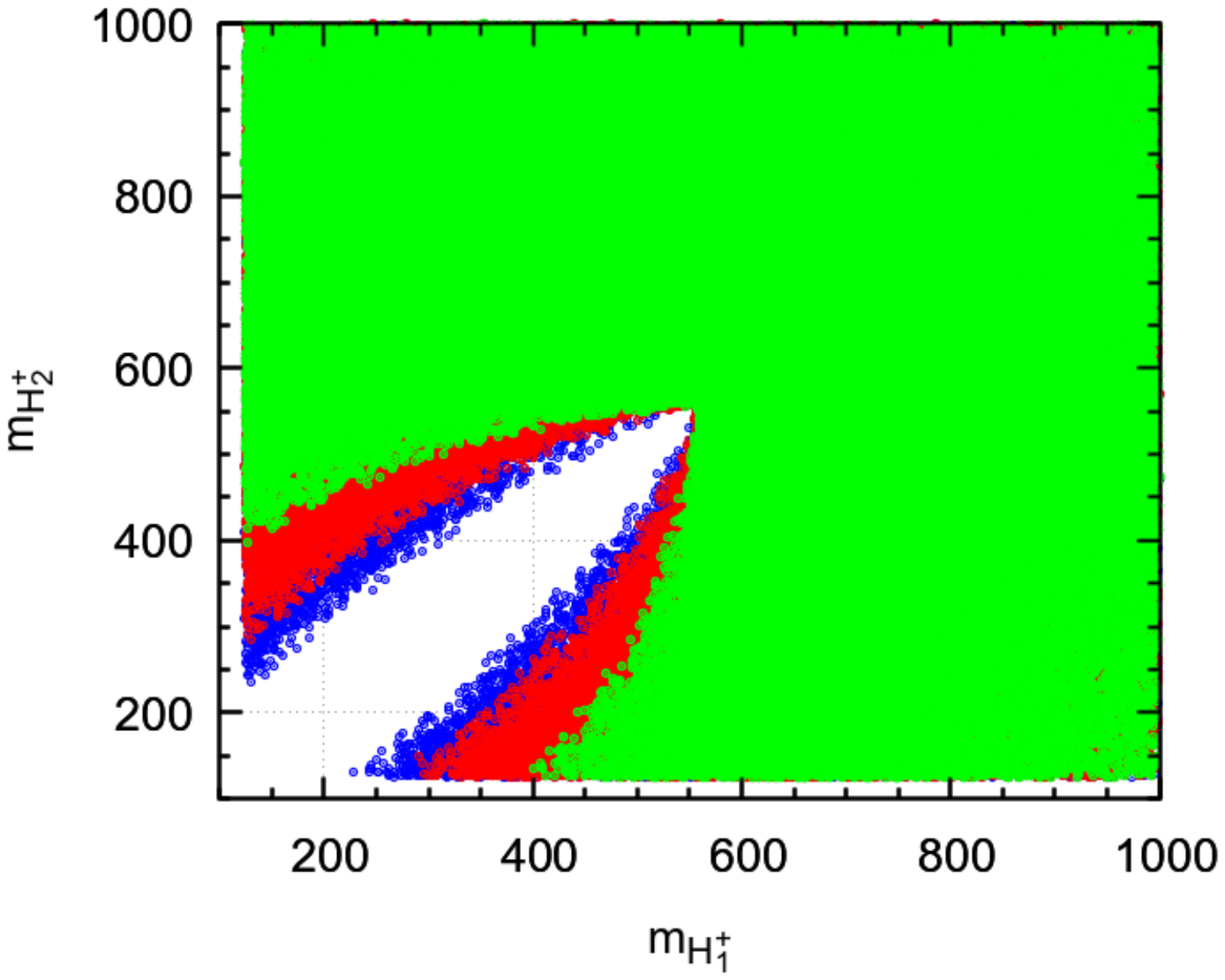}
      &
      \includegraphics[width=0.48\textwidth]{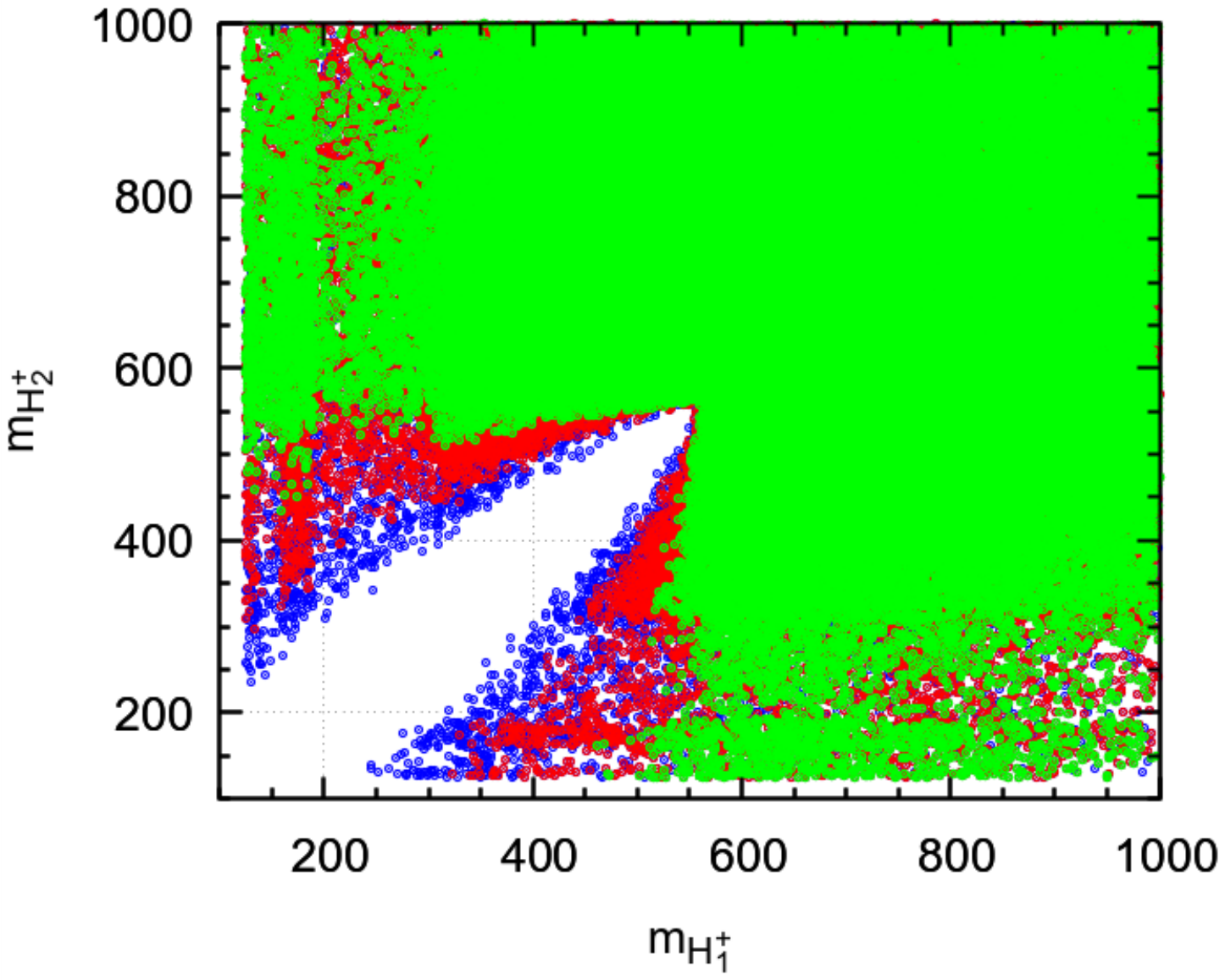}
   \end{tabular}
  \caption{All points are within 1\% of the perfect alignment of
    \eqs{eq:alignment}{eq:alignment2}.
    Left panel: All points passed all constraints except for \texttt{HiggsBounds-5.9.1}.
    The blue
    points satisfy \eq{eq:tanbetaeffect}. The red points are for
    $\tan\beta_{1,2}>0.5$ and the green points
        are for $\tan\beta_{1,2}>1$. Right panel: same color code as in
    the left panel but only showing points surviving after requiring \texttt{HiggsBounds-5.9.1}.}
  \label{fig:6}
\end{figure}

In Fig.~\ref{fig:6} we show the results for the case when we allow the
parameters 
to differ 1\% from the perfect alignment limit.

\begin{figure}[htb]
  \centering
  \begin{tabular}{cc}
      \includegraphics[width=0.48\textwidth]{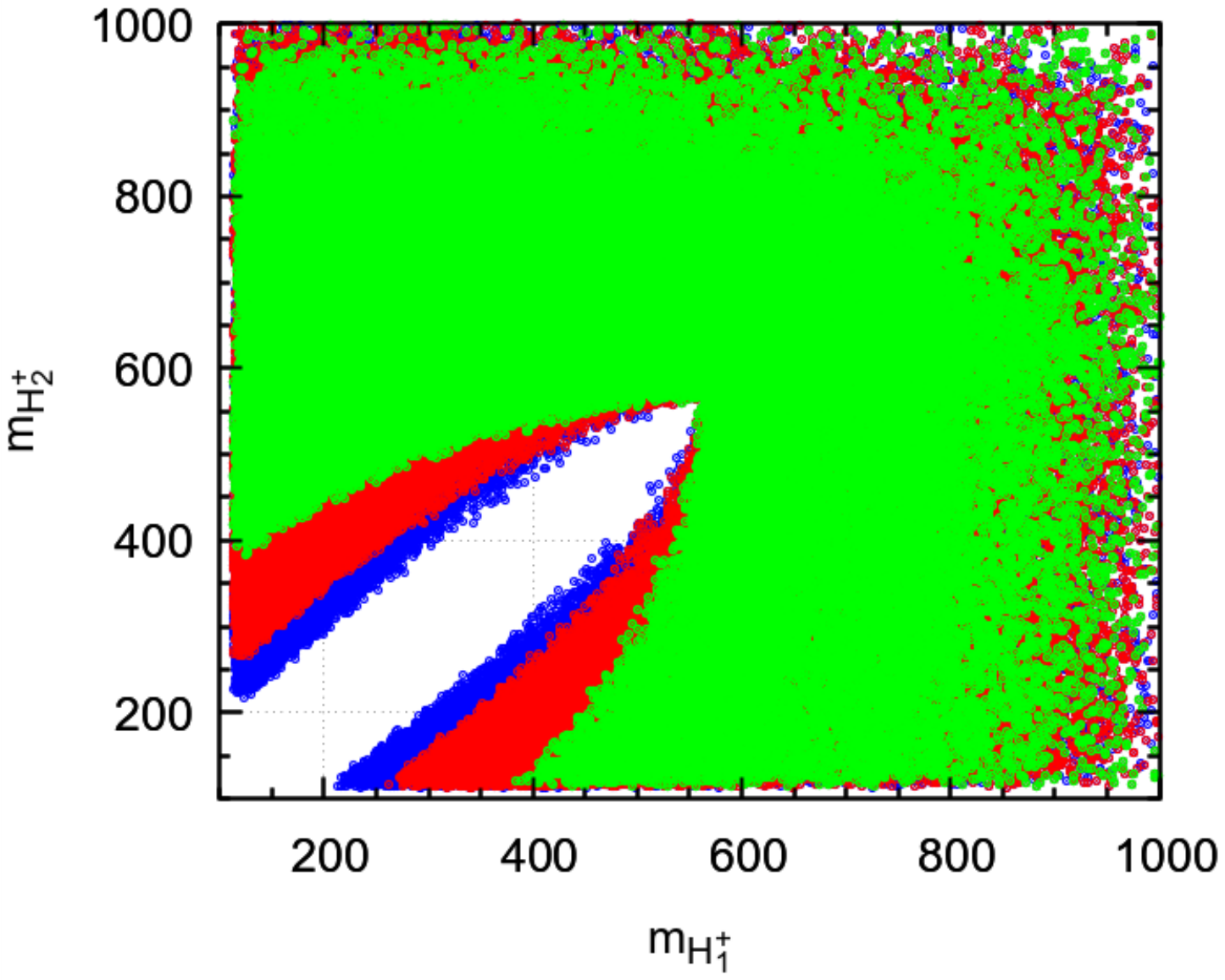}
      &
      \includegraphics[width=0.48\textwidth]{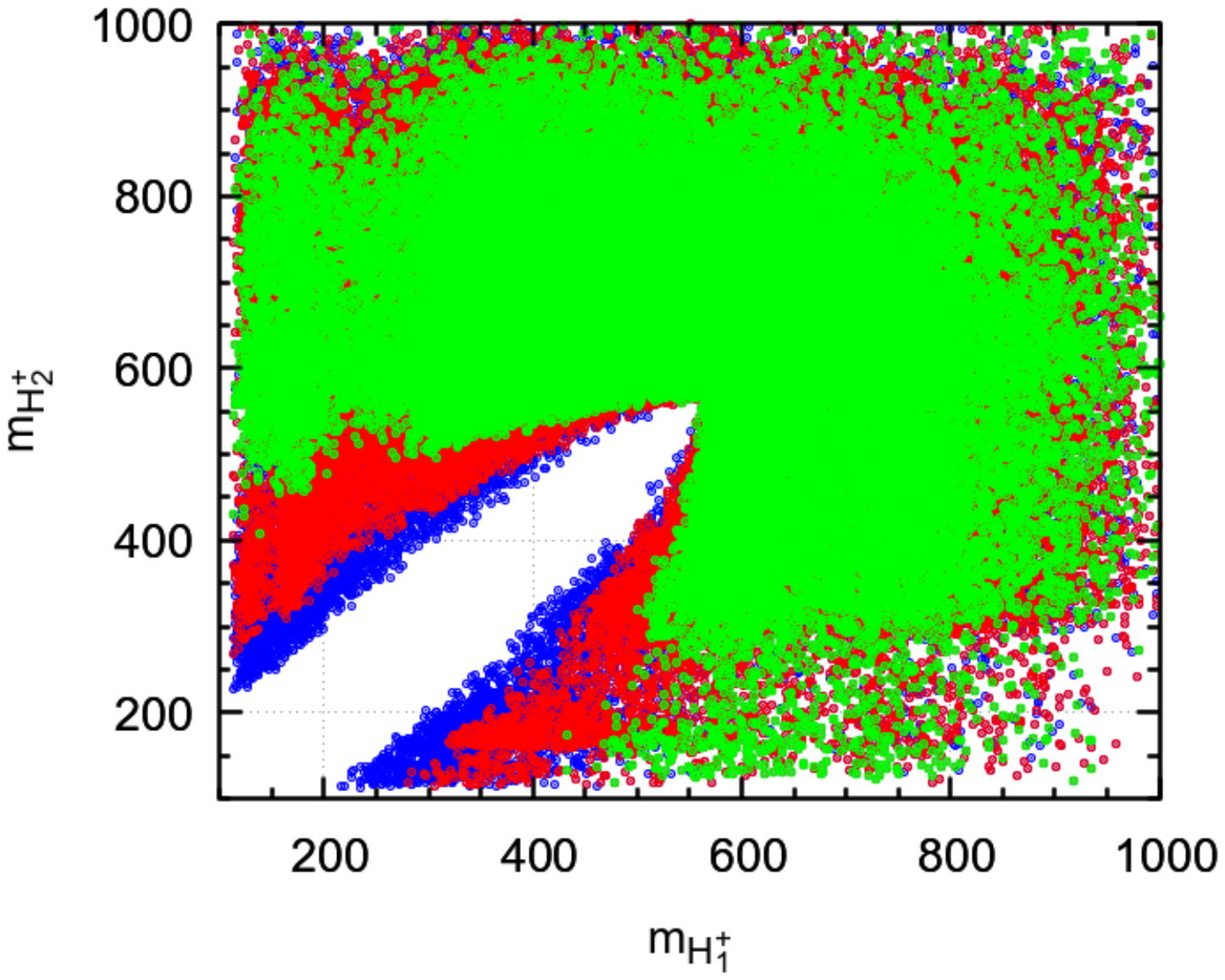}
   \end{tabular}
  \caption{All points are within 10\% of the perfect alignment of
   \eqs{eq:alignment}{eq:alignment2}.
    Left panel: All points passed all constraints except for \texttt{HiggsBounds-5.9.1}.
    The blue
    points satisfy \eq{eq:tanbetaeffect}. The red points are for
    $\tan\beta_{1,2}>0.5$ and the green points
        are for $\tan\beta_{1,2}>1$. Right panel: same color code as in
    the left panel but only showing points surviving after requiring \texttt{HiggsBounds-5.9.1}.}
  \label{fig:7}
\end{figure}

Next we considered the case when the difference for perfect alignment
was 10\%. This is shown in Fig.~\ref{fig:7}.
\noindent
We see that the acceptable points which differ more from perfect alignment
are less frequent, as expected.\footnote{To be more specific,
for the same number of points generated with the constraint of alignment
within 10\% or 1\%, fewer of the former are obtained which pass all
requirements.} 
Nevertheless,
one can still find many points which differ from exact alignment by
as much as 10\%, while satisfying all experimental and theoretical constraints.
And such points do allow for qualitatively different predictions,
as we saw when looking at the charged scalar masses consistent with
$b \rightarrow s \gamma$.
We conclude that imposing perfect alignment is too constraining and
does not cover 
all the interesting features of the $\Z3$ 3HDM.

\subsection{\label{subsec:unusual}Unusual signals of charged scalars}

\begin{figure}[htb]
\centering
\includegraphics[width=0.48\textwidth]{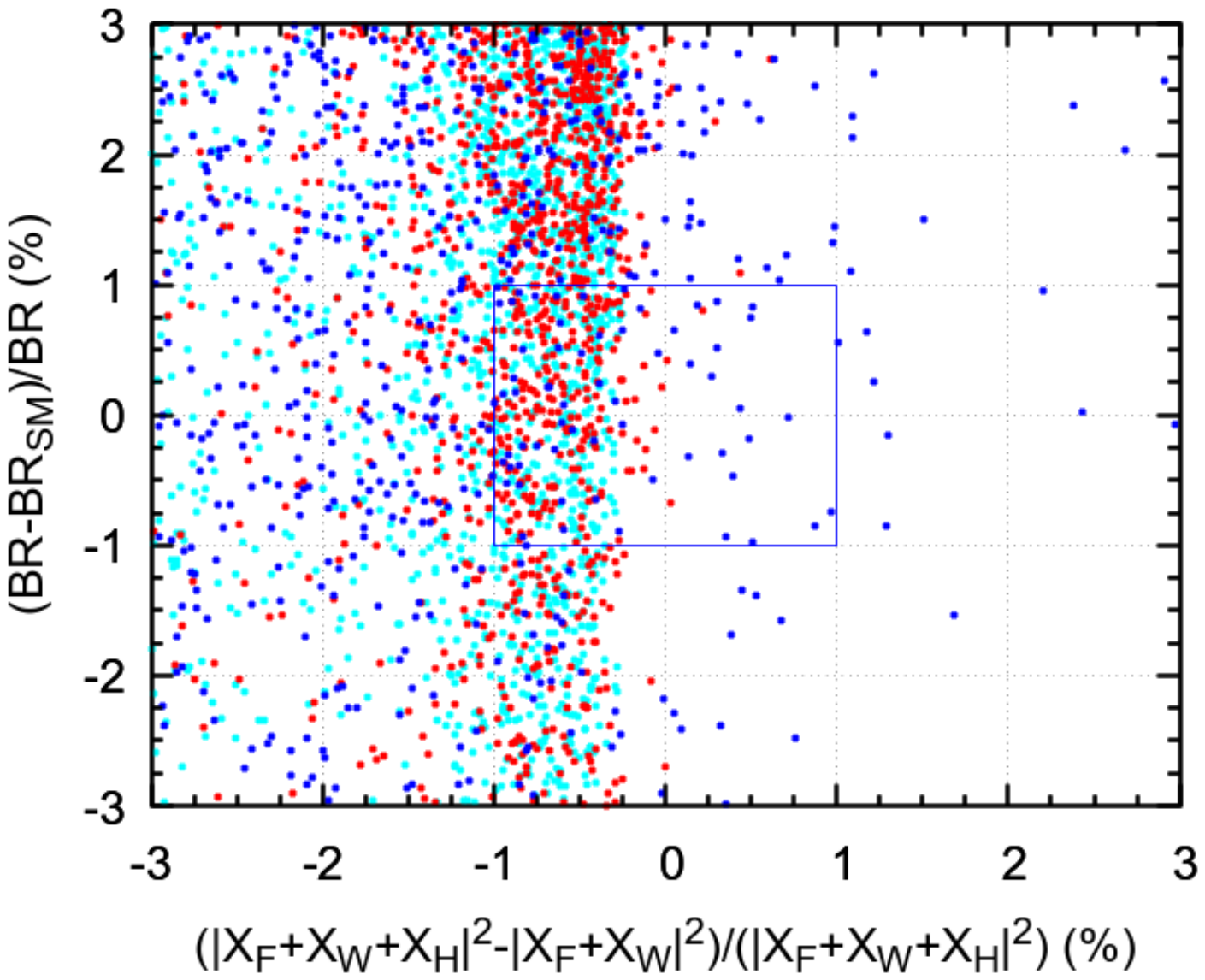}
\caption{\label{fig:cancel} Points with significant approximate
cancellation in both $h \to \gamma \gamma$ (horizontal axis) and
$B \to X_s \gamma$ (vertical),
which pass all theoretical and experimental bounds, including \texttt{HiggsBounds-5.9.1}.
Color code: cyan is perfect alignment, red means alignment
within 1\%, and blue means alignment within 10\%.
The blue box guides the eye to those points closest to (0,0).}
\end{figure}

The contributions of the two charged scalars can exhibit
large cancellations in the decays $h \rightarrow \gamma \gamma$
and $B \rightarrow X_s \gamma$.\footnote{For 3HDMs,
the cancellation
can be exact in $B \rightarrow X_s \gamma$ because there are two
charged components of Higgs doublets feeding the two physical
charged Higgs states. This is no longer the case in the Zee model,
with two Higgs doublets and one charged scalar
singlet~\cite{Florentino:2021ybj}.}
For some choices of parameter space, it is even possible that
there are cancellations in both decays simultaneously.
This is illustrated in Fig.~\ref{fig:cancel}.
\noindent
Such charged scalars would, thus, be difficult to probe indirectly.
Notice that points with exact alignment,
in cyan in Fig.~\ref{fig:cancel},
do not allow for cancellation in $h \to \gamma \gamma$;
but alignment with 1\% already does.

Most points within the blue box close to (0,0) have $H_2^+$
decays into quarks or leptons, which  are being sought at LHC.
But there are points which could also be difficult to probe directly with
such common searches, even tough one or both charged scalars
might have relatively small masses.
Indeed,
one can find fine-tuned points in parameter space
where the $H_2^+$ does not decay primordially into quarks or leptons,
but rather as $H_2^+ \rightarrow H_1^+ h_j$ with $h_j=h_1,h_2,A_1$.
We proposed that such decays are actively searched for at LHC's next run.

There has been a recent interest in the literature for unusual decays
of the charged Higgs~\cite{Bahl:2021str}, specially those in which the
charged Higgs decays to $W^+ h_i$ where $h_i$ is any of the scalars or
pseudo scalars in the model. We have performed a search in our large data sets and found many points where BR($H_1^+\to W^+ + h_{125}$) was larger than 80\%. From
those we selected three benchmark points (BP) that we list in
Table~\ref{tab:benchZ3}. 
\begin{table}[H]
  \centering
  \vspace{-2mm}
  \begin{tabular}{lccc}
    \toprule
    Type-Z & BP1 & BP2 & BP3\\
    \midrule
    $m_{h_2}$ & 419.00 & 494.60  & 486.26 \\
    $m_{h_3}$ & 799.60 & 850.88   & 694.44 \\
    $m_{A_1}$ & 413.80 & 483.96  & 513.46 \\
    $m_{A_2}$ & 763.15 &  806.44 & 647.56 \\
    $m_{H_1^\pm}$ & 396.13 & 477.63  & 506.36 \\
    $m_{H_2^\pm}$ & 752.81 &  843.034  & 654.77 \\
    $(m_{12}^2)$ & -8350 & -31768 & -19562   \\
    $(m_{13}^2)$ & -83278   & -80800 & -63134  \\
    $(m_{23}^2)$ & -231428 & -232361 & -197019  \\
    $\alpha_1$ &   1.289   & 1.343 & 1.328 \\
    $\alpha_2$ & 0.5419 & 0.4406 & 0.7119 \\
    $\alpha_3$ &  0.00543  & -0.00299 &  0.01136 \\
    $\gamma_1$ & -0.00503 & 0.00322  & -0.01078 \\
    $\gamma_2$ & -0.00504 & 0.00301 & -0.01011  \\
    $\beta_1$ &  1.192  & 1.263 & 1.231 \\
    $\beta_2$ & 0.5077   & 0.4311  & 0.7351 \\
    \midrule   
    BR($H_1^+\rightarrow\nu_\tau+\tau^+$) & 0.0688 & 0.0790 &    0.0784 \\
    BR($H_1^+\rightarrow t+\bar{b}$) & 0.0383 & 0.0197 & 0.0358 \\ 
    BR($H_1^+\rightarrow W^+h_1$) & 0.8926 & 0.9011 & 0.8855 \\ 
    \midrule
    BR($H_2^+\rightarrow t+\bar{b}$) & 0.9970 & 0.9995 & 0.9965 \\
    BR($H_2^+\rightarrow W^+h_1$) & 0.0012 & 0.0001 & 0.0009 \\
    BR($H_2^+\rightarrow W^+h_2$) & 0.0007 & 0.0003 & 0.0006 \\
    \bottomrule
  \end{tabular}
  \caption{Benchmark points for the Type Z $\Z3$-3HDM.}
  \label{tab:benchZ3}
\end{table}

\section{Distinguishing the Type-Z models}

There are usually two different ways in which a Type-Z Yukawa structure is realized.
The first method employs a $\Z3$ symmetry~\cite{Das:2019yad}, studied in detail in Section~\ref{sec:Z3_constraints}, whereas the second option uses
a $\Z2\times \Z2$ symmetry~\cite{Akeroyd:2016ssd}.
In both versions of the Type-Z real 3HDM the scalar potential of \eq{3hdmpotential}
contains a total of 18 parameters (6 bilinear parameters and 12 quartic parameters), traded to the equivalent set in the physical basis, as defined in Section~\ref{sec:3hdmphys}.


\begin{table}[htb]
  \centering
  \makebox[0pt][c]{\parbox{0.96\textwidth}{%
      \hskip -2mm
      \begin{minipage}[b]{0.49\textwidth}\centering
        \begin{tabular}{|c|c|c|c|c|}\hline
          \multicolumn{5}{|c|}{$ \Z2 \times  \Z2$\  ($\textrm{Al-2}-10\%$) }\\\hline\hline
          Check   &    N   &  Y     & 100*p   & 100*$\delta_p$\\\hline\hline
          STU  &   500000  &   407162  &   81.432  &    0.172\\*[1mm]
          BFB  &  5000000 &    380066  &    7.601  &    0.013\\*[1mm]
          Unitarity  &   500000  &    26386  &    5.277 &     0.033\\*[1mm]
          $b \rightarrow s \gamma$  &    50000 &     22198  &   44.396 &     0.358\\*[1mm]
           $\mu$'s  &   50000 &  4168     &  8.336  &  0.282   \\\hline
        \end{tabular}
      \end{minipage}
      \hfill
      \begin{minipage}[b]{0.49\textwidth}\centering
        \begin{tabular}{|c|c|c|c|c|}\hline
          \multicolumn{5}{|c|}{$ \Z3$\  ($\textrm{Al-2}-10\%$) }\\\hline\hline
          Check    &   N  &  Y   & 100*p   & 100*$\delta_p$\\\hline\hline
          STU  &   500000  &   407176  &   81.435  &    0.172\\*[1mm]
          BFB &   5000000  &    42703  &    0.854  &    0.004\\*[1mm]
          Unitarity  &   500000  &    18424  &    3.685  &    0.027\\*[1mm]
          $b \rightarrow s \gamma$  &    50000  &    21810  &    43.62  &    0.354\\*[1mm]
          $\mu$'s &     50000  &   4141  & 8.282  & 0.271    \\\hline
        \end{tabular}
      \end{minipage}
    }}
\caption{Impact of individual constraints for the two Type-Z models
  while the scanning is done within $10\%$ of the alignment
	condition of \eq{Al-2}.}
  \label{tab:k10}
\end{table}

To explicitly compare the symmetries with our scanning method, we
display in Table~\ref{tab:k10} how restrictive the
individual constraints of Chapter~\ref{ch:constraints_real} can be. In these tables,
$N$ represents the number of initial input points and $Y$ stands for
the number of output points that can successfully pass through a given
constraint labeled appropriately.
Thus $p = Y/N$ gives an estimate for the
probability of successfully negotiating a particular constraint. The
quantity $\delta_p$ represents the typical uncertainty associated with
the estimate of $p$ and is calculated using the formula for the
propagation of errors.
From Table~\ref{tab:k10} it should be evident that the
BFB constraints have a very low acceptance ratio for the input
points.
We should point out that
our choice of scanning around \eqs{eq:alignment}{eq:alignment2} does
definitely increase the number of output points that pass
through all the constraints.
This is seen in Table~\ref{tab:k10},
where we show equivalent numbers for a run generating points within
50\% of the alignment limit. Comparing Table~\ref{tab:k10} with Table~\ref{tab:k16},
we notice that, away from alignment, the unitarity and $\mu$ constraints cut
most of the allowed parameter space.

\begin{table}[htb]
\centering
\makebox[0pt][c]{\parbox{0.97\textwidth}{%
    \begin{minipage}[b]{0.47\textwidth}\centering
      \begin{tabular}{|c|c|c|c|c|}\hline
        \multicolumn{5}{|c|}{$ \Z2 \times  \Z2$\ ($\textrm{Al-2}-50\%$)  }\\\hline\hline
        Check    &     N       &  Y     & 100*p   & 100*$\delta_p$\\\hline\hline
        STU   &   500000  &     46179   &      9.236  &     0.045\\*[1mm]
        BFB  &   5000000  &    277802  &     5.556 &      0.011\\*[1mm]
        Unitarity   &   500000  &      3397  &     0.679 &      0.012\\*[1mm]
        $b \rightarrow s \gamma$   &    50000  &     19340  &    38.680  &     0.328\\*[1mm]
        $\mu$'s   &    50000   &    217   & 0.434    &    0.077 \\\hline
      \end{tabular}
    \end{minipage}
    \hfill
    \begin{minipage}[b]{0.47\textwidth}\centering
      \begin{tabular}{|c|c|c|c|c|}\hline
        \multicolumn{5}{|c|}{$ \Z3$\  ($\textrm{Al-2}-50\%$) }\\\hline\hline
        Check    &     N       &  Y     & 100*p   & 100*$\delta_p$\\\hline\hline
        STU   &   500000  &     46296  &     9.259  &     0.045\\*[1mm]
        BFB  &   5000000  &     25673  &     0.513  &     0.003\\*[1mm]
        Unitarity  &    500000  &      2639  &     0.528  &     0.010\\*[1mm]
        $b \rightarrow s \gamma$  &     50000  &     19193  &    38.386  &     0.326\\*[1mm]
        $\mu$'s  &     50000  &    178   & 0.356     & 0.069    \\\hline
      \end{tabular}
    \end{minipage}
  }}
\caption{Impact of individual constraints for the two Type-Z models
  while the scanning is done within $50\%$ of the alignment
	condition of \eq{Al-2}.}
\label{tab:k16}
\end{table}
Having compared the constraints with our scan strategy clearly, we continue with results from our numerical studies that pass all constraints. We will do this in
two stages. At first we will demonstrate the results from the general
scans and point out features that may distinguish between the two
variants of Type-Z 3HDMs.  In the second part we will presume that
some nonstandard scalars have been discovered and therefore we will
work with some illustrative benchmark points in the hope of making the
distinction between the two models more pronounced.

\begin{figure}[htb]
  \centering
\includegraphics[width=0.47\textwidth]{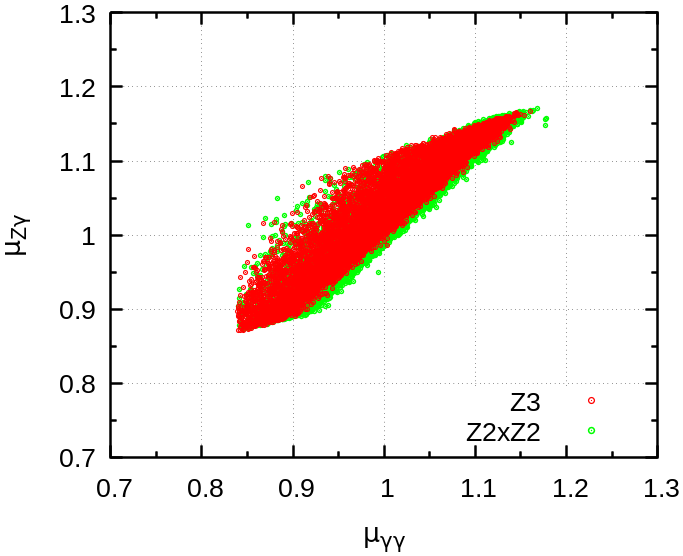}
\caption{Output points that pass all the constraints are plotted in
  $\mu_{\gamma\gamma}$ vs $\mu_{Z\gamma}$ 
  plane for the gluon fusion production channel. The scanning is
  done assuming the Al2-10\% condition of \eq{Al-2}.
  The red and the green points correspond to the
  $\Z3$ and $ \Z2 \times \Z2$ models respectively.}
\label{fig:mugagak10}
\end{figure}
We have to always keep in mind that the difference between the two versions of
Type-Z 3HDM is marked by the scalar potential. Therefore, we focus on the
measurements that involve the scalar self-couplings. Quite naturally, our first
choice will be to study $\mu_{\gamma\gamma}$ and $\mu_{Z\gamma}$ (Higgs signal
strengths in the two photon and $Z$-photon channels respectively) which pick up
extra contributions from charged scalar loops that depend on couplings of the
form $hH_i^+ H_i^-$ ($i=1,2$). However, as we have displayed
in Fig.~\ref{fig:mugagak10}, the points that pass through all the constraints
span very similar regions in the $\mu_{\gamma\gamma}$ vs $\mu_{Z\gamma}$
plane for both versions of Type-Z 3HDM. Thus no significant distinction between
the two models can be made from $\mu_{\gamma\gamma}$ and $\mu_{Z\gamma}$.


\begin{figure}[htb]
	\centering
	\includegraphics[width=0.31\textwidth]{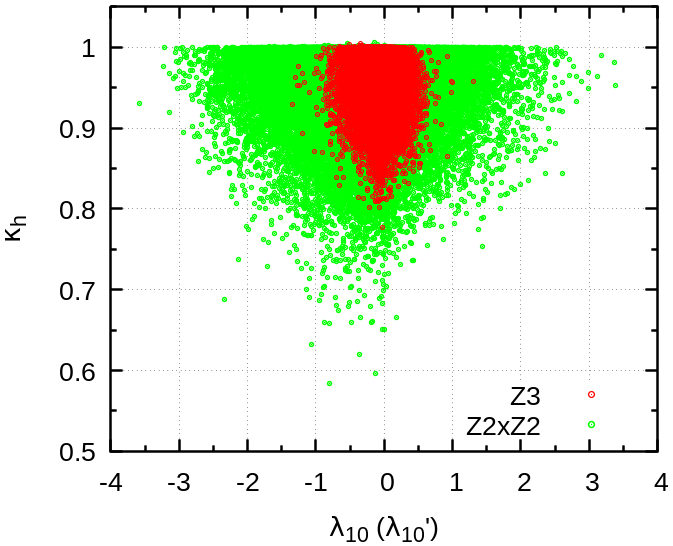}
	\includegraphics[width=0.31\textwidth]{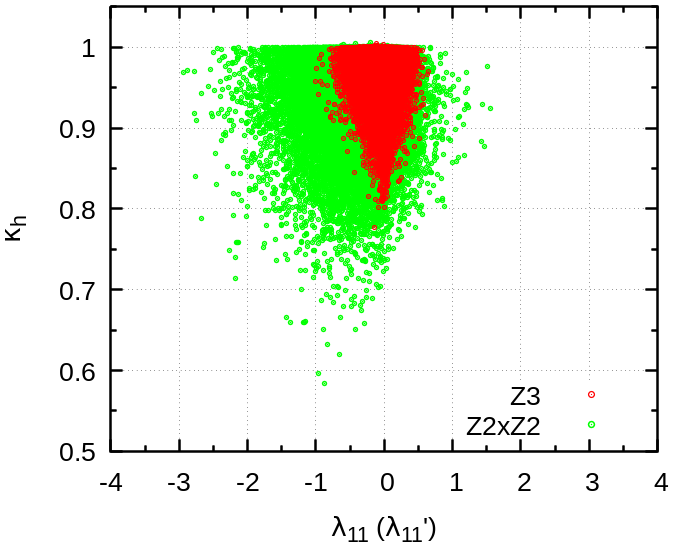}
	\includegraphics[width=0.31\textwidth]{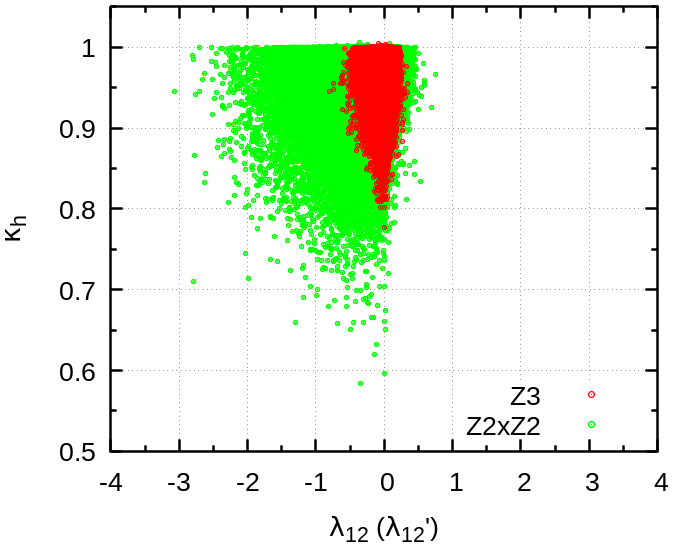}
	\caption{Points that pass through all the constraints are plotted in the
	$\kappa_h$ vs $\lambda_{10}^{(\prime)}, \lambda_{11}^{(\prime)}, \lambda_{12}^{(\prime)}$ plane. The scanning is
	done assuming the $\textrm{Al-2}-10\%$  condition of \eq{Al-2}.
	The red and the green points correspond to the
	$\Z3$ and $ \Z2 \times \Z2$ models respectively.}
\label{f:khlam}
\end{figure}


\begin{figure}[htb]
  \centering
  \includegraphics[width=0.45\textwidth]{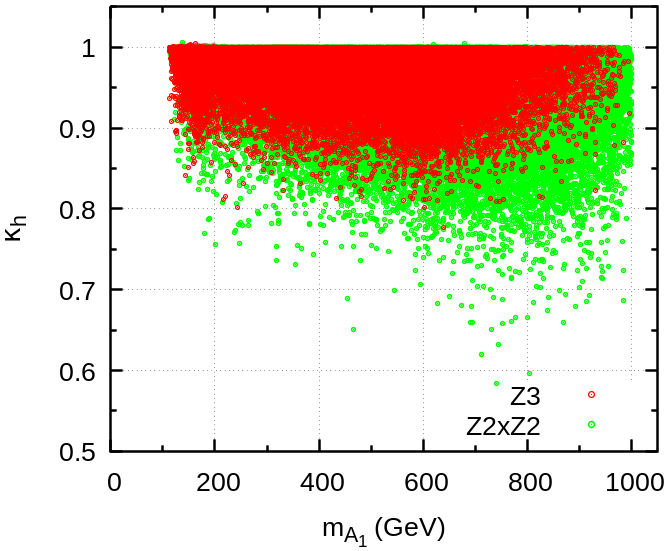}
  \includegraphics[width=0.45\textwidth]{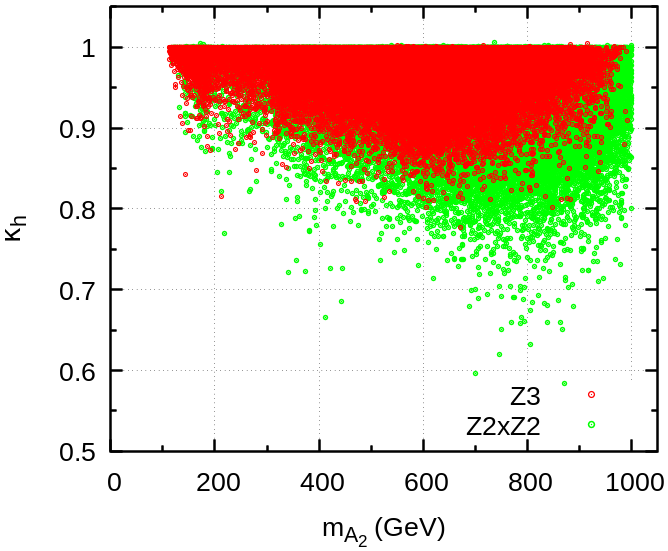}
  \caption{Points that pass through all the constraints are plotted in the
    $\kappa_h$ vs $m_{A1}, m_{A2}$ plane. The scanning is
    done assuming the $\textrm{Al-2}-10\%$  condition of \eq{Al-2}.
    The red and the green points correspond to the
    $\Z3$ and $ \Z2 \times \Z2$ models respectively.
     }
\label{f:khmass}
\end{figure}

Next we turn our attention to the trilinear Higgs self-coupling of the
following form:
\begin{eqnarray}
	{\mathscr L}_{hhh} = g_{hhh} h^3 \,.
\end{eqnarray}	
In the SM we have $g_{hhh}^{\rm SM} = -m_h^2/(2v)$.
Thus we define the following coupling modifier
\begin{eqnarray}
	\kappa_h = \frac{g_{hhh}}{g_{hhh}^{\rm SM}}
\end{eqnarray}
which is already being measured experimentally and some preliminary values
have been reported in Refs.~\cite{CMS:2022cpr,ATLAS:2021ifb}. We have checked that for both models, $\kappa_h = 1$ in the alignment limit defined by
\eq{eq:alignment}, as expected. Therefore we have to hope that the LHC Higgs data will
eventually settle for some nonstandard values away from exact alignment 
so that some distinguishing features can be found. To this end we recall
that the quartic parameters $\lambda_{10-12}$ in \eq{eq:z2z2part} and \eq{Z3quartic} mark the essential difference
between the two models. It should also be noted that in the limit
$\lambda_{10}^{(\prime)}, \lambda_{11}^{(\prime)}, \lambda_{12}^{(\prime)} = 0$, the
quartic part of the potential possesses a $U(1)\times U(1)$ symmetry
(independent from the $U(1)_Y$ hypercharge symmetry). Consequently,
$\lambda_{10}^{(\prime)}$, $\lambda_{11}^{(\prime)}$ and $\lambda_{12}^{(\prime)}$ are
the only quartic parameters that get involved in the expressions of the
pseudoscalar masses, $m_{A1}$ and $m_{A2}$. Keeping these in mind we
exhibit in Fig.~\ref{f:khlam} the scatter plot of the points that pass through
all the constraints in the $\kappa_h$ vs $\lambda_{10}^{(\prime)}, \lambda_{11}^{(\prime)}, \lambda_{12}^{(\prime)}$ plane. There we observe
that values of $\kappa_h$ in the ballpark $0.8$ or lower will definitely
favor the $\Z2\times \Z2$ scenario over the $\Z3$ version of Type-Z 3HDM.
To give these results a better physical context, in Fig.~\ref{f:khmass},
 we plot the same points in
the $\kappa_h$ vs pseudoscalar mass planes. This figure clearly indicates
that unlike the $\Z3$ model, the $\Z2\times \Z2$ model can still allow
$\kappa_h$ values as low as $0.7$.
In passing, we also note that values of $\kappa_h$ around $1.1$ or higher
 will disfavor both versions of Type-Z 3HDMs.

%
\begin{table}[htb]
  \centering
  \begin{tabular}{|c|c|c|c|c|c|c|}
    \hline
    & $m_{H1}$ & $m_{H2}$ & $m_{A1}$ &  $m_{A2}$ & $m_{C1}$ & $m_{C2}$ \\
    \hline\hline
    Benchmark 1 & 365  &    450  &  340 &    470 &  335 &   465  \\*[1mm]
    Benchmark 2 &  530 &  645 &  515 &  610 &   540 &     610   \\*[1mm]
    Benchmark 3 &  641  &  775 &  615 &     745 &    645  &   770 \\
    \hline
  \end{tabular}
  \caption{Benchmark values for the nonstandard masses (in ${\rm GeV}$) used in Fig.~\ref{f:bench}.
     }
  \label{t:bench}
\end{table}

\begin{figure}[htb]
  \centering
  \includegraphics[width=0.44\textwidth]{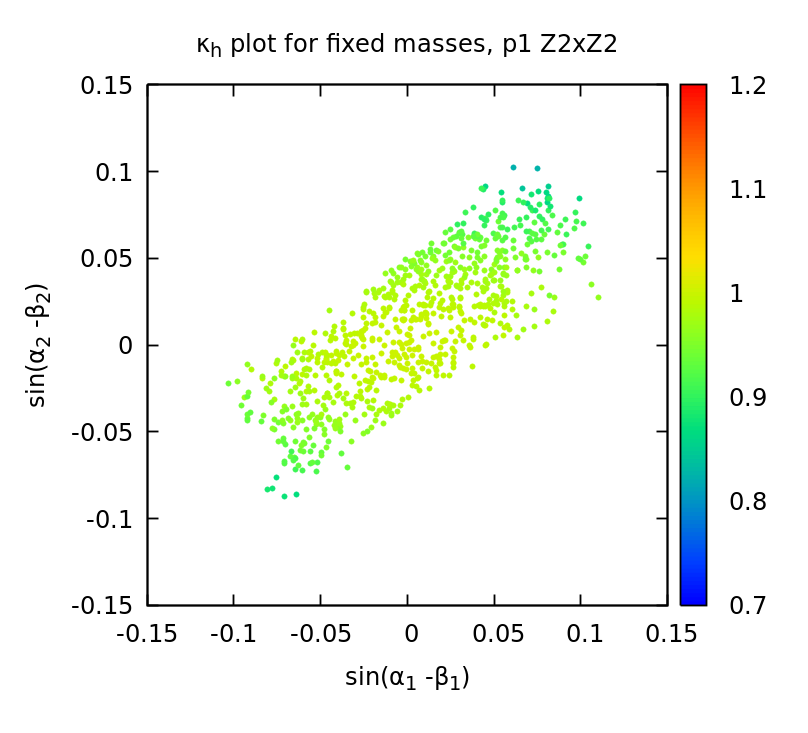} ~~
  \includegraphics[width=0.44\textwidth]{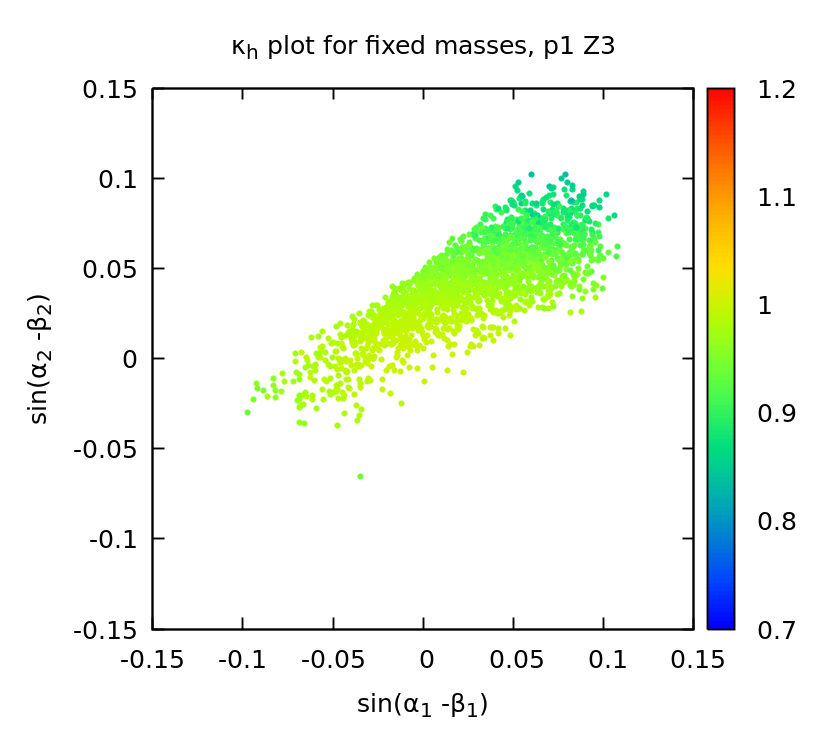} \\
  \includegraphics[width=0.44\textwidth]{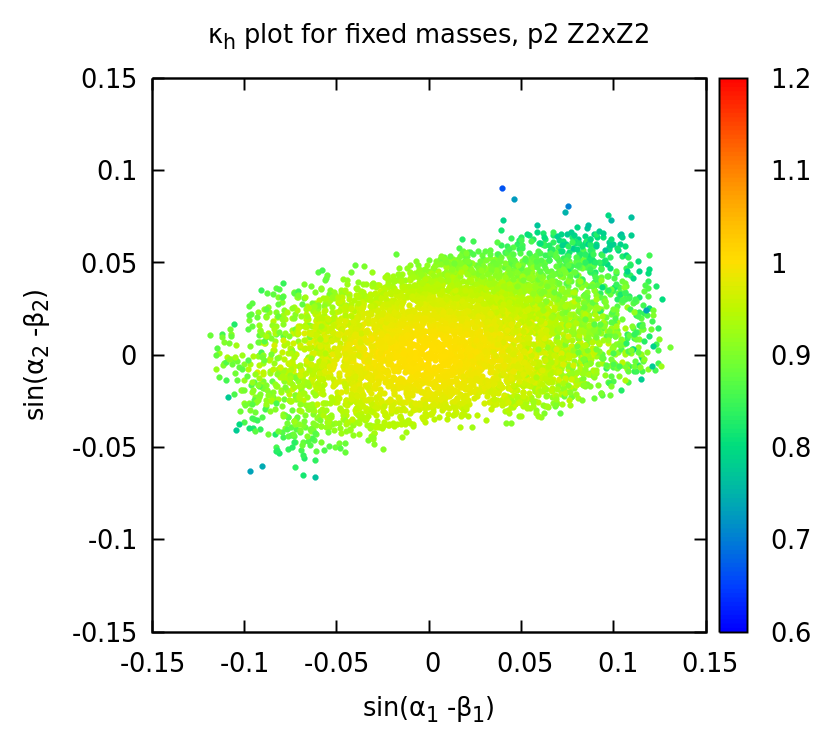} ~~
  \includegraphics[width=0.44\textwidth]{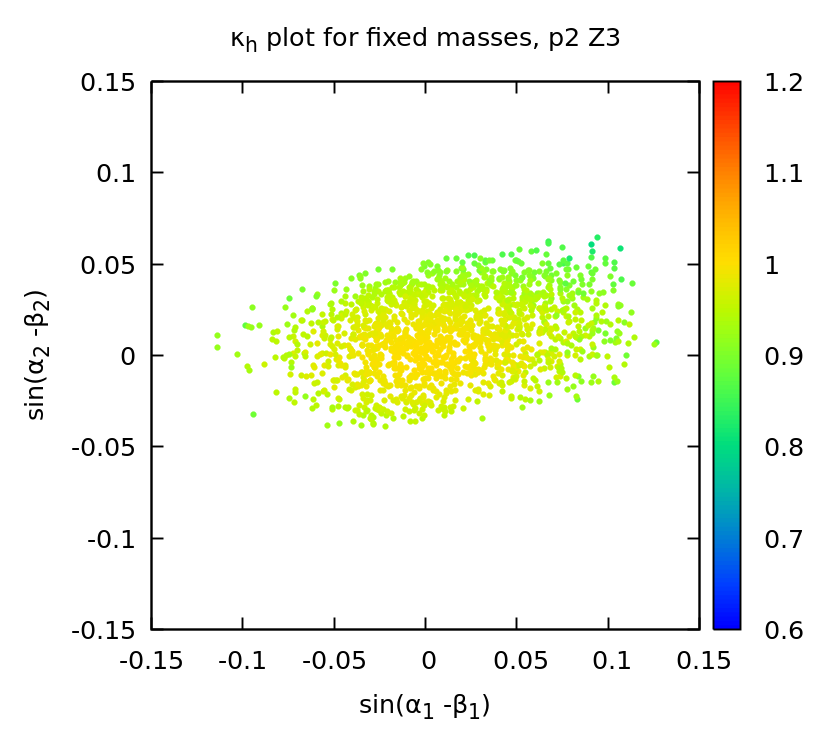} \\
  \includegraphics[width=0.44\textwidth]{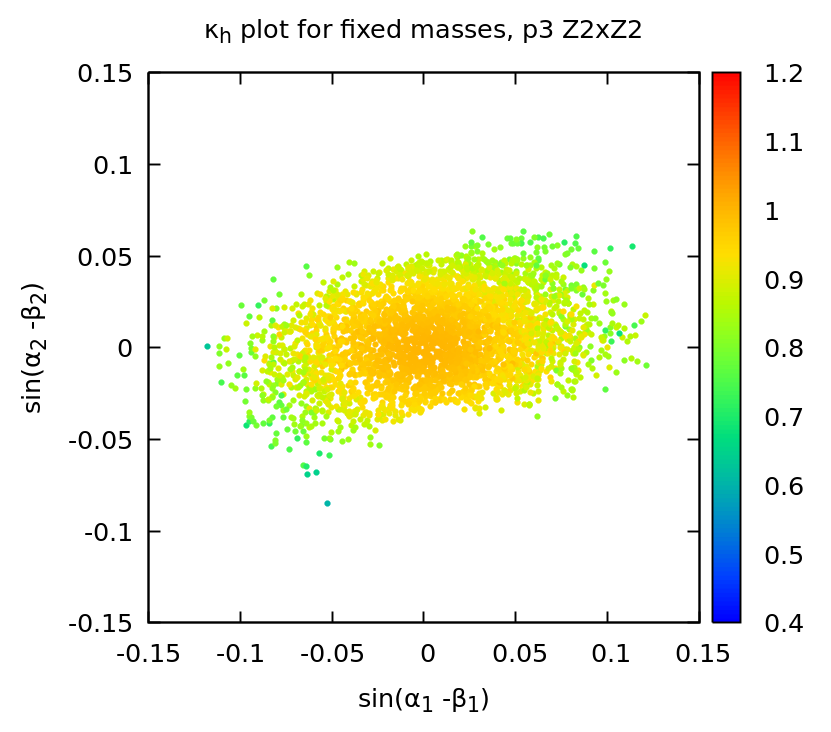} ~~
  \includegraphics[width=0.44\textwidth]{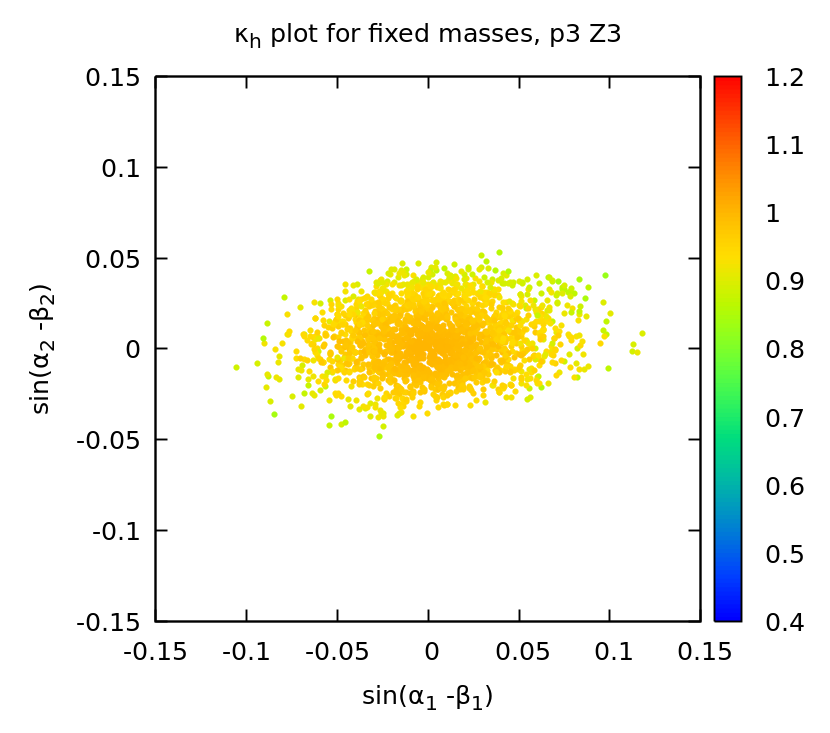}
  \caption{Plot in the $\sin{(\alpha_1-\beta_1)}$ vs
  $\sin{(\alpha_1-\beta_1)}$ plane for the benchmark values
(labeled appropriately) of Table~\ref{t:bench}. The
color bar associated with each plot marks the gradient of
values taken by $\kappa_h$. The plots in the left panel
correspond to the $\Z2\times \Z2$ model whereas the plots in
the right panel correspond to the $\Z3$ model.
Clearly, the distinguishability between the two models depends
on the benchmark point (mass region) chosen.
}
\label{f:bench}
\end{figure}

In a final and more optimistic effort, we presume that some nonstandard scalars have
already been observed and we try to ascertain whether, in view of the set of
nonstandard parameters, one of the Type-Z 3HDMs can be preferred over the other.
Our benchmark values for the nonstandard masses appear in Table~\ref{t:bench}.
The remaining parameters are scanned following $\textrm{Al-2}-20\%$ of \eq{Al-2}. 
For these benchmark values we have plotted all the points that pass through the
constraints in the $\sin(\alpha_1-\beta_1)$ vs $\sin(\alpha_2-\beta_2)$ plane.
The results have been displayed in Fig.~\ref{f:bench} where we have also color
coded the value of $\kappa_h$ for each point. There we can see that the points
span a relatively larger region for the $\Z2\times \Z2$ model. Therefore, if both
$\sin(\alpha_1-\beta_1)$ and $\sin(\alpha_2-\beta_2)$ are measured to be close to
$0.1$ along with $\kappa_h$ to be around $0.7$, then
it would definitely point towards the $\Z2\times \Z2$ model.
Thus, again, we have found that although we can find corners in the parameter space
that can isolate the $\Z2\times \Z2$ model, it seems to be very difficult to point
out exclusive features characterizing the $\Z3$ version of the Type-Z 3HDM.


\section{Summary}
\label{s:summary_alignment}
Multi Higgs models with $N \geq 3$ allow for the possibility that all
fermions of a given charge couple exclusively to one dedicated
scalar. These are known as Type-Z models, and constitute a fifth
alternative beyond the four natural flavour conservation models
allowed in the 2HDM.  We investigate the current bounds on the Type-Z
3HDM imposed by a $\Z3$ symmetry and compare with the $\Z2\times\Z2$ implementation. 
We stress the fact that interesting physical observables may
differ significantly when one considers situations close to the
alignment limit, versus adopting the exact alignment limit.   Indeed,
current LHC bounds on the productions and branching ratios of the
$125~{\rm GeV}$ neutral scalar force the measured couplings to lie close to
those obtained for the SM Higgs.  Nevertheless, forcing those
couplings to match \textit{exactly} those in the SM is too
constraining on the parameter space and precludes much of the
interesting new features present in the models.

%

We look at the constraints allowed by current data on the $125~{\rm GeV}$ Higgs
decays, including a detailed look at $h \to \gamma \gamma$ and its
correlations with the other decays.  We point out the possibility that
the contributions from the two charged scalars might cancel in $h \to
\gamma \gamma$.  This is also possible in $B \to X_s \gamma$, and we
explore explicitly how this allows for lower masses for the charged
scalars.  We provide illustrative benchmark points to aid in
experimental searches, proposed at LHC meetings.  By comparing the constraints from
\texttt{HiggsBounds-5.7.1} and the newer \texttt{HiggsBounds-5.9.1} we
highlight the importance that the next LHC run will have in further
constraining this model, or perhaps, finally uncovering new physics in
the scalar sector.

We point out that the difference between these two implementations of the Type-Z Yukawa couplings is captured
by certain quartic terms in the scalar potential.
Then we proceed to uncover the effects of these quartic terms in creating
distinctions between the two Type-Z models. In doing so we have performed exhaustive scans over the set of free parameters in these models and show numerical results on the impact of individual constraints. We have found that although $\mu_{\gamma\gamma}$ and
$\mu_{Z\gamma}$ are not the best discriminators, the trilinear Higgs 
self coupling modifier $(\kappa_h)$ has the potential to distinguish between
the two models. We have concluded that relatively lower values of $\kappa_h$
will favor the $\Z2\times \Z2$ version of Type-Z 3HDM. We have also emphasized
that some nonstandard physics need to be discovered in the LHC Higgs data
for us to be able to discriminate between the two Type-Z 3HDMs. Our study
underscores the importance of the ongoing effort to measure the trilinear
Higgs self coupling with increased precision.
%

%

%% file: chapters/c3hdm.tex
\chapter{The complex 3HDM}
\label{chapter:C3HDM}
\hspace*{0.3cm}
Up till now we have studied scalar extensions of the SM with the simplification of all the parameters of the potential considered real. In the real 2HDM, this simplification has been argued as not being theoretically sound~\cite{Fontes:2021znm,deLima:2024hnk}. The existing CP violation in the quark mixing matrix can leak divergent CP-violation effects, at higher order in perturbation theory,  into the scalar potential lacking the counterterms to absorb them, spoiling renormalizability and overall consistency of the theory. In this Chapter we do not study such effects but motivate the extension of our previous studies of the softly-broken $\Z2\times\Z2$ 3HDM to allow for complex parameters in the scalar potential. 

The general $\Z2\times\Z2$ symmetric 3HDM, referred to as Weinberg's 3HDM~\cite{Weinberg:1976hu}, was originally introduced with the characteristic of having CP-violation in the charged scalar sector. We consider a softly-broken version of this model, which we dub the complex 3HDM (\gls{C3HDM}). In this Section we present a
parameterization of the rotations needed to go from the initial fields to the mass eigenstates, a task
complicated by the fact that the neutral scalars have no longer a definite CP parity. We complete with a derivation of the alignment and real limits.

\section{The scalar potential}

The scalar potential obeying the $\Z2\times \Z2'$ symmetry defined in \eq{eq:z2z2_def},
including soft breaking terms, can alternatively be written as~\cite{Botella:1994cs}
\begin{equation} \label{VNHDM}
	V = V_2 + V_4 = \mu_{ij} (\Phi_i^\dag \Phi_j) + z_{ijkl} (\Phi_i^\dag \Phi_j)(\Phi_k^\dag \Phi_l) \;,
\end{equation}
with
\begin{equation}
	V_2 =\, \mu_{11} (\Phi_1^\dag \Phi_1) + \mu_{22} (\Phi_2^\dag \Phi_2) + \mu_{33} (\Phi_3^\dag \Phi_3) + \left( \mu_{12}  (\Phi_1^\dag \Phi_2) + \mu_{13}  (\Phi_1^\dag \Phi_3) + \mu_{23}  (\Phi_2^\dag \Phi_3) + {\rm h.c.}\right)
	\;, \\
\end{equation}
where $\mu_{11}$, $\mu_{22}$, $\mu_{33}$ are real, and
the complex $\mu_{12}$, $\mu_{13}$, $\mu_{23}$ parameters break the
$\Z2\times \Z2'$ symmetry softly. Moreover, similar to \eq{Z2Z2quartic},
\begin{equation}
\begin{split}
		V_4 =\,& V_{RI} + V_{\Z2\times \Z2'} \;,\\
		V_{RI} =\,& \lambda_1 (\Phi_1^\dag \Phi_1)^2 + \lambda_2 (\Phi_2^\dag \Phi_2)^2 + \lambda_3 (\Phi_3^\dag \Phi_3)^2  + \lambda_4 (\Phi_1^\dag \Phi_1)(\Phi_2^\dag \Phi_2) + \lambda_5 (\Phi_1^\dag \Phi_1)(\Phi_3^\dag \Phi_3) \\&+ \lambda_6 (\Phi_2^\dag \Phi_2)(\Phi_3^\dag \Phi_3)  + \lambda_7 (\Phi_1^\dag \Phi_2)(\Phi_2^\dag \Phi_1) + \lambda_8 (\Phi_1^\dag \Phi_3)(\Phi_3^\dag \Phi_1) + \lambda_9 (\Phi_2^\dag \Phi_3)(\Phi_3^\dag \Phi_2) \;, \\
		V_{\Z2\times \Z2'} =\,& \lambda_{10} (\Phi_1^\dag \Phi_2)^2 + \lambda_{11} (\Phi_1^\dag \Phi_3)^2 + \lambda_{12} (\Phi_2^\dag \Phi_3)^2 + {\rm h.c.}
		\;,
\end{split} 
\end{equation} 
where $\lambda_1,\, ..., \lambda_9$ are real parameters and $\lambda_{10}$, $\lambda_{11}$, $\lambda_{12}$ can now be potentially complex. We write our doublets around a charge preserving vev as
\begin{equation}
	\Phi_i = 
	\begin{pmatrix}
		w_i^+ \\ (v_i + x_i + i\ z_i) /\sqrt{2}
	\end{pmatrix} 
	=
	\begin{pmatrix}
		w_i^+ \\ (v_i + \varphi_i) /\sqrt{2}
	\end{pmatrix} 
	\;.
\end{equation}
The existence of complex quadratic and quartic parameters enables explicit CP violation. The vev components $v_i=|v_i| e^{-i \alpha_i}$ can be made real if
we perform a basis change that rephases each doublet by the
opposite phase of its vev component,
i.e. $\,\Phi_i' = U_{ij} \Phi_j\,$,
with $\, U = \text{diag}(e^{i\alpha_1},e^{i\alpha_2},e^{i\alpha_3})\,$. A general basis change comes with an associated transformation in the parameters:
\begin{equation}\label{e:doub_repar}
	\mu^\prime_{ab} = U_{ai} \mu_{ij} U^\dag_{jb} 
	\qquad\text{and}\qquad 
	z^\prime_{abcd} = U_{ai} U_{ck} z_{ijkl} U^\dag_{ld} U^\dag_{jb} 
	\;,
\end{equation}
that in this case simply rephases the complex parameters:
\begin{equation}\begin{split}
		\lambda_{10} \rightarrow \lambda_{10} \cdot e^{ i\cdot 2\alpha_{12}} \;,\qquad
		\lambda_{11} \rightarrow \lambda_{11} \cdot e^{ i\cdot 2\alpha_{13}} \;,\qquad
		\lambda_{12} \rightarrow \lambda_{12} \cdot e^{ i\cdot 2\alpha_{23}} \;,\\
		\mu_{12} \rightarrow \mu_{12} \cdot e^{i\cdot \alpha_{12}} \;,\qquad
		\mu_{13} \rightarrow \mu_{13} \cdot e^{i\cdot \alpha_{13}} \;,\qquad
		\mu_{23} \rightarrow \mu_{23} \cdot e^{i\cdot \alpha_{23}} \;,
\end{split}\end{equation}
where $\alpha_{ij} = \alpha_i - \alpha_j$. 
We can therefore choose a real vev without loss of generality.
 With the scalar minimum established, the Higgs basis~\cite{Georgi:1978ri,Donoghue:1978cj,Botella:1994cs} can be introduced  by the following rotation,
\begin{equation}\label{higgsbasisc3hdm}
     \begin{pmatrix} H_0 \\ R_1 \\ R_2 \end{pmatrix}
     =
    R_H
     \begin{pmatrix} x_1 \\ x_2 \\ x_3 \end{pmatrix}
     =
     \begin{pmatrix} c_{\beta_2} c_{\beta_1} & c_{\beta_2} s_{\beta_1} & s_{\beta_2} \\
                    -s_{\beta_1} & c_{\beta_1} & 0 \\ 
                    -c_{\beta_1} s_{\beta_2} & -s_{\beta_1} s_{\beta_2} & c_{\beta_2}\end{pmatrix}
     \begin{pmatrix} x_1 \\ x_2 \\ x_3 \end{pmatrix},
\end{equation}
which, similar to \eq{higgsbasisZ3}, corresponds to parameterizing the vevs as,
\begin{equation}\label{c3hdmvevs}
     v_1=v c _{\beta_1} c _{\beta_2}\,,\qquad v_2=v s _{\beta_1} c _{\beta_2}\, ,\qquad v_3=v s _{\beta_2},
\end{equation}
where $c_{\beta_i}$ and $s_{\beta_i}$ account for $\cos(\beta_i)$ and $\sin(\beta_i)$, respectively.

We have 24 parameters in the scalar potential,
besides the three real vevs. These are constrained by five stationarity conditions, derived in \eq{STAT}, yielding 22 parameters. Of these, two will be reserved for $v$ and the mass of the $125~{\rm GeV}$
scalar, leaving 20 free parameters.

\section{Physical basis\label{sec:physbasis}}
To obtain the physical quantities, we cannot follow the procedure of Section~\ref{sec:3hdmphys}, as the $5\times 5$ neutral Higgs mass matrix now combines components on the $x_i$ fields with $z_i$ fields~\cite{Luis:2023}. To relate the potential parameters with the vev and masses we need to consider
the linear and quadratic terms of the potential,
after spontaneous symmetry breaking (SSB).
We can write these in general as\footnote{Here and henceforth, bold symbols
(such as $ \boldsymbol{v}$)
refer to
vectors made out of the corresponding components (such as $v_i$).}
\begin{equation}\label{eq:V1}
	V^{(1)} = \Re(\boldsymbol{\varphi}^\dagger A \boldsymbol{v})
= \Re(\boldsymbol{x}^T A \boldsymbol{v} - i \boldsymbol{z}^T A \boldsymbol{v)}
= \boldsymbol{x}^T \Re(A\boldsymbol{v}) + \boldsymbol{z}^T \Im(A\boldsymbol{v})  \;,
\end{equation}
introducing the hermitian matrix
\begin{equation}\label{e:A}
	A_{ij} = \mu_{ij} + z_{ijkl} v_k^* v_l  \;,
\end{equation}
and as
\begin{equation}\label{eq:V2}
	\begin{split}
		V^{(2)} = 
		V^{(2)}_\text{ch} + V^{(2)}_\text{n} &= 
		( \boldsymbol{w}^+)^\dag A  \boldsymbol{w}^+
+ \frac{1}{2} \boldsymbol{\varphi}^\dag A \boldsymbol{\varphi}
+ \frac{1}{2} z_{ijkl} \varphi_i^* v_j v_k^* \varphi_l + \frac{1}{2} \Re(z_{ijkl} v_i^* \varphi_j v_k^* \varphi_l ) = \\
		&= (\boldsymbol{w}^+)^\dagger M^2_{ch} \boldsymbol{w}^+ +  
		\frac{1}{2} \begin{pmatrix}
			\boldsymbol{x}^T & \boldsymbol{z}^T
		\end{pmatrix}
		\begin{pmatrix}
			M^2_x & M^2_{xz} \\
			(M^2_{xz})^T & M^2_z 
		\end{pmatrix}
		\begin{pmatrix}
			\boldsymbol{x} \\ \boldsymbol{z}
		\end{pmatrix} = 
		\\
		&= 
		(\boldsymbol{w}^+)^\dagger M^2_{ch} \boldsymbol{w}^+ +  
		\frac{1}{2} \begin{pmatrix}
			\boldsymbol{x}^T & \boldsymbol{z}^T
		\end{pmatrix}
		M^2_n
		\begin{pmatrix}
			\boldsymbol{x} \\ \boldsymbol{z}
		\end{pmatrix}
		\;,
	\end{split}
\end{equation}
where the mass matrices are given by
\begin{equation}\label{eq:mass_matrices}\begin{split}
		M^2_{ch} &= A \;, \\
		M^2_x &= \Re(A+B + C) \;, \\
            M^2_z &= \Re(A+B - C) \;, \\
		M^2_{xz} &= -\Im(A+B+C) \;,
\end{split}\end{equation}
introducing two new matrices (hermitian and symmetric respectively)
\begin{equation}\label{e:matBC}
	B_{ij} =  z_{iklj} v_l^* v_k 
	\quad \text{and} \quad
	C_{ij} =  z_{kilj} v_l^* v_k^* 
	\;.
\end{equation}
As a result, $M^2_{ch}$ is hermitian, $M^2_x$ and $ M^2_z$ are real and symmetric,
and $M^2_{xz}$ is real, so that $M^2_n$ is a real symmetric matrix.
We include in Appendix~\ref{app:parametrization} the specific forms of the
$z_{ijkl}$ tensor and the $A$, $B$, and $C$ matrices
in terms of the potential parameters. Imposing stationarity conditions, i.e. ensuring the vev is a
stationary point of the potential, is equivalent to the tadpole condition $V^{(1)}=0$, or
\begin{equation}\label{e:stat_conditions}
	A \boldsymbol{v}=0 \,; 
\end{equation}
leading to five independent equations,
\small
\begin{subequations}
	\label{STAT}
	\begin{eqnarray}
		&\mu_{11} v_1 = -\Re(\mu_{12}) v_2 -\Re(\mu_{13}) v_3 - v_1 \left( \lambda_1 v_1^2 + \left(\Re(\lambda_{10}) + \frac{1}{2}\lambda_4 + \frac{1}{2}\lambda_7\right) v_2^2 + \left(\Re(\lambda_{11}) + \frac{1}{2} \lambda_5 + \frac{1}{2}\lambda_8\right) v_3^2 \right ) 
		,\qquad \\*[1mm]
		&\mu_{22} v_2 =  -\Re(\mu_{12}) v_1 -\Re(\mu_{23}) v_3 - v_2 \left( \lambda_2 v_2^2 + \left(\Re(\lambda_{10}) + \frac{1}{2}\lambda_4 + \frac{1}{2}\lambda_7\right) v_1^2 + \left(\Re(\lambda_{12}) + \frac{1}{2}\lambda_6 + \frac{1}{2}\lambda_9\right) v_3^2 \right) 
		,\qquad\\*[1mm]
		&\mu_{33} v_3 =  -\Re(\mu_{13}) v_1 -\Re(\mu_{23}) v_2 - v_3 \left( \lambda_3 v_3^2 + \left(\Re(\lambda_{11}) + \frac{1}{2} \lambda_5 + \frac{1}{2}\lambda_8\right) v_1^2 + \left(\Re(\lambda_{12}) + \frac{1}{2}\lambda_6 + \frac{1}{2}\lambda_9\right) v_2^2\right) 
		,\qquad\\*[1mm]
		&\Im(\mu_{13}) v_3 =  -v_1\left( \Im(\lambda_{10}) v_2^2 + \Im(\lambda_{11}) v_3^2 \right) - \Im(\mu_{12}) v_2 
		\;, \\*[1mm]
		&\Im(\mu_{23}) v_3 = v_2\left( \Im(\lambda_{10}) v_1^2 - \Im(\lambda_{12})v_3^2 \right) + \Im(\mu_{12}) v_1
		\;.
	\end{eqnarray}
\end{subequations}
\normalsize
We proceed by changing $w^+_i$ and $z_i$ into the Higgs basis, by
\begin{equation}\label{e:field_transf}
	\begin{pmatrix}
		x \\ z^\prime
	\end{pmatrix}
	= \begin{pmatrix}
		\mathbbm{1} & 0 \\
		0 & R_H
	\end{pmatrix}
	\begin{pmatrix}
		x \\ z
	\end{pmatrix}
	\;,
	\qquad \qquad
	w^{+\prime} = R_H w^+
	\; .
\end{equation}
\begin{equation}
 R_H M_{ch}^2 R_H^T =\begin{pmatrix}
     0 & 0 & 0  \\
    0 &\multicolumn{2}{c}{\multirow{2}{*}{ $ M_{ch}^{2\prime}$}}\\ 
    0  &
    \end{pmatrix}\;, 
\qquad  \quad  
\begin{pmatrix}
		\one & 0 \\
		0 & R_H
	\end{pmatrix}M_{n}^2\begin{pmatrix}
		\mathbbm{1} & 0 \\
		0 & R_H^T
	\end{pmatrix}=\begin{pmatrix}
\multicolumn{3}{c}{\multirow{3}{*}{ $ M_{x}^{2}$}}  & 0 & \multicolumn{2}{c}{\multirow{3}{*}{ $ M_{xz}^{2\prime}$}}\\ 
      &  &  & 0 &  &   \\
           &  &  & 0 &  &   \\
     0 & 0 & 0 & 0 & 0 & 0 \\
    \multicolumn{3}{c}{\multirow{2}{*}{ $ M_{xz}^{2\prime T}$}}  & 0 & \multicolumn{2}{c}{\multirow{2}{*}{ $ M_z^{2\prime}$}}\\ 
     &  &  & 0 &  &   
    \end{pmatrix}
		\;.
\end{equation}
This isolates the would be Goldstone bosons, $G^0$ and $G^\pm$. Thus, only a $2\times 2$ charged Higgs mass matrix and a $5\times 5$ neutral Higgs mass matrix remain to be diagonalized. This is performed with orthogonal $W$ and $R$ matrices (respectively) according to 
\begin{equation}\label{mass_eigenstates}
	\Big(H_1^+ \; H_2^+\Big)^T = W\;\Big(w_2^{+\prime}\; w_3^{+\prime}\Big)^T  \;, \qquad \quad \Big(h_1\;h_2\;h_3\;h_4\;h_5\Big)^T = R 
	\;\Big(x_1\;x_2\;x_3\;z_2^\prime\; z_3^\prime\Big)^T \;,
\end{equation}
such that
\begin{equation} \label{e:mass_equations}
	M_{ch}^{2\prime} = W^\dag 
	\begin{pmatrix}
		m_{H_1^\pm}^2 & 0 \\ 0 & m_{H_2^\pm}^2
	\end{pmatrix}
	W
	\;,
	\quad 
	\begin{pmatrix}
\left(M_{x}^{2}\right)_{\textbf{3x3}} &   \left(M_{xz}^{2\prime}\right)_{\textbf{3x2}}\\
\vspace{2pt} \\
   \left(M_{xz}^{2\prime T}\right)_{\textbf{3x2}} &  \left(M_z^{2\prime}\right)_{\textbf{2x2}}\\   
    \end{pmatrix}= R^T 
	\begin{pmatrix}
		m_{h_1}^2 & 0 & 0 & 0 & 0 \\ 
		0 & m_{h_2}^2 & 0 & 0 & 0 \\ 
		0 & 0 & m_{h_3}^2 & 0 & 0 \\ 
		0 & 0 & 0 & m_{h_4}^2 & 0 \\ 
		0 & 0 & 0 & 0 & m_{h_5}^2 
	\end{pmatrix}
	R
	\;.
\end{equation}
We parameterize $W$ using two angles $\theta$ and $\phi$ as
\begin{equation}
	W = 
	\begin{pmatrix}
		c_\theta e^{i \varphi} & s_\theta e^{-i \varphi} \\*[1mm]
		-s_\theta e^{i \varphi} & c_\theta e^{-i \varphi}
	\end{pmatrix},
\end{equation}
and $R$ as some product of the ten rotations $R_{ij}$ in five dimensions. For our choice, we start by defining the $O_{ij}$ $5 \times 5$ matrices which are like the identity matrix in all entries
except the entries $ii$ and $jj$, given by $\cos{\alpha_{ij}}$,
the entry $ij$, given by $\sin{\alpha_{ij}}$,
and the entry $ji$ given by $-\sin{\alpha_{ij}}$.
For our studies we made the possible choice
\begin{subequations}
	\label{eq:Rspecific}
	\begin{eqnarray}
R_x &=& O_{23} O_{13}O_{12} \;,
\\
R_z &=& O_{45} \;,
\\
R_{CPV} &=& O_{35}O_{34}O_{25}O_{24}O_{15}O_{14}\, ,\\
R &=& R_{CPV} R_x R_z = R_{CPV} R_z R_x \; .
	\end{eqnarray}
\end{subequations}
With these definitions, taking the real limit corresponds to setting the $R_{CPV}$ as the identity matrix and setting $\varphi=0$.
For later use, we define the full set of neutral mass eigenstates as,
\begin{equation}\label{matrixQ_c3hdm}
    \begin{split}
    \begin{pmatrix}
			\xi_1 \\  \xi_2 \\ \xi_3 \\ \xi_4 \\ \xi_5 \\ \xi_6
		\end{pmatrix}
        \equiv
		\begin{pmatrix}
			G^0 \\ h_1 \\ h_2 \\ h_3 \\ h_4 \\ h_5
		\end{pmatrix}
        		&=
		Q
		\begin{pmatrix}
			x_1 \\ x_2 \\ x_3 \\ z_1 \\ z_2 \\ z_3
		\end{pmatrix}
\\
	&= 		\left(\begin{matrix}0 & 0 & 0 & c_{\beta_1} c_{\beta_2} & c_{\beta_2} s_{\beta_1} & s_{\beta_2}\\R_{11} & R_{12} & R_{13} & - R_{14} s_{\beta_1} - R_{15} c_{\beta_1} s_{\beta_2} & R_{14} c_{\beta_1} - R_{15} s_{\beta_1} s_{\beta_2} & R_{15} c_{\beta_2}\\R_{21} & R_{22} & R_{23} & - R_{24} s_{\beta_1} - R_{25} c_{\beta_1} s_{\beta_2} & R_{24} c_{\beta_1} - R_{25} s_{\beta_1} s_{\beta_2} & R_{25} c_{\beta_2}\\R_{31} & R_{32} & R_{33} & - R_{34} s_{\beta_1} - R_{35} c_{\beta_1} s_{\beta_2} & R_{34} c_{\beta_1} - R_{35} s_{\beta_1} s_{\beta_2} & R_{35} c_{\beta_2}\\R_{41} & R_{42} & R_{43} & - R_{44} s_{\beta_1} - R_{45} c_{\beta_1} s_{\beta_2} & R_{44} c_{\beta_1} - R_{45} s_{\beta_1} s_{\beta_2} & R_{45} c_{\beta_2}\\R_{51} & R_{52} & R_{53} & - R_{54} s_{\beta_1} - R_{55} c_{\beta_1} s_{\beta_2} & R_{54} c_{\beta_1} - R_{55} s_{\beta_1} s_{\beta_2} & R_{55} c_{\beta_2}\end{matrix}\right)
		\begin{pmatrix}
			x_1 \\ x_2 \\ x_3 \\ z_1 \\ z_2 \\ z_3
		\end{pmatrix} .
	\end{split}
\end{equation}

Taking $\xi_2\equiv h_1$ as the  $125~{\rm GeV}$ Higgs, it is easy to identify the
conditions that guarantee that $h_1$ couples as the SM Higgs.
These conditions may be written as
\begin{equation} \label{cond_alignment_c3hdm}
R_{1k}=\left(R_H\right)_{1k}\,,\quad (k=1,2,3)\,,\qquad \, R_{14}=R_{15}=0\,,
\end{equation}
or, which is the same, as
\begin{equation}\label{alignment_c3hdm}
\alpha_{12}=\beta_1\,,\quad \alpha_{13}=\beta_2\, ,\quad \alpha_{14}=\alpha_{15}=0\,.
\end{equation}
These conditions align the $125~{\rm GeV}$ Higgs regardless of the exact values of $ \alpha_{23},\, \alpha_{24},\,\alpha_{25},\,
\alpha_{34},\, \alpha_{35},\,\alpha_{45}$.
Indeed, one can show that the $hVV$ coupling divided
by its SM value ($k_V$) may be written as
\begin{equation}
    \frac{k_V}{\cos{\alpha_{14}}\cos{\alpha_{15}}} = 1-2\sin{\left(\frac{\alpha_{12}-\beta_1}{2}\right)^2}\cos{\left(\frac{\alpha_{13}+\beta_2}{2}\right)^2}-2\sin{\left(\frac{\alpha_{13}-\beta_2}{2}\right)^2}\cos{\left(\frac{\alpha_{12}-\beta_1}{2}\right)^2}\,.\label{eq:k_V}
\end{equation}
\section{Independent parameters\label{sec:parameters}}

For a phenomenological study, one would like to perform
an extensive scan of the parameter space.
Our fixed inputs are $v = 246~{\rm GeV}~{\rm GeV}$ and $m_{h_1} = 125~{\rm GeV}$. We then would take random values in the
ranges:
\begin{subequations}
	\label{eq:scanparameters_c3hdm}
	\begin{eqnarray}
  &\theta,\,\varphi,\, \in
\left[-\pi,\pi\right]; \\[8pt]
& \alpha_{12},\, \alpha_{13},\, \alpha_{14},\,
\alpha_{15},\, \alpha_{23},\, \alpha_{24},\,\alpha_{25},\,
\alpha_{34},\, \alpha_{35},\,\alpha_{45}\, \in
\left[-\pi,\pi\right];\\[8pt]
&\tan{\beta_1},\,\tan{\beta_2}\,\in \left[0.3,10\right]; 
\\[8pt]
& m_{h_2}\, 
\in \left[126,1000\right]~{\rm GeV},\,
m_{H_1^\pm},\,m_{H_2^\pm}\,
\in \left[100,1000\right]~{\rm GeV};\\[8pt]
&
\text{Re}(m^2_{12}),\text{Re}(m^2_{13}),
\text{Re}(m^2_{23}) \in  \left[\pm 10^{-1},\pm 10^{7}\right]\,
~{\rm GeV}^2\, .
	\end{eqnarray}
\end{subequations}
These 20 free parameters (in addition to $v$ and $m_{h_1}$), fully determine the point in parameter space, as we had found from the Lagrangian. The masses $m_{h_3},\, m_{h_4},\, m_{h_5}$ are noticeably absent from the previous list. This occurs because they are not independent from the parameters in 
\eq{eq:scanparameters_c3hdm}; indeed this model has the constraint equations
\begin{equation}\label{e:X1i}
	m_{h_i}^2 \left[ R_{i5} c_{\beta_2}(R_{i1}s_{\beta_1}-R_{i2}c_{\beta_1}) - R_{i3}R_{i4} \right] = X_{1i} m_{h_i}^2 = 0
	\;,
\end{equation}
\begin{equation}\label{e:X2i}
	m_{h_i}^2 R_{i5} \frac{ c_{\beta_2}(R_{i1}c_{\beta_1}+R_{i2}s_{\beta_1}) - R_{i3} s_{\beta_2} }{s_{\beta_2}}  = X_{2i} m_{h_i}^2  = 0
	\;,
\end{equation}
\begin{equation}\label{e:X3i}
	m_{h_i}^2 \frac{ R_{i4} (R_{i1}c_{\beta_1}-R_{i2}s_{\beta_1}) - R_{i5} s_{\beta_2}(R_{i1}s_{\beta_1}+R_{i2}c_{\beta_1}) }{s_{\beta_1}}  = X_{3i} m_{h_i}^2  = 0
	\;,
\end{equation}
where we have implicitly defined a $3\times 5$ matrix $X$. Inverting the system and assuming $m_{h_1}^2$ and $m_{h_2}^2$ are given, 
\begin{equation}
	\begin{split}
		m_{h_3}^2 = 
		-\frac{ \sum_{i=1}^2 (X_{1i}X_{24}X_{35}-X_{1i}X_{25}X_{34}-X_{14}X_{2i}X_{35}+X_{14}X_{25}X_{3i}+X_{15}X_{2i}X_{34}-X_{15}X_{24}X_{3i}) m_{h_i}^2}{X_{13}X_{24}X_{35}-X_{13}X_{25}X_{34}-X_{14}X_{23}X_{35}+X_{14}X_{25}X_{33}+X_{15}X_{23}X_{34}-X_{15}X_{24}X_{33}} 
		\;,
	\end{split}\label{mh3}
\end{equation}
\begin{equation}
	m_{h_4}^2 = - \frac{1}{X_{24}X_{35}-X_{25}X_{34}} \sum_{i=1}^3 (X_{2i}X_{35}- X_{3i} X_{25}) m_{h_i}^2
	\;,\label{mh4}
\end{equation}
\begin{equation}
	m_{h_5}^2 = -\frac{1}{X_{35}}\sum_{i=1}^4  X_{3i}  m_{h_i}^2
	\;.\label{mh5}
\end{equation}

Notice that Eqs.~\eqref{mh3}-\eqref{mh5} do \textit{not} guarantee that the left-hand sides are indeed positive. Thus, points in parameter space yielding negative squared masses are to be discarded \footnote{The non-generality of the mass matrices is similar to what happens
in the C2HDM, where there is a single constraint.
However, the presence of one single constraint on the C2HDM makes it
easily solvable for one of the rotation angles, unlike in our C3HDM,
where the situation is much more complicated.}.

Explicit expressions for the quartic potential parameters $\lambda_{1-12}$ and the remaining quadratic terms, when written in terms of the parameters in \eq{eq:scanparameters_c3hdm}, can be found in Appendix~\ref{app:lambdas_c3hdm}.

\section{The Yukawa Lagrangian} \label{ssub:yukawa}

 The Yukawa sector is as described in Section~\ref{sec:yukawa} for the Type-Z, imposing a $\Z2\times \Z2'$ symmetry that acts on the scalars fields, right-handed down quarks ($d_R$) and right-handed charged-leptons ($\ell_R$) as 
\begin{equation}
    \begin{aligned}
     \label{eq:Symmetry}
     \Z2 :&\ \Phi_1 \to -\Phi_1,\quad \ell_R \rightarrow -\ell_R, \\
      \Z2' :&\ \Phi_2 \to - \Phi_2,\quad d_R \rightarrow -d_R ,
    \end{aligned}
\end{equation}
resulting in the Type-Z couplings of the model, in which each scalar couples to a different family of fermions. In this configuration, one may assume the right-handed fermions of each electric-charge to couple only to one of the Higgs doublets, which will be dubbed $\Phi_u$, $\Phi_d$ and $\Phi_\ell$ going forward. To understand how the neutral scalars couple to fermions, we need to look at the terms of the Yukawa Lagrangian, given by, 
%
\begin{equation}\label{LY}
- \mathcal{L}_\text{Yukawa} = \overline{Q}_L Y^u \Phi_d n_R
+\overline{Q}_L  Y^d \tilde{\Phi}_u p_R
+\overline{L}_L Y^\ell \Phi_\ell \ell_R + {\rm h.c.} 
\;,
\end{equation}
where $Q_L = (p_L \; n_L)^T$, $L_L = (\nu_L \; \ell_L)^T$,
while $n_R$, $p_R$, and $\ell_R$ are,
respectively,
right-handed down-type, up-type, and charged lepton fields,
written in a weak basis. The Yukawa matrices $Y^u, Y^d$ and $Y^\ell$ are $3\times 3$ in the respective fermionic sectors.
After Spontaneous Symmetry Breaking (SSB), the scalars acquire vevs leading to mass terms for the fermions. In this work, we do not consider right-handed neutrino fields. As a result, the neutrinos will be massless, and we can choose the charged lepton basis such that the $Y^\ell$ matrix is already diagonal:
\begin{equation}
    \frac{v_\ell}{\sqrt{2}}Y^\ell = D_\ell \equiv \text{diag}(m_e, m_\mu, m_\tau).
\end{equation}
As for the quarks, to perform the required basis change from flavour to the diagonal mass basis, one may rotate the quarks fields by an unitary transformation of the form
\begin{equation}
    f_{L/R} = U^f_{L/R}\tilde{f}_{L/R}.
\end{equation}
As a result, 
%
%
\begin{equation}
      (U^{f}_L)^\dagger  \frac{v_f}{\sqrt2} Y^f U^{f}_R \equiv D_{f} ,
\end{equation}
where $f = u, d$, $D_u=\text{diag}(m_u,m_c,m_t)$, and $D_d=\text{diag}(m_d,m_s,m_b)$. By changing to the quark mass basis, we obtain the Cabibbo-Kobayashi-Maskawa (CKM) matrix~\cite{Cabibbo:1963yz,Kobayashi:1973fv}, $V=(U^p_L)^\dagger U^n_L$ . 

The Lagrangian is now composed of fermion mass terms and scalar-fermion interactions, for which the CKM matrix is involved in the terms with charged scalars. The expression for the interactions with the neutral scalars, $\xi_{1-6}$, are
\begin{equation}
    -\mathcal{L}_{\xi ff} = \sum_f \sum^{2N}_{j=1} \frac{m_f}{v} \Bar{f} \left( c^e_{\xi_j ff} + i\gamma_5 c^o_{\xi_j ff} \right) f \xi_j ,
\end{equation}
where,
\begin{equation}\label{h0ff_couplings}
	\;\; c^e_{\xi_j ff} + i\gamma_5 c^o_{\xi_j ff} = \frac{v}{v_f}\left( Q_{jf} \pm i \gamma_5 Q_{j,N+f}\right)
	\;,
\end{equation}
while $f$ refers to the doublet that couples to a given fermion, the $``+''$ is applied to leptons or down-type quarks, and the $``-''$ for up-type quarks. In particular, for the $125~{\rm GeV}$ Higgs found at LHC,
\begin{equation}\label{yuk:c3hdm}
    c^e_{h_{125}ff} + i\gamma_5c^o_{h_{125}ff} = \frac{v}{v_f} (Q_{2,f} \pm i\gamma_5 Q_{2,3+f}),
\end{equation}
or, in terms of the rotation angles,
\begin{equation}
	c^e_{ff} \equiv
	c^e_{h_{125}ff} =  
	\frac{R_{11}}{c_{\beta_2} c_{\beta_1}} \;,\;
	\frac{R_{12}}{c_{\beta_2} s_{\beta_1}} \;,\;
	\frac{R_{13}}{s_{\beta_2} } \;,\;
	\qquad \text{for} \quad f=\ell,d,u  \;,
\end{equation}
and
\begin{equation}
	c^o_{ff} \equiv
	c^o_{h_{125}ff} = 
	\frac{-R_{14}s_{\beta_1}-R_{15}c_{\beta_1}s_{\beta_2}}{c_{\beta_2} c_{\beta_1}} \;, \;
	\frac{R_{14}c_{\beta_1}-R_{15}s_{\beta_1}s_{\beta_2}}{c_{\beta_2} s_{\beta_1}} \;,\;
	-  \frac{R_{15} c_{\beta_2}}{s_{\beta_2}}\, ,
	\qquad \text{for} \quad f=\ell,d,u \;. \label{eq:cff_odd}
\end{equation}

If we take the SM-like limit for the couplings in \eq{cond_alignment_c3hdm},
the even components and odd components are determined to be
$c^e_{ff} = 1$ and $c^o_{ff}=0$, respectively.

There are limits on the CP-odd coupling of the $125~{\rm GeV}$ Higgs coupling
with the top, arising from $tth$ production
~\cite{ATLAS:2020ior, CMS:2020cga, ATLAS:2023cbt}. Taking into account that $Q$ is an orthogonal matrix, with  $Q_{2 i} Q_{2 i} + Q_{2,3+i} Q_{2,3+i} = 1 $, and $Q_{2,3+i} \hat{v}_i = 0$, we can obtain from \eq{yuk:c3hdm} and the couplings of gauge
bosons to neutral scalars that 
\begin{align}
	c^e_i \hat{v}_i^2 &= \kappa_V \,, \label{e:e3c3hdm} \\
	c^o_i \hat{v}_i^2 &= 0 \,, \label{e:e4c3hdm} 
\end{align}
for $i=\ell,d,u $. If we take $c^o_{tt} = 0$, we find
an analytical relation between the odd coupling of the $\tau$ and $b$ quarks:
\begin{equation}
    \frac{c^o_{\tau\tau}}{c^o_{bb}} = - \frac{\hat{v}_2^2}{\hat{v}_1^2} = - \tan^2\beta_1 \, .
\label{notthere}
\end{equation}
This relation means that,
despite the apparent uncorrelated nature of the scalar coupling to the different fermion types in this Type-Z model, there exists a remnant correlation among the odd couplings in this limit.
We will test the assumption $c^o_{tt} \sim 0$ and \eq{notthere} with different numerical simulations.
As we will see in the next Chapter,
one dramatic result from better sampling methods is that we can
produce points consistent with all known data where both these
statements are far from realized.

\section{Summary}

In this Chapter we studied the structure of the mass matrices and couplings in the softly-broken complex $\Z2\times\Z2$ 3HDM (C3HDM). This has never been studied before, and
presents the possibility of addressing two important issues. 
Firstly, due to the fact that the Yukawa Lagrangian is already complex
because of the (complex) CKM matrix, the renormalization will require that the scalar
potential is also complex~\cite{Fontes:2021znm,deLima:2024hnk}.
Secondly, the C3HDM can address the possibility of having large pseudoscalar
couplings for the fermions.
Recently, this possibility has been almost
ruled out in the C2HDM~\cite{Biekotter:2024ykp},
due to the recent data on the eEDM and the searches at LHC.
One important constraint came from the measurement
of the decay $h_{125}\to \tau\bar{\tau}$. In the Type-Z C3HDM, as each right-handed fermion of a given charge
couples with only one Higgs doublet, we have more freedom. We will follow with a full phenomenological study in the next Chapter, with a focus on showing this possibility.

We have introduced a parameterization for the rotation matrices needed
for the diagonalization of the neutral and charged scalars. This is
not trivial, as now all the five neutral scalars lack definite CP
properties and mix. 
This new formalism holds for Higgs-fermion couplings in any generic NFC 3HDM. We include the definition of the alignment limit and real limits. 

%% file: chapters/ai.tex

\chapter{Machine learning Exploration}\label{chapter:ml}
\hspace*{0.3cm}
There has been a great increase in the use of Machine Learning (ML)
for a variety of physical applications, recognized by the 2024 Nobel Prize in Physics
~\cite{Hopfield:1982pe,Rumelhart:1986gxv,LeCun:2015pmr}.
These techniques have found application in experimental particle physics searches
as, for example:
in ATLAS searches
for the Higgs ($H$)~\cite{ATLAS:2022vkf},
the single top ($t$) channel~\cite{ATLAS:2012byx},
and vector-like quarks~\cite{ATLAS:2018ziw};
in CMS boosted decision trees for Higgs detection~\cite{CMS:2012qbp}
and ttH couplings~\cite{CMS:2018uxb};
as well as the combined ATLAS+CMS determination of the Higgs
mass~\cite{ATLAS:2015yey}.

A different application has been used in theoretical particle physics.
Many physical models, including most particle physics models
beyond the Standard Model (SM),
involve a large number of parameters.
Exploring their parameter space becomes increasingly challenging, due to
stringent experimental constraints and the difficulty in identifying
viable regions in high-dimensional spaces. As shown in the previous Chapter~\ref{chapter:Alignment},  the traditional methods exhibit
low efficiency and often require assumptions that may leave significant
regions unexplored, such as starting close to the alignment limit.
The ML increase in efficiency is documented in,
for example, Refs.~\cite{Hollingsworth:2021sii,Baretz:2023mra,deSouza:2022uhk}.

In this Chapter, we also examine in depth the second advantage of ML relative to traditional
scanning methods. 
By employing a loss function with focused Novelty Reward,
we identify observables corresponding to clear signs of New Physics,
which would otherwise seem impossible to attain. The important result of this study is that
traditional scanning methods gave a completely wrong physical picture of the model. 

Although we apply this here to a study of CP-odd Higgs couplings in the context of the C3HDM defined in Chapter~\ref{chapter:C3HDM}, the method explained here finds applicability in all theoretical physics models that require a sampling; being especially impactful in models with large numbers of parameters.

This Chapter is organized as follows. In Section~\ref{sec:method_ml}, we describe the machine learning techniques employed to explore the parameter space of the model. Section~\ref{sec:constraints_ml} completes the constraints in Chapter~\ref{chapter:Constraints} for real scalar potentials with the bounds for models with explicit CP violation. 
 Section~\ref{sec:strategy_ml} defines the simulation strategies taken. Section~\ref{sec:results_ml} presents and discusses the results obtained, followed by a summary in Section~\ref{sec:summary_ml}.

\section{The method} \label{sec:method_ml}
\subsection{Black-box Machine Learning Optimisation}
\begin{figure}[H]
\centering
\includegraphics[height=7cm]{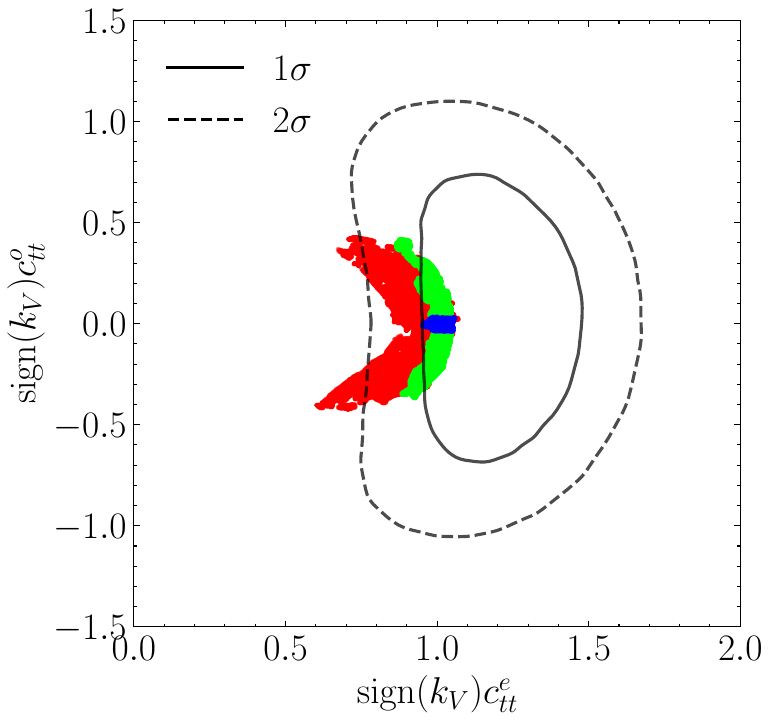}
\caption{\label{fig-top}Combined seeded plots with CMA-ES,
novelty detection and \textbf{including focus on top-top couplings}, in red and green. Points in green and blue include agreement at $3\sigma$ with $\Delta \chi^2$, as calculated with \texttt{HiggsSignals-2}.
The points in blue are obtained starting from the near-alignment real 3HDM explained in Section~\ref{sec:classic_c3hdm}. The experimental data from $tth$ production data is taken from~\cite{ATLAS:2020ior}. }
\end{figure}

To explore the parameter space, we adapted the AI black box optimisation approach first presented in~\cite{deSouza:2022uhk} and applied to a real 3HDM in~\cite{Romao:2024gjx}.
The procedure starts with defining the constraint function, $C(\mathcal{O})$, as
\begin{equation}
	C(\mathcal{O}) = \max(0, -\mathcal{O} + \mathcal{O}_{LB}, \mathcal{O} - \mathcal{O}_{UB}),
\end{equation}
where $\mathcal{O}$ is the value of an observable or a constrained quantity,
$\mathcal{O}_{LB}$ is its lower bound, and $\mathcal{O}_{UB}$ its upper bound.
The constraint function $C(\mathcal{O})$ returns 0 only if $\mathcal{O}$
lies within the specified interval; otherwise, it returns a positive
number quantifying \textit{how far} the value of the observable is
from the specified bounds.
Importantly, points that fail to satisfy constraints yield useful information.
The values $\mathcal{O}$ are obtained by a black box computational routine that
takes a vector $\theta$ in the parameter space as inputs to calculate all relevant physical quantities $\mathcal{O}$.
Finding points that satisfy a given constraint is equivalent to minimizing the respective  $C(\mathcal{O})$ function. To address multiple constraints simultaneously, we make the choice of optmising a \textit{single} loss function that encapsulates all the constraints of the model as
\begin{equation} \label{eq:lossf}
	L(\theta) = \sum_{i=1}^{N_c} C(\mathcal{O}_i(\theta)),
\end{equation}
where the sum runs over all the $N_c$ constraints,
ensuring $L \geq 0$ for all parameters $\theta$,
with $L = 0$ only when all constraints are
satisfied \footnote{In terms of implementation details, we perform two further 
transformations before using \eq{eq:lossf}  as the target function. First, we compute
the logarithm of the different $\mathcal{C}$ contributions to set them all to equivalent
nominal values and to endow us with a notion of numerical infinite necessary to
weed out unphysical points. Secondly, each $\mathcal{C}$  contribution is min-max normalised
within each generation so that no constraint wins over any other.}.
Notably, the quantity $\mathcal{O}_i$ does not necessarily need to be an experimental observable.
Mixing measurements, theoretical constraints and cuts in the same loss
function is a key strength of this methodology. While $\mathcal{O}_i$
may be referred to as ``observables'' in subsequent discussions,
we emphasize the flexibility to select diverse constraint types,
which allowed the exploration of exotic couplings.
All desired bounds must be specified, and the method efficiently
attempts to find valid points in parameter space, without knowledge
of the form of the likelihood. A $\Delta \chi^2$ test is performed
by calculating a single quantity and imposing a numerical bound
corresponding to $3\sigma$.
We used the \texttt{HiggsSignals} module of \texttt{HiggsTools-1.1.3} for
this calculation involving the $125~{\rm GeV}$ Higgs couplings.The dramatic impact can be seen in Fig.~\ref{fig-top}, 

\hspace*{-0.5cm}which shows the prediction for CP-violating couplings of the
top quark with the $125\textrm{GeV}$ Higgs.
The points in blue correspond to a traditional scan starting near the alignment of the real 3HDM, as in Chapter~\ref{chapter:Alignment};
while the points in red and green correspond to points
found using the new technique. All details are explained below; but the salient point here is that
traditional scanning methods gave a completely wrong physical picture of the model.
They implied that large CP-violating top couplings  were impossible.
In contrast,
the new approach decisively allows much larger couplings consistent with
current experimental data \cite{ATLAS:2020ior} -
shown as solid ($1\sigma$) and dashed ($2\sigma$) black lines.

\subsection{Optimisation with an Evolutionary Strategy}

Evolutionary Strategies are powerful numerical optimisation algorithms, which are characterized by an iterative process that samples the best points as candidate solutions for a problem, extracting them from a distribution and using them to generate the new distribution, from which the next generation is generated.

The sampling algorithm is the
Covariant Matrix Adaptation Evolutionary Strategy (CMA-ES)
~\cite{Hansen2001,hansen2023cmaevolutionstrategytutorial},
implemented in a python package~\cite{nomura2024cmaessimplepractical}.
In CMA-ES, the distribution is a highly localized multivariate normal,
initialized with its mean at a random point in parameter space and its
covariance matrix set to the identity matrix, $\one$, scaled by a constant,
$\sigma$. A generation of candidate solutions is sampled from this distribution,
and their losses are evaluated using \eq{eq:lossf}.
The candidates are ranked from best to worst, meaning how close
they are to fulfilling the constraints, with the best candidates used to
compute a new mean and approximate the covariance matrix.

This change in mean directs the algorithm towards the steepest descent of the loss function, similar to a gradient descent as an example of a first-order method, while the covariance matrix approximates the local Hessian, similar to a second-order method. However, CMA-ES does not compute derivatives, making it suitable for not so well behaved loss functions, which allows it to converge rapidly across a variety of optimisation problems.

\subsection{Novelty Reward} \label{ssub:NoveltyReward}

Although CMA-ES converges quickly, its reliance on a single multivariate normal exhibits limited discovery capacity due to the localized nature of its solution distribution. To address this,~\cite{Romao:2024gjx} introduced a \textit{novelty reward} into the loss function, penalizing solutions based on the density of previously identified valid regions. This density is obtained using a Histogram-Based Outlier Score (HBOS), a univariate anomaly detection algorithm~\cite{HBOS}. HBOS constructs a histogram for each parameter dimension in the parameter space $P$  and/or observable space $\mathcal{O}$, dividing the numerical data into equal-width bins over the range of allowed values. The frequency of samples that fall into each bin is used as an estimate for the density, given by the height of the bins. The penalties are normalised to $p \in [0,1]$, such that novelty points receive a penalty of value 0, corresponding to a lower density, and points similar to previously explored approach a maximal penalty of 1.

This adjustment encourages CMA-ES to explore novel regions by penalising the loss function when constraints are satisfied but solutions are near already discovered areas.

To ensure proper minimization with these penalties, the loss function is shifted accordingly,
\begin{equation}
	\Tilde{L}(\theta) =
	\left\{
	\begin{array}{ll}
		1 + L(\theta)   & \text{if } L(\theta) > 0 \\
		\text{ } \\
		0               & \text{if } L(\theta) = 0
	\end{array}
	\right..
\end{equation}
We can then add the penalties and obtain the final version of the loss function,
\begin{equation}
	L_T(\theta) = \Tilde{L}(\theta) + \frac{1}{2} \left( \frac{1}{N^{\mathcal{P}}_p} \sum_{i=1}^{N^{\mathcal{P}}_p} p^{\mathcal{P}}_i(\theta^i) + \frac{1}{N^{\mathcal{O}}_p}\sum_{i=1}^{N^{\mathcal{O}}_p}p^{\mathcal{O}}_i [\mathcal{O}^{i}(\theta)]\right),
\end{equation}
where $p^{\mathcal{P}}_i(\theta^i)$ is the density penalty of the parameter
space $\mathcal{P}$, normalised by the number of parameter penalties
considered, $N^{\mathcal{P}}_p$, and $p^{\mathcal{O}}_i [\mathcal{O}^{i}(\theta)]$
is the density penalty of the observable space $\mathcal{O}$,
also normalised by the number of observables considered to be penalised,
given by $N^{\mathcal{O}}_p$.
For a valid point,
$\sum_i C(\mathcal{O}_i) = 0 \Rightarrow  L_T \in [0, 1]$,
and for invalid points $L_T > 1$.
We stress that penalties do not need to apply to all parameters $\theta$. One may choose a subset of parameters of interest, ${\theta^i}$, and perform \textit{focused runs} with density penalty on those specific parameters. The same applies to the space of observables $\mathcal{O}(\theta)$, with the additional benefit of achieving a direct improvement in the exploration of novel phenomenological features of a model.

By design, each run is independent, as they are initialized with new values for the CMA-ES mean and covariant matrix parameters and trained solely on points from that run. The information from previous runs can however be also used to guide new runs using seeds. We may choose valid points from previous runs, store their parameter values and start new runs initialized already in that region.
CMA-ES allows initialization with a specific mean of the multivariate Gaussian and an overall scale of the covariant matrix, $\sigma$. Unseeded runs start with $\sigma=1$, while seeded runs with  $\sigma=0.01$ are confirmed to already start at the minimum of the constraint loss function, with increased up to $\sigma=0.1$  in an attempt to uncover new features.

\section{Implementing constraints} \label{sec:constraints_ml}

The model consistency is equivalent to requiring a set of
theoretical and experimental constraints.
For this analysis we choose the first neutral scalar as the
$125~{\rm GeV}$ Higgs. As described in Sec.~\ref{sec:parameters},
the parametrization leads to the squared mass of three other neutral
scalars to be derived parameters.
The requirement that all three are positive quantities is set as a
constraint that must be satisfied before the fitting procedure
is initiated~\cite{deSouza:2025uxb}.
During the simulation, all other constraints are considered on equal footing, both theoretical and experimental,  with the techniques described below.
This is a distinguishing feature of this method; (after the initial
squared mass step) there is no hierarchical sequence of constraints.
This is a very positive asset for the approach.
Indeed, for most models, one does not know ahead of the simulation which particular observables will be easier to obey, and which will turn out to be very difficult and require a dedicated analysis.
With this method one can remain agnostic to such foresight.

A more philosophical issue concerns the concept of ``fine-tuning''.
Imagine that we need to obey constraints $A$ and $B$.
By starting with $A$, we may find an easily populated region in
parameter space and some protrusion where one is less likely
to find good points.
Let us also imagine that observable $B$ is only satisfied precisely in that protruded region. One would be tempted to think of it as a fine-tuned region.
Conversely, had one started with constraint $B$, whatever region appeared would feel natural; especially if it automatically allowed a good fit to $A$. In the method described here the two observables are treated simultaneously. None is more important than the other; none drives our urge to find fine-tuning where it does not exist. All the constraints applied to the real 3HDM case are also to be applied: boundedness from below in Section~\ref{sec:3hdm_pot}, all the restrictions in Chapter~\ref{chapter:Constraints} and $b \to s \gamma$ in Section~\ref{sec:bsgamma}. In addition we have two specific constraints for models with explicit CP violation: 

\begin{itemize}
\item 
\textbf{eEDM:}  The calculation of the non-zero electric dipole moment (\gls{EDM}) of the electron
(eEDM) follows the formulae in
~\cite{Barr:1990vd,Yamanaka:2013pfn,Abe:2013qla,Inoue:2014nva,Altmannshofer:2020shb}, with significant experimental constraints ~\cite{ACME:2018yjb,Roussy:2022cmp}.
\item
\textbf{Direct searches for CP-violation:}
There are CP-violation constraints
from the decays of the $125~{\rm GeV}$ Higgs into $\tau \bar{\tau}$
~\cite{CMS:2021sdq,ATLAS:2022akr}:
$|\theta_{\tau}| = |\arctan(c^o_{\tau} / c^e_{\tau})| < 34^\circ$.
There are also limits on the CP-odd coupling of the $125~{\rm GeV}$ Higgs coupling
with the top, arising from $tth$ production
~\cite{ATLAS:2020ior, CMS:2020cga, ATLAS:2023cbt}.
We show the $1\sigma$ and $2\sigma$ contour plots in the relevant figures.
\end{itemize}

\section{Simulation strategies}\label{sec:strategy_ml}

\subsection{Random scan}
Ideally one would like to perform
an extensive scan of the parameter space.
Our fixed inputs are $v = 246~{\rm GeV}$ and $m_{h_1} = 125~{\rm GeV}$. We then would take random values in the
ranges in \eq{eq:scanparameters_c3hdm}.
These 20 free parameters (in addition to $v$ and $m_{h_1}$), fully determine the point in parameter space, as we had found from the Lagrangian. We sampled $10^{13}$ points using
    2100 CPU hours (on an AMD Ryzen 9 7950X 16-Core processor) and did not find any good point.

\subsection{Importing from the near-alignment real 3HDM}\label{sec:classic_c3hdm}

With the definitions for the rotation matrix $R$ in \eq{eq:Rspecific} taking the real limit corresponds to setting the $R_{CPV}$ as the identity matrix and setting $\varphi=0$, or equivalently setting,
\begin{align}
  \label{eq:enlarging1}
  &\alpha_{12}=\alpha_{1},\ \ 
  \alpha_{13}=\alpha_{2},\ \ \alpha_{23}=\alpha_{3},\ \ 
  \alpha_{45}=-\gamma_1,\ \ \theta=-\gamma_2 \nonumber\\[8pt]
  &\varphi=\alpha_{14}=\alpha_{15} = \alpha_{24}=\alpha_{25}=
  \alpha_{34}=\alpha_{35}=0\, ,
\end{align}
where $\alpha_1,\,\alpha_2,\,\alpha_3,\,\gamma_1,\,\gamma_2$ are the
variables defined in the real 3HDM parameterization of \eq{CPevenDiag},
and $\beta_1,\,\beta_2$ retain their meaning in both cases.
So we are able to produce sets of points in the real $\Z2\times\Z2$ 3HDM, following a strategy
of closeness to the alignment limit, as in Chapter~\ref{chapter:Alignment}. These points
should also be good C3HDM points. The next step is to scan around
these points. This was done within the ranges
\begin{align}
  \alpha_{14},&\, \alpha_{15} \in [-0.01,0.01];\nonumber\\[8pt]
  \varphi,\,\alpha_{24},\,&\alpha_{25},\,\alpha_{34},\,\alpha_{35} \in
  [-0.1,0.1] . \label{eq:enlarging2}
\end{align}
These ranges were chosen to be close to the real case. $\alpha_{14},\, \alpha_{15}$ are involved with the alignment condition for the C3HDM in \eq{alignment_c3hdm} and therefore have to be smaller to comply with
the experimental limits, especially of the eEDM. 

\begin{figure}[H]
  \centering
  \begin{tabular}{cc}
      \includegraphics[width=0.47\textwidth]{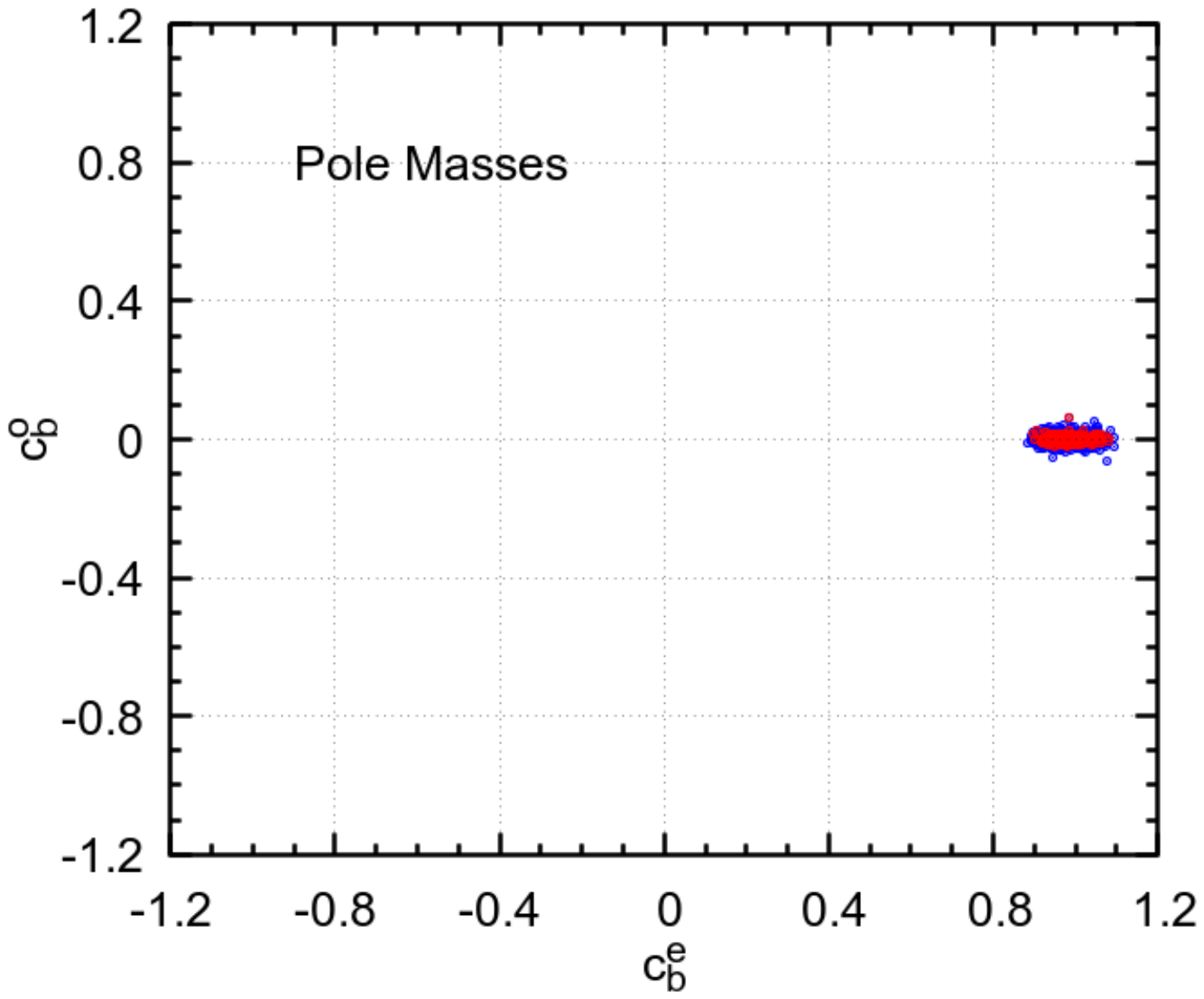}
    &
      \includegraphics[width=0.47\textwidth]{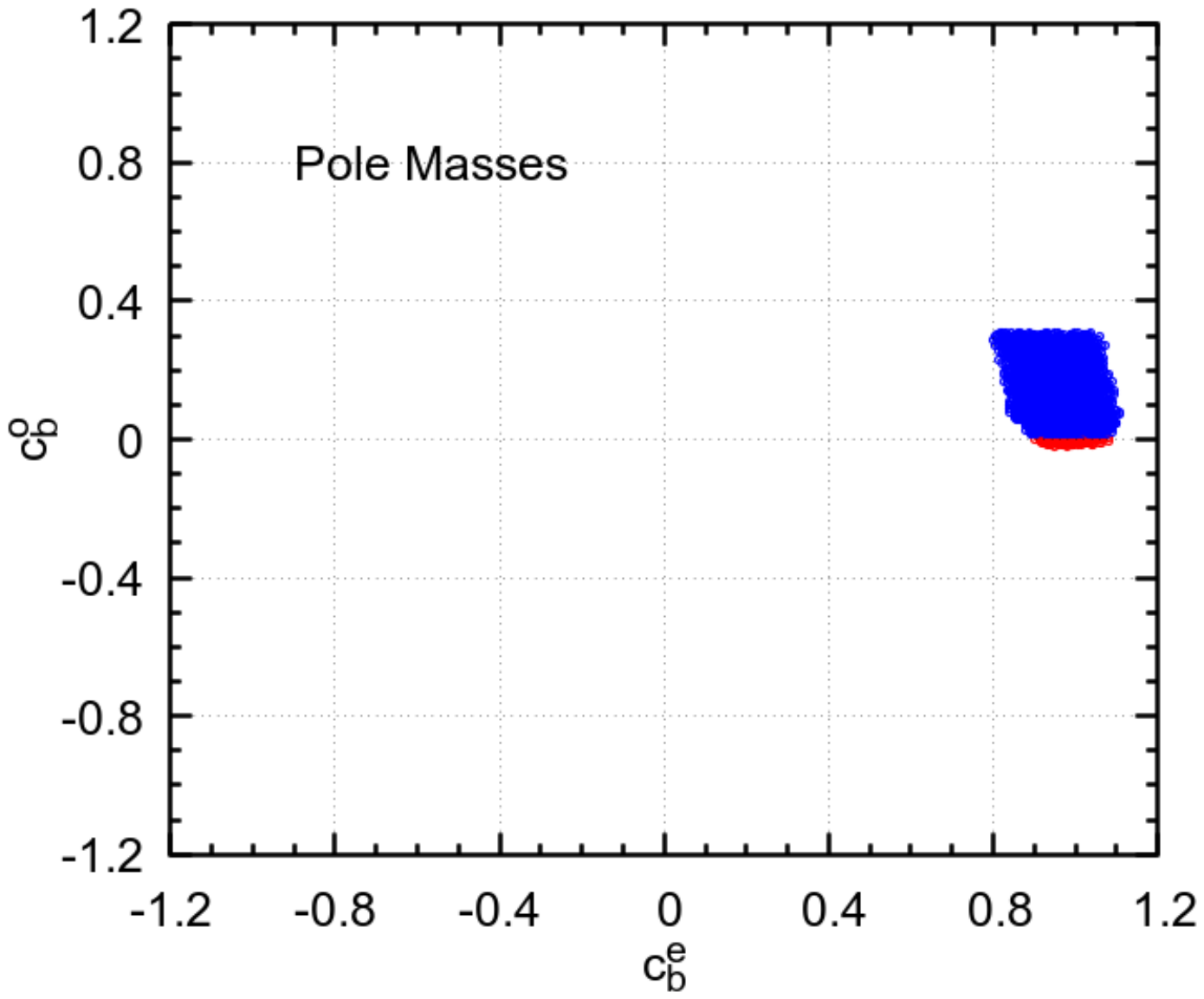}
  \end{tabular}
  \caption{Enlarging the pseudoscalar component around solutions of the real
    3HDM. Left Panel: The blue (red) points are before (after) the
    \texttt{\rm HiggsTools} constraints. Right panel: the points that pass \texttt{\rm HiggsTools} are used iteratively in an attempt to enlarge the pseudoscalar component.} 
  \label{fig:3}
\end{figure}

 We started from 81386 points 
from the real 3HDM, already after passing the
\texttt{HiggsTools}~\cite{Bahl:2022igd} constraints. From these, only
2893 passed all the C3HDM constraints (in blue in Fig.~\ref{fig:3} Left)
and after \texttt{HiggsTools} only 372 remained (in red). This so
large difference is easy to understand. Once we generate the C3HDM
parameters, the masses $m_{h_3},\, m_{h_4},\, m_{h_5}$, are derived
quantities, which have to satisfy being in the interval $[125,1000]~{\rm GeV}$ and the STU constraints (these could have been true for the real 3HDM
points, but we have no control over these masses when we import the
points into the C3HDM). But most importantly, the calculation of the
parameters of the 
potential also changes and many points are discarded due to BFB or
perturbative unitarity constraints.

The points shown in Fig.~\ref{fig:3} Left have non zero pseudoscalar
components, but those are small because of the way in which they were generated,
coming from the real 3HDM case. So the question arises: can one
enlarge the pseudoscalar component? To answer this question we
developed the following strategy. We start with the good C3HDM set of points
after \texttt{HiggsTools}, in red in Fig.~\ref{fig:3} Left. We then
consider random points \textit{around} those points. From this run, we select
only the points with the largest pseudoscalar component (supposing,
for the moment that we are going in the positive direction, the same
can be applied going into the negative direction, of course). We
consider these as input and scan around these points, keeping only the
points with the largest pseudoscalar component. We repeat this
procedure until we cannot go any higher. Much of the success depends,
clearly, on the meaning of \textit{around}, and of course, this is a
time consuming method. We present in Fig.~\ref{fig:3} Right the situation after a few
iterations. The points in red are those of Fig.~\ref{fig:3} Left, generated from the
initial points from the real 3HDM. Points in blue were generated by
the method explained above\footnote{As mentioned above, this is a time
  consuming method because it is difficult to automatize.}.

\begin{figure}[h!]
  \centering
  \begin{tabular}{cc}
    \includegraphics[width=0.48\textwidth]{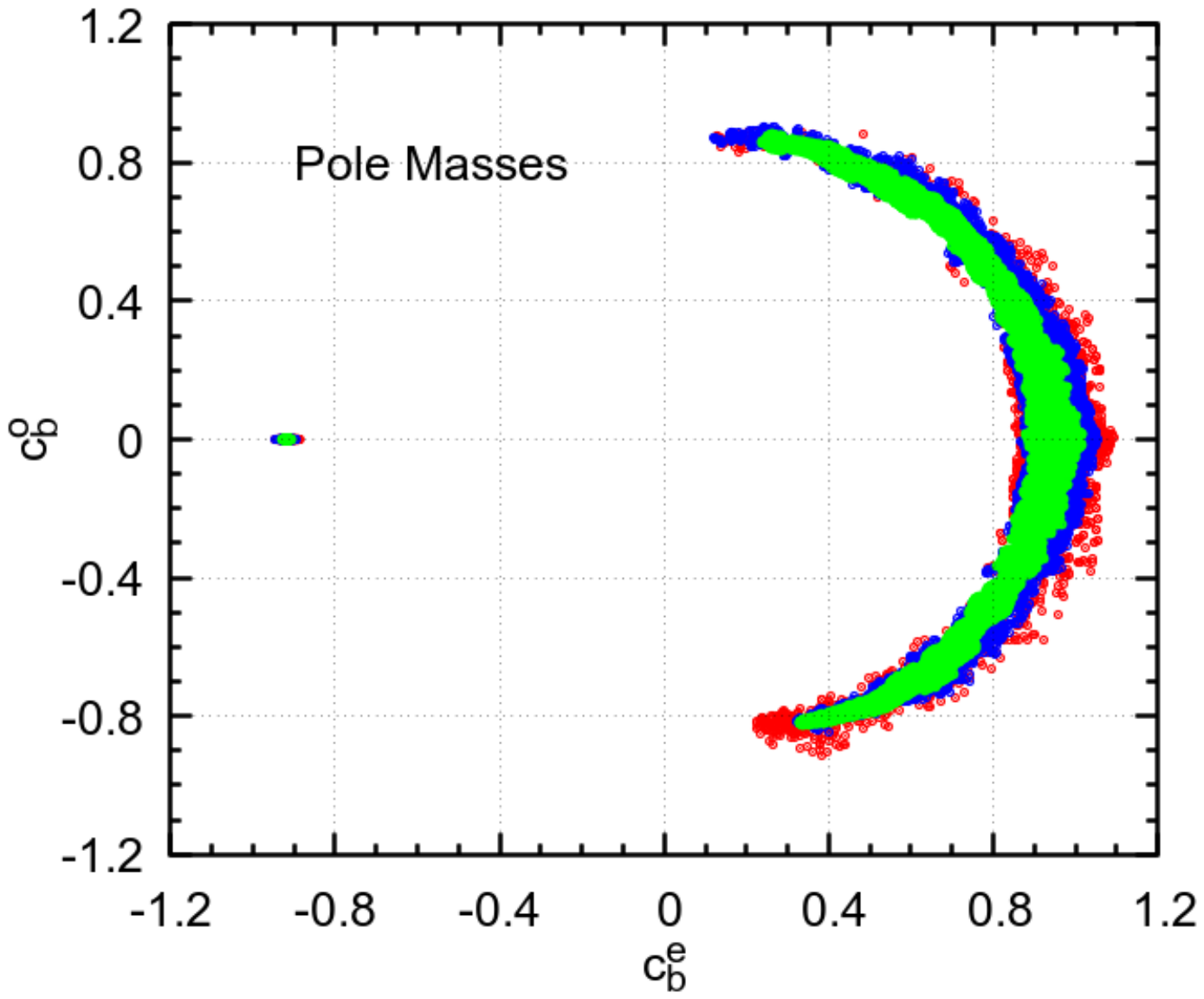}
    &
    \includegraphics[width=0.48\textwidth]{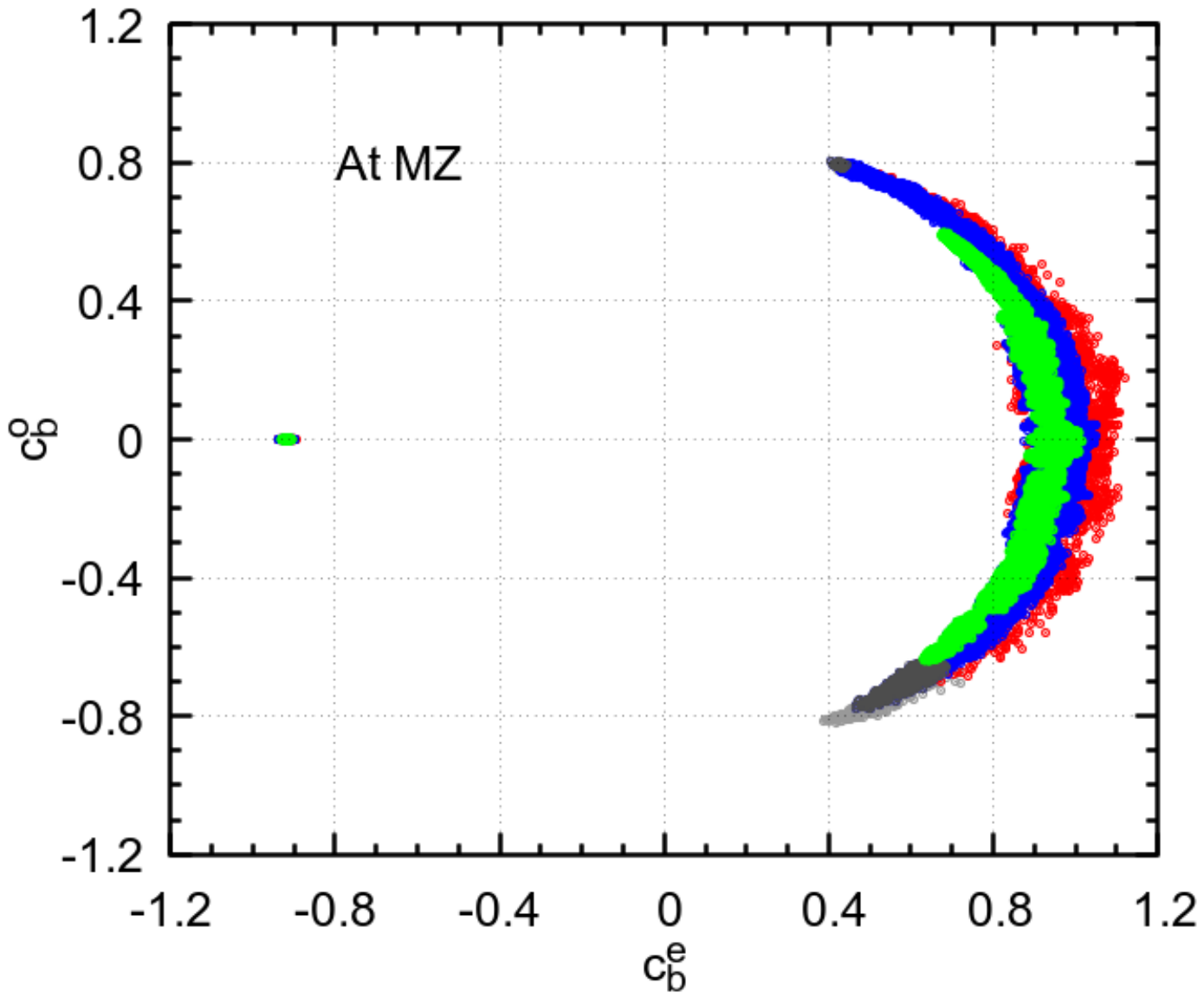}
  \end{tabular}
  \caption{In the left panel we plot $c^o_b$ vs. $c^e_b$  for the pole
    masses and right panel for 
    masses at the $M_Z$ scale. The constraints on $h_{125}\to
    \tau\bar{\tau}$ do not affect the former but slightly affect the
    latter (gray). The
points in red correspond to a $2 \sigma$ agreement with each individual
signal strength measurement.
The points in blue and green correspond to the method where
the signal strengths are constrained using the \texttt{HiggsSignals-2}
module in \texttt{HiggsTools-1.1.3},
for $\Delta \chi^2$ corresponding to $3 \sigma$ and $2 \sigma$, respectively.}
  \label{fig:mz_mpole_c3hdm}
\end{figure}

 In Fig.~\ref{fig:mz_mpole_c3hdm}, we have the results for the scalar and pseudoscalar coupling of the down quarks after iteratively enlarging the pseudoscalar component. We see in Fig.~\ref{fig:mz_mpole_c3hdm} that we can obtain
large pseudoscalar couplings, although maximal values are
excluded. As
explained in Ref.~\cite{Biekotter:2024ykp}, the cancellations needed to
obey the present eEDM limit
of $4.1 \times 10^{-30}\, \textrm{e.cm}$ reported by
the JILA collaboration~\cite{Roussy:2022cmp},
are quite large and depend
strongly on the precise value of the constants and at which scale they
are taken. In Ref.~\cite{Biekotter:2024ykp} two cases were studied:
one case where the masses were taken as pole masses;
and another case where everything was considered at the $M_Z$ scale.
This had a strong impact; for instance,
the possibility of large pseudoscalar couplings for the Type-II C2HDM
disappeared when taking the constants at the $M_Z$ scale.
The implication is that the scale dependence and, thus,
the assumed theoretical errors in estimating the eEDM warrants a full
independent and thorough analysis. In the remaining figures of this Chapter, we always take the constants at the $M_Z$ scale, as the focus is set on comparing the sampling methods.

\subsection{Focused ML exploration with CMA-ES}
We considered yet a very different strategy, taking advantage of the ML
methods described in Sections~\ref{sec:method_ml} and~\ref{sec:constraints_ml}. When we perform a run, CMA-ES varies the parameter
values slightly in the direction of valid points. This may converge to valid regions,
but it does not necessarily give us a broad exploration in specific
parameters of interest.
To force the algorithm to increase exploration for these particular parameters,
we can add them to the penalized quantities as explained in
Section~\ref{ssub:NoveltyReward}, which will then focus
the search on the chosen parameters and help populate plots.
After a successful run that uncovered promising points,
we generate seeds from them to guide the next point distribution.
These seeds help steer subsequent runs toward the desired regions,
improving both convergence and exploration.\\
Both features, focused runs and seeds, can and should be used
simultaneously when appropriate to boost performance.
In the results section below, we present only seeded and
focused runs.
We also specify the plane variables that are being focused on,
which generally include the plotted variables.

We should stress that sets of points generated in the fashion proposed
do \textbf{not} have a final statistical interpretation; neither a frequencist
nor a posterior interpretation.
These points were generated by \textit{focusing} on certain sets of observables
(in the technical sense explained above). What is sought in this method is a thorough exploration
of the possible phenomenological consequences of a given model, and not an attribution
of likelihood of any kind.

\section{Results and Comparison}\label{sec:results_ml}

In this Section, we present the final combination of our scans with the implemented techniques.
In several subsequent figures, we consider the parameters at scale $M_Z$ and adhere to the following color code:
\begin{itemize}
\item
The points in \textbf{red} pass all the theoretical and experimental constraints, including a $2\sigma$ agreement with
each individual $125~{\rm GeV}$ signal strength, and the new Higgs searches in
the \texttt{HiggsBounds} module found in \texttt{HiggsTools-1.1.3}.
In this way of generating points with \textbf{machine learning}, there may be a large number of observables
which lie at the edge of their allowed regions, yielding a large overall $\Delta \chi^2$.
\item
The points in \textbf{green} combine points originally in red that are later found to
also satisfy $\Delta \chi^2$ at $3\sigma$, calculated using \texttt{HiggsSignals-2}
module in \texttt{HiggsTools-1.1.3}, with points generated with
$\Delta \chi^2$ at $3\sigma$ as a constraint in the \textbf{machine learning} algorithm,
with an appropriate  $C(\mathcal{O})$ contribution to \eq{eq:lossf}.
Thus, points in green imply the usual attribution of significance to the
$\Delta \chi^2$ of points on the theoretical parameter space.
\item
The points in \textbf{blue} correspond to the standard scanning technique described in Section~\ref{sec:classic_c3hdm}, 
without requirements on $\Delta \chi^2$,
to draw comparisons between the two methods considered.
\end{itemize}
We see that using ML techniques, the traditional iterative method are improved upon dramatically for all plots shown.

The plots below consist of order $5\times 10^8$ parameter points in different projection planes,
and combine scans where the novelty reward was aimed at different sets of parameters.
The first highlight of our technique was the ability to generate valid parameter
points within the full parameter domain, and at high efficiency, with order $10^5$
points in less than 100 CPU hours.
This should be compared with the painstaking job
of having to devise a strategy
of scanning around solutions of the real 3HDM and slowly
increasing the CP-violating phases, obtaining order $10^6$ parameter points.

As the signs of $c^e_{ff}$ and $c^o_{ff}$ have no meaning without taking into consideration
the sign of  $k_V \equiv c(h_{125}V V )$, in the following plots we always
present quantities in the form  $\text{sgn}(k_V)c^e_{ff}\,\text{vs. sgn}(k_V)c^o_{ff}$,
with $k_V$ given by \eq{eq:k_V}.

\subsection{\texorpdfstring{$b$ quark couplings}{b quark couplings}}
\begin{figure}[H]
	\centering
	\includegraphics[height=7cm]{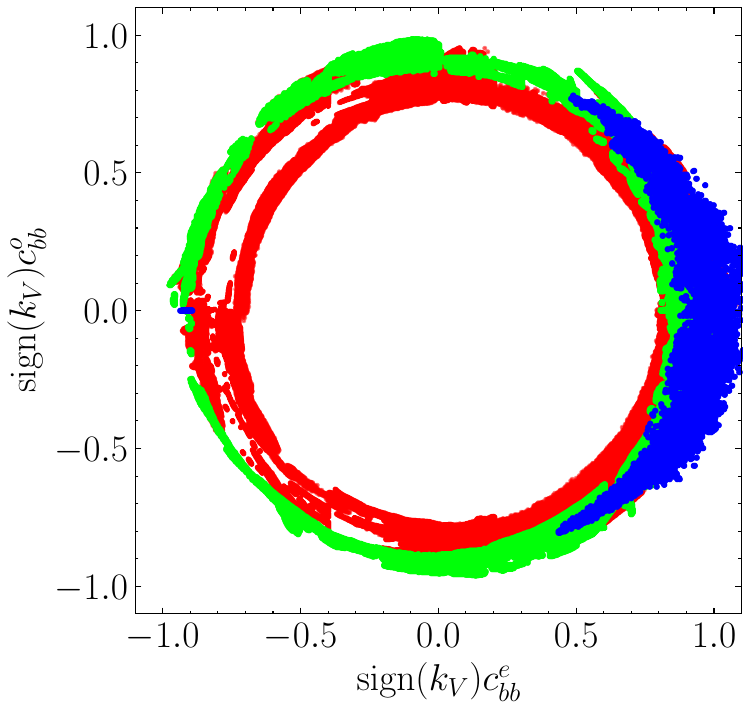}
	\hfill
	\includegraphics[height=7cm]{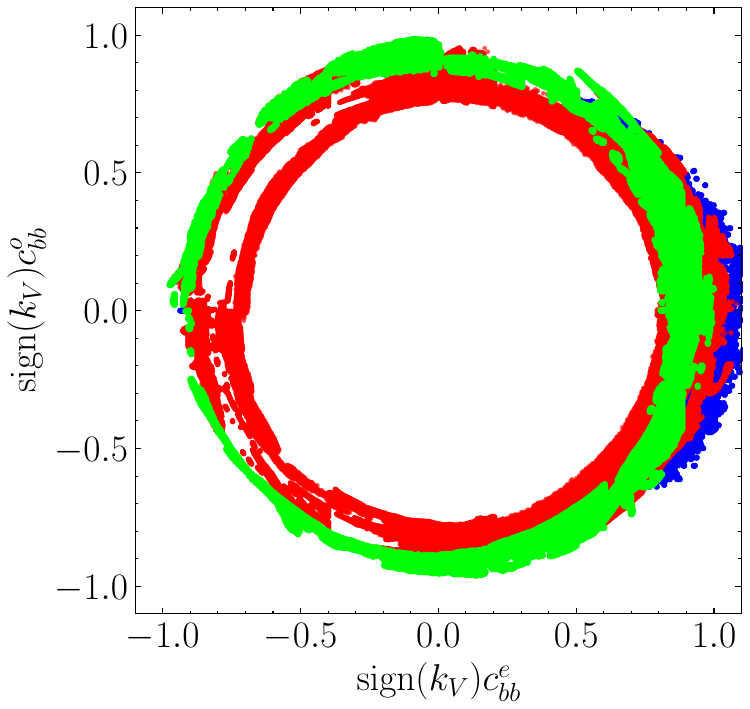}
	\caption{Combined seeded plots with CMA-ES, novelty detection and \textbf{including focus on $b\bar{b}$ couplings}. The blue coloured points obtained with the method in Section~\ref{sec:classic_c3hdm} are shown above (Left plot) or below (Right plot) the points produced with the ML evolutionary technique. }
	\label{fig-bb-vs-paper}
\end{figure}

In Fig.~\ref{fig-bb-vs-paper}, we present our results for the scalar and
pseudoscalar coupling of the down quarks.

The first striking difference
between sampling methods
is that with ML we are now able to find maximal values of the pseudoscalar
couplings, even when making the stronger requirement of meeting
$\Delta \chi^2$ at $3\sigma$.
The algorithm is able to efficiently find the cancellations
needed to obey the latest eEDM limit,
while simultaneously meeting every other constraint.
We are also able to completely populate the wrong sign region\footnote{The wrong sign regime is defined by a relative sign of the Higgs-fermion Yukawa coupling compared to the Higgs-gauge coupling. It was shown that the Type Z in the real 3HDM~\cite{Das:2022gbm} offers more flexibility than the C2HDM \cite{Ferreira:2014naa,Fontes:2017zfn} with respect to the possibility of the wrong sign. It was shown that while keeping $\kappa_V \approx \kappa_U \approx 1$ we can have $\kappa_{D,L}\approx \pm 1$, which we obtain.}. 

In broad strokes, the strategy to fill the plots was done in three main steps.
First, runs were performed with novelty reward on the parameters $|c_{bb}^o|$,
until a significant amount of converged runs that can be used for seeded runs.
This only filled the right-hand side of the plot,
with some exceptions for wrong sign with small $|c_{bb}^o|$.
Second, we added a constraint that forced $|c_{bb}^o|$ to be large,
with appropriate $C(\mathcal{O})$, and repeated the same setup of runs
with novelty reward followed by seeded runs.
Lastly, we forced $c_{bb}^e$ to have increasingly negative values, adding this
as a constraint to force the convergence in the wrong sign case
(again running seeded runs based on valid runs with novelty reward).
The last two steps each took around $10^5$ CPU hours with the first taking a fraction.
This method results in having some cuts as remnants of the fractured scans.
An improvement would be to employ novelty reward that remembers all
the previous scans, which could consist of a large amount of data. However, this significantly
increases HBOS computational time, hindering its usefulness. 
\begin{figure}[H]
	\centering
	\includegraphics[height=6cm]{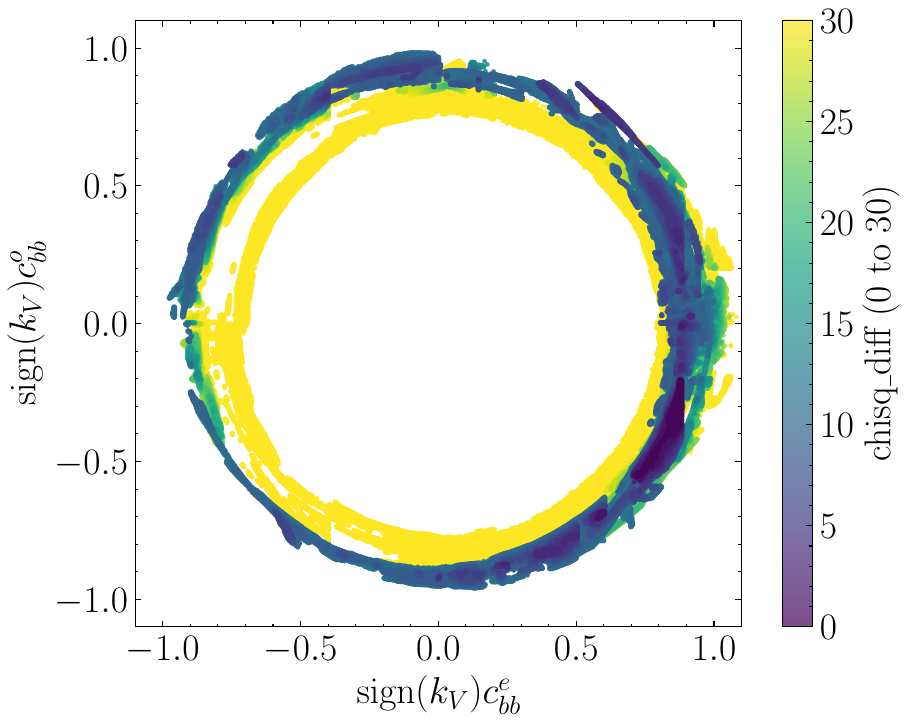} 
	\hfill
	\includegraphics[height=6cm]{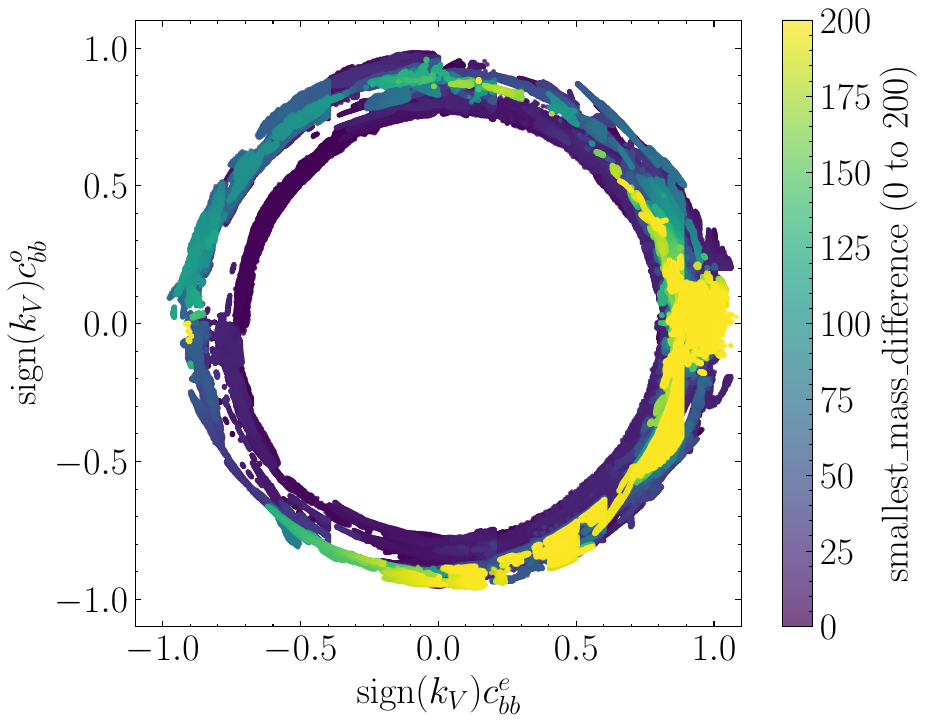}
	\caption{Combined seeded plots with CMA-ES, novelty detection and \textbf{including focus on $b\bar{b}$ couplings}. The points combine runs that required agreement with $\Delta \chi^2$ at $3\sigma$ and ones that did not. The left figure has a colour scale showing the lowest found $\Delta \chi^2$ for a given point in 2D and the right figure the highest found difference between the mass for the lightest scalar and the $125~{\rm GeV}$ Higgs. }
	\label{fig-bb-chisq}
\end{figure}
On the plot on the left (right) of Fig.~\ref{fig-bb-vs-paper},
the blue points with the method in Section~\ref{sec:classic_c3hdm} are drawn last (first).
We notice two features.
First,
because the (blue) points were drawn starting from
the real 3HDM limit, they are concentrated in a region extending away from
the $(1,0)$ point.
And, in that way of searching, the farthest away, the more difficult it is to find a point.
The fact that we now have green points across the full circle shows the true impact of this
novel search technique.
Indeed,
by ``focusing''
(in the technical simulation sense mentioned above) on this $hbb$ plane,
we can generate allowed points on this plane with relative ease.

Second,
points with 
$\text{sgn}(k_V)c^o_{ff} > 1.0$,
found to be possible starting from the real limit,
were difficult to generate here, and, by comparing green and red regions,
they require larger values of $\Delta \chi^2$.
Points with lower radius $\sqrt{|c^e_{bb}|^2 + |c^o_{bb}|^2}$
also require larger values of  $\Delta \chi^2$.
This is also apparent on the left panel of Fig.~\ref{fig-bb-chisq},
as we explain next.

In Fig.~\ref{fig-bb-chisq} we use the red points from Fig.~\ref{fig-bb-vs-paper};
that is, both points that obeyed $\Delta \chi^2$ at $3\sigma$ calculated
using \texttt{HiggsSignals-2}, and points that did not.
On the left panel, we show the lowest found $\Delta \chi^2$ for a given point in 2D;
the yellow points represent $\Delta \chi^2$ equal or above $30$.
The right panel of Fig.~\ref{fig-bb-chisq} presents a colouring based on the highest
found difference between the mass for the lightest scalar and
the $125~{\rm GeV}$ Higgs.
We see that points with lower radius would require a second Higgs almost
degenerate with the $125~{\rm GeV}$ Higgs.

\subsection{\texorpdfstring{$\tau$ lepton couplings}{tau lepton couplings}}
\begin{figure}[H]
	\centering
	\includegraphics[height=6cm]{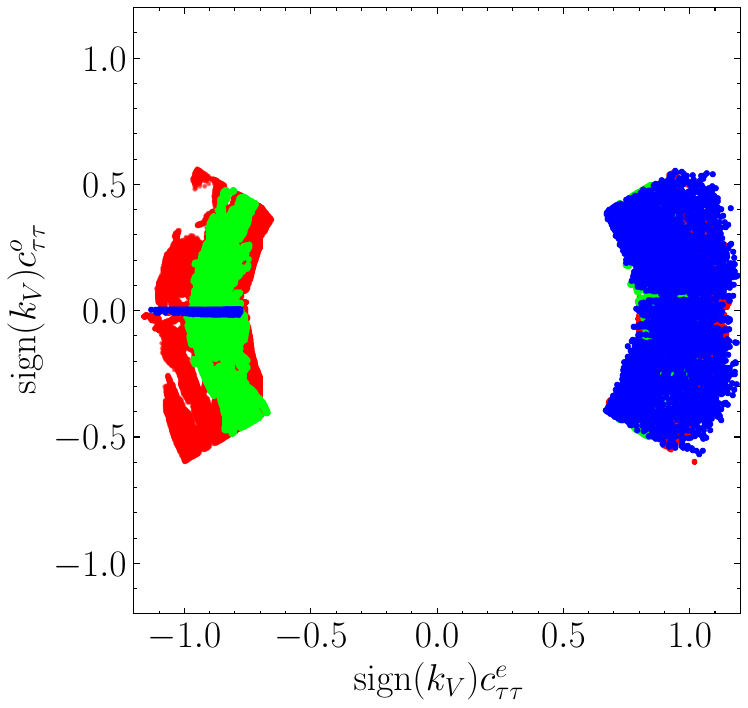} 
	\hfill
	\includegraphics[height=6cm]{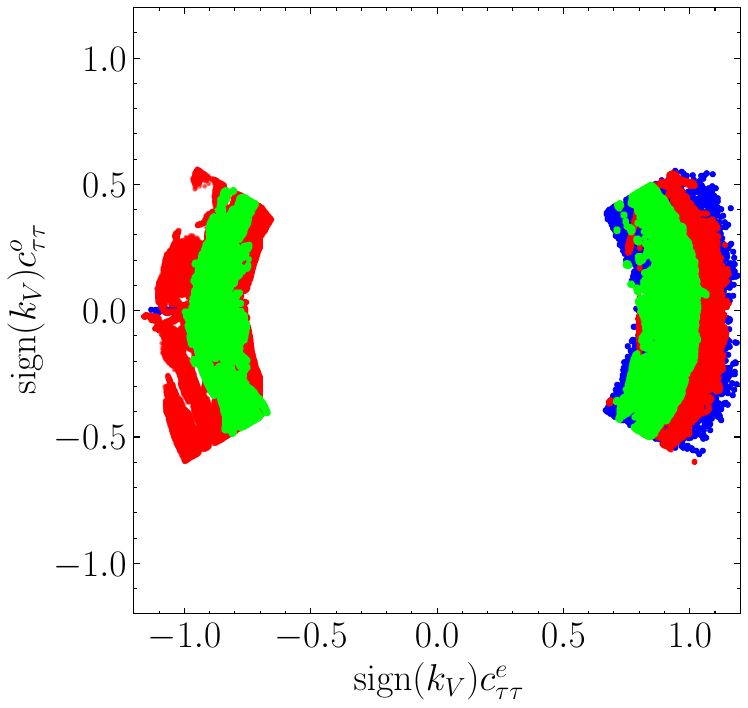}
	\caption{Combined seeded plots with CMA-ES, novelty detection and \textbf{including focus on $\tau\bar{\tau}$ couplings}. The blue coloured points obtained with the method in Section~\ref{sec:classic_c3hdm}  are shown above (Left plot) or below (Right plot) the points produced with the evolutionary technique.}
	\label{fig-tautau-vs-paper}
\end{figure}

In Fig.~\ref{fig-tautau-vs-paper}, we show our results with respect to the charged leptons.
The latest data from direct searches for CP-violation in decay planes of $\tau$ leptons
places an upper limit on $c^o_{\tau\tau}/c^e_{\tau\tau} \equiv \tan{(\alpha_{h\tau\tau})}$,
yielding $|\alpha_{h\tau\tau}|<34^{\circ}$~\cite{CMS:2021sdq,ATLAS:2022akr}.
We include this as a strict constraint.
These experimental bounds are also included in \texttt{HiggsSignals-2}.

\begin{figure}[H]
	\centering
	\includegraphics[height=6cm]{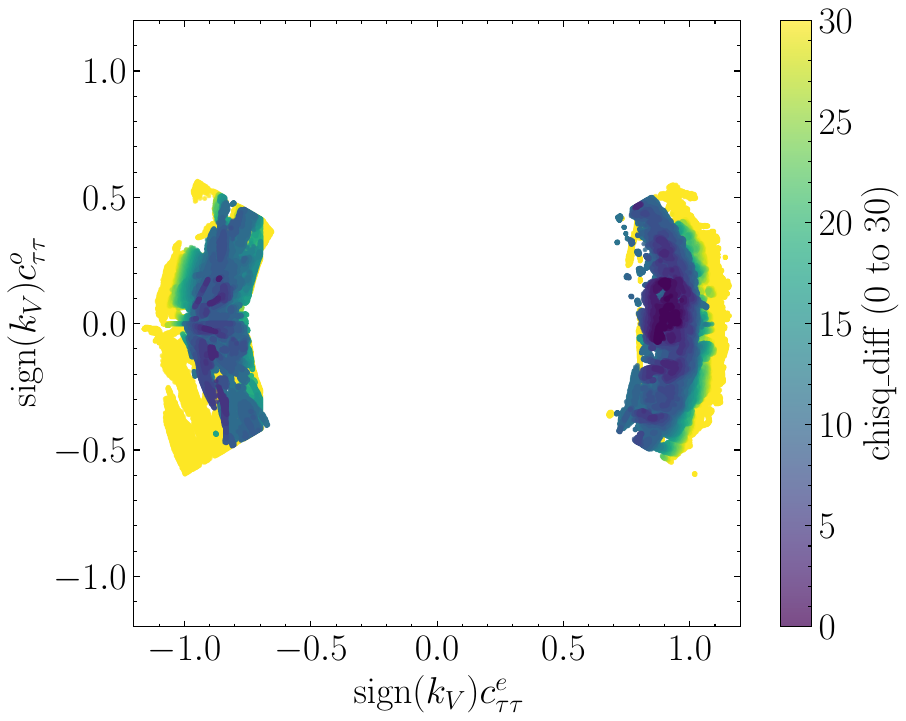}
	\hfill
	\includegraphics[height=6cm]{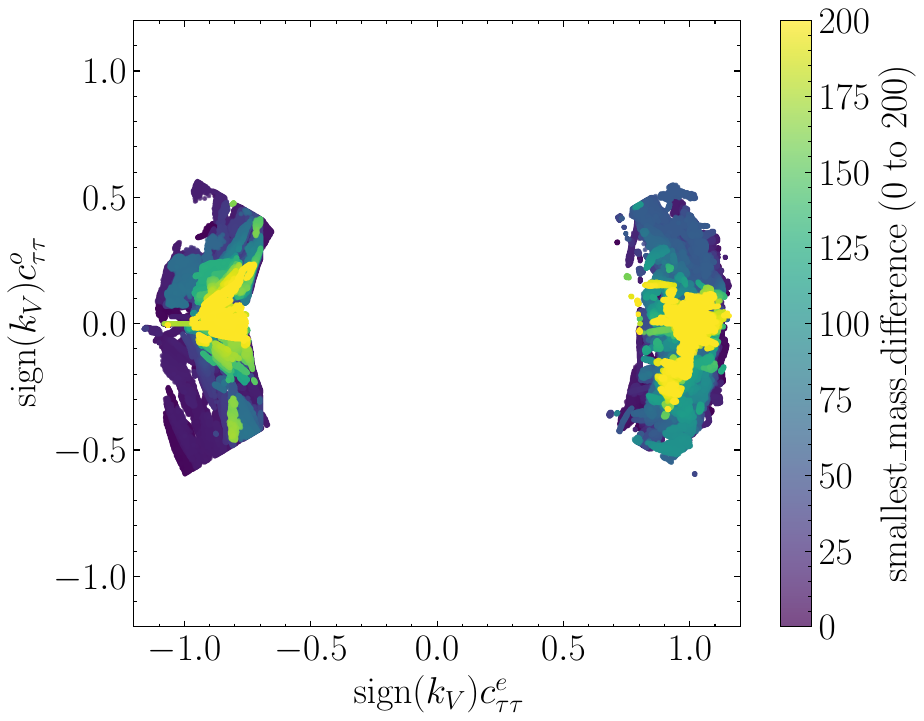}
	\caption{Combined seeded plots with CMA-ES, novelty detection and \textbf{including focus on $\tau\bar{\tau}$ couplings}. The left figure has a colour scale showing the lowest found $\Delta \chi^2$ for a given point in 2D and the right figure the highest found difference between the mass for the lightest scalar and the $125~{\rm GeV}$ Higgs.}
	\label{fig-tautau-chisq}
\end{figure}

The angles that meet this criteria can coexist with all other constraints.
Requiring agreement with the $\Delta \chi^2$ test at $3\sigma$ lowers the
possible radius $\sqrt{|c^e_{\tau\tau}|^2 + |c^o_{\tau\tau}|^2}$,
as also shown on the left panel of Fig.~\ref{fig-tautau-chisq}.
We are able to populate the region with negative sign for the lepton coupling,
even for large values of $|c^0_{\tau\tau}| \sim 0.5$.
As seen clearly from the line of blue points on the wrong-sign region of
Fig.~\ref{fig-tautau-chisq}-Left, this seemed completely impossible when using the
scanning method of Section~\ref{sec:classic_c3hdm}.

\subsection{\texorpdfstring{Uncorrelated $\tau$ and $b$ CP-odd couplings}{Uncorrelated tau and b CP-odd couplings}}

\begin{figure}[H]
\centering
\includegraphics[height=7cm]{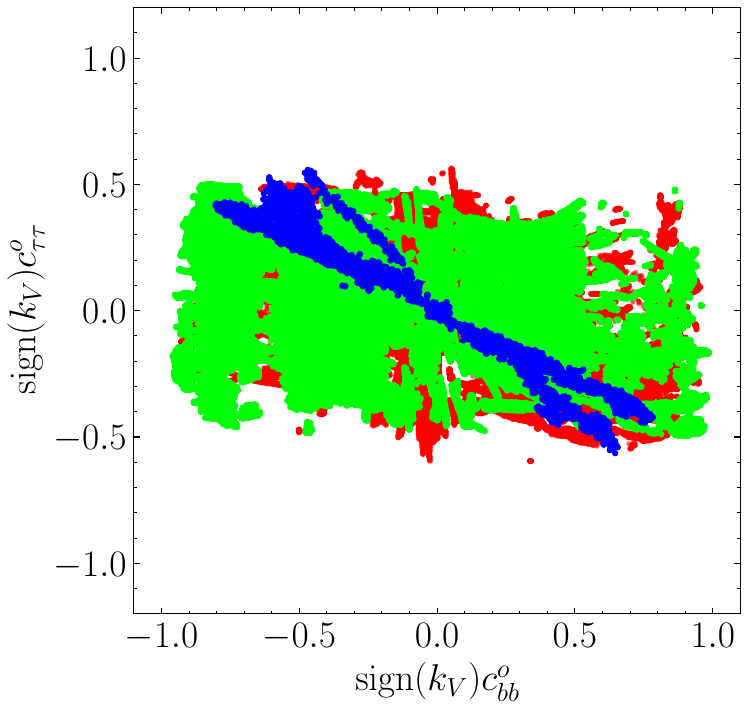}
\hfill
\includegraphics[height=7cm]{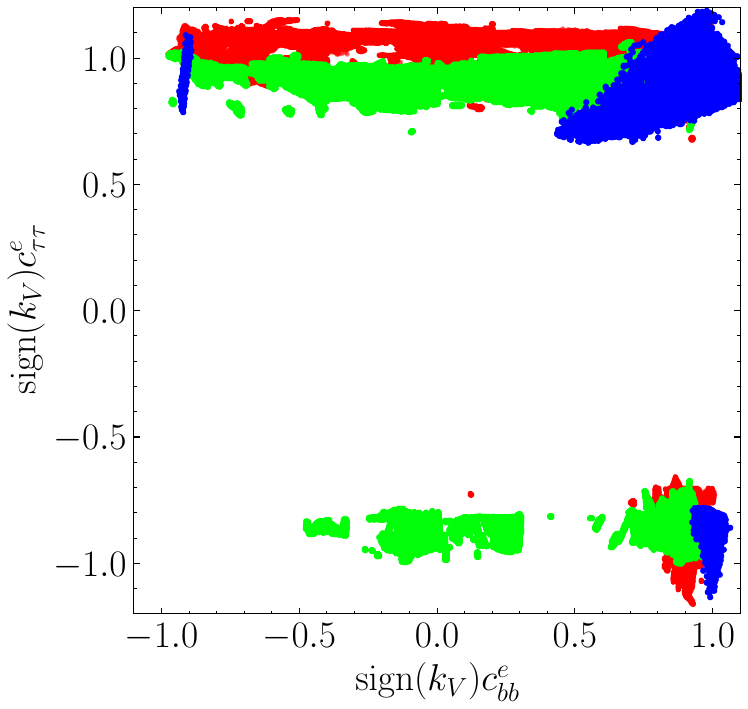} 
\caption{Combined seeded plots with CMA-ES, novelty detection
and \textbf{including focus on $\tau\bar{\tau}-b\bar{b}$ pseudoscalar couplings}.
The color code is the same as before, with blue points drawn last.}
\label{fig-bothwrong}
\end{figure}

We now turn to the second ``money-plot'' showing the full power of the
new search method, presented on the left panel of Fig.~\ref{fig-bothwrong},
where we compare $c^o_{\tau\tau}$ with $c^o_{bb}$.
This shows a very dramatic difference between the search 
performed starting from points near the aligned real 3HDM followed by 
increasing $c^o_{ff}$ (blue points),
and the search performed here using ML techniques (red and green points).
Indeed,
the blue points follow roughly the relation in \eq{notthere},
which, as pointed out in Section~\ref{ssub:yukawa}, would arise from taking $c^o_{tt}=0$.
This seemed to be confirmed by the simulations with the method of Section~\ref{sec:classic_c3hdm}.
Using the ML method,
we were able to access new regions of parameter
space where \eq{notthere} is far from valid.
Said otherwise, we can now access regions where $c^o_{tt}$ is very different from zero
(we will come back to $c^o_{tt}$ in Sec.~\ref{ssub:top} below).
We notice from Fig.~\ref{fig-bothwrong}-Left that one can fill the whole plane,
roughly for $|c^e_{\tau\tau}|<0.5$, as limited by~\cite{CMS:2021sdq,ATLAS:2022akr}, and $|c^e_{bb}|<1.0$.

Figure~\ref{fig-bothwrong}-Right shows that we are able to populate the region 
that have simultaneous wrong-sign couplings for the down quarks and for the leptons,
but that, with the ranges in \eq{eq:scanparameters_c3hdm},
we cannot find $\textrm{sgn}(k_V)c^e_{bb} \sim -1$.

\subsection{top couplings} \label{ssub:top}

The relation in \eq{notthere} was derived
assuming $c_{tt}^o=0$.
The scanning method used in Section~\ref{sec:classic_c3hdm}, based on starting with points for
the real case in the alignment limit,
supported this relation and assumption.
However, the ML approach finds no anticorrelation between
$c_{\tau\tau}^o$ and $c_{bb}^o$,
as shown on the left panel of Fig.~\ref{fig-bothwrong}.
This is confirmed in Fig.~\ref{fig-top}, where $|c_{tt}| > 0.3$ is a clear possibility,
when simulations cover properly the full range of parameters,
as achieved with our new method.

It turns out that the CP nature of the $125~{\rm GeV}$ Higgs coupling to the top quark
has been probed experimentally in $tth$ production data~\cite{ATLAS:2020ior}.
For ease of comparison,
the choice of scale in Fig.~\ref{fig-top} mimics that in~\cite{ATLAS:2020ior}.
The experimental results are shown as $1\sigma$ (solid) and $2\sigma$ contour lines
~\cite{ATLAS:2020ior},
indicate consistency with $c_{tt}^o \neq0$.
%
%
%

\begin{figure}[htb]
\centering
\includegraphics[height=7cm]{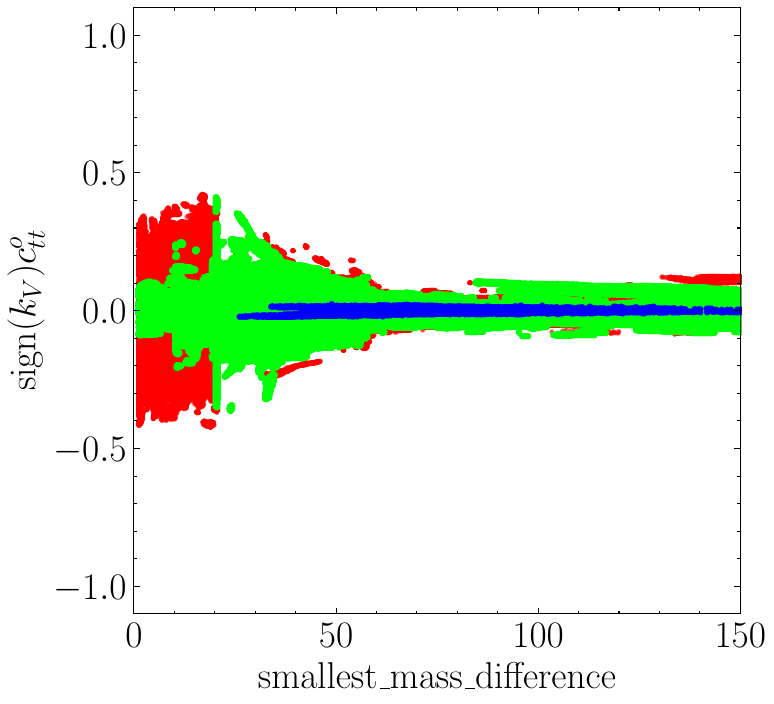}
\caption{Combined seeded plots with CMA-ES, novelty detection
and \textbf{including focus on top coupling and mass difference}.
The ``mass difference'' refers to the smallest mass difference
between every neutral scalar and $m_h=125~{\rm GeV}$.
The color code for points is the same as before.}
\label{fig-top-massdiff}
\end{figure}

Figure~\ref{fig-top-massdiff}, shows that  increasing $c_{tt}^o$ implies the existence of a second Higgs with mass closer
to the $125~{\rm GeV}$ Higgs.

Spurred by this plot of the C3HDM,
we have reanalyzed the C2HDM~\cite{Biekotter:2024ykp} using the
same ML scheme described in this Chapter,
hoping that values of $|c_{tt}^o| > 0.1$ would also be possible for that simpler model.
Indeed,  
we found points where $|c_{tt}^o| > 0.1$,
but only when the lightest scalar’s mass lies below $126~{\rm GeV}$,
and in situations very close to degeneracy.
As pointed out in Section~\ref{sec:parameters},
we chose our parameter range in \eq{eq:scanparameters_c3hdm} precisely to
exclude closely degenerate cases, thus precluding this region of the C2HDM.

\subsection{The real limit}

\begin{figure}[htb]
	\centering
	\includegraphics[height=7cm]{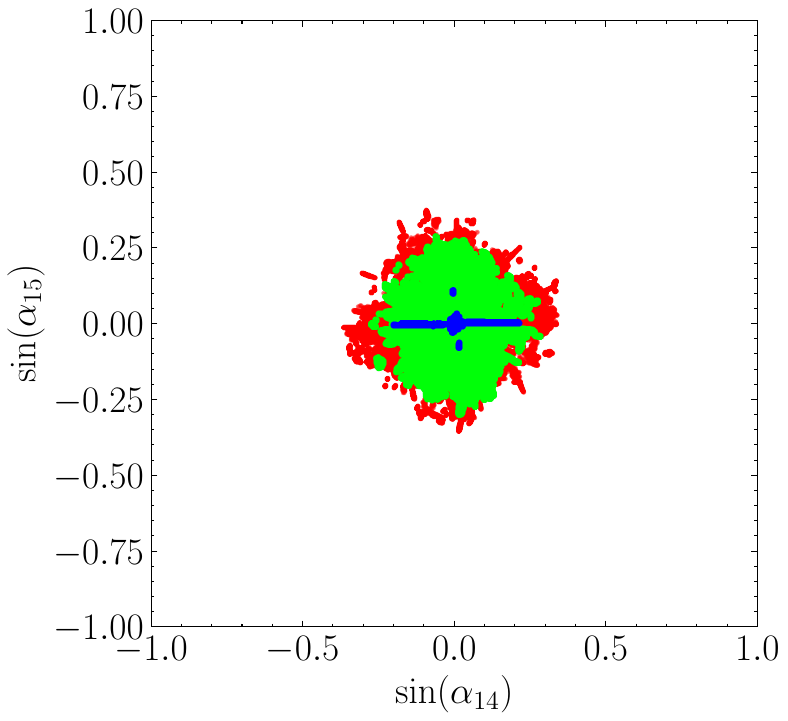}
	\hfill
	\includegraphics[height=7cm]{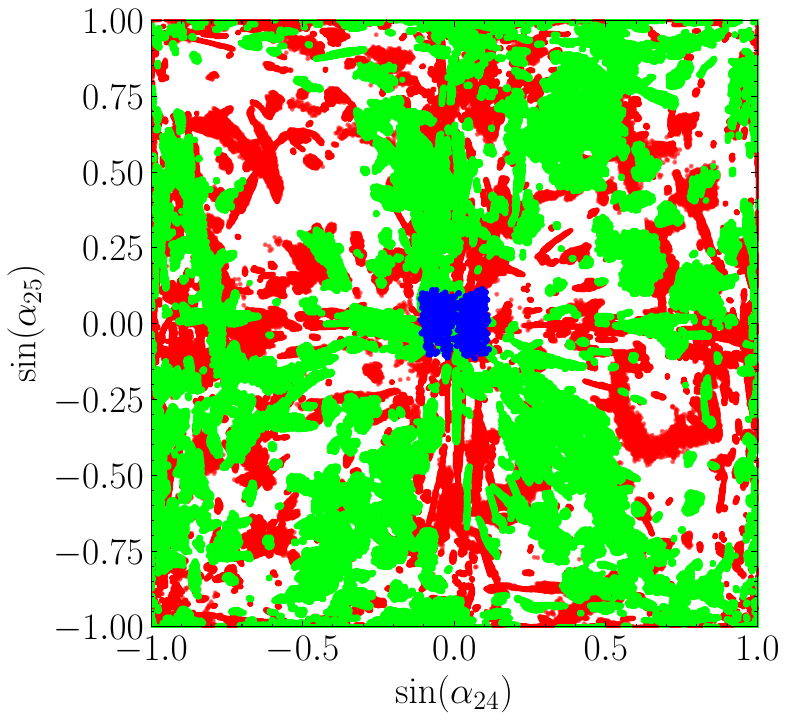}
	\caption{  Combined seeded plots with CMA-ES. The points shown \textbf{did not include dedicated scans with novelty reward focused on the parameters shown}. Blue the points obtained with the method in Section~\ref{sec:classic_c3hdm}. }
	\label{fig-14-15}
\end{figure}

\begin{figure}[htb]
	\centering
	\includegraphics[height=7cm]{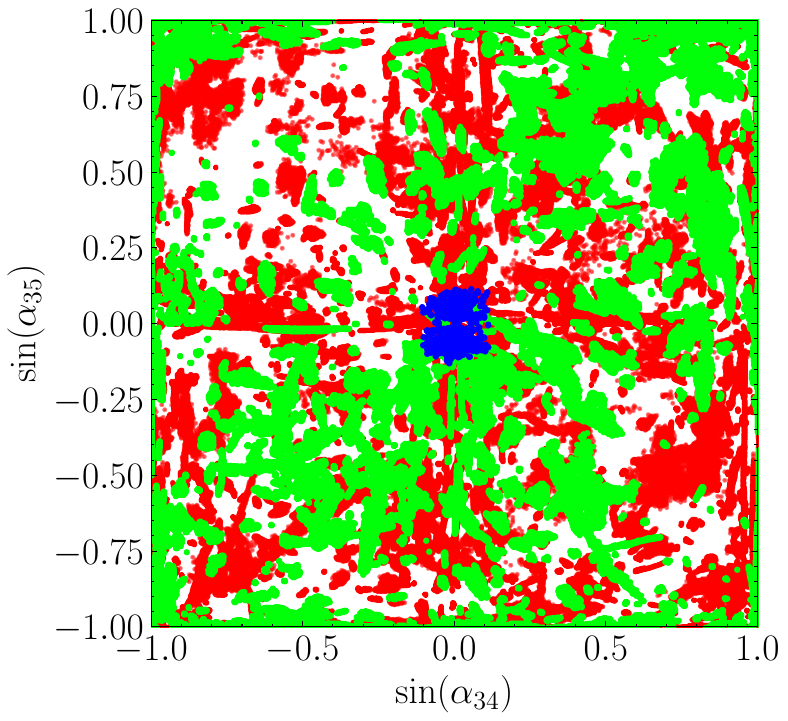}
	\hfill
	\includegraphics[height=7cm]{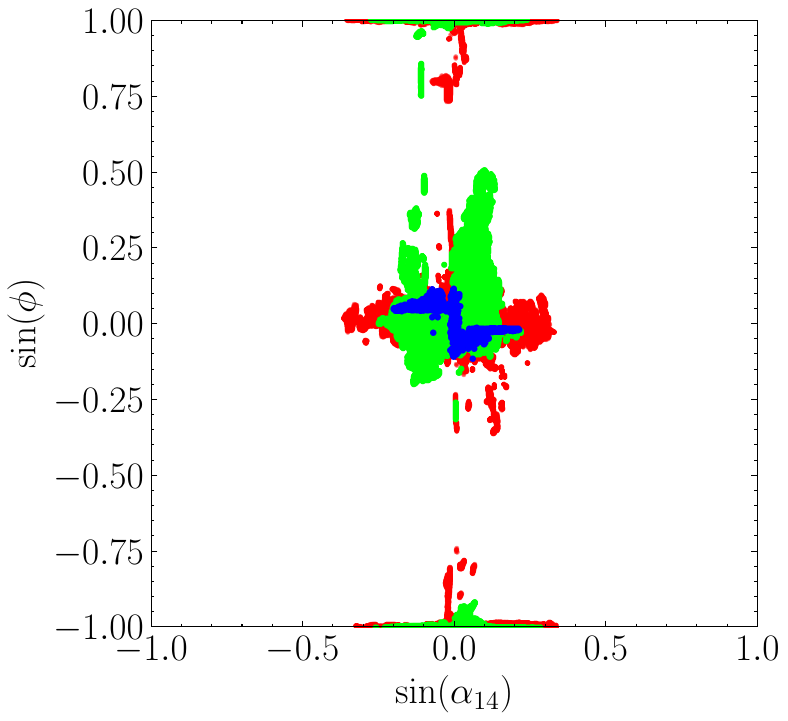}
	\caption{  Combined seeded plots with CMA-ES. The points shown \textbf{did not include dedicated scans with novelty reward focused on the parameters shown}.  Blue the points obtained with the method in Section~\ref{sec:classic_c3hdm}. }
	\label{fig-34-15}
\end{figure}

As introduced in Section~\ref{sec:physbasis},
the real limit of the model is equivalent to setting the $R_{CPV}$
as the identity matrix and $\varphi =0$.
With the iterative sampling of enlarging the pseudoscalar component, the intervals in \eq{eq:enlarging2} were found to allow compatibility with the eEDM measurement.
With the sampling using CMA-ES,  we are able to scan all the parameters in
\eq{eq:scanparameters_c3hdm} for the full $[-\pi, \pi]$ domain.
The set of phases corresponding to this deviation from the real limit
is plotted in Figs.~\ref{fig-14-15}-\ref{fig-34-15}.
We are able to find agreement with the statement that
$\alpha_{14}$ and $\alpha_{15}$ must be small.

Recall that
the $125~{\rm GeV}$ Higgs can be written as the combination
\begin{equation}
    h_1 = R_{11} x_1 +  R_{12} x_2 +  R_{13} x_3 - (R_{14} s_{\beta_1} + R_{15} c_{\beta_1} s_{\beta_2}) z_1 + (R_{14} c_{\beta_1} - R_{15} s_{\beta_1} s_{\beta_2})z_2 + R_{15} c_{\beta_2} z_3,
\end{equation}
leading to the couplings with the fermions as in \eq{eq:cff_odd}.
Setting $\alpha_{14}$ and $\alpha_{15}$ to be small,
reduces the (CP-odd) contributions of $z_k (k=1,2,3)$ into $h_1$,
as required. 
In contrast,
we find that the other 5 parameters can take any possible values
in the full chosen domain. This has
a deep physical implication, since $\alpha_{24}-\alpha_{25}$,  $\alpha_{34}-\alpha_{35}$ control CP violating effect in heavier scalar states. The traditional
simulation forces these to be small; the new technique
allows for any value.

Notice that the points in Figs.~\ref{fig-14-15}-\ref{fig-34-15} were obtained not by
focusing on the parameters shown on those plots (no attempt was made to cover the particular parameter
planes shown). These points were obtained by focusing on \textit{observables}, such as
$c^e_{tt}$ and $c^o_{tt}$. The fact that focusing on \textit{observables} implies automatically
very wide scanning regions for the model \textit{parameters} is a very interesting feature,
which we have also found in preparatory studies using other specific models.

\section{Summary}\label{sec:summary_ml}

We have looked at the impact of ML techniques in the full exploration of the parameter space
and physical consequences of the C3HDM.
Specifically,
we have used an Evolutionary Strategy algorithm
integrated with an anomaly detection-based Novelty Reward mechanism,
to ensure robust exploration not only of the model’s parameter space but,
crucially, of its physical implications. 

We have demonstrated the effectiveness of this new method,
by pitting it against the traditional scanning approach starting from the real and alignment limit.
This allowed for the efficient identification of valid points within the model parameter space.
What in the traditional sampling approach was a painstaking, slow search, requiring constant intervention
in the quest for new features, becomes here incomparably faster and requiring no
user supervision whatsoever.
Most importantly,
because the method converges rapidly to points with novel characteristics,
one is able to fully explore the phenomenological implications of the model.
In this respect, the differences between the blue (iterative method) and green (ML method)
regions in Figs.~\ref{fig-top} and~\ref{fig-bothwrong}-Left are impressive.  We have shown that the maximal CP-odd case
($c^e_{bb}=0,c^o_{bb}=\pm 1$) is not excluded. For the case of the
$\tau$ lepton, the measurement of the CP properties of the $\tau$
lepton at the LHC~\cite{CMS:2021sdq,ATLAS:2022akr} imposes stronger
constraints. 
Thus, the prospect of the CP-even/CP-odd $t$/$b$ possibility deserves continued
experimental exploration.

Of course, this is a particular application of the Machine Learning with novelty reward tecnhique,
as the method can be used to great advantage
in any model with large numbers of parameters.
The approach introduced here offers a promising direction for systematically uncovering
novel and physically meaningful solutions in complex theoretical frameworks.

%% file: chapters/DarkMatter_problem.tex
\chapter{The Dark Matter problem}
\label{chapter:Darkmatter_problem}
\hspace*{0.3cm}
The quest to unravel the nature of Dark Matter (DM) remains one of the most compelling challenges in modern physics, as it eludes unique incorporation into a complete model of particle physics. Despite the remarkable precision of the Standard Model (SM), which successfully describes approximately $15\%$ of the Universe's matter content, primarily baryonic matter, the remaining $85\%$, attributed to Dark Matter, continues to defy explanation~\cite{Planck:2015sxf}. 
In astrophysical systems, from galactic to cosmological scales, observed dynamical anomalies cannot be explained by visible matter alone. These discrepancies can be reconciled either by postulating the existence of a significant amount of unseen Dark Matter or by proposing modifications to the known laws of gravitation and general relativity. In this work, we focus on particle physics solutions, exploring Dark Matter candidates that satisfy key cosmological and experimental constraints, including consistency with the observed relic abundance~\cite{Planck:2018vyg},  over cosmological timescales~\cite{Audren:2014bca, MAGIC:2018tuz}, only gravitational weakly interacting with ordinary matter~\cite{Dimopoulos:1989hk, McDermott:2010pa}, cold and non-relativistic properties for large-scale structure formation, and limits on self-interactions from cluster collisions like the Bullet Cluster~\cite{Clowe:2006eq}.

 Such candidates are  typically produced in the early Universe at high temperatures via mechanisms such as the thermal freeze-out~\cite{Lee:1977zp}. The production occurs within the context of an expanding universe, ruled at first order by the Hubble constant:
\begin{equation}
    H_0\coloneqq\frac{\dot{r}}{r}\,;\quad \text{with value } h=\frac{H_0}{100 \frac{\text{km}}{\text{s Mpc}}}\approx 0.7\, .
\end{equation}
Dark Matter today is analyzed based on the one very precisely measured property: the relic density $\Omega_{DM} h^2\approx 0.1200\pm0.0012$~\cite{Planck:2018vyg}. The definition of this parameter comes from solving the Friedmann equations, a particular form of the Einstein equations, and is defined as a fraction of the critical density $\rho_c$, which separates a collapsing and expanding Universe with $\Lambda = 0$,
\begin{equation}
    \rho_c=3 H(t)^2 M_{PI}^2\,.
\end{equation}
 The relevant densities to be obtained from experiment are
\begin{equation}
    \quad\Omega_m \coloneqq \frac{\rho_m}{\rho_c}\text{ , }\quad\Omega_r\coloneqq\frac{\rho_r}{\rho_c}\text{ , }\quad\Omega_b\coloneqq\frac{\rho_b}{\rho_c}\quad\text{ and }\quad\Omega_\chi\coloneqq\Omega_m-\Omega_b ;
\end{equation}
for matter, radiation, baryonic matter and Dark Matter, respectively.

While the Standard Model does not include a valid DM candidate, the weakly interacting massive particle (\gls{WIMP}) has been introduced in Beyond the SM (\gls{BSM}) scenarios to
account for the DM relic density in our universe. Based on this hypothetical framework and others, three main experimental approaches were established to search for DM particles. The first is indirectly searching for signals from the annihilation of DM into SM particles in astroparticle experiments. The
second is directly looking for DM-nucleus scattering in large volume detection experiments.   The third is using colliders such as the Large Hadron Collider (\gls{LHC}) as a production factory for DM particles.

 In this text, we review current evidence for the existence of Dark Matter. We concentrate on the hypothesis of particle solutions with a freeze-out mechanism of production and follow with the calculation of their relic abundance, discussing the possible relationship between Dark Matter and, in the topic of this thesis, scalar extensions of the Standard Model of particle physics. An overview of the various Dark Matter experimental searches and the constraints obtained is then given.

\section{Present Evidence}

\subsection{Data and measurements of rotation curves}

\begin{figure}[H]
\centering
\includegraphics[width = 0.5\textwidth]{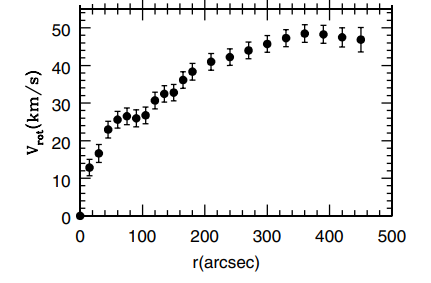}
\caption{\label{NGC3741} Rotation curve for NGC 3741~\cite{Gentile:2006hv}.}
\end{figure}

The pioneering observations for DM were made by Jan Oort~\cite{Oort:436532} and Fritz Zwicky~\cite{Zwicky:1933gu}. Oort pointed out that the velocity of nearby stars were too high to be explained by the luminous mass present in the Galaxy. Zwicky took information from redshift measurements on the velocity of galaxy clusters to apply the virial theorem, deriving a cluster mass much larger than the luminous mass present.  The velocity of galaxy clusters exceeded the predicted velocity based on total mass of the galaxy using classical Newtonian physics. Rotation curves have since then been tools for relating departures from the expected form to the amount and distribution of Dark Matter. The data is obtained through multiple different methods~\cite{Yegorova:2007}:
\begin{itemize}
    \item \textbf{Long-slit spectroscopy} is used to measure the Doppler shift of spectral lines at various positions within a galaxy. This technique captures the spectrum along a narrow slit, typically aligned with the major axis of a galaxy. The rotational motion of the galaxy induces a Doppler shift, causing spectral lines to shift toward shorter (blue) or longer (red) wavelengths, enabling the mapping of rotational velocities across the galaxy.
    
    \item \textbf{Radio emission lines} originate from transitions at high energy levels in ionized hydrogen gas, typically in H\,II regions energized by nearby hot stars. Key lines include the hydrogen H$\alpha$ line (6562.8 \AA), nitrogen [N\,II] lines (6548.03 \AA, 6583.41 \AA), and sulfur [S\,II] lines (6716.47 \AA, 6730.84 \AA). These lines are critical for studying ionized gas kinematics in galaxies.
    
    \item \textbf{CO rotational transition lines} in the millimeter wavelength range are employed to probe the kinematics of molecular gas in the inner disks and central regions of spiral galaxies. 
    
    \item \textbf{Maser lines} in the microwave range, such as those from SiO, OH, and H$_2$O, provide precise radial velocity measurements. These lines, emitted from circumstellar dust and gas clouds, allow for detailed kinematic studies of stellar components in the disks and bulges of nearby galaxies.
\end{itemize}
The observed flux of a galaxy can be used to derive its luminosity by integrating over its projected area. This approach is particularly effective for constructing luminosity profiles of elliptical galaxies, where the light density decreases rapidly in the halo. Assuming that the mass distribution follows the light distribution, most of the galaxy’s mass is interior to a given stellar orbit in the halo. Using Newton's law, the expected velocity in this halo would then be:
\begin{equation}
    \frac{V_c^2}{R}=\frac{G M}{R^2} \implies V_c(R) \propto R^{-1/2},
\end{equation}
where $G$ is the gravitational constant and $M$ is the mass interior to radius $R$. However, observed rotation curves of spiral galaxies deviate significantly from this prediction. Rather than declining as $ R^{-1/2}$ beyond the radius of visible matter, the rotation curves typically exhibit a plateau at distances of several kiloparsecs from the galactic center, indicating velocities higher than expected from models based solely on visible matter and Newtonian mechanics. This discrepancy suggests the presence of additional, non-luminous mass, commonly attributed to Dark Matter. The rotation curve of the galaxy NGC 3741, shown in Figure~\ref{NGC3741}, exemplifies this property.

To account for the observed flat rotation curves, additional mass in the galactic halo is required. The density profile of this mass can be approximated as:
 \begin{subequations}
    \label{eq:velocity_profile}
	\begin{eqnarray}
    \rho &\approx& \rho_0 \left(\frac{R_0}{R}\right)^\gamma ,\\
    M(<R)&=&\int_0^R 4\pi \rho(R)R^2 dR \quad	\propto\quad R^{3-\gamma} ,\\
    v_c &\approx& \sqrt{\frac{G M(<R)}{R}} \quad\xrightarrow[\text{flat curve}]{} \quad\gamma=2;
	\end{eqnarray}
\end{subequations}
where $\rho_0$ is the density at radius $R_0$, $\gamma$ describes the power-law of the density profile, and $M(<R)$ is the mass enclosed within radius $R$. For a flat rotation curve, \eq{eq:velocity_profile} implies a mass profile $M(<R) \propto R$, indicating that the enclosed mass increases linearly with radius. Given that the light density in the halo of galaxies decreases rapidly, the flat rotation curves suggest the presence of ``invisible'' gravitational mass, commonly attributed to Dark Matter. This Dark Matter component dominates the gravitational potential at large radii, compensating for the steep decline in luminous matter. The inner rotation curve is reproduced by just the stellar disk with suitable mass/light ratio.

\subsection{Weak gravitational lensing}
One of the most compelling pieces of evidence for the existence of dark matter (DM) comes from gravitational lensing, with the \textit{Bullet Cluster} (catalog name 1E0657-558) being an iconic example~\cite{Clowe:2006eq}. First studied in 2006, the \textit{Bullet Cluster} consists of two galaxy clusters that collided approximately 150 million years ago. During this collision, the baryonic matter, primarily hot X-ray emitting plasma, experienced a collisional shock wave and gathered between the clusters, as mapped by its X-ray emission profile.  In contrast, the dark matter, experienced negligible collisions with itself and normal matter, moving through and separating from the visible matter. This separation cannot be accounted for by altering gravitational laws, providing clear observational support for the existence of dark matter. Similar systems have been studied and a study in 2015~\cite{Harvey:2015hha} combined results on 72 collisions to conclude the existence of Dark Matter at $7.6\,\sigma$ significance.

\begin{figure}[htb]
	\centering
	\includegraphics[height=6cm]{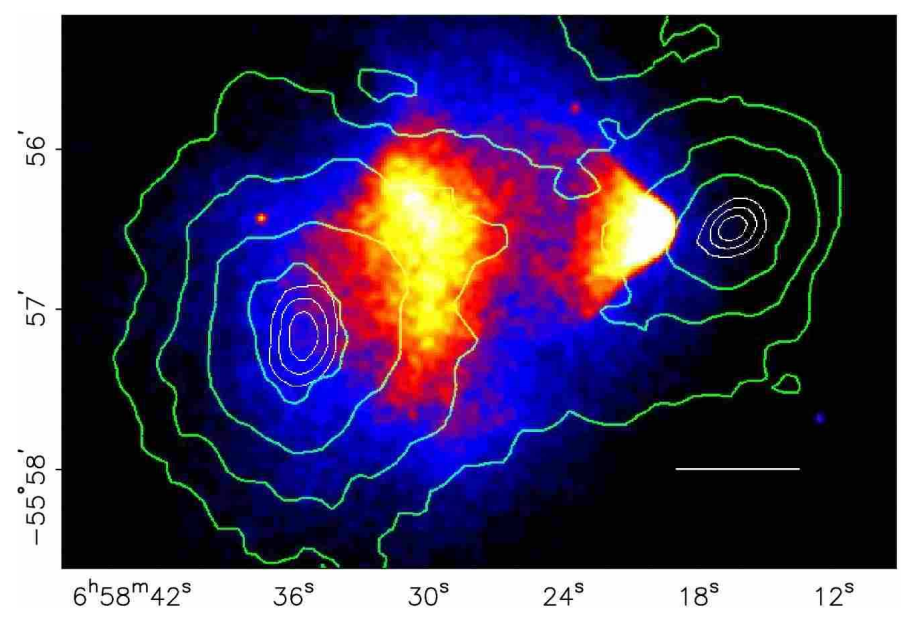}
\caption{\label{bullet} X-ray image of the Bullet Cluster obtained by the NASA Chandra X-ray Observatory. The colored map represents the X-ray image of the hot baryonic gas. The distribution of the total mass reconstructed through weak lensing is shown with green contours. Taken from~\cite{Clowe:2006eq}.}
\end{figure}

\subsection{Cosmic Microwave Background}

Evidence for Dark Matter emerges from observations on the scale of galaxies and galaxy clusters. These measurements indicate the presence of non-luminous mass. However, these observations do not provide constraints on the total amount of Dark Matter in the Universe. 
To estimate the total Dark Matter content, analysis of the Cosmic Microwave Background (\gls{CMB}) is required. The existence of background radiation originating from the propagation of photons in the early Universe, once decoupled from matter, was predicted
by George Gamow~\cite{Gamow:1946eb} and his collaborators Ralph Alpher and Robert Herman predicted the radiation to be at the temperature of $5\,K$~\cite{Alpher:1948srz}.

The first measurement of the microwave background was done by Arno Penzias and Robert Wilson in 1965 and awarded with the 1978 Nobel Prize in Physics~\cite{Wilson:1978}. Their instrument, a Dicke radiometer, observed an excess $4.2\,K$ antenna temperature which they could not account for. 
In the following years, it was estimated that there should be residual fluctuations $\delta T / T$ of order $10^{-4}$ to $10^{-5}$ in the early universe~\cite{Harrison:1970sb,Peebles:1970ag}. Experiments then started to place limits on the anisotropy of the CMB. The first CMB satellite, \textit{RELIKT-1}, launched in 1983 and observed the entire galactic plane in 6 months using a Dicke-type modulation radiometer. 


The following satellite, COBE by NASA, included an instrument named Differential Microwave Radiometer (DMR) that was able to create full sky maps of the CMB by subtracting out galactic emissions and dipole at various frequencies. 
The maps obtained detected and quantified large scale anisotropies at the limit of its detection capabilities and confirmed the blackbody form of the CMB. The results were first published in 1992~\cite{Smoot:1992td} and members of the team received the 2006 Nobel Prize in Physics for the discovery.

\begin{figure}[htb]
	\centering
	\includegraphics[height=5.3cm]{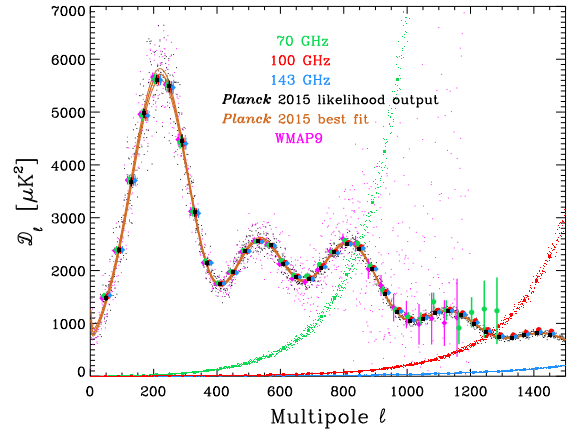}
	\includegraphics[height=5.1cm]{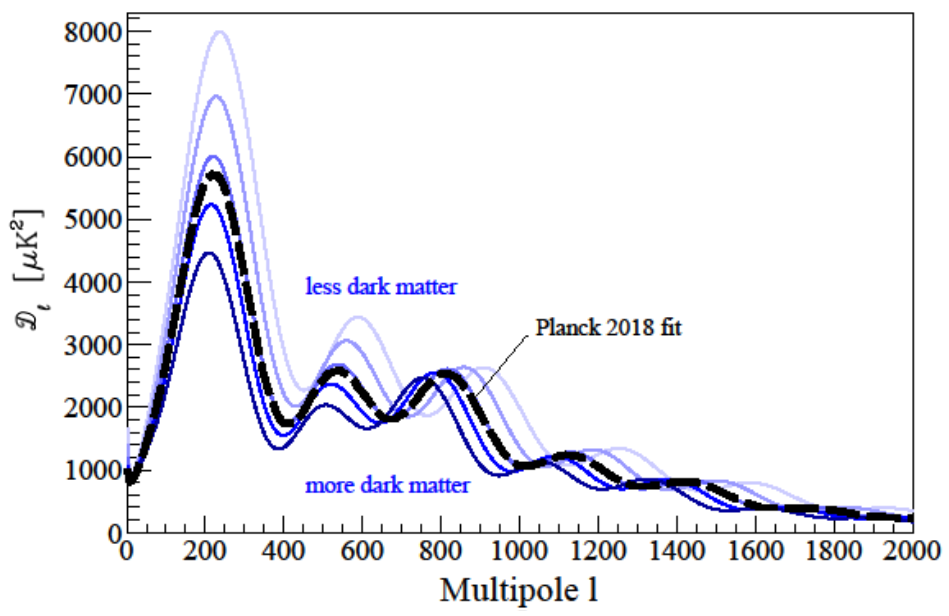}
\caption{\label{planck} Left: Direct comparison of Planck and WMAP-9 CMB power spectra. Taken from~\cite{Planck:2015sxf}. \\
Right: Temperature power spectrum of the CMB
for a DM density $\Omega_{DM}$ varying between 0.11 and 0.43 (blue lines). All other input paramters are kept constant. The dashed black line shows the best fit to Planck data from the 2018 release~\cite{Planck:2018vyg}. Taken from~\cite{Schumann:2019eaa}.}
\end{figure}

COBE was then followed by two more advanced spacecraft: NASA's Wilkinson Microwave Anisotropy Probe (WMAP) operated from 2001 to 2010 and the ESA Planck spacecraft from 2009 to 2013. 
The Planck probe had the angular resolution to measure the smaller scale fluctuations and corresponds to the most complete temperature sky currently available. Today, the CMB is known to be isotropic at the $10^{-5}$ level and to follow the spectrum of a black body corresponding to a temperature $T=2.7255\pm 0.0006\,K$~\cite{Fixsen_2009}. The analysis of CMB anisotropies is then performed by decomposing the map of the sky into spherical harmonics, 
\begin{equation}
    \frac{\Delta T}{T}(\theta, \phi)= \sum_{lm} a_{lm} Y_{lm} (\theta, \phi),
\end{equation}
where the $a_{lm}$ term measures the mean temperature and $Y_{lm}(\theta,\phi)$ the fluctuations, with $l$ the multipole number while $m$ is the azimuthal number. The angular power spectrum $C_l$ of $a_{lm}$ is given by,
\begin{equation}
 <a_{lm}a^*_{l'm'}> \equiv C_l \delta_{ll'}\delta_{m m'}.
\end{equation}
This leads to the average of square temperature fluctuations as,
\begin{equation}
    <|\Delta T/T|^2>=\sum_{ll'mm'} <a_{lm}a_{l'm'}*> \int d \cos{\theta} \int d\phi  Y_{lm} (\theta, \phi)  Y^*_{l'm'} (\theta, \phi) = \sum_l^\infty \frac{2l+1}{4\pi} C_l \equiv \sum_l^\infty \mathcal{D}_l ,
\end{equation}
meaning that $l$ provides information regarding the temperature variations on a certain scale. The Planck experiment provided the CMB spectrum, $\mathcal{D}_l$. The comparison with the WMAP data in Fig.~\ref{planck} clearly shows the improvement made at smaller scales, that is at larger $l$.

The temperature fluctuations can be analyzed in parts based on individual physical processes that dominate at a given scale. These are a part of the $\Lambda$CDM Standard model of Cosmology with six parameters that is fitted to the data obtained. The effect at large angular scales, the Sachs–Wolfe effect, is caused by changing gravitational fields affecting the propagation of the photons originally emitted at the time of decoupling, due to fluctuations in matter/energy density.
If there is a higher density region than the average, photons have to climb the potential well and are redshifted. 

The next effect and the most useful comes from the acoustic oscillations of the photon-baryon fluid. Since photons, protons and electrons are coupled through Compton and Coulomb interactions before photon decoupling, they are treated as a mixed fluid with density fluctuations in the form of acoustic waves. The radiation pressure from the photons resists the gravitational compression of the fluid creating oscillations, similar to sound waves in the air due to pressure differences. The peaks correspond, roughly, to resonances in which the photons decouple when a particular mode is at its peak amplitude. The odd numbered acoustic peaks are associated with how much the plasma compresses, so they are enhanced by an increase in the amount of baryons in the universe. The even numbered peaks are associated with how much the plasma rarefies due to the radiation pressure.  Thus, with the addition of baryons the odd peaks are enhanced over the even peaks. The ratio of the first peak to the second peak determines the baryon density. Since Dark Matter does not interact with light, it is not affected by the radiation pressure while still experiencing the same gravitational compression as baryons. The third peak, related to compression, can then be used to get accurate information about the Dark Matter density. 

For the smaller scales, there is a damping effect, known as Silk damping~\cite{Silk:1967kq}, due to diffusion of the photons from the overdense and hot regions to the cold regions, causing the temperatures to be averaged. Therefore, the CMB power spectrum declines as $l$ increases.

With an appropriate fitting including the effects mentioned previously, the analysis of CMB anisotropies enables accurate testing of cosmological models and puts constraints on cosmological parameters. The parameters quantified by the Planck experiment~\cite{Planck:2018vyg} are:
\begin{align}
    \Omega_{DM} h^2 = 0.1200 \pm 0.0012;\label{relicdensity}\\
    \Omega_b h^2 = 0.02237 \pm 0.00015,\\
        \Omega_\Lambda  = 0.6847 \pm 0.0073,\\
            H_0 = 67.36 \pm 0.54 \frac{\text{km}}{\text{s Mpc}}\, .
\end{align}
  The relic density $\Omega_{DM} h^2$ is the parameter most relevant for BSM physics. The power spectrum of temperature fluctuations is now measured up to the third acoustic peak by WMAP and $l\approx 2500$ by Planck, and even smaller scales by ground-based experiments such as the Atacama Cosmology Telescope (ACT)~\cite{Choi:2020ccd} in Chile and the South Pole Telescope (SPT)~\cite{Reichardt:2020jmt}. While the existence of acoustic oscillations with diffusion damping has been confirmed and obtains values for the six parameters of the current $\Lambda$CDM model, there is still mention of some tension between the parameters derived from different measurements. These could be indicative of yet unidentified systematics or of missing components of the model. These are a part of the ground-based CMB Stage-4 experiment currently operating.

Next-generation satellite experiments that are planned to launch in the coming years, such as PIXIE, LiteBIRD (in 2028) and COrE (in 2028 and 2034), will complement the CMB-S4 measurements and be a probe of the thermal history of the universe. In terms of next-generation ground telescope there is the Square Kilometre Array (SKA)~\cite{Maartens:2015hva}, the world's largest radio telescope with thousands of small antennas, that started construction activities in 2021 and is projected to have first observations in 2027. The other main observatory under construction is the  Extremely Large Telescope (ELT) in Chile with first observations also planned for 2027. The cosmological measurements enabled by the SKA and the ELT will be part of the leading force in cosmological discovery in coming years, including the testing of models of Dark Matter and of the evolution of the Universe.

\section{Weakly interacting massive particles (WIMP)}

This Section aims to summarize the considerations used for the class of dark matter (DM) candidates known as Weakly Interacting Massive Particles (WIMPs), as considered in this thesis~\footnote{See, e.g., Refs.~\cite{Kolb:1990vq,Bauer:2017qwy} for complete discussions on the topic.}.

A given particle species existed in thermal equilibrium in the very early stages of the history of the Universe. As its interaction rate drops below the expansion rate of the Universe, the equilibrium is no longer maintained and the particle is considered to be decoupled (\textit{freeze-out}) from the primordial plasma. In this scenario, assuming Standard model of Cosmology, the DM relic density is determined by a single Particle Physics input: the thermally averaged annihilation cross-section of the DM $\langle \sigma v \rangle$. We consider a hypothetical massive, weakly interacting Dark Matter candidate. Assuming masses of at least the ${\rm GeV}$ scale, DM is non-relativistic at the time of decoupling.   

\subsection{Relic Density}
A statistical description  is considered at the Early Universe such that the evolution of the phase space distribution function, $f(t,\vec{x},\vec{p})$, is obtained by the Boltzmann equation,
\beq
\mathbf{L}[f]=\mathbf{C}[f]\,,
\eeq
where $\mathbf{L}[f]$ is the Liouville operator for the propagation of a particle, and $\mathbf{C}[f]$ is the collision operator, describing the interactions of the particle species considered. Assuming homogeneity and isotropy at large scales in the Early Universe, a Friedmann--Lema\^{i}tre--Robertson--Walker metric is considered and, after some manipulation, the Boltzmann equation for a particle species $\chi$ can be written as an equation for the number density, $n_{\chi}$, as:
\beq
\frac{dn_{\chi}}{dt} +3 H(T) n_\chi = -\langle \sigma_{\text{ann}} v \rangle (n_{\chi}^2 -n_{\chi, eq}^2), \label{Boltzmanneq}
\eeq
with $n_{\chi, eq}$ the number density at thermal equilibrium of the species $\chi$, $H(T)$ the Hubble expansion parameter, and $\langle \sigma v \rangle$ is the thermally averaged annihilation cross-section multiplied by the relative velocity of the particles in collision. It is assumed that the long lived DM candidates interact with a pair of SM states via $2\to\,2$ annihilation processes, with the possibility of coannihilation in scenarios with multiple DM candidates.  
For massive particles in the non-relativistic limit, the Maxwell-Boltzmann approximation gives, 
\beq
n_{\chi, eq} (T) = g_{\chi} \left(\frac{m_{\chi}T}{2\pi}\right)^{3/2}e^{-m_{\chi}/T}\,, \label{n_Boltzmann}
\eeq
where $m_\chi$ represents the DM mass and $g_{\chi}$ the number of internal degrees of freedom of $\chi$. 
The DM annihilation rate is related to the thermally averaged cross-section through $\Gamma=n_\chi\langle \sigma v \rangle$ . With a dependence on a power $T$, in a scenario at very high temperatures we have $\Gamma >> H$, as the annihilation is highly efficient. As the temperature of the Universe drops, pair production of the particle species $\chi$ becomes Boltzmann suppressed and, consequently, annihilation of $\chi$ becomes inefficient. When $\Gamma (T_f)\approx H(T_f)$,  the annihilation process stops occuring and the particle number of $\chi$ per co-moving volume is conserved, defining $T_f$ as the freeze-out temperature.
From that moment, $\Gamma << H$ so the DM particles have not interacted since the decoupling.  The ratio $x_f=m_\chi/T_f$ appears in the exponential of \eq{n_Boltzmann} for $T=T_f$ and comes to order $\sim 23$ for WIMP masses $0.1-1\,{\rm TeV}$. The abundance today is 
\beq
\Omega_\chi h^2 =\frac{\rho_\chi (T_0)}{\rho_c} h^2= \frac{m_\chi n_\chi(T_0)}{\rho_c} h^2\, , \label{omegaeq}
\eeq
where $T_0=2.73\,K$ is the temperature today at which both $\rho_\chi$ and $\rho_c$ are calculated. The density today then relates with the one at decoupling by,
\beq
m_{\chi}n_{\chi}(T_0) = m_{\chi}n_{\chi}(T_f) \frac{g_{\text{eff}}(T_0) \, T_0^3}{g_{\text{eff}}(T_f) \, T_f^3}\, , \label{conservation_entropy}
\eeq
where $g_{\text{eff}}$  is the an effective total number of spin states of the particle species. This result comes from the evolution of radiation/matter with particle number conserved. The DM relic density can be calculated numerically with great precision for arbitrary models by public packages, solving \eq{Boltzmanneq}. All the results in this thesis used \texttt{micrOMEGAs}~\cite{Alguero:2023zol}. Other avaliable codes include DARKSUSY~\cite{Bringmann:2018lay}, MadDM~\cite{Ambrogi:2018jqj}, RelExt~\cite{Capucha:2025iml}.

A rough analysis is possible and instructive. The velocity can be estimated based on the kinetic energy as $v\sim \sqrt{2T_f/m_\chi}$.
From $\Gamma(T_f)=H(T_f)$ we find the condition
\begin{equation}
n_{\chi}(T_f) = \frac{H(T_f)}{\sigma(T_f) v}, 
\end{equation}
which, assuming freeze-out to have occured during the radiation-dominated era, can be replaced with the Hubble law for Bose-Einstein statistics~\cite{Bauer:2017qwy}
\begin{equation}
g_{\chi}(T_f) \left( \frac{m_{\chi} T_f}{2\pi} \right)^{3/2} e^{-m_{\chi}/T_f}  = \frac{\frac{\pi \sqrt{g_{\text{eff}}(T_f)}}{\sqrt{90}} \frac{T_f^2}{M_{PI}}}{\sigma_{\chi} \sqrt{\frac{2T_f}{m_\chi}}} \, .
\end{equation}
We choose $g_{\chi}(T_f) = 2$ arbitrarily. Then, we get
\begin{equation}
e^{-m_{\chi}/T_f} = \frac{\pi^{5/2}}{3\sqrt{10}} \frac{\sqrt{g_{\text{eff}}(T_f)}}{m_{\chi} M_{PI} \sigma}.
\end{equation}
To calculate a relic density we need to estimate the cross section of the process $\chi\chi \to f \bar{f}$, which we will consider to be given by a generic annihilation cross section with the scale of the electroweak interaction. Using the example in~\cite{Bauer:2017qwy},
\begin{equation}
\sigma = \frac{\pi \alpha^2 m_\chi^2}{c_w^4 m_Z^4}.
\end{equation}
The computation of \eq{n_Boltzmann} leads to,
\begin{equation}
\rho_{\chi}(T_0) = m_\chi n_{\chi}(T_f) \frac{g_{\text{eff}}(T_0) T_0^3}{g_{\text{eff}}(T_f) T_f^3} = T_0^3 \frac{n_\chi(T_f) x_f^3}{28 m_\chi^2}\,,
\end{equation}
where the approximation to the ratio $\frac{g_{\text{eff}}(T_0)}{g_{\text{eff}}(T_f)}\sim 1/28$ was based on the counting of degrees of freedom with $g_{\text{eff}}(T_0)=3.6$ today and $\approx\,100$ for $g_{\text{eff}}(T_f)$. The other ingredient still missing is the decoupling temperature. As an estimate, we use $x_f \approx 23$. Computing $\Omega_{\chi} h^2$ yields,
\begin{equation}
\Omega_{\chi} h^2 \approx 0.12 \left( \frac{13 ~{\rm GeV}}{m_{\chi}} \right)^2. 
\end{equation}
Thus predicting that for a DM candidate with annihilation into SM mediated by the weak interaction, the required mass has to be of the electroweak scale in order to obtain the measured relic density. This conclusion is known as the WIMP miracle and also occurs with better approximations.

\section{Indirect Searches}

While current model-independent constraints on DM self-interaction and its coupling with Standard Model particles are loose, lacking the precision to pinpoint specific mechanisms, numerous experimental strategies have been developed to probe signals from specific DM particle models. The most advanced efforts focus on WIMPs, with indirect searches referring to the observation of annihilation or decay products of WIMPs in the Universe, $\chi+\chi\to SM+SM$. In regions with a clump of gravitational matter, such as the center of galaxies or the Sun, the DM should annihilate efficiently even at present, with products including neutrinos, positrons, anti-protons and gamma-rays. The results are presented as upper limits on $\langle \sigma v \rangle$ in terms of the WIMP mass.

To detect cosmic gamma-rays directly, observations must be conducted from space. This is because, at the energies of interest (${\rm GeVs}$), photons interact with matter via pair production, leading to an interaction length far shorter than the thickness of Earth's atmosphere, preventing sufficient transmission to ground-based telescopes. There remains the possibility of observing the results of the electromagnetic cascade produced and the consequent shower of secondary particles. Notable ground-based telescopes include MAGIC, VERITAS, HESS and HAWC. The Cherenkov Telescope Array (CTA) is the next generation observatory for gamma-rays at high energies.

\begin{figure}[htb]
	\centering
	\includegraphics[height=5.3cm]{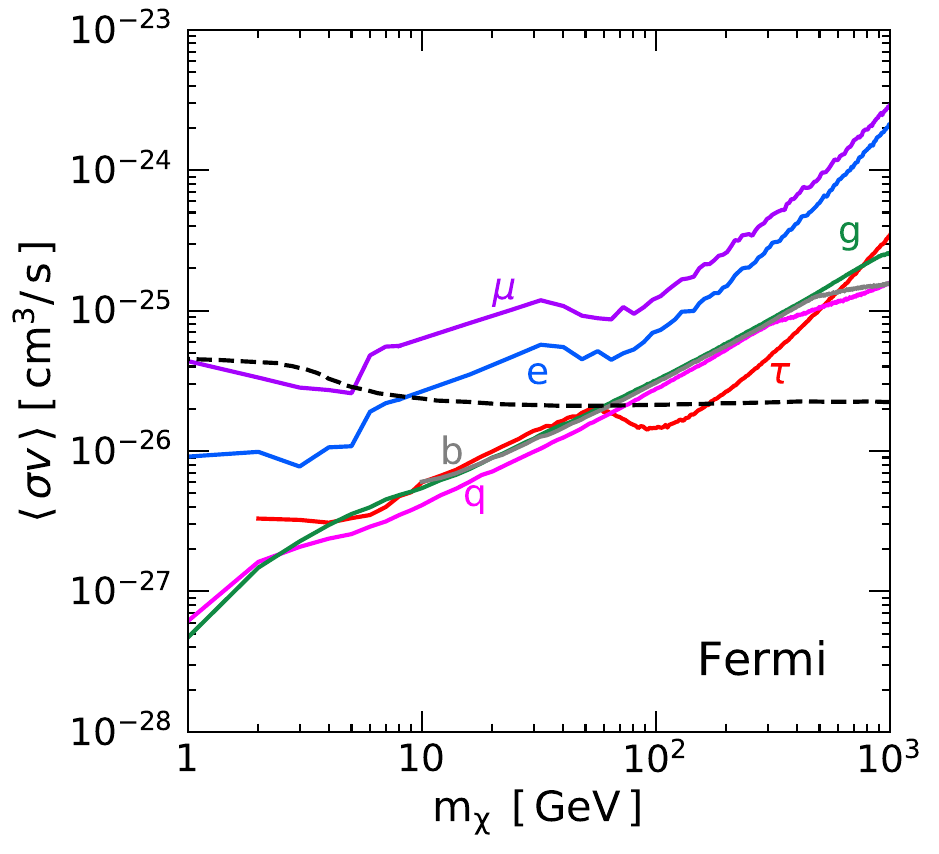}
	\includegraphics[height=5.1cm]{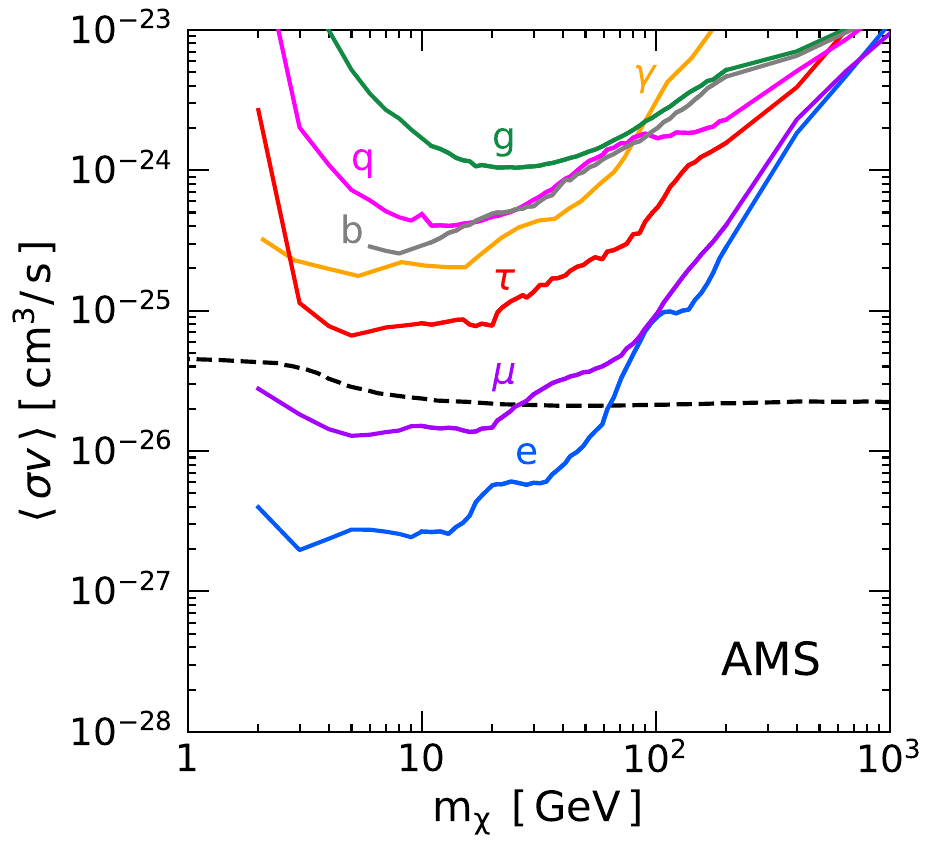}
\caption{\label{IDlimits} \textit{Fermi}-LAT (left) and conservative AMS limits limits at 95$\%$ C.L. for DM annihilation 100$\%$ to individual channels: electrons (blue), muons (purple), taus (red), $b$-quarks (gray), gluons (green), and light quarks $q=u,d,s,c$ (magenta). Taken from Ref.~\cite{Leane:2018kjk} with the black dashed line indicating the expected WIMP cross-section calculated in Ref.~\cite{Steigman:2012nb}.}
\end{figure}

\begin{figure}[htb]
	\centering
	\includegraphics[height=6cm]{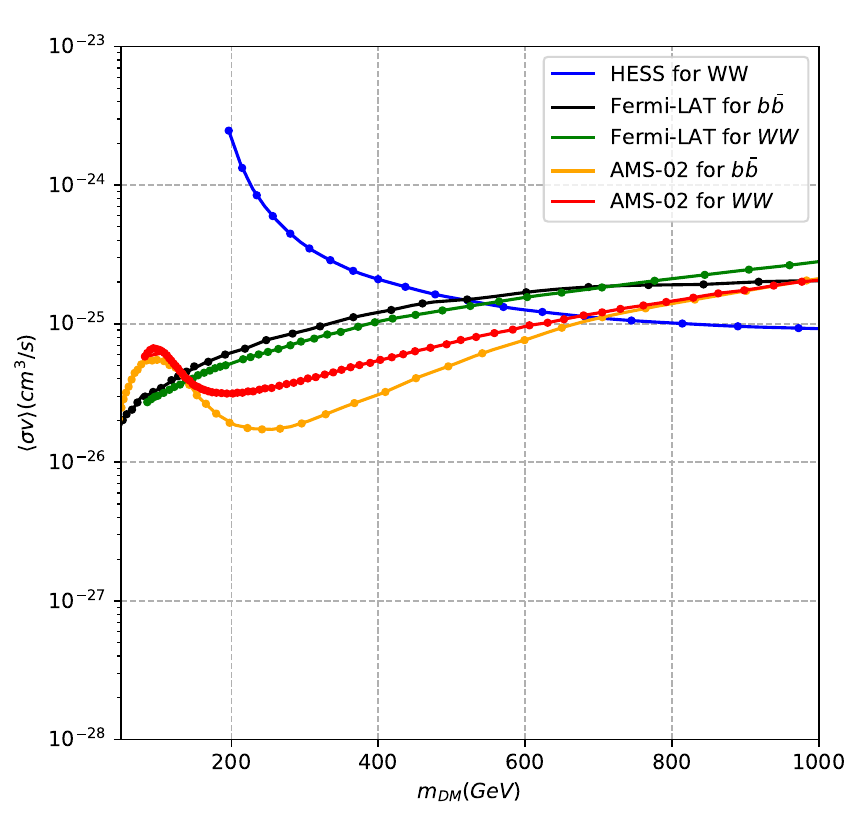}
\caption{\label{indirect_lines} Combined relevant limits from indirect searches on the total
$\langle\sigma v\rangle$ as a function of the mass of the DM candidate $m_{DM}$. The lines coming from
Fermi-LAT~\cite{Fermi-LAT:2015att} and \gls{H.E.S.S.}~\cite{HESS:2022ygk} assume a Navarro-Frenk-White (\gls{NFW})
DM density profile and the \gls{AMS}-02~\cite{AMS:2016oqu} lines correspond to the conservative approach derived in Ref.~\cite{Reinert:2017aga}. }
\end{figure}

The first high energy (above ${\rm GeV}$) gamma-ray space telescope was EGRET launched in 1991. The Fermi-\gls{LAT} gamma ray telescope, launched in 2008, looks at Milky Way dwarf spheroidal galaxies and has current limits for indirect DM detection through photons~\cite{Fermi-LAT:2015att}
ranging from $\langle\sigma v\rangle\approx 3\times 10^{-26}\, \text{cm}^3/\text{s}$,
for light DM,
to around $\langle\sigma v\rangle\approx  10^{-25}\, \text{cm}^3/\text{s}$,
for heavier DM. The searches for anti-protons with AMS-02~\cite{AMS:2016oqu,AMS:2016brs} in the International Space Station yield competing constraints for WIMP DM in this mass range. The limits from HESS~\cite{HESS:2022ygk},
a Cherenkov gamma ray telescope,
which points to the central region of the Milky Way, give the strongest indirect detection constraints for DM masses
from $\sim 500~{\rm GeV}$ upwards.  The strongest limits
coming from these experiments for several annihilation channels are shown in Fig.~\ref{IDlimits} from Ref.~\cite{Leane:2018kjk}. As in~\cite{Belanger:2021lwd}, we find that in the scalar extensions studied  the $\langle \sigma v \rangle$
which are in reach of Fermi-LAT arise mainly from
annihilation decays into $VV$.
So we sum only the  $WW$ and $ZZ$ final states,
assuming a similar spectrum, which we dub $\langle\sigma v\rangle_{VV}$. Below the $W$ threshold, the annihilation proceeds mostly into
$b \bar{b}$. In Fig.~\ref{indirect_lines} we show the experimental lines used to set constraints on the models considered.

The flux measured from the annihilation of two particles depends on the Dark Matter density squared. A steeper DM halo profile with the density increasing rapidly in the center of the galaxy results in stronger bounds on DM annihilation. The main issue with this interpretation of measured fluxes in terms of DM bounds is that we cannot measure the Dark Matter distribution, $\rho_{DM}$, in the galaxy directly. The density profiles are obtained by hydrodynamic simulations assuming a cold DM model, having to solve the collision-less Boltzmann equation coupled
with Poisson’s equation, while also modelling gas dynamics using the Euler equations for fluid dynamics~\cite{Vogelsberger:2019ynw}. Out of the existing profiles that fall within the range of simulations in Fig.~2 of Ref.~\cite{Hussein:2025xwm}, we will use the Navarro-Frenk-White (NFW) DM density profile. We note that the uncertainty due to this choice is large, and does not exist for measurements of the CMB~\cite{Planck:2018vyg}, applied for the low DM mass range (below $10~{\rm GeV}$).

Another type of indirect detection method is through neutrino detectors. Their detection involves a large volume of water or ice and detecting the Cherenkov light produced in the detector as neutrino interactions occur in it. The IceCube neutrino observatory has been a leading force in this area through it's multiple phases, with an energy threshold of a few ${\rm GeV}$.  

Several current and planned missions also aim to measure the flux of charged cosmic rays from space. 
The Pierre Auger Observatory (Auger) is a ground-based detector with an array of water surface detectors able to detect particles coming from cosmic-ray initiated particle showers. These detectors are sensitive to cosmic rays of very high energy, opening up the possibility of detecting signatures of superheavy Dark Matter annihilation. 










\section{Direct Searches}

If the galaxy is filled with weakly interacting massive particles, then many of them should pass through the Earth, making it possible to look for the interaction of such particles with matter, through measuring the recoil energy of nuclei. The key requirements for the calculation of the signal in direct detection experiments are the density and the velocity distribution of WIMPs that come to Earth and the DM-nucleon scattering cross section. It is then possible to evaluate the rate of events expected in an experiment per unit time, per unit detector material mass. The rate is given by~\cite{Bertone:2004pz},

\begin{equation}
    R \approx \sum_i N_i n_\chi <\sigma_{i\chi}>,
\end{equation}
with the index, i, running through the nuclei species present in the detector.
\begin{equation}
    N_i=\frac{\text{Detector mass}}{\text{Atomic mass of species i}},
\end{equation}
is the number of target nuclei in the detector. The local WIMP density is given as,
\begin{equation}
    n_\chi=\frac{\text{WIMP energy density}}{\text{WIMP mass}},
\end{equation}
and $<\sigma_{i\chi}>$ is the cross section for the scattering of the WIMP particle $\chi$ off the nuclei $i$, averaged over the relative WIMP velocity. The WIMPs as a class of Dark Matter candidates have masses in the 1 to $10^5~{\rm GeV}$ range and interaction cross sections from $10^{-41}$ to $10^{-51}$ ${\rm cm}^2$.

The detectors used vary greatly in the target materials used and techniques to mitigate background noise. The first detectors used Germanium semiconductor crystals to look for Dark Matter induced signals, and later also silicium. A current state-of-the-art experiment using Ge crystals is the CDEX-10. Another type of experiments uses arrays of high-purity scintillator crystals, which can operate in large areas and for a long time at a cost of high background signal. The DAMA/LIBRA experiment falls under this category, aiming to find a modulation in the annual signal.

A different type of detector named the cryogenic detector aims to measure the temperature increase following a particle interaction. To optimize sensitivity they are operated at very low temperatures, $\leq 50mK$, and aim to reduce the heat capacity $C$. Leading experiments using cryogenic detectors include CREST-II/III and SuperCDMS.

The last main type of detectors is based on noble gases such as argon and xenon that are excellent scintillators and can be ionized easily. Both of them can be liquefied to build dense Dark Matter targets. The first type of configuration used is that of a single-phase detector, where the spherical noble liquid target is surrounded by photomultipliers to collect the most scintillation signal possible. The DEAP-3600 detector operating since 2016 fits this category.  The second type, the dual phase time projection chamber (TPC), has a cylindrical noble liquid target with two arrays of photomultiplier tubes installed above and below. The simultaneous measurement of both signals aims to allow identification of multiple scatters and reconstruction of the interaction events. Darkside-50 and DEAP-3600 with argon and XENONnT, PandaX-4T and LUX-ZEPLIN (\gls{LZ}) using xenon consist of a TPC and are at the frontier of the current experimental limits.

\begin{figure}[htb]
	\centering
	\includegraphics[height=5.2cm]{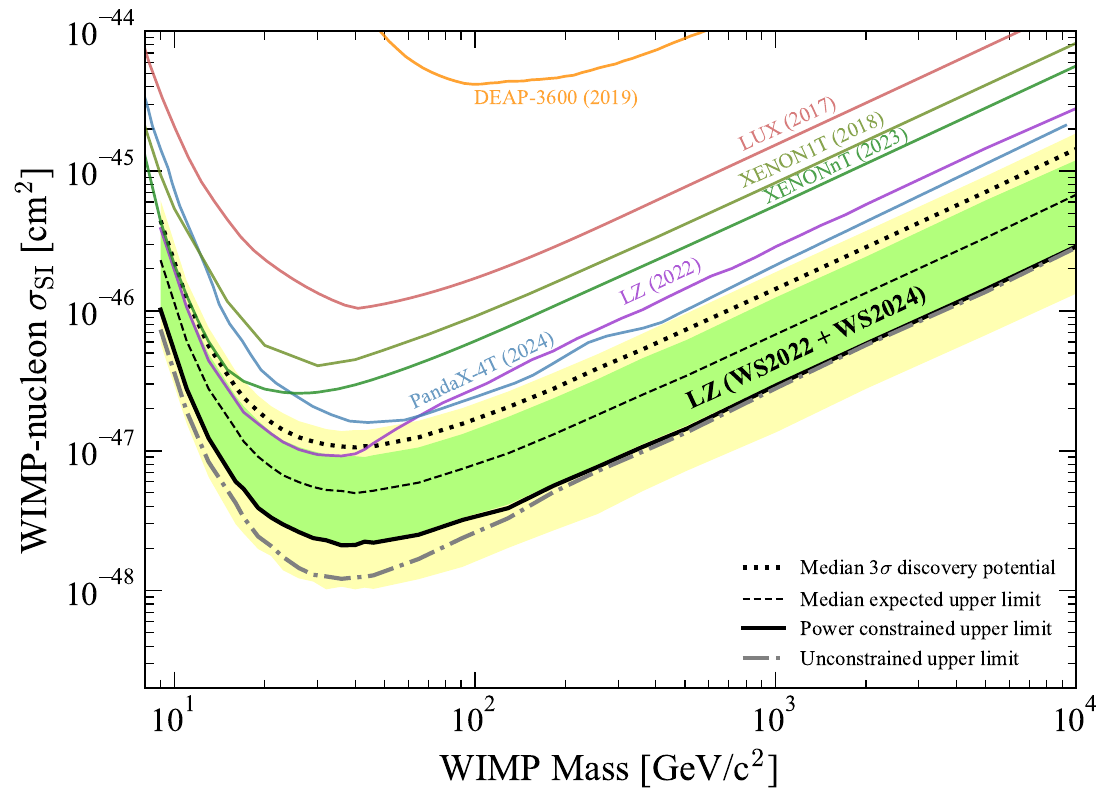}
	\includegraphics[height=5.2cm]{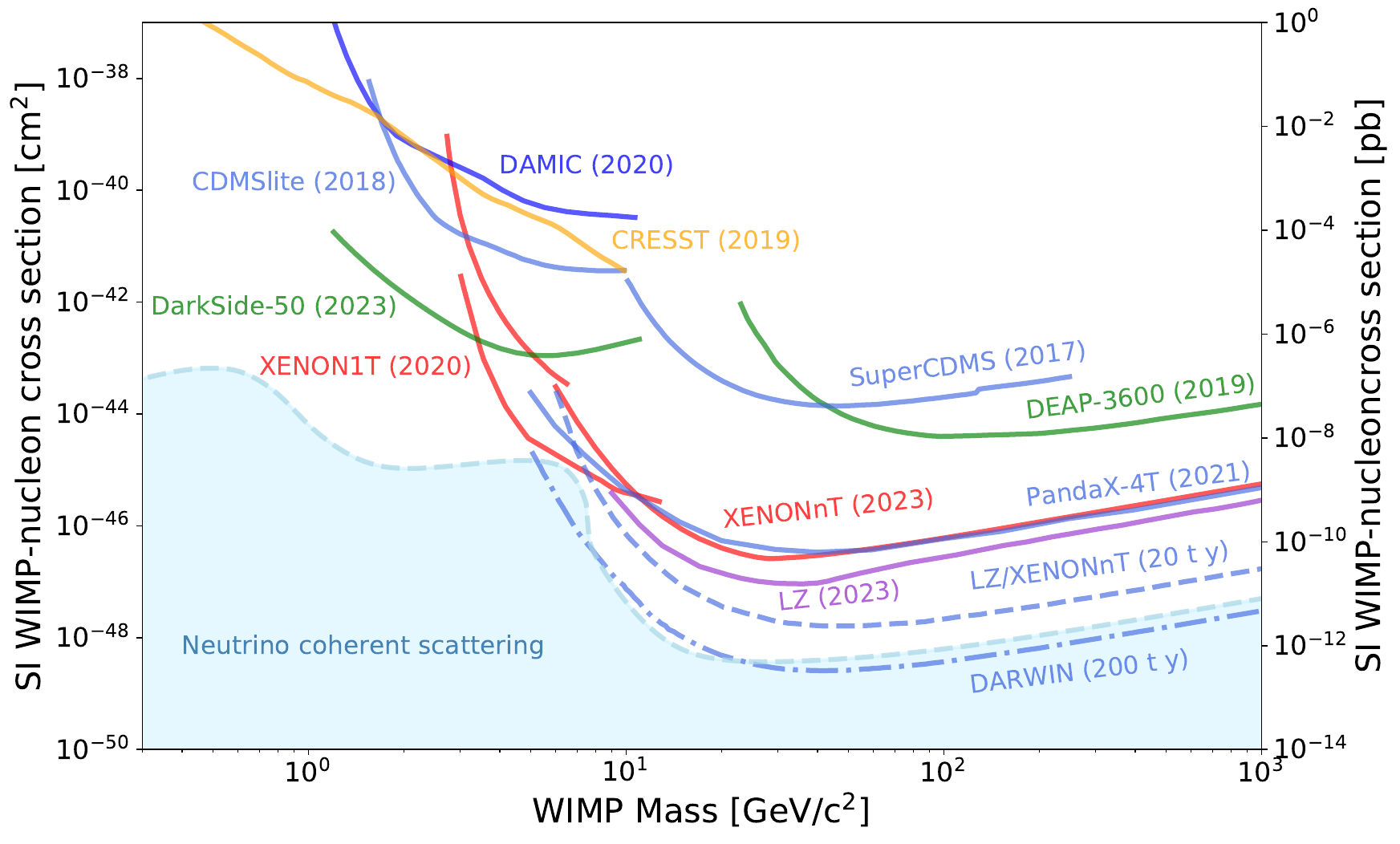}
\caption{\label{alldirect} Exclusion limits (solid) for spin-independent (\gls{SI}) WIMP-nucleon cross sections. Left panel: LZ 2024 data release of Ref.~\cite{LZ:2024zvo}, with the space above the lines excluded at $90\%$ CL. Right panel: From Ref.~\cite{Baudis:2024jnk}, the projections for LZ and XENONnT (dashed) and for DARWIN/XLZD (dashed-dotted) are also shown. The region known as the neutrino fog~\cite{OHare:2021utq} is shown in blue. }
\end{figure}

The current status of the searches with WIMP-nucleon scattering are shown in Fig.~\ref{alldirect}-Left and the strongest limits above WIMP mass of $5~{\rm GeV}$ are indeed those using a xenon TPC. In the interval between $1.8~{\rm GeV}$ and $5~{\rm GeV}$ the best limitis that of DarkSide-50 using argon. Other types of detectors have the sensitivity to explore the low mass ranges for WIMPs such as the SENSEI experiment built at the Fermilab with a Si-based sensor. No experiment so far observed a statistically significant excess above its background expectation. Future sensitivity of direct detection experiments are expected to reach the high mass section of the neutrino fog~\cite{OHare:2021utq}, where a distinction between a DM signal and a background of astrophysical neutrinos is challenging, thus requiring complementary probes. Fig.~\ref{alldirect}-Right includes the projections for LZ and XENONnT (dashed) and for DARWIN/XLZD (dashed-dotted).

\section{Collider Searches}

Although it remains uncertain whether Dark Matter (DM) particles interact strongly enough with ordinary matter to enable their production via collisions of Standard Model (SM) particles, collider searches for DM particles have attracted significant interest from the particle physics community due to their experimental feasibility. SM background signals are now understood with sufficient precision that even small deviations in the missing energy spectrum could be detectable and used to constrain DM models that predict such signatures. The fundamental requirement for a model to predict missing energy signals is that of a mechanism for DM production in the early Universe able to reproduce the one well-known property of DM, the relic density in \eq{relicdensity}. Such mechanism turns out to work for a wide range of masses and couplings and the constraints which can be placed on a DM candidate from collider experiments become highly model dependent. In this Section we review some of the most important collider searches which have been carried out. To note that these all consider models that predict large enough production cross sections so that a signal can be detected. Notably, other well-motivated DM candidates, such as axions or neutrinos, often predict signatures that are not detectable at colliders like the LHC.

\begin{figure}[htb]
	\centering
	\includegraphics[height=5cm]{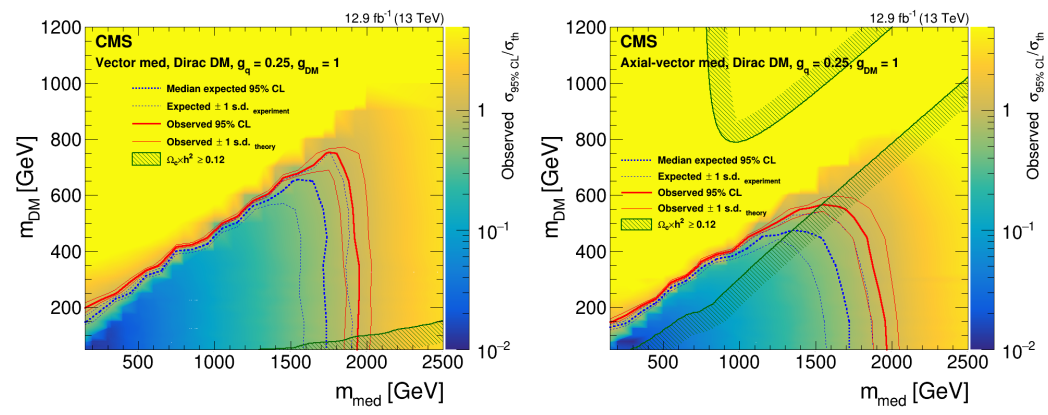}
\caption{\label{cmsexclusion1} Exclusion limits at $95\%$ CL on the ratio of the signal cross section to the predicted cross section, $\mu=\sigma/\sigma_{th}$, in the $m_{med}-m_{DM}$ plane assuming vector (left) and axial-vector (right) mediators. Taken from~\cite{CMS:2017tbk}.}
\end{figure}

\begin{figure}[htb]
	\centering
	\includegraphics[height=5cm]{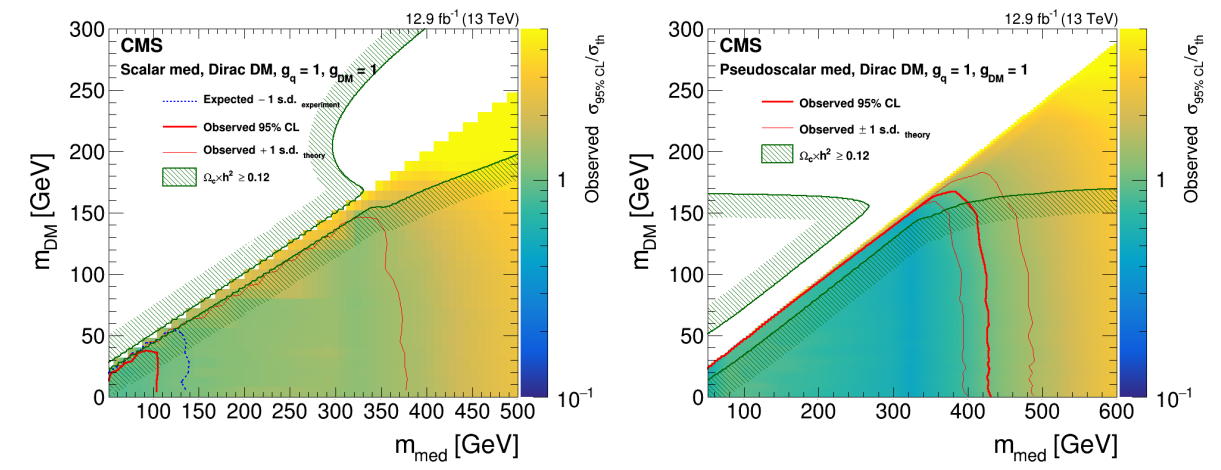}
\caption{\label{cmsexclusion2} Exclusion limits at $95\%$ CL on the ratio of the signal cross section to the predicted cross section, $\mu=\sigma/\sigma_{th}$, in the $m_{med}-m_{DM}$ plane assuming scalar (left) and pseudo-scalar (right) mediators. Taken from~\cite{CMS:2017tbk}.}
\end{figure}

\textbf{\textit{Mono-X Searches}}
This detection strategy is the most common and involves the identification in the scattering event of isolated objects from the initial state, accompanied by large missing momentum. The object produced in association with the DM candidate can be one or more QCD jet or other Standard Model particles, such as $\gamma$, $h$, $Z$, leptons, etc. Results from the Run-2 for the CMS and ATLAS collaborations using proton-proton collision data set on mono-X searches can be found in~\cite{CMS:2017tbk,ATLAS:2016tsc}. Exclusion limits are computed using simplified models in which DM production is mediated by either spin-1 or spin-0 particles. Fig.~\ref{cmsexclusion1} shows the exclusion contours in the $m_{med}-m_{DM}$ plane for the vector and axial-vector mediators, and Fig.~\ref{cmsexclusion2} for scalar and pseudoscalar mediators. These limits are computed using the monojet and mono-V, $V=\gamma, W, Z$, signal except for the pseudoscalar case which only considers the monojet process. A recent review of dark sector searches with the CMS~\cite{CMS:2024zqs} continues to report that no significant excess is observed.

Searches for a SM-like Higgs boson in association with missing transverse momentum have also been proposed and resulted in searches at colliders, in the $\gamma \gamma$~\cite{ATLAS:2015xoa} and in the $b \Bar{b}$~\cite{ATLAS:2016oyy} final state. These searches can make use of the high number of techniques that have been developed in the past years to identify the Higgs boson but, due to how faint the signal is for the Higgs, are limited by statistics. 

\textbf{\textit{Invisible Higgs Decays}}--- 
If weakly interacting massive particles (WIMPs) are lighter than half the Higgs boson mass, $m_h=125~{\rm GeV}$, the Higgs might invisibly decay into
WIMP pairs. In this case, one can use bounds from LHC on the invisible branching ratio
of the Higgs, BR($h\rightarrow \text{inv}$) $\leq$ 0.107 at $95\%$ C.L.~\cite{ATLAS:2023tkt}, to set strong constraints on WIMP models.

\textbf{\textit{Invisible Z Decays}}--- 
There is the possibility of a model with the DM particle coupling to the Z boson, thus having the feature that when the mass of the DM candidate is less than half of the Z particle, $m_Z= 80.38~{\rm GeV}$, there are strong constraints coming from the precise measurements of the Z boson decay width. The current limit is $\Gamma (Z\rightarrow \text{inv}) \leq 499 \pm 1.5~{\rm MeV}$~\cite{ParticleDataGroup:2020ssz}, which is about $20\%$ of the total width.

\textbf{\textit{Top quarks}}--- 
If the DM particle couples dominantly to heavy quark flavors, a search for a top-quark pair in association with missing energy would be a promising way to discover a DM signal. Analyses have been done in the semileptonic and fully hadronic channels~\cite{CMS-PAS-EXO-16-005}, or in the fully leptonic channel~\cite{CMS-PAS-EXO-16-028}. The constraints obtained are stronger in the first case.

\section{Summary}
In this Chapter we reviewed the current status of particle dark matter, including present experimental evidence. We introduced the class of dark matter (DM) candidates we will consider, known as Weakly Interacting Massive Particles (WIMPs). We followed with a description of the most recent experimental searches for WIMPs, with direct direction, indirect detection and collider searches. The most stringent bounds are to be considered as constraints for the phenomenological studies of the models of DM considered.

\begin{figure}[htb]
	\centering
	\includegraphics[height=5.8cm]{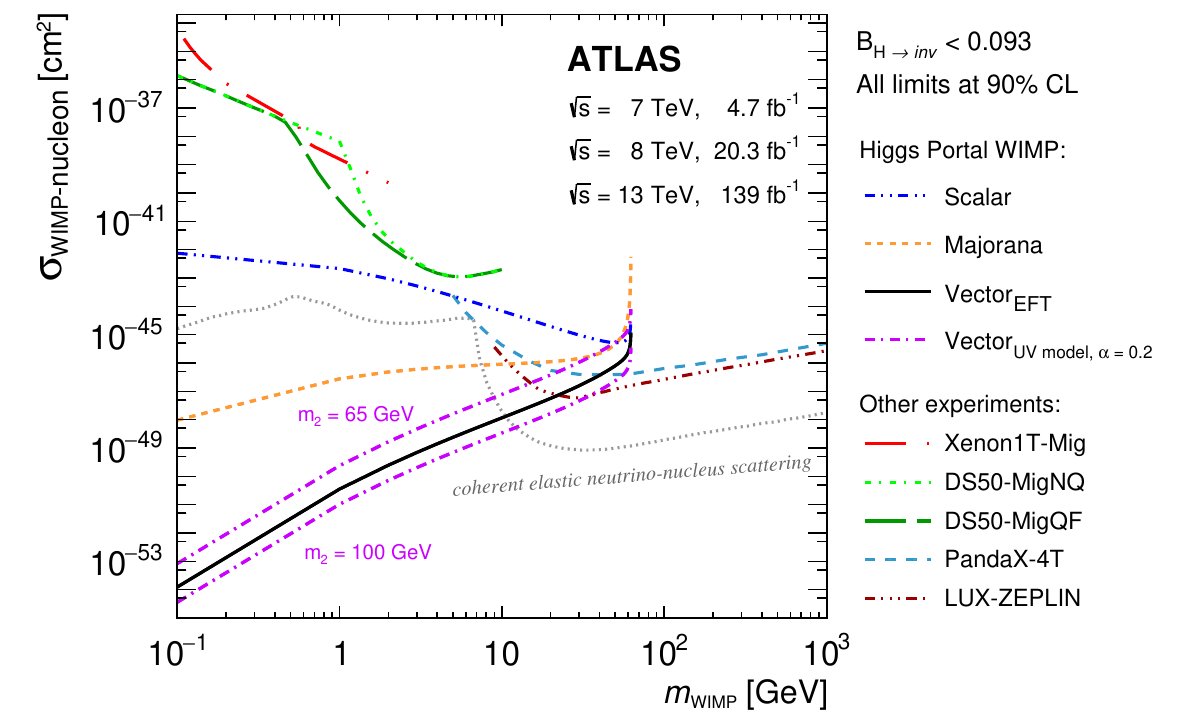}
	\includegraphics[height=5.8cm]{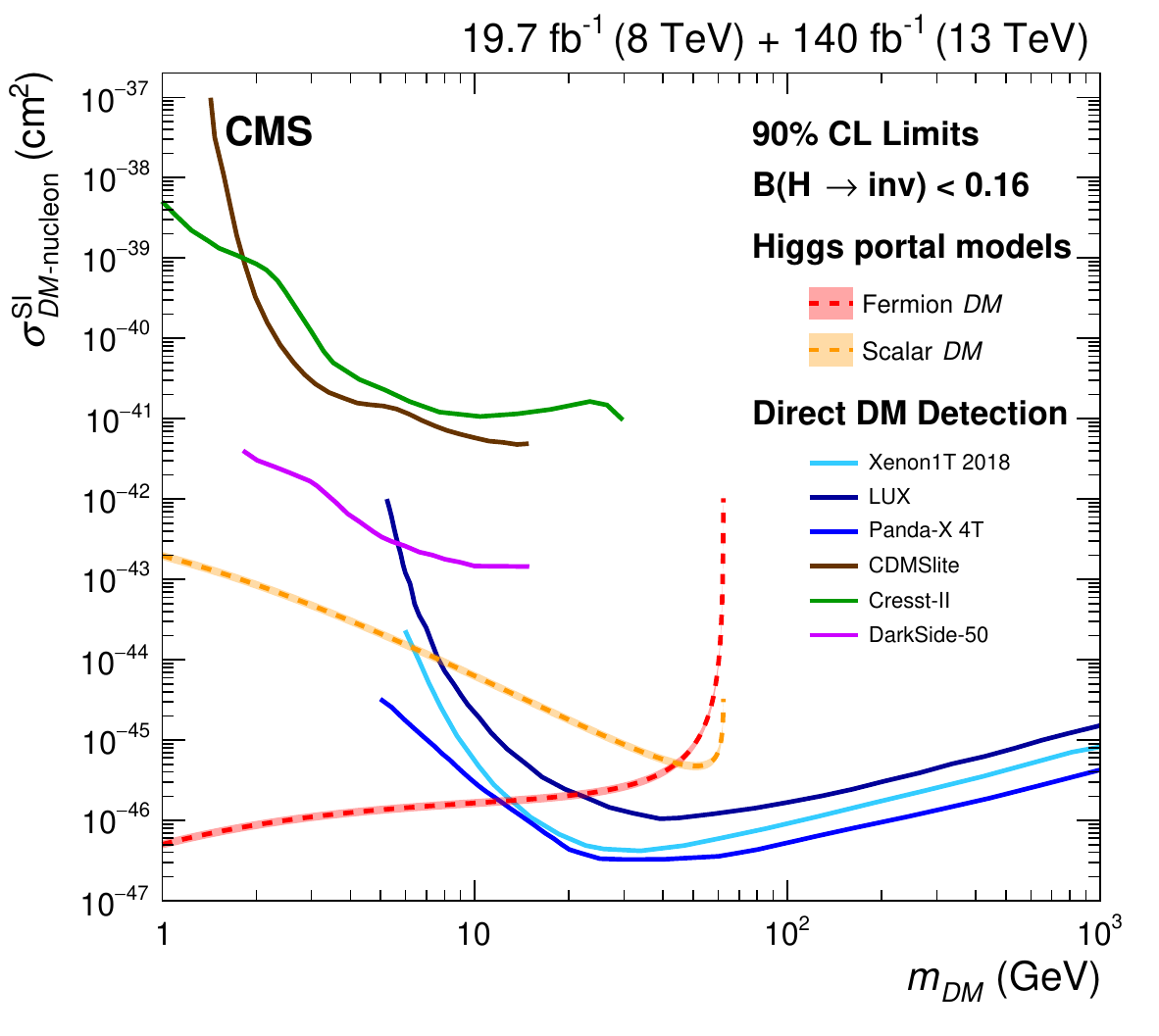}
\caption{\label{Complementarity} Upper limit at the $90\%$ CL on the SI- WIMP-nucleon scattering cross-section as a function of
the DM mass for direct detection experiments and the interpretation of the $H\to$invisible collider results in
the context of Higgs portal models. In the Left panel, the ATLAS~\cite{ATLAS:2023tkt} figure shows the combination results considering scalar, Majorana and vector WIMP hypotheses. In the Right panel, the CMS~\cite{CMS:2022qva} results consider an Higgs boson produced via vector boson fusion and assuming either a scalar or fermionic DM candidate.}
\end{figure}

We finish with an example of the existing complementarity between direct detection experiments and collider searches in the context of Higgs-portal DM models, shown in Fig.~\ref{Complementarity} for the ATLAS~\cite{ATLAS:2023tkt} (2023) and CMS~\cite{CMS:2022qva} (2022) collaborations at the LHC.

%% file: chapters/Darkmatter_model.tex
\chapter{Building a Dark Matter Model}
\label{chapter:Darkmatter_model}
\section{Stabilizing dark matter: symmetry groups vs. individual symmetries}
In order to explain Dark Matter, one popular idea is to assume that the scalar sector is not as minimal as in the SM but contains new scalar fields.
Equipped with a new global symmetry that remains unbroken in the vacuum, 
this scalar sector leads to one or more scalar DM candidates. 

A prototypical illustration of this scenario is the inert doublet model (IDM)~\cite{Deshpande:1977rw,Ma:2006km,Barbieri:2006dq,LopezHonorez:2006gr}. 
This is a version of the two-Higgs-doublet model (2HDM) with an exact $\Z2$ symmetry 
that acts trivially on the first Higgs doublet $\phi_1$ and the SM fields 
and flips the sign of the second Higgs doublet $\phi_2$.
If the vacuum expectation value (vev) of the second doublet is zero, $\lr{\phi_2} = 0$,
$\Z2$ remains unbroken and stabilizes the lightest scalar from $\phi_2$ against decay.
As the model introduces very few free parameters and leads to numerous collider and cosmological predictions, 
it has become a popular playground, see for example 
~\cite{Arhrib:2013ela,Ilnicka:2015jba,Belyaev:2016lok,Kalinowski:2018ylg,Belyaev:2018ext} and references therein.

Using $\Z2$ as a DM stabilizing group is not the only option.\footnote{We mention in passing that 
there exist examples of scalar DM that are not based on any exact symmetry~\cite{Kajiyama:2011gu}.}
One can easily construct multi-Higgs models in which DM candidates are protected by the unbroken $\Z{n}$, $n > 2$, see 
~\cite{Ma:2007gq,Batell:2010bp,Ivanov:2012hc,Belanger:2012vp,Belanger:2014bga,Yaguna:2019cvp,Belanger:2020hyh,Yaguna:2021vhb,Yaguna:2021rds,Belanger:2021lwd,Belanger:2022esk} and references therein.
Such models often contain additional processes that affect the DM evolution 
or feature several DM components with distinct interaction preferences.
One can even build a 3HDM model in which a pair of mass-degenerate DM candidates is stabilized 
by an exotic $CP$ symmetry of higher order rather than a Higgs family transformations~\cite{Ivanov:2015mwl,Ivanov:2018srm}.
Another idea considered in the literature is to replicate the IDM, that is, 
to consider a 3HDM with two independent $\Z2$ symmetries flipping the signs of 
$\phi_2$ and $\phi_3$, respectively. 
The model possesses two DM candidates coming from the two inert doublets 
with much freedom to adjust the properties of the two dark sectors
~\cite{Belanger:2011ww,Aoki:2012ub,Bhattacharya:2016ysw,Hernandez-Sanchez:2020aop}. This is the model we will use for a full phenomenological study in the next Chapter.

The main objective of this Chapter is to explore yet another possibility, with scalar DM candidates stabilized not by an abelian symmetry 
but by a non-abelian symmetry group, in a special situation where a {\em remnant} non-abelian global symmetry group $G_v \subset G$ 
survives after the minimization of the scalar potential invariant under $G$. In doing so, we will find characteristic features of the DM 
in multi-Higgs-doublet models highly constrained by large global symmetry groups and chart the limits 
of what multi-Higgs-doublet models can in principle accommodate.
We will see that it is not an easy task to arrange for stable DM candidates that could account for at least a part 
of the total relic density without running into conflict with direct detection and collider constraints.

This idea is not new. The first---and the simplest---example based on the residual group $G_v = S_3$
was presented in~\cite{Adulpravitchai:2011ei}. In that work, a simplified DM model was constructed,
which featured, in addition to the SM Higgs doublet, two electroweak (EW) singlet fields, the real $\eta$ 
and the complex $\chi$.
The interaction potential was designed in such a way that the model possesses $S_3$
and that the vev alignment preserves this group, making $\chi$ and $\chi^*$ the two mass degenerate DM candidates. 
The model contained enough freedom to introduce new channels
that affect the DM thermal evolution and to satisfy observational constraints.
Another attempt to build a DM model based on a non-abelian group, the quaternion group $Q_4$, 
was described in~\cite{Lovrekovic:2012bz}.

We are interested in a different twist of the non-abelian DM scenario.
In the above papers, as well as in many other DM models, one postulates a dark sector which, 
by construction, does not participate in the EW interactions and couples to the visible sector
either through portal-like interactions or flavor physics connections. 
Since the visible and the dark sectors are governed by different Lagrangians, 
it is no wonder that this scenario offers enough freedom
to adjust the DM sector separately from the SM fields. However, we want to explore the same scenario in the multi-Higgs-doublet models.
Here, the scalar DM protected by a non-abelian group appears from the same self-interaction
potential as the SM-like Higgs boson itself. As a result, the connections and constraints become much tighter,
and it is not clear a priori whether such a DM scenario is tenable. 

We explore this scenario and its DM implications with the example of the three-Higgs-doublet model (3HDM)
whose scalar sector is invariant under the global symmetry group $G=\Sigma(36)$. In the next Section, we review the concept of structural rigidity and its importance in building symmetry-based multi-Higgs models.  
In Section~\ref{subsection:potential}, we will review the scalar sector of the $\Sigma(36)$ symmetric 3HDM,
list its minima and the physical scalars.
We will also highlight the features that arise due to stabilization of DM candidates
by a non-abelian group, as contrasted with stabilization by a single symmetry.
Then, in Section~\ref{section-realistic} we attempt to complete the model
with a suitable Yukawa sector.
We first show in Subsection~\ref{section-Yukawa-exact} that $\Sigma(36)$ can be, 
in principle, extended to the quark sector but leads to unphysical quark properties and unstable DM candidates.

We then relax our assumptions and, in Subsection~\ref{section-Yukawa-S3}, we construct a Yukawa sector that
is invariant not under the full $\Sigma(36)$ but under its subgroup $G_v = S_3$,
the residual symmetry group at a chosen minimum.
This can be easily done in a fully realistic way.
Moreover, since $\Sigma(36)$ is not the symmetry of the full Lagrangian,
we also take the liberty of introducing soft breaking terms respecting the same $S_3$,
which adds only one additional parameter.
This is done in a consistent way, so that the minimum we initially selected
becomes the unique global minimum of the resulting potential.

We incorporate the model in \texttt{micrOMEGAs 6.0.5}~\cite{Alguero:2023zol} and present, 
in Section~\ref{section-DM}, the results of our study.
First, we focus on the model without any soft breaking and
explore how the relic density and direct detection signals depend on the free parameters.
We compare our predictions with the Planck observations~\cite{Planck:2018vyg}
and the upper limits from the direct detection experiments Xenon1T~\cite{XENON:2018voc} and LZ~\cite{LZ:2022lsv},
including the very recent LZ results~\cite{LZ:2024zvo}. 
We find that, due to a strong correlation among the signals, the $\Sigma(36)$-symmetric scalar sector 
is unable to account even for a part of DM density. 
This strong conclusion is the second lesson emerging from our study.

We then proceed to the softly broken $\Sigma(36)$ 3HDM, 
complemented with the $S_3$ symmetric Yukawa sector,
and discuss in Subsection~\ref{subsection-DM-soft} the DM relic density and 
direct detection signals.
As the tight correlation between the scalar masses and couplings is now relaxed,
we find it possible to match the DM abundance or to remain in the sub-dominant DM scenario
without violating observational constraints.
However, even with these relaxed assumptions, we see a serious clash between theoretical constraints 
and the LZ-2024 direct detection results, which is the third lesson of our study.
We place a lower bound on the magnitude of the soft breaking terms
that allow us to reach minimally viable models, and comment on the origin of these persistent conflict 
between different experimental constraints.

\section{Structural rigidity and its consequences}\label{subsection:Delta27}

Structural rigidity is the central theme of the symmetry-based multi-Higgs model building activity.
Here, ``rigidity'' means that, for a sufficiently large finite group, the phenomenological properties of the model
follow robust patterns insensitive to the exact numerical values of the parameters.
The key goal is then to identify which pattern seems to (approximately) fit the observables
and which multi-Higgs models lend support to such a pattern.

We find it instructive to remind the reader of the situation in which the particle physics found itself
at the end of the 1970s. With the third generation of fermions just discovered, 
and their masses and mixing parameters measured, 
the hierarchical structure of the fermion masses and the Cabibbo-Kobayashi-Maskawa (CKM) 
matrix $\VCKM$ called for explanations. The problem was optimistically attacked by constructing 
$N$-Higgs-doublet models with several generations of Higgs doublets. The scalar doublets were assumed to couple to fermions and among themselves
in such a way that the Lagrangian stayed invariant under a group $G$ of global transformations acting both on the scalar and fermions
fields. In this early activity, several options for the group $G$ were studied, such as the groups of sign flips, 
rephasing transformations, permutations, or their combinations; for a brief historical overview see~\cite{Ivanov:2017dad}.
For example, the early suggestion~\cite{Segre:1978ji} 
was based on the group of $\Delta(27) \subset SU(3)$ generated by the following order-3
rephasing and permutation transformations: 
\begin{equation}
	a = \mtrx{1&0&0\\ 0&\omega&0\\ 0&0&\omega^2}\,, \quad
	b = \mtrx{0&1&0\\ 0&0&1\\ 1&0&0}\,,\quad\mbox{where}\quad \omega = e^{2\pi i /3}, \ \omega^3 = 1\,.
	\label{Delta27-generators}
\end{equation}
The idea of Ref.~\cite{Segre:1978ji} was to ``naturally'' explain the mass and mixing hierarchies by assuming that 
the three Higgs doublets acquire hierarchical vevs $|v_1| \ll |v_2| \ll |v_3|$.
However, as it was pointed out already in~\cite{Segre:1978ji}, the $\Delta(27)$ invariant scalar potential,
with its limited structures and very few free parameters, was unable to yield such a vev alignment. 

The failure of the $\Delta(27)$-invariant 3HDM to reproduce the desired vevs was later understood as a particular example of a general phenomenon.
Multi-Higgs potentials based on large finite symmetry groups possess a rigid structure of their minima~\cite{Degee:2012sk,Ivanov:2014doa}
and even of their saddle points~\cite{Yang:2024bys}. Their vevs scale as simple ratios 
such as $1:1:1$ and remain stable against smooth variation of the coefficients.
Another manifestation of a rigid vev structure is the so-called geometric $CP$ violation;
the calculable relative phase between vevs,
which was first noticed in~\cite{Branco:1983tn} for the same $\Delta(27)$ 3HDM and 
later used in other multi-Higgs models~\cite{deMedeirosVarzielas:2011zw,deMedeirosVarzielas:2012rxf,Ivanov:2013nla}.

Structural rigidity has far-reaching consequences for the fermion sector of the model.
As was first pointed out in~\cite{Leurer:1992wg}, 
structural rigidity may render the quark sector unphysical at any minimum of a symmetry-constrained potential. 
A no-go theorem was formulated in~\cite{Leurer:1992wg} and later refined in~\cite{GonzalezFelipe:2014mcf}, 
which states that the only way to obtain a non-block-diagonal CKM mixing matrix and, simultaneously, non-degenerate
and non-zero quark masses is to make sure that vev alignment breaks the group $G$ completely, except possibly for a symmetry
belonging to the baryon number or a symmetry located purely within the inert sector (i.e. for the doublets which do not couple to quarks).
But in potentials with a large symmetry group, there are always residual symmetries left at each of the possible minima.
One popular way out is to add soft breaking terms, which can in many cases provide enough freedom to adjust vevs
at least within some limits.
A nice illustration of the above situation was given for the $A_4$-invariant 3HDM:
while the model with an exact $A_4$ symmetry always leads to pathological fermion sectors
~\cite{GonzalezFelipe:2013xok}, adding soft breaking terms was enough
to remove the unwanted residual symmetries and render the model compatible with experiment~\cite{Bree:2023ojl}.
For the $\Delta(27)$ symmetric 3HDM, numerous attempts were undertaken 
to incorporate it into realistic models~\cite{deMedeirosVarzielas:2011zw, Varzielas:2012nn, Varzielas:2013eta, Kalinowski:2021lvw},
but all of them had to resort to additional non-SM fields, going beyond the pure multi-Higgs-doublet models,
or even had to accept the massless fermions as a viable option as in~\cite{Kalinowski:2021lvw}.

Stabilization of scalar DM candidates is another manifestation of structural rigidity,
as certain vevs can remain exactly zero in broad regions of the parameter space.
Thus, rigidity can be either a welcome feature, as in the case of DM candidates, 
or a nuisance factor, as for the fermion sector.
Whether one can keep its positive features and avoid its downsides remains to be analyzed in specific models.

\section{\texorpdfstring{The $\Sigma(36)$ 3HDM potential}{The Sigma(36) 3HDM potential}\label{subsection:potential}}

Let us now describe and analyze the $\Sigma(36)$ 3HDM scalar sector. 
First, a technical remark is in order. The group $\Sigma(36)$, of order $36$, derived in the 3HDM symmetry classification 
~\cite{Ivanov:2011ae,Ivanov:2012ry,Ivanov:2012fp} was, by construction, considered as a subgroup of $PSU(3) \simeq SU(3)/Z(SU(3))$,
where  $Z(SU(3)) \simeq \Z3 = 
\{\id_3, \, \omega\cdot \id_3, \, \omega^2\cdot \id_3\}$ is the center of $SU(3)$.
However, since it is convenient to work with unitary, not projective representations, 
we will work inside $SU(3)$. In this group space, the finite symmetry group we deal with is denoted as $\Sigma(36\varphi)$, or $\Sigma(36\times 3)$, 
see e.g.~\cite{Grimus:2010ak,Hagedorn:2013nra}, and is of order 108. 
Although we will occasionally write $\Sigma(36)$ as in ``$\Sigma(36)$-based 3HDM,'' we will actually deal with $\Sigma(36\varphi)$.

The group $\Sigma(36\varphi)$, which can also be represented as $\Delta(27)\rtimes \Z4$, is generated by the same $a$ and $b$ as 
in \eq{Delta27-generators} and a new transformation $d$ of order four:
\begin{equation}
	d = \frac{i}{\sqrt{3}} \left(\begin{array}{ccc} 1 & 1 & 1 \\ 1 & \omega^2 & \omega \\ 1 & \omega & \omega^2 \end{array}\right)\,,
	\quad
	\mbox{so that}\  d^2 = -\mtrx{1&0&0\\ 0&0&1\\ 0&1&0}\,.
	\label{Sigma36-generators}
\end{equation}
The 3HDM scalar potential invariant under this group is
\begin{eqnarray}
	V & = &  - m^2 \left(\phi_1^\dagger \phi_1+ \phi_2^\dagger \phi_2+\phi_3^\dagger \phi_3\right)
	+ \lambda_1 \left(\phi_1^\dagger \phi_1+ \phi_2^\dagger \phi_2+\phi_3^\dagger \phi_3\right)^2 \nonumber\\
	&&
	- \lambda_2 \left[|\phi_1^\dagger \phi_2|^2 + |\phi_2^\dagger \phi_3|^2 + |\phi_3^\dagger \phi_1|^2
	- (\phi_1^\dagger \phi_1)(\phi_2^\dagger \phi_2) - (\phi_2^\dagger \phi_2)(\phi_3^\dagger \phi_3)
	- (\phi_3^\dagger \phi_3)(\phi_1^\dagger \phi_1)\right] \nonumber\\
	&&
	+ \lambda_3 \left(
	|\phi_1^\dagger \phi_2 - \phi_2^\dagger \phi_3|^2 +
	|\phi_2^\dagger \phi_3 - \phi_3^\dagger \phi_1|^2 +
	|\phi_3^\dagger \phi_1 - \phi_1^\dagger \phi_2	|^2\right)\, .
	\label{Vexact}
\end{eqnarray}
Here, the first two lines are invariant under the entire $SU(3)$ group of family rotations of the three Higgs doublets,
while the $\lambda_3$ term selects the finite group $\Sigma(36\varphi)$.
This potential is also $CP$ invariant~\cite{Ivanov:2011ae,Ivanov:2014doa}, following Fig.~\ref{f:symmetrybreaking};
in fact, the $\Z4$ symmetry group within the 3HDM scalar sector automatically forbids 
any form of $CP$ violation in the scalar sector, be it explicit or spontaneous \cite{Ivanov:2014doa}.

The potential in \eq{Vexact} contains very few free parameters.
The coefficients $m^2$ and $\lambda_1$ fix the overall scales 
of the vev $v$ and the SM-like Higgs mass,
while the locations of the global minima depend on $\lambda_2$ and $\lambda_3$.
The conditions for the potential to be bounded from below (BFB) were derived in~\cite{Yang:2024bys}:
\begin{equation}
	\lambda_1 > 0\,, \quad \lambda_1 + \lambda_3 > 0\,,\quad
	\lambda_1 + \frac{1}{4}\lambda_2 > 0\,,\quad \lambda_1 + \frac{1}{4}(\lambda_2 + \lambda_3) > 0\,.
	\label{BFB}
\end{equation}
All these inequalities are strict, meaning we must impose the BFB conditions in the strong sense~\cite{Maniatis:2006fs}.

\subsection{The global minima and residual symmetries}

Depending on the values of the quartic coefficients, the global minimum of the $\Sigma(36)$ 3HDM can be either neutral or charge-breaking. 
The latter option must be avoided at zero temperature, although existence of a charge-breaking phase
at intermediate temperatures is an interesting and insufficiently explored opportunity
for the hot early Universe evolution~\cite{Aoki:2023lbz,Yang:2024bys}. 
The regions on the $(\lambda_2,\lambda_3)$ plane that guarantee the neutral minimum of the tree-level potential 
were established in~\cite{Degee:2012sk} and are listed below.
The vev alignment corresponding to a neutral global minimum can be written as
\begin{equation}
	\lr{\phi_1} = \frac{1}{\sqrt{2}}\doublet{0}{v_1}\,,\quad  
	\lr{\phi_2} = \frac{1}{\sqrt{2}}\doublet{0}{v_2}\,,\quad 
	\lr{\phi_3} = \frac{1}{\sqrt{2}}\doublet{0}{v_3}\,,
	\label{neutral-vevs}
\end{equation}
with $v_i$ being, in general, complex. To simplify the notation, we will take the overall scale out, 
implicitly assuming that $|v_1|^2 + |v_2|^2 + |v_3|^2 = v^2$, where $v = 246~{\rm GeV}$, 
and indicate the ratios between the individual vevs. 
For example, if a minimum corresponds to $v_1 = v_2 = v_3$, we write 
\begin{equation}
	(\lr{\phi_1^0},\, \lr{\phi_2^0},\, \lr{\phi_3^0}) = 
	\left(\frac{v_C}{\sqrt{2}},\, \frac{v_C}{\sqrt{2}},\, \frac{v_C}{\sqrt{2}}\right) = \frac{v_C}{\sqrt{2}}\left(1, 1, 1\right)\,,
	\quad
	\mbox{with}\quad v_C = \frac{v}{\sqrt{3}}\,,
\end{equation}
and label this vev alignment as $(1,1,1)$.

Depending on the free parameters, $\Sigma(36)$ 3HDM allows for two scenarios for neutral minima~\cite{Ivanov:2014doa,Yang:2024bys}.
\begin{itemize}
	\item 
	$BC$-scenario: if $\lambda_2 > 0$ and $\lambda_3 > 0$, then the global minima correspond to the following vev alignments: 
	\begin{eqnarray}
		&&	\mbox{alignment $B$:}\quad B_1 = (1,\,0,\,0)\,, \quad B_2 = (0,\,1,\,0),\, \quad B_3 = (0,\,0,\,1)\, ;\label{points-B}\\
		&&\mbox{alignment $C$:}\quad C_1 = (1,\,1,\,1)\,,\quad C_2 = (1,\,\omega,\,\omega^2)\,,\quad C_3 = (1,\,\omega^2,\,\omega)\label{points-C}\,.
	\end{eqnarray}
	Other configurations such as $(\omega,\,\omega^2,\, 1)$ can be 
	reduced to those already listed by an overall phase rotation of the three doublets.
	For example, $(\omega,\,\omega^2,\, 1) = \omega (1,\,\omega,\,\omega^2)$ corresponds to the alignment $C_2$.
	\item 
	$AA'$-scenario: if $\lambda_3 < 0$, while $\lambda_2$ satisfies $\lambda_2 - 3 \lambda_3 > 0$,
then the global minima correspond to
	\begin{eqnarray}
		&&\mbox{alignment $A$:}\quad A_1 = (\omega,\,1,\,1)\,, \quad A_2 = (1,\,\omega,\,1),\, \quad A_3 = (1,\,1,\,\omega)\, ;\label{points-A}\\
		&& \mbox{alignment $A'$:}\quad A'_1 = (\omega^2,\,1,\,1)\,, \quad A'_2 = (1,\,\omega^2,\,1),\, \quad A'_3 = (1,\,1,\,\omega^2)\, .\label{points-Ap}
	\end{eqnarray}
\end{itemize}
In either case, we observe a similar picture: there are six degenerate global minima, which are all linked by the transformations from the group $\Sigma(36)$
broken by the vev alignment. For example, the broken transformation $d$ links $B_i$ and $C_i$.
Also, no higher-lying local minima exists for any of these cases~\cite{Yang:2024bys}.

For any choice of the global minimum, the symmetry group $G = \Sigma(36)$ is spontaneously 
broken to the residual subgroup $G_v = S_3$. 
It is well known that any $S_3$ can be generated by two generators $g_2$ and $g_3$ satisfying $g_2^2 = g_3^3 = e$, $g_2^{-1} g_3 g_2 = g_3^{-1}$.  
The specific transformations $g_2$ and $g_3$ from $\Sigma(36)$ that remain unbroken and generate this residual $S_3$ differ for each minimum.
Here are several examples of the residual symmetry generators:
\begin{eqnarray}
	&&\mbox{alignment}\ B_1 = (1,\,0,\,0):\quad g_2 = d^2\,, \quad g_3 = a\,.\label{residual-B1}\\
	&&\mbox{alignment}\ C_1 = (1,\,1,\,1):\quad  g_2 = d^2\,, \quad g_3 = b\,.\label{residual-C1}\\
	&&\mbox{alignment}\ C_2 = (1,\,\omega,\,\omega^2):\quad g_2 = d^2a\,, \quad g_3 = b\,.\label{residual-C2}\\
	&&\mbox{alignment}\ A_1 = (\omega,\,1,\,1):\quad g_2 = d^2\,, \quad g_3 = ba^2\,.\label{residual-A1}
\end{eqnarray}
Besides, each vev alignment conserves $CP$, either the canonical $CP_0: \phi_a \mapsto \phi_a^*$
or a generalized $CP$ symmetry which combines $CP_0$ with a transformation from $\Sigma(36)$.
For example, the alignment $A_1$ is invariant under $CP_0$ followed by $d$.

\subsection{The physical scalars and DM candidates}

Since any choice of the global minimum leads to residual symmetries, we expect scalar DM candidates stabilized by these symmetries.
The fact that the residual symmetry group is $S_3$ hints at a two-dimensional space 
of mass-degenerate DM candidates. This is the feature which we want to explore with our choice of the symmetry group $\Sigma(36)$.

In this subsection, we give the spectrum of physical scalars and indicate their conserved quantum numbers which protect 
them against decay. We remind the reader that, at this stage, we are working with the bosonic degrees of freedom only.

Let us begin with the $BC$-scenario and choose the vev alignment $B_1$ as a representative case. 
We expand the Higgs doublets around the minimum as
\begin{equation}
	\phi_1 = \frac{1}{\sqrt{2}}\doublet{\sqrt{2}h_1^+}{v + h_1 + i a_1}\,,\quad  
	\phi_2 = \frac{1}{\sqrt{2}}\doublet{\sqrt{2}h_2^+}{h_2 + i a_2}\,,\quad 
	\phi_3 = \frac{1}{\sqrt{2}}\doublet{\sqrt{2}h_3^+}{h_3 + i a_3}\,.
	\label{field-expansion}
\end{equation}
The fields $h_1^+$ and $a_1$ are the would-be Goldstone bosons which disappear in the unitary gauge.
The remaining fields are physical scalars with the following masses:
\begin{eqnarray}
	h_{SM} = h_1: &\qquad & m_{SM}^2 = 2\lambda_1 v^2 = 2 m^2\,,\label{B1-spectrum-SM}\\
	H_2^+ = h_2^+\,, \ H_3^+ = h_3^+: &\qquad & m_{H^+}^2 = \frac{\lambda_2 v^2}{2}\quad \mbox{(double degenerate)}\,,\label{B1-spectrum-charged}\\
	h = \frac{1}{\sqrt{2}}(h_2 + h_3)\,, \quad a = \frac{1}{\sqrt{2}}(a_2 - a_3): &\qquad &
	m_{h}^2 = \frac{\lambda_3 v^2}{2}\quad \mbox{(double degenerate)}\,,\label{B1-spectrum}\\
	H = \frac{1}{\sqrt{2}}(h_2 - h_3)\,, \quad A = \frac{1}{\sqrt{2}}(a_2 + a_3): &\qquad &
	m_{H}^2 = \frac{3 \lambda_3 v^2}{2}\quad \mbox{(double degenerate)}\,.\label{B1-spectrum-HA}
\end{eqnarray}
Assuming $\lambda_2 > \lambda_3$, the neutral fields $h$ and $a$ indicated in the third line
are the DM candidates.
Since they are mass degenerate, one can alternatively choose any linear combination of 
$h$ and $a$, which can also be considered as the DM candidate.
Indeed, the residual symmetry group $S_3$ given in \eq{residual-B1}
acts on $h$ and $a$ by $2\pi/3$ rotations and reflections in the $(h,a)$ plane.
Thus, the pair $(h,a)$ forms a real 2D irreducible representation of $S_3$, and it is this 2D space of spaces
which is stabilized by the residual group.

For the $AA'$-scenario, we choose the alignment $A_3 = (1,1,\omega)$ and parametrize the doublets as
\begin{equation}
	\phi_1 = \frac{1}{\sqrt{2}}\doublet{\sqrt{2}h_1^+}{v_A + h_1 + i a_1}\,,\quad  
	\phi_2 = \frac{1}{\sqrt{2}}\doublet{\sqrt{2}h_2^+}{v_A + h_2 + i a_2}\,,\quad 
	\phi_3 = \frac{\omega}{\sqrt{2}}\doublet{\sqrt{2}h_3^+}{v_A + h_3 + i a_3}\,,
	\label{field-expansion-2}
\end{equation}
where $v_A = v/\sqrt{3}$. Note that we factorized the phase factor $\omega$ on the third doublet.
The mass of the SM-like Higgs is again
\begin{equation}
	\hSM = \frac{1}{\sqrt{3}}(h_1 + h_2 + h_3): \quad \mSM^2 = 2(\lambda_1 + \lambda_3) v^2 = 2m^2\,. 
\end{equation}
The charged Goldstone is $G^+ = (h_1^+ + h_2^+ + h_3^+)/\sqrt{3}$, while the two charged Higgses orthogonal to it
are again mass degenerate:
\begin{equation}
	H_2^+, \, H_3^+ \ \perp \ G^+: \quad m_{H^+}^2 = \frac{v^2}{2}(\lambda_2 - 3\lambda_3)\,. 
\end{equation}
To find the remaining four neutral mass eigenstates, we define
\begin{equation}
	h_- = \frac{1}{\sqrt{2}}(h_1 - h_2)\,, \quad 
	h_+ = \frac{1}{\sqrt{6}}(h_1 + h_2 - 2 h_3)\,, \quad
	a_- = \frac{1}{\sqrt{2}}(a_1 - a_2)\,, \quad 
	a_+ = \frac{1}{\sqrt{6}}(a_1 + a_2 - 2 a_3)\,,
\end{equation}
and observe that the neutral mass matrix splits into two identical $2\times 2$ blocks ${\cal M}_2$ acting within the spaces
$(h_-,a_-)$ and $(h_+, a_+)$:
\begin{equation}
	{\cal M}_2 = -\frac{\lambda_3 v^2}{4} \mmatrix{5}{\sqrt{3}}{\sqrt{3}}{3}\,.
\end{equation}
Inside $(h_-,a_-)$, the eigenvectors and eigenvalues are
\begin{eqnarray}
	h = -\frac{1}{2}h_- + \frac{\sqrt{3}}{2}a_-\,: &\quad& m_h^2 = -\frac{\lambda_3 v^2}{2}\nonumber\\
	H = \frac{\sqrt{3}}{2}h_- + \frac{1}{2}a_-\,: &\quad& m_H^2 = -\frac{3\lambda_3 v^2}{2}\,.
\end{eqnarray}
Note that $A,A'$ can be the minima only for $\lambda_3 < 0$. 
We observe the same mass hierarchy $m_H^2 = 3 m_h^2$ as in \eqs{B1-spectrum}{B1-spectrum-HA}.
The construction of the physical states in the space $(h_+, a_+)$ proceeds in the same way,
yielding $a$ and $A$. Thus, we again have two mass degenerate DM candidates, $h$ and $a$.

\subsection{DM stabilization by a group vs. by a single symmetry}

In the above discussion, we state that the DM candidates are stabilized by the residual symmetry group $S_3$
rather than by individual residual symmetries.
This language may be somewhat unconventional but it is fully justified. 
Suppose that, trying to stick to the conventional language,
we would say that $a$ is stabilized by the residual $\Z2$ symmetry $-g_2$, which flips the sign of $a$
but keeps $h$ unchanged.
Then we would not be able to identify the symmetry which stabilizes $h$ alone.
Alternatively, we could switch to the complex fields
\begin{equation}
	\chi = \frac{1}{\sqrt{2}}(h + i a)\,, \quad \chi^* = \frac{1}{\sqrt{2}}(h - i a)\label{psipsi}
\end{equation}
and say that they are stabilized by the residual $\Z3$ symmetry, under which they transform with the charges 
$+1$ and $-1$. Although such a representation may be useful to identify the semi-annihilation channel
$\psi \psi \to \psi^*+\mbox{SM}$, it still does not capture the full picture.
The full information on DM stabilization is provided by indicating that it is the group $S_3$
that protects the two real component scalar DM candidates.


Stabilization by a non-abelian group leads to another important feature, 
which prevents an immediate clash with experiment that other models 
with two mass-degenerate scalar DM candidates may possess.
Consider again the IDM, the 2HDM with the residual symmetry group $\Z2$.
In this model, $H$ is the lightest DM candidate stabilized by $\Z2$, 
and $A$ is a heavier state, which is also $\Z2$ odd.
Being members of the same doublet and possessing the opposite $CP$ parities, 
the two scalars can interact through the $ZHA$ interaction:
\begin{equation}
	{\cal L} \supset i \frac{g}{2 c_W}\, Z^\mu\left(H\partial_\mu A - A\partial_\mu H\right)\,.\label{ZHA}
\end{equation}
Then, in the mass-degenerate limit of $m_A = m_H$, we find that the DM particles
can elastically scatter off a nucleus via the transitions $H \to A$ and $A\to H$
mediated by the $Z$-exchange.
This scattering would lead to an unacceptable large direct detection (DD) signal
and is definitely ruled out by the negative results of DD searches.

One may wonder if the two degenerate DM candidates $h$ and $a$, which our model contains,  
can also couple to the $Z$ bosons in a similar way.
Fortunately, they do not; we observe only the $ZhA$ and $ZHa$ vertices but not $Zha$. 
The reason is again group-theoretical. Since $(h, a)$ form a doublet under $S_3$,
their hypothetical interaction terms with $Z$ 
must be of the type $\bm{2} \otimes \bm{2} = \bm{1} \oplus \bm{1}' \oplus \bm{2}$.
Since the conserved $S_3$ does not act on the $Z$, we must select the trivial singlet, 
which involves only the diagonal combinations.
But the $Zhh$ and $Zaa$ couplings are impossible by construction,
and the non-symmetric $Zha$ does not enter the trivial $S_3$ singlet.
We conclude that the residual group $S_3$ protects the model against the $Zha$ interactions.

Finally, stabilization of DM by $S_3$ differs from the case of $\Z2\times\Z2$ in that it allows 
for semi-annihilation channels such as $hh \to a+\mbox{SM}$.
This is due to the fact that the $S_3$ decomposition 
of $\bm{2} \otimes \bm{2}$, which corresponds to the $S_3$ quantum numbers of the initial state, 
contains a $\bm{2}$, which matches the final state quantum numbers.


\section{\texorpdfstring{Building realistic $\Sigma(36)$-based dark matter models}{Building realistic Sigma(36)-based dark matter models}\label{section-realistic}}

\subsection{\texorpdfstring{A toy model: exact $\Sigma(36)$ in the Yukawa sector}{A toy model: exact Sigma(36) in the Yukawa sector}\label{section-Yukawa-exact}}

In order to track the freeze-out evolution of DM density and explore the interplay of two DM candidates, 
one must complement the scalar interactions with a minimally realistic Yukawa sector.
In this Section, we describe our attempt to extend the full symmetry group $\Sigma(36)$ to the quark Yukawa.
The no-go theorem proved~\cite{Leurer:1992wg,GonzalezFelipe:2014mcf} makes it clear that,
even if such a sector can be constructed, the resulting quark properties 
cannot fully match the experimentally measured values and will unavoidably display pathologies 
such as massless quarks and insufficient mixing parameters.
Nevertheless, since the main goal of our study is to check the evolution of scalar DM stabilized 
by a non-abelian residual group, we find it appropriate to embrace this situation, considering it as a toy model,
provided the scalar DM candidates remain stable.

To set up notation, we write the quark Yukawa sector of the 3HDM as
\begin{equation}
	-{\cal L}_Y = \overline{Q}^0_{Li} \Gamma_{a,ij} \phi_a d_{Rj}^0 +
	\overline{Q}^0_{Li} \Delta_{a,ij} \tilde\phi_a u_{Rj}^0 + {\rm h.c.}\label{Yukawa-general}
\end{equation}
Here, the indices $i,j = 1,2,3$ refer to the quark generations, while $a = 1,2,3$ label the Higgs doublets. 
The superscript $0$ for the quark fields indicates that these are the starting quark fields;
when we pass to the physical quarks by diagonalizing their mass matrices, we will remove this superscript.

To construct the $\Sigma(36)$-invariant quark Yukawa sector, we need to assign the quark fields $Q_L$,
$d_R$, and $u_R$ to specific group representations, follow the rules for decomposition of their products,
and extract the trivial singlet.
To do this, we first briefly review the irreducible representations (irreps) of $\Sigma(36\varphi)$, 
reproducing some of the results of~\cite{Grimus:2010ak,Hagedorn:2013nra}.

The finite group $\Sigma(36\varphi)$ has four 1D irreps labeled $\bm{1}^{(p)}$, $p = 0, 1, 2, 3$,
which trivially represent the generators $a$ and $b$, $\rho_{1^{(p)}}(a) = \rho_{1^{(p)}}(b) = 1$,
and differ only by the representing value of the generator $d$: $\rho_{1^{(p)}}(d) = i^p$.
For shorthand notation, we will denote the trivial representation $\bm{1}^{(0)}$ as $\bm{1}$.
Next, it has four complex 3D irreps $\bm{3}^{(p)}$, whose $\rho_3(a)$ and $\rho_3(b)$ are given by \eq{Delta27-generators}
and whose $\rho_3(d)$ is as in \eq{Sigma36-generators} with the same extra factor $i^{p}$.
These are complemented by the four conjugate triplets $(\bm{3}^{(p)})^*$. 
Finally, the group possesses two real 4D irreps, $\bm{4}$ and $\bm{4}'$, which we will not encounter in our construction.
The squares of the irrep dimensions sum up to the group order as $\sum_i d_i^2 = 4\times 1 + 8\times 3^2 + 2\times 4^2 = 108$.

In Appendix~\ref{appendix-group}, we collect the irrep decomposition rules for the products of a triplet with
another triplet or an antitriplet. These rules follow a uniform pattern, which allows us, without losing generality, 
to assign the Higgs doublets to $\bm{3}^{(0)}$, which we shorten as $\bm{3}$.
Then, the two most relevant decomposition rules are
\begin{equation}
	\bm{3} \otimes \bm{3}^* = \bm{1} \oplus \bm{4} \oplus \bm{4}'\,, \qquad
	\bm{3} \otimes \bm{3} = \bm{3}^* \oplus (\bm{3}^{(1)})^* \oplus (\bm{3}^{(3)})^*\,.\label{irrep-products}
\end{equation}
It is instructive to compare this situation with the irreps of $\Delta(27)$ and their product decomposition,
which can be found, for example, in~\cite{deMedeirosVarzielas:2015amz}.
The group $\Delta(27)$ has nine distinct 1D irreps $\bm{1}_{ij}$,
which are labeled by the powers of $\omega$ in their $\rho_1(a)$ and $\rho_1(b)$,
and two complex 3D irreps, $\bm{3}$ and $\bm{3}^*$.
As a result, $\bm{3} \otimes \bm{3}^*$ decomposes into the sum of all nine distinct singlets.
This feature makes the group convenient for flavor model building. Indeed, assuming that $\phi_a$ and $Q_{Li}$
transform as $\bm{3}$, one can assign each individual $d_{Rj}$ to a different singlet, 
which introduces flexibility in building the Yukawa sector~\cite{Kalinowski:2021lvw}.
In the case of $\Sigma(36\varphi)$, this freedom is strongly reduced, because 
the product $\bm{3} \otimes \bm{3}^*$ involves now only one singlet. 

In Appendix~\ref{appendix-group}, we list all classes of irrep assignments for the quark fields.
Although the Yukawa matrices $\Gamma_a$ and $\Delta_a$ depend on cases,
the consequences are similar: in each sector, we obtain massless or mass degenerate quarks and insufficient mixing.
For example, in the case where $Q_L$ are trivial singlets, $d_R$ transform as $\bm{3}^*$, 
and $u_R$ transform as $\bm{3}$, 
the three Yukawa matrices $\Gamma_a$ take the following form:
\begin{equation}
	\Gamma_1 = \mtrx{g_1&\cdot&\cdot \\ g_2&\cdot&\cdot \\ g_3&\cdot&\cdot}\,, \quad 
	\Gamma_2 = \mtrx{\cdot&g_1&\cdot \\ \cdot&g_2&\cdot \\ \cdot&g_3&\cdot}\,, \quad 
	\Gamma_3 = \mtrx{\cdot&\cdot&g_1 \\ \cdot&\cdot&g_2 \\ \cdot&\cdot&g_3}\,,\label{case1333-Gamma}
\end{equation}
where dots correspond to the zero entries. The three independent coefficients $g_i$ are in general complex.
In the up-quark sector, the matrices $\Delta_a$ have the same structure,
bearing their own parameters $d_i$. The quark mass matrices become rank-1 matrices: 
$(M_d)_{ij} = g_i v_j/\sqrt{2}$, $(M_u)_{ij} = d_i v^*_j/\sqrt{2}$.
Diagonalizing these matrices, we find two generations of massless quarks in the down and up-quark sectors
and one generation of massive ones, with $m_b^2 = v^2 |\vec g|^2/2$ and $m_t^2 = v^2 |\vec d|^2/2$.
These masses do not depend on the vev alignment.

Since the mass matrices can be diagonalized analytically, 
we write the quark rotation matrices explicitly, insert them back in the Yukawa Lagrangian,
and, for a each choice of the vev alignment, establish how physical scalars interact with quark pairs.  
Taking the vev alignment $B_1$ as a reference,
we find the following interaction terms for the DM candidates $h$ and $a$:
\begin{equation}
	-{\cal L}_Y \supset \frac{m_b}{v}\, \bar b_L \left[(d_R + s_R) h + i (d_R - s_R) a \right] 
	+ \frac{m_t}{v}\, \bar t_L \left[(u_R + c_R) h - i (u_R - c_R) a \right] 
	+ {\rm h.c.}\label{h-a-quark}
\end{equation}
We arrive at a very concrete conclusion:
the scalars $h$ and $a$, which we considered as DM candidates in the previous Section,
cannot be stable as they unavoidable decay to quark pairs with unsuppressed couplings.

We checked that this key result remains valid for all other vev alignments as well as for all 
the irrep choices that can lead to $\Sigma(36)$-symmetric Yukawa sectors, even though
the patterns of the Yukawa matrices and the Higgs-quark couplings can be distinct.
For completeness, in Appendix~\ref{appendix-Yukawa-exact} we provide more details 
for each case. 

In summary, when building an exactly $\Sigma(36)$-invariant Yukawa sector,
there is no way to avoid tree-level decays of the anticipated DM candidates to quark pairs.
These scalars are intrinsically unstable and cannot play the role of dark matter.
The main obstacle is that quarks also carry the conserved quantum numbers with respect 
to the same group $S_3$ that was used to stabilize the scalars $h$ and $a$.
Thus, these scalars are no longer forbidden to decay into quark pairs.

\subsection{\texorpdfstring{A realistic model: $S_3$-symmetric Yukawa sector and softly broken $\Sigma(36)$}{A realistic model: S3-symmetric Yukawa sector and softly broken Sigma(36)}\label{section-Yukawa-S3}}

Since the exact $\Sigma(36)$ in the Yukawa sector leads not only to the pathological quark sector
but also to unavoidable decays of the anticipated DM candidates,
we need to relax our assumptions. 
We can construct a realistic model using the same $\Sigma(36)$-invariant scalar potential \eqref{Vexact} 
and complementing it with a Yukawa sector that is invariant not under the full $\Sigma(36)$
but under a $S_3$ subgroup.
Moreover, we require this $S_3$ to be exactly the same residual $S_3$ that 
remains unbroken at the minimum.

The specific example we consider is the vev alignment $B_1 = (1, 0, 0)$,
with the residual symmetry group $S_3$ given in \eq{residual-B1}.
The doublet $\phi_1$ is now the trivial singlet of $S_3$, while $\phi_2$ and $\phi_3$ transform
as a doublet irreducible representation.
In the Yukawa sector, we assume that all quarks transform trivially under $S_3$, 
so that they couple only to the first doublet with the SM Yukawa matrices. 
In this way, the doublets $\phi_2$ and $\phi_3$ become truly inert,
and their lightest states are indeed stable.

Of course, the full $\Sigma(36)$ is no longer a conserved symmetry group of the entire Lagrangian.
As a result, $\Sigma(36)$-breaking terms will leak, via quark loops, from the Yukawa sector back into the scalar sector.
Instead of calculating these terms, 
we take a different strategy and add soft breaking terms to the scalar potential 
that violate the full $\Sigma(36)$ but preserve $S_3$.
Various soft breaking options for the $\Sigma(36)$ 3HDM were discussed 
in~\cite{deMedeirosVarzielas:2021zqs,Yang:2024bys}. If we fix $v$ and $m_h$, 
then there is only one type of $S_3$-preserving soft breaking terms,
\begin{equation}
	V_{\scr{soft}} = \mu^2(\phi_2^\dagger\phi_2 + \phi_3^\dagger\phi_3)\,,\label{soft}
\end{equation}
which must be added to the potential $V_0$ given in \eq{Vexact}.
The requirement that the vev alignment $B_1$ is the global minimum corresponds to $\mu^2 > 0$.
Since the symmetry group $S_3$ is unbroken,
we still have the two DM candidates $h$ and $a$ with the bosonic interactions
coming from the scalar potential and the kinetic terms.
Compared to Eqs.~\eqref{B1-spectrum-charged}--\eqref{B1-spectrum-HA}, the soft breaking 
term leads to a uniform shift 
of the masses of all the scalars from the second and third doublets
keeping the physical scalar doublet degenerate:
\begin{eqnarray}
	(H_2^+, H_3^+): \,\, m_{H^+}^2 = \mu^2 + \frac{\lambda_2 v^2}{2}\,,\quad 
	(h, a): \,\, m_{h}^2 = \mu^2 + \frac{\lambda_3 v^2}{2}\,,\quad
	(H, A): \,\, m_{H}^2 = \mu^2 + \frac{3 \lambda_3 v^2}{2}\,.\quad \label{B1-spectrum-soft}
\end{eqnarray}
We then choose $m_h$, $m_{H^+} > m_h$, and $\mu^2$ as the three independent free parameters\footnote{Another possible choice
of the three free parameters is $m_h$, $m_H$, and $m_{H^+} > m_h$.}
and express the quartic couplings $\lambda_2$ and $\lambda_3$ by inverting the first two relations
from \eq{B1-spectrum-soft}, while the heavy inert Higgs mass is given by 
$m_H^2 = 3 m_h^2 - 2\mu^2$.
Note that all three quantities $\mu^2$, $\lambda_2$, $\lambda_3$ must be positive for the global minimum to reside at $B_1$.
Also, due to the present of $\mu^2$, the model now has a well-defined decoupling limit.

We still call this model the $\Sigma(36)$-based DM model.
Although the exact symmetry group is $S_3$, the quartic potential is the same as in the $\Sigma(36)$-symmetric 3HDM,
which constrains in a significant way the types of interactions and imposes correlations among the couplings.
This model ``inherits'' certain features from the $\Sigma(36)$-symmetric 3HDM~\cite{deMedeirosVarzielas:2021zqs},
and this is why we distinguish it from a generic $S_3$-based 3HDM DM models~\cite{Khater:2021wcx,Kuncinas:2022whn}.


\section{\texorpdfstring{Dark matter properties in the $\Sigma(36)$-based 3HDM}{Dark matter properties in the Sigma(36)-based 3HDM}\label{section-DM}}

\subsection{Constraints on the model and comparison with the IDM}

We implemented the $\Sigma(36)$-based DM model just described in \texttt{micrOMEGAs 6.0.5}~\cite{Alguero:2023zol}, 
which allowed us to explore the relic density and direct detection (DD) signals
as functions of the free parameters of the model.
Note that the model features an exact scalar alignment limit, which allows us to evade most of the phenomenological
constrains on the SM-like Higgs measurements.
We must, however, take into account the important constraint on the model coming from the SM-like Higgs decay to the DM candidates, 
which exists for sufficiently light $h$ and $a$.
The negative results of the LHC searches for an invisible Higgs decay~\cite{ATLAS:2023tkt} lead 
to the upper limit on its branching ratio, $B_{\scr{inv}} < 0.107$,
which translates into $\Gamma(\hSM \to \mbox{inv.}) < 0.42$~MeV.

To discuss the resulting constraints on the free parameters of our model,
we find it instructive to compare it with the IDM in the notation of~\cite{Ilnicka:2015jba,Belyaev:2016lok,Kalinowski:2018ylg}.
The scalar potential of the IDM contains seven free parameters. After fixing $v$ and $\mSM$,
we still have five real parameters to adjust, which affect the IDM phenomenology.
Among them, the key role is played by the trilinear $\hSM$-DM-DM
coupling divided by $v$, which is traditionally labeled as $\lambda_{345}$. This coefficient sets the magnitude
of several important processes, such as DM annihilation through the $s$-channel Higgs resonance and the invisible Higgs decay to light DM candidates.
It also contributes to the quartic vertices $hhW^+_LW^-_L$ and $hhZ_LZ_L$ to be discussed below.
The key feature is that, within the IDM, this all-important $\lambda_{345}$ is an independent parameter, 
not directly related to the scalar masses. 
It can be chosen sufficiently small, $|\lambda_{345}| \lsim 0.04$, 
so that, for the light DM candidates, the Higgs invisible decay width stays suppressed.

However, in our model, the coefficient in the vertices $\hSM hh$ and $\hSM aa$ is not an independent parameter but is expressed
via masses of the $\hSM$ and the SM candidates:
\begin{equation}
	\hSM hh,\ \hSM aa: \quad \bar\lambda v \equiv (2\lambda_1 + \lambda_3)v = \frac{\mSM^2 + m_H^2 - m_h^2}{v}
	= \frac{\mSM^2 + 2 m_h^2 - 2\mu^2}{v}\,,\label{hh-aa}
\end{equation}
which is never too small. It is at least as large as $v/4$ and grows further as the inert sector mass splitting increases.
This coefficient leads to a huge invisible decay width of the SM-like Higgs 
\begin{equation}
	\Gamma(\hSM \to hh, aa) = \frac{(\mSM^2 + 2 m_h^2 - 2\mu^2)^2}{16\pi \mSM v^2} \beta\,, \quad
	\mbox{where}\ \beta = \sqrt{1 - \frac{4 m_h^2}{\mSM^2}}\,.\label{invisible}
\end{equation}
For $m_h^2 = \mu^2$, it yields $640\,\mbox{MeV}\cdot \beta$
and exceeds the experimental upper limit by orders of magnitude.
Even the nominal threshold $\mSM = 2 m_h$ is excluded: a slightly off-shell Higgs boson 
with the invariant mass exceeding the nominal value $\mSM$ by a few $\Gamma_{\scr{SM}}$ would already generate an unacceptably large signal.
This analysis implies that in our model, unlike in the IDM, the light DM region $m_h \le \mSM/2$ is altogether ruled out.

\begin{figure}[htb]
	\centering
	\includegraphics[height=6cm]{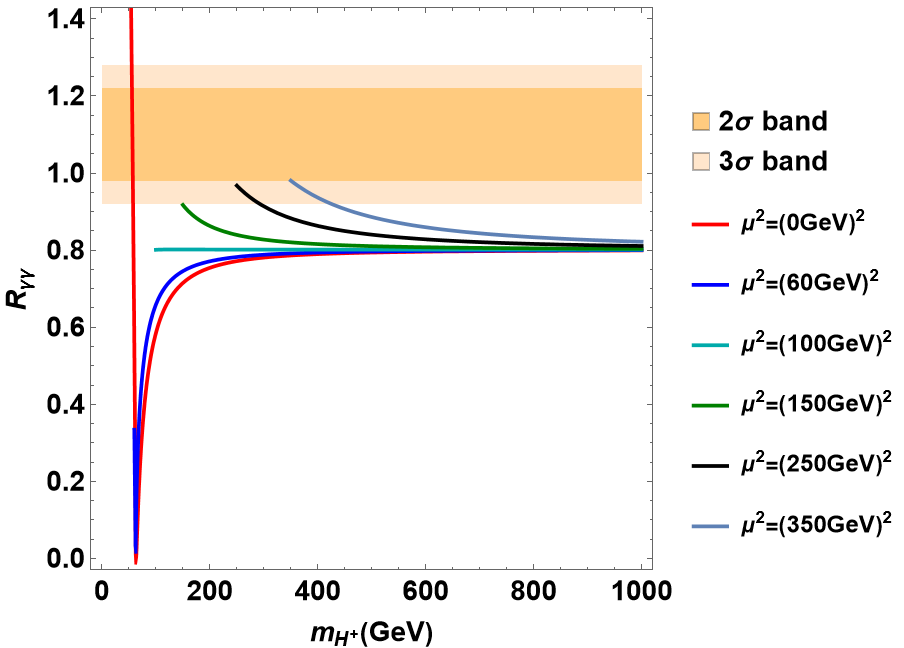}
	\caption{The ratio $R_{\gamma\gamma}$ defined in \eq{Rgamgam} as a function of he charged Higgs mass for several values of $\mu^2$ compared with 
		the LHC results~\cite{ParticleDataGroup:2024cfk} for the $2\sigma$ and $3\sigma$ bands around the central value.}
	\label{fig-Rgamgam}
\end{figure}

The presence of inert scalars also modifies the loop-induced decays of the SM-like Higgs, in particular, $\hSM \to \gamma\gamma$. 
Let us denote the modification of this decay width with respect to the SM prediction as
\begin{equation}
	R_{\gamma\gamma} = \frac{\Gamma(\hSM \to \gamma\gamma)}{\Gamma(\hSM \to \gamma\gamma)^{\mbox{\scriptsize SM}}}\,.\label{Rgamgam}
\end{equation}
The expression for this decay width can be recovered from the generic 2HDM result~\cite{Djouadi:2005gj}:
\begin{eqnarray}
	\Gamma(\hSM \to \gamma\gamma)=\frac{G_F\alpha^2 \mSM^3}{128\sqrt{2}\pi^3}\bigg | 
	\sum_f N_c Q_f^2 A_{1/2}(\tau_f) + A_1(\tau_W)
	+2\times \frac{\mSM^2 + 2 m_{H^+}^2 - 2\mu^2}{2 m_{H^+}^2}A_0(\tau_+)\bigg |^2,\label{gamgam}
\end{eqnarray}
while for the SM expression we simply omit the last contribution.
The factor 2 in the last term shows that both $H_2^+$ and $H_3^+$ give equal contributions.
Following~\cite{Djouadi:2005gj}, for a particle with mass $m_i$, we define the mass ratio as $\tau_i = \mSM^2/(4m_i^2)$
and use the formfactors 
\begin{eqnarray}
	A_0(\tau) = -\frac{\tau-f(\tau)}{\tau^{2}},\quad
	A_{1/2}(\tau) = 2\frac{\tau+(\tau-1)f(\tau)}{\tau^{2}},\quad
	A_1(\tau) = -\frac{2\tau^2+3\tau+3(2\tau-1)f(\tau)}{\tau^{2}}\,,
\end{eqnarray}
where
\begin{equation}
	f(\tau)= \left\{\arcsin^2\sqrt{\tau} \ \ \mbox{for $\tau\le 1$}, \quad
		-\frac{1}{4}\left[\log\left(\frac{\displaystyle 1+\sqrt{1-\tau^{-1}}}{ \displaystyle  1-\sqrt{1-\tau^{-1}}}\right)-i\pi\right]^2  
		\ \mbox{for $\tau > 1$} \right\}.
\end{equation}
The structure of the charged Higgs contribution closely resembles the IDM counterpart~\cite{Cao:2007rm,Arhrib:2012ia,Swiezewska:2012eh,Krawczyk:2013jta}.
However, unlike in the IDM, the sign of the coefficient is fixed, which leads to $R_{\gamma\gamma} < 1$ for a heavy charged Higgs. 

The coefficient in the last term of \eq{gamgam} makes it clear that, without the soft breaking terms, 
the contribution of the charged Higgses is not negligible. 
We find that, for $\mu^2 = 0$ and heavy charged Higgses, $R_{\gamma\gamma}$ is as low as $0.8$, which is already ruled out 
by the LHC measurements~\cite{ParticleDataGroup:2024cfk}: $R^{\textrm{exp.}}_{\gamma\gamma} = 1.10 \pm 0.06$.
However, a non-zero $\mu^2$ reduces the tension, provided the charged Higgs mass stays close to $\mu$.
In Fig.~\ref{fig-Rgamgam}, we show the behavior of $R_{\gamma\gamma}$ as a function of $m_{H^+}^2$
for several values of $\mu^2$.
As we can see, the $R_{\gamma\gamma}$ curves can reach $0.92$, which lies within $3\sigma$ of the central experimental value,
only for  $\mu^2 > (150~{\rm GeV})^2$, and can exceed $0.98$ ($2\sigma$ band) only for $\mu^2 > (350~{\rm GeV})^2$.

We arrive at the conclusion that the charged Higgs contributions to $\hSM \to \gamma\gamma$ are very sizable
and, in the absence of any additional contribution to compensate their effect,
rule out the $\Sigma(36)$ symmetric scalar sector. 
The only way to bring $\hSM \to \gamma\gamma$ back within experimental limits is to assume a sufficiently large soft breaking term.

\subsection{DM properties: no soft breaking}

\begin{figure}[htb]
	\centering
	\includegraphics[height=5.2cm]{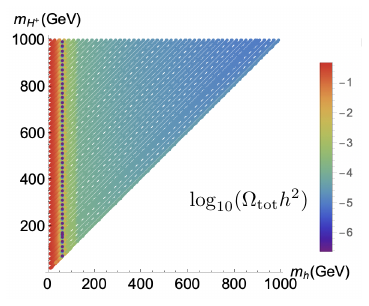}
	\hfill
	\includegraphics[height=5cm]{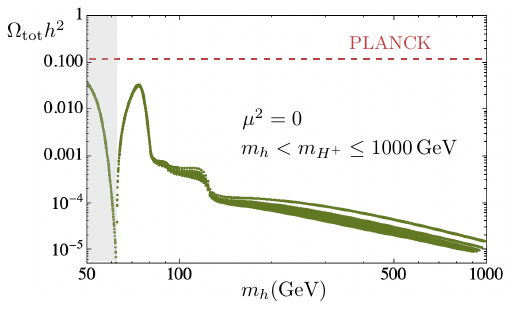}
	\caption{Combined relic density of $h$ and $a$ in the 3HDM with an exact $\Sigma(36)$. 
		Left: the relic density encoded in color on the parameter plane $(m_h, m_{H^+})$.
	Right: the relic density as a function of $m_h$ for several values of $m_{H^+}$.}
	\label{fig-exact-relic}
\end{figure}

Keeping the above results in mind, we turn now to the study of the relic density as a function of the free parameters.
We begin the presentation of our results 
with the case of the $\Sigma(36)$-symmetric scalar sector, that is, the model without soft breaking terms, $\mu^2 = 0$.
Although the measurements of the two photon decay width effectively rule out this scenario,
we still want to see whether additional problems arise from the DM sector alone.

When exploring the relic density evolution, we always observe that $h$ and $a$ give equal contributions.
This is of course expected due to the symmetry that links $h$ and $a$, the two components of the $S_3$ doublet.
Thus, when showing the values for $\Omega h^2$, we will always present the sum of their contributions.

In Fig.~\ref{fig-exact-relic}, we show the resulting relic density $\Omega h^2$ as a function of the two remaining free parameters
$m_h$ and $m_{H^+}$. The left plot shows the results of a general scan, with the relic density encoded in color,
while the left plot reveals finer details of $\Omega h^2$ as a function of $m_h$ for several representative values of $m_{H^+}$
that go up to 1 TeV.
The shaded region in the right plot corresponds to $m_h < \mSM/2$ and is excluded by the invisible decay width.

A salient feature of these plots is that, for $m_h > 45~{\rm GeV}$, the relic density is always
below the Planck result~\cite{Planck:2018vyg}: $(\Omega h^2)_{\scr{Planck}} = 0.1200 \pm 0.0012$.
This is due to the rather large annihilation, coannihilation, and semi-annihilation cross sections.
The only region where the predicted relic density matches the Planck result is the narrow band around $m_h \approx 40~{\rm GeV}$,
but this low mass region is already excluded by the invisible Higgs decay constraint.

The low relic density in the high-mass region can also be understood by comparing our model with the IDM.
In the IDM with heavy DM candidates, annihilation mainly proceeds into the longitudinally polarized $W^+W^-$ and $ZZ$ pairs
~\cite{Belyaev:2016lok}, which are dominated by the quartic vertices 
\begin{equation}
	hhW^+_LW^-_L\mbox{(IDM):}\quad \lambda_{345} + \frac{2(m_H^2-m_h^2)}{v^2}\,, \qquad
	hhZ_LZ_L\mbox{(IDM):}\quad \lambda_{345} + \frac{2(m_{H^+}^2-m_h^2)}{v^2}\,.\label{IDM-quartic}
\end{equation} 
As mentioned earlier, $\lambda_{345}$ is an independent parameter and can be chosen small. 
If the mass splitting in the inert sector is also small, the annihilation cross section is suppressed, 
and the relic density can match the observed value.
This is the mechanism behind the high-mass region of the IDM, which still remains viable.

However, in the $\Sigma(36)$ model, these couplings are
\begin{equation}
	hhZ_LZ_L: \quad 2\lambda_1 + 3\lambda_3 = \bar\lambda + \frac{2(m_H^2-m_h^2)}{v^2}\,,\qquad
	hhW^+_LW^-_L: \quad 2\lambda_1 + \lambda_2 = \bar\lambda + \frac{2(m_{H^+}^2-m_h^2)}{v^2}\,,\label{hhVV}
\end{equation}
where $\bar\lambda$ is given in \eq{hh-aa}.
Both expressions are never small. The minimal value for the $hhZ_LZ_L$ coupling in the exactly $\Sigma(36)$-symmetric scalar sector 
is $5\mSM^2/2v^2 \approx 0.6$ and further grows approximately as $m_h^2$.
This explains why, for large DM masses, we always obtain the relic density significantly below the observed value.

We conclude that the model can account only for a small fraction of total dark matter.
Although it still leaves open the question of the origin of the dominant DM component, 
this sub-dominant DM scenario is, in principle, viable.

Next, we confront this model with the direct detection limits.
Let us denote by $\xi_\Omega = \Omega/\Omega_{\scr{Planck}}$ the fraction of the total relic density constituted by $h$ and $a$. 
Then, we can compute the DD signal expected in this case, taking into account the sub-dominant DM scenario we work in,
and compare it with the experimental upper limits on the spin-independent cross section $\sigma_{SI}$ established by the DD searches.
Instead of up-scaling the published upper limits on $\sigma_{\scr{SI}}$ for such a comparison, 
we downscale our predicted cross section by defining 
\begin{equation}
\hat\sigma_{SI} = \sigma_{SI}\cdot \xi_\Omega\,,\label{hat-sigma}
\end{equation}
and directly compare it with the published results.

\begin{figure}[htb]
	\centering
	\includegraphics[height=6cm]{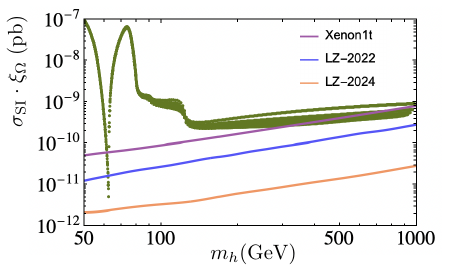}
	\caption{The rescaled cross section for the direct detection signal $\hat\sigma_{SI} = \sigma_{SI}\cdot \xi_\Omega$ obtained in our model,
		compared with the upper limits from recent DD experiments~\cite{XENON:2017vdw,LZ:2022lsv,LZ:2024zvo}.}
	\label{fig-exact-DD}
\end{figure}

In Fig.~\ref{fig-exact-DD} we present this comparison. Even with the scaling down taken into account, the model predicts rather large DD signals. 
This is unavoidable because the DM-Higgs interaction is governed by the same coupling $\bar\lambda$ as in \eq{hh-aa}.
Still, there remains a sizable part of the random scan which passes the limits obtained by the Xenon1T experiment.
However, these predictions are in a clear conflict with the upper limits announced by the LZ experiment 
both in 2022~\cite{LZ:2022lsv} and especially in 2024~\cite{LZ:2024zvo}, definitely ruling out the model we consider.

The bottom line is: with the latest LZ-2024 result, the scenario featuring two DM candidates arising 
within the 3HDM with an exact $\Sigma(36)$ symmetry in the scalar sector is ruled out.

\subsection{DM properties with soft breaking terms}\label{subsection-DM-soft}

\begin{figure}[htb]
	\centering
	\includegraphics[height=4.5cm]{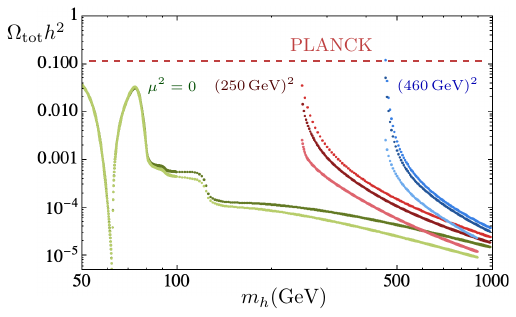}
	\hfill
	\includegraphics[height=4.5cm]{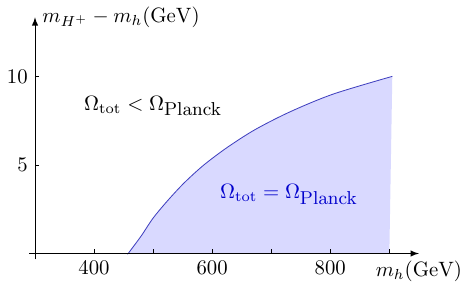}
	\caption{Left: The total relic density in the 3HDM for three values of the soft breaking parameter $\mu^2$ and
		several options for $m_{H^+} - m_h$, see the main text. 
		Right: The region in the $(m_{H^+} - m_h, \,m_h)$ parameter space 
		in which the relic density computed with a suitable $\mu^2$ can match the Planck results.}
	\label{fig-soft-relic}
\end{figure}

%

Next, we introduce the soft breaking terms \eqref{soft} and explore the consequences for the DM observables.
In Fig.~\ref{fig-soft-relic}, left, we show how the DM relic density depends on $\mu^2 > 0$.
The light and dark shaded green points correspond to the exactly $\Sigma(36)$-symmetric scalar sector 
with $m_{H^+} - m_h = 10~{\rm GeV}$ and $100~{\rm GeV}$;
they agree with Fig.~\ref{fig-exact-relic}, right.
The red and blue families of points correspond to $\mu^2 = (250~{\rm GeV})^2$ and $(460~{\rm GeV})^2$,
respectively. In both cases, the three sequences of points, from top to bottom, refer to 
$m_{H^+} - m_h = 0.1~{\rm GeV}$, $10~{\rm GeV}$, and $100~{\rm GeV}$. 
As expected, if $m_h$ is fixed but $\mu$ increases, the annihilation cross section goes down, 
and as a result the relic density rises.

Fig.~\ref{fig-soft-relic}, left, indicates that, for $\mu^2 > (460~{\rm GeV})^2$ and with a small mass splitting,
a region of parameters opens up, in which the calculated relic density matches the Planck measurements.
For a better visualization, we show this region in Fig.~\ref{fig-soft-relic}, right, on the 
plane of the mass splitting $m_{H^+} - m_h$ vs. $m_h$.
For any choice of masses $m_{H^+}$ and $m_h$ that fall inside the blue region, 
it is possible to find a suitable $\mu^2$ that leads to the total relic density
that matches the Planck result. For masses $m_{H^+}$ and $m_h$ outside this region,
one always obtains subdominant scalar DM for any choice of $\mu^2$.
It is interesting to note that this borderline value of $\mu$ 
is very similar to the value of the DM mass in the IDM, $m_{DM} \sim 500~{\rm GeV}$, above which the so-called heavy mass region
with a matching relic density opens up~\cite{Arhrib:2013ela,Ilnicka:2015jba,Belyaev:2016lok}.

\begin{figure}[htb]
	\centering
	\includegraphics[height=4.5cm]{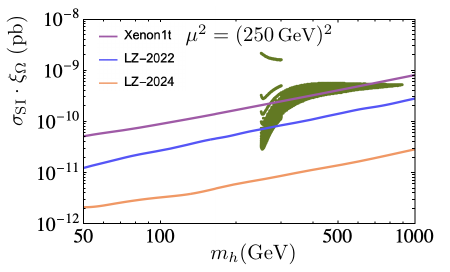}
	\hfill
	\includegraphics[height=4.5cm]{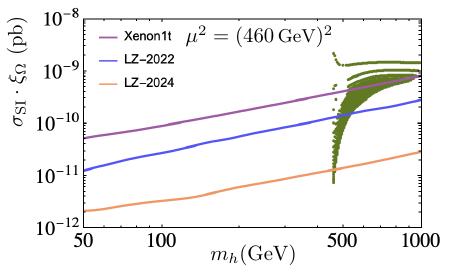}
	\caption{The rescaled cross section for the direct detection signal $\hat\sigma_{SI} = \sigma_{SI}\cdot \xi_\Omega$ for 
		$\mu^2 = (250~{\rm GeV})^2$ and $(460~{\rm GeV})^2$,
		compared with the experimental upper limits~\cite{XENON:2017vdw,LZ:2022lsv,LZ:2024zvo}.}
	\label{fig-soft-DD}
\end{figure}

In Fig.~\ref{fig-soft-DD}, we show how the rescaled cross section for the direct detection signal $\hat\sigma_{SI}$ 
changes as $\mu^2$ grows. We show the predictions for the same soft breaking parameter values:
$\mu^2 = (250~{\rm GeV})^2$ and $(460~{\rm GeV})^2$.
For DM mass well above $\mu$, the rescaled DD signals still hover above the upper limits from LZ,
similarly to the case $\mu^2=0$, and are not significantly suppressed with $\mu^2$ variation.
This behavior can be understood by noticing once more that the same $\bar\lambda$, which is never too small, 
both sets the DM scattering process as in \eq{hh-aa} and drives the annihilation cross section, \eq{hhVV}.
If we fix a large $m_h$ and increase $\mu^2$, the DD cross section goes down, but the relic density fraction goes up,
and the two effects approximately compensate each other in the rescaled DD signal.

The only way to keep DD cross section low and, at the same time, further suppress relic density
is to keep $\bar\lambda$ small and increase the charged Higgs mass.
This is what we observe for $m_h$ very close to $\mu$. In particular, we find that the model with 
$\mu^2 = (460~{\rm GeV})^2$ is, in principle, capable of bringing the DD signal
below not only LZ-2022 but also the latest LZ-2024 upper limits~\cite{LZ:2024zvo}.
However, this is done at the expense of a large $\lambda_2$;
we checked that the lowest lying points in Fig.~\eqref{fig-soft-DD}, right, correspond to 
$m_{H^+} > 720~{\rm GeV}$, which translates into $\lambda_2$ above 10.
Also, the large mass splitting between $H, A$ and $H_{2,3}^+$, which arises in this case,
drives the electroweak precision observables far beyond the experimental limits.

We conclude that the tight correlations arising in our model even in the softly broken regime, 
once more, lead the predictions to severe clashes with theoretical and experimental constraints.
It may prove useful to extend the parameter $\mu^2$ to even larger values in a quest for a fully viable DM model.
But we believe that the present study clearly indicates the persistent problems
that arise in this and other similar DM models based on a multi-Higgs-doublet with very large finite symmetry groups.

\section{Summary}

In this Chapter we addressed the question whether scalar DM candidates stabilized by a non-abelian residual group 
is a viable option within multi-Higgs-doublet models. 
A conserved non-abelian group not only protects the multiplet of DM candidates against decay 
but also gives rise to additional DM evolution channels. 
In contrast to previous works, in which a non-abelian group protected dark sector, by construction, 
was nearly decoupled from the SM fields, 
we considered a multiplet of DM candidates emerging from additional inert electroweak scalar doublets 
that enter the same Higgs potential that leads to the SM-like Higgs boson. 
As the benchmark example, we used the $\Sigma(36)$-based 3HDM~\cite{Ivanov:2012ry,Ivanov:2012fp}
and took the advantage of its property that, for any choice of its free parameters,
the global minimum of the potential always preserves a non-abelian subgroup $S_3 \subset \Sigma(36)$ 
and therefore always stabilizes a pair of mass degenerate DM candidates~\cite{Ivanov:2014doa}.
 
	Due to very tight symmetry-based relations between the SM-like Higgs, the quark sector, and the dark sector,
we found that this idea repeatedly runs into conflict with observations.
\begin{itemize}
	\item 
The 3HDM Yukawa sector invariant under the exact $\Sigma(36)$
not only leads to non-physical quark properties but also destabilizes all the scalars,
rendering a realistic quark sector incompatible with DM candidates.
To avoid this problem, we relaxed our assumptions and built the Yukawa sector invariant not under the full $\Sigma(36)$
but under $S_3$, the same subgroup that remains unbroken after minimization.
	\item
The extra charged Higgses significantly reduce the $\hSM \to \gamma\gamma$ decay width, which conflicts the LHC measurements.
As the $\hSM$ coupling with DM is never suppressed, the unsuppressed invisible Higgs decay rules out this model
for $m_h < \mSM/2$.
	\item
If the scalar sector respects the full $\Sigma(36)$ symmetry, the scalar DM is unavoidably subdominant and, 
at the same time, enters in conflict with the direct detection results from the LZ experiment~\cite{LZ:2024zvo}. 
The bottom line is that the model with an exactly $\Sigma(36)$-symmetric scalar sector 
complemented with $S_3$-invariant Yukawa interactions is ruled out.
	\item
The soft breaking terms governed by the single new free parameter $\mu^2$ can relax the tension.
These soft breaking terms violate $\Sigma(36)$ but preserve the same $S_3$ that remains as a residual symmetry group at the minimum,
thus, supporting the same pair of DM candidates stabilized by the residual non-abelian group.
Still, the quartic potential inherits many features of the exact $\Sigma(36)$ 3HDM and remains very constrained. 
A moderately large $\mu > 350~{\rm GeV}$ can bring the $\hSM \to \gamma\gamma$ within the experimental limits.
For $\mu > 460~{\rm GeV}$, a high-mass, compressed-spectrum region opens up, in which the relic density predictions
can match the Planck result. However, the DD signals in this region are way too high, 
above not only LZ~\cite{LZ:2022lsv,LZ:2024zvo} but even Xenon-1T~\cite{XENON:2018voc} results.
	\item
Alternatively, there is a region for $\mu > 460~{\rm GeV}$ with a very sub-dominant DM scenario, in which the DD signal
dives below the latest LZ limits~\cite{LZ:2024zvo}. However, it requires a large charged vs. neutral Higgs mass splitting,
and an extreme value of $\lambda_2$, violating experimental and theoretical constraints.
\end{itemize}
The origin of these repeated conflicts is clear.
Our starting idea of a large finite global symmetry group strongly constraints the shape of the scalar sector.
Looking back at the Higgs potential \eqref{Vexact}, we see that it is the term 
$\lambda_1(\phi_1^\dagger \phi_1+ \phi_2^\dagger \phi_2+\phi_3^\dagger \phi_3)^2$
that forces the effective DM-SM coupling $\bar\lambda$ in \eq{hh-aa} to always remain
unsuppressed. It is this unsuppressed coupling that drives the large invisible Higgs decay for light DM,
the large contribution of the charged Higgs loops to $\hSM \to \gamma\gamma$ for heavy DM,
as well as the large DD signal.
Thus, the large DM effects are tightly connected with our starting assumption of a large symmetry group.

It may happen that extending our study to $\mu^2$ in excess of $1\,\mbox{TeV}^2$ would reveal a model that would be 
technically within the existing limits. But judging from the present experience, it will always be at the brink of exclusion
and never be a natural candidate.
But the main goal of this work was not to construct yet another DM model but 
to explore the limits of what multi-Higgs-doublet models can in principle accommodate.
We met this objective: through a concrete example, we learned several lessons 
that further highlight the conflicts between symmetry assumptions,
the LHC results, and DM properties that arise if we try to derive a DM multiplet stabilized by a non-abelian group
from the multi-Higgs-doublet model potential.
We believe that these lessons are of interest to the community and help clarify what can and what cannot 
be achieved with several Higgs doublets alone, without additional New Physics assumptions.

%% file: chapters/DarkMatter_multi.tex
\chapter{Two particle candidates for dark matter}
\label{chapter:Darkmatter_multi}
\hspace*{0.3cm}
Having attempted to construct a viable Dark Matter model in the previous Chapter, we now continue with a full phenomenological study of a model known to be viable in the literature~\cite{Belanger:2011ww,Aoki:2012ub,Bhattacharya:2016ysw,Hernandez-Sanchez:2020aop}. We consider the real $\Z2\times\Z2$ symmetric 3HDM of Chapter ~\ref{sec:3hdm_pot} in a situation where, upon spontaneous symmetry breaking,
both new fields do not develop a vacuum expectation value, allowing for two DM candidates. The model is motivated as a follow-up to the IDM, a model with one additional Higgs doublets odd under a $\Z2$ symmetry that leaves all the SM fields unchanged. 
The upshot of all theoretical, collider, astrophysical and cosmological constraints
is that the DM candidate can only have its mass restricted to two
regions; one region around $m_h/2$ and another region with mass above
around $500~{\rm GeV}$.
The exclusion of the intermediate region comes from an interplay between the requirements
from relic density and the constraints from direct detection (DD) experiments.
In addition, a variety of indirect detection (ID) constraints can arise
from DM annihilation into
photons~\cite{Modak:2015uda,Queiroz:2015utg,Garcia-Cely:2015khw}
cosmic rays~\cite{Nezri:2009jd},
or neutrinos~\cite{Agrawal:2008xz,Andreas:2009hj}.

The DM (intermediate) mass region can be extended in theories
that have more than one DM component~\cite{Boehm:2003ha};
the so-called multi-component DM models
~\cite{Zurek:2008qg,Batell:2009vb,Profumo:2009tb}.
This is due to the possibility that the various components contribute to the
relic density and also due to the new processes of
co-annihilation~\cite{Mizuta:1992qp,Ellis:1998kh},
DM conversion between sectors~\cite{Liu:2011aa},
and/or semi-annihilation~\cite{DEramo:2010keq}.
Examples involving exclusively new scalars include
models based on $\Z4$
~\cite{Belanger:2012vp,Belanger:2021lwd}
or the $\Z2\times\Z2$
~\cite{Belanger:2011ww,Aoki:2012ub,Bhattacharya:2016ysw,Hernandez-Sanchez:2020aop}.
An interesting feature of multi-component DM models is that they can be used
to explain putative anomalies in uncorrelated DM signals which require different
DM mass scales.
And they may yield signal detectable at the FCC; see for example
~\cite{Bhattacharya:2022wtr,Bhattacharya:2022qck}.

In this Chapter, we study a three Higgs doublet model (3HDM),
with a $\Z2\times\Z2$ symmetry, where two scalar doublets are inert.
This leads to two separate DM sectors and two natural DM candidates.
Whilst in models based on $\Z4$
~\cite{Belanger:2012vp,Belanger:2021lwd} the existence of two DM candidates
hinges on a suitable choice of masses such that decays of one sector
into the other are  kinematically forbidden. This is not the case here,
where such decays are symmetry forbidden.

In Sections~\ref{sec:vacua} and ~\ref{sec:mineq},
we discuss in detail all possible vacua, and the conditions guaranteeing that
the double inert vacuum is indeed the global minimum,
improving on the conditions in~\cite{Hernandez-Sanchez:2020aop}.
In Sections~\ref{sec:scan} we set up the parameter ranges which we
will use in our simulations.
This model has a number of interesting processes relevant for DM studies,
including DM conversion and co-annihilation, which we discuss in
section~\ref{sec:processes}.
The results of our scan, and their discussion and implications are presented
in Section~\ref{sec:results}.
We outline our conclusions in Section~\ref{sec:concl}.
For completeness,
we include in Appendix~\ref{app:masses} the formulas for the scalar masses
in the various vacua.

\section{The possible vacua}
\label{sec:vacua}

This Section is devoted to the identification of
the possible vacua of the 3HDM $\Z2\times\Z2$ model
and the criteria ensuring that the inert vacuum corresponds to the global minimum.
We start by describing the various vacua.
Then, we explain how numerical and analytical explorations show that 
not all vacua are contained in Ref.~\cite{Hernandez-Sanchez:2020aop}.

\subsection{Neutral vacua}
\label{subsec:neutral-vacua}

The most general neutral vacuum configuration may be parametrized as
\begin{equation}
  \label{eq:1n}
  \langle \phi_1 \rangle  =
  \begin{pmatrix}
    0\\
    v_1 e^{i\xi_1}
  \end{pmatrix},\quad
  \langle \phi_2 \rangle = 
  \begin{pmatrix}
    0\\
    v_2 e^{i\xi_2}
  \end{pmatrix},\quad
  \langle \phi_3 \rangle =
  \begin{pmatrix}
    0\\
    v_3
  \end{pmatrix} .
\end{equation}
Various of its distinct incarnations were studied in Ref.~\cite{Hernandez-Sanchez:2020aop}.
We follow their notation for the classification, which is shown in
the upper part of Table~\ref{tab:1}.
\begin{table}[htb]
  \centering
  \begin{tabular}{|c|c|c|c|}\hline
    Name &vevs &Symmetry& Properties\\
         &&of vacuum & \\\hline
   \texttt{EWs}&(0,0,0)&$\Z2\times\Z2^\prime$ &EW Symmetry\\\hline
  \texttt{2-Inert}&$(0,0,v_3)$&$\Z2\times\Z2^\prime$ &SM + 2 DM candidates\\\hline
    \texttt{DM1}&$(0,v_2,v_3)$&$\Z2$ &2HDM + 1 DM candidates\\\hline
\texttt{DM2}&$(v_1,0,v_3)$&$\Z2^\prime$ &2HDM + 1 DM candidates\\\hline
\texttt{F0DM1}&$(0,v_2,0)$&$\Z2$ &1 DM candidates + massless fermions
    \\\hline
\texttt{F0DM2}&$(v_1,0,0)$&$\Z2^\prime$ &1 DM candidates + massless fermions
    \\\hline
\texttt{F0DM0}&$(v_1,v_2,0)$&None &No DM candidate + massless fermions
    \\\hline
\texttt{N}&$(v_1,v_2,v_3)$&None &3HDM no DM candidate
    \\\hline
\texttt{sCPv}&$(v_1e^{i \xi_1},v_2e^{i \xi_2},v_3)$&None &Spontaneous
CP violation\\\hline
\multicolumn{4}{c}{ }\\\hline
\texttt{F0DM0'}&$(v_1,i v_2,0)$&None &No DM candidate + massless fermions
\\\hline
  \end{tabular}
  \caption{Possible neutral vacua. The top of the table has all vacua found in
    Ref.~\cite{Hernandez-Sanchez:2020aop}; the last line corresponds to a new vacuum. (See text for explanation.)}
  \label{tab:1}
\end{table}

We want to get the conditions where the \texttt{2-Inert} minimum lies below
all other neutral minima.
Ref.~\cite{Hernandez-Sanchez:2020aop} shows that \texttt{DM1} and \texttt{DM2}
are always above \texttt{2-Inert}.
As will be explained below, we found that guaranteeing that the \texttt{2-Inert} minimum lies below
the other minima on the first part of Table~\ref{tab:1} does \textit{not} guarantee that
it lies below our new minimum.
This is easy to see with the following argument.
The difference between the new \texttt{F0DM0'} case and the \texttt{F0MD0} case in
Ref.~\cite{Hernandez-Sanchez:2020aop} is that the former can be obtained from
the latter with the substitution $\phi_2 \rightarrow i \phi_2$.
As can be seen from \eqs{3hdmquadratic}{Z2Z2quartic},
this corresponds to $\lambda_{10}'' \rightarrow - \lambda_{10}''$
and
$\lambda_{12}'' \rightarrow - \lambda_{12}''$.

Now, in the \texttt{F0MD0} case, ensuring that \texttt{2-Inert} lies below involves
$\lambda_7 + 2 \lambda_{10}''$.
But, since the vev $(v_1, i v_ 2, 0)$ is allowed, and since it is obtainable
through $\lambda_{10}'' \rightarrow - \lambda_{10}''$, we must also
study $\lambda_7 - 2 \lambda_{10}''$.
This plausibility argument will be fully proved both analytically
and numerically below.

\subsection{Charge breaking vacua}
\label{subsec:CB-vacua}

In addition to normal vacua, where the photon is massless,
there are also charge breaking (CB) vacua,
and one must also guarantee stability against then.
Those discussed in Ref.~\cite{Hernandez-Sanchez:2020aop} can be found
in the upper part of Table~\ref{tab:2}.
\begin{table}[htb]
  \centering
  \begin{tabular}{|c|c|}\hline
    Name &vevs \\ \hline
   \texttt{CB1}&
$
\left( \begin{array}{c}
u_1\\
c_1
\end{array} \right)
\ \ 
\left( \begin{array}{c}
u_2\\
c_2
\end{array} \right)
\ \ 
\left( \begin{array}{c}
0\\
c_3
\end{array} \right)
$
 \\ \hline
  \texttt{CB2}&
$
\left( \begin{array}{c}
u_1\\
0
\end{array} \right)
\ \ 
\left( \begin{array}{c}
u_2\\
c_2
\end{array} \right)
\ \ 
\left( \begin{array}{c}
0\\
c_3
\end{array} \right)
$
 \\\hline
    \texttt{CB3}&
$
\left( \begin{array}{c}
u_1\\
c_1
\end{array} \right)
\ \ 
\left( \begin{array}{c}
u_2\\
0
\end{array} \right)
\ \ 
\left( \begin{array}{c}
0\\
c_3
\end{array} \right)
$
\\\hline
\texttt{CB4}&
$
\left( \begin{array}{c}
u_1\\
c_1
\end{array} \right)
\ \ 
\left( \begin{array}{c}
u_2\\
c_2
\end{array} \right)
\ \ 
\left( \begin{array}{c}
0\\
0
\end{array} \right)
$
 \\\hline
\texttt{CB5}&
$
\left( \begin{array}{c}
0\\
c_1
\end{array} \right)
\ \ 
\left( \begin{array}{c}
u_2\\
c_2
\end{array} \right)
\ \ 
\left( \begin{array}{c}
0\\
c_3
\end{array} \right)
$
 
    \\\hline
\texttt{CB6}&
$
\left( \begin{array}{c}
u_1\\
c_1
\end{array} \right)
\ \ 
\left( \begin{array}{c}
0\\
c_2
\end{array} \right)
\ \ 
\left( \begin{array}{c}
0\\
c_3
\end{array} \right)
$
 
    \\\hline
\texttt{CB7}&
$
\left( \begin{array}{c}
u_1\\
0
\end{array} \right)
\ \ 
\left( \begin{array}{c}
u_2\\
0
\end{array} \right)
\ \ 
\left( \begin{array}{c}
0\\
c_3
\end{array} \right)
$

    \\\hline
\texttt{CB8}&
$
\left( \begin{array}{c}
u_1\\
0
\end{array} \right)
\ \ 
\left( \begin{array}{c}
0\\
0
\end{array} \right)
\ \ 
\left( \begin{array}{c}
0\\
c_3
\end{array} \right)
$

    \\\hline
\texttt{CB9}&
$
\left( \begin{array}{c}
0\\
0
\end{array} \right)
\ \ 
\left( \begin{array}{c}
u_2\\
0
\end{array} \right)
\ \ 
\left( \begin{array}{c}
0\\
c_3
\end{array} \right)
$
 \\\hline
\multicolumn{2}{c}{ }\\\hline
\texttt{F0CB}&
$
\left( \begin{array}{c}
u_1\\
c_1
\end{array} \right)
\ \ 
\left( \begin{array}{c}
u_2\\
-\frac{u_1^* u_2}{c_1^*}
\end{array} \right)
\ \ 
\left( \begin{array}{c}
0\\
0
\end{array} \right)
$

\\\hline
  \end{tabular}
  \caption{Possible charge breaking (CB) vacua.
The top of the table has all vacua found in Ref.~\cite{Hernandez-Sanchez:2020aop};
the last line corresponds to a new vacuum. (See text for explanation.)
In all cases, the vacua shown explicitly are assumed to be non-vanishing and unrelated;
except on the last line, where the explicit relation $u_1^* u_2 + c_1^* c_2 = 0$ holds. }
  \label{tab:2}
\end{table}

Ref.~\cite{Hernandez-Sanchez:2020aop} studies the tapdole (stationarity) 
equations for CB vacua in their equations (3.34)-(3.38).
Typically, those equations yield two solutions for the quartic parameters $m_{11}^2$
and/or $m_{22}^2$.
And, forcing their equality gives a constraint on  
the quartic couplings which, when used in the value of the potential at the minimum,
imposes that the CB vacua 
\texttt{CB1}-\texttt{CB9} always lie above the \texttt{2-Inert} vacuum~\cite{Hernandez-Sanchez:2020aop}.

As we will show below, we have found both analytically and numerically
that there is, however, a further CB vacuum,
which we dub \texttt{F0CB},
corresponding to the last line
of Table~\ref{tab:2}.
In hindsight,
this arises because, under those very specific conditions
($c_3=0$ and $u_1^* u_2 + c_1^* c_2 = 0$),
both $m_{11}^2$ and $m_{22}^2$ are unequivocally determined from the tadpole
equations, with no further constraint on the quartic parameters.
Again, the existence of this new and independent vacuum will be proved
numerically and analytically below.

\subsection{Numerical  minimization}
\label{subsec:numerical}

The most general vacuum to be compared with the  \texttt{2-Inert} minimum
may be parametrized
as~\cite{Faro:2019vcd}
\begin{equation}
  \label{eq:vacua_param}
  \phi_1=\sqrt{r_1}
  \begin{pmatrix}
    \sin\alpha_1\\
    \cos\alpha_1\ e^{i \beta_1}
  \end{pmatrix},\quad
  \phi_2= \sqrt{r_2}
  e^{i \gamma}
  \begin{pmatrix}
    \sin\alpha_2\\
    \cos\alpha_2\ e^{i \beta_2}
  \end{pmatrix},\quad
  \phi_3=\sqrt{r_3}
  \begin{pmatrix}
    0\\
    1
  \end{pmatrix} .
\end{equation}
The main result of Ref.~\cite{Hernandez-Sanchez:2020aop}
is that they only worry about the \texttt{F0DM1}, \texttt{F0DM2} and
\texttt{F0DM0} cases to make sure that the \texttt{2-Inert} is the
global minimum.  They give conditions on the parameters of the
potential, but for \texttt{F0DM0} it is simpler to make a numerical
comparison of the value of the potential in the two situations.

To cross check their results we used the following method.
We start by choosing some point in the $(m^2_{ii}, \lambda_k)$
parameter space of \eqs{3hdmquadratic}{Z2Z2quartic}.
We ensure that the point satisfies the BFB conditions explained in
Section~\ref{sec:BFB-Z2xZ2}.
Then, we impose
\begin{equation}
  \label{eq:3}
  V_{\texttt{2Inert}} < V_{\texttt{X}}\, ,
\end{equation}
for all the $\texttt{X}$ mentioned in Ref.~\cite{Hernandez-Sanchez:2020aop}.
Next we use the multi-step procedure explained in
Ref.~\cite{Bree:2023ojl}, utilizing CERN's Minuit
library~\cite{James:1975dr} in order to minimize the potential.
In this method,
we minimize the potential starting from a large number of random
initial conditions for the parameters in \eq{eq:vacua_param}.
We found that out of 10000 points satisfying BFB and \eq{eq:3},
there were still 161 (1.6\%) that had a lower minima than
$V_{\texttt{2Inert}}$.
Note that this is not a numerical precision problem,
because for those cases where $V_{\texttt{2Inert}}$ was indeed the global
minimum we got precisely the value of the potential, and also
$r_1=r_2=0$ in the notation of \eq{eq:vacua_param}.
After adding the constraint
\begin{equation}
  \label{eq:f0dm0_cond}
  V_{\texttt{2Inert}} < V_{\texttt{F0DM0'}}\, ,
\end{equation}
we still find points whose minimum lies below \texttt{2Inert}.
But, after adding the constraint
\begin{equation}
  \label{eq:8888}
  V_{\texttt{2Inert}} < V_{\texttt{F0CB}}\, ,
\end{equation}
our extensive minimization procedure no longer finds any
global minimum below $V_{\texttt{2Inert}}$.

It turns out that all the new points correspond to $r_3=0$ and,
thus, we study next that case analytically in detail.

\section{Solving the minimization equations when \texorpdfstring{$r_3=0$}{}}
\label{sec:mineq}

In this Section we perform an analytical study of the minimization in the situation where we have found new minima,
which agrees with the numerical results discussed previously.

\subsection{The potential for \texorpdfstring{$r_3=0$}{}}

\begin{figure}[htpb!]
\centering
\includegraphics[width = 0.4\textwidth]{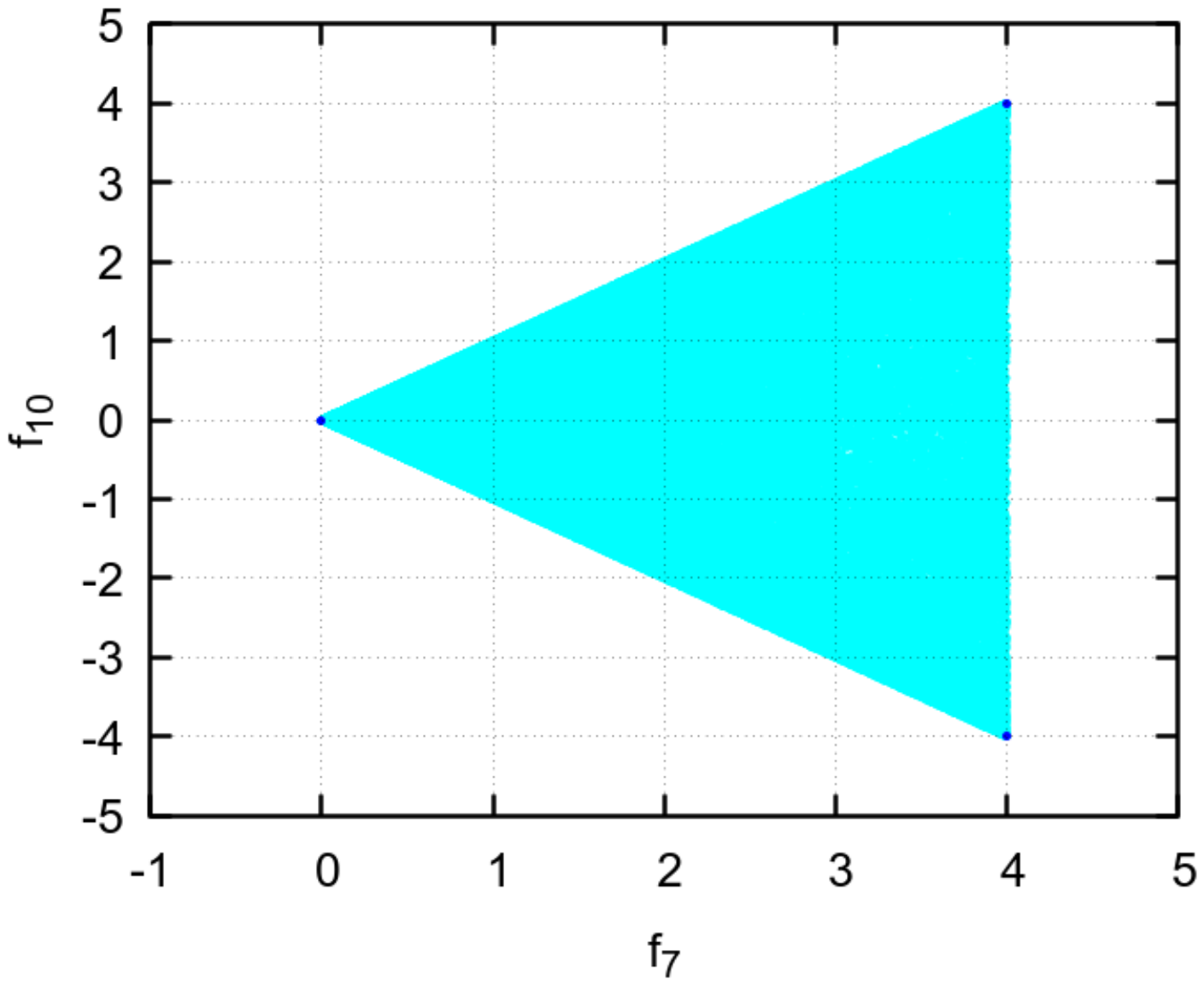}
\caption{Possible values of the functions $f_7$ and $f_{10}$ required for the minimization of the angular part.\label{fig:f7Vsf10}}
\end{figure}

For both new cases we have found numerically,
\texttt{F0DM0'} and \texttt{F0CB},
one has $r_3=0$.
So,
to have a better understanding of the situation,
we consider the parameterization of \eq{eq:vacua_param} with $r_3=0$.
Certainly, there are redundant angles,
as we shall see in a moment.
With these conditions the potential reads,
\begin{equation}
V
= V_1
+\frac{1}{4}\lambda_7 r_1 r_2 f_7(\alpha_{+},\alpha_{-},\beta)
+ \frac{1}{2}\lambda''_{10} r_1 r_2
f_{10}(\alpha_{+},\alpha_{-},\beta,\gamma),
  \label{eq:pot_r30}
\end{equation}
where
\begin{equation}
V_1 = m^2_{11} r_1 + m^2_{22} r_2+ \lambda_1 r_1^2   + \lambda_2
r_2^2+ \lambda_4 r_1 r_2\, ,
\end{equation}
and we have defined
\begin{equation}
  \label{eq:r3_def}
  \alpha_{+}=\alpha_1+\alpha_2,\quad
  \alpha_{-}=\alpha_1-\alpha_2,\quad
  \beta=\beta_1-\beta_2,
\end{equation}
already indicating that we need less angles to describe this
situation, and where we also have defined,
\begin{align}
  \label{eq:r3_fs}
  f_7(\alpha_{+},\alpha_{-},\beta)=&
  2 - \cos(2 \alpha_{+}) (-1 + \cos\beta) + \cos(2 \alpha_{-}) (1 + \cos\beta),
\nonumber\\[+2mm]
f_{10}(\alpha_{+},\alpha_{-},\beta,\gamma)=&
\left[\cos(2 \alpha_{+}) (-1 + \cos\beta) + 2 \cos\beta
+ \cos(2 \alpha_{-}) (1 + \cos\beta)\right]
\cos(\beta - 2 \gamma)\nonumber\\
&
- 4 \cos(\alpha_{-}) \cos(\alpha_{+}) \sin\beta \sin(\beta - 2 \gamma).
 \end{align}
In Fig.~\ref{fig:f7Vsf10}, we plot $f_7$ versus $f_{10}$, for random values of
$\alpha_+$, $\alpha_-$, $\beta$, and $\gamma$.
We see that $x=f_7 \in (0, 4)$,
while $y=f_{10} \in (-4,4)$, lying between the lines
$y = - x$ and $y=+x$.
The angular part of our potential is of the form
$g(x,y) = a x + b y$, with $a= \lambda_7/4$ and $b= \lambda_{10}''/2$.
It is easy to show that $g(x,y)$ cannot have extrema in the interior of the triangle
in Fig.~\ref{fig:f7Vsf10}.
And, assuming $\lambda_7 \neq \pm 2 \lambda_{10}''$,
it must have its extrema at the vertices of the triangle.
The possibilities are, thus,
\begin{eqnarray}
(f_7, f_{10}) = (0,0)
&\ \ \Longrightarrow \ \ &
g(0,0)=0\, ,
\label{0,0}
\\
(f_7, f_{10}) = (4,4)
&\ \ \Longrightarrow \ \ &
g(4,4) = \lambda_7 + 2 \lambda_{10}''\, ,
\label{4,4}
\\
(f_7, f_{10}) = (4,-4)
&\ \ \Longrightarrow \ \ &
g(4,-4) = \lambda_7 - 2 \lambda_{10}''\, .
\label{4,-4}
\end{eqnarray}
The true minimum depends on which of Eqs.~\eqref{0,0}-\eqref{4,-4} lies the lowest.
Notice that this is the only part that depends on $\lambda_7$ and/or $\lambda_{10}''$;
the candidates for extrema do \textit{not} depend on $\lambda_7$ and/or $\lambda_{10}''$,
for they can only lie at the vertices of the triangle regardless.

We conclude that
\begin{eqnarray}
\lambda_7 > 2 |\lambda_{10}''| \hspace{7ex}
&\ \ \Longrightarrow \ \ &
V_\textrm{min} = V_1\, ,
\label{case:0}
\\
\lambda_7 < 2 |\lambda_{10}''| \ \ \ \textrm{and}\ \ \ 
\lambda_{10}'' < 0
&\ \ \Longrightarrow \ \ &
V_\textrm{min} = V_1 + r_1 r_2 (\lambda_7 + 2 \lambda_{10}'')\, ,
\label{case:+}
\\
\lambda_7 < 2 |\lambda_{10}''| \ \ \ \textrm{and}\ \ \ 
\lambda_{10}'' > 0
&\ \ \Longrightarrow \ \ &
V_\textrm{min} = V_1 + r_1 r_2 (\lambda_7 - 2 \lambda_{10}'')\, .
\label{case:-}
\end{eqnarray}
Equations~\eqref{case:0}-\eqref{case:-} correspond, respectively,
to the values obtained for the potential:
i) in our new charge breaking case \texttt{F0CB};
ii) in the case \texttt{F0DM0} of  Ref.~\cite{Hernandez-Sanchez:2020aop};
and
iii) in our new case \texttt{F0DM0'}.

By looking at the extrema conditions for $f_7$ and $f_{10}$
one can show that
\eqref{case:0} occurs for one of the following angle combinations:
\begin{eqnarray}
\textit{i)}
&\ \ &
\beta =0 \ \ \textrm{and}\ \  \cos{\alpha_-}=0\, ,
\\
\textit{ii)}
&\ \ &
\beta =\pi \ \ \textrm{and}\ \   \cos{\alpha_+}=0\, ,
\\
\textit{iii)}
&\ \ &
\sin\beta \neq 0 \ \ \textrm{and}\ \ \cos(2 \alpha_+) = \cos(2 \alpha_-) = -1\, .
\end{eqnarray}

Similarly,
by looking at the extrema conditions for $f_7$ and $f_{10}$, one can show that
\eqref{case:+} and \eqref{case:-} occur, respectively, for one of the following angle combinations:
\begin{eqnarray}
\textit{i)}
&\ \ &
\beta =0 \ \ \textrm{and}\ \  \sin{\alpha_-}=0 \ \ \textrm{and}\ \  \cos(2 \gamma) = \pm 1\, ,
\\
\textit{ii)}
&\ \ &
\beta =\pi \ \ \textrm{and}\ \  \sin{\alpha_+}=0 \ \ \textrm{and}\ \  \cos(2 \gamma) = \pm 1\, ,
\\
\textit{iii)}
&\ \ &
\sin\beta \neq 0 \ \ \textrm{and}\ \ (\alpha_+,\alpha_-) = (0,0), (\pi,\pi)
\ \ \textrm{and}\ \ \cos(2 \beta - 2 \gamma) = \pm 1\, ,
\\
\textit{iv)}
&\ \ &
\sin\beta \neq 0 \ \ \textrm{and}\ \ (\alpha_+,\alpha_-) = (0,\pi), (\pi,0)
\ \ \textrm{and}\ \ \cos(2 \gamma) = \pm 1\, .
\end{eqnarray}

We have crossed checked our results with our numerical method of
finding the global minimum and they completely agree.

\section{Setting up the scan}
\label{sec:scan}

In the previous Sections we discussed in detail how to ensure that
our \texttt{2-Inert} is indeed the global minimum.
We used implicitly the relations found for this minimum.
Namely, 
we take $v_1=v_2=0$, $v_3=v$ and find
\begin{equation}
  \label{eq:minim_h_z2z2}
  m_h^2=2 \lambda_3 v^2, \quad v^2=-\frac{m_{33}^2}{\lambda_3}\, ,
\end{equation}
requiring $\lambda_3 >0$ (already needed for BFB) and $m_{33}^2 <0$.
The fields can be parametrized as
\begin{equation}
  \label{eq:fields}
  \phi_1=
  \begin{pmatrix}
    H_1^+\\
    \tfrac{1}{\sqrt{2}} \left( H_ 1 + i A_ 1\right)
  \end{pmatrix},\quad
  \phi_2= 
 \begin{pmatrix}
    H_2^+\\
    \tfrac{1}{\sqrt{2}} \left( H_ 2 + i A_ 2\right)
  \end{pmatrix},\quad
  \phi_3=
  \begin{pmatrix}
    G^+\\
    \tfrac{1}{\sqrt{2}} \left( v + h + i G_0 \right)
  \end{pmatrix} .
\end{equation}
Since the vacuum does not break the 
$\Z2\times\Z2$ symmetry,
all states are unmixed; they are already in the mass basis.
Moreover, $G^0$ and $G^+$ are the would-be Goldstone bosons,
which, in the unitary gauge, become the longitudinal components
of the $Z^0$ and $W^+$ gauge bosons, respectively.

It proves useful\footnote{Our definition of
$\Lambda_1$ agrees with the vertex in Fig.~9b of
~\cite{Hernandez-Sanchez:2020aop}, but not with their Eqs.~(2.39),
(5.7)-(5.8).}
to define~\cite{Hernandez-Sanchez:2020aop}
\begin{eqnarray}
\Lambda_1 =
\tfrac{1}{2} \left( \lambda_4 + \lambda_7 + 2 \lambda''_{10} \right)\, ,
&\hspace{4ex}&
\bar\Lambda_1 =
\tfrac{1}{2} \left( \lambda_4 + \lambda_7 - 2 \lambda''_{10} \right)\, ,
\\
\Lambda_2 =
\tfrac{1}{2} \left( \lambda_6 + \lambda_9 + 2 \lambda''_{12} \right)\, ,
&\hspace{4ex}&
\bar\Lambda_2 =
\tfrac{1}{2} \left( \lambda_6 + \lambda_9 - 2 \lambda''_{12} \right)\, ,
\\
\Lambda_3 =
\tfrac{1}{2} \left( \lambda_5 + \lambda_8 + 2 \lambda''_{11} \right)\, ,
&\hspace{4ex}&
\bar\Lambda_3 =
\tfrac{1}{2} \left( \lambda_5 + \lambda_8 - 2 \lambda''_{11} \right)\, .
\end{eqnarray}
The other masses are given by
\begin{align}
  m_{H_{1}}^2=&\ m_{11}^2 + \frac{1}{2}\left(\lambda_5 + \lambda_8 + 2
    \lambda''_{11} \right) v^2 \equiv m_{11}^2 + \Lambda_3 v^2\, ,\label{m_Hpm_2_1}\\ 
  m_{A_{1}}^2=&\ m_{11}^2 + \frac{1}{2}\left(\lambda_5 + \lambda_8 - 2
    \lambda''_{11} \right)v^2 \equiv m_{11}^2 + \bar{\Lambda}_3 v^2\, ,\\ 
  m_{H^\pm_{1}}^2=&\ m_{11}^2 + \frac{1}{2} \lambda_5 v^2\, ,\\[+2mm]
  m_{H_{2}}^2=&\ m_{22}^2 + \frac{1}{2}\left(\lambda_6 + \lambda_9 + 2
    \lambda''_{12} \right) v^2 \equiv m_{22}^2 + \Lambda_2 v^2\, ,\\ 
  m_{A_{2}}^2=&\ m_{22}^2 + \frac{1}{2}\left(\lambda_6 + \lambda_9 - 2
    \lambda''_{12} \right)v^2 \equiv m_{22}^2 + \bar{\Lambda}_2 v^2\, ,\\ 
  m_{H^\pm_{2}}^2=&\ m_{22}^2 + \frac{1}{2} \lambda_6 v^2\, .
\label{m_Hpm_2_6}
\end{align}
The conditions for a local minimum are
\begin{align}
  v^2=-\frac{m_{33}^2}{\lambda_3}>0\, ,
\quad
\Lambda_2 > -m_{22}^2/v^2\, ,
\quad
\Lambda_3 > - m_{11}^2/v^2\, ,
\end{align}
but, if we take the masses as input parameters, these will be
automatically satisfied. The value of the potential at the minimum is
\begin{align}
  V_{\texttt{2Inert}} = - \frac{m_{33}^4}{4\lambda_3}\, .
\label{V_2Inert}
\end{align}
As mentioned, we ensure that $ V_{\texttt{2Inert}}$ lies below
the value of the potential at the other local extrema,
whose explicit expressions
can be found in Appendix~\ref{app:masses}. We take
\begin{equation}
v \approx 246.2~{\rm GeV}\, , \ \ \  m_h = 125~{\rm GeV}\, ,
\end{equation}
as fixed inputs.
We follow Ref.~\cite{Hernandez-Sanchez:2020aop} and choose as our free parameters
\begin{equation}
m_{H_1}^2, m_{H_2}^2, m_{A_1}^2,m_{A_2}^2,
m_{H_1^\pm}^2, m_{H_2^\pm}^2,
\Lambda_1, \Lambda_2, \Lambda_3,
\lambda_1, \lambda_2, \lambda_4, \lambda_7\, .
\label{input_param}
\end{equation}
All other parameters of the scalar potential can be extracted from
\eq{m_Hpm_2_1}-\eqref{m_Hpm_2_6}.
We choose random values for the remaining parameters in the
set of \eq{input_param},
in the ranges
\begin{align}
&\Lambda_1,\, \Lambda_2,\, \Lambda_3,\, \lambda_1,\, \lambda_2,\, \lambda_4,\, \lambda_7  \in \pm\left[10^{-3},10\right];
\nonumber\\[8pt]
&m_{H_1},\, m_{H_2},\, m_{A_1},\,m_{A_2}\,
\in \left[50,1000\right]~{\rm GeV};
\nonumber\\[8pt]
&
m_{H_1^\pm},\,m_{H_2^\pm}\,
\in \left[70,1000\right]~{\rm GeV},
\label{eq:scanparameters_DM}
\end{align}
with the chosen condition that $m_{H_1}<m_{H_2}$, without loss of generality.
The lower limit on the mass of the charged scalars comes from Ref.~\cite{Pierce:2007ut}.
Although this bound has not been established within the context of the
current model of two component DM, we take it as a conservative
lower bound on the masses of all charged scalars.

For the interactions with fermions and gauge bosons,
it is assumed that all such SM fields transform into themselves under
$\Z2\times\Z2$.
Thus,
Table~\ref{tab:Types} implies that \textit{all} fermion fields only couple with $\phi_3$.
This is a so-called Type-I model.
Since the Yukawa couplings are identical to the SM ones,
there are no FCNCs at tree-level and flavour bounds
such as $B \to X_s \gamma$ are trivially satisfied.
As the fields in \eq{eq:fields} are already in the mass basis,
there are thus three sectors: the dark-$\Z2$ sector,
constituted by the fields in $\phi_1$;
the dark-$\Z2^\prime$ sector,
constituted by the fields in $\phi_2$;
and the active or SM sector,
constituted by the fields in $\phi_3$
and all SM fermions.
Connections among different sectors can only occur due to gauge bosons or due
to the cubic and quartic interactions of the Higgs potential.


Our numerical scan proceeds in the following fashion.
We start by taking a random value for the parameters \eqref{input_param} within
the intervals \eqref{eq:scanparameters_DM}.
As mentioned above,
we apply the constraints from BFB and global minimum.
Then, we confirm numerically that indeed no lower-lying minimum is found. 
Next,
we impose all the constraints described in Chapter~\ref{chapter:Constraints}. In addition to \texttt{HiggsTools} 1.1.3~\cite{Bahl:2022igd} for current bounds from searches for additional
scalars, we speed up our scan range by initially forbidding decays of SM gauge bosons into the new scalars by enforcing:
\begin{equation}
m_{H_i}+m_{H_i^\pm} \geq m_W^\pm\, , \hspace{2ex}
m_{A_i}+m_{H_i^\pm} \geq m_W^\pm\, , \hspace{2ex}
m_{H_i}+m_{A_i} \geq m_Z\, , \hspace{2ex}
2 m_{H_i^\pm}\geq m_Z.
\end{equation}
Taking into account the LEP 2 results re-interpreted for the
I(1+1)HDM, we exclude the region of masses where the following
conditions are simultaneously satisfied
\cite{Lundstrom:2008ai} $(i=1,2)$:
\begin{equation}
m_{A_i} \leq 100~{\rm GeV},\quad
m_{H_i} \leq 80~{\rm GeV},\quad
|m_{A_i}-m_{H_i}| \geq 8 ~{\rm GeV}.
\end{equation}
In order to evade the bounds from long-lived charged particle searches given in Ref.~\cite{Heisig:2018kfq}, we set
the upper limit on the charged scalar lifetime of $\tau \leq 10^{-7}\text{s}$.

We also have to consider the Dark Matter relic density measurement and direct, indirect and collider searches, described in Chapter~\ref{chapter:Darkmatter_problem}. Our results for the relic density, scattering amplitudes and
annihilation cross section are obtained using the implementation
of this model in \texttt{micrOMEGAs 6.0.5}~\cite{Alguero:2023zol},
which we have constructed.

We calculate the dark matter relic density as the sum of the contributions
of each DM candidate:
\begin{equation}
    \Omega_T h^2 = \Omega_1 h^2 + \Omega_2 h^2\,,
\end{equation}
and impose the range very precisely determined value by the
by the Planck collaboration
~\cite{Planck:2018vyg}\footnote{Alternatively,
one could adopt a more permissive range. Indeed, some models have been studied
where loop effects can induce corrections to the relic density which are
of the order of 10\%~\cite{Banerjee:2019luv,Banerjee:2021hal}; this lead the authors of
~\cite{Belanger:2021lwd} to consider instead the augmented range
$0.094 < \Omega_T h^2 < 0.142$. We will keep to the
Planck constraint in \eqref{Planck}.}:
\begin{equation}
    \Omega_{T} h^2= 0.1200\pm 0.0012\, .
\label{Planck}
\end{equation}

The direct detection limit from dark matter-nucleon
scattering considered is by LZ in 2022~\cite{LZ:2022lsv}.
To compare directly with the experimental limit, we follow the
method presented in ~\cite{Belanger:2014bga} of computing the
normalized cross section of DM on a point-like nucleus (taken to be Xenon)
\begin{equation}
\sigma_{\text{SI}}^{\text{Xe},k}=
\frac{4 \mu_k^2}{\pi}\frac{\left(Z f_p + (A-Z)f_n\right)^2}{A^2},
\end{equation}
with $\mu_k$ the reduced mass of the DM candidate and
$f_p$, $f_n$ the amplitudes for protons and neutrons.
As there are two dark matter candidates, we rescale
the obtained cross section for each DM candidate by the
relative density of the component:
\begin{equation}
\sigma_{\text{SI}}^{\text{r},k} =
\sigma_{\text{SI}}^{\text{Xe},k}\ \xi_k\, , 
\end{equation}
where
\begin{equation}
\xi_k = \frac{\Omega_k}{\Omega_T}\, .
\end{equation}

To deal with indirect detection constraints, we follow closely
the strategy adopted in~\cite{Belanger:2021lwd}.
We start by using our \texttt{micrOMEGAs 6.0.5} model implementation
in order to calculate the thermally averaged cross section for DM annihilation
(or co-annihilation, or DM conversion)
times velocity $\langle \sigma v \rangle$.
We find that the $\langle \sigma v \rangle$ detectable arises mainly from
annihilation decays into $VV$, and sum only the  $WW$ and $ZZ$ final states,
assuming a similar spectrum, which we dub $\langle\sigma v\rangle_{VV}$,
and we apply the exclusion bounds in Fig.~\ref{indirect_lines}.

\section{Some interesting processes}
\label{sec:processes}

\begin{figure}[htbp!]
\centering
\includegraphics[width=0.75\textwidth]{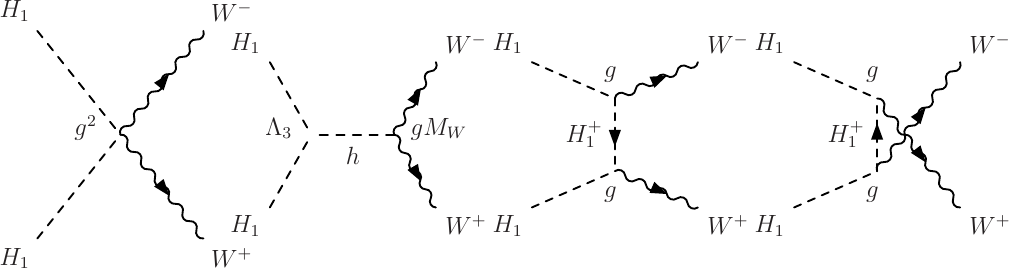}
\caption{Feynman diagrams for $H_1 H_1 \rightarrow WW$.
}
\label{H1H1TOWW}
\end{figure}
%
%
Before we proceed, it is interesting to describe some of the classes of processes achievable
in this very rich model.
Indeed, we have:
\begin{itemize}
\item (vanilla) annihilation: e.g. $H_1 H_1 \rightarrow b \bar{b}$,
\item co-annihilation: e.g. $A_1 H_1 \rightarrow b \bar{b}$,
\item co-scattering: e.g. $H_1^+ W^- \rightarrow A_1 Z$,
\item DM conversion: e.g. $H_2 H_2 \rightarrow H_1 H_1$.
\end{itemize}
There are, however, no semi-annihilations processes $x_i x_j \rightarrow x_k \textrm{SM}$,
such as appears in~\cite{DEramo:2010keq} and the model discussed in Chapter~\ref{chapter:Darkmatter_model}.
Fig.~\ref{H1H1TOWW}
shows the Feynman diagrams for $H_1 H_1 \rightarrow WW$.
The Feynman diagrams for $H_1 H_1 \rightarrow ZZ$ are obtained
from Fig.~\ref{H1H1TOWW} with the substitutions $W^\pm \rightarrow Z$,
$H_1^\pm \rightarrow A_1$.

\begin{figure}[htbp!]
\centering
\includegraphics[width=0.40\textwidth]{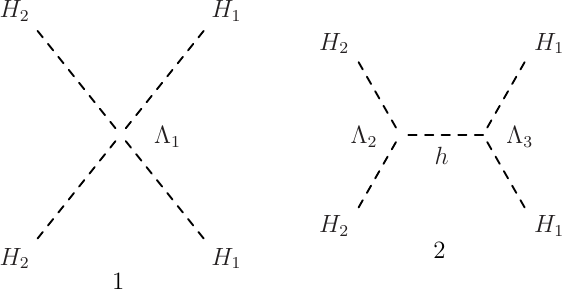}
\caption{Feynman diagrams for the DM conversion processes
$H_2 H_2 \rightarrow H_1 H_1$.
}
\label{22TO11}
\end{figure}

\begin{figure}[htbp!]
\centering
\includegraphics[width=0.4\textwidth]{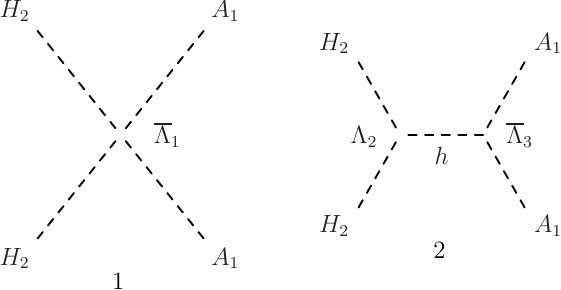}
\caption{Feynman diagrams for the DM conversion processes
$H_2 H_2 \rightarrow A_1 A_1$.
}
\label{22TOA1A1}
\end{figure}

\begin{figure}[htbp!]
\centering
\includegraphics[width=0.40\textwidth]{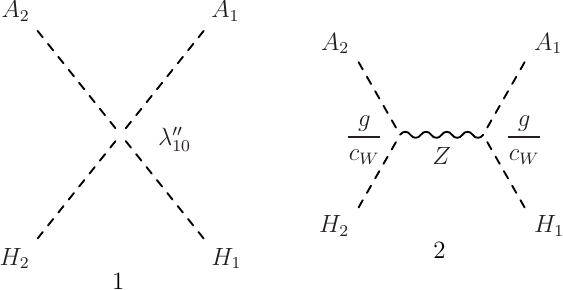}
\caption{Feynman diagrams for the DM conversion processes
$H_2 A_2 \rightarrow H_1 A_1$.
}
\label{2A2TO1A1}
\end{figure}

This model has DM conversion processes between the DM sector 1 and the
DM sector 2:
\begin{itemize}
\item $H_2 H_2 \rightarrow H_1 H_1$: quartic ($\sim \Lambda_1$), and
through $h$ ($\sim \Lambda_2 \Lambda_3$);
\item $H_2 H_2 \rightarrow A_1 A_1$: quartic ($\sim \bar{\Lambda}_1$), and
through $h$ $(\sim \Lambda_2 \bar{\Lambda}_3)$;
\item  $H_2 A_2 \rightarrow H_1 A_1$: quartic ($\sim \lambda''_{10}$),
and through $Z$ ($\sim g^2$).
\end{itemize}
We show the Feynman diagrams for these DM conversion processes
in Figs.~\ref{22TO11},~\ref{22TOA1A1}, and~\ref{2A2TO1A1}.

The numerical results to be discussed below include all processes, which arise out of
our implementation of the model in \texttt{micrOMEGAs 6.0.5}.
The authors of Ref.~\cite{Hernandez-Sanchez:2020aop} choose to concentrate on the mass region
$m_h/2 < m_{H_1} < 80 ~{\rm GeV}$ and $ m_{H_2} \simeq 100 ~{\rm GeV}$.
We extend significantly the analysis by considering all available parameter space.
Not surprisingly, we find different conclusions.
In particular, we find many situations in which both DM components can contribute equally
to the relic density.
We also find wide regions of parameter space where
the possibility that the lighter DM component is mainly probed through direct nuclear recoil
while the heavier DM component is probed in indirect DM detection does not hold.
This will be discussed next.

\section{Results and discussion}
\label{sec:results}

In the IDM, the DM mass is constrained to two regions.
The reason is the following.
The annihilation of the DM particle into $WW$ (equivalent to that in Fig.~\ref{H1H1TOWW})
is controlled by a gauge coupling and, thus, it is not tunable.
For the most part, it leads to a decay rate so high that it depletes the
IDM DM candidate, making its relic density under-abundant.
This is avoided for $m_\textrm{dm} \gtrsim 500 ~{\rm GeV}$ if all ``dark'' scalars
($H$, $A$, and $H^\pm$) have similar masses.
Below the $W$ threshold, the annihilation proceeds mostly into
$b \bar{b}$; this depends on the DM-Higgs coupling, which is tunable to comply
with the relic density.
However, that requires large couplings, which are precluded by direct detection.
The exception occurs around $m_\textrm{dm} \simeq m_h/2$ where the annihilation
has a resonance, allowing for a fit to the relic density with a coupling low
enough to comply with direct detection constraints.

\begin{figure}[htbp!]
\centering
\includegraphics[width=0.55\textwidth]{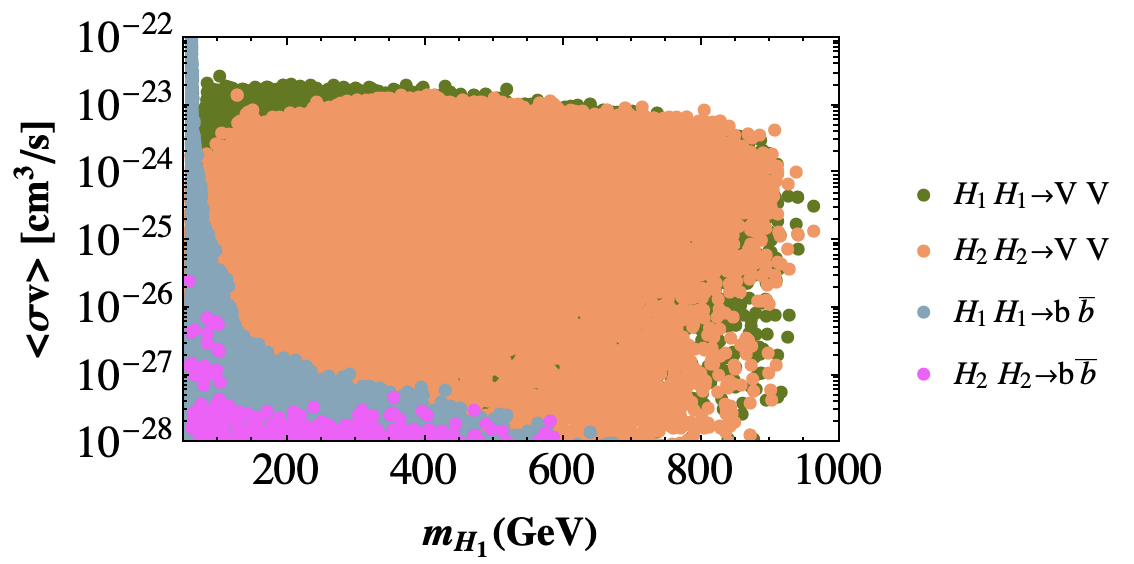}
\caption{$\langle\sigma v\rangle$ for
$H_1$ ($H_2$) annihilation into $b \bar{b}$ and $VV$ as a function of $m_{H_1}$.
The colour codes are in the figure.
}
\label{sigv_H1_and_H2}
\end{figure}

\begin{figure}[htbp!]
\centering
\includegraphics[width=0.45\textwidth]{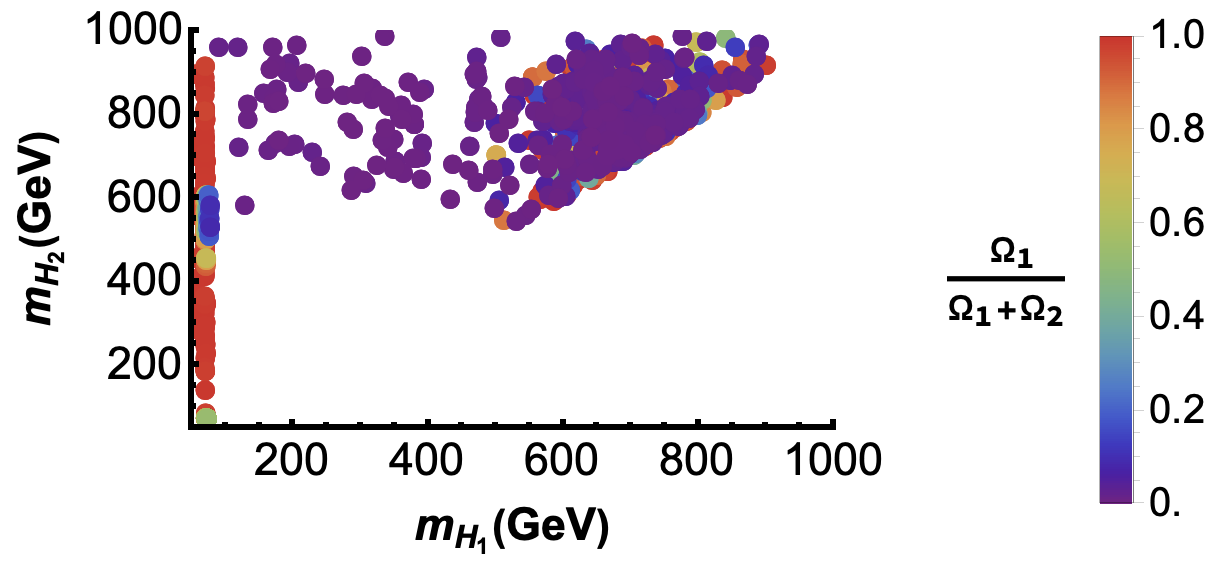}
\hspace{5mm}
\includegraphics[width=0.45\textwidth]{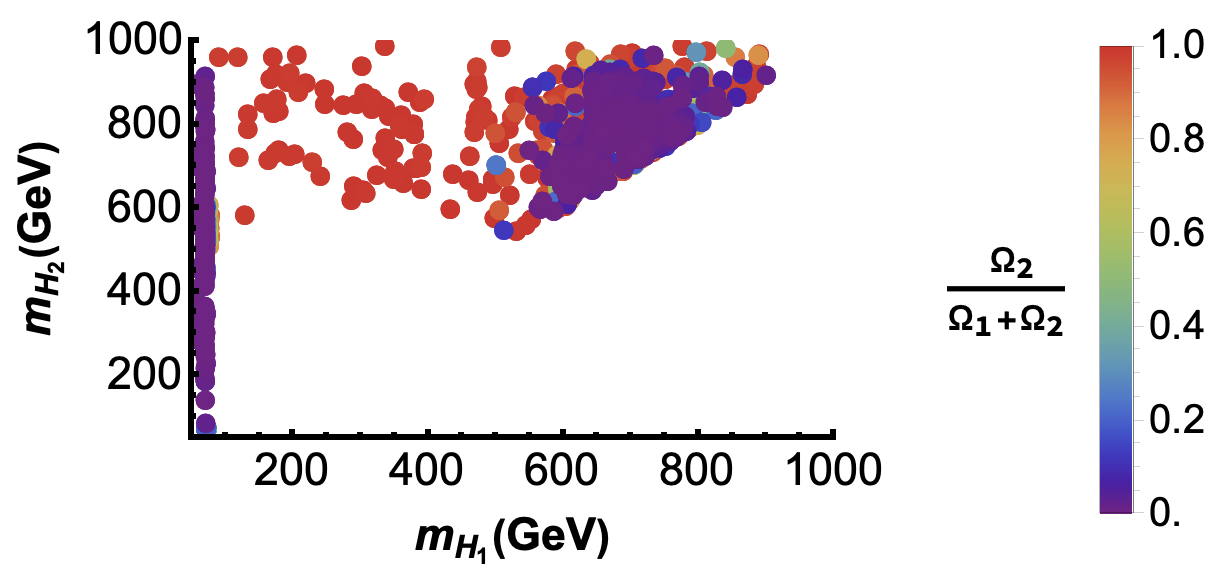}
\caption{Range of allowed ($m_{H_1}, m_{H_2}$) masses with a ``temperature''
colour code for $\Omega_1/\Omega_T$ ($\Omega_2/\Omega_T$) on the left (right).
Both plots have the same information but were included to aid the eye.
These points have passed all theory, collider and astrophysical constraints.
}
\label{mH1mH2_relic}
\end{figure}

The situation is both similar and different in our two component DM
$\Z2\times\Z2$ 3HDM.
This is best seen with the help of Fig.~\ref{sigv_H1_and_H2},
where we show the values of $\langle\sigma v\rangle$ for
$H_1$ ($H_2$) annihilation into $b \bar{b}$ and $VV$ as a function of $m_{H_1}$.
First, we note that above the $W$ threshold
$\langle\sigma v\rangle_{VV} \gg \langle\sigma v\rangle_{bb}$ for both $H_1$ and $H_2$.
Second, for $H_1$, $\langle\sigma v\rangle_{bb}$ can be very large
for low values of $m_{H_1}$, as in the IDM.
Thus, if $H_1$ were the only DM component, combining relic density
and direct detection constraints would lead to the same mass regions
as in the IDM.
However, we can force $m_{H_1}$ into the intermediate mass range, by requiring that
it is $H_2$ which is mostly responsible for the relic density.
This can be seen in Fig.~\ref{mH1mH2_relic}.
Notice also that for small $m_{H_1}$ there are two possibilities.
For any value of $m_{H_2}$,
one can have $H_1$ be the major relic density component.
On the other hand,
for $ 500~{\rm GeV}  \lesssim m_{H_2} \lesssim  600 ~{\rm GeV}$,
one can have $H_2$ be the major relic density component.
Moreover,
one can find quite a number of interesting points
where $\Omega_1 \simeq \Omega_2$.
These occur for the $H_1$ mass regions which would be allowed in the IDM.
The reason is simple; $H_1$ would be able to allow for all the relic density,
and one can tune it down for 50\%, while tuning the $H_2$ parameters to account
for the remainder 50\%.
Finally, we note that if a DM component has a mass between $\sim 100 ~{\rm GeV}$ and
$\sim 500 ~{\rm GeV}$,
then it is guaranteed to give a very suppressed contribution to the
total relic density.

There is one relevant issue concerning Fig.~\ref{sigv_H1_and_H2}.
One might worry that there are significant contributions from
$\langle\sigma v\rangle_{hh}$.
We have checked that this is the case if we do not impose that $\Omega_T$
must equal the measured relic density.
However, once we impose the Planck limit in \eq{Planck},
the dominant contributions are those shown in Fig.~\ref{sigv_H1_and_H2}.
Albeit in a different model, this is also what was found in Ref.~\cite{Belanger:2021lwd}.
This will be important to the discussion on indirect detection below.

\begin{figure}[htbp!]
\centering
\includegraphics[width=0.37\textwidth]{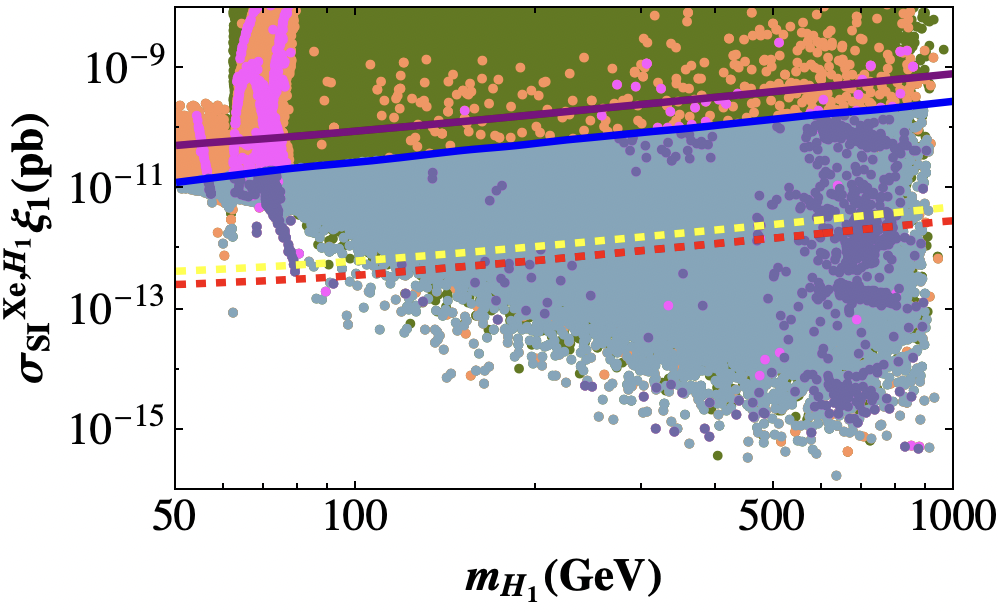}
\hspace{3mm}
\includegraphics[width=0.14\textwidth]{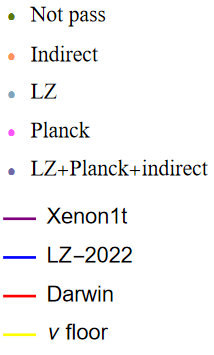}
\hspace{3mm}
\includegraphics[width=0.37\textwidth]{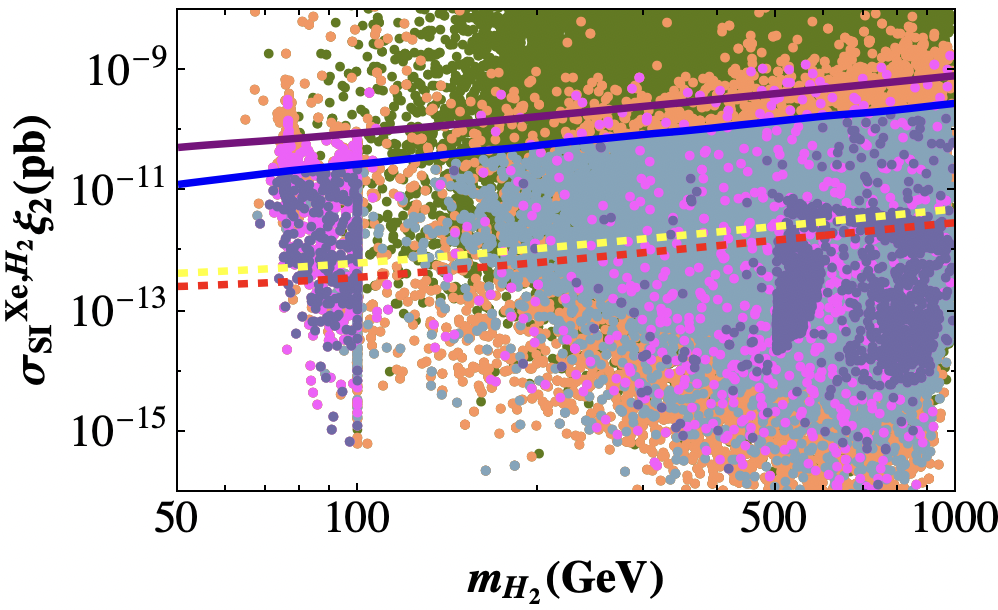}
\caption{Direct detection constraints on $H_1$ ($H_2$) on the left (right) figure.
See text for details.
}
\label{direct}
\end{figure}

But first, we turn to the constraints from direct detection shown in Fig.~\ref{direct}.
The lines shown have the following origins:
i) the solid purple line refers to the XENON1T~\cite{XENON:2018voc},
which is included as recast limits inside \texttt{micrOMEGAs 6.0.5};
ii) the solid blue line refers to the current LUX-ZEPLIN (LZ) bound of 2022
~\cite{LZ:2022lsv};
iii) the dashed red line corresponds to the expected reach
of the DARWIN experiment~\cite{DARWIN:2016hyl};
iv) the dashed yellow line corresponds to the neutrino floor,
as presented in~\cite{OHare:2021utq}.
As for the points shown,
all have passed theory and collider constraints.
The colour code refers to the additional astrophysical constraints as follows:
i) the green points have passed  theory and collider constraints,
but failed all of the relic density, direct and indirect limits;
ii) the orange points have agreement with Fermi-LAT's indirect detection
bounds~\cite{Fermi-LAT:2015att}, but fail both Planck and LZ;
iii) the gray points Pass LZ, but fail Planck (some pass Fermi-LAT;
some do not);
iv) the pink points achieve the correct relic density,
but failed either LZ or Fermi-LAT;
finally, v) the dark-purple points pass all current astrophysical
constraints.

\begin{figure}[htbp!]
\centering
\includegraphics[width=0.4\textwidth]{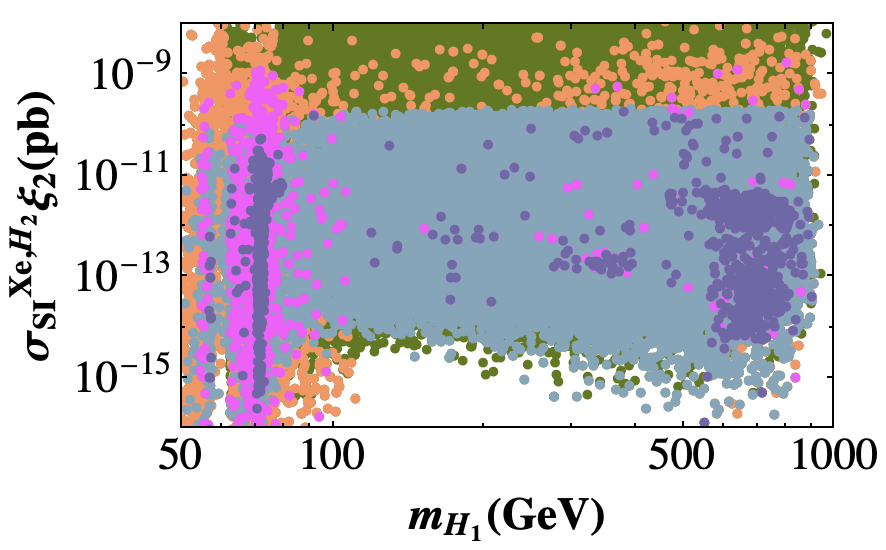}
\caption{Constraints on $\sigma_{\text{SI}}^{\text{Xe}2k}\ \xi_2$  as a function of $m_{H_1}$.
The colours of points have the same meaning as in Fig.~\ref{direct}.
Notice that the curves for experimental constraints are not appropriate
for this graph.
}
\label{direct2mh1}
\end{figure}

Let us concentrate first on Fig.~\ref{direct}-left.
The presence of red and orange points above the LZ line shows that,
in agreement with our previous discussion,
this constraint from direct detection is relevant for low $H_1$ masses.
Remarkably, for such low masses the DARWIN experiment will be able to
exclude many points.
In contrast,
many points with high $H_1$ masses will not be invalidated by DARWIN.
And, since this exclusion is expected to lie below the neutrino floor,
other collider and/or astrophysical probes must be used.
Turning to Fig.~\ref{direct}-right, we see that direct detection
may also constrain $H_2$.
This is more clearly seen in a plot as a function of
$m_{H_1}$, as in Fig.~\ref{direct2mh1}.
However, looking at the pink and dark-purple points in Fig.~\ref{direct2mh1}
for low values of $m_{H_1}$ it is possible that direct
detection probes $H_1$, while it does not affect $H_2$,
in accordance with the special case discussed in
~\cite{Hernandez-Sanchez:2020aop}.

Note that the appearance of green points bellow all exclusion lines in one of the
plots in Fig.~\ref{direct} is due to the fact that, although the point passes
the direct detection for the corresponding DM component,
it does not pass it for the other DM component.

We now turn to the constraints arising from indirect detection.
As mentioned, our strategy was to prove that $\langle\sigma v\rangle_{VV}$ dominates for
most of the parameter space, and use the lines determined by the authors of Ref.~\cite{Reinert:2017aga} together with gamma ray searches for dark matter~\cite{Fermi-LAT:2015att,HESS:2022ygk}.
The exception occurs for small masses, where we apply the $\langle\sigma v\rangle_{bb}$ lines also obtained by the authors of Ref.~\cite{Reinert:2017aga} and Fermi-LAT experiment~\cite{Fermi-LAT:2015att}.

\begin{figure}[htbp!]
\centering
\includegraphics[width=0.38\textwidth]{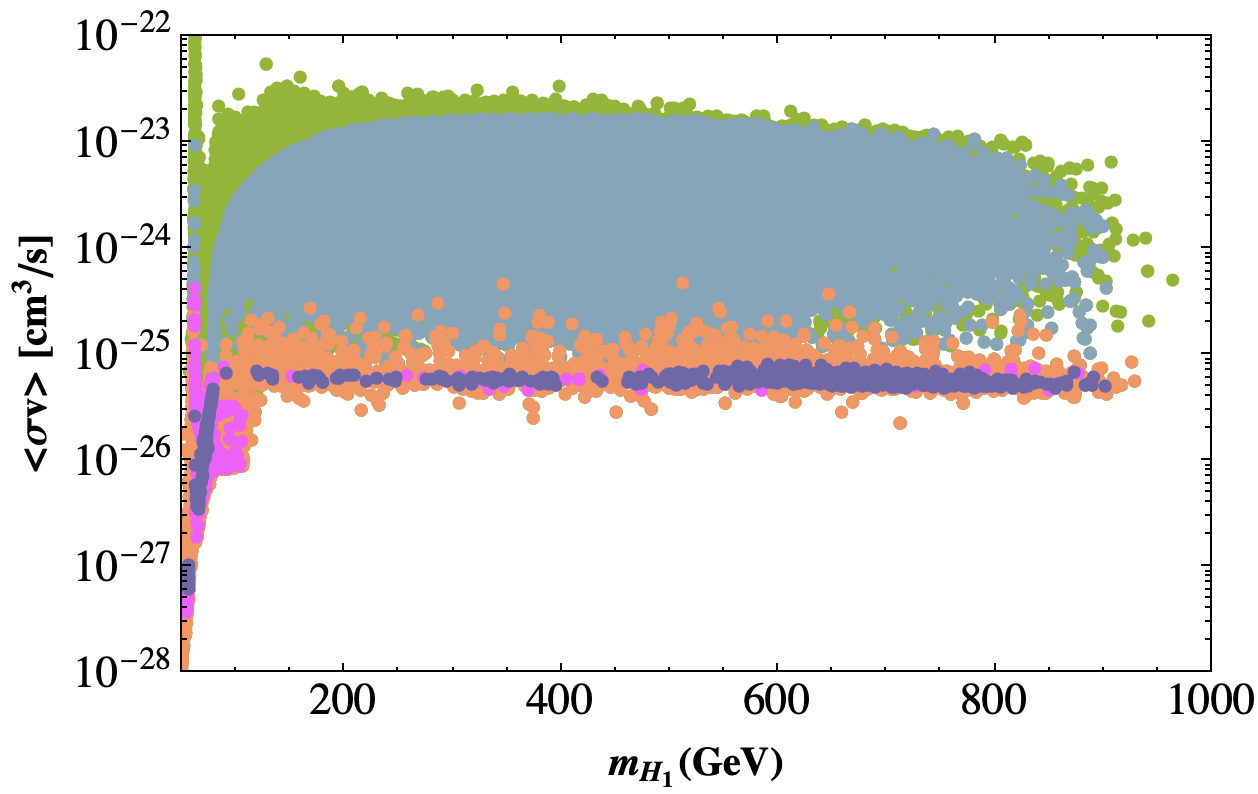}
\hspace{3mm}
\includegraphics[width=0.14\textwidth]{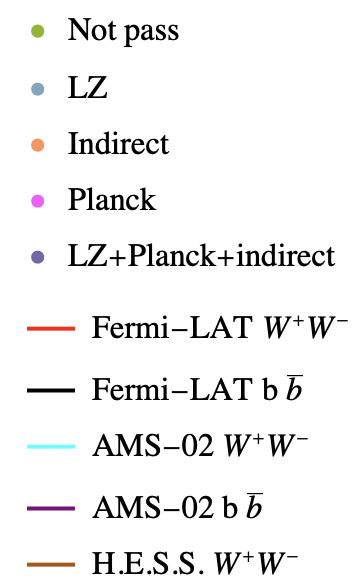}
\hspace{3mm}
\includegraphics[width=0.38\textwidth]{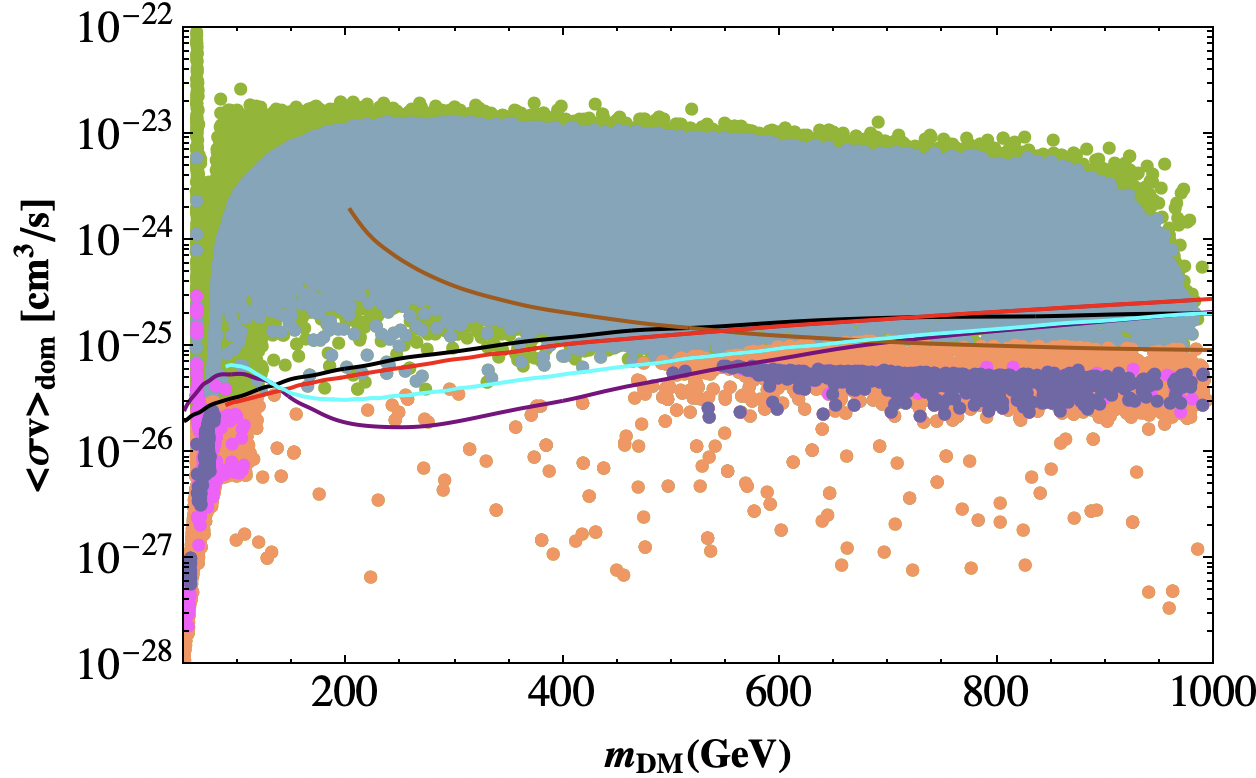}
\caption{The colours of the points have the same meaning as in Fig.~\ref{direct}.
The left figure shows the total
$\langle\sigma v\rangle$ as a function of $m_{H_1}$. 
The right figure shows the dominant contribution to
$\langle\sigma v\rangle$ as a function of the mass of the DM candidate, $m_\textrm{DM}$,
which corresponds to the $\langle\sigma v\rangle$ plotted on the vertical axis.
The lines coming from
Fermi-LAT~\cite{Fermi-LAT:2015att} and H.E.S.S.~\cite{HESS:2022ygk} assume a Navarro-Frenk-White (NFW)
DM density profile and the AMS-02~\cite{AMS:2016oqu} lines correspond to the conservative approach derived in Ref.~\cite{Reinert:2017aga},
with the colour codes also shown in the figure.
}
\label{indirect1}
\end{figure}

Fig.~\ref{indirect1}-left shows the total $\langle\sigma v\rangle$.
The plots have 200000 points (in fact, we generated 1 million points, but the conclusions
are not altered).
The colours of the points have the same meaning as in Fig.~\ref{direct}.
The red points have passed Planck bounds but may, or not, have passed
the bounds from indirect detection.
However, we found that, out of 2761 points that pass Planck, only 36 are ruled out by
indirect detection, and that this occurs only for masses of $H_1$ between
$\sim 62~{\rm GeV}$ and $\sim 66~{\rm GeV}$, where the dominant contribution is
$\langle\sigma v\rangle_{bb}$.
This is also seen on Fig.~\ref{indirect1}-right,
where we plot the dominant contribution to
$\langle\sigma v\rangle$ as a function of the mass of the DM candidate, $m_\textrm{DM}$,
which corresponds to the (dominant) $\langle\sigma v\rangle$ plotted on the vertical axis.
This is to be compared with the exclusion
lines for $\langle\sigma v\rangle_{VV}$ and $\langle\sigma v\rangle_{bb}$
corresponding to $H_1$ or $H_2$; whichever yields the dominant  $\langle\sigma v\rangle$.
The lines come from Ref.~\cite{Reinert:2017aga} and refer to exclusions extrapolated from
Fermi-LAT~\cite{Fermi-LAT:2015att} (in red for $VV$ and in black for $bb$),
and from AMS-02~\cite{AMS:2016oqu} (in light-blue for $VV$ and in purple for $bb$).
We also include (in a solid brown line) a limit coming from
H.E.S.S.~\cite{HESS:2022ygk},
which is important for DM masses above $\sim 500~{\rm GeV}$.
As a result, we can conclude that, except for those very specific 36 points,
the Planck constraints (almost) guarantee that the indirect detection will be
ineffectual.
Again, the dark purple points pass every constraint.

\begin{figure}[htbp!]
\centering
\includegraphics[width=0.45\textwidth]{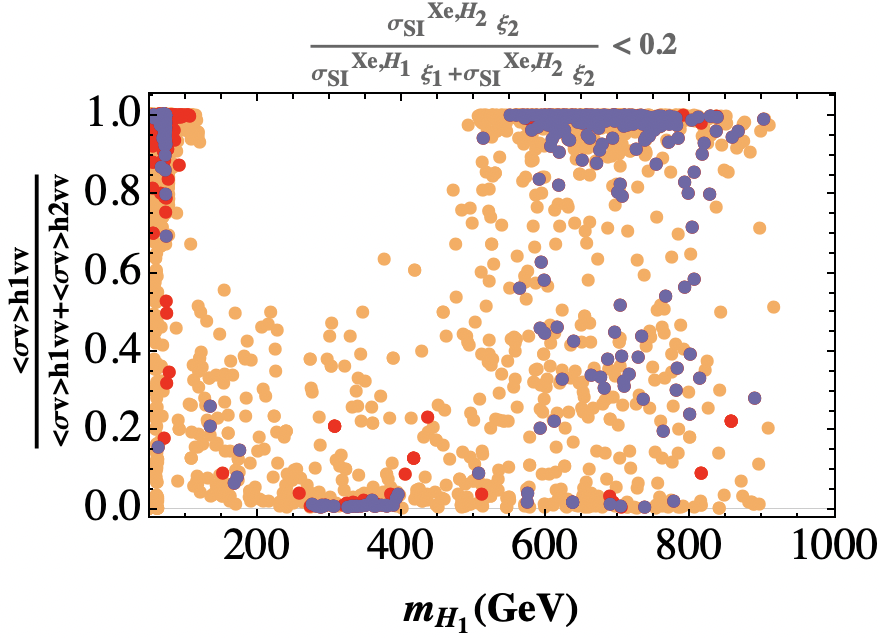}
\hspace{5mm}
\includegraphics[width=0.45\textwidth]{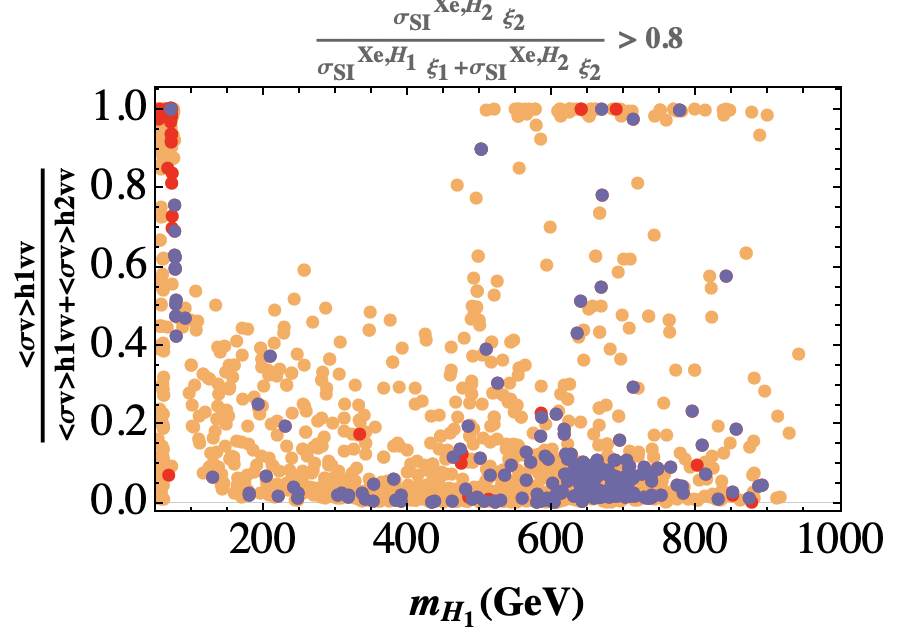}
\caption{The orange points pass the indirect detection bounds; the red ones
achieve the correct density; and the dark-purple points pass all astrophysical constraints.
On the left (right) we show points where direct detection is dominated by $H_1$ ($H_2$).}
\label{indirect2}
\end{figure}

We now turn to the interplay between direct and indirect detection.
Fig.~\ref{indirect2}-left (-right) contains points for which the dominant
direct detection cross-section is due to $H_1$ ($H_2$).
In both, we plot  $\langle\sigma v\rangle_{VV}$ for $H_1$ over the sum of
$\langle\sigma v\rangle_{VV}$ for $H_1$ and $H_2$.
Concentrating on Fig.~\ref{indirect2}-left,
we learn that, whilst direct detection is dominated by $H_1$,
$H_1$ can, in some cases, give the dominant contribution to
indirect detection, while, in other cases, it is $H_2$ which gives
the dominant contribution to indirect detection.
Conversely,
on Fig.~\ref{indirect2}-right,
we learn that, whilst direct detection is dominated by $H_2$,
$H_1$ can, in some cases, give the dominant contribution to
indirect detection, while, in other cases, it is $H_2$ which gives
the dominant contribution to indirect detection.
That is, depending on the parameters of the model,
including $m_{H_1}$, we can have all four possible combinations.

Notice the following feature on both plots in Fig.~\ref{indirect2}.
When $m_{H_1}$ lies roughly between $100~{\rm GeV}$ and $500~{\rm GeV}$,
it can never be the dominant contribution to indirect detection signals.
This is also the region where $H_1$ cannot be the dominant contribution to
the relic density. This is the previously referred relation between
relic density and indirect detection.

\begin{figure}[htbp!]
\centering
\includegraphics[width=0.4\textwidth]{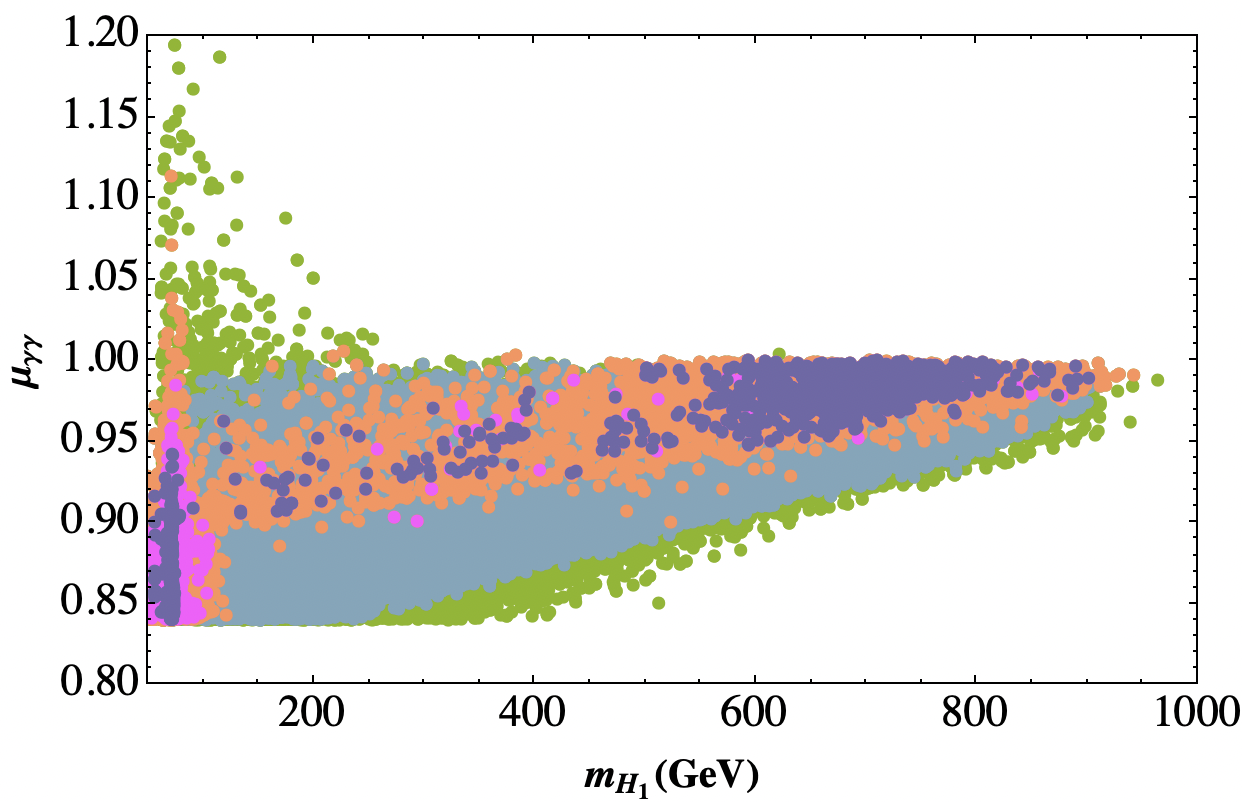}
\caption{Constraints on $\mu_{\gamma\gamma}$ as a function of $m_{H_1}$.
The colours of points have the same meaning as in Fig.~\ref{direct}.
}
\label{hgg}
\end{figure}

Recall,
that since we have only one active scalar, all tree level $125~{\rm GeV}$ Higgs
couplings are as in the SM model.
It is only in loop mediated (or loop corrections to tree level) decays
that we are sensitive to the dark sectors.
As an example,
we show results for the $h \rightarrow \gamma\gamma$ decay in Fig.~\ref{hgg}.
We start by noticing that, after theoretical and LHC constraints,
there is a region where $\mu_{\gamma\gamma}$ could exceed unity.
Although not completely apparent from the figure,
this region only starts after $M_{H_1} > m_h/2$, and decreases sharply until around
$m_{H_1} \sim 300~{\rm GeV}$.
This is exactly the same that one finds in the IDM;
compare, for example, with figure 3-left of~\cite{Krawczyk:2015vka};
see also~\cite{Krawczyk:2013jta}.
However,
this region is already excluded by the LZ direct detection bound.
Thus, in the $\Z2\times\Z2$ 3HDM, current astrophysical bounds force
$\mu_{\gamma\gamma} \lesssim 1$.

\begin{figure}[htbp!]
\centering
\includegraphics[width=0.4\textwidth]{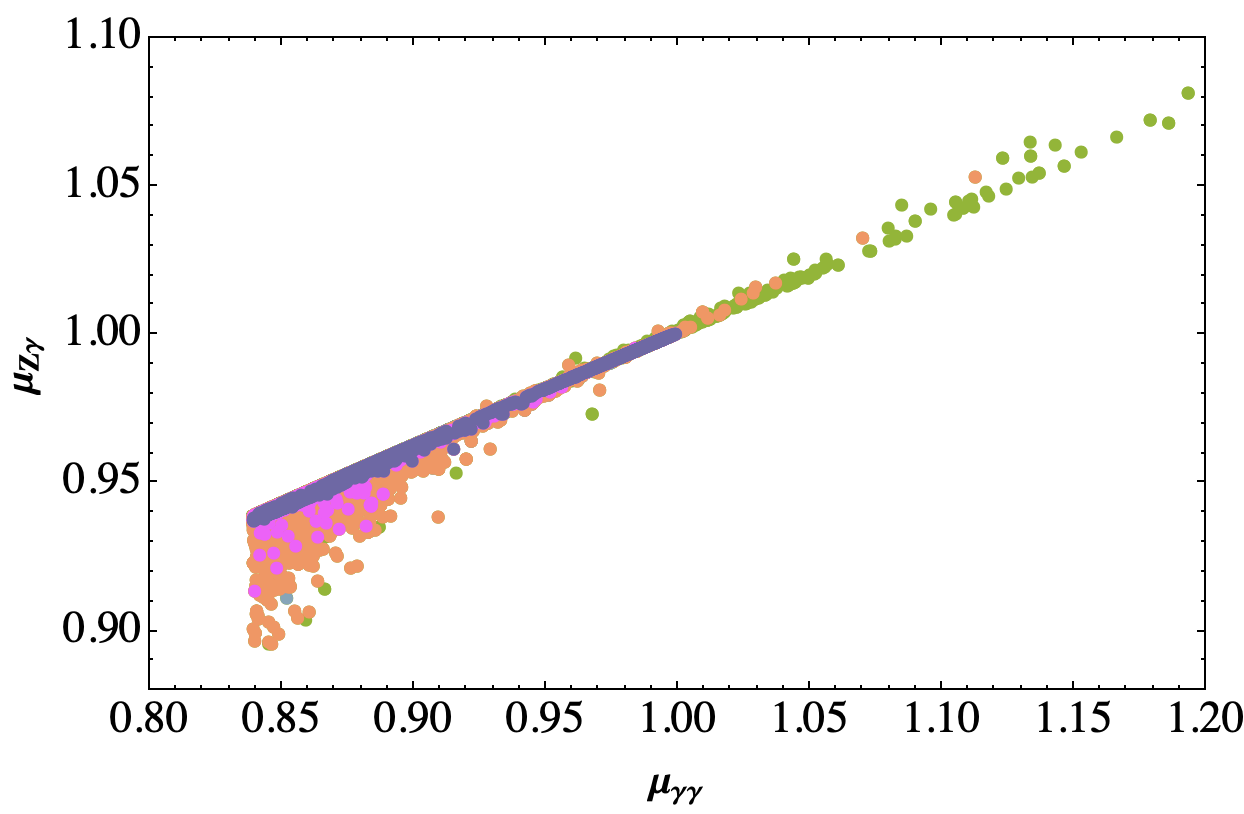}
\caption{Constraints on the $\mu_{\gamma\gamma} - \mu_{Z\gamma}$ plane.
The colours of points have the same meaning as in Fig.~\ref{direct}.
}
\label{hzg}
\end{figure}

As for $h \rightarrow Z \gamma$, 
the current measurement of $ \mu_{Z \gamma} = 2.2 \pm 0.7$
~\cite{ATLAS:2020qcv,ATLAS:2023yqk,CMS:2022ahq}, still has large errors.
Nonetheless,
if its central value remained with shrinking errors,
it would constitute a definite sign of new Physics.
We show the results for the $\mu_{\gamma\gamma}$ versus $\mu_{Z \gamma}$
within the $\Z2\times\Z2$ 3HDM in Fig.~\ref{hzg}.
One sees a very strong correlation between
$\mu_{\gamma\gamma}$ and $\mu_{Z\gamma}$,
due to the fact that very similar new virtual charge Higgs diagrams
are involved in both cases.\footnote{It is viable to uncorrelate
$\mu_{Z\gamma}$ from $\mu_{\gamma\gamma}$ in multi-Higgs models
where the $Z$ couples to two different charged Higgses, as shown with our study in Section~\ref{studyhzgamma}.
}
As a result, should a more precise measurement of
$\mu_{Z \gamma}$ uncover new physics,
the $\Z2\times\Z2$ 3HDM would be ruled out together with the SM.

\section{Summary}
\label{sec:concl}

Recently there has been renewed interest in multi-component DM models.
In this Chapter we focus our attention on a  $\Z2\times\Z2$ symmetric 3HDM with a double
inert vacuum.
We start by reassessing the possible solutions of the stationarity equations,
making sure that ours is the absolute minimum.
We found new relevant minima, which we dub \texttt{F0DM0}'
and \texttt{F0CB}.
To be certain that we keep all the points that are global minima, we
have to compare not only with the \texttt{F0DM1}, \texttt{F0DM2} and
\texttt{F0DM0}, but also with the new cases, \texttt{F0DM0}' and 
\texttt{F0CB}.
We obtained explicit expressions for all the cases and
therefore it is easy to compare.

After this step, we subject the parameter space of our model to all
current theoretical, collider and DM constraints, including limits in the $125~{\rm GeV}$ couplings,
searches for extra scalars and 
flavour observables.
We concentrate on the implications from relic density, direct and indirect detection 
of DM. By performing a wide scan, we found that simple implications obtained
when concentrating on small regions of parameter space cease to be valid,
and a much richer pallet of possibilities emerges.
In particular, we found regions where one can have two DM candidates contributing
equally to the relic density. The whole mass range for a given component can be populated in the $\Z2\times\Z2$ model, even for intermediate mass regions that require the other component to dominate the relic density calculation. We include the future sensitivity of DD experiments that are expected to reach the high mass section of the neutrino fog without being able to invalidate the model, thus requiring complementary probes.


%% file: chapters/conclusion.tex

\chapter{Conclusion}
\label{chapter:conclusion}
\hspace*{0.3cm}
After the discovery at LHC by ATLAS~\cite{ATLAS:2012yve} and
CMS~\cite{CMS:2012qbp} of the $125~{\rm GeV}$ scalar ($h_{125}$),
we have entered the epoch of the precise determination of its properties.
And, given that one such scalar exists, one is also eager to learn
whether additional fundamental scalars exist in Nature. These two ideas cross-fertilize: precise determination of $h_{125}$
couplings constrain the properties of theories with extra scalars;
theories with extra scalars point out avenues for exploration
of peculiar $h_{125}$ properties. In this thesis, we contributed to the study of three Higgs Doublets Models (3HDM), starting from their formulation and consistency, to their physical implications after taking into account all existing constraints. We consider new sources of Charge Parity (CP) violation and viable Dark Matter (DM) candidates.

In Chapter~\ref{chapter:3HDM} we derive sufficient conditions for the scalar potential to be bounded from below and complete with Appendix~\ref{a:comparison_bfb}, where we argue for the necessity of considering charge breaking directions and that our derived sufficient conditions do not show physical differences from using necessary \textit{and} sufficient conditions in available cases. 

In Chapter~\ref{chapter:Constraints} and~\ref{chapter:Alignment}, we consider all theoretical and experimental bounds on real 3HDMs to perform full phenomenological studies for different discrete symmetries. We devise a method of sampling away from the alignment limit in the literature, to study in detail the individual impact of constraints and attempt to distinguish models. We highlight the impact of the decays: $h\to \gamma\gamma$ , $h\to Z\gamma$, $b \to s \gamma$; and continue with the importance of the precise measurement of the trilinear Higgs self coupling modifier $(\kappa_h)$, as it can distinguish models. 

We follow with a new 3HDM where (some) parameters
in the potential are complex with the possibility that all fermions of a given charge couple to a different scalar doublet, which we named the C3HDM for the specific softly-broken $\Z2\times\Z2$ symmetry chosen. In Chapter~\ref{chapter:C3HDM}, we introduced a parameterization for the rotation matrices needed for the diagonalization
of the neutral and charged scalars, defining the alignment and real limits. In Chapter~\ref{chapter:ml}, we consider all bounds, including the electron's Electric Dipole Moment (EDM) present in models that explicitly violate CP symmetry, and find the possibility of having large
pseudoscalar couplings for the fermions.

We have looked at the impact of Machine Learning techniques in the full exploration of the parameter space
and physical consequences of the C3HDM.
Specifically,
we have used an Evolutionary Strategy algorithm
integrated with an anomaly detection-based Novelty Reward mechanism,
to ensure robust exploration not only of the model’s parameter space but,
crucially, of its physical implications. 

We have demonstrated the effectiveness of this new method,
by pitting it against the traditional scanning approach we also developed.
This allowed for the efficient identification of valid points within the model parameter space.
Importantly,
because the method converges rapidly to points with novel characteristics,
one is able to fully explore the phenomenological implications of the model.
In this respect, the differences between the blue (old method) and green (new method)
regions in Figs.~\ref{fig-bothwrong}-Left and~\ref{fig-top} are impressive. This is a particular application,
as the method can be used to great advantage
in any model with large numbers of parameters.

We then turn to the possibility of the 3HDM having viable Dark Matter candidates. In Chapter~\ref{chapter:Darkmatter_model} we attempt to have scalar DM candidates stabilized by a non-abelian
residual group as a viable option within multi-Higgs-doublet model, with the example of the global $\Sigma(36)$ symmetry.  This symmetry leads to very tight relations between the $125~{\rm GeV}$ Higgs, the quark sector and the dark sector, leading to conflict with observations. We study in detail the areas of tension and explore the limits of what multi-Higgs doublet models can accommodate.

In Chapter~\ref{chapter:Darkmatter_multi} we consider a viable model with two Dark Matter (DM) candidates. We found new solutions of the stationarity equations that may be below the desired minimum, and need to be compared to. We consider all theoretical, collider and Dark Matter constraints and study the implications on the obtained relic density and experimental searches for DM, in the present and future.

In the course of this thesis, eight articles were produced, as described in the preface. We hope the work produced offers a promising direction for uncovering novel and physically meaningful solutions in scalar extensions of the SM, with contributions to building models, considering experimental searches and methodology for sampling the allowed parameter space. 

%% file: appendix/main.tex

\include{appendix/lambdas}
\include{appendix/yukawa_types}

\include{appendix/bfb_comp}
\include{appendix/offdiagonal}

\include{appendix/hzgamma}
\include{appendix/param_c3hdm}
\include{appendix/sigma36_reps}
\include{appendix/sigma36_yukawa}

\include{appendix/z2z2_minima}

%% file: appendix/lambdas.tex
\chapter{\label{app:lambdas}Potential parameters in terms of physical variables}

We list here the relation of the parameters of the potential and
masses and angles for the the 3HDM potentials considered. We follow the parameterization described in Section~\ref{sec:3hdmphys} for the real potentials and the procedure in Chapter~\ref{chapter:C3HDM} for the complex $\Z2\times\Z2$. 

\section{\texorpdfstring{The $\U1\times\U1$ potential}{The U1XU1 potential}}

As there are no $\lambda''_{10}$, $\lambda''_{11}$,  and $\lambda''_{12}$,
we can also solve for the three  soft-breaking terms in \eq{3hdmquadratic}.  The expressions are
\begin{align} 
\label{eq:41} 
  \lambda_1=& 
\frac{1}{2 c_{\beta_{1}}^3
  c_{\beta_{2}}^3 v^2}
\left[c_{\alpha_{1}}^2 c_{\alpha_{2}}^2 c_{\beta_{1}} c_{\beta_{2}}
m_{h}^2+c_{\alpha_{1}}^2 c_{\alpha_{3}}^2
   c_{\beta_{1}} c_{\beta_{2}} m_{H_{2}}^2 s_{\alpha_{2}}^2+c_{\alpha_{1}}^2
   c_{\beta_{1}} c_{\beta_{2}} m_{H_{1}}^2 s_{\alpha_{2}}^2
   s_{\alpha_{3}}^2\right.\nonumber\\
&\left.
   +2
   c_{\alpha_{1}} c_{\alpha_{3}} c_{\beta_{1}} c_{\beta_{2}} m_{H_{1}}^2 s_{\alpha_{1}}
   s_{\alpha_{2}} s_{\alpha_{3}}-2 c_{\alpha_{1}} c_{\alpha_{3}} c_{\beta_{1}} c_{\beta_{2}}
   m_{H_{2}}^2 s_{\alpha_{1}} s_{\alpha_{2}} s_{\alpha_{3}}+c_{\alpha_{3}}^2
   c_{\beta_{1}} c_{\beta_{2}} m_{H_{1}}^2 s_{\alpha_{1}}^2\right.\nonumber\\
&\left.
   +c_{\beta_{1}}
   c_{\beta_{2}} m_{H_{2}}^2 s_{\alpha_{1}}^2 s_{\alpha_{3}}^2+c_{\beta_{2}}
   m_{12}^2 s_{\beta_{1}}+m_{13}^2 s_{\beta_{2}}\right]\, ,
  \\[+2mm]
  \lambda_2=&\frac{1}{2
    c_{\beta_{2}}^3 s_{\beta_{1}}^3 v^2}
  \left[c_{\alpha_{1}}^2 c_{\alpha_{3}}^2 c_{\beta_{2}} m_{H_{1}}^2
  s_{\beta_{1}}+c_{\alpha_{1}}^2 c_{\beta_{2}}
   m_{H_{2}}^2 s_{\alpha_{3}}^2 s_{\beta_{1}}-2 c_{\alpha_{1}} c_{\alpha_{3}}
   c_{\beta_{2}} m_{H_{1}}^2 s_{\alpha_{1}} s_{\alpha_{2}} s_{\alpha_{3}}
   s_{\beta_{1}}\right.\nonumber\\
&\left.\hskip 15mm
   +2 c_{\alpha_{1}} c_{\alpha_{3}} c_{\beta_{2}} m_{H_{2}}^2
   s_{\alpha_{1}} s_{\alpha_{2}} s_{\alpha_{3}} s_{\beta_{1}}
   +c_{\alpha_{2}}^2 c_{\beta_{2}} m_{h}^2 s_{\alpha_{1}}^2
   s_{\beta_{1}}+c_{\alpha_{3}}^2 c_{\beta_{2}} 
   m_{H_{2}}^2 s_{\alpha_{1}}^2 s_{\alpha_{2}}^2 s_{\beta_{1}}
 \right.\nonumber\\
 &\left.\hskip 15mm
   +c_{\beta_{1}}
   c_{\beta_{2}} m_{12}^2+c_{\beta_{2}} m_{H_{1}}^2 s_{\alpha_{1}}^2
   s_{\alpha_{2}}^2 s_{\alpha_{3}}^2 s_{\beta_{1}}+m_{23}^2 s_{\beta_{2}}\right]\, ,
  \\[+2mm]
  \lambda_3=&\frac{1}{2
   s_{\beta_{2}}^3 v^2}\left[c_{\alpha_{2}}^2 c_{\alpha_{3}}^2 m_{H_{2}}^2
  s_{\beta_{2}}+c_{\alpha_{2}}^2 m_{H_{1}}^2 s_{\alpha_{3}}^2
   s_{\beta_{2}}+c_{\beta_{1}} c_{\beta_{2}} m_{13}^2+c_{\beta_{2}}
   m_{23}^2 s_{\beta_{1}}+m_{h}^2 s_{\alpha_{2}}^2 s_{\beta_{2}}\right]
  \\[+2mm]
  \lambda_4=&\frac{1}{c_{\beta_{1}} c_{\beta_{2}}^2 s_{\beta_{1}} v^2}
  \left[-c_{\alpha_{1}}^2 c_{\alpha_{3}} m_{H_{1}}^2 s_{\alpha_{2}}
  s_{\alpha_{3}}+c_{\alpha_{1}}^2 c_{\alpha_{3}}
   m_{H_{2}}^2 s_{\alpha_{2}} s_{\alpha_{3}}+c_{\alpha_{1}} c_{\alpha_{2}}^2
   m_{h}^2 s_{\alpha_{1}}\right.\nonumber\\
&\left.\hskip 20mm
   -c_{\alpha_{1}} c_{\alpha_{3}}^2 m_{H_{1}}^2
   s_{\alpha_{1}}+c_{\alpha_{1}} c_{\alpha_{3}}^2 m_{H_{2}}^2 s_{\alpha_{1}}
   s_{\alpha_{2}}^2+c_{\alpha_{1}} m_{H_{1}}^2 s_{\alpha_{1}} s_{\alpha_{2}}^2
   s_{\alpha_{3}}^2 \right.\nonumber\\
&\left.\hskip 20mm
 -c_{\alpha_{1}} m_{H_{2}}^2 s_{\alpha_{1}}
   s_{\alpha_{3}}^2+c_{\alpha_{3}} m_{H_{1}}^2 s_{\alpha_{1}}^2 s_{\alpha_{2}}
   s_{\alpha_{3}}-c_{\alpha_{3}} m_{H_{2}}^2 s_{\alpha_{1}}^2 s_{\alpha_{2}}
   s_{\alpha_{3}}\right.\nonumber\\
&\left.\hskip 20mm
   -c_{\beta_{1}} c_{\beta_{2}}^2 \lambda_{7} s_{\beta_{1}}
   v^2-m_{12}^2 \right]\, ,
  \\[+2mm]
  \lambda_5=&\frac{1}{c_{\beta_{1}} c_{\beta_{2}} s_{\beta_{2}} v^2}
  \left[-c_{\alpha_{1}} c_{\alpha_{2}} c_{\alpha_{3}}^2 m_{H_{2}}^2
  s_{\alpha_{2}}+c_{\alpha_{1}} c_{\alpha_{2}}
   m_{h}^2 s_{\alpha_{2}}-c_{\alpha_{1}} c_{\alpha_{2}} m_{H_{1}}^2
   s_{\alpha_{2}} s_{\alpha_{3}}^2\right.\nonumber\\
&\left.\hskip 20mm
   -c_{\alpha_{2}} c_{\alpha_{3}} m_{H_{1}}^2
   s_{\alpha_{1}} s_{\alpha_{3}}+c_{\alpha_{2}} c_{\alpha_{3}} m_{H_{2}}^2 s_{\alpha_{1}}
   s_{\alpha_{3}}-c_{\beta_{1}} c_{\beta_{2}} \lambda_{8} s_{\beta_{2}}
   v^2-m_{13}^2\right]\, ,
  \\[+2mm]
  \lambda_6=&\frac{1}{c_{\beta_{2}} s_{\beta_{1}} s_{\beta_{2}} v^2}
  \left[c_{\alpha_{1}} c_{\alpha_{2}} c_{\alpha_{3}} m_{H_{1}}^2
  s_{\alpha_{3}}-c_{\alpha_{1}} c_{\alpha_{2}} c_{\alpha_{3}}
   m_{H_{2}}^2 s_{\alpha_{3}}-c_{\alpha_{2}} c_{\alpha_{3}}^2 m_{H_{2}}^2
   s_{\alpha_{1}} s_{\alpha_{2}}\right.\nonumber\\
&\left.\hskip 20mm
   +c_{\alpha_{2}} m_{h}^2 s_{\alpha_{1}}
   s_{\alpha_{2}}-c_{\alpha_{2}} m_{H_{1}}^2 s_{\alpha_{1}} s_{\alpha_{2}}
   s_{\alpha_{3}}^2-c_{\beta_{2}} \lambda_{9} s_{\beta_{1}} s_{\beta_{2}}
   v^2-m_{23}^2 \right]\, ,
  \\[+2mm]
  \lambda_7=&-\frac{2}{c_{\beta_{1}}
    c_{\beta_{2}}^2 s_{\beta_{1}} v^2 }
  \left[ c_{\beta_{1}}^2 \left(c_{\gamma_{2}} s_{\beta_{2}}
  s_{\gamma_{2}}
   \left(m_{H_2^\pm}^2-m_{H_1^\pm}^2\right)+m_{12}^2\right)+c_{\beta_{1}}
   s_{\beta_{1}} \left(c_{\gamma_{2}}^2 \left(m_{H_1^\pm}^2-m_{H_2^\pm}^2
       s_{\beta_{2}}^2\right)\right.\right.\nonumber\\
&\left.\left.\hskip 20mm
     +s_{\gamma_{2}}^2 \left(m_{H_2^\pm}^2-m_{H_1^\pm}^2
   s_{\beta_{2}}^2\right)\right)+s_{\beta_{1}}^2 \left(c_{\gamma_{2}} s_{\beta_{2}}
   s_{\gamma_{2}}
   \left(m_{H_1^\pm}^2-m_{H_2^\pm}^2\right)+m_{12}^2\right) \right]\, ,
  \\[+2mm]
  \lambda_8=&-\frac{2}{c_{\beta_{1}}
   c_{\beta_{2}} s_{\beta_{2}} v^2} \left[c_{\beta_{1}} c_{\beta_{2}} s_{\beta_{2}}
  \left(c_{\gamma_{2}}^2 m_{H_2^\pm}^2+m_{H_1^\pm}^2
   s_{\gamma_{2}}^2\right)+c_{\beta_{2}} c_{\gamma_{2}} s_{\beta_{1}} s_{\gamma_{2}}
 \left(m_{H_2^\pm}^2-m_{H_1^\pm}^2\right)\right.\nonumber\\
&\left.\hskip 25mm
 +m_{13}^2\right]\, ,
  \\[+2mm]
  \lambda_9=&-\frac{2}{c_{\beta_{2}} s_{\beta_{1}}
    s_{\beta_{2}} v^2}
  \left[c_{\beta_{2}} \left(c_{\beta_{1}} c_{\gamma_{2}}
  s_{\gamma_{2}}
   \left(m_{H_1^\pm}^2-m_{H_2^\pm}^2\right)+c_{\gamma_{2}}^2 m_{H_2^\pm}^2
   s_{\beta_{1}} s_{\beta_{2}}+m_{H_1^\pm}^2 s_{\beta_{1}} s_{\beta_{2}}
   s_{\gamma_{2}}^2\right)\right.\nonumber\\
&\left.\hskip 25mm
 +m_{23}^2\right]\, ,
  \\[+2mm]
  m^2_{12}=&c_{\beta_{1}}^2 c_{\gamma_{1}} s_{\beta_{2}} s_{\gamma_{1}}
   \left(m_{A_{1}}^2-m_{A_{2}}^2\right)+c_{\beta_{1}} s_{\beta_{1}}
   \left(c_{\gamma_{1}}^2 \left(m_{A_{2}}^2
   s_{\beta_{2}}^2-m_{A_{1}}^2\right)+s_{\gamma_{1}}^2 \left(m_{A_{1}}^2
   s_{\beta_{2}}^2-m_{A_{2}}^2\right)\right)\nonumber\\
&\hskip 2mm
+c_{\gamma_{1}} s_{\beta_{1}}^2
   s_{\beta_{2}} s_{\gamma_{1}}
   \left(m_{A_{2}}^2-m_{A_{1}}^2\right)\, ,
  \\[+2mm]
  m^2_{13}=&-c_{\beta_{2}} \left(c_{\beta_{1}} s_{\beta_{2}} \left(c_{\gamma_{1}}^2
  m_{A_{2}}^2+m_{A_{1}}^2
   s_{\gamma_{1}}^2\right)+c_{\gamma_{1}} s_{\beta_{1}} s_{\gamma_{1}}
   \left(m_{A_{2}}^2-m_{A_{1}}^2\right)\right)\, ,
  \\[+2mm]
  m^2_{23}=&-c_{\beta_{2}} \left(c_{\beta_{1}} c_{\gamma_{1}} s_{\gamma_{1}}
   \left(m_{A_{1}}^2-m_{A_{2}}^2\right)+c_{\gamma_{1}}^2 m_{A_{2}}^2
   s_{\beta_{1}} s_{\beta_{2}}+m_{A_{1}}^2 s_{\beta_{1}} s_{\beta_{2}}
   s_{\gamma_{1}}^2\right)\, .
\end{align}

\section{\texorpdfstring{The $\U1\times\Z2$ potential}{The U1XZ2 potential}}

As there are no $\lambda''_{11}$ and $\lambda''_{12}$,
we can also solve for two of the soft-breaking terms in \eq{3hdmquadratic}. We choose
$m^2_{13},m^2_{23}$ leaving $m^2_{12}$ as independent. The expressions are
\begin{align}
\label{eq:42}
\lambda_1=&\frac{1}{2 c_{\beta_{1}}^3 c_{\beta_{2}}^3 v^2}
\left[c_{\alpha_{1}}^2 c_{\beta_{1}} c_{\beta_{2}}
  \left(c_{\alpha_{2}}^2
   m_{h}^2+s_{\alpha_{2}}^2 \left(c_{\alpha_{3}}^2
     m_{H_{2}}^2+m_{H_{1}}^2 s_{\alpha_{3}}^2\right)\right)\right.\nonumber\\
&\left.\hskip 20mm
 +2 c_{\alpha_{1}}
   c_{\alpha_{3}} c_{\beta_{1}} c_{\beta_{2}} s_{\alpha_{1}} s_{\alpha_{2}} s_{\alpha_{3}}
   \left(m_{H_{1}}^2-m_{H_{2}}^2\right)+c_{\alpha_{3}}^2 c_{\beta_{1}}
   c_{\beta_{2}} m_{H_{1}}^2 s_{\alpha_{1}}^2\right.\nonumber\\
&\left.\hskip 20mm
   +c_{\beta_{1}} c_{\beta_{2}}
   m_{H_{2}}^2 s_{\alpha_{1}}^2 s_{\alpha_{3}}^2+c_{\beta_{2}} m_{12}^2
   s_{\beta_{1}}+m_{13}^2 s_{\beta_{2}}\right]\, ,
  \\[+2mm]
  \lambda_2=&\frac{1}{2 c_{\beta_{2}}^3
    s_{\beta_{1}}^3 v^2}
  \left[c_{\alpha_{1}}^2 c_{\beta_{2}} s_{\beta_{1}} \left(c_{\alpha_{3}}^2
   m_{H_{1}}^2+m_{H_{2}}^2 s_{\alpha_{3}}^2\right)+2 c_{\alpha_{1}} c_{\alpha_{3}}
   c_{\beta_{2}} s_{\alpha_{1}} s_{\alpha_{2}} s_{\alpha_{3}} s_{\beta_{1}}
   \left(m_{H_{2}}^2-m_{H_{1}}^2\right)\right.\nonumber\\
 &\left.\hskip 20mm
   +c_{\alpha_{2}}^2 c_{\beta_{2}}
   m_{h}^2 s_{\alpha_{1}}^2 s_{\beta_{1}}+c_{\alpha_{3}}^2 c_{\beta_{2}}
   m_{H_{2}}^2 s_{\alpha_{1}}^2 s_{\alpha_{2}}^2 s_{\beta_{1}}+c_{\beta_{1}} c_{\beta_{2}}
   m_{12}^2\right.\nonumber\\
 &\left.\hskip 20mm
   +c_{\beta_{2}} m_{H_{1}}^2 s_{\alpha_{1}}^2 s_{\alpha_{2}}^2
   s_{\alpha_{3}}^2 s_{\beta_{1}}+m_{23}^2 s_{\beta_{2}}\right]\, ,
  \\[+2mm]
  \lambda_3=&\frac{1}{2 s_{\beta_{2}}^3 v^2}
  \left[c_{\alpha_{2}}^2 c_{\alpha_{3}}^2 m_{H_{2}}^2 s_{\beta_{2}}+c_{\alpha_{2}}^2
   m_{H_{1}}^2 s_{\alpha_{3}}^2 s_{\beta_{2}}+c_{\beta_{1}} c_{\beta_{2}}
   m_{13}^2+c_{\beta_{2}} m_{23}^2 s_{\beta_{1}}+m_{h}^2
   s_{\alpha_{2}}^2 s_{\beta_{2}}\right]\, ,
  \\[+2mm]
  \lambda_4=&\frac{1}{c_{\beta_{1}} c_{\beta_{2}}^2 s_{\beta_{1}} v^2}
  \left[c_{\alpha_{1}}^2 c_{\alpha_{3}} s_{\alpha_{2}} s_{\alpha_{3}}
   \left(m_{H_{2}}^2-m_{H_{1}}^2\right)+c_{\alpha_{1}} s_{\alpha_{1}}
   \left(c_{\alpha_{2}}^2 m_{h}^2+c_{\alpha_{3}}^2 \left(m_{H_{2}}^2
       s_{\alpha_{2}}^2-m_{H_{1}}^2\right)\right.\right.\nonumber\\
 &\left.\left.\hskip 20mm
     +s_{\alpha_{3}}^2 \left(m_{H_{1}}^2
   s_{\alpha_{2}}^2-m_{H_{2}}^2\right)\right)+c_{\alpha_{3}} m_{H_{1}}^2
   s_{\alpha_{1}}^2 s_{\alpha_{2}} s_{\alpha_{3}}-c_{\alpha_{3}} m_{H_{2}}^2 s_{\alpha_{1}}^2
   s_{\alpha_{2}} s_{\alpha_{3}}\right.\nonumber\\
 &\left.\hskip 20mm
   -c_{\beta_{1}} c_{\beta_{2}}^2 \lambda_{7} s_{\beta_{1}} v^2-2
   c_{\beta_{1}} c_{\beta_{2}}^2 \lambda''_{10} s_{\beta_{1}}
   v^2-m_{12}^2 \right]\, ,
  \\[+2mm]
  \lambda_5=&-\frac{1}{c_{\beta_{1}} c_{\beta_{2}} s_{\beta_{2}} v^2}
  \left[c_{\alpha_{1}} c_{\alpha_{2}} s_{\alpha_{2}} \left(c_{\alpha_{3}}^2
   m_{H_{2}}^2-m_{h}^2+m_{H_{1}}^2 s_{\alpha_{3}}^2\right)+c_{\alpha_{2}}
 c_{\alpha_{3}} m_{H_{1}}^2 s_{\alpha_{1}} s_{\alpha_{3}}\right.\nonumber\\
&\left.\hskip 25mm
 -c_{\alpha_{2}} c_{\alpha_{3}}
   m_{H_{2}}^2 s_{\alpha_{1}} s_{\alpha_{3}}+c_{\beta_{1}} c_{\beta_{2}} \lambda_{8}
   s_{\beta_{2}} v^2+m_{13}^2 \right]\, ,
  \\[+2mm]
  \lambda_6=&-\frac{1}{c_{\beta_{2}} s_{\beta_{1}} s_{\beta_{2}}
    v^2}
  \left[c_{\alpha_{2}} \left(c_{\alpha_{1}} c_{\alpha_{3}} s_{\alpha_{3}}
   \left(m_{H_{2}}^2-m_{H_{1}}^2\right)+c_{\alpha_{3}}^2 m_{H_{2}}^2
   s_{\alpha_{1}} s_{\alpha_{2}}+m_{h}^2 (-s_{\alpha_{1}})
   s_{\alpha_{2}}\right.\right.\nonumber\\
&\left.\left.\hskip 25mm
   +m_{H_{1}}^2
   s_{\alpha_{1}} s_{\alpha_{2}} s_{\alpha_{3}}^2\right)+c_{\beta_{2}} \lambda_{9}
   s_{\beta_{1}} s_{\beta_{2}} v^2+m_{23}^2 \right]\, ,
  \\[+2mm]
  \lambda_7=&-\frac{2}{
    c_{\beta_{1}} c_{\beta_{2}}^2 s_{\beta_{1}} v^2}
  \left[c_{\beta_{1}}^3 c_{\beta_{2}}^2 \lambda''_{10} s_{\beta_{1}}
   v^2+c_{\beta_{1}}^2 \left(c_{\gamma_{2}} s_{\beta_{2}} s_{\gamma_{2}}
     \left(m_{H_2^\pm}^2-m_{H_1^\pm}^2\right)+m_{12}^2\right)
 \right.\nonumber\\ 
&\left.\hskip 25mm
   +c_{\beta_{1}}
   s_{\beta_{1}} \left(c_{\beta_{2}}^2 \lambda''_{10} s_{\beta_{1}}^2 v^2+c_{\gamma_{2}}^2
   \left(m_{H_1^\pm}^2-m_{H_2^\pm}^2 s_{\beta_{2}}^2\right)-m_{H_1^\pm}^2
   s_{\beta_{2}}^2 s_{\gamma_{2}}^2+m_{H_2^\pm}^2
   s_{\gamma_{2}}^2\right) \right.\nonumber\\ 
&\left.\hskip 25mm
 +s_{\beta_{1}}^2
   \left(c_{\gamma_{2}} s_{\beta_{2}} s_{\gamma_{2}}
   \left(m_{H_1^\pm}^2-m_{H_2^\pm}^2\right)+m_{12}^2\right)\right]\, ,
  \\[+2mm]
  \lambda_8=&-\frac{2}{c_{\beta_{1}}
    c_{\beta_{2}} s_{\beta_{2}} v^2}
  \left[c_{\beta_{1}} c_{\beta_{2}} s_{\beta_{2}} \left(c_{\gamma_{2}}^2
   m_{H_2^\pm}^2+m_{H_1^\pm}^2 s_{\gamma_{2}}^2\right)+c_{\beta_{2}} c_{\gamma_{2}}
   s_{\beta_{1}} s_{\gamma_{2}}
   \left(m_{H_2^\pm}^2-m_{H_1^\pm}^2\right)+m_{13}^2\right]\, ,
  \\[+2mm]
  \lambda_9=&-\frac{2}{c_{\beta_{2}} s_{\beta_{1}}
    s_{\beta_{2}} v^2}
  \left[c_{\beta_{2}} \left(c_{\beta_{1}} c_{\gamma_{2}} s_{\gamma_{2z}}
   \left(m_{H_1^\pm}^2-m_{H_2^\pm}^2\right)+c_{\gamma_{2}}^2 m_{H_2^\pm}^2
   s_{\beta_{1}} s_{\beta_{2}}+m_{H_1^\pm}^2 s_{\beta_{1}} s_{\beta_{2}}
   s_{\gamma_{2}}^2\right)+m_{23}^2\right]\, ,
  \\[+2mm]
  \lambda''_{10}=&-\frac{1}{2
    c_{\beta_{1}} c_{\beta_{2}}^2 s_{\beta_{1}} v^2}
  \left[c_{\beta_{1}}^2 \left(c_{\gamma_{1}} s_{\beta_{2}} s_{\gamma_{1}}
   \left(m_{A_{2}}^2-m_{A_{1}}^2\right)+m_{12}^2\right)+c_{\beta_{1}}
   s_{\beta_{1}} \left(c_{\gamma_{1}}^2 \left(m_{A_{1}}^2-m_{A_{2}}^2
       s_{\beta_{2}}^2\right)\right.\right.\nonumber\\
&\left.\left.\hskip 25mm
     +s_{\gamma_{1}}^2 \left(m_{A_{2}}^2-m_{A_{1}}^2
   s_{\beta_{2}}^2\right)\right)+s_{\beta_{1}}^2 \left(c_{\gamma_{1}} s_{\beta_{2}}
   s_{\gamma_{1}} \left(m_{A_{1}}^2-m_{A_{2}}^2\right)+m_{12}^2\right)\right]\, ,
  \\[+2mm]
  m^2_{13}=&-c_{\beta_{2}} \left[c_{\beta_{1}} s_{\beta_{2}} \left(c_{\gamma_{1}}^2
   m_{A_{2}}^2+m_{A_{1}}^2 s_{\gamma_{1}}^2\right)+c_{\gamma_{1}} s_{\beta_{1}}
 s_{\gamma_{1}} \left(m_{A_{2}}^2-m_{A_{1}}^2\right)\right]\, ,
\\[+2mm]
  m^2_{23}=&-c_{\beta_{2}} \left[c_{\beta_{1}} c_{\gamma_{1}} s_{\gamma_{1}}
   \left(m_{A_{1}}^2-m_{A_{2}}^2\right)+c_{\gamma_{1}}^2 m_{A_{2}}^2
   s_{\beta_{1}} s_{\beta_{2}}+m_{A_{1}}^2 s_{\beta_{1}} s_{\beta_{2}}
   s_{\gamma_{1}}^2\right]\, .
\end{align}

\section{\texorpdfstring{The real $\Z2\times\Z2$ potential}{The real Z2XZ2 potential}}
\begin{align}
  \lambda_{1}=&
  \frac{1}{2 c_{\beta_{1}}^3
    c_{\beta_{2}}^3 v^2}
  \left[c_{\alpha_{1}}^2 c_{\beta_{1}} c_{\beta_{2}}
    \left(c_{\alpha_{2}}^2 m_{h}^2+s_{\alpha_{2}}^2 
   \left(c_{\alpha_{3}}^2 m_{H_{2}}^2+m_{H_{1}}^2
     s_{\alpha_{3}}^2\right)\right)\right.\nonumber\\
 &\left.\hskip 20mm
 +2 c_{\alpha_{1}} 
   c_{\alpha_{3}} c_{\beta_{1}} c_{\beta_{2}} s_{\alpha_{1}}
   s_{\alpha_{2}} s_{\alpha_{3}} 
   \left(m_{H_{1}}^2-m_{H_{2}}^2\right)
   +c_{\alpha_{3}}^2 c_{\beta_{1}}
   c_{\beta_{2}} m_{H_{1}}^2 
   s_{\alpha_{1}}^2\right.\nonumber\\
 &\left.\hskip 20mm
   +c_{\beta_{1}} c_{\beta_{2}} m_{H_{2}}^2
   s_{\alpha_{1}}^2 s_{\alpha_{3}}^2+c_{\beta_{2}} 
   m^2_{12} s_{\beta_{1}}+m^2_{13} s_{\beta_{2}}\right]\, ,
\\[+2mm]
 \lambda_{2}=&
\frac{1}{2 c_{\beta_{2}}^3 s_{\beta_{1}}^3 
   v^2}\left[c_{\alpha_{1}}^2 c_{\beta_{2}} s_{\beta_{1}}
  \left(c_{\alpha_{3}}^2 m_{H_{1}}^2+m_{H_{2}}^2 
   s_{\alpha_{3}}^2\right)+2 c_{\alpha_{1}} c_{\alpha_{3}}
 c_{\beta_{2}} s_{\alpha_{1}} s_{\alpha_{2}} s_{\alpha_{3}} 
 s_{\beta_{1}} \left(m_{H_{2}}^2-m_{H_{1}}^2\right) \right. \nonumber\\
&\left.\hskip 20mm
 +c_{\alpha_{2}}^2
   c_{\beta_{2}} m_{h}^2 
   s_{\alpha_{1}}^2 s_{\beta_{1}}+c_{\alpha_{3}}^2 c_{\beta_{2}}
   m_{H_{2}}^2 s_{\alpha_{1}}^2 s_{\alpha_{2}}^2 
   s_{\beta_{1}}\right. \nonumber\\
&\left.\hskip 20mm
   +c_{\beta_{1}} c_{\beta_{2}} m^2_{12}+c_{\beta_{2}}
   m_{H_{1}}^2 s_{\alpha_{1}}^2 
   s_{\alpha_{2}}^2 s_{\alpha_{3}}^2 s_{\beta_{1}}+m^2_{23}
   s_{\beta_{2}} \right]\, ,
\\[+2mm]
 \lambda_{3}=&
 \frac{1}{2 s_{\beta_{2}}^3 v^2}
 \left[c_{\alpha_{2}}^2 c_{\alpha_{3}}^2 m_{H_{2}}^2
  s_{\beta_{2}}+c_{\alpha_{2}}^2 m_{H_{1}}^2 s_{\alpha_{3}}^2 
   s_{\beta_{2}}+c_{\beta_{1}} c_{\beta_{2}} m^2_{13}+c_{\beta_{2}} m^2_{23}
   s_{\beta_{1}}+m_{h}^2 s_{\alpha_{2}}^2 s_{\beta_{2}}\right]\, ,
 \\[+2mm]
 \lambda_{4}=&
\frac{1}{c_{\beta_{1}}
  c_{\beta_{2}}^2 s_{\beta_{1}} v^2}
\left[c_{\alpha_{1}}^2 c_{\alpha_{3}} s_{\alpha_{2}} s_{\alpha_{3}}
   \left(m_{H_{2}}^2-m_{H_{1}}^2\right)+c_{\alpha_{1}} s_{\alpha_{1}}
   \left(c_{\alpha_{2}}^2 
   m_{h}^2+c_{\alpha_{3}}^2 \left(m_{H_{2}}^2
     s_{\alpha_{2}}^2-m_{H_{1}}^2\right)\right.\right.\nonumber\\
&\left.\left.\hskip 20mm
   +s_{\alpha_{3}}^2 
   \left(m_{H_{1}}^2
     s_{\alpha_{2}}^2-m_{H_{2}}^2\right)\right)+c_{\alpha_{3}}
 m_{H_{1}}^2 
   s_{\alpha_{1}}^2 s_{\alpha_{2}} s_{\alpha_{3}}-c_{\alpha_{3}}
   m_{H_{2}}^2 s_{\alpha_{1}}^2 s_{\alpha_{2}} 
   s_{\alpha_{3}}\right.\nonumber\\
 &\left.\hskip 20mm
   -c_{\beta_{1}} c_{\beta_{2}}^2 \lambda_{7}
   s_{\beta_{1}} v^2
   -2 c_{\beta_{1}} c_{\beta_{2}}^2 
   \lambda''_{10} s_{\beta_{1}} v^2-m^2_{12}\right]\, ,
 \\[+2mm]
 \lambda_{5}=&
-\frac{1}{c_{\beta_{1}}
  c_{\beta_{2}} s_{\beta_{2}} v^2}
\left[c_{\alpha_{1}} c_{\alpha_{2}} s_{\alpha_{2}} \left(c_{\alpha_{3}}^2
   m_{H_{2}}^2-m_{h}^2+m_{H_{1}}^2
   s_{\alpha_{3}}^2\right)+c_{\alpha_{2}} c_{\alpha_{3}} 
 m_{H_{1}}^2 s_{\alpha_{1}} s_{\alpha_{3}}\right.\nonumber\\
&\left.\hskip 25mm
 -c_{\alpha_{2}}
   c_{\alpha_{3}} m_{H_{2}}^2 s_{\alpha_{1}} 
   s_{\alpha_{3}}+c_{\beta_{1}} c_{\beta_{2}} \lambda_{8}
   s_{\beta_{2}} v^2+2 c_{\beta_{1}} c_{\beta_{2}} 
   \lambda''_{11} s_{\beta_{2}} v^2+m^2_{13}\right]\, ,
 \\[+2mm]
 \lambda_{6}=&
 -\frac{1}{c_{\beta_{2}} s_{\beta_{1}} s_{\beta_{2}} v^2}
 \left[c_{\alpha_{2}} \left(c_{\alpha_{1}} c_{\alpha_{3}} s_{\alpha_{3}}
   \left(m_{H_{2}}^2-m_{H_{1}}^2\right)+c_{\alpha_{3}}^2 m_{H_{2}}^2
   s_{\alpha_{1}} 
   s_{\alpha_{2}}+m_{h}^2 (-s_{\alpha_{1}}) s_{\alpha_{2}}\right.\right.
 \nonumber\\
&\left.\left. \hskip 25mm
   +m_{H_{1}}^2
   s_{\alpha_{1}} s_{\alpha_{2}} 
   s_{\alpha_{3}}^2\right)+c_{\beta_{2}} s_{\beta_{1}} s_{\beta_{2}}
 v^2 (\lambda_{9}+2 
   \lambda''_{12})+m^2_{23}\right]\, ,
 \\[+2mm]
 \lambda_{7}=&
-\frac{2}{c_{\beta_{1}} 
  c_{\beta_{2}}^2 s_{\beta_{1}} v^2}
\left[c_{\beta_{1}}^3 c_{\beta_{2}}^2 \lambda''_{10}
    s_{\beta_{1}} v^2+c_{\beta_{1}}^2 
   \left(c_{\gamma_{2}} s_{\beta_{2}} s_{\gamma_{2}}
     \left(m_{H_2^\pm}^2-m_{H_1^\pm}^2\right)+m^2_{12}\right)\right.
 \nonumber\\
 &\left.\hskip 25mm
   +c_{\beta_{1}}
 s_{\beta_{1}} 
   \left(c_{\beta_{2}}^2 \lambda''_{10} s_{\beta_{1}}^2 v^2+c_{\gamma_{2}}^2
   \left(m_{H_1^\pm}^2-m_{H_2^\pm}^2 s_{\beta_{2}}^2\right)-m_{H_1^\pm}^2
   s_{\beta_{2}}^2 
   s_{\gamma_{2}}^2+m_{H_2^\pm}^2
   s_{\gamma_{2}}^2\right)\right.\nonumber\\
&\left.\hskip 25mm
 +s_{\beta_{1}}^2 \left(c_{\gamma_{2}}
   s_{\beta_{2}} 
   s_{\gamma_{2}}
   \left(m_{H_1^\pm}^2-m_{H_2^\pm}^2\right)+m^2_{12}\right)\right]\, ,
 \\[+2mm]
 \lambda_{8}=&
-\frac{2}{c_{\beta_{1}}
  c_{\beta_{2}}    s_{\beta_{2}} v^2}
\left[c_{\beta_{1}} c_{\beta_{2}} s_{\beta_{2}}
    \left(c_{\gamma_{2}}^2 m_{H_2^\pm}^2+\lambda''_{11} 
   v^2+m_{H_1^\pm}^2 s_{\gamma_{2}}^2\right)+c_{\beta_{2}}
 c_{\gamma_{2}} s_{\beta_{1}} s_{\gamma_{2}} 
   \left(m_{H_2^\pm}^2-m_{H_1^\pm}^2\right)\right.\nonumber\\
 &\left.\hskip 25mm
   +m^2_{13}\right]\, ,
 \\[+2mm]
 \lambda_{9}=&
-\frac{2}{c_{\beta_{2}}
  s_{\beta_{1}} s_{\beta_{2}} v^2}
\left[c_{\beta_{2}} \left(c_{\beta_{1}} c_{\gamma_{2}} s_{\gamma_{2}}
   \left(m_{H_1^\pm}^2-m_{H_2^\pm}^2\right)+c_{\gamma_{2}}^2 m_{H_2^\pm}^2
   s_{\beta_{1}} 
   s_{\beta_{2}}\right.\right.\nonumber\\
 &\left.\left.\hskip 25mm
   +s_{\beta_{1}} s_{\beta_{2}} \left(\lambda''_{12} v^2+m_{H_1^\pm}^2
   s_{\gamma_{2}}^2\right)\right)+m^2_{23}\right]\, ,
 \\[+2mm]
 \lambda''_{10}=&
-\frac{1}{2  c_{\beta_{1}} c_{\beta_{2}}^2 s_{\beta_{1}} v^2}
\left[c_{\beta_{1}}^2 \left(c_{\gamma_{1}} s_{\beta_{2}} s_{\gamma_{1}}
   \left(m_{A_{2}}^2-m_{A_{1}}^2\right)+m^2_{12}\right)+c_{\beta_{1}}
 s_{\beta_{1}} 
   \left(c_{\gamma_{1}}^2 \left(m_{A_{1}}^2-m_{A_{2}}^2
       s_{\beta_{2}}^2\right)\right.\right.\nonumber\\
&\left.\left.\hskip 25mm
     +s_{\gamma_{1}}^2 
   \left(m_{A_{2}}^2-m_{A_{1}}^2
     s_{\beta_{2}}^2\right)\right)+s_{\beta_{1}}^2
 \left(c_{\gamma_{1}} 
   s_{\beta_{2}} s_{\gamma_{1}}
   \left(m_{A_{1}}^2-m_{A_{2}}^2\right)+m^2_{12}\right)\right] \, ,
 \\[+2mm]
 \lambda''_{11}=&
-\frac{c_{\beta_{1}} c_{\beta_{2}} s_{\beta_{2}}
  \left(c_{\gamma_{1}}^2 m_{A_{2}}^2+m_{A_{1}}^2 
   s_{\gamma_{1}}^2\right)+c_{\beta_{2}} c_{\gamma_{1}} s_{\beta_{1}}
 s_{\gamma_{1}} 
   \left(m_{A_{2}}^2-m_{A_{1}}^2\right)+m^2_{13}}{2 c_{\beta_{1}}
   c_{\beta_{2}} s_{\beta_{2}}
   v^2}\, ,
 \\[+2mm]
 \lambda''_{12}=&
 -\frac{c_{\beta_{2}} \left(c_{\beta_{1}} c_{\gamma_{1}} s_{\gamma_{1}}
   \left(m_{A_{1}}^2-m_{A_{2}}^2\right)+c_{\gamma_{1}}^2 m_{A_{2}}^2
   s_{\beta_{1}} 
   s_{\beta_{2}}+m_{A_{1}}^2 s_{\beta_{1}} s_{\beta_{2}}
   s_{\gamma_{1}}^2\right)+m^2_{23}}{2 
   c_{\beta_{2}} s_{\beta_{1}} s_{\beta_{2}} v^2}\, .
\end{align}

\section{\texorpdfstring{The real $\Z3$ potential}{The real Z3 potential}}

\ba
\lambda_{1}&=&\frac{m_h^2}{2v^2}\frac{\cai^2\caii^2}{\cbi^2\cbii^2}+\frac{m_{H_1}^2}{2v^2\cbi^2\cbii^2}(\cai\saii\saiii+\sai\caiii)^2+\frac{m_{H_2}^2}{2v^2\cbi^2\cbii^2}(\cai\saii\caiii-\sai\saiii)^2\nonumber\\[8pt]
&&
+\frac{\tan\beta_1\tan\beta_2}{4\cbi^2}(\lambda_{11}\sbi+\lambda_{12}\tan\beta_2)+\frac{1}{2\cbi^3\cbii^3v^2}(m_{12}^2 \cbii  \sbi + m_{13}^2 \sbii) ,\label{3hdm-l1}\\[8pt]
\lambda_{2}&=&\frac{m_h^2}{2v^2}\frac{\sai^2\caii^2}{\sbi^2\cbii^2}+\frac{m_{H_1}^2}{2v^2\sbi^2\cbii^2}(\cai\caiii-\sai\saii\saiii)^2+\frac{m_{H_2}^2}{2v^2\sbi^2\cbii^2}(\cai\saiii+\sai\saii\caiii)^2 \nonumber\\[8pt]
&&
+\frac{\tan\beta_2}{4\sbi^2\tan\beta_1}(\lambda_{10}\cbi+\lambda_{12}\tan\beta_2)+\frac{1}{2\cbii^3\sbi^3v^2}(m_{12}^2 \cbi  \cbii + m_{23}^2 \sbii) ,\label{3hdm-l2}\\[8pt]
\lambda_3&=&\frac{m_h^2}{2v^2}\frac{\saii^2}{\sbii^2}+\frac{m_{H_1}^2\caii^2\saiii^2}{2v^2\sbii^2}+\frac{m_{H_2}^2\caii^2\caiii^2}{2v^2\sbii^2}+\frac{s_{2\beta_1}}{8\tan^3\beta_2}(\lambda_{10}\cbi+\lambda_{11}\sbi)\nonumber\\[8pt]
&&
+\frac{\cbii}{2\sbii^3v^2}(m_{13}^2 \cbi  + m_{23}^2 \sbi) ,\label{3hdm-l3} \\[8pt]
   \lambda_4&=&\frac{1}{4v^2s_{2\beta_1}\cbii^2}\left[(m_{H_1}^2-m_{H_2}^2)\left\{(-3+c_{2\alpha_2})s_{2\alpha_1}c_{2\alpha_3}-4c_{2\alpha_1}\saii s_{2\alpha_3}\right\}-2(m_{H_1}^2+m_{H_2}^2)s_{2\alpha_1}\caii^2\right]\nonumber\\[8pt]
&&
+\frac{m_h^2}{v^2}\frac{s_{2\alpha_1}\caii^2}{s_{2\beta_1}\cbii^2}-\frac{\tan\beta_2}{s_{2\beta_1}}(2\lambda_{10}\cbi+2\lambda_{11}\sbi+\lambda_{12}\tan\beta_2)-\lambda_7-\frac{m_{12}^2}{\cbi\cbii^2\sbi v^2} ,\label{3hdm-l4}\\[8pt]
\lambda_5 &=& \frac{m_h^2}{v^2}\frac{\cai s_{2\alpha_2}}{\cbi s_{2\beta_2}}-\frac{m_{H_1}^2}{v^2\cbi s_{2\beta_2}}(\cai s_{2\alpha_2} \saiii^2+\sai\caii s_{2\alpha_3})+\frac{m_{H_2}^2}{v^2\cbi s_{2\beta_2}}(\sai\caii s_{2\alpha_3}-\cai s_{2\alpha_2} \caiii^2)\nonumber\\[8pt]
&&
-\frac{\sbi}{2\tan\beta_2}(2\lambda_{10}+\lambda_{11}\tan\beta_1)-\lambda_{12}\tan\beta_1-\lambda_8-\frac{m_{13}^2}{\cbi\cbii\sbii v^2} ,\label{3hdm-l5}\\[8pt]
\lambda_6 &=& \frac{m_h^2}{v^2}\frac{\sai s_{2\alpha_2}}{\sbi s_{2\beta_2}}+\frac{m_{H_1}^2}{v^2}\frac{\caii}{\sbi s_{2\beta_2}}(-2\sai\saii\saiii^2+\cai s_{2\alpha_3})-\frac{m_{H_2}^2}{v^2}\frac{\caii}{\sbi s_{2\beta_2}}(2\sai\saii\caiii^2+\cai s_{2\alpha_3})\nonumber\\[8pt]
&&
-\frac{\cbi}{2\tan\beta_2}(\lambda_{10}\cot\beta_1+2\lambda_{11})-\lambda_{12}\cot\beta_1-\lambda_9-\frac{m_{23}^2}{\cbii\sbi\sbii v^2} ,\label{3hdm-l6}\\[8pt]
\lambda_{7} &=& \frac{(m_{C1}^2-m_{C_2}^2)}{2v^2}\left[(-3+c_{2\beta_2})\frac{c_{2\gamma_2}}{\cbii^2}+\frac{4\tan\beta_2}{\tan2\beta_1}\frac{s_{2\gamma_2}}{\cbii}\right]-\frac{(m_{C_1}^2+m_{C2}^2)}{v^2}-\lambda_{10}\frac{\tan\beta_2}{\sbi}\nonumber\\[8pt]
&&
-\lambda_{11}\frac{\tan\beta_2}{\cbi}- \frac{2 m_{12}^2}{\cbi\cbii^2\sbi v^2} ,\label{3hdm-l7}\\[8pt]
\lambda_{8} &=& \frac{m_{C1}^2}{v^2}\left(-2\sgii^2+\tan\beta_1\frac{s_{2\gamma_2}}{\sbii}\right)-\frac{m_{C2}^2}{v^2}\left(2\cgii^2+\tan\beta_1\frac{s_{2\gamma_2}}{\sbii}\right)-\lambda_{10}\sbi\cot\beta_2\nonumber\\[8pt]
&&
-\lambda_{12}\tan\beta_1 - \frac{2 m_{13}^2}{\cbi\cbii\sbii v^2}
 ,\label{3hdm-l8}\\[8pt]
\lambda_9 &=& -\frac{m_{C1}^2}{v^2}\left(2\sgii^2+\cot\beta_1\frac{s_{2\gamma_2}}{\sbii}\right)+\frac{m_{C2}^2}{v^2}\left(-2\cgii^2+\cot\beta_1\frac{s_{2\gamma_2}}{\sbii}\right)-\lambda_{11}\cbi\cot\beta_2\nonumber\\[8pt]
&&
-\lambda_{12}\cot\beta_1- \frac{2 m_{23}^2}{\cbii \sbi \sbii v^2}
 ,\label{3hdm-l9}
  \ea
 
   \ba
\lambda_{10} &=& \frac{2m_{A1}^2}{9v^2}\left[\frac{s_{2\gamma_1}}{\cbi\cbii}-\frac{2\sbi\cgi^2}{\sbii\cbii}+\frac{s_{3\beta_3}\sgi\cgi}{\sbi\cbi\cbii}+\tan\beta_2\sgi^2\left\{\frac{\tan\beta_1}{\cbi}-2\cbi\cot\beta_1\right\}\right]\nonumber\\[8pt]
&&
-\frac{m_{A2}^2}{9v^2}\left[(2c_{2\beta_1}+3)\frac{s_{2\gamma_1}}{\cbi\cbii}+4\frac{\sbi\sgi^2}{\sbii\cbii}-2\tan\beta_2\cgi^2\left\{\frac{\tan\beta_1}{\cbi}-2\cbi\cot\beta_1\right\}\right]\nonumber\\[8pt]
&&
+\frac{1}{9 \cbi^2 \cbii^2 \sbi \sbii v^2}\left[2 m_{23}^2 \sbi\sbii -4\cbi(\cbii m_{12}^2 \sbi + m_{13}^2 \sbii)\right]
    ,  \label{3hdm-l10}\\[8pt]
\lambda_{11} &=& \frac{m_{A1}^2}{9v^2}\left[-\frac{4\cbi\cgi^2}{\sbii\cbii}+\frac{(-3+2c_{2\beta_1})}{\sbi\cbii}s_{2\gamma_1}+2(\cot^4\beta_1+\cot^2\beta_1-2)\sbi\sgi^2\tan\beta_1\tan\beta_2\right]\nonumber\\[8pt]
&&
+\frac{m_{A2}^2}{9v^2}\left[-\frac{4\cbi\sgi^2}{\sbii\cbii}+\frac{(5+\cot^2\beta_1)}{\cbii}s_{2\gamma_1}\sbi+2(\cot^4\beta_1+\cot^2\beta_1-2)\sbi\cgi^2\tan\beta_1\tan\beta_2\right]\nonumber\\[8pt]
&&
-\frac{1}{9 \cbi\cbii^2 \sbi^2 \sbii v^2}(4 \cbi \cbii m_{12}^2 \sbi - 2 \cbi m_{13}^2 \sbii +4 m_{23}^2 \sbi \sbii) ,\label{3hdm-l11}\\[8pt]
\lambda_{12} &=& \frac{m_{A1}^2}{36v^2}\left[\frac{4s_{2\beta_1}\cgi^2}{\sbii^2}-\frac{4c_{2\beta_1}s_{2\gamma_1}}{\sbii}+(c_{4\beta_1}-17)\frac{\sgi^2}{\sbi\cbi}\right]\nonumber\\[8pt]
&&
+\frac{m_{A2}^2}{36v^2}\left[\frac{4s_{2\beta_1}\sgi^2}{\sbii^2}+\frac{4c_{2\beta_1}s_{2\gamma_1}}{\sbii}+(c_{4\beta_1}-17)\frac{c_{\gamma_1}^2}{\sbi\cbi}\right]\nonumber\\[8pt]
&&
+\frac{2}{9 \cbi\cbii\sbi\sbii^2 v^2}(\cbi\cbii m_{12}^2 \sbi - 2 \cbi m_{13}^2 \sbii -2 m_{23}^2 \sbi \sbii) .\label{3hdm-l12}
 \ea

\section{\label{app:lambdas_c3hdm}\texorpdfstring{The complex $\Z2\times\Z2$ potential}{The complex Z2XZ2 potential}}

The equality of matrices in \eqref{e:mass_equations} contains 19 independent real equations, linear in the squared masses, and in the potential parameters $\mu_{ij}$ and $\lambda_i v^2$, which we solve for,
\begin{equation}\label{eq:L1}
	\lambda_1 v^2 = \frac{1}{2 c_{\beta_1}^2 c_{\beta_2}^2  }\left[ t_{\beta_1} \Re(\mu_{12}) + \frac{t_{\beta_2}}{c_{\beta_1}} \Re(\mu_{13}) + R_{i1}^2 m_{h_i}^2 \right]
	\;,
\end{equation}
\begin{equation}\label{eq:L2}
	\lambda_2 v^2 = \frac{1}{2 s_{\beta_1}^2 c_{\beta_2}^2  }\left[ \frac{1}{t_{\beta_1}} \Re(\mu_{12}) + \frac{t_{\beta_2}}{s_{\beta_1}} \Re(\mu_{23}) + R_{i2}^2 m_{h_i}^2 \right]
	\;,
\end{equation}
\begin{equation}\label{eq:L3}
	\lambda_3 v^2 = \frac{1}{2 s_{\beta_2}^2  }\left[ \frac{c_{\beta_1}}{t_{\beta_2}} \Re(\mu_{13}) + \frac{s_{\beta_1}}{t_{\beta_2}} \Re(\mu_{23}) + R_{i3}^2 m_{h_i}^2 \right]
	\;,
\end{equation}
\begin{equation}\label{eq:L4}
	\lambda_4 v^2 = -\lambda_7 v^2 - 2 \Re(\lambda_{10}) v^2 + \frac{1}{c_{\beta_1} s_{\beta_1} c_{\beta_2}^2  }\left[ -\Re(\mu_{12}) + R_{i1} R_{i2} m_{h_i}^2 \right]
	\;,
\end{equation}
\begin{equation}\label{eq:L5}
	\lambda_5 v^2 = -\lambda_8 v^2 - 2 \Re(\lambda_{11}) v^2 + \frac{1}{c_{\beta_1} c_{\beta_2} s_{\beta_2}  }\left[ -\Re(\mu_{13}) + R_{i1} R_{i3} m_{h_i}^2 \right]
	\;,
\end{equation}
\begin{equation}
	\lambda_6 v^2 = -\lambda_9 v^2 - 2 \Re(\lambda_{12}) v^2+ \frac{1}{ s_{\beta_1} c_{\beta_2} s_{\beta_2}  }\left[ -\Re(\mu_{23}) + R_{i2} R_{i3} m_{h_i}^2 \right]
	\;,
\end{equation}
\begin{equation}
	\begin{split}
		\lambda_7 v^2 = - 2 \Re(\lambda_{10}) v^2 + \frac{1}{c_{\beta_1} s_{\beta_1} c_{\beta_2}^2  }\Big[ -2\Re(\mu_{12}) 
		+2\Big( 
		-c_{\beta_1}s_{\beta_1}(c_\theta^2 m_{H_1^\pm}^2 + s_\theta^2 m_{H_2^\pm}^2) \\
		+ c_{\beta_1}s_{\beta_1} s_{\beta_2}^2 (s_\theta^2 m_{H_1^\pm}^2 + c_\theta^2 m_{H_2^\pm}^2)
		+c_\theta s_\theta s_{\beta_2} (1-2 s_\varphi^2)(1-2 s_{\beta_1}^2)(m_{H_2^\pm}^2-m_{H_1^\pm}^2)
		\Big) \Big]
		\;,
	\end{split}
\end{equation}
\normalsize
\begin{equation}
	\begin{split}
		\lambda_8 v^2 = - 2 \Re(\lambda_{11}) v^2& + \frac{1}{c_{\beta_1} s_{\beta_2} c_{\beta_2}  }\Big[ -2\Re(\mu_{13}) \\
		+2 c_{\beta_2}\Big( 
		-c_{\beta_1}s_{\beta_2}&(s_\theta^2 m_{H_1^\pm}^2 + c_\theta^2 m_{H_2^\pm}^2)
		+c_\theta s_\theta s_{\beta_1} (1-2 s_\varphi^2)(m_{H_2^\pm}^2-m_{H_1^\pm}^2)
		\Big) \Big]
		\;,
	\end{split}
\end{equation}
\normalsize
\begin{equation}
	\begin{split}
		\lambda_9 v^2 = - 2 \Re(\lambda_{12}) v^2& + \frac{1}{s_{\beta_1} s_{\beta_2} c_{\beta_2}  }\Big[ -2\Re(\mu_{23}) \\
		+2 c_{\beta_2}\Big( 
		-s_{\beta_1}s_{\beta_2}&(s_\theta^2 m_{H_1^\pm}^2 + c_\theta^2 m_{H_2^\pm}^2)
		-c_\theta s_\theta c_{\beta_1} (1-2 s_\varphi^2)(m_{H_2^\pm}^2-m_{H_1^\pm}^2)
		\Big) \Big]
		\;,
	\end{split}
\end{equation}
\normalsize
\begin{equation}
	\Re(\lambda_{10}) v^2 = \frac{-1}{2 c_{\beta_2}^2} \left[
	\frac{1}{s_{\beta_1}c_{\beta_1}} \Re(\mu_{12}) + m_{h_i}^2\left( R_{i4}^2 + s_{\beta_2}\frac{2 c_{\beta_1}^2-1}{s_{\beta_1}c_{\beta_1}} R_{i4}R_{i5} - R_{i5}^2 s_{\beta_2}^2 \right)
	\right]
	\;,
\end{equation}
\begin{equation}
	\Re(\lambda_{11}) v^2 = \frac{-1}{2 c_{\beta_1} s_{\beta_2}} \left[
	\frac{1}{c_{\beta_2}} \Re(\mu_{13}) + m_{h_i}^2 R_{i5}\left( R_{i4} s_{\beta_1} + R_{i5} c_{\beta_1} s_{\beta_2} \right)
	\right]
	\;,
\end{equation}
\begin{equation}
	\Re(\lambda_{12}) v^2 = \frac{-1}{2 s_{\beta_1} s_{\beta_2}} \left[
	\frac{1}{c_{\beta_2}} \Re(\mu_{23}) + m_{h_i}^2 R_{i5}\left( R_{i5} s_{\beta_1}s_{\beta_2} - R_{i4} c_{\beta_1} \right)
	\right]
	\;,
\end{equation}
\begin{equation}\label{eq:L10I}
	\Im(\lambda_{10}) v^2 = \frac{1}{2 c_{\beta_1}s_{\beta_1}c_{\beta_2}^2} \left[
	- \Im(\mu_{12}) + m_{h_i}^2 R_{i1} \left( R_{i5} s_{\beta_1}s_{\beta_2} - R_{i4}c_{\beta_1}  \right)
	\right]
	\;,
\end{equation}
\begin{equation}\label{eq:L11I}
	\Im(\lambda_{11}) v^2 = \frac{s_{\beta_1}}{2 c_{\beta_1}s_{\beta_2}^2} \left[
	\Im(\mu_{12}) - m_{h_i}^2 R_{i1} \left( R_{i4} c_{\beta_1} + R_{i5} \frac{s_{\beta_2}(1+c_{\beta_1}^2)}{s_{\beta_1}}  \right)
	\right]
	\;,
\end{equation}
\begin{equation}\label{eq:L12I}
	\Im(\lambda_{12}) v^2 = \frac{c_{\beta_1}}{2 s_{\beta_1}s_{\beta_2}^2} \left[
	- \Im(\mu_{12}) + m_{h_i}^2 \left( R_{i1}(R_{i4}c_{\beta_1} -R_{i5}s_{\beta_1}s_{\beta_2} ) - \frac{2 s_{\beta_2}}{c_{\beta_1}} R_{i2}R_{i5} \right)
	\right]
	\;,
\end{equation}
\begin{equation}\label{eq:mu12I}
	\Im(\mu_{12}) = 4 c_\varphi s_\varphi c_\theta s_\theta s_{\beta_2} (m_{H_2^\pm}^2-m_{H_1^\pm}^2) + m_{h_i}^2 R_{i1}(R_{i4}c_{\beta_1}-R_{i5}s_{\beta_1}s_{\beta_2})
	\;,
\end{equation}
with 3 additional constraints on the neutral mass matrix:
\begin{equation}\label{e:X1iapp}
	m_{h_i}^2 \left[ R_{i5} c_{\beta_2}(R_{i1}s_{\beta_1}-R_{i2}c_{\beta_1}) - R_{i3}R_{i4} \right] = X_{1i} m_{h_i}^2 = 0
	\;,
\end{equation}
\begin{equation}\label{e:X2iapp}
	m_{h_i}^2 R_{i5} \frac{ c_{\beta_2}(R_{i1}c_{\beta_1}+R_{i2}s_{\beta_1}) - R_{i3} s_{\beta_2} }{s_{\beta_2}}  = X_{2i} m_{h_i}^2  = 0
	\;,
\end{equation}
\begin{equation}\label{e:X3iapp}
	m_{h_i}^2 \frac{ R_{i4} (R_{i1}c_{\beta_1}-R_{i2}s_{\beta_1}) - R_{i5} s_{\beta_2}(R_{i1}s_{\beta_1}+R_{i2}c_{\beta_1}) }{s_{\beta_1}}  = X_{3i} m_{h_i}^2  = 0
	\;.
\end{equation}

%% file: appendix/yukawa_types.tex
\chapter{\label{app:Types}Yukawa interactions in the mass basis}
\section{Type-I}

 For this case we assume that under the group all the fermion
fields are unaffected. Therefore they can only couple to $\phi_3$. With the conventions of
\eqs{eq:couplingNeutralFerm}{eq:couplingChargedFerm} we have, 
\begin{align}
a_j^f \to&
\frac{\textbf{R}_{j,3}}{\hat{v_3}},
\qquad\qquad j=1,2,3\qquad \text{for all leptons} ,\nonumber\\[2pt]
b_j^f \to&
\frac{\textbf{P}_{j-2,3}}{\hat{v_3}},
\qquad\quad j=4,5\quad\qquad \text{for all leptons} ,\nonumber\\[2pt]
a_j^f \to&
\frac{\textbf{R}_{j,3}}{\hat{v_3}},
\qquad\qquad j=1,2,3\qquad \text{for all up quarks} ,\nonumber\\[2pt]
b_j^f \to&
-\frac{\textbf{P}_{j-2,3}}{\hat{v_3}},
\quad\quad j=4,5\quad\qquad \text{for all up quarks} ,\nonumber\\[2pt]
a_j^f \to&
\frac{\textbf{R}_{j,3}}{\hat{v_3}},
\qquad\qquad j=1,2,3\qquad \text{for all down quarks} ,\nonumber\\[2pt]
b_j^f \to&
\frac{\textbf{P}_{j-2,3}}{\hat{v_3}},
\qquad\quad j=4,5\quad\qquad \text{for all down quarks} ,
\label{eq:coeffNeutralFerm-Type-I}
\end{align}
and
\begin{equation}
\label{eq:coeffChargedFerm-Type-I}  
\eta_k^{\ell\,L}=-\frac{\textbf{Q}_{k+1,3}}{\hat{v_3}}\,,\quad\eta_k^{\ell\,R}=
0\,,\quad\eta_k^{q\,L} =-\frac{\textbf{Q}_{k+1,3}}{\hat{v_3}}\,,
\quad\eta_k^{q\,R}=\frac{\textbf{Q}_{k+1,3}}{\hat{v_3}}\,,\quad
\text{k=1,2}\, .
\end{equation}

\section{Type-II}

For this case we assume that under the group,
\begin{equation}
\label{eq:Type-IIa}
n_R\to (+,e^{-i \theta'})\,n_R,\qquad \ell_R\to (+,e^{-i \theta'})\,\ell_R ,
\end{equation}
the other fermion fields remain unaffected,
where we have used the notation of
Table~\ref{tab:Types}, only altered by using ``+'' for
invariance.\footnote{We used a space instead of ``+''
for an invariance in Table~\ref{tab:Types},
in order not to clutter the notation.}    
Therefore, up quarks couple to $\phi_3$ and down quarks and leptons
couple only to $\phi_2$. With the conventions of
\eqs{eq:couplingNeutralFerm}{eq:couplingChargedFerm} we have,
\begin{align}
a_j^f \to&
\frac{\textbf{R}_{j,2}}{\hat{v_2}},
\qquad\qquad j=1,2,3\qquad \text{for all leptons} ,\nonumber\\[2pt]
b_j^f \to&
\frac{\textbf{P}_{j-2,2}}{\hat{v_2}},
\qquad\quad j=4,5\quad\qquad \text{for all leptons} ,\nonumber\\[2pt]
a_j^f \to&
\frac{\textbf{R}_{j,3}}{\hat{v_3}},
\qquad\qquad j=1,2,3\qquad \text{for all up quarks} ,\nonumber\\[2pt]
b_j^f \to&
-\frac{\textbf{P}_{j-2,3}}{\hat{v_3}},
\quad\quad j=4,5\quad\qquad \text{for all up quarks} ,\nonumber\\[2pt]
a_j^f \to&
\frac{\textbf{R}_{j,2}}{\hat{v_2}},
\qquad\qquad j=1,2,3\qquad \text{for all down quarks} ,\nonumber\\[2pt]
b_j^f \to&
\frac{\textbf{P}_{j-2,2}}{\hat{v_2}},
\qquad\quad j=4,5\quad\qquad \text{for all down quarks} ,
\label{eq:coeffNeutralFerm-Type-II}
\end{align}
and
\begin{equation}
\label{eq:coeffChargedFerm-Type-II}  
\eta_k^{\ell\,L}=-\frac{\textbf{Q}_{k+1,2}}{\hat{v_2}}\,,
\quad\eta_k^{\ell\,R}=0\,,\quad\eta_k^{q\,L}
=-\frac{\textbf{Q}_{k+1,2}}{\hat{v_2}}\,,\quad\eta_k^{q\,R}
=\frac{\textbf{Q}_{k+1,3}}{\hat{v_3}}\,,\quad \text{k=1,2} ,
\end{equation}

\section{Type-X}

For this case we assume that under the group,
\begin{equation}
  \label{eq:Type-IIb}
    \ell_R\to (+,e^{-i \theta'})\,\ell_R ,
  \end{equation}
the other fermion fields remain unaffected, where we have used the notation of
Table~\ref{tab:Types}.   
Therefore up and down quarks couple to $\phi_3$ and leptons
couple only to $\phi_2$. With the conventions of
\eqs{eq:couplingNeutralFerm}{eq:couplingChargedFerm} we have,
\begin{align}
a_j^f \to&
\frac{\textbf{R}_{j,2}}{\hat{v_2}},
\qquad\qquad j=1,2,3\qquad \text{for all leptons} ,\nonumber\\[2pt]
b_j^f \to&
\frac{\textbf{P}_{j-2,2}}{\hat{v_2}},
\qquad\quad j=4,5\quad\qquad \text{for all leptons} ,\nonumber\\[2pt]
a_j^f \to&
\frac{\textbf{R}_{j,3}}{\hat{v_3}},
\qquad\qquad j=1,2,3\qquad \text{for all up quarks} ,\nonumber\\[2pt]
b_j^f \to&
-\frac{\textbf{P}_{j-2,3}}{\hat{v_3}},
\quad\quad j=4,5\quad\qquad \text{for all up quarks} ,\nonumber\\[2pt]
a_j^f \to&
\frac{\textbf{R}_{j,3}}{\hat{v_3}},
\qquad\qquad j=1,2,3\qquad \text{for all down quarks} ,\nonumber\\[2pt]
b_j^f \to&
\frac{\textbf{P}_{j-2,3}}{\hat{v_3}},
\qquad\quad j=4,5\quad\qquad \text{for all down quarks} ,
\label{eq:coeffNeutralFerm-Type-X}
\end{align}
and
\begin{equation}
\label{eq:coeffChargedFerm-Type-X}  
\eta_k^{\ell\,L}=-\frac{\textbf{Q}_{k+1,2}}{\hat{v_2}}\,,
\quad\eta_k^{\ell\,R}=0\,,\quad\eta_k^{q\,L}
=-\frac{\textbf{Q}_{k+1,3}}{\hat{v_3}}\,,\quad\eta_k^{q\,R}
=\frac{\textbf{Q}_{k+1,3}}{\hat{v_3}}\,,\quad \text{k=1,2} ,
\end{equation}

\section{Type-Y}

For this case we assume that under the group,
\begin{equation}
  \label{eq:Type-IIc}
    n_R\to (+,e^{-i \theta'})\,n_R,
  \end{equation}
the other fermion fields remain unaffected, where we have used the notation of
Table~\ref{tab:Types}.
Therefore up quarks and leptons couple to $\phi_3$ and down quarks
couple only to $\phi_2$. With the conventions of
\eqs{eq:couplingNeutralFerm}{eq:couplingChargedFerm} we have,
\begin{align}
a_j^f \to&
\frac{\textbf{R}_{j,3}}{\hat{v_3}},
\qquad\qquad j=1,2,3\qquad \text{for all leptons} ,\nonumber\\[2pt]
b_j^f \to&
\frac{\textbf{P}_{j-2,3}}{\hat{v_3}},
\qquad\quad j=4,5\quad\qquad \text{for all leptons} ,\nonumber\\[2pt]
a_j^f \to&
\frac{\textbf{R}_{j,3}}{\hat{v_3}},
\qquad\qquad j=1,2,3\qquad \text{for all up quarks} ,\nonumber\\[2pt]
b_j^f \to&
-\frac{\textbf{P}_{j-2,3}}{\hat{v_3}},
\quad\quad j=4,5\quad\qquad \text{for all up quarks} ,\nonumber\\[2pt]
a_j^f \to&
\frac{\textbf{R}_{j,2}}{\hat{v_2}},
\qquad\qquad j=1,2,3\qquad \text{for all down quarks} ,\nonumber\\[2pt]
b_j^f \to&
\frac{\textbf{P}_{j-2,2}}{\hat{v_2}},
\qquad\quad j=4,5\quad\qquad \text{for all down quarks} ,
\label{eq:coeffNeutralFerm-Type-Y}
\end{align}
and
\begin{equation}
\label{eq:coeffChargedFerm-Type-Y}  
\eta_k^{\ell\,L}=-\frac{\textbf{Q}_{k+1,3}}{\hat{v_3}}\,,
\quad\eta_k^{\ell\,R}=0\,,\quad\eta_k^{q\,L}
=-\frac{\textbf{Q}_{k+1,2}}{\hat{v_2}}\,,
\quad\eta_k^{q\,R}=\frac{\textbf{Q}_{k+1,3}}{\hat{v_3}}\,,\quad \text{k=1,2} ,
\end{equation}

%% file: appendix/bfb_comp.tex
\chapter{Comparison of the different BFB conditions}
\label{a:comparison_bfb}

For three symmetries - $U(1) \times U(1)$, $U(1) \times\Z2$,
and $\Z2\times\Z2$ - 
we have generated a large set of points which are consistent
with $\textrm{Al-1}-50\%$ in \eqref{Al-1},
and which pass all current constraints from
$B$-physics, measurements of the $125~{\rm GeV}$ Higgs properties,
and searches for extra scalars.
We repeated the process for $\textrm{Al-2}-50\%$ in \eqref{Al-2}.

Within each group, we denote by different colours those points
which pass different BFB conditions,
using the following notation:
\begin{itemize}
\item{BFB1} (red points): those points which pass BFB-n but do not pass
BFB-c;
\item{BFB2} (green points): those points which pass BFB-n and also
pass the necessary and sufficient conditions for BFB-c.
These conditions are only known for $U(1)\times U(1)$,
shown in subsection~\ref{subsec:U1U1_BFB_NS},
and for $U(1)\times \Z2$,
shown in subsection~\ref{subsec:U1Z2_BFB_NS}.
The necessary and sufficient conditions for BFB-c are unknown
in the case of $\Z2\times\Z2$ and, thus,
there will be no green points in the corresponding plots.
\item{BFB3} (orange points): those points which pass BFB-c and also
pass the sufficient conditions for BFB-c derived in this article,
but do not pass BFB4 below.
\item{BFB4} (blue points): those points which pass BFB-c and also
pass the sufficient conditions for BFB-c adapted from
those of the $\Z2\times\Z2$ case presented in Ref.~\cite{Grzadkowski:2009bt},
but do not pass BFB3 above.
\item{BFB3+4} (grey points): those points which pass BFB-c and also
pass the sufficient conditions for BFB-c derived in this article,
and in addition also
pass the sufficient conditions for BFB-c adapted from
those of the $\Z2\times\Z2$ case presented in Ref.~\cite{Grzadkowski:2009bt}.
Grey points are the would-be overlap between orange and blue points.
\end{itemize}

%

\begin{figure}[htb] 
  \centering
  \begin{tabular}{cc}
\includegraphics[width=0.47\textwidth]{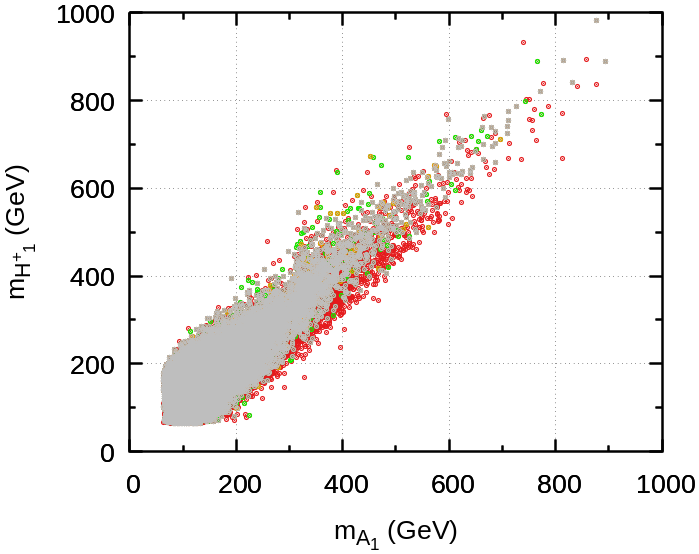}
    &
\includegraphics[width=0.47\textwidth]{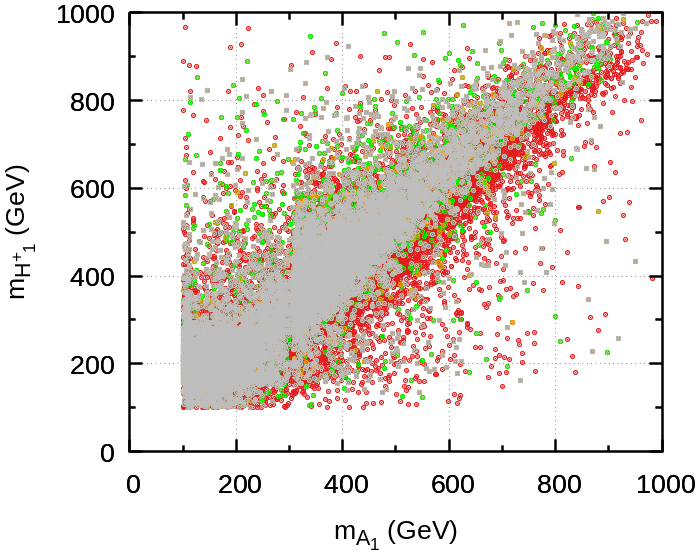}
  \end{tabular}
   \caption{$U(1)\times U(1)$: $(m_{A_1}, m_{H_1^\pm})$ for Type-1 with the
$\textrm{Al-2}-50\%$ conditions (left panel) and loose $\textrm{Al-1}-50\%$ conditions (right panel).
BFB1=red, BFB2=green, BFB3=orange, BFB4=blue, BFB3+4=gray}
    \label{fig:ma1-mc1_t1_U1U1}
\end{figure}

We first comment on the difference between the two alignment conditions:
$\textrm{Al-1}-50\%$ and $\textrm{Al-2}-50\%$.
In Figure~\ref{fig:ma1-mc1_t1_U1U1}a,
we show in the $m_{A_1} - m_{H_1^\pm}$ plane
the points generated by a 50\% range around \eq{Al-2}.
Figure~\ref{fig:ma1-mc1_t1_U1U1}b repeats the exercise
for the much looser alignment constraints in \eq{Al-1}.
Naturally, points which obey $\textrm{Al-1}$ but do not obey $\textrm{Al-2}$ are much
more difficult to generate than points which obey $\textrm{Al-2}$.
However, as Figure~\ref{fig:ma1-mc1_t1_U1U1}b illustrates,
such points are allowed and correspond to physically interesting
regions of parameter space.
Namely,
and contrary to popular belief,
the oblique parameters do not require degeneracy
within each scalar family.
This confirms and extends results mentioned
in Ref.~\cite{Hernandez-Sanchez:2020aop}
for the case of a very specific DM implementation of $\Z2\times\Z2$.

\begin{figure}[htb]
  \centering
  \begin{tabular}{cc}
\includegraphics[width=0.47\textwidth]{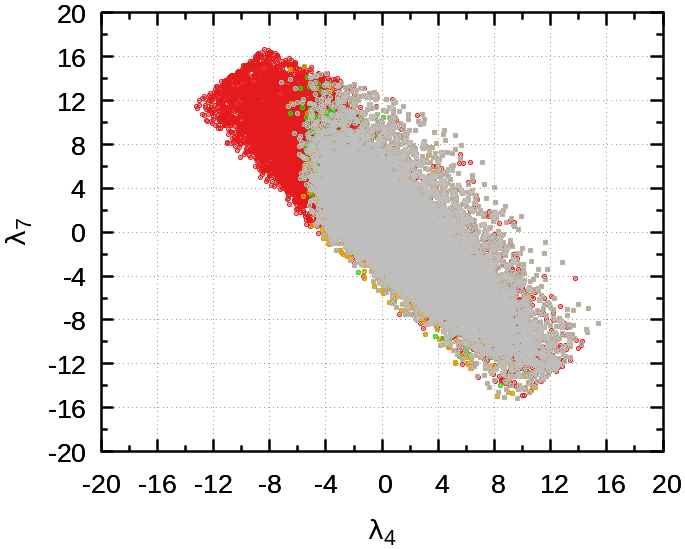}
    &
\includegraphics[width=0.47\textwidth]{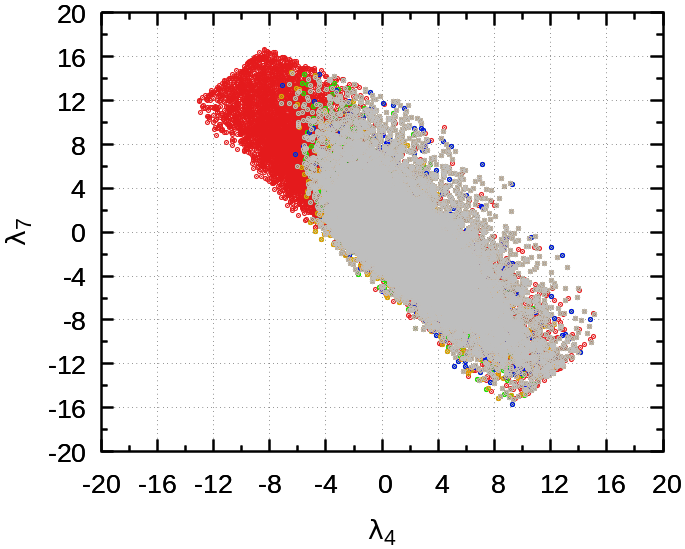}
  \end{tabular}
   \caption{Left panel:    $U(1)\times U(1)$: $(\lambda_4, \lambda_7)$ for Type-1 with the Al-2 conditions.\\ Right panel: $U(1)\times \Z2$: $(\lambda_4, \lambda_7)$ for Type-1 with the $\textrm{Al-2}-50\%$  conditions.}
    \label{fig:L4-L7_t1_U1U1_U1Z2_k16}
\end{figure}

\begin{figure}[htb]
\centering
\includegraphics[width = 0.5\textwidth]{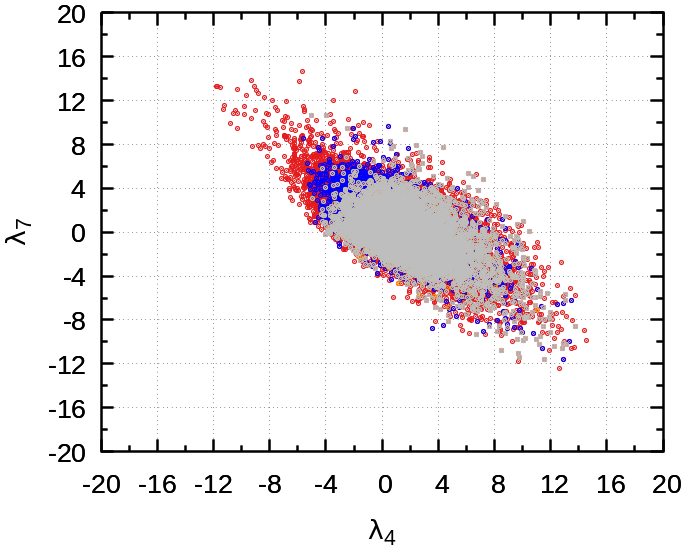}
\caption{$\Z2\times \Z2$: $(\lambda_4, \lambda_7)$ for Type-1 with the
$\textrm{Al-2}-50\%$  conditions.}
    \label{fig:L4-L7_t1_Z2Z2_k16}
\end{figure}

Now we turn to a second important issue.
Could it be that by using only BFB-n, without concern
about BFB-c one is led into wrong physical conclusions?
After all, it could be that points which are BFB-n
but not BFB-c do not differ in their physical
consequences from points which obey both BFB-n
and BFB-c.
This is not the case, as we illustrate in 
Figures~\ref{fig:L4-L7_t1_U1U1_U1Z2_k16}a,
\ref{fig:L4-L7_t1_U1U1_U1Z2_k16}b,
and \ref{fig:L4-L7_t1_Z2Z2_k16}.
We see that points which are BFB-n but not BFB-c
(BFB1-red points) do allow for a negative and large $\lambda_4$,
together with a positive and large $\lambda_7$.
The same type of features appear in the
$\lambda_5-\lambda_8$ plane.
And this occurs for all symmetries studied in this article:
$U(1)\times U(1)$ in Figure~\ref{fig:L4-L7_t1_U1U1_U1Z2_k16}a;
$U(1)\times \Z2$ in Figure~\ref{fig:L4-L7_t1_U1U1_U1Z2_k16}b;
and $\Z2\times \Z2$ in Figure~\ref{fig:L4-L7_t1_Z2Z2_k16}.
We thus conclude that ignoring BFB-c does lead
to wrong physical conclusions.
Dealing with the charge breaking directions is not an option;
it is a must.

A third and curious conclusion arises from the implementation
of the different BFB constraints.
In the $U(1)\times U(1)$ and $U(1)\times \Z2$ cases there
are three BFB conditions of interest.
The true necessary and sufficient BFB conditions
in subsection 
\ref{subsec:U1U1_BFB_NS},
and \ref{subsec:U1Z2_BFB_NS},
respectively;
the sufficient conditions derived in this thesis;
and the adaptation of the
sufficient conditions for BFB-c presented for the $\Z2\times\Z2$
case in Ref.~\cite{Grzadkowski:2009bt}.

We start by noticing that the green points in
Figures~\ref{fig:ma1-mc1_t1_U1U1}a,
\ref{fig:ma1-mc1_t1_U1U1}b,
\ref{fig:L4-L7_t1_U1U1_U1Z2_k16}a,
and \ref{fig:L4-L7_t1_U1U1_U1Z2_k16}b,
do not seem to occupy regions of parameter space
far different from those allowed by the more stringent
sufficient conditions BFB3, BFB4, and BFB3+4.
This is a first hint that maybe using sufficient conditions
does not skew the physical interpretation of the models.
Interestingly,
there seem to be no single blue point in the
Figures \ref{fig:ma1-mc1_t1_U1U1}a,
\ref{fig:ma1-mc1_t1_U1U1}b,
and
\ref{fig:L4-L7_t1_U1U1_U1Z2_k16}a,
corresponding to the $U(1) \times U(1)$ case.
Indeed, we have found numerically that all
points which obey the adaptation to
$U(1) \times U(1)$ of the 
sufficient BFB-c conditions presented for the $\Z2\times\Z2$
case in Ref.~\cite{Grzadkowski:2009bt}
also obey the $U(1)\times U(1)$ sufficient BFB-c
conditions proposed by us in subsection~\ref{subsec:U1U1_BFB_us}.
This is illustrated by the gray points.
The converse is not true.
Thus we find in Figures \ref{fig:ma1-mc1_t1_U1U1}a,
\ref{fig:ma1-mc1_t1_U1U1}b,
and
\ref{fig:L4-L7_t1_U1U1_U1Z2_k16}a orange points,
which correspond to points which pass the sufficient BFB-c
conditions of subsection~\ref{subsec:U1U1_BFB_us} but do
\textit{not} pass the sufficient BFB-c
conditions of subsection~\ref{subsec:U1U1_BFB_Grz}.
In contrast,
Figure~\ref{fig:L4-L7_t1_U1U1_U1Z2_k16}b,
which correspond to the $U(1)\times \Z2$ case,
contain: i) points in gray which pass both sets of bounds;
ii) points in orange which pass the conditions of 
subsection~\ref{subsec:U1U1_BFB_us} but do
\textit{not} pass the conditions of subsection~\ref{subsec:U1U1_BFB_Grz};
but also iii) points in blue which pass the conditions of 
subsection~\ref{subsec:U1U1_BFB_Grz} but do
\textit{not} pass the conditions of subsection~\ref{subsec:U1U1_BFB_us}.
In fact,
we find the quite curious result that our simulation
consistently generated more blue points than orange points.
This is even more apparent in
Figure~\ref{fig:L4-L7_t1_Z2Z2_k16} concerning the $\Z2\times\Z2$ case.
That figure is based on one simulation where we found 
6977 BFB3 orange points,
10087 BFB4 blue points,
and 6842 BFB3+4 gray overlap points.
This could be due to the following.
The sufficient conditions for the $\Z2\times\Z2$ case were found in
Ref.~\cite{Grzadkowski:2009bt} by a careful study of the
charge breaking directions of that specific potential.
Thus, it is not surprising that they are more helpful in that case
than in the $U(1)\times U(1)$ or $U(1)\times \Z2$ cases.

\begin{figure}[htb]
  \centering
  \begin{tabular}{cc}
\includegraphics[width=0.47\textwidth]{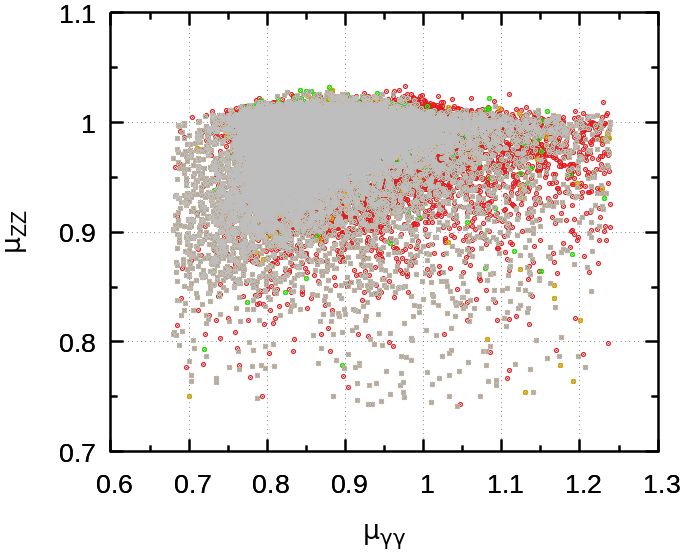}
    &
\includegraphics[width=0.47\textwidth]{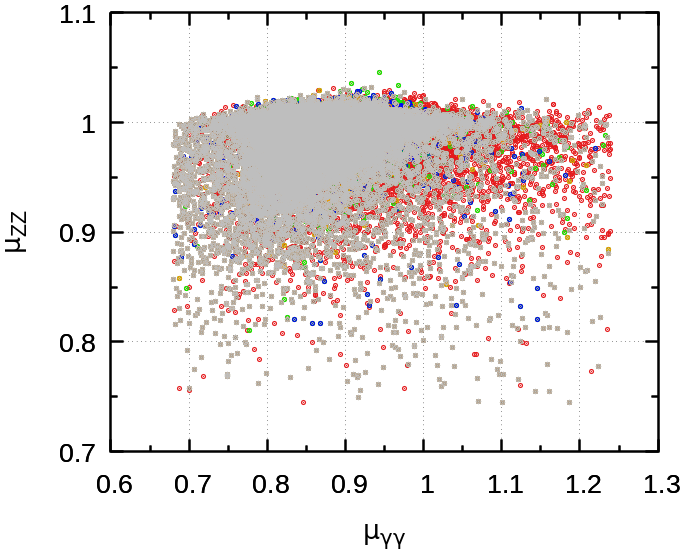}
  \end{tabular}
   \caption{Left panel:    $U(1) \times U(1)$: $\mu_{ZZ}-\mu_{\gamma\gamma}$
    plane for the gluon fusion production channel.
    For Type-1 with Al-2 conditions.\\ Right panel: $U(1) \times\Z2$: $\mu_{ZZ}-\mu_{\gamma\gamma}$
    plane for the gluon fusion production channel.
    For Type-1 with Al-2 conditions.}
    \label{fig:gaga-zz_t1_U1U1_U1Z2_k16}
\end{figure}

\begin{figure}[htb]
\centering
\includegraphics[width = 0.5\textwidth]{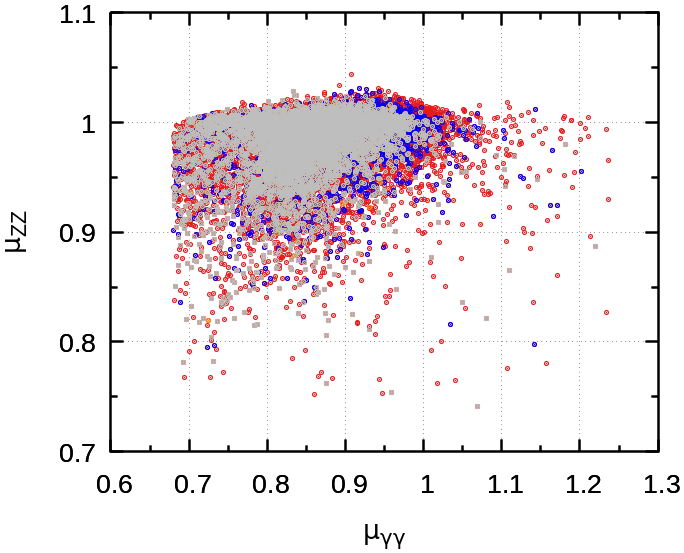}
\caption{$\Z2 \times \Z2$: $\mu_{ZZ}-\mu_{\gamma\gamma}$
    plane for the gluon fusion production channel.
    For Type-1 with Al-2 conditions.}
    \label{fig:gaga-zz_t1_Z2Z2_k16}
\end{figure}

\begin{figure}[htb]
  \centering
  \begin{tabular}{cc}
\includegraphics[width=0.47\textwidth]{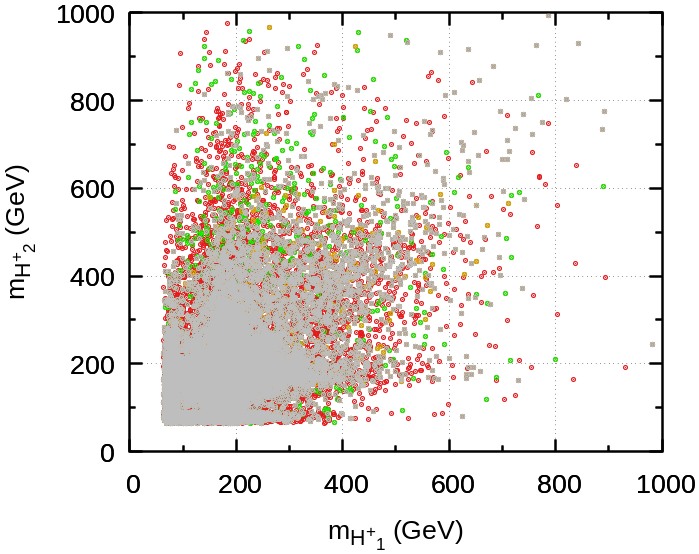}
    &
\includegraphics[width=0.47\textwidth]{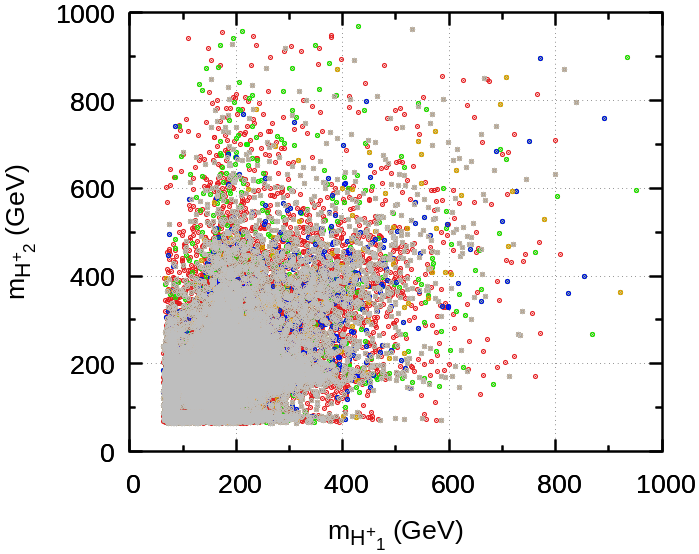}
  \end{tabular}
   \caption{Left panel:    $U(1) \times U(1)$: charged scalar masses for Type-1 with Al-2 conditions.\\ Right panel: $U(1) \times \Z2$: charged scalar masses for Type-1 with Al-2 conditions.}
    \label{fig:mc1-mc2_t1_U1U1_U1Z2_k16}
\end{figure}

\begin{figure}[htb]
\centering
\includegraphics[width = 0.5\textwidth]{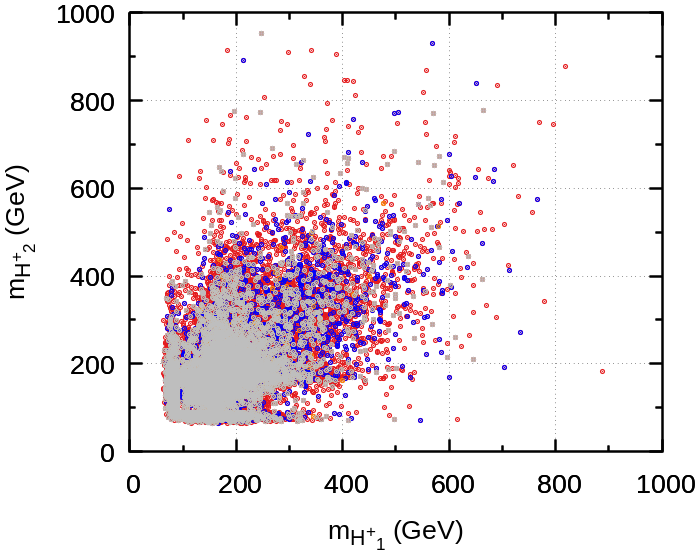}
\caption{$\Z2 \times \Z2$: charged scalar masses for Type-1 with Al-2 conditions.}
    \label{fig:mc1-mc2_t1_Z2Z2_k16}
\end{figure}

We now turn to the question of whether using sufficient BFB-c
instead of the correct necessary and sufficient BFB-c conditions
does (or not) constrain unduly the physical quantities.
We have calculated all consequences of the various points found
for the $125~{\rm GeV}$ scalar and for searches into heavier scalars.
Next we plotted all pairs of observables in the respective
planes, looking for physical differences between the placement
of the green points versus points with sufficient BFB-c conditions.
We have found no evidence of a difference.
We illustrate such searches below.
We show the $\mu_{\gamma\gamma} - \mu_{ZZ}$ plane
in Figures~\ref{fig:gaga-zz_t1_U1U1_U1Z2_k16}a,
\ref{fig:gaga-zz_t1_U1U1_U1Z2_k16}b,
and \ref{fig:gaga-zz_t1_Z2Z2_k16},
for the $U(1)\times U(1)$,
$U(1)\times \Z2$,
and $\Z2\times \Z2$ symmetries, respectively.
The exercise is repeated for the $m_{H_1^\pm} - m_{H_2^\pm}$
plane in Figures \ref{fig:mc1-mc2_t1_U1U1_U1Z2_k16}a,
\ref{fig:mc1-mc2_t1_U1U1_U1Z2_k16}b,
and \ref{fig:mc1-mc2_t1_Z2Z2_k16}.

There is no physical difference that can be considered statistically
significant; there are only minor differences between placement of colours,
due to the sparse placement of a (necessarily) limited numerical
simulation.
By looking at hundreds of such plots we conclude that:
\begin{enumerate}
\item Using BFB-n bounds while ignoring BFB-c considerations does lead
to wrong physical conclusions.
\item In contrast, using safe sufficient BFB-c bounds versus using
(when available) the exact necessary and sufficient BFB-c conditions
does not seem to introduce a bias in the physical observables.
\item Moreover, using different safe BFB-c bounds does affect the
number of points generated (for equal running time) but it does not
seem to introduce a bias in the analysis.
\end{enumerate}

%% file: appendix/offdiagonal.tex
\chapter{\texorpdfstring{Theoretical conditions for off-diagonal scalar $Z$-boson couplings}{Theoretical conditions for off-diagonal scalar Z-boson couplings}}
\label{a:appoffdiag}

%
%
Let $H_1^{+Q}$ and $H_2^{+Q}$ be two (unphysical) charged scalars originating
from two different $SU(2)_L$ multiplets and therefore having well defined
$T_3$ eigenvalues denoted by $T_3^{(1)}$ and $T_3^{(2)}$ respectively.
Thus, their interactions with the $Z$-boson may be parametrized as
\begin{eqnarray}
	\label{e:quartic}
	{\mathscr L}_H^{Z} &=&  i R_{z_1} \left \{ \left(\partial_\mu H_1^{+Q}\right)H_1^{-Q} - \left(\partial_\mu H_1^{-Q}\right)H_1^{+Q}\right\}Z^\mu + R_{z_1}^2 (H_1^{+Q} H_1^{-Q})(Z^\mu Z_\mu)  
	\nonumber \\
	&& + i R_{z_2} \left \{ \left(\partial_\mu H_2^{+Q}\right)H_2^{-Q} - \left(\partial_\mu H_2^{-Q}\right)H_2^{+Q}\right\}Z^\mu + R_{z_2}^2 (H_2^{+Q} H_2^{-Q})(Z^\mu Z_\mu) 
	 \,, 
\end{eqnarray}
where,
\begin{eqnarray}
	R_{z_i} = \frac{g}{c_w}\left(T_3^{(i)} - Qs_w^2\right) \qquad i=1,2 \,.
\end{eqnarray}
Here $g$ denotes the $SU(2)_L$ gauge coupling, $s_w$ and $c_w$ are the sine and the cosine of the weak mixing angle,
respectively. The physical charged scalars, $S_1^{+Q}$ and $S_2^{+Q}$, should
be obtained by the following rotation:   
\begin{eqnarray}
	\label{e:basisrot}
	\begin{pmatrix}
		H_1^{+Q} \\
		H_2^{+Q} 
	\end{pmatrix}
	&=& \begin{pmatrix}
		\cos \zeta & \sin \zeta  \\
		-\sin \zeta & \cos \zeta \end{pmatrix}
	\begin{pmatrix}
		S_1^{+Q} \\
		S_2^{+Q} 
	\end{pmatrix}\, .
\end{eqnarray}
Substituting \eq{e:basisrot} into \eq{e:quartic}, we find,
\begin{eqnarray}
	{\mathscr L}_S^{Z} &=&  i R_{z_1}  \left[  \cos^2 \zeta \left \{  \left(\partial_\mu S_1^{+Q}\right)S_1^{-Q} - \left(\partial_\mu S_1^{-Q}\right)S_1^{+Q}\right\} + \sin^2 \zeta \left \{  \left(\partial_\mu S_2^{+Q}\right)S_2^{-Q} - \left(\partial_\mu S_2^{-Q}\right)S_2^{+Q}\right\} \right.
	\nonumber \\
 && \left. + \sin \zeta \cos \zeta \left \{  \left(\partial_\mu S_1^{+Q}\right)S_2^{-Q} - \left(\partial_\mu S_1^{-Q}\right)S_2^{+Q} + \left(\partial_\mu S_2^{+Q}\right)S_1^{-Q} - \left(\partial_\mu S_2^{-Q}\right)S_1^{+Q}\right\} \right]Z^\mu \nonumber \\
	&& +  R_{z_1}^2 \left[ \cos^2 \zeta \, S_1^{+Q} S_1^{-Q} + \sin^2 \zeta \,  S_2^{+Q} S_2^{-Q} + \cos \zeta \sin \zeta \,  \left\{S_1^{+Q} S_2^{-Q} + S_2^{+Q} S_1^{-Q} \right\}\right](Z^\mu Z_\mu)  	\nonumber \\
	&&  + i R_{z_2}  \left[ \sin^2 \zeta  \left \{  \left(\partial_\mu S_1^{+Q}\right)S_1^{-Q} - \left(\partial_\mu S_1^{-Q}\right)S_1^{+Q}\right\} + \cos^2 \zeta \left \{  \left(\partial_\mu S_2^{+Q}\right)S_2^{-Q} - \left(\partial_\mu S_2^{-Q}\right)S_2^{+Q}\right\} \right.
	\nonumber \\
 && \left. - \sin \zeta \cos \zeta  \left \{  \left(\partial_\mu S_1^{+Q}\right)S_2^{-Q} - \left(\partial_\mu S_1^{-Q}\right)S_2^{+Q} + \left(\partial_\mu S_2^{+Q}\right)S_1^{-Q} - \left(\partial_\mu S_2^{-Q}\right)S_1^{+Q}\right\} \right]Z^\mu \nonumber \\
	&& +  R_{z_2}^2 \left[ \sin^2 \zeta \, S_1^{+Q} S_1^{-Q} + \cos^2 \zeta \,  S_2^{+Q} S_2^{-Q} - \cos \zeta \sin \zeta \,  \left\{S_1^{+Q} S_2^{-Q} + S_2^{+Q} S_1^{-Q} \right\}\right](Z^\mu Z_\mu)  	
	 \,, 
\end{eqnarray}
such that the diagonal couplings are given by,
\begin{eqnarray}
g_{zs_1s_1}&=&\cos^2 \zeta \, R_{z_1}+\sin^2 \zeta \, R_{z_2} = R_{z_1}- \sin^2 \zeta \, \left( R_{z_1} - R_{z_2} \right)   \,, \nonumber\\
g_{zzs_1s_1}&=&\cos^2 \zeta \, R^2_{z_1}+\sin^2 \zeta \, R^2_{z_2} \,, \nonumber\\
g_{zs_2s_2}&=&\sin^2 \zeta \, R_{z_1}+\cos^2 \zeta \, R_{z_2} = \sin^2 \zeta \, \left( R_{z_1} - R_{z_2} \right)  +  R_{z_2}  \,, \nonumber\\
g_{zzs_2s_2}&=&\sin^2 \zeta \, R^2_{z_1}+\cos^2 \zeta \, R^2_{z_2} 
 \,, 
\end{eqnarray}
and off-diagonal,
\begin{eqnarray}
g_{zs_1s_2}&=&\tfrac{1}{2} \left(   R_{z_1}- R_{z_2} \right) \sin 2\zeta  \,, \nonumber\\
 g_{zzs_1s_2}&=&\tfrac{1}{2} \left(   R^2_{z_1}- R^2_{z_2} \right) \sin 2\zeta = g_{zs_1s_2} \left( R_{z_1} + R_{z_2}\right)
 \,, 
\end{eqnarray}
Both trilinear diagonal couplings, $g_{zs_1s_1}$ and $g_{zs_2s_2}$, will
vanish simultaneously if the following conditions are satisfied
\begin{subequations}
	\label{e:offconds}
	\begin{eqnarray}
	&& R_{z_1}=-R_{z_2} \,, \quad \Rightarrow \, T_3^{(1)} + T_3^{(2)} = 2Q s_w^2 \,,
	\label{e:offcond1} \\
	{\rm and,} && \sin \zeta = \frac{1}{\sqrt{2}} \,.
	\label{e:offcond2}
	\end{eqnarray}
\end{subequations}
Under these conditions the remaining couplings take the form,
\begin{subequations}
	\label{e:othercoup}
	\begin{eqnarray}
	g_{zzs_1s_2}  &=&  0  \,, \\
	g_{zs_1s_2} &=& R_{z_1} \,,  \\
	g_{zzs_1s_1}  &=&   R_{z_1}^2 = g^2_{zs_1s_2}  \,, \\
	g_{zzs_2s_2}  &=&   R_{z_1}^2 = g^2_{zs_1s_2} \,.
	\end{eqnarray}
\end{subequations}
Because of the numerical value~\cite{ParticleDataGroup:2022pth} of $s_w^2 \approx 0.23$, \eq{e:offcond1} can 
be approximately satisfied when two singly charged scalars ($Q=1$) arise from
a mixing between an $SU(2)_L$ singlet ($T_3^{(1)}=0$) and an 
$SU(2)_L$ doublet ($T_3^{(2)}=1/2$). The Zee-Type model~\cite{Zee:1980ai} constitutes a prototypical
example of such a scenario. 
We have verified the existence of allowed points in the parameter space of such a model~\cite{Florentino:2021ybj}, which conform to maximal mixing as in
\eq{e:offcond2}\footnote{The angle $\zeta$  corresponds to $\gamma$ in the Zee-Type model~\cite{Zee:1980ai}. } and lead to very suppressed trilinear diagonal couplings with the Higgs and the $Z$-boson as compared to the corresponding off-diagonal couplings. 
Furthermore, the Zee-Type model admits
an ‘alignment limit’ which ensures that the lightest CP-even scalar mimics an
SM-like Higgs boson. Therefore, one can achieve the SM-like $hXX$ couplings 
($X$ denotes a massive SM-particle)
by staying in the proximity of `alignment limit', while independently
realizing the maximal mixing ($\zeta \approx 45^\circ$) between the charged scalars
by adjusting the parameters in the scalar potential. 
%

%% file: appendix/hzgamma.tex

\chapter{Explicit calculations of the scattering amplitudes}
\label{a:appScattering}
\begin{figure}[htbp!]
	\centering
	\includegraphics[scale=0.15]{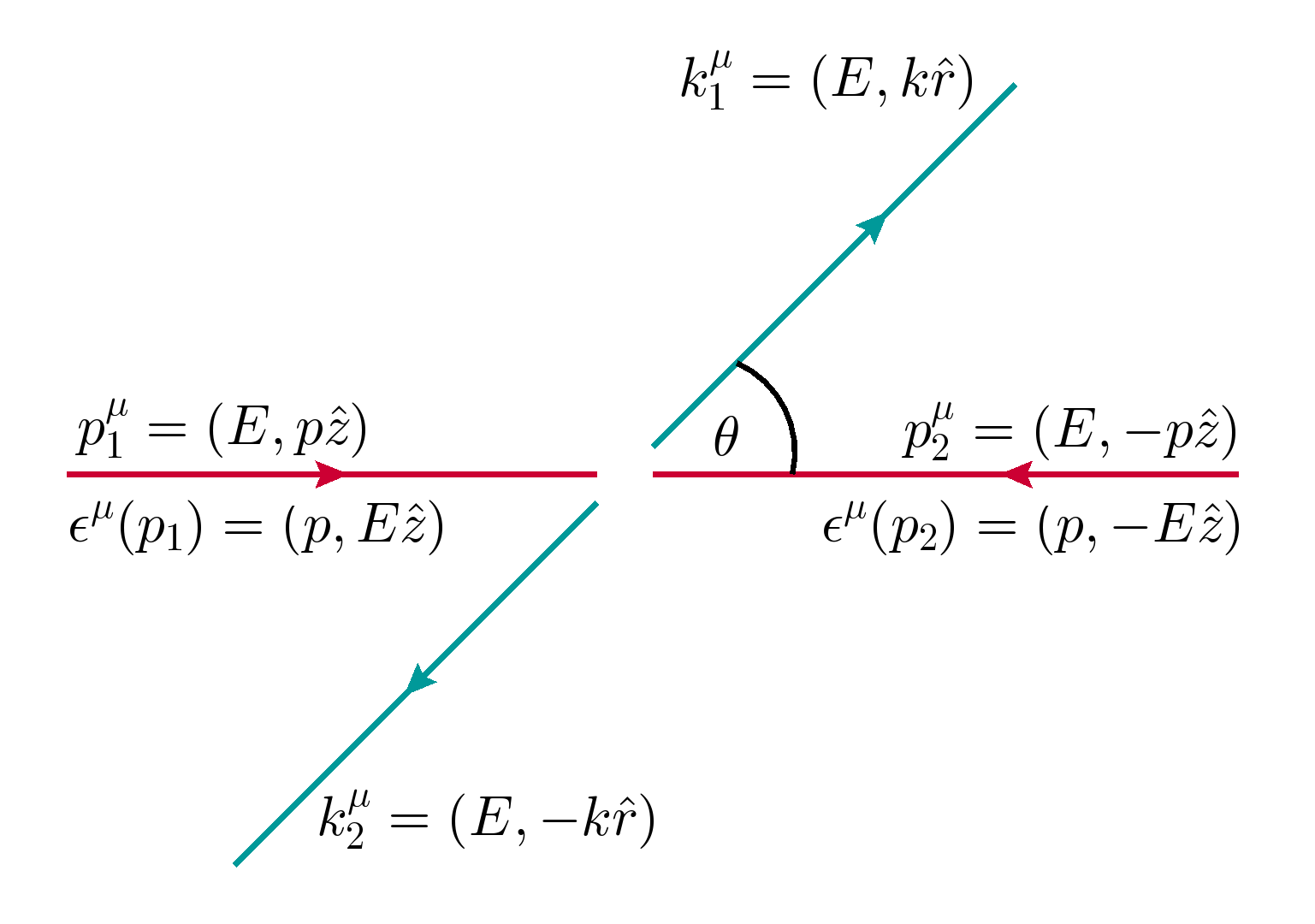}
	\caption{Kinematics in the CM frame for the process in \eq{e:P1}.}
	\label{f:kine}
\end{figure}

In this Appendix, we show the explicit calculations of the scattering amplitudes discussed in the
main text. First we consider the process
\begin{eqnarray}
	\centering
	Z_L(p_1) + Z_L(p_2) \to S_1^+(k_1) + S_1^-(k_2) \,.
	\label{e:P1}
\end{eqnarray}
%
\begin{figure}[ht!]
	\centering
	\includegraphics[scale=0.17]{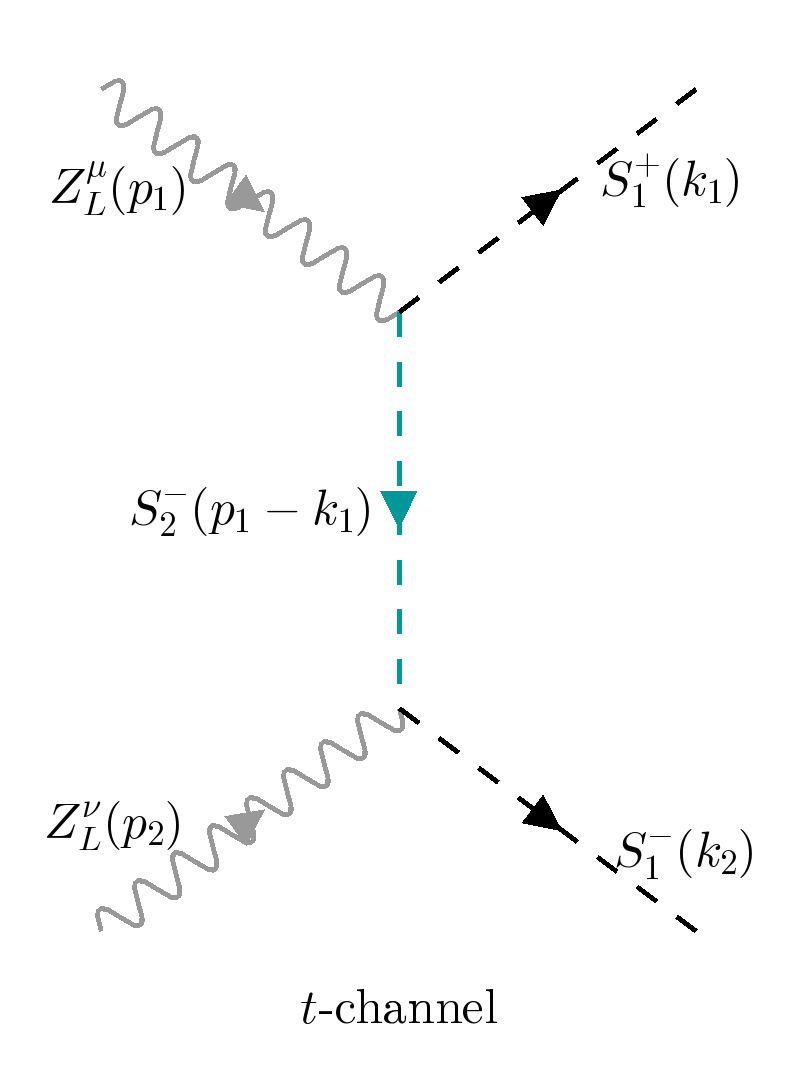}
	\includegraphics[scale=0.17]{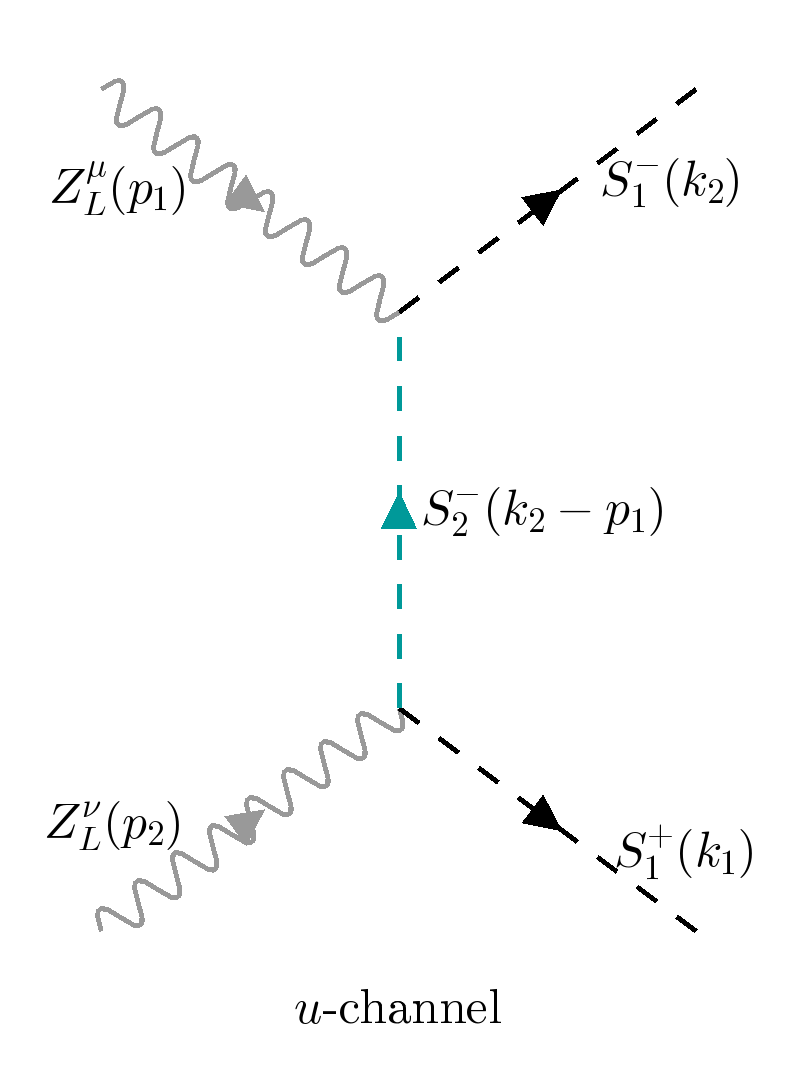}
		\includegraphics[scale=0.15]{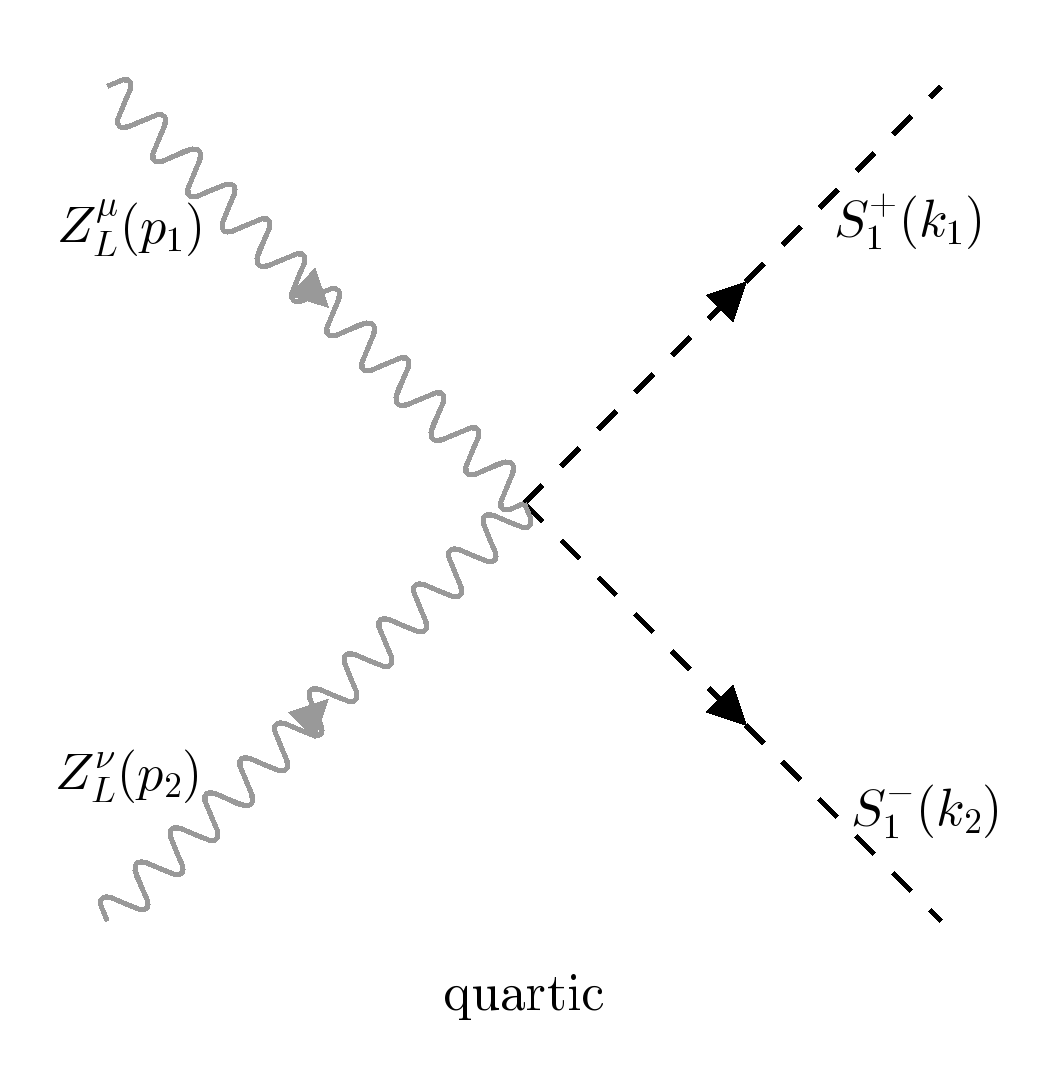}
	\caption{Feynman diagrams for $Z_L Z_L \to S_1^+ S_1^-$.}
	\label{f:FD1}
\end{figure}
The vertex factor for the interaction $Z_\mu S_1^+(p) S_2^-(p^\prime)$ is written as $i g_{z s_1 s_2} (p - p^\prime)_\mu $ where, $p$ and $p^\prime$ 
are the momenta of the incoming charged scalars. 
Considering the momentum assignment of the initial and final states particles as in \eq{e:P1}, we can write
\begin{eqnarray}
	p_1 + p_2 = k_1 + k_2 \,.
	\label{eq:momcons}
\end{eqnarray}
Now, the Feynman amplitude for the $t$-channel diagram is given by,
\begin{eqnarray}
	i{\cal M}_t &=& (i g_{z s_1 s_2})^2 \left[-(p_1 - k_1)+k_1\right]_\mu \frac{i}{t - m_{C_2}^2} \left[-k_2 - \left(p_1 - k_1\right)\right]_\nu \epsilon^\mu(p_1) \epsilon^\nu(p_2) \,, \nonumber \\
	&=& \frac{i g_{z s_1 s_2}^2}{t - m_{C_2}^2} \left(p_1 - 2k_1\right)_\mu \epsilon^\mu(p_1) \left(p_2 - 2 k_2 \right)_\nu \epsilon^\nu(p_2) \,,
	\label{eq:mattchan}
\end{eqnarray}
where we used \eq{eq:momcons} in the last step. In a similar manner, we write down the 
matrix element for the $u$-channel diagram as,
\begin{eqnarray}
	i{\cal M}_u &=& (i g_{z s_1 s_2})^2 \left[-k_2 - \left(k_2 - p_1\right)\right]_\mu \frac{i}{u - m_{C_2}^2} \left[k_1 - \left(k_2 - p_1\right)\right]_\nu \epsilon^\mu(p_1) \epsilon^\nu(p_2) \,, \nonumber \\
	&=& \frac{i g_{z s_1 s_2}^2}{u - m_{C_2}^2} \left(p_1 - 2k_2\right)_\mu \epsilon^\mu(p_1) \left(p_2 - 2 k_1 \right)_\nu \epsilon^\nu(p_2) \,.
	\label{eq:matuchan}
\end{eqnarray}
Next, we express the longitudinal polarization vector for the $Z$-boson as $\epsilon_L^\mu(p) \equiv {\epsilon^\mu(p)}/{M_Z}$, 
with the understanding that $\epsilon^\mu(p)\epsilon_\mu(p)= -M_Z^2$ and $p_\mu \epsilon^\mu(p)= 0$. 
The kinematics for the process in the CM frame has been schematically depicted in fig.~\ref{f:kine}.
Following this, we may write
\begin{subequations}
	\label{eq:pola}
	\begin{eqnarray}
	&&	k_1 \cdot \epsilon_L(p_1) = \frac{E}{M_Z}(p - k\cos\theta) = k_2 \cdot \epsilon_L(p_2)\,; \\
	&& k_2 \cdot \epsilon_L(p_1) = \frac{E}{M_Z}(p + k\cos\theta) = k_1 \cdot \epsilon_L(p_2)\,.
	\end{eqnarray}
\end{subequations}
Using the relations given in \eq{eq:pola}, we now rewrite the matrix elements as,
\begin{subequations}
	\label{eq:mat2}
	\begin{eqnarray}
	{\cal M}_t &=& 4  g_{z s_1 s_2}^2\frac{E^2}{M_Z^2} \frac{(p -k \cos\theta)^2}{t - m_{C_2}^2} \,,  \\
	{\cal M}_u &=& 4 g_{z s_1 s_2}^2\frac{E^2}{M_Z^2} \frac{(p + k \cos\theta)^2}{u - m_{C_2}^2} \,.
	\end{eqnarray}
\end{subequations}
Now, using $t = (p_1 - k_1)^2$ and $u = (p_2 - k_1)^2$ in combination with \eq{eq:momcons}, we can show that
\begin{subequations}
\label{eq:delm}
	\begin{eqnarray}
	\left(p - k \cos \theta \right) &=& \frac{\Delta m^2 - t}{2 p} \,, \\
	\left(p + k \cos \theta \right) &=& \frac{\Delta m^2 - u}{2 p} \,,
	\end{eqnarray}
\end{subequations}
where, we have defined $\Delta m^2 = m_{C_1}^2 - M_Z^2$.
Thus, substituting \eq{eq:delm} into \eq{eq:mat2}, we find,
\begin{subequations}
	\begin{eqnarray}
	{\cal M}_t &=& \frac{g_{z s_1 s_2}^2 E^2}{p^2 M_Z^2} \left(\Delta m^2 - t\right)^2 \frac{1}{t}\left(1 - \frac{m_{C_2}^2}{t}\right)^{-1} \,, \\	
	\Rightarrow	{\cal M}_t	&\simeq& \frac{g_{z s_1 s_2}^2 E^2}{p^2 M_Z^2} \left(t - 2 \Delta m^2 + m_{C_2}^2 \right) \,,	\\
	{\rm and,~ similarly} \quad {\cal M}_u	&\simeq& \frac{g_{z s_1 s_2}^2 E^2}{p^2 M_Z^2} \left(u - 2 \Delta m^2 + m_{C_2}^2 \right) \,,
	\end{eqnarray}
\end{subequations}
where, we have neglected the terms ${\cal O}(M^2/E^2)$ in the high-energy limit. 
Furthermore, using the identity
\begin{eqnarray}
	\frac{E^2}{p^2} = E^2 (E^2 - M_Z^2)^{-1} \simeq \left(1 + \frac{M_Z^2}{E^2} \right) \,,
\end{eqnarray}
we reduce the above amplitudes into,
\begin{subequations}
	\begin{eqnarray}
	{\cal M}_t &=&  \frac{g_{z s_1 s_2}^2}{M_Z^2} t + g_{z s_1 s_2}^2 \left[\frac{t}{E^2} - \frac{\Lambda^2}{M_Z^2} \right] + {\cal O}\left(\frac{M^2}{E^2}\right) \,,  \\
	{\rm and,}\quad		{\cal M}_u &=&  \frac{g_{z s_1 s_2}^2}{M_Z^2} u + g_{z s_1 s_2}^2 \left[\frac{u}{E^2} - \frac{\Lambda^2}{M_Z^2} \right] + {\cal O}\left(\frac{M^2}{E^2}\right) \,,
	\end{eqnarray}
\end{subequations}
where we defined $\Lambda^2 = (2 \Delta m^2 - m_{C_2}^2)$.
Next, the Feynamn amplitude for the quartic diagram is,
\begin{eqnarray}
	i{\cal M}_Q &=& 2 i(g_{z s_1 s_2})^2 \epsilon^\mu_L(p_1) \epsilon_{\mu L}(p_2) \,. 
	\label{eq:matquart}
\end{eqnarray}
Replacing the longitudinal polarization vector for the $Z$-boson as $\epsilon_L^\mu(p) \equiv {\epsilon^\mu(p)}/{M_Z}$
we get
	\begin{eqnarray}
		&&	\epsilon_L(p_1)  \cdot \epsilon_L(p_2) =  \frac{1}{M_Z^2}\epsilon(p_1)  \cdot \epsilon(p_2)  = \left(\frac{2E^2}{M_Z^2} - 1\right) \,.
	\end{eqnarray}
Thus the total amplitude will be given by,
{\small
\begin{eqnarray}
{\cal M}_{Z_L Z_L \to S_1^+ S_1^-} &=& ({\cal M}_t + {\cal M}_u + {\cal M}_Q)
\nonumber \\ 
&=& \frac{g_{z s_1 s_2}^2}{M_Z^2} (t + u) + g_{z s_1 s_2}^2 \left[\frac{(t+u)}{E^2} - \frac{2\Lambda^2}{M_Z^2} \right] + 2g_{z s_1 s_2}^2 \left(\frac{2E^2}{M_Z^2} - 1\right)  +{\cal O}\left(\frac{M^2}{E^2}\right) \,.
\end{eqnarray}
}
Substituting $t+u = 2(M_Z^2 + m_{C_1}^2) - 4 E^2 $ we obtain,
{\small
\begin{subequations}
	\begin{eqnarray}
	{\cal M}_{Z_L Z_L \to S_1^+ S_1^-} &=& -\frac{4 g_{z s_1 s_2}^2}{M_Z^2} E^2 + \frac{g_{z s_1 s_2}^2}{M_Z^2} \left[2(M_Z^2 + m_{C_1}^2) - 2\Lambda^2 - 4 M_Z^2 \right]  \nonumber \\
	&& + 2g_{z s_1 s_2}^2 \left(\frac{2E^2}{M_Z^2} - 1\right) +{\cal O}\left(\frac{M^2}{E^2}\right) \,.
 	\end{eqnarray}
\normalsize Now we can use the definition of $\Lambda$ to write
\small
	\begin{eqnarray} 
	\Rightarrow {\cal M}_{Z_L Z_L \to S_1^+ S_1^-}   &\approx&  -\frac{4 g_{z s_1 s_2}^2}{M_Z^2} E^2 + \frac{2 g_{z s_1 s_2}^2}{M_Z^2} \left[ M_Z^2 + \left( m_{C_1}^2 - m_{C_2}^2 \right) \right] + 2g_{z s_1 s_2}^2 \left(\frac{2E^2}{M_Z^2} - 1\right) \,,  \\
	\Rightarrow {\cal M}_{Z_L Z_L \to S_1^+ S_1^-}  &\approx&  \frac{2 g_{z s_1 s_2}^2}{M_Z^2}\left( m_{C_1}^2 - m_{C_2}^2 \right) \,.
	\label{eq:mtot}
	\end{eqnarray}
\end{subequations}
}
%
It is quite interesting to note that the energy growths of ${\cal O}(E^2)$
arising from the $t$ and $u$ channel diagrams get exactly cancelled by the
quartic diagram, as expected in spontaneously broken gauge theories.

Next we consider the process
\begin{eqnarray}
	\centering
	Z_L(p_1) + Z_L(p_2) \to S_1^+(k_1) + S_2^-(k_2) \,.
\end{eqnarray}
Since we are assuming the presence of off-diagonal couplings only, with the Higgs and
the $Z$-boson, the above process can only proceed via the $s$-channel Higgs exchange.
The corresponding amplitude will be given by
\begin{subequations}
	\begin{eqnarray}
	i{\cal M}_{Z_L Z_L \to S_1^+ S_2^-} &=& \left(\frac{i g M_Z}{c_w}\right) \frac{i}{s - m_h^2} \left(i \lambda_{hs_1s_2} M_W \right) \epsilon_L^\mu(p_1)\epsilon_{L\mu}(p_2) \,, \\
	\Rightarrow {\cal M}_{Z_L Z_L \to S_1^+ S_2^-}  &=& \left(- g M_Z^2 \lambda_{hs_1s_2}\right) \frac{1}{s - m_h^2} \frac{\left(p^2 + E^2\right)}{M_Z^2}  \,,
	\end{eqnarray}
\end{subequations}
where we used $p_1^\mu = (E, p\hat{z})$ and $p_2^\mu = (E, -p\hat{z})$ in the CM frame. 
Furthermore, using $p^2 + E^2 = s/2 - M_Z^2$ we get,
\begin{eqnarray}
	{\cal M}_{Z_L Z_L \to S_1^+ S_2^-} \approx -\frac{1}{2}g \lambda_{hs_1s_2} + {\cal O}\left(\frac{M^2}{E^2}\right) \,. 
	\label{eq:math}
\end{eqnarray}

\begin{figure}[htbp!]
	\centering
	\includegraphics[scale=0.115]{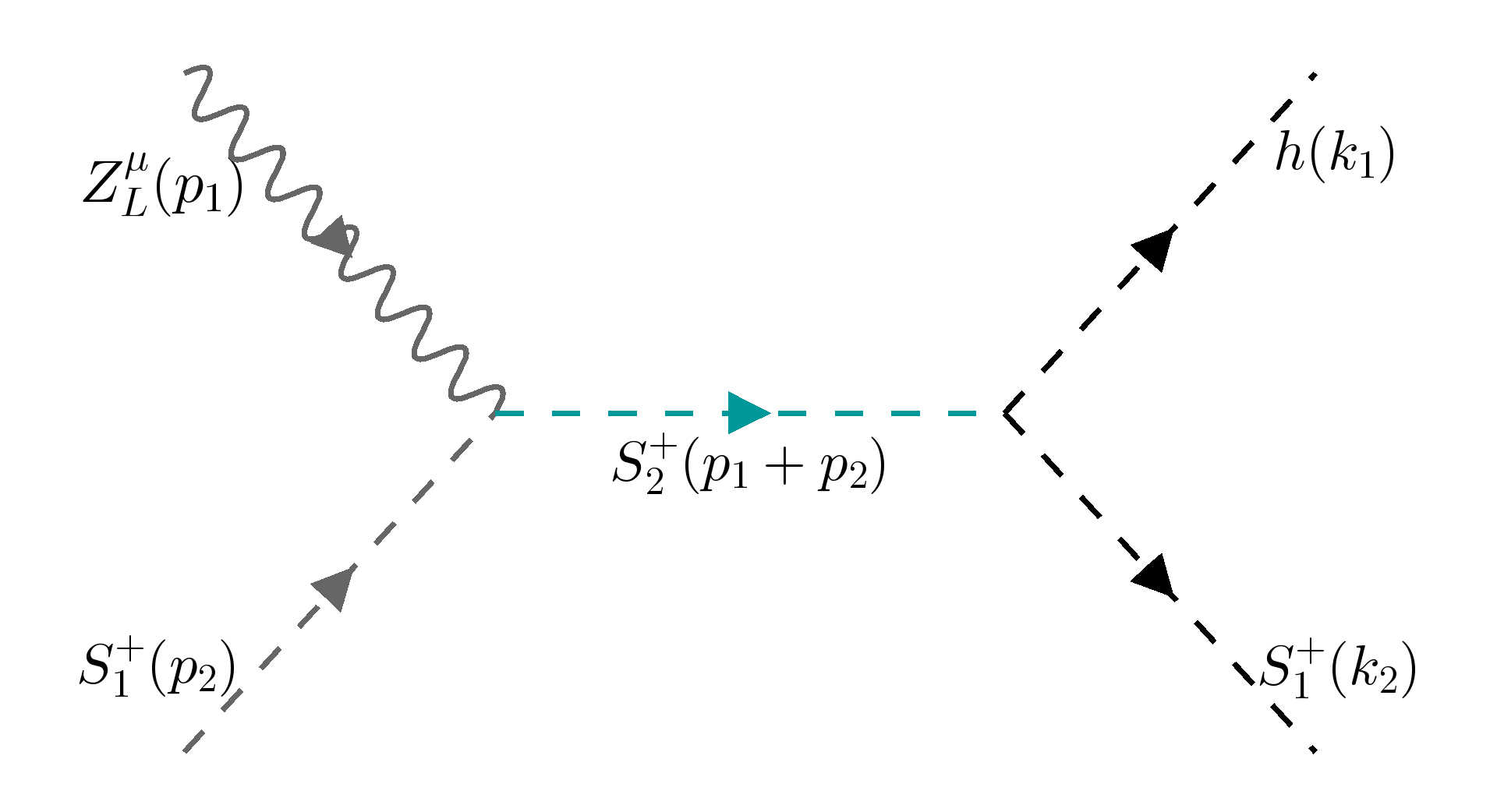} ~~
		\includegraphics[scale=0.125]{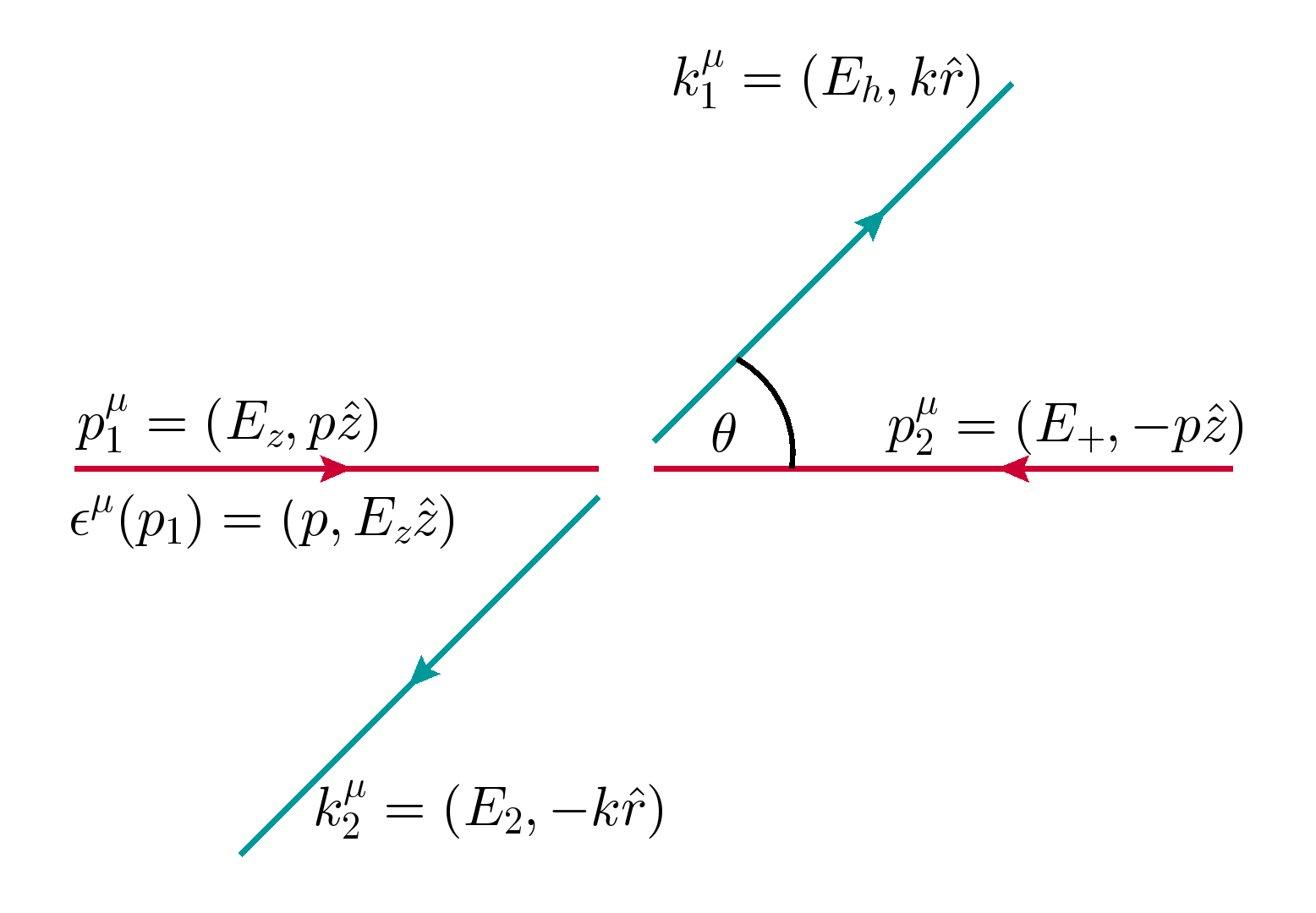}
	\caption{Feynman diagram for the process appearing in \eq{e:p3} and the corresponding kinematics
		in the CM frame.}
	\label{f:FD3}
\end{figure}

Finally, we consider the process
\begin{eqnarray}
	\centering
	Z_L(p_1) + S^+_1(p_2) \to h (k_1) + S_1^+(k_2) \,.
	\label{e:p3}
\end{eqnarray}
The Feynman diagram along with the kinematics in the CM frame have been depicted in Fig.~\ref{f:FD3}.
%
The amplitude for the process may be written as,
\begin{subequations}
	\begin{eqnarray}
	&&	i{\cal M}_{Z_L S_1^+ \to h S_1^+} = ig_{zs_1s_2} \left\{p_2 +(p_1 + p_2)\right\}_\mu \frac{i}{s - m_{C_2}^2} (i\lambda_{hs_1s_2}M_W)
	\epsilon_{L}^\mu(p_1) \,, \\
	\Rightarrow && {\cal M}_{Z_L S_1^+ \to h S_1^+} = -  g_{zs_1s_2}\lambda_{hs_1s_2} \frac{M_W}{M_Z} \frac{1}{s - m_{C_2}^2} \left(2 p_2 \cdot \epsilon(p_1) \right) \,.
	\label{e:MZShS}
	\end{eqnarray}
\end{subequations}
Now, following the kinematics in Fig.~\ref{f:FD3} we have the following relations,
\begin{subequations}
	\begin{eqnarray}
	&&	p_2 \cdot \epsilon (p_1)=  (E_+ p + E_z p) = p(E_+ + E_z) \equiv p \sqrt{s} \,, \\
	{\rm and,} && p^2 = E_z^2 - M_Z^2 = E_+^2 - m_{C_1}^2 \,, \\
	\Rightarrow &&  M_Z^2 - m_{C_1}^2 = E_z^2 - E_+^2 \,, \\
	\Rightarrow && (E_z - E_+) = \frac{M_Z^2 - m_{C_1}^2}{\sqrt{s}} \,.
	\end{eqnarray}
\end{subequations}
Alternatively, one can also write,
\begin{subequations}
	\begin{eqnarray}	
	2p^2 &=& E_z^2 + E_+^2 - (M_Z^2 + m_{C_1}^2) \,, \nonumber\\
	&=& \frac{1}{2}\left[(E_+ + E_z)^2 + (E_+ - E_z)^2\right] - (M_Z^2 + m_{C_1}^2) \,, \nonumber \\
	&=& \frac{1}{2}\left[ s + \frac{(M_Z^2 - m_{C_1}^2)^2}{s} - 2(M_Z^2 + m_{C_1}^2)\right] \,, \\
	\Rightarrow	p^2 &=& \frac{1}{4}\left[ s + \frac{(M_Z^2 - m_{C_1}^2)^2}{s} - 2(M_Z^2 + m_{C_1}^2)\right] \,, \\
	\Rightarrow p &=& 	\frac{\sqrt{s}}{2} \left[ 1 - 2\frac{(M_Z^2 + m_{C_1}^2)}{s} + \frac{(M_Z^2 - m_{C_1}^2)^2}{s^2} \right]^{\frac{1}{2}} \,, \\ 
	p &\approx& \frac{\sqrt{s}}{2} \left[ 1 - \frac{(M_Z^2 + m_{C_1}^2)}{s} + {\cal O}\left(\frac{M^4}{s^2}\right) \right]  \,.
	\end{eqnarray}
\end{subequations}
Thus, one may write
\begin{eqnarray}
p_2\cdot\epsilon(p_1) &=& \frac{s}{2} \left[ 1 - \frac{(M_Z^2 + m_{C_1}^2)}{s} + {\cal O}\left(\frac{M^4}{s^2}\right) \right] \,.
\end{eqnarray}
Using this in \eq{e:MZShS}, the final expression for the amplitude, in the high-energy limit, can be written as
\begin{eqnarray}
	{\cal M}_{Z_L S_1^+ \to h S_1^+} \approx - g_{zs_1s_2} \lambda_{hs_1s_2} \frac{M_W}{M_Z} + {\cal O}\left(\frac{M^2}{s}\right) \,.
\end{eqnarray}

%% file: appendix/param_c3hdm.tex
\chapter{Matrices for parameterizing the C3HDM}\label{app:parametrization}

The z-tensor of quartic parameters is
\footnote{In this representation of $z_{ijkl}$,
one of the pairs of indices, $ij$/$kl$, refers to the entries of
the larger $3\times 3$ matrix,
while the other refers to the smaller $3\times 3$ matrices;
the order is not relevant due to the symmetry $z_{ijkl}=z_{klij}$.}
\begin{equation}\label{e:ztensor}
	z = 
	\begin{pmatrix}
		\begin{pmatrix}
			\lambda_1 & 0 & 0 \\
			0 & \lambda_4/2 & 0 \\
			0 & 0 & \lambda_5/2 \\
		\end{pmatrix}
		&
		\begin{pmatrix}
			0 & \lambda_{10} & 0 \\
			\lambda_7/2 & 0 & 0 \\
			0 & 0 & 0 \\
		\end{pmatrix}
		&
		\begin{pmatrix}
			0 & 0 & \lambda_{11} \\
			0 & 0 & 0 \\
			\lambda_8/2 & 0 & 0 \\
		\end{pmatrix}
		\\
		\begin{pmatrix}
			0 & \lambda_7/2 & 0 \\
			\lambda_{10}^* & 0 & 0 \\
			0 & 0 & 0 \\
		\end{pmatrix}
		&
		\begin{pmatrix}
			\lambda_4/2 & 0 & 0 \\
			0 & \lambda_2 & 0 \\
			0 & 0 & \lambda_6/2 \\
		\end{pmatrix}
		&
		\begin{pmatrix}
			0 & 0 & 0 \\
			0 & 0 & \lambda_{12} \\
			0 & \lambda_9/2 & 0 \\
		\end{pmatrix}
		\\
		\begin{pmatrix}
			0 & 0 & \lambda_8/2 \\
			0 & 0 & 0 \\
			\lambda_{11}^* & 0 & 0 \\
		\end{pmatrix}
		&
		\begin{pmatrix}
			0 & 0 & 0 \\
			0 & 0 & \lambda_9/2 \\
			0 & \lambda_{12}^* & 0 \\
		\end{pmatrix}
		&
		\begin{pmatrix}
			\lambda_5/2 & 0 & 0 \\
			0 & \lambda_6/2 & 0 \\
			0 & 0 & \lambda_3 \\
		\end{pmatrix}
	\end{pmatrix}
	\;,
\end{equation}
giving us the matrices 
from which we can obtain the matrices $A$, $B$, $C$ for our $\Z2\times\Z2$ model
\begin{equation}\label{e:matA}
	A = \boldsymbol{\mu}  + 
	\left(\begin{matrix}\lambda_{1} v_{1}^{2} + \frac{\lambda_{4} v_{2}^{2}}{2} + \frac{\lambda_{5} v_{3}^{2}}{2} & \frac{\lambda_{7} v_{1} v_{2}}{2} + \lambda_{10} v_{1} v_{2} & \frac{\lambda_{8} v_{1} v_{3}}{2} + \lambda_{11} v_{1} v_{3}\\\frac{\lambda_{7} v_{1} v_{2}}{2} + v_{1} v_{2} \lambda_{10}^* & \lambda_{2} v_{2}^{2} + \frac{\lambda_{4} v_{1}^{2}}{2} + \frac{\lambda_{6} v_{3}^{2}}{2} & \frac{\lambda_{9} v_{2} v_{3}}{2} + \lambda_{12} v_{2} v_{3}\\\frac{\lambda_{8} v_{1} v_{3}}{2} + v_{1} v_{3} \lambda_{11}^* & \frac{\lambda_{9} v_{2} v_{3}}{2} + v_{2} v_{3} \lambda_{12}^* & \lambda_{3} v_{3}^{2} + \frac{\lambda_{5} v_{1}^{2}}{2} + \frac{\lambda_{6} v_{2}^{2}}{2}\end{matrix}\right) = M^2_{ch}
	\;,
\end{equation}
\begin{equation}\label{e:matB}
	B =
	\left(\begin{matrix}\lambda_{1} v_{1}^{2} + \frac{\lambda_{7} v_{2}^{2}}{2} + \frac{\lambda_{8} v_{3}^{2}}{2} & \frac{\lambda_{4} v_{1} v_{2}}{2} + \lambda_{10} v_{1} v_{2} & \frac{\lambda_{5} v_{1} v_{3}}{2} + \lambda_{11} v_{1} v_{3}\\\frac{\lambda_{4} v_{1} v_{2}}{2} + v_{1} v_{2} {\lambda_{10}^*} & \lambda_{2} v_{2}^{2} + \frac{\lambda_{7} v_{1}^{2}}{2} + \frac{\lambda_{9} v_{3}^{2}}{2} & \frac{\lambda_{6} v_{2} v_{3}}{2} + \lambda_{12} v_{2} v_{3}\\\frac{\lambda_{5} v_{1} v_{3}}{2} + v_{1} v_{3} {\lambda_{11}^*} & \frac{\lambda_{6} v_{2} v_{3}}{2} + v_{2} v_{3} {\lambda_{12}^*} & \lambda_{3} v_{3}^{2} + \frac{\lambda_{8} v_{1}^{2}}{2} + \frac{\lambda_{9} v_{2}^{2}}{2}\end{matrix}\right)
	\;,
\end{equation}
\begin{equation}\label{e:matC}
	C =
	\left(\begin{matrix}\lambda_{1} v_{1}^{2} + v_{2}^{2} {\lambda_{10}^*} + v_{3}^{2} {\lambda_{11}^*} & \frac{\lambda_{4} v_{1} v_{2}}{2} + \frac{\lambda_{7} v_{1} v_{2}}{2} & \frac{\lambda_{5} v_{1} v_{3}}{2} + \frac{\lambda_{8} v_{1} v_{3}}{2}\\\frac{\lambda_{4} v_{1} v_{2}}{2} + \frac{\lambda_{7} v_{1} v_{2}}{2} & \lambda_{2} v_{2}^{2} + \lambda_{10} v_{1}^{2} + v_{3}^{2} {\lambda_{12}^*} & \frac{\lambda_{6} v_{2} v_{3}}{2} + \frac{\lambda_{9} v_{2} v_{3}}{2}\\\frac{\lambda_{5} v_{1} v_{3}}{2} + \frac{\lambda_{8} v_{1} v_{3}}{2} & \frac{\lambda_{6} v_{2} v_{3}}{2} + \frac{\lambda_{9} v_{2} v_{3}}{2} & \lambda_{3} v_{3}^{2} + \lambda_{11} v_{1}^{2} + \lambda_{12} v_{2}^{2}\end{matrix}\right)
	\;,
\end{equation}
from which the mass matrices can be written.

%% file: appendix/sigma36_reps.tex
\chapter{\texorpdfstring{Representations of the group $\Sigma(36\varphi)$}{Representations of the group Sigma(36phi)}\label{appendix-group}}

Irreducible representations of the group $\Sigma(36\varphi)$ together with the character table 
were listed in~\cite{Grimus:2010ak,Hagedorn:2013nra}.
They include four 1D irreps $\bm{1}^{(p)}$, which differ by the representing value of $\rho_1(d) = i^p$,
four complex 3D irreps $\bm{3}^{(p)}$, whose $\rho_3(a)$ and $\rho_3(b)$ are given in \eq{Delta27-generators}
and $\rho_3(d)$ is the same as in \eq{Sigma36-generators} with an extra factor $i^{p}$,
four conjugate triplets $(\bm{3}^{(p)})^*$, and finally two real 4D irreps.
As in the main text, we shorten $\bm{3}^{(0)}$ to just $\bm{3}$.
The sum of the squares of the irrep dimensions is $\sum_i d_i^2 = 4\times 1 + 8\times 3^2 + 2\times 4^2 = 108$,
which is the order of $\Sigma(36\varphi)$.

The decomposition of various products of irreps were derived in~\cite{Grimus:2010ak}, Appendix C.
We are interested in products of $\bm{3}$, which is how the Higgs doublets transform,
with $\bm{3}^{(p)}$ or $(\bm{3}^{(p)})^*$.
First,
\begin{equation}
	\bm{3} \otimes \bm{3}^* = \bm{1}^{(0)} \oplus \bm{4} \oplus \bm{4}'\,.\label{CG-3030*}
\end{equation}
If $a \sim \bm{3}$ and $b^* \sim \bm{3}^*$, then the trivial singlet is obtained as
\begin{equation}
	[ab^*]_1 = a_1 b_1^* + a_2 b_2^* + a_3 b_3^*\,. 
\end{equation}
\eq{CG-3030*} can be generalized to other triplets-antitriplet products as
\begin{equation}
	\bm{3}^{(p)} \otimes (\bm{3}^{(p')})^* = \bm{1}^{(p-p')} \oplus \bm{4} \oplus \bm{4}'\,,\label{CG-3p3p*}
\end{equation}
where $p-p'$ is understood as taken mod $4$.

Next, the product of two triplets is decomposed as 
\begin{equation}
	\bm{3} \otimes \bm{3} = \bm{3}^* \oplus (\bm{3}^{(1)})^* \oplus (\bm{3}^{(3)})^*\,.\label{CG-3030}
\end{equation}
If $a \sim \bm{3}$ and $b \sim \bm{3}$, then the explicit expression for the antitriplet $\bm{3}^*$, which is antisymmetric under $a\leftrightarrow b$, 
is inherited from the $SU(3)$ irrep products:
\begin{equation}
	\frac{1}{\sqrt{2}}\epsilon_{ijk}a_jb_k = \frac{1}{\sqrt{2}}\triplet{a_2 b_3 - a_3 b_2}{a_3 b_1 - a_1 b_3}{a_1 b_2 - a_2 b_1} \sim \bm{3}^*\,.
	\label{CG3030-30*}
\end{equation}
The six-dimensional symmetric subspace splits into the two invariant subspaces: 
\begin{equation}
	\frac{1}{\sqrt{24}}\triplet{\sqrt{2}\,\tau_- a_1 b_1 - \tau_+(a_2 b_3 + a_3 b_2)}%
							   {\sqrt{2}\,\tau_- a_2 b_2 - \tau_+(a_3 b_1 + a_1 b_3)}%
							   {\sqrt{2}\,\tau_- a_3 b_3 - \tau_+(a_1 b_2 + a_2 b_1)} \sim (\bm{3}^{(1)})^*\,, \quad
	\frac{1}{\sqrt{24}}\triplet{\sqrt{2}\,\tau_+ a_1 b_1 + \tau_-(a_2 b_3 + a_3 b_2)}%
							   {\sqrt{2}\,\tau_+ a_2 b_2 + \tau_-(a_3 b_1 + a_1 b_3)}%
							   {\sqrt{2}\,\tau_+ a_3 b_3 + \tau_-(a_1 b_2 + a_2 b_1)} \sim (\bm{3}^{(3)})^*\,.
	\label{CG3030-3p*}
\end{equation}
Here, we used the shorthand notation
\begin{equation}
	\tau_- = \sqrt{2(3-\sqrt{3})}\,, \quad 	\tau_+ = \sqrt{2(3+\sqrt{3})}\,, \quad \tau_-\tau_+ = \sqrt{24}\,.
\end{equation}
Again, the decomposition \eq{CG-3030} can be generalized to the product of any two triplets:
\begin{equation}
	\bm{3}^{(p)} \otimes \bm{3}^{(p')} = (\bm{3}^{(-p-p')})^* \oplus (\bm{3}^{(1-p-p')})^* \oplus (\bm{3}^{(3-p-p')})^*\,,\label{CG-3p3p}
\end{equation}
where all indices are computed mod $4$. However, the decomposition rules remain the same as in \eqs{CG3030-30*}{CG3030-3p*}.


%% file: appendix/sigma36_yukawa.tex
\chapter{\texorpdfstring{Yukawa sectors for the exact $\Sigma(36)$ symmetry}{Yukawa sectors for the exact Sigma(36) symmetry}\label{appendix-Yukawa-exact}}

We present here the details of extending the symmetry group $\Sigma(36)$ to the quark Yukawa sector.
With the list of irreps of the group $\Sigma(36\varphi)$ and their product composition rules,
we present in Table~\ref{table-options} the main options 
for the irrep assignments in a $\Sigma(36\varphi)$-invariant quark sector.
In principle, one can also choose different triplets; for example, 
in the second line, $\overline{Q_L}$ could be chosen to transform as $\bm{3}^{(1)}$.
But with a simultaneous relabeling the irrep assignments for $d_R$ and $u_R$, 
this choice would still lead to the same Yukawa structures.
Let us now build the Yukawa matrices for each of the cases listed in Table~\ref{table-options}. 

\begin{table}[!h]
	\centering
	\begin{tabular}[t]{ccccc}
		\toprule
		label & $\overline{Q_L}$ & $\phi$ & $d_R$ & $u_R$\\
		\midrule
		case 1-3-3-3\quad & $\quad\bm{1}\quad$ & $\quad\bm{3}\quad$ & $\quad\bm{3}^*\quad$ & $\quad\bm{3}\quad$ \\
		case 3-3-1-3\quad & $\bm{3}^*$ & $\bm{3}$ & $\bm{1}$ & $\bm{3}^*$ or $(\bm{3}^{(1)})^*$ or $(\bm{3}^{(3)})^*$ \\
		case 3-3-3-1\quad & $\bm{3}$ & $\bm{3}$ & $\bm{3}$ or $\bm{3}^{(1)}$ or $\bm{3}^{(3)}$ & $\bm{1}$ \\
		\bottomrule
	\end{tabular}
	\caption{The main options for $\Sigma(36\varphi)$-invariant Yukawa sector.}
	\label{table-options}
\end{table}

{\bf Case 1-3-3-3}: $Q_L$ are trivial singlets, $d_R \sim \bm{3}^*$, $u_R \sim \bm{3}$, the first line of Table~\ref{table-options}.
The three Yukawa matrices $\Gamma_a$ were given in \eq{case1333-Gamma},
and similar matrices $\Delta_a$ with their own parameters $d_i$ arise for the up sector.
Starting from the mass matrices $(M_d)_{ij} = g_i v_j/\sqrt{2}$ and $(M_u)_{ij} = d_i v^*_j/\sqrt{2}$,
we define, as usual, their hermitean squares $H_d = M_d M_d^\dagger$ and $H_u = M_u M_u^\dagger$: 
\begin{equation}
	(H_d)_{ij} = \frac{v^2}{2}g_i g^*_j\,, \quad 
	(H_u)_{ij} = \frac{v^2}{2}d_i d^*_j\,. \label{case1333-HdHu}
\end{equation}
Both in the down and up-quark sectors, we have two generations of massless quarks 
and one generation of massive ones, with $m_b^2 = v^2 |\vec g|^2/2$ and $m_t^2 = v^2 |\vec d|^2/2$.

In general, the mass matrices are diagonalized through the rotations
\begin{equation}
	d_L^0 = V_{dL} d_L\,, \quad d_R^0 = V_{dR} d_R\,, \quad
	u_L^0 = V_{uL} u_L\,, \quad u_R^0 = V_{uR} u_R\,.\label{quark-rotations}
\end{equation}
which lead to the CKM matrix $\VCKM = V_{uL}^\dagger V_{dL}$ and 
the diagonal matrices $D_d = V_{dL}^\dagger M_d V_{dR}$ and $D_u = V_{uL}^\dagger M_u V_{uR}$. 
Let us assume that the non-zero masses correspond to the third eigenvector. 
Then the matrices $V_{dL}$ and $V_{uL}$ take the following generic form:
\begin{equation}
	V_{dL} = \frac{1}{\sqrt{|\vec g|^2}}\mtrx{\uparrow& \uparrow & \uparrow\\ \vec x_1&\vec x_2&\vec g\\ \downarrow& \downarrow & \downarrow}\,,\quad
	V_{uL} = \frac{1}{\sqrt{|\vec d|^2}}\mtrx{\uparrow& \uparrow & \uparrow\\ \vec y_1&\vec y_2&\vec d\\ \downarrow& \downarrow & \downarrow}\,.
	\label{case1333-VdLVuL}
\end{equation}
Here, for the down-quark sector, $\vec g = (g_1, g_2, g_3)$ and $\vec x_1$, $\vec x_2$ are the other two eigenvectors corresponding to zero eigenvalue, 
which cannot be uniquely defined. The same construction holds for the up-sector in terms of the vectors
$\vec d = (d_1, d_2, d_3)$ and $\vec y_1$, $\vec y_2$. 

Proceeding in a similar way with $G_d = M_d^\dagger M_d$ and $G_u = M_u^\dagger M_u$,
one finds $(G_d)_{ij} = |\vec g|^2 v_i^* v_j$ and $(G_u)_{ij} = |\vec d|^2 v_i v_j^*$.
One can then find the rotation matrices $V_{dR}$ and $V_{uR}$, inserting them back in the Yukawa lagrangian
and obtain the interaction patterns of the neutral scalar fields $\phi_a^0 = (h_a + i a_a)/\sqrt{2}$
with quark pairs:
\begin{equation}
	-{\cal L}_Y \supset \frac{\sqrt{2}m_b}{v^2} \bar{d}_{L} 
	\left( \begin{array}{ccc}0& 0 & 0\\ 0& 0 & 0 \\ 
		\vec{\phi^0}\cdot \vec u_1& \vec{\phi^0}\cdot \vec u_2 & \vec{\phi^0}\cdot \vec v\end{array}\right)  
	d_{R} + h.c.\,,\label{couplings-down}
\end{equation}
and a similar matrix for the up-quark sector.
Here, the two vectors $\vec u_1$ and $\vec u_2$, of norm $v$, are orthogonal to $\vec v$ and with each other.
Although the three vectors $\vec u_1$, $\vec u_2$, $\vec v$ are different for different vev alignments, 
they all share the key properties: the SM-like Higgs couples only to $\bar b b$, 
while all the additional neutral Higgses couple to $b\bar{d}$ and $b\bar{s}$.
In particular, for the vev alignment $B_1$ we recover the coupling patterns given in \eq{h-a-quark}.

{\bf Case 3-3-1-3}: $d_R$ are trivial singlets, $Q_L \sim \bm{3}$, $u_R \sim \bm{3}^*$.
The Yukawa matrices $\Gamma_a$ again come from the simple product of $\bm{3}^* \otimes \bm{3}$, 
which is then coupled to the three singlets $d_R$, each with its own coefficient:
\begin{equation}
	\Gamma_1 = \mtrx{g_1&g_2&g_3 \\ \cdot&\cdot&\cdot \\ \cdot&\cdot&\cdot}\,, \quad 
	\Gamma_2 = \mtrx{\cdot&\cdot&\cdot \\ g_1&g_2&g_3 \\ \cdot&\cdot&\cdot}\,, \quad 
	\Gamma_3 = \mtrx{\cdot&\cdot&\cdot \\ \cdot&\cdot&\cdot \\ g_1&g_2&g_3}\,,\label{case3313-Gamma}
\end{equation}
with the dots being null entries. We arrive at $(H_d)_{ij} = |\vec g|^2v_i v^*_j/2$. 
again yielding two massless generations and a massive $b$-quark, with the same mass squared as before: $m_b^2 = v^2 |\vec g|^2/2$.
However, the quark rotation matrix is now governed by the vector of vevs rather than the coefficients:
\begin{equation}
	V_{dL} = \frac{1}{v}\mtrx{\uparrow& \uparrow & \uparrow\\ \vec u_1&\vec u_2&\vec v\\ \downarrow& \downarrow & \downarrow}\,, \label{Vdl-3313}
\end{equation}
while $V_{dR}$ takes a form similar to \eq{case1333-VdLVuL}.
The interactions of the neutral Higgses with quark pairs is now described by the matrix 
\begin{eqnarray}
	-{\cal L}_Y \supset \frac{\sqrt{2}m_b}{v^2} \bar{d}_{L} \left( \begin{array}{ccc}0& 0 & \vec{\phi^0}\cdot \vec u_1\\ 0& 0 & \vec{\phi^0}\cdot \vec u_2 \\ 0& 0 & \vec{\phi^0}\cdot \vec v\end{array}\right)  d_{R} + h.c.\,.\label{couplings-down-3313}
\end{eqnarray}
We again observe that the decays of $h$ and $a$ to $\bar b d$ and $\bar b s$ are unavoidable.

In the up-quark sector, we encounter the product $\bm{3}^* \otimes \bm{3}^* \otimes \bm{3}^*$,
which contains only one singlet. As a result, the entire up-quark sector now involves only one free parameters $d$.
Using the representation product rules listed in Appendix~\ref{appendix-group}, 
we obtain the following Yukawa matrices:
\begin{equation}
	\Delta_1 = d\mtrx{\cdot&\cdot&\cdot \\ \cdot&\cdot& -1 \\ \cdot& 1 &\cdot}\,, \quad 
	\Delta_2 = d\mtrx{\cdot&\cdot& 1 \\ \cdot&\cdot&\cdot \\  -1 &\cdot&\cdot}\,, \quad 
	\Delta_3 = d\mtrx{\cdot& -1 &\cdot \\  1 &\cdot&\cdot \\ \cdot&\cdot&\cdot}\,.\label{case3313-Delta}
\end{equation}
The mass matrix and its hermitean square are
\begin{equation}
	(M_u)_{ij} = -\frac{d}{\sqrt{2}}\epsilon_{ijk} v^*_k\,, \quad 
	(H_{u})_{ij} = \frac{|d|^2}{2}\left(v^2 \delta_{ij} - v^*_i v_j\right)\,.\label{up-Hd-3313}
\end{equation}
The mass spectrum is now different: one zero eigenvalue, with the eigenvector $(v_1^*, v_2^*, v_3^*)$
and two non-zero degenerate eigenvalues $m^2 = |d|^2 v^2/2$.
The Higgs-quark couplings now show a different pattern but the overall conclusion remains unchanged: 
$h$ and $a$ unavoidably decay to quark pairs.

{\bf Case 3-3-3-1}, with $u_R$ being trivial singlets, while $Q_L \sim \bm{3}^*$, $d_R \sim \bm{3}$,
is similar to the previous one, with the structures in the up and down-quark sectors swapped.

Finally, {\bf case 3-3-1-3$^{(1)}$} and {\bf case 3-3-1-3$^{(3)}$}, 
with $Q_L \sim \bm{3}$ and $u_R \sim (\bm{3}^{(1)})^*$ or $u_R \sim (\bm{3}^{(3)})^*$,
involve new structures. 
The down sector remains the same as in case 3-3-1-3, while the up sector is now constructed 
from a different contraction of triplets:
\begin{equation}
	\Delta_1 = d\mtrx{\sqrt{2}\tau_\mp&\cdot&\cdot \\ \cdot&\cdot& \mp\tau_\pm \\ \cdot& \mp\tau_\pm &\cdot}\,, \quad 
	\Delta_2 = d\mtrx{\cdot&\cdot& \mp\tau_\pm \\ \cdot&\sqrt{2}\tau_\mp&\cdot \\  \mp\tau_\pm &\cdot&\cdot}\,, \quad 
	\Delta_3 = d\mtrx{\cdot& \mp\tau_\pm &\cdot \\  \mp\tau_\pm &\cdot&\cdot \\ \cdot&\cdot&\sqrt{2}\tau_\mp}\,,\label{case3313'-Delta}
\end{equation}
where $\tau_- = \sqrt{2(3-\sqrt{3})}$, $\tau_+ = \sqrt{2(3+\sqrt{3})}$, so that $\tau_-\tau_+ = \sqrt{24}$.
We now obtain a mass matrix with three non-zero masses, two of them being degenerate.
Although the analytic diagonalization for a generic vev alignment is cumbersome,
it is easily done for the alignments which are possible for the $\Sigma(36)$ potential.
The bottom line is the same: all new neutral Higgses couple to quark pairs, leading to decays of $h$ and $a$.

%% file: appendix/z2z2_minima.tex
\chapter{\texorpdfstring{Local minima in the $\Z2\times\Z2$ 3HDM and mass formulas}{Local minima in the Z2xZ2 3HDM and mass formulas}\label{app:masses}}

In this Appendix we present the formulas for the scalar masses under the
various minima of interest. Requiring that the mass squared are positive is akin
to guaranteeing that the corresponding extreme is indeed a local minimum.

\section{\texttt{2-Inert}}

This case can be found in Eq.~\eqref{m_Hpm_2_1}-\eqref{m_Hpm_2_6} of
Section~\ref{sec:scan}.

%

\section{\texttt{F0DM1}}

In this case we have $v_1=0$, $v_2\not=0, v_3=0$. The minimization gives
\begin{align}
  v_2^2=-\frac{m_{22}^2}{\lambda_2} \, ,
\end{align}
implying $m_{22}^2 <0$ as $\lambda_2>0$ from BFB. For the masses we
have
\begin{align}
  m_{H_{1}}^2=& m_{11}^2 + \frac{1}{2}\left(\lambda_4 + \lambda_7 + 2
    \lambda''_{10} \right) v_2^2 \equiv m_{11}^2 + \Lambda_1 v_2^2\, ,\\ 
  m_{A_{1}}^2=& m_{11}^2 + \frac{1}{2}\left(\lambda_4 + \lambda_7 - 2
    \lambda''_{10} \right)v_2^2 \equiv m_{11}^2 + \bar{\Lambda}_1 v_2^2\, ,\\ 
  m_{H^\pm_{1}}^2=& m_{11}^2 + \frac{1}{2} \lambda_4 v_2^2\, ,\\[+2mm]
  m_{H_{2}}^2=& 2 v_2^2 \lambda_2\, ,\\[+2mm]
  m_{H_{3}}^2=& m_{33}^2 + \frac{1}{2}\left(\lambda_6 + \lambda_9 + 2
    \lambda''_{12} \right) v_2^2 \equiv m_{33}^2 + \Lambda_2 v_2^2\, ,\\ 
  m_{A_{3}}^2=& m_{33}^2 + \frac{1}{2}\left(\lambda_6 + \lambda_9 - 2
    \lambda''_{12} \right)v_2^2 \equiv m_{33}^2 + \bar{\Lambda}_2 v_2^2\, ,\\ 
  m_{H^\pm_{3}}^2=& m_{33}^2 + \frac{1}{2} \lambda_6 v_2^2 \, .
\end{align}
We have to require all these masses squared to be positive in order to
have a local minimum. This is easier than finding conditions on the
parameters. The value of the potential at the minimum is
\begin{align}
  V_{\texttt{F0DM1}} = - \frac{m_{22}^4}{4\lambda_2}\, ,
\end{align}
and this has to be compared with $V_{\texttt{2Inert}}$.

\section{\texttt{F0DM2}}

In this case we have $v_1\not=0$, $v_2=0, v_3=0$. The minimization gives
\begin{align}
  v_1^2=-\frac{m_{11}^2}{\lambda_1}\, ,
\end{align}
implying $m_{11}^2 <0$ as $\lambda_1>0$ from BFB. For the masses we
have
\begin{align}
  m_{H_{1}}^2=& 2 v_1^2 \lambda_1\, ,\\[+2mm]
  m_{H_{2}}^2=& m_{22}^2 + \frac{1}{2}\left(\lambda_4 + \lambda_7 + 2
    \lambda''_{10} \right) v_1^2 \equiv m_{22}^2 + \Lambda_1 v_1^2\, ,\\ 
  m_{A_{2}}^2=& m_{22}^2 + \frac{1}{2}\left(\lambda_4 + \lambda_7 - 2
    \lambda''_{10} \right)v_1^2 \equiv m_{22}^2 + \bar{\Lambda}_1 v_1^2\, ,\\ 
  m_{H^\pm_{2}}^2=& m_{22}^2 + \frac{1}{2} \lambda_4 v_1^2\, ,\\[+2mm]
  m_{H_{3}}^2=& m_{33}^2 + \frac{1}{2}\left(\lambda_5 + \lambda_8 + 2
    \lambda''_{11} \right) v_1^2 \equiv m_{33}^2 + \Lambda_3 v_1^2\, ,\\ 
  m_{A_{3}}^2=& m_{33}^2 + \frac{1}{2}\left(\lambda_5 + \lambda_8 - 2
    \lambda''_{11} \right)v_1^2 \equiv m_{33}^2 + \bar{\Lambda}_3 v_1^2\, ,\\ 
  m_{H^\pm_{3}}^2=& m_{33}^2 + \frac{1}{2} \lambda_5 v_1^2\, .
\end{align}
We have to require all these masses squared to be positive in order to
have a local minimum. This is easier than finding conditions on the
parameters.
The value of the potential at the minimum is
\begin{align}
  V_{\texttt{F0DM2}} = - \frac{m_{11}^4}{4\lambda_1}.
\end{align}
and this has to be compared with $V_{\texttt{2Inert}}$.

\section{\texttt{F0DM0}}

In this case we have $v_1\not=0$, $v_2\not=0, v_3=0$. The minimization gives
\begin{align}
  v_1^2=\frac{\lambda_2 m_{11}^2-\Lambda_1
    m_{22}^2}{\Lambda_1^2-\lambda_1 \lambda_2},\
  v_2^2=\frac{\lambda_1 m_{22}^2-\Lambda_1
    m_{11}^2}{\Lambda_1^2-\lambda_1 \lambda_2}\, ,
\end{align}
requiring $v_{1}^2,v_2^2 >0$. For the masses we
have
\begin{align}
  m_{H_{1}}^2=&\lambda_1 v_1^2+\lambda_2 v_2^2
  -\sqrt{4 \Lambda_1^2 v_1^2 v_2^2 + (\lambda_1 v_1^2 - \lambda_2
    v_2^2)^2} \, ,\\[+2mm] 
  m_{H_{2}}^2=&\lambda_1 v_1^2+\lambda_2 v_2^2
  +\sqrt{4 \Lambda_1^2 v_1^2 v_2^2 + (\lambda_1 v_1^2 - \lambda_2 v_2^2)^2}\, ,\\ 
  m_{H_{3}}^2=& m_{33}^2 + \frac{1}{2}\left(\lambda_5 + \lambda_8 + 2
    \lambda''_{11} \right) v_1^2 + \frac{1}{2}\left(\lambda_6 + \lambda_9 + 2
    \lambda''_{12} \right)v_2^2 \, ,\\ 
  m_{A_{1}}^2=&-2 \lambda''_{10}\left( v_1^2 + v_2^2\right)\, ,\\ 
  m_{A_{2}}^2=& m_{33}^2 + \frac{1}{2}\left(\lambda_5 + \lambda_8 - 2
    \lambda''_{11} \right)v_1^2 + \frac{1}{2}\left(\lambda_6 + \lambda_9 - 2
    \lambda''_{12} \right)v_2^2  \, ,\\
  m_{H^\pm_{1}}^2=& - \frac{1}{2} \left(\lambda_7 +2
    \lambda''_{10}\right)(v_1^2+  v_2^2)\, ,\\[+2mm]
  m_{H^\pm_{2}}^2=& m_{33}^2 + \frac{1}{2}
  \left(\lambda_5 v_1^2+\lambda_6 v_2^2\right)\, .
\end{align}
We have to require all these masses squared to be positive in order to
have a local minimum. This is easier than finding conditions on the
parameters.
The value of the potential at the minimum is
\begin{align}
  V_{\texttt{F0DM0}} = \frac{\lambda_1 m_{22}^4+\lambda_2 m_{11}^4-2
    m_{11}^2m_{22}^2 \Lambda_1}{4(\Lambda_1^2-\lambda_1\lambda_2)}\, ,
\end{align}
and this has to be compared with $V_{\texttt{2Inert}}$.

\section{\texttt{F0DM0'}}

Let us take
\begin{equation}
  \label{eq:f0dm0_rel}
  \sqrt{r_1}=\frac{v_1}{\sqrt{2}},
  \sqrt{r_2}=\frac{v_2}{\sqrt{2}},r_3=0,\quad
  \alpha_1=\alpha_2=\beta_1=\gamma=0,\quad
  \beta_2=\frac{\pi}{2}.
\end{equation}
This is still along the neutral directions, but in comparison with
\texttt{F0DM0} corresponds to making\footnote{Notice
that this case is not contradiction with Eq.(3.32) of
Ref.~\cite{Hernandez-Sanchez:2020aop}.
Although it looks like a particular case of
\texttt{sCPv}, we checked explicitly by calculating the
full $6 \times 6$ mass matrix for the neutral scalars that indeed that matrix
separates into CP even and CP odd blocks, ensuring CP conservation.} $v_2\to i\, v_2$.
The stationary conditions give,
\begin{align}
  v_1^2=\frac{\lambda_2 m_{11}^2-\bar{\Lambda}_1
    m_{22}^2}{\bar{\Lambda}_1^2-\lambda_1 \lambda_2},\
  v_2^2=\frac{\lambda_1 m_{22}^2-\bar{\Lambda}_1
    m_{11}^2}{\bar{\Lambda}_1^2-\lambda_1 \lambda_2},\
\end{align}
and
\begin{equation}
  \label{eq:f0dm0_pot}
  V_{\texttt{F0DM0}'} = \frac{\lambda_1 m_{22}^4+\lambda_2 m_{11}^4-2
    m_{11}^2m_{22}^2 \bar{\Lambda}_1}{4(\bar{\Lambda}_1^2-\lambda_1\lambda_2)}\, ,
\end{equation}
where
\begin{equation}
  \label{eq:f0dm0_l1}
  \bar{\Lambda}_1= \frac{1}{2}\left( \lambda_4+\lambda_7 -2
    \lambda_{10}'' \right)\, ,
\end{equation}
was defined before. We require that $v_1^2,v_2^2$ are positive and
also check for the positiveness of the masses. This is not absolutely
necessary, because if $ V_{\texttt{F0DM0}'} <  V_{\texttt{2Inert}}$,
even if it is a saddle-point, it indicates that there should be a
minimum below and the point is not a good point. But our statement is
stronger if we also identify the local minima.
It should be stressed that such a minimum should
exist, as our sufficient BFB conditions were checked for all cases in
our numerical simulation.
For the masses we
have
\begin{align}
  m_{H_{1}}^2=&
2 \lambda_{1} v_{1}^2
\, ,\\[+2mm] 
m_{H_{2}}^2=&
2 \lambda''_{10} v_{1}^2
\, ,\\ 
m_{H_{3}}^2=&
\frac{1}{2} \left(\lambda_{5} v_{1}^2+\lambda_{6}
   v_{2}^2+\lambda_{8} v_{1}^2+\lambda_{9} v_{2}^2+2
   \lambda''_{11} v_{1}^2-2 \lambda''_{12} v_{2}^2+2
   m^2_{33}\right)
\, ,\\ 
m_{A_{1}}^2=&
2 \lambda_{2} v_{2}^2
\, ,\\ 
m_{A_{2}}^2=&
2 \lambda''_{10} v_{2}^2
\, ,\\
m_{A_{3}}^2=&
\frac{1}{2} \left(\lambda_{5} v_{1}^2+\lambda_{6}
   v_{2}^2+\lambda_{8} v_{1}^2+\lambda_{9} v_{2}^2-2
   \lambda''_{11} v_{1}^2+2 \lambda''_{12} v_{2}^2+2
   m^2_{33}\right)
\, ,\\
m_{H^\pm_{1}}^2=&
-\frac{1}{2} (\lambda_{7}-2 \lambda''_{10})
   \left(v_{1}^2+v_{2}^2\right)
\, ,\\[+2mm]
m_{H^\pm_{2}}^2=&
\frac{1}{2} \left(\lambda_{5} v_{1}^2+\lambda_{6} v_{2}^2+2
   m^2_{33}\right)
\, .
\end{align}

\section{\texttt{F0CB}}

Looking at the results for the new minima found by our numerical simulation,
we realized that there is another particular case.
It is somewhat hidden because, for $v_3=0$, we
are using too many angles. We identify a new situation, that we call
\texttt{F0CB}, and that can be defined by
\begin{equation}
 \label{eq:f0cb_rel}
  \sqrt{r_1}=\frac{v_1}{\sqrt{2}},
  \sqrt{r_2}=\frac{v_2}{\sqrt{2}},r_3=0,\quad
  \beta_1=\beta_2=\gamma=0,\quad
  \alpha_1\not=0,\alpha_2\not=0 .
\end{equation}
This is, in principle, a charge breaking minimum. We get the
stationary equations,
\begin{align}
 0=& \frac{1}{4} \left[4 m^2_{11} + 4 \lambda_{1} v_1^2 + 2 \lambda_{4} v_2^2
 + \lambda_{7} v_2^2 +  
 2 \lambda''_{10} v_2^2 + (\lambda_{7} + 2 \lambda''_{10}) v_2^2
 \cos(2 (\alpha_1 - \alpha_2))\right]\,,\\ 
 0=&
 \frac{1}{4} \left[4 m^2_{22} + 2 \lambda_{4} v_1^2 + \lambda_{7} v_1^2 + 2 \lambda''_{10} v_1^2 + 
 4 \lambda_{2} v_2^2 + (\lambda_{7} + 2 \lambda''_{10}) v_1^2 \cos(2
 (\alpha_1 - \alpha_2))\right]\,,\\ 
 0=& -\frac{1}{4} (\lambda_{7} + 2 \lambda''_{10}) v_1^2 v_2^2
 \sin(2 (\alpha_1 - \alpha_2))\,,\\ 
 0=&\frac{1}{4} (\lambda_{7} + 2 \lambda''_{10}) v_1^2 v_2^2 \sin(2
 (\alpha_1 - \alpha_2)) \,.
\end{align}
These equations have many solutions.
As we showed analytically,
they are equivalent, giving the same value at the
minimum.
As an example, we take
\begin{align}
  \alpha_1=& \frac{\pi}{2}, \quad
  &&\alpha_2=0\,,\\
  v_1^2=&-\frac{2 (2 \lambda_2 m^2_{11} - \lambda_4 m^2_{22})}{4 \lambda_1
    \lambda_2 - \lambda_4^2},\quad 
  &&v_2^2=-\frac{2 (-\lambda_4 m^2_{11} + 2 \lambda_1 m^2_{22})}{4 \lambda_1
    \lambda_2 - \lambda_4^2} \,,
\end{align}
and
\begin{equation}
  \label{eq:9}
  V_{\texttt{F0CB}}=-\frac{\lambda_{1} m^4_{22}+\lambda_{2}
      m^4_{11}-\lambda_{4} m^2_{11} 
   m^2_{22}}{4 \lambda_{1} \lambda_{2}-\lambda^2_{4}}
 \end{equation}

To have a local minimum, we take $v_1^2,v_2^2 >0$, as well as the
squared masses to be positive,
\begin{align} 
  m_{H_{1}}^2=&
\frac{4 \lambda_2 (\lambda_4 m^2_{11}-2 \lambda_1
   m^2_{22})}{4 \lambda_1 \lambda_2-\lambda_4^2}
\, ,\\[+2mm] 
m_{H_{2}}^2=&
\frac{(\lambda_7+2 \lambda''_{10}) (\lambda_4 m^2_{11}-2
   \lambda_1 m^2_{22})}{4 \lambda_1 \lambda_2-\lambda_4^2}
\, ,\\ 
m_{H_{3}}^2=&
\frac{1}{4 \lambda_1
  \lambda_2-\lambda_4^2}\left[
4 \lambda_1 \lambda_2 m^2_{33}-2 \lambda_1 \lambda_6
   m^2_{22}-2 \lambda_1 \lambda_9 m^2_{22}-4 \lambda_1
   \lambda''_{12} m^2_{22}-2 \lambda_2 \lambda_5
   m^2_{11}\right.\nonumber\\
 &\left. \hskip 20mm
   +\lambda_4^2 (-m^2_{33})
   +\lambda_4
   \lambda_5 m^2_{22}+\lambda_4 \lambda_6
   m^2_{11}+\lambda_4 \lambda_9 m^2_{11}+2 \lambda_4
   \lambda''_{12} m^2_{11} \right]
\, ,\\ 
m_{A_{1}}^2=&
\frac{(\lambda_7-2 \lambda''_{10}) (\lambda_4 m^2_{11}-2
   \lambda_1 m^2_{22})}{4 \lambda_1 \lambda_2-\lambda_4^2}
\, ,\\ 
m_{A_{2}}^2=&
\frac{1}{4 \lambda_1
  \lambda_2-\lambda_4^2}\left[
4 \lambda_1 \lambda_2 m^2_{33}-2 \lambda_1 \lambda_6
   m^2_{22}-2 \lambda_1 \lambda_9 m^2_{22}+4 \lambda_1
   \lambda''_{12} m^2_{22}-2 \lambda_2 \lambda_5
   m^2_{11}
\right.\nonumber\\
 &\left. \hskip 20mm
   +\lambda_4^2 (-m^2_{33})+\lambda_4
   \lambda_5 m^2_{22}+\lambda_4 \lambda_6
   m^2_{11}+\lambda_4 \lambda_9 m^2_{11}-2 \lambda_4
   \lambda''_{12} m^2_{11}\right]
\, ,\\
m_{H^\pm_{1}}^2=&
-\frac{2 \lambda_1 (2 \lambda_2 m^2_{11}-\lambda_4
   m^2_{22})}{4 \lambda_1 \lambda_2-\lambda_4^2}
\, ,\\[+2mm]
m_{H^\pm_{2}}^2=&
\frac{\lambda_4 \lambda_7 m^2_{22}-2 \lambda_2 \lambda_7
   m^2_{11}}{4 \lambda_1 \lambda_2-\lambda_4^2}
\, ,\\[+2mm]
m_{H^\pm_{3}}^2=&
\frac{1}{4
  \lambda_1 \lambda_2-\lambda_4^2}\left[
4 \lambda_1 \lambda_2 m^2_{33}-2 \lambda_1 \lambda_6
   m^2_{22}-2 \lambda_2 \lambda_5 m^2_{11}-2 \lambda_2
   \lambda_8 m^2_{11}
   \right.\nonumber\\
 &\left. \hskip 20mm
   +\lambda_4^2
   (-m^2_{33})+\lambda_4 \lambda_5 m^2_{22}+\lambda_4
   \lambda_6 m^2_{11}+\lambda_4 \lambda_8 m^2_{22}\right]
\, .
\end{align}